\documentclass{article}

\usepackage[english]{babel}
\usepackage{placeins}
\usepackage[letterpaper,top=2cm,bottom=2cm,left=3cm,right=3cm,marginparwidth=1.75cm]{geometry}

\usepackage{amsmath}
\usepackage{graphicx}
\usepackage{ dsfont }
\usepackage{amssymb}
\usepackage{chngcntr}
\usepackage{diagbox}
\usepackage{multirow}
\usepackage{xcolor}

\newcommand{\jakvf}{\textit{JAK2$^{V617F}$}}
\usepackage{bm}
\newcommand{\vect}[1]{\bm{#1}}
\newcommand{\matr}[1]{\bm{#1}}

\newcommand{\rev}[1]{\textcolor{black}{#1}}

\usepackage{caption}
\usepackage{subcaption}
\graphicspath{{fig/}{fig_identif/}{fig/}{fig_identif/patients/}{fig_res_annexe/}{fig_res_annexe/dyn/}{fig_sel/}{fig_synthetic_patients/}{fig_resAIC_synth/}{fig_inference_synth/}{fig_inference_synth/dose/}{fig_inference_synth/Delta/}{fig_inference_synth/parameters/}{fig_inference_synth/R/}}

\title{Mathematical modelling, selection, and hierarchical inference to determine the minimal dose in IFN$\alpha$ therapy against 
Myeloproliferative Neoplasms}
\usepackage{authblk}

\author[1]{\small Gurvan Hermange}
\author[2, 3, 4]{\small William Vainchenker}
\author[2, 3, 4]{\small  Isabelle Plo}
\author[1*]{\small Paul-Henry Courn\`ede}

\affil[1]{\footnotesize Universit{\'e} Paris-Saclay, CentraleSup{\'e}lec, Laboratory of Mathematics and Informatics (MICS), Gif-sur-Yvette, France.}
\affil[2]{\footnotesize  INSERM U1287 (INSERM, Gustave Roussy, Universit{\'e} Paris-Saclay), Villejuif, France}
\affil[3]{\footnotesize Gustave Roussy, Villejuif, France}
\affil[4]{\footnotesize Universit{\'e} Paris-Saclay, Villejuif, France}
\affil[*]{\footnotesize corresponding author: paul-henry.cournede@centralesupelec.fr}

\date{}
\begin{document}
\maketitle

\section*{Abstract}
Myeloproliferative Neoplasms (MPN) are blood cancers that appear after acquiring a driver mutation in a hematopoietic stem cell. These hematological malignancies result in the overproduction of mature blood cells and, if not treated, induce a risk of cardiovascular events and thrombosis. Pegylated IFN$\alpha$ is commonly used to treat MPN, but no clear guidelines exist concerning the dose prescribed to patients. We applied a model selection procedure and ran a hierarchical Bayesian inference method to decipher how dose variations impact the response to the therapy. We inferred that IFN$\alpha$ acts on mutated stem cells by inducing their differentiation into progenitor cells; the higher the dose, the higher the effect. We found that the treatment can induce long-term remission when a sufficient (patient-dependent) dose is reached. We determined this minimal dose for individuals in a cohort of patients and estimated the most suitable starting dose to give to a new patient to increase the chances of being cured. 

\section{Introduction}
Classical Myeloproliferative Neoplasms (MPN) are malignant hematological pathologies resulting in the overproduction of mature blood cells and the deregulation of hematopoiesis. 
MPN are classified into three diseases based on the type of mature cells overproduced~\cite{tefferi2008classification}: Essential Thrombocythemia (ET) characterized by an overproduction of platelets, Polycythemia Vera (PV) characterized by an overproduction of red blood cells, and Primary Myelofibrosis (PMF) characterized by deregulation in the production of megakaryocytes and granulocytes, along with debilitating collagen fibres in the bone marrow. MPN is a rare disease, with an estimated global incidence of 1-4 cases per 100,000 per year, which increases with age~\cite{mehta2014epidemiology,titmarsh2014common, maynadie2011twenty}.
A study conducted across Europe estimates a prevalence (per 100,000 individuals) ranging from 4.96 to 30.0 for PV, 4.0 to 24.0 for ET, and 0.51 to 2.7 for PMF~\cite{moulard2013epidemiology}.
The MPN disease is often detected belatedly - at median ages of around 65 years for ET and PV, and 70 years for PMF~\cite{rohrbacher2009clinical} - after complications such as thrombosis or cardiovascular events and can progress to acute leukemia. These blood cancers occur following the acquisition of a specific somatic mutation in a hematopoietic stem cell (HSC). The most frequent driver mutation of the MPN disease affects the JAK2 protein (mutation \jakvf) that plays a crucial role in cell signaling~\cite{vainchenker2017genetic}. Following homologous recombination, homozygous malignant subclones can develop in parallel to heterozygous subclones.
The apparition of homozygous subclones is often linked to the progression of the disease from ET to PV~\cite{scott2006progenitors} and is associated with a more severe disease phenotype. This underscores the importance of having mathematical models that account for homozygous mutated cells.
Although the disease prevalence is low, the \jakvf mutation is found in the general population at a higher frequency, estimated in a Danish cohort at 3.1\%~\cite{cordua2019prevalence}.
Advances in understanding this hematological malignancy are crucial to developing treatments that will lead to the cure of the disease. Interferon alpha (IFN$\alpha$), a natural inflammatory cytokine that has long been used to treat many diseases, has shown promising results in MPN. 
Indeed, pegylated IFN$\alpha$ induces a hematological response, i.e., a normalization in blood cell counts, and a molecular response, i.e., a reduction of the number of mutated cells~\cite{yacoub2019pegylated,Gisslinger2020}.
The reduction in the number of mutated hematopoietic cells could be explained by the effect of IFN$\alpha$ on HSCs, promoting both their exit from quiescence and their differentiation, as has been hypothesized from mouse model studies~\cite{hasan2013jak2v617f,mullally2013depletion}.
\\
Understanding and quantifying the impact of this treatment on (mutated) hematopoietic cells is 
essential for developing personalized medicine. With their potential power to predict cell population dynamics,
mathematical models are promising tools to guide clinical decisions~\cite{hoffmann2020integration}.
Several mathematical models have been proposed to model the effect of a given treatment against hematological malignancies~\cite{roeder2006dynamic, michor2005dynamics, bunimovich2019optimization, ottesen2020mathematical, haeno2009progenitor, zhang2017determining, ottesen2019bridging, andersen2020global}, but few of them study the impact of dose variations during long-term therapies, while in clinical routine, physicians rarely prescribe a constant posology over several years. Instead, they often continuously increase the dose until reaching a maximum tolerated dose (or reaching a sufficient hematological response) and then proceed to a dose de-escalation. Such dosage strategies were observed by Mosca et al.~\cite{mosca2021}. 
To study the effect of IFN$\alpha$ on hematopoietic stem cells, Mosca et al. proposed a hierarchical model they calibrated based on data from a cohort of MPN patients followed over several years.
Better molecular responses were obtained in \jakvf MPN patients treated on average with higher IFN$\alpha$ doses.  However, the model did not account for dose variations over time and, consequently, the minimal IFN$\alpha$ dose to administer to the patients to improve their chance of getting long-term molecular remission could not be determined. Pedersen et al.~\cite{pedersen2021dose} proposed a dose-dependent mathematical modelling of the action of IFN$\alpha$ on MPN, but they did not account for differences between heterozygous and homozygous mutated cells. \\
Here, we aim to accurately model the changes of posology in IFN$\alpha$ therapy and derive appropriate dose-response relationships.
We start from the model by Mosca et al.~\cite{mosca2021} which does not account for dose variations. We refer to it as the baseline model and explore various ways to make it more complex by considering different dose-response relationships, leading to a set of different models that we will confront to the data published by Mosca et al. to select the best one.
We propose a two-step selection procedure to determine the best model, starting with a very large set of potential models.
Considering the high number of models that we choose to compare (225), it is computationally too expensive to estimate the parameters of these models in a hierarchical framework as done in~\cite{mosca2021}. Therefore, our two-step procedure is designed to first eliminate most models with a coarser but computationally feasible strategy and then conduct a more refined statistical analysis for the few selected models. First, the estimation considers all individuals independently, and models are compared based on the Akaike Information Criterion (AIC). In a second step, we perform a hierarchical Bayesian estimation on a subselection of models to find the best one according to the Deviance Information Criterion (DIC), as defined by Spiegelhalter et al.~\cite{spiegelhalter2002bayesian}. Finally, we analyze the selected model to characterize better how the dose impacts the response to the IFN$\alpha$ therapy for MPN patients and to determine for each of them which would be the minimal dose under which a remission might not be reached. 
In addition, taking advantage of the hierarchical framework that allows us to estimate the population effect, we can determine the most suitable starting IFN$\alpha$ dose for a new MPN patient. Altogether, this work aims to improve clinicians' decision-making regarding the doses of IFN$\alpha$ to be prescribed, be it for the initial dose or the lower limit when de-escalating the dose.

\section{Methods}

\subsection{Modelling the dose-response to IFN$\alpha$}

\begin{figure}
    \centering
    \includegraphics[width=0.9\textwidth]{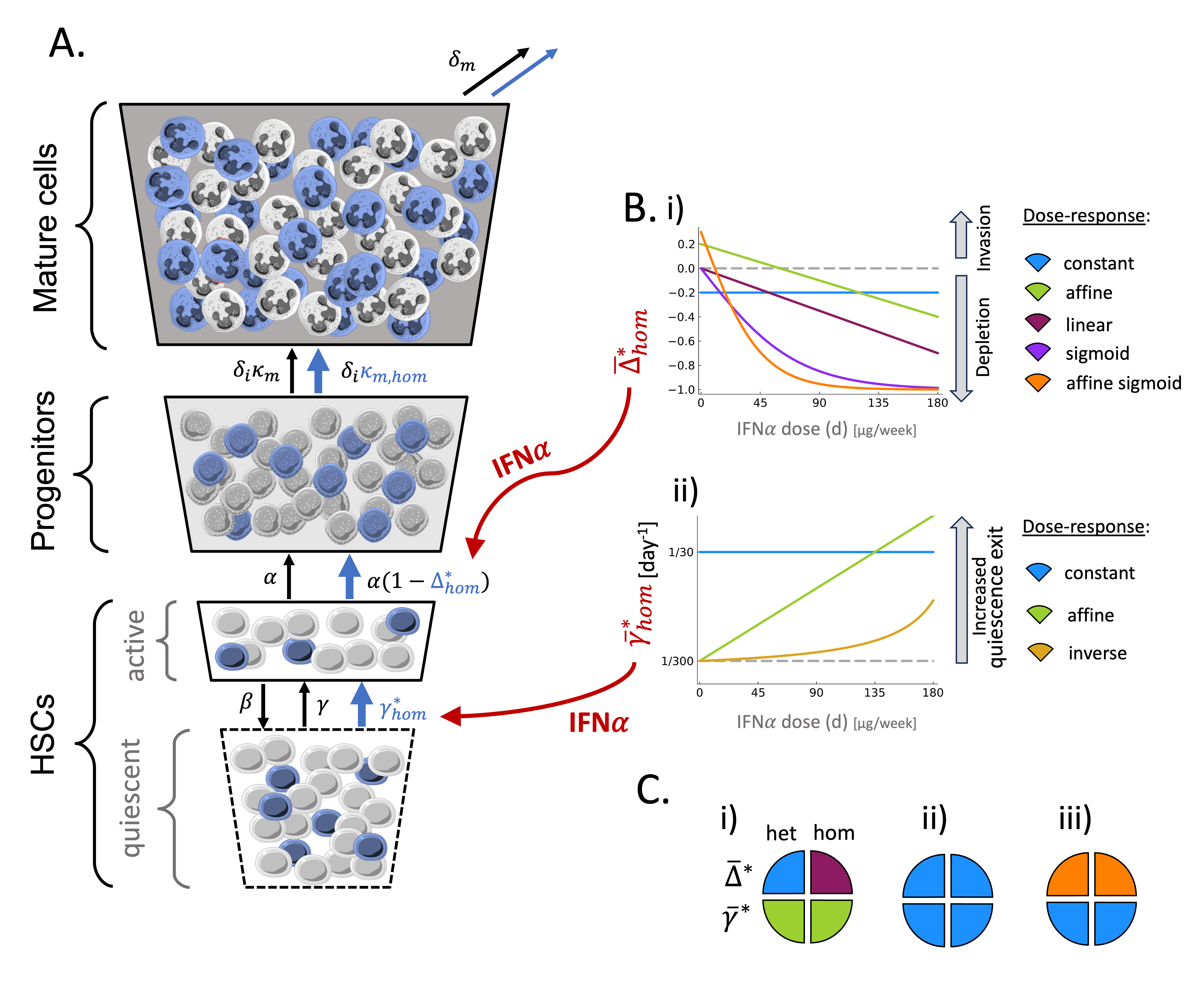}
    \caption{A. In the baseline model, our cell populations are described by a compartmental model consisting of four compartments: quiescent or active HSC that give rise to progenitors and, ultimately, to mature cells. Within each compartment, we can find WT cells (in black) and mutated \jakvf cells, either homozygous (in blue) or heterozygous (not represented for clarity), either before ($t<0$) or under IFN$\alpha$ treatment ($t \geq 0$). WT cells are considered to be in a steady state, while IFN$\alpha$ will target mutated HSCs by potentially favouring their differentiation into progenitors (impact on $\Delta^*$) and/or increasing their quiescent exit (impact on $\gamma^*$). \\
    B. The impact of the treatment on parameters $\Delta^*$ (i) and $\gamma^*$ (ii) might depend on the weekly IFN$\alpha$ dose $d$ administered to the patient. Several dose-response relationships are considered (involving parameters that would have to be estimated). The horizontal dashed line represents the baseline level associated with WT HSCs. \\
    C. i) According to how we combine the different dose-response relationships for $\bar{\Delta}^*_{het}$, $\bar{\Delta}^*_{hom}$, $\bar{\gamma}^*_{hom}$, and $\bar{\gamma}^*_{het}$, we can end up with 225 different configurations, that is, different models from which we want to select the best. Each configuration can be schematized by the circle where each quadrant colour is associated with a given dose-response relationship (in agreement with the colour legend in B). Configuration (ii)  corresponds to the baseline model, where each dose-response relationship is constant. Configuration (iii) corresponds to the model where the effect of IFN$\alpha$ on the quiescent exit of both heterozygous and homozygous mutated HSCs does not depend on the dose $d$ and where IFN$\alpha$ impacts the differentiation of the heterozygous and homozygous active HSCs according to the affine sigmoid relation.
}
    \label{fig:schema_model}
\end{figure}

\subsubsection{Baseline model}
\label{sec:based_model}

To understand how variations of pegylated IFN$\alpha$ (Pegasys) doses precisely impact mutated HSCs for MPN patients, we extend the model of Mosca et al.~\cite{mosca2021} that described the dynamics of mutated hematopoietic cells under treatment, yet without considering a differential impact of the posology over the therapy.\\
Briefly, this baseline model considers quiescent HSCs (compartment 1) that can become active at a rate $\gamma$, active HSCs (compartment 2) that can return to quiescence at a rate $\beta$ or be recruited to divide at a rate $\alpha$. In this latter case, according to the type of division the HSC will encounter, the cell generates 0, 1, or 2 progenitor cells (compartment $i$) (Fig.~\ref{fig:schema_model}.A). Parameter $\Delta$ models the balance between differentiated and symmetrical division; that is, $\Delta$ is equal to the probability that an HSC generates two HSCs minus the probability that it generates two progenitor cells. $\Delta \in [-1,1]$, and under homeostatic conditions we have $\Delta = 0$.
Progenitor cells exit their compartment at rate $\delta_i$ to become, after expansion at a rate $\kappa_m$, mature cells (compartment $m$). Mature cells are fully differentiated cells that die at a rate $\delta_m$. In this work, only granulocytes are considered, i.e., we study granulopoiesis. Yet, the model itself remains general and valid for other types of mature hematopoietic cells (depending mainly on the value we set for $\delta_m$). Differentiated cells, like platelets, erythrocytes, or lymphocytes, are also produced during hematopoiesis. Some mathematical models have focused on describing the production of a given mature cell type (megakaryopoiesis~\cite{boullu2019model}, erythropoiesis~\cite{crauste2008adding}, lymphopoiesis~\cite{chulian2021dynamical}) when others have modeled multiple cell lineages~\cite{colijn2005mathematical,xu2019statistical}.
In this study, we focus on granulocytes since the \jakvf Variant Allele Frequency (VAF) used for model estimations was measured in these mature cells. Note that the VAF in erythroblasts, megakaryocytes and granulocytes are generally very similar~\cite{anand2011effects}, such that the measurement of the VAF in granulocytes could be viewed as a surrogate measurement of the VAF in mature cells. \\
In our case, the hematopoietic dynamics is described by the following system of (linear) ordinary differential equations (ODEs):
\begin{equation}
\label{eq:syst_comp}
\left\{
\begin{array}{ll}
  \frac{dN_1(t)}{dt} &= -\gamma N_1(t) + \beta N_2(t)  \\
  \frac{dN_2(t)}{dt} & = \gamma N_1(t) + (\alpha \Delta - \beta) N_2(t) \\
   \frac{dN_i(t)}{dt} &= \alpha ( 1-\Delta) \kappa_i N_2(t) - \delta_i N_i(t) \\
   \frac{dN_{m}(t)}{dt} &= \delta_i \kappa_m N_i(t) - \delta_{m} N_{m}(t) \\		
\end{array}
\right.
\end{equation}
where $N_1(t)$, $N_2(t)$, $N_i(t)$, and $N_m(t)$ describe the numbers of inactive HSCs, active HSCs, progenitors, and mature cells respectively. To avoid a numerical resolution of these equations, which would result in an additional computational cost when dealing with the parameter estimation procedure, we derive an analytical solution for this ODE system (see Appendix A.1).\\
Actually, several genotypes are considered according to whether the cells are wild-type (subscript wt) or have the \jakvf mutation in one (heterozygous, subscript het) or two (homozygous, subscript hom) alleles. A system of ODEs is derived for each genotype, as presented previously. In terms of notation, corresponding subscripts refer to wt, het, or hom quantities, and superscript * indicates parameters impacted by IFN$\alpha$. 
Table~\ref{tab:tab_assumptions_params} synthesises those different conditions and the model parameters.
To note that, if we denote by $N_{HSC, het}(t) = N_{1,het}(t) + N_{2, het}(t)$ the number of mutated heterozygous HSCs, we have $\frac{dN_{HSC, het}(t)}{dt}  = \alpha_{het}^* \Delta_{het}^* N_{2, het}(t)$. Then, $\Delta_{het}^* < 0$ implies that there will be, in the long-term, an exhaustion of the pool of mutated heterozygous HSCs, that is, a molecular remission. The same goes for homozygous mutated cells.\\
More details of the baseline model, its simplifications, and its parameters to estimate are presented in Appendix A. After such simplifications, Mosca et al. ended up with 7 parameters to estimate.\\ Furthermore, we also verify the practical identifiability of the model. As highlighted by Duchesne et al.~\cite{duchesne2019calibration}, testing the identifiability cannot be avoided if we aim to further use the model for predicting purposes.
In Appendix C, we use a classical approach based on model identification from synthetic data to verify identifiability. 

\begin{table}[t]
\centering
\begin{tabular}{|c|c c|c c|}
\hline
\multirow{2}{*}{} & \multicolumn{2}{c|}{Before treatment and initial conditions} & \multicolumn{2}{c|}{Under IFN$\alpha$ (t $\geq$ 0)} \\
\cline{2-5}
& Parameter & Value & Parameter & Value \\
\hline
\multirow{10}{*}{WT} & $\alpha$ & $1/30$ & $\alpha^*$ & $= \alpha$ \\
& $\Delta$ & $0$ & $\Delta^*$ & $= \Delta$ \\
& $\gamma$ & $1/300$ & $\gamma^*$ & $= \gamma$ \\
& $\beta$ & $= \gamma(1-\chi)/\chi$ & $\beta^*$ & $= \beta$ \\
& $\kappa_i$ & NR & $k_i^* = \kappa_i^*/\kappa_i$ & NR \\
& $\delta_i$ & $1/6$ & $\delta_i^*$ & $= \delta_i$ \\
& $\kappa_m$ & NR & $k_m^* = \kappa_m^*/\kappa_m$ & NR \\
& $\delta_m$ & $1$ & $\delta_m^*$ & $= \delta_m$ \\
& $\chi$ & $0.1$ &  &  \\
& $N_{HSC}$ & NR &  &  \\\hline
\multirow{10}{*}{het} &$\alpha_{het}$ & $= \alpha$ & $\alpha_{het}^*$ & $= \alpha$ \\
& $\Delta_{het}$ & $0$ & $\Delta_{het}^*$ & To estimate  \\
& $\gamma_{het}$ & NR & $\gamma_{het}^*$ & To estimate \\
& $\beta_{het}$ & NR & $\beta_{het}^*$ & $=\beta$ \\
& $k_{i,het} = \kappa_{i,het}/ \kappa_i$ & 1 & $k_{i,het}^*= \kappa_{i,het}^* / \kappa_{i,het}$ & $=k_i^*$ \\
& $\delta_{i,het}$ & $= \delta_{i}$ & $\delta_{i,het}^*$ & $= \delta_{i}$ \\
& $k_{m,het}= \kappa_{m,het} / \kappa_m$ & To estimate & $k_{m,het}^* = \kappa_{m,het}^* / \kappa_{m,het}$ & $=k_m^*$ \\
& $\delta_{m,het}$ & $= \delta_{m}$ & $\delta_{m,het}^*$ & $= \delta_{m}$ \\
& $\chi_{het}$ & $= \chi$ &  &  \\
& $\eta_{het}$ & To estimate &   &   \\
\hline
\multirow{10}{*}{hom} &$\alpha_{hom}$ & $= \alpha$ & $\alpha_{hom}^*$ & $= \alpha$ \\
& $\Delta_{hom}$ & $0$ & $\Delta_{hom}^*$ & To estimate  \\
& $\gamma_{hom}$ & NR & $\gamma_{hom}^*$ & To estimate \\
& $\beta_{hom}$ & NR & $\beta_{hom}^*$ & $=\beta$ \\
& $k_{i,hom} = \kappa_{i,hom}/ \kappa_i$ & 1 & $k_{i,hom}^*= \kappa_{i,hom}^* / \kappa_{i,hom}$ & $=k_i^*$ \\
& $\delta_{i,hom}$ & $= \delta_{i}$ & $\delta_{i,hom}^*$ & $= \delta_{i}$ \\
& $k_{m,hom}= \kappa_{m,hom} / \kappa_m$ & $=k_{m,het}$ & $k_{m,hom}^* = \kappa_{m,hom}^* / \kappa_{m,hom}$ & $=k_m^*$ \\
& $\delta_{m,hom}$ & $= \delta_{m}$ & $\delta_{m,hom}^*$ & $= \delta_{m}$ \\
& $\chi_{hom}$ & $= \chi$ &  &  \\
& $\eta_{hom}$ & To estimate &   &   \\
\hline
\end{tabular}
\caption{Summary table of the parameters used in the baseline model (from Mosca et al.~\cite{mosca2021}). NR is for 'Not Relevant', i.e., after normalization, these quantities simplify and no longer intervene in the model output. $N_{HSC}$ corresponds to the number of WT HSCs. We define $\eta_{het} = \frac{N_{1,het}(0) + N_{2,het}(0)}{N_{HSC} }$ and $\chi_{het} = \frac{N_{2,het}(0)}{N_{1,het}(0) + N_{2,het}(0) }$ to express the initial conditions for heterozygous cells (the same goes for homozygous cells). Appendix A.2 provides additional details about the initial conditions and any simplifications.}
\label{tab:tab_assumptions_params}
\end{table}

\FloatBarrier
\subsubsection{Dose-response relationships}
\label{sec:dose-effect}

Mosca et al.~\cite{mosca2021} found that IFN$\alpha$ might increase the exit of quiescence of homozygous and heterozygous HSCs and increase their propensity to differentiate into progenitors. In other terms, a potential dose effect was identified on parameters $\Delta^*_{het}$, $\Delta^*_{hom}$, $\gamma^*_{het}$ and $\gamma^*_{hom}$. In their baseline model, following the idea from Michor et al.~\cite{michor2005dynamics}, it was only considered that the treatment acted by modifying parameters values from the start of the therapy, without considering further variations of posology, which corresponds to a constant dose-response relationship.
However, patients under therapy generally undergo many variations of dosage, and sometimes even temporary treatment interruptions (as we can see, for example, in Fig.~\ref{fig:ex_dyn}). To describe the effect of IFN$\alpha$ on MPN patients more precisely, we should incorporate these variations of posology as model inputs and derive appropriate dose-response relationships. For that purpose, we introduce the variable $d(t) \in [0,1]$, which describes the weekly IFN$\alpha$ dose administered at time $t$ ($t=0$ corresponds to the start of the therapy) normalized by the maximal observed dose equal to 180~$\mu$g/week. This latter is the maximum dose administered to some of the patients studied in this article (see Supplemental Tab.~B.1), but it is also a consensus of the clinical community; we are unaware of any higher doses ever being administered.
$\bar{\Delta}^*_{het}$, $\bar{\Delta}^*_{hom}$, $\bar{\gamma}^*_{het}$ and $\bar{\gamma}^*_{hom}$ are now functions of the variable $d$ -  a model input - and no longer parameters. We make the distinction between functions and parameters by using the bar symbol $\bar{\circ}$. 
Note that $d: t\longmapsto d(t)$ is a piece-wise constant function. It implies that we can still obtain an analytical solution of the ODE system~\eqref{eq:syst_comp}. To get smoother inferred dynamics at the changes of dosage, we could have used a pharmacokinetics equation as done by Ottesen et al.~\cite{ottesen2020mathematical} or Pedersen et al.~\cite{pedersen2021dose} when modelling the uptake of IFN$\alpha$. However, this model, while being more complex, did not improve the results in our case, as we show in Appendix H.\\

We adopt a model-based approach to study the dose response to IFN$\alpha$ and consider several potential relationships (Fig.~\ref{fig:schema_model}.B).\\
For $\bar{\Delta}^*_{het}$ (and equivalently for homozygous cells) that models the propensity of mutated heterozygous HSCs to differentiate into progenitors, in addition to the constant relationship $\bar{\Delta}^*_{het}: d \longmapsto \Delta_{het}^*$ used in the baseline model, we consider:
\begin{itemize}
    \item a linear relation:
\begin{equation}
    \label{eq:Delta_linear}
    \bar{\Delta}^*_{het}: d \longmapsto \Delta_{het}^* \cdot d,
\end{equation}
\item an affine relation:
\begin{equation}
    \label{eq:Delta_affine}
    \bar{\Delta}^*_{het}: d \longmapsto \Delta^*_{het} \cdot d + \Delta_{het},
\end{equation}
\item a sigmoid relation:
\begin{equation}
\label{eq:Delta_sigmo}
\bar{\Delta}_{het}^{*}: d \longmapsto \frac{-2}{1+e^{-\Delta^*_{het} \cdot d}}+1,
\end{equation}
\item and an affine sigmoid relation:
\begin{equation}
\label{eq:Delta_sigmo_affine}
\bar{\Delta}_{het}^{*}: d \longmapsto-2\left(\frac{1}{1+e^{-\Delta^*_{het} \cdot d}}-0.5\right) \cdot\left(1+\Delta_{het}\right)+\Delta_{het}.
\end{equation}
\end{itemize}
In eq.~\eqref{eq:Delta_affine} and~\eqref{eq:Delta_sigmo_affine}, we get $\bar{\Delta}_{het}^{*}(d=0) = \Delta_{het}$; that is, we study the possibility that mutated \jakvf HSCs naturally invade the stem cell pool (assuming $\Delta_{het}>0$) in the absence of treatment, as evidenced in~\cite{hermange2021, van2021reconstructing}. In eq.~\eqref{eq:Delta_linear} and~\eqref{eq:Delta_sigmo} on the opposite, we explicitly force $\Delta_{het}$ to be equal to zero, as done in~\cite{mosca2021}, with thus one less degree of freedom. We must have $\forall d \in [0,1], \bar{\Delta}_{het}^{*}(d) \in [-1,1]$. This condition is ensured by a proper choice of prior distributions (more precisely, by the choice of the lower and upper bound for the support of the prior distributions) in the case of the constant, linear, and affine relations and is automatically verified in both sigmoid relations.\\

For $\bar{\gamma}^*_{het}$ (and equivalently for  $\bar{\gamma}^*_{hom}$) that models the quiescence exit of mutated HSCs, in addition to the constant relationship $\bar{\gamma}^*_{het}: d \longmapsto \gamma_{het}^*$  used in the baseline model, we consider an affine relation:
\begin{equation}
    \label{eq:gamma_linear}
    \bar{\gamma}^*_{het}: d \longmapsto \gamma_{het}^* \cdot d + \gamma_{het},
\end{equation}
and an inverse relation:
\begin{equation}
    \label{eq:gamma_inverse}
    \bar{\gamma}^*_{het}: d \longmapsto \frac{1}{\tau_{het}^* \cdot d + 1/\gamma_{het}},
\end{equation}
which corresponds to an affine relation for the inverse of $ \bar{\gamma}^*_{het}$. Actually, there is no particular reason for preferring the parameter $\gamma$, which corresponds to a quiescence exit rate, to its inverse $\tau=1/\gamma$, which would correspond to an average time of quiescence exit. \\
Following Mosca et al.~\cite{mosca2021}, we consider that $\bar{\gamma}_{het}^{*}(d=0) = \gamma_{het} = 1/300$ [days$^{-1}$], and thus, we do not study linear relations as we do for $\bar{\Delta}^*_{het}$. Here again, proper prior distributions are chosen to ensure that $d \longmapsto \bar{\gamma}_{het}^{*}(d)$ is a decreasing and positive function.\\

Many other dose-response relationships could have been studied, as done, for example, in~\cite{bretz2005combining}, but since we combine the dose-response relationships for four different parameters in our dynamic model, it would result in a very large set of models to calibrate. Thus, we have chosen to restrict ourselves to some standard relations.

\subsection{Two-step model selection procedure}

\subsubsection{Data}
The data used in this study come from the previous study by Mosca et al.~\cite{mosca2021}.
We denote the data by $\mathcal{D}=\{\mathcal{D}_i\}_{i \in \{1,\cdots, N\}}$ with $N=19$ the number of \jakvf MPN patients.
Among them, 5 were diagnosed with an ET, 13 with a PV, and 1 with a PMF.
For a patient $i$, the data consist of observations at different times $t_k^{(i)}$, from the beginning of the therapy ($t=0$) and up to 5 years of treatment, of the VAF $\hat{y}_k^{(i)}$ among granulocytes (which correspond to compartment $m$ in the dynamic model)
and of the Clonal Fraction (CF) $\hat{z}_{het,k}^{(i)}$ (respectively, $\hat{z}_{hom, k}^{(i)}$) of mutated heterozygous (respectively, homozygous) progenitor cells (corresponding to compartment $i$).
More precisely, the measured VAF $\hat{y}_k^{(i)}$ should be compared to the theoretical VAF $y(t_k^{(i)})$ obtained from the model at time $t_k^{(i)}$:
\begin{equation*}
y(t_k^{(i)}) = \frac{N_{m,hom}(t_k^{(i)})+0.5N_{m,het}(t_k^{(i)})}{N_{m,hom}(t_k^{(i)})+N_{m,het}(t_k^{(i)})+N_{m,wt}(t_k^{(i)})}
\end{equation*}
and so for the heterozygous (or homozygous) CF:
\begin{equation*}
z_{het}(t_k^{(i)}) = \frac{N_{i,het}(t_k^{(i)})}{N_{i,hom}(t_k^{(i)})+N_{i,het}(t_k^{(i)})+N_{i,wt}(t_k^{(i)})}
\end{equation*}
We provide additional information regarding the observation model in Appendix B. \\
Observations are measured on average every 156 days ([72, 368] for the 5 and 95 percentiles, see Supplemental Fig. B.1), with a tendency to have a higher frequency of data acquisition at the beginning of the follow-up. 
For a given patient, we can expect to have observed it 12 times (at least 7 times for patient \#22, at most 15 for patients \#25 and 29).
The information per patient is provided in the Supplemental Tab. B.1. 
\begin{figure}
    \centering
    \includegraphics[width=0.8\textwidth]{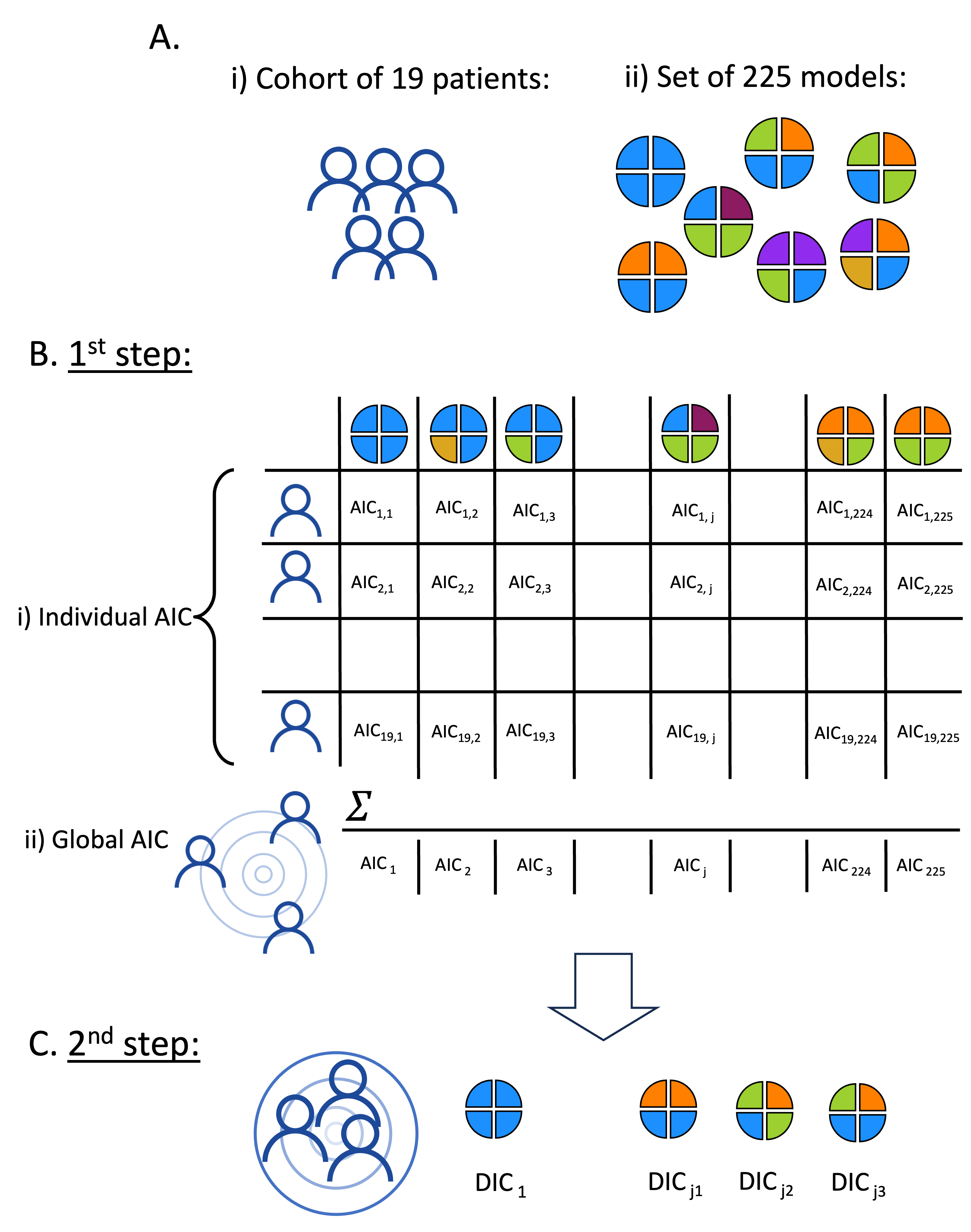}
    \caption{Schema of the two-step model selection procedure.
    A. The method is applied considering i) a cohort of 19 \jakvf MPN patients from Mosca et al.~\cite{mosca2021} and ii) a set of 225 models (see Fig.~\ref{fig:schema_model}). 
    B. First, we confront each potential model (index $j$) to the experimental observations of each patient (index $i$) to compute individual AIC$_{i,j}$ values (i). Considering the patients are independent of each other, for each model $j$, we sum up the contribution of all patients to get a global $AIC_j$ value. Then, we select the models leading to the best global AIC values. C. For each of them and the baseline model, we then apply a model selection based on a hierarchical Bayesian inference method. The patients are not considered independent anymore; their parameters are now assumed to be found in a common population distribution. The DIC of each preselected model is computed to select the one leading to the best (the lowest) value.
    }
    \label{fig:schema_selection}
\end{figure}

\subsubsection{Selection based on AIC}

Given the dose-response relationships in §~\ref{sec:dose-effect}, we end up with a large set of different mathematical models to compare. Indeed, we study 5 relations for $\bar{\Delta}_{het}^{*}$ and $\bar{\Delta}_{hom}^{*}$, and 3 for $ \bar{\gamma}^*_{het}$ and $\bar{\gamma}^*_{hom}$. Since it would be possible to have different dose-response relationships according to the genotype (het or hom), we end up with $5^2 \times 3^2 = 225$ models to compare. For $j \in \{1, \cdots, 225\}$, model $j$ ($\mathcal{M}_j$) corresponds to the dynamic model (described by three independent ODE systems~\eqref{eq:syst_comp}, one for each genotype) with a particular combination of dose-response relationships for the four previous parameters (Fig.~\ref{fig:schema_model}.C and~\ref{fig:schema_selection}.A).\\
In the first step, given the large number of models that we want to compare, we first use a coarse but quick method based on the Akaike Information Criterion (AIC)~\cite{akaike1974new, burnham2002practical}, to compare the different models and select the most adequate (Fig.~\ref{fig:schema_selection}.B). 
\\
For each patient $i$ and model $j$, we compute an AIC (Fig.~\ref{fig:AIC_zoom}):
\begin{equation}
    \label{eq:AIC_loc}
    AIC_{i,j} = -2\log( \mathcal{L}_{i,j}) + 2k_j
\end{equation}
with $\mathcal{L}_{i,j}$ the maximum of likelihood and $k_j$ the number of parameters to estimate with model $j$. 
Note that we could also have used the Bayesian Information Criterion~\cite{gideon1978estimating}, which would have led, in our case, to the same conclusions.
The statistical model used for expressing the likelihood is the same as in the baseline model (see Appendix C).
The maximum likelihood is computed using the CMA-ES (Covariance Matrix Adaptation - Evolution Strategy) algorithm~\cite{hansen2016cma}. The CMA-ES algorithm is a stochastic method for optimization that gives good results in a wide range of problems, including problems that are non-linear, non-separable, and in high dimension~\cite{hansen2006cma}. Briefly, this algorithm searches for the maximum of a function (here, the likelihood) over generations. At each generation, a sample of $\lambda$ individuals (i.e., parameter vectors) is generated according to a multidimensional normal distribution whose mean and variance-covariance matrix is computed from the selected individuals of the previous generation. Among these $\lambda$ offsprings, $\mu$ of them are selected (those which give
the highest likelihood values) and used for the next generation. It continues until convergence.\\
Finally, to compare how the different models perform on the whole cohort, we compute a global AIC (Suppl. Fig.~D.2) that sums the contribution of all patients:
\begin{equation}
    \label{eq:AIC_global}
    AIC_{j} = \sum_{1 \leq i \leq N} AIC_{i,j}
\end{equation}
It corresponds to the likelihood of a model considering all the patients altogether but as independent observations and with independent individual parameters. Based on this global criterion, we can sort the different models and select the ones that give the best results (i.e., the smaller values for the global AIC).

\subsubsection{Model selection based on a hierarchical Bayesian estimation}

One drawback with the statistical model upon which criterion~\eqref{eq:AIC_global} is built is that all patients are considered independently. No population effect is considered, with a risk of overfitting \cite{llamosi2016population}. 
Therefore, in the second step, we apply a hierarchical Bayesian estimation method to infer the distributions of the parameters for each patient as well as population parameters (referred to as hyper-parameters) (Fig.~\ref{fig:schema_selection}.C). Besides, compared to standard Bayesian methods, hierarchical models tend to improve the robustness of the estimations by reducing variance between individuals~\cite{gelman2004bayesian,llamosi2016population}.\\
Briefly, if we consider a population $\mathcal{P}=\{1, \cdots, N\}$ of $N$ patients, whose hematopoietic dynamics are described according to model $\mathcal{M}_j$, $\vect{\theta}=\left\{\vect{\theta}^{(i)}\right\}_{i \in \mathcal{P}}$ denotes the set of all patient parameters with:
\begin{equation*}
\begin{array}{c}
\vect{\theta}^{(1)}=\left(\theta_{1}^{(1)}, \cdots, \theta_{P}^{(1)}\right) \\
\vdots \\
\vect{\theta}^{(N)}=\left(\theta_{1}^{(N)}, \cdots, \theta_{P}^{(N)}\right)
\end{array}
\end{equation*}
where $P$ is the number of parameters to estimate for model $\mathcal{M}_j$. 
With the hierarchical inference method, instead of estimating each $\vect{\theta}^{(i)}$ independently, we assume that all individual parameter vectors are realizations of the same random variable of unknown distribution in a statistical model. Thus, the hierarchical model (also known as a random-effect model) can account for inter-individual variability but also for similarity between patients. In practice, we consider here:
\begin{equation}
\label{eq:prior_HP}
\forall i \in \mathcal{P}, \forall k \in\{1, \cdots, P\}, \theta_{k}^{(i)} \mid \tau_{k}, \sigma_{k}^{2} \sim \mathcal{N}_{c, k}\left(\tau_{k}, \sigma_{k}^{2}\right)
\end{equation}
where the population distribution for each component is a truncated Gaussian distribution $\mathcal{N}_{c, k}$ (over a range that depends on the considered parameter $k$), and $\vect{\tau}=(\tau_1,\cdots,\tau_P)$ and $\vect{\sigma^2}=(\sigma^2_1,\cdots,\sigma^2_P)$ are the hyper-parameters.

We can thus estimate the joint posterior distributions of $\vect{\theta}$ and hyper-parameters $\vect{\tau}$ and $\vect{\sigma^2}$ (as done for the baseline model in~\cite{mosca2021} and detailed in Appendix C.2):
\begin{align}
\mathbb{P}\left[\vect{\theta}, \vect{\tau}, \vect {\sigma^{2}} \mid \mathcal{D}\right] & \propto \mathbb{P}\left[\mathcal{D} \mid \vect{\theta}^{(1)}, \cdots, \vect {\theta}^{(N)}, \vect{\tau}, \vect{\sigma^{2}}\right]\mathbb{P}\left[\vect{\theta}^{(1)}, \cdots, \vect {\theta}^{(N)}, \vect{\tau}, \vect{\sigma^{2}} \right] \nonumber  \\
& \propto \prod_{i \in \mathcal{P}}\left(\mathbb{P}\left[\mathcal{D}_{i} \mid \vect{\theta}^{(i)}\right]\mathbb{P}\left[\vect{\theta}^{(i)} \mid \vect{\tau}, \vect{\sigma^{2}}\right]\right)\mathbb{P}[\vect{\tau}]\mathbb{P}\left[\vect{\sigma^{2}}\right]
\end{align}
Then, we sample from the posterior distribution using a Markov Chain Monte Carlo (MCMC) method, namely the Metropolis-Hastings within Gibbs algorithm~\cite{hastings1970monte, geman1984stochastic}. Conditionally on the hyper-parameters, patients are independent and their parameters can be sampled using a standard Metropolis-Hasting scheme.
When increasing the parameter space dimension, the Metropolis-Hastings algorithm can become inefficient: too many iterations might be required to reach convergence. Indeed, the algorithm relies on the choice of a proposal, usually a multivariate distribution whose covariance matrix $\matr{\Sigma_i}$ (related to patient $i$) has to be finely tuned before running the algorithm to get proper acceptance rates. Adaptive algorithms~\cite{andrieu2008tutorial} can be used to circumvent this difficulty. Here, we adopt another approach and choose for $\matr{\Sigma_i}$ the covariance matrix (up to a multiplication factor) learned from the CMA-ES algorithm. The rationale behind this approach is that learning the covariance matrix in CMA-ES is analogous to learning the inverse Hessian matrix in a quasi-Newton method~\cite{hansen2016cma}. Concerning the hyper-parameters, they are sampled using the Gibbs method, which consists in sampling from the marginal conditional posterior distribution of the hyper-parameters.
Details of the calculations are presented in Appendix C.2.
Model calibration was achieved by implementing the previous methods in the Julia programming language.  
The framework used for parameter estimation (and that can be used for a wide range of problems) is available at: 
\begin{verbatim}
https://gitlab-research.centralesupelec.fr/2012hermangeg/bayesian-inference
\end{verbatim}

Because of the high computational cost of the hierarchical inference method, we only use it for the comparison of a limited number of models: the baseline model and those that we first selected based on the AIC.\\

After running the parameter estimation procedure until convergence (assessed using standard diagnostic tools such as trace plots, evaluation of the autocorrelation, computation of the ergodic means and variances), we compute the Deviance Information Criterion (DIC) of each model to select the best one~\cite{spiegelhalter2002bayesian} (Tab.~\ref{tab:best_config}). For model $\mathcal{M}_j$, DIC$_j$ is defined by:
\begin{equation}
    \label{eq:DIC}
    DIC_j = D\left( \mathbb{E}[\vect{\theta} | \mathcal{D}, \mathcal{M}_j] \right) +2 p_{D_j}
\end{equation}
With the deviance defined by $D(\vect{\theta}) = -2 \log\left(\mathbb{P}[\mathcal{D} | \vect{\theta}, \mathcal{M}_j] \right)$ and $p_{D_j}$ the effective number of parameters defined, following Gelman et al.~\cite{gelman2004bayesian}, by $p_{D_j} = 0.5 \mathbb{V}[D(\vect{\theta})]$.
Finally, we select the model with the lowest DIC value.

\subsection{Inferring the impact of IFN$\alpha$ and determining its minimal dose in MPN}

Once we have run the model selection procedure and selected the best model (that is, the most appropriate dose-response relationships for $\bar{\Delta}_{het}^{*}$, $\bar{\Delta}_{hom}^{*}$, $ \bar{\gamma}^*_{het}$ and $\bar{\gamma}^*_{hom}$), we can analyze the results of this procedure in more details (section~\ref{sec:res_model_selection}). We can study if heterozygous and homozygous malignant clones respond differently to variations of IFN$\alpha$ doses. Indeed, our method allows the comparison of models with dose-response relationships that might be different depending on the genotype, following Tong et al.~\cite{tong2021hematopoietic} who suggested that heterozygous HSCs might respond to IFN$\alpha$ differently from homozygous cells. Besides, for the best model (that we will denote by $\mathcal{M}$), we can compare the posterior distributions of hyper-parameters $\vect{\tau}=\left(\tau_{1}, \cdots, \tau_{P}\right)$ and study how they differ between heterozygous or homozygous cells (Fig.~\ref{fig:posterior_HP_genotype}).\\

Moreover, from the results of model $\mathcal{M}$ calibration, we can study how the patients individually respond to the treatment (section~\ref{sec:ind_responses}). We can first compare the posterior distributions of their individual parameters (Fig.~\ref{fig:param_Delta}).
Then, by sampling from these distributions using a Monte-Carlo method, we can propagate the uncertainty from the input parameters to the output of the models (that is, the dynamics of the CF in each hematopoietic compartment), and display the inferred dynamics and a 95\% credibility interval for each patient (Fig.~\ref{fig:ex_dyn}).\\

More interestingly, we can study for each patient how $\bar{\Delta}_{het}^*$ and $\bar{\Delta}_{hom}^*$ depend on the IFN$\alpha$ dose $d$ (section~\ref{sec:ind_min_dose}). $\bar{\Delta}_{het}^*$ and $\bar{\Delta}_{hom}^*$ are actually key quantities of our model. Indeed, a molecular remission can only be achieved if negative values are reached for both quantities, meaning, according to our model, that \jakvf HSCs will encounter more differentiated divisions than self-renewal ones, leading to a depletion of the mutant stem cell pool.
Since IFN$\alpha$ might potentially induce some side effects, notably major depression~\cite{ lotrich2007depression, trask2000psychiatric}, it would be key progress in therapy to know the minimal dose that is necessary and sufficient to eliminate the malignant clones, i.e., such that $\bar{\Delta}_{het}^*$ and $\bar{\Delta}_{hom}^*$ are negative. A too-high dose might increase the toxicity of the therapy, while a too-low dose might not be able to induce a (major) molecular response in the patient.
Thus, we introduce $d_{min}$ as the minimal value of dose $d$, for a given patient, above which remission is possible:

\begin{equation}
\label{eq:def_dmin}
d_{min} = \arg\min_{d} \left( \bar{\Delta}_{het}^*(d) \leq 0
\ ; \
\bar{\Delta}_{hom}^*(d) \leq 0 \right)
\end{equation}

In addition, taking into account the uncertainty about $d_{min} $, we can introduce the probability that a dose $d$ might induce long-term remission: 
\begin{equation}
\label{eq:P_rem}
    P_{rem}(d) = \mathbb{P}[ d \geq d_{min} \mid \mathcal{M}, \mathcal{D} ]  \end{equation}
    
According to the selected dose-response relations, we can express $d_{min}$ as a function of the model parameters (Appendix G) and, therefore, infer its posterior distribution by sampling from the parameter posterior distributions.

We can then estimate for each patient of the cohort the lower limit of IFN$\alpha$ dose to be given by computing the posterior mean of $d_{min}$ as well as 95\% credibility intervals (Fig.~\ref{fig:minimal_dose_true_cohort}). Since these patients are still under treatment, this information might be helpful for clinicians. \\

One advantage of the hierarchical Bayesian framework is that we not only estimate individual parameters but also infer a population effect. Assuming that the patients in our cohort are representative of the population of MPN patients, especially regarding the range of IFN$\alpha$ doses that were given, we can consider that the population effect we inferred (through the estimation of the posterior distribution of the hyper-parameters) can be generalized for other MPN patients outside our cohort.
More precisely, we assumed that the individual model parameters $\theta_k^{(i)}$ (parameter $k$, patient $i\in \{1, \cdots, N \}$) followed a (truncated) Gaussian distribution~\eqref{eq:prior_HP} of mean and variance $\tau_k$ and $\sigma^2_k$ respectively: $\theta_k^{(i)}\sim \mathcal{N}_{c, k}\left(\tau_{k}, \sigma_{k}^{2}\right)$, where $\tau_k$ and $\sigma^2_k$ were hyper-parameters to estimate and for which we had only vague priors. After the estimation from data $\mathcal{D}$, we get the posterior distribution of $\tau_k$ and $\sigma^2_k$ to update our knowledge at the population level. For a new patient $N+1$, we can now consider as new prior:
\begin{equation}
\label{eq:prior_new_patient}
   \theta_k^{(N+1)}\sim \mathcal{N}_{c, k}\left(\mathbb{E}[\tau_{k} |\mathcal{D}] , \mathbb{E}[\sigma_{k}^{2}|\mathcal{D}]\right) 
\end{equation}
By sampling from the previous distribution, we can infer the general (that is, at the population level) dose-response relationships for $\bar{\Delta}_{het}^{*}$ and $\bar{\Delta}_{hom}^{*}$
(section~\ref{sec:starting_dose}, Fig.~\ref{fig:pop_Delta}) and estimate the minimal dose $d_{min}$ that should be given to a new patient having either heterozygous, homozygous, or both mutated HSCs (and prior to any other relevant medical information or clinical observations) to maximize his chances of getting a long-term molecular remission (Fig.~\ref{fig:pop_prob_rem}). 
We can also evaluate, for increasing posologies, the proportion of patients that might be ultimately cured and confront our results to standard dose escalation strategies starting from 45 $\mu$g/week up to 135 $\mu$g/week~\cite{knudsen2021genomic, yacoub2019pegylated, Gisslinger2020, barbui2021ropeginterferon}.

\subsection{Validation based on a synthetic study}

In this section, we test our methodology on synthetic data to study to which extent we can draw robust conclusions from our two-step model selection procedure and the further post-selection inferences - especially regarding determining a minimal dose. Indeed, as highlighted by Berk et al.~\cite{berk2013valid}, making inferences from a model we selected from data rather than from prior knowledge might lead to some biases.  We use our synthetic study to verify that this is not the case.   \\
First, we generate synthetic data similar to the study by Mosca et al. We randomly choose 20 models among the 225 ones that we consider and, for each of them, simulate a synthetic cohort consisting of noisy observations for 19 virtual patients. Each virtual patient will be associated with a real one; that is, the $i^{th}$ virtual patient from each cohort will have the exact observation times and same dose variations as the $i^{th}$ patient from the cohort of Mosca et al. In a virtual cohort, the patient’s virtual dynamics are simulated by randomly sampling the model parameter values from some population distributions chosen beforehand. In total, we simulate noisy observations for 380 virtual patients. More details about the simulation of the synthetic data are provided in Appendix F.1. \\
Then, we study each synthetic cohort separately, with the objectives to 1) retrieve the model used to simulate the data and 2) infer the impact of IFN$\alpha$, that is, accurately estimate the model parameters, the dose-response relationships, and the minimal IFN$\alpha$ dose.\\

We begin with the first step consisting of computing a global AIC value for each virtual cohort and keeping only the models with the best ones (Appendix F.2). At this stage, we find that the true model is among the top three models (over 225) in 65\% of the cases (that is, for 13 over the 20 synthetic cohorts), and is the best one in 45\% of the cases.\\
Given the high number of models considered, these results are satisfactory. However, since we will only consider the subsets of the three best models for the second step of the model selection procedure, it also means that, in 35\% of the cases, the true model is discarded and could not be the one that would be selected. On closer inspection, when the true model is not selected, we show that its dynamics are very close to those of the preselected models in most cases. The dose-response relationships are usually correctly retrieved (Appendix F.2.2).\\

Secondly, we run our hierarchical Bayesian inference method for the three preselected models of each virtual cohort, that is, 60 times in total (Appendix F.3). For each synthetic cohort, the selected model is the one with the lowest DIC value among the three considered.
At this stage, we only slightly improved the results after the first step, with now 50\% of the true models being correctly retrieved. It means that if the true data were actually coming from one of the 225 considered models, we might not correctly find it. However, we show that we would still select a model very close to the true one regarding dose-response relationships. It is the key expectation in terms of method robustness, with the final purpose of making accurate inferences about biological quantities of interest.\\
Therefore, we pursued our validation study to see if we could make accurate inferences from the observations and the selected models, even if they were not the true ones (Appendix F.4). We show that we could estimate with accuracy the model parameters (Appendix F.4.1) and long-term response of the therapy (F.4.2). 
Among the 380 virtual patients, there is for 247 of them a minimal dose $d_{min}^{(i)}$ - under which no remission is possible - that we aim to accurately estimate.  
From our synthetic study, we show that
we get overall accurate estimations of the minimal dose, even if the selected model is not the true one, with a median error (L1-norm) of 0.046 (corresponding to $\sim 8$~$\mu$g/week). There are still a few situations for which we poorly estimate the minimal dose, essentially when the selected model involves a constant dose-response relationship, either associated with the heterozygous or homozygous malignant clone (Appendix F.4.4).\\

\FloatBarrier

\section{Results}

\subsection{Differences between genotypes: from the model selection procedure to the analysis of  hyper-parameter posterior distributions}
\label{sec:res_model_selection}

\begin{figure}[ht]
\centering
\includegraphics[width=0.85\textwidth]{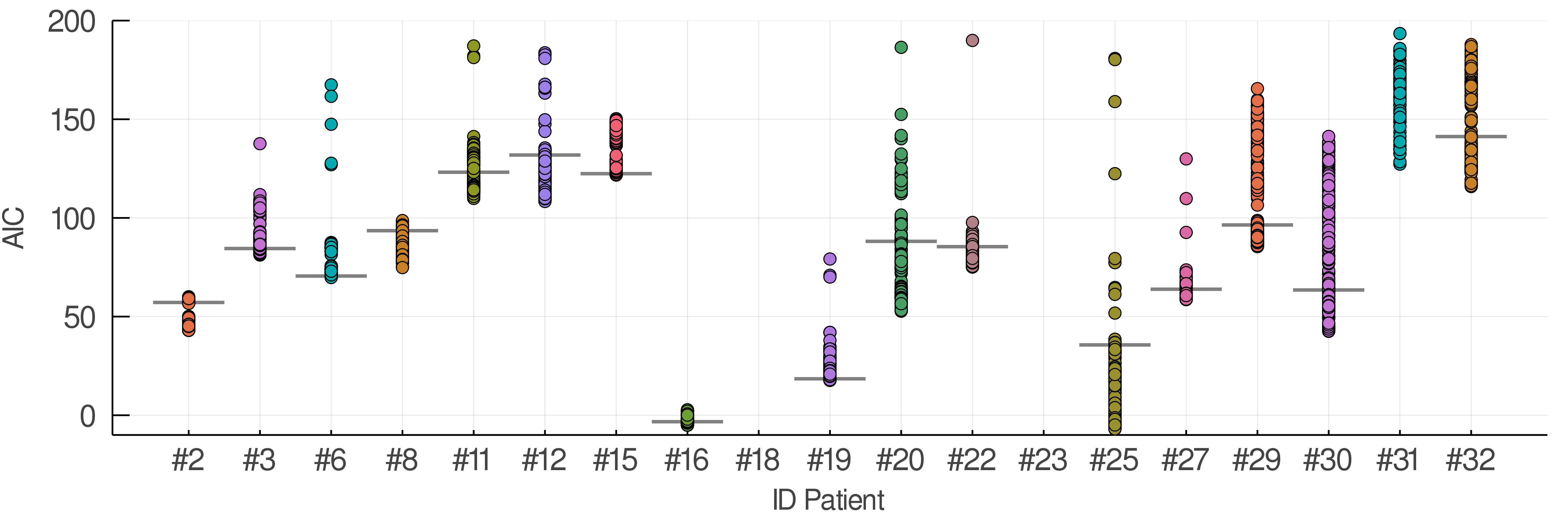}
\caption{AIC values for each model $j$ (represented by a dot) and patient $i$. For each patient, the horizontal line displays the AIC value of the baseline model. The y-axis is truncated for clarity (see complete figure~D.1 in Appendix).}
\label{fig:AIC_zoom}
\end{figure}

To better decipher how IFN$\alpha$ impacts the dynamics of mutated HSCs in MPN patients, and particularly understand its impact depending on the genotype (het vs hom), we apply our two-step model selection procedure.\\ 
We first compute the AIC (eq.~\eqref{eq:AIC_loc}) for each model and each patient. Results are displayed in Figure~\ref{fig:AIC_zoom}. For almost all patients, the baseline model is improved by considering more complex dose-response relationships than the constant ones. 

Actually, the best model for a given patient is not necessarily the best one for another. Therefore, to compare the different models on the whole cohort, we compute the global AIC (eq.~\eqref{eq:AIC_global}).
Results are displayed in Suppl. Fig.~D.2. Three models (presented in Tab.~\ref{tab:best_config}) stand out with a global AIC of around 2,000 when the global AIC of the baseline model is about 2,700.\\
For each of them (as well as for the baseline model), we run a hierarchical Bayesian estimation procedure and compute the DIC value (see eq.~\eqref{eq:DIC}).
Results are presented in Table~\ref{tab:best_config}. Both AIC and DIC are in agreement and lead us to select the model with a constant dose-effect relationship for both $\bar{\gamma}^*_{het}$ and $\bar{\gamma}^*_{hom}$ (same as in the baseline model) and an affine sigmoid (eq.~\eqref{eq:Delta_sigmo_affine}) relation for both $\bar{\Delta}_{het}^*$ and $\bar{\Delta}^*_{hom}$. \\

\begin{table}[ht]
\centering
   \begin{tabular}{ | c | c | c | c | c |c | c | }
     \hline
      \rule{0pt}{10pt}$\bar{\Delta}^*_{hom}$ & $\bar{\Delta}^*_{het}$ & $\bar{\gamma}^*_{hom}$ & $\bar{\gamma}^*_{het}$ & global AIC & DIC\\ \hline
	affine sigmoid&	affine sigmoid&	constant&		constant&	2,016&	2,292\\ 
	affine sigmoid&	affine&				affine&	constant	&	2,040&	2,423\\ 
	affine sigmoid&	affine&				constant&		constant&	2,060&	2,309\\ \hline 
	constant	&	constant&				constant&	constant&	2,659&	2,878\\ 
     \hline
   \end{tabular}
\caption{Best models based on AIC and DIC criteria and comparison with the baseline model (constant dose-response relationships).}
\label{tab:best_config}
 \end{table}

\begin{figure}[h]
\centering
    \begin{subfigure}[b]{0.32\textwidth}
        \includegraphics[width=\textwidth]{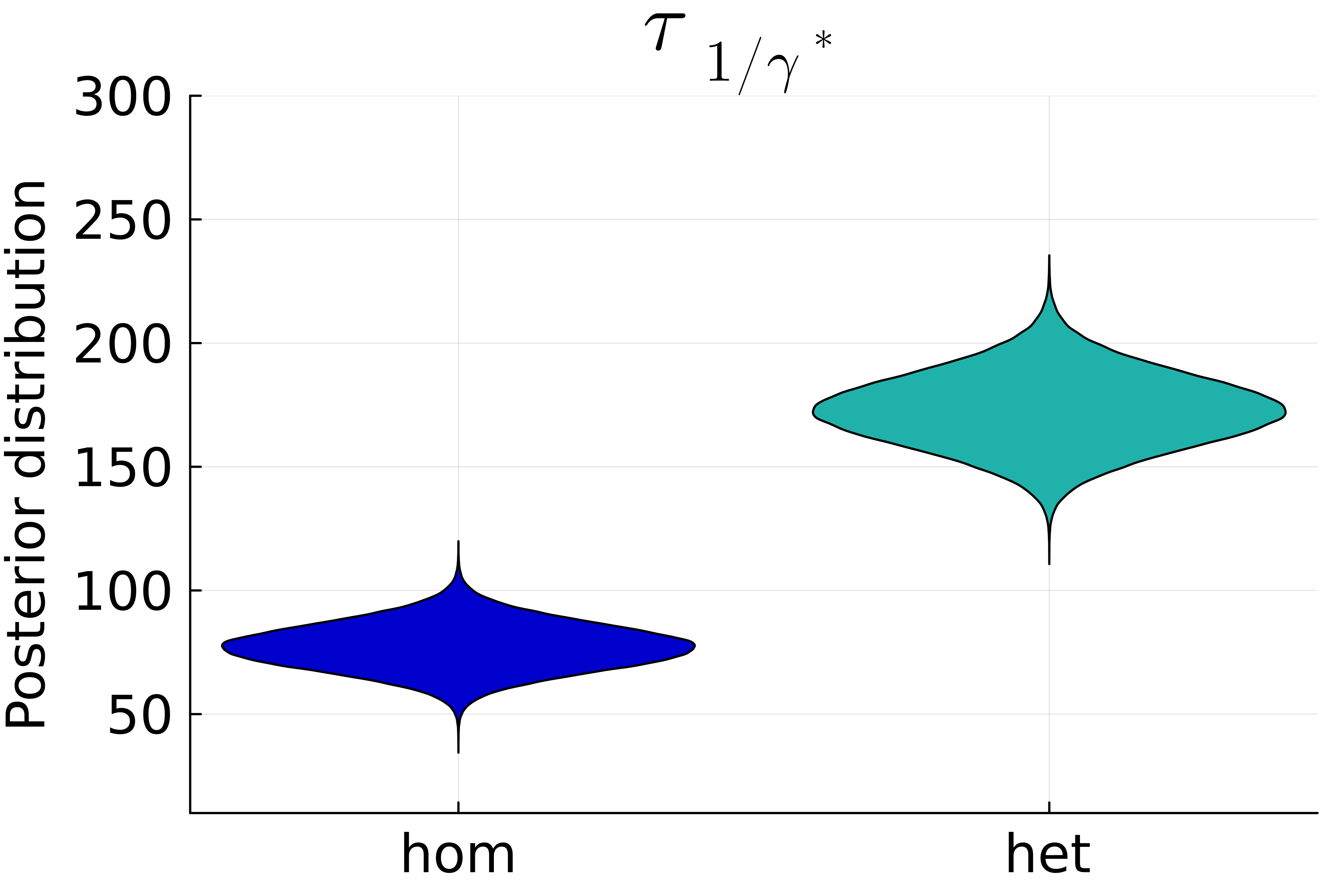}
    \end{subfigure}
    \begin{subfigure}[b]{0.32\textwidth}
        \includegraphics[width=\textwidth]{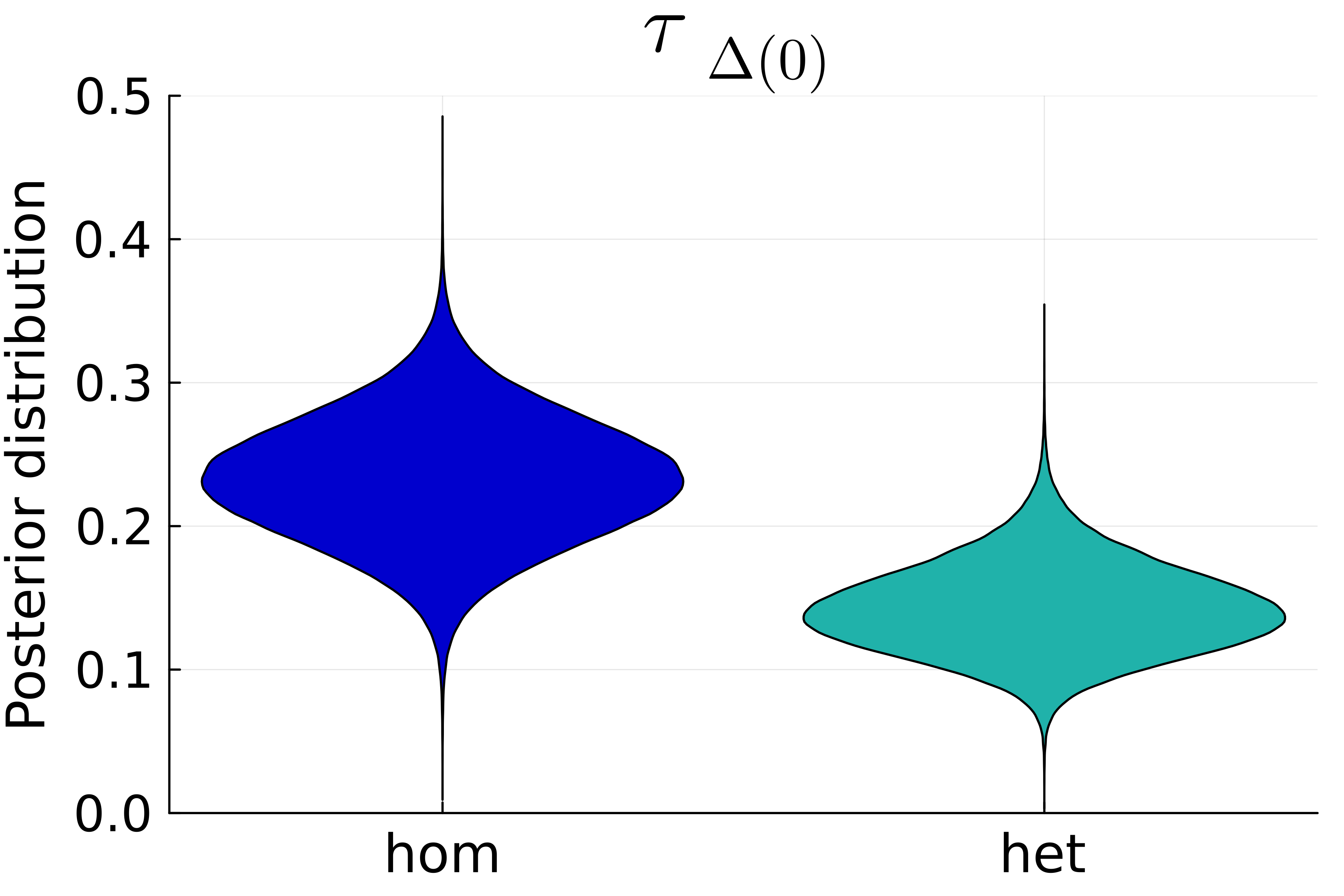}
    \end{subfigure}
    \begin{subfigure}[b]{0.32\textwidth}
\includegraphics[width=\textwidth]{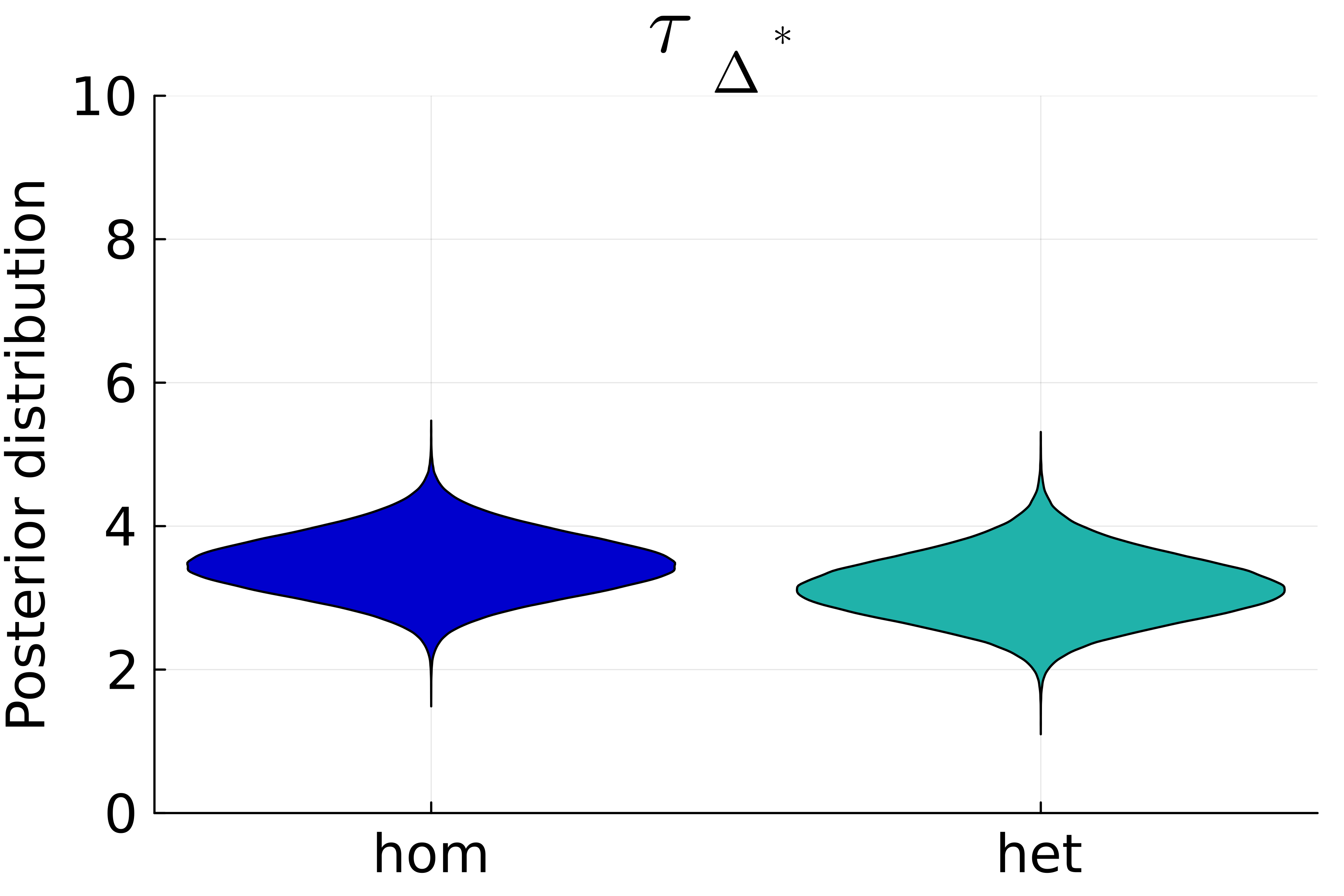}
    \end{subfigure}

    \caption{Comparison of the posterior distributions of the population hyper-parameters (HP) $\vect{\tau}$ for heterozygous (green) and homozygous (blue) clones. Left: HP related to the quiescence exit ($1/\tau^*$). Lower values indicate a higher rate of exit of quiescence. Middle: HP related to the initial (i.e., without treatment) propensity of mutated HSCs to invade the stem cell pool ($\Delta(0)$). Higher values indicate a higher propensity to invade the stem cell pool, that is, more self-renewal divisions than differentiated ones. Right: HP related to the effect of the dose in dose-response relationship~\eqref{eq:Delta_sigmo_affine} ($\Delta^*$). Higher values indicate a stronger depletion of the stem cell pool. }
    \label{fig:posterior_HP_genotype}
\end{figure}

Interestingly, the results of this model selection procedure suggest that IFN$\alpha$ has a similar mechanism of action against heterozygous and homozygous malignant subclones because we end up with dose-response relationships that do not depend on the genotype.
By comparing in Fig.~\ref{fig:posterior_HP_genotype} the posterior distributions of hyper-parameters $\vect{\tau}=\left(\tau_{1}, \cdots, \tau_{P}\right)$, which corresponds to the means of the population distributions~\eqref{eq:prior_HP} associated to $1/\gamma^*_{hom}$ and $1/\gamma^*_{het}$ (related to quiescence exit), $\Delta_{hom}(0)$ and $\Delta_{het}(0)$ (related to the initial propensity of mutated HSCs to invade the stem cell pool), and $\Delta^*_{hom}$ and $\Delta^*_{het}$ (related to the slope of the decreasing self-renewal capacity over IFN$\alpha$ dose), we find that the magnitude of the response to IFN$\alpha$ differs between heterozygous and homozygous mutated cells. As already evidenced by Mosca et al.~\cite{mosca2021}, overall, the quiescence exit under IFN$\alpha$ is increased in homozygous subclones compared to in heterozygous ones (Fig.~\ref{fig:posterior_HP_genotype} left). At the population level, the dose-related response concerning the differentiation of HSCs into progenitor cells does not significantly differ between genotypes (Fig.~\ref{fig:posterior_HP_genotype} right). Finally, we find a higher propensity of homozygous mutated cells to encounter self-renewal divisions compared to heterozygous HSCs (Fig.~\ref{fig:posterior_HP_genotype} middle). This result is biologically consistent since homozygous HSCs present the mutation on two alleles. Therefore, the selective advantage conferred by the mutation $\jakvf$ should be increased compared to heterozygous HSCs.

\FloatBarrier

\subsection{Individual responses to the therapy}
\label{sec:ind_responses}

\begin{figure}[h]
\centering
    \begin{subfigure}[b]{0.49\textwidth}
        \includegraphics[width=\textwidth]{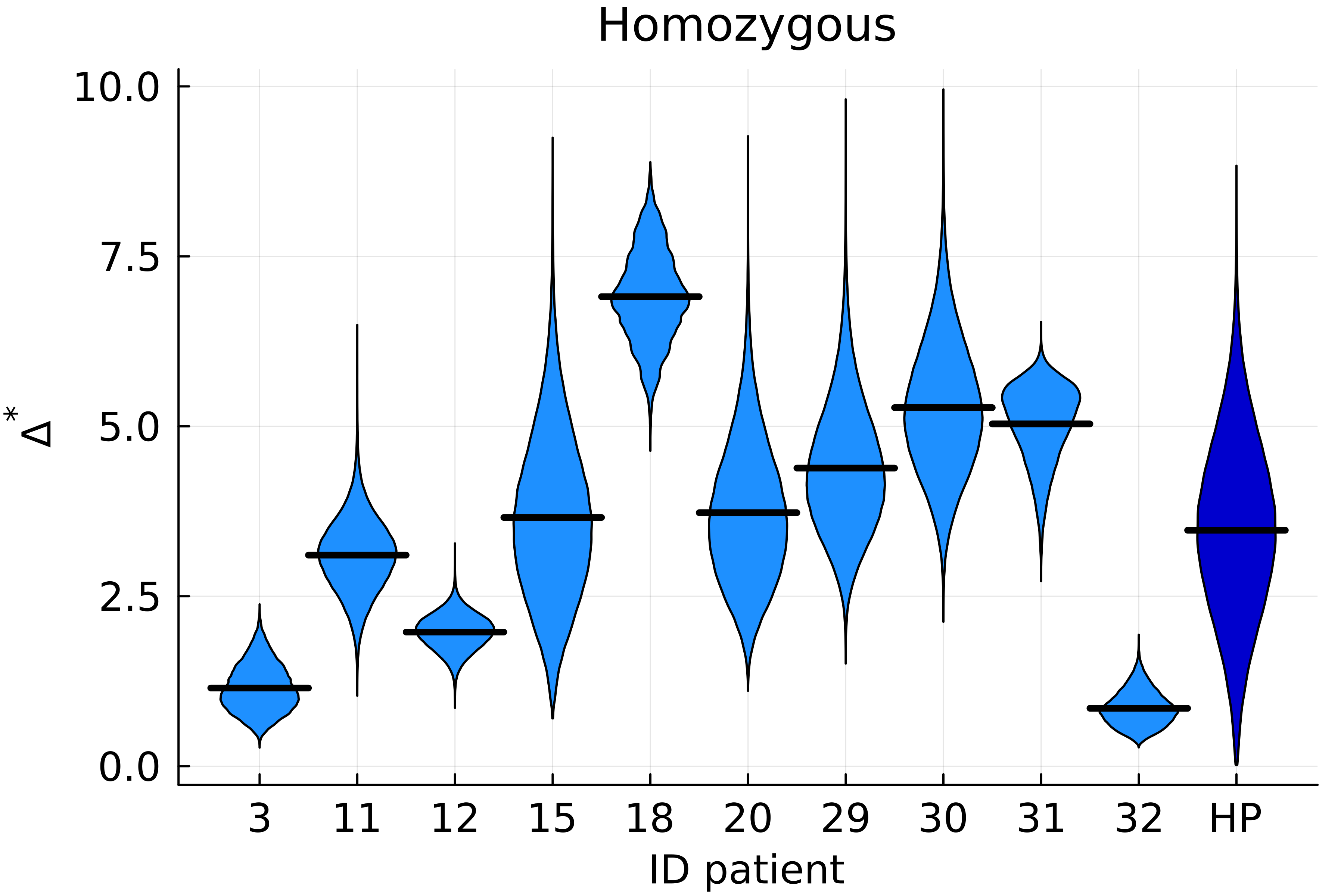}
    \end{subfigure}
    \begin{subfigure}[b]{0.49\textwidth}
        \includegraphics[width=\textwidth]{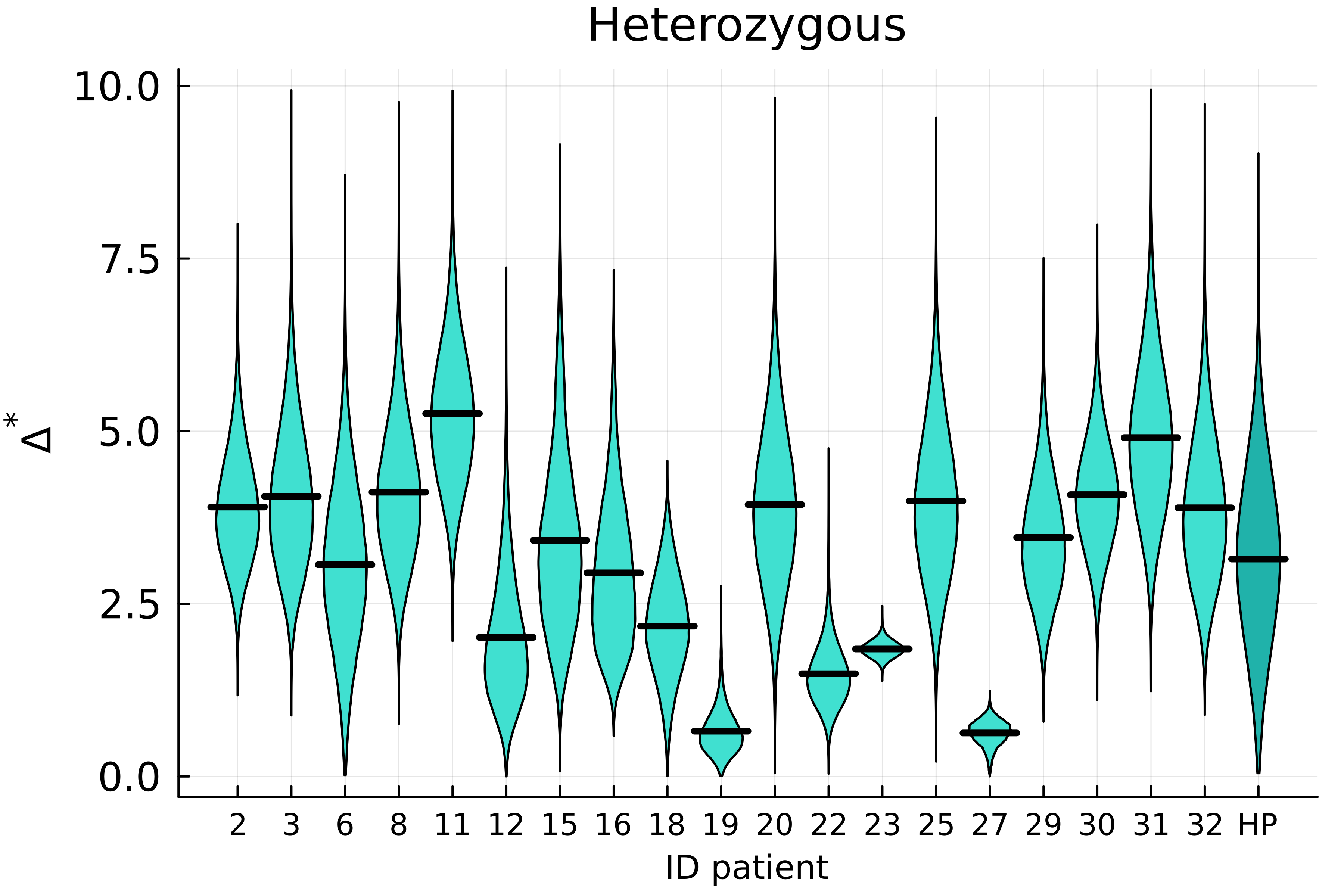}
    \end{subfigure}

\begin{subfigure}[b]{0.49\textwidth}
        \includegraphics[width=\textwidth]{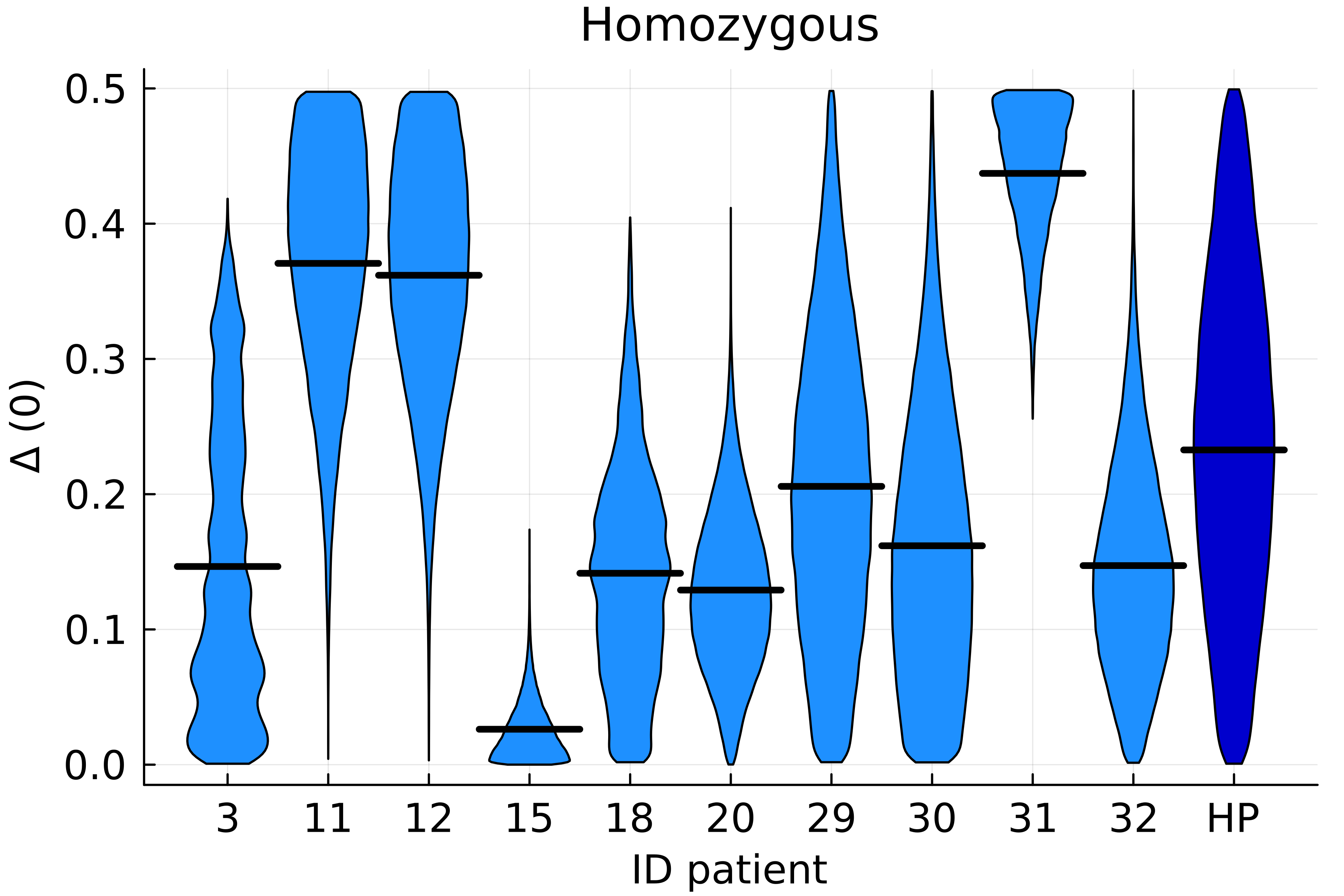}
    \end{subfigure}
    \begin{subfigure}[b]{0.49\textwidth}
        \includegraphics[width=\textwidth]{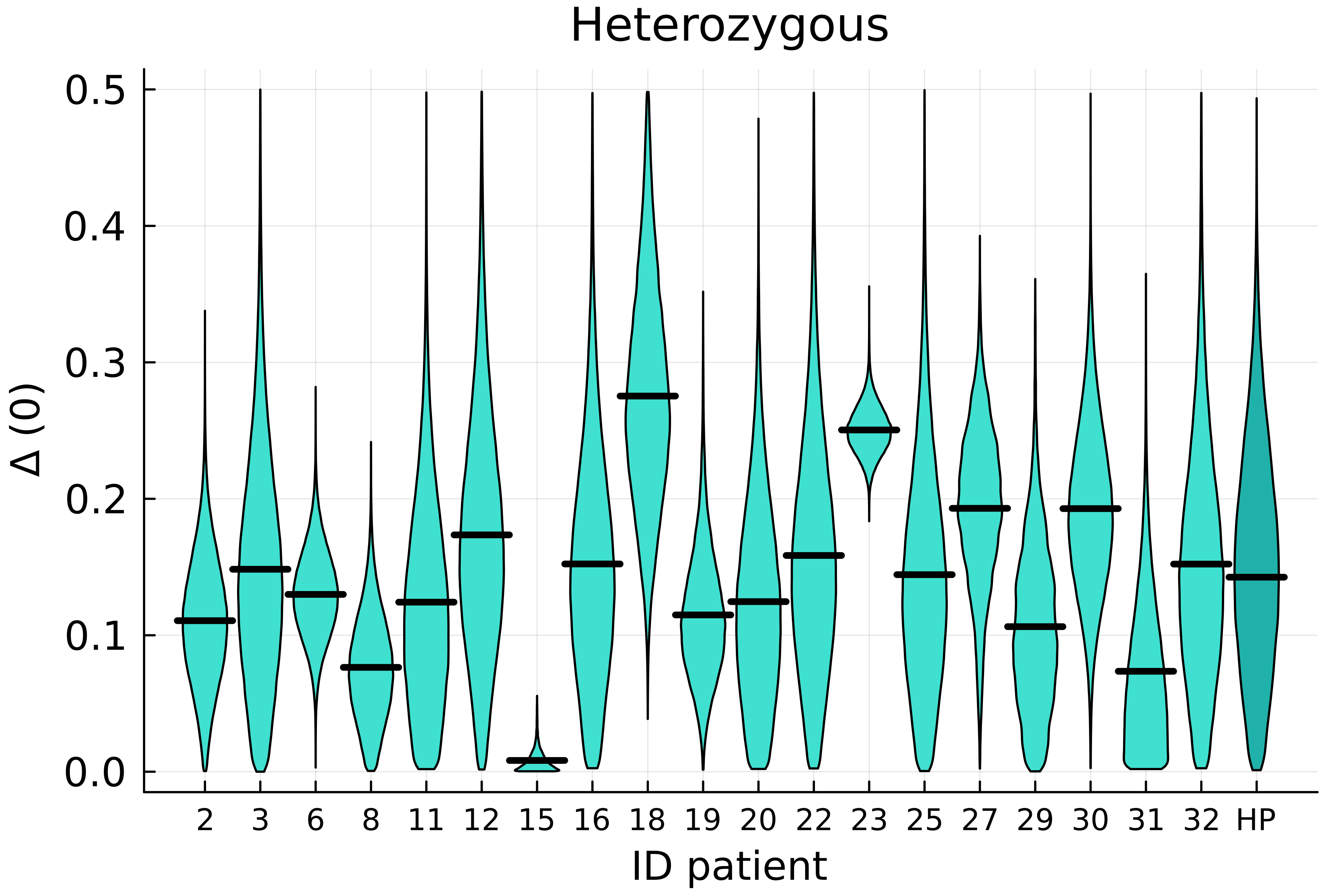}
    \end{subfigure}
    
    \caption{Posterior distributions of the parameters involved in the dose-response relationships of $\bar{\Delta}_{hom}^*$ (left) and $\bar{\Delta}_{het}^*$ (right) estimated for each patient. HP indicates the population distribution, described by a (truncated) Gaussian distribution~\eqref{eq:prior_HP} with mean $\mathbb{E}[\vect{\tau} |\mathcal{D}]$ and variance $\mathbb{E}[\vect{\sigma^2} |\mathcal{D}]$. 
    At the top, we display the distributions for parameter $\Delta^*$ that can be seen as the slope in the affine sigmoid relationships~\eqref{eq:Delta_sigmo_affine}, and at the bottom, the initial propensity of the mutated cells to invade the stem cell pool ($\Delta (0)$). For parameters related to homozygous cells, only patients that exhibit homozygous clones are presented. Horizontal lines indicate mean values. }
    \label{fig:param_Delta}
\end{figure}

In the selected model $\mathcal{M}$, variations of IFN$\alpha$ impact $\bar{\Delta}_{het}^*$ and $\bar{\Delta}^*_{hom}$ through the affine sigmoid relationship~\eqref{eq:Delta_sigmo_affine}. It involves two parameters, the initial value $\Delta$ (without IFN$\alpha$, which we also write for clarity: $\Delta(0)$) and $\Delta^*$ that can be interpreted as a slope for the dose-response relationship. The posterior distributions of these parameters, both for heterozygous and homozygous malignant subclones, were estimated for each patient and at the population level through the hierarchical Bayesian inference framework. Those distributions are displayed in Fig.~\ref{fig:param_Delta}.
Note that not all patients in our cohort present homozygous HSCs.
We observe some inter-individual heterogeneity; patients respond differently to the therapy, as already evidenced by Mosca et al.~\cite{mosca2021}. The inferred population effect, described by a (truncated) Normal law (see eq.~\eqref{eq:prior_HP}) with (posterior) mean $\mathbb{E}[\vect{\tau} |\mathcal{D}]$ and variance $\mathbb{E}[\vect{\sigma^2} |\mathcal{D}]$, is also displayed in Fig.~\ref{fig:param_Delta}.
The estimations for the other parameters involved in model $\mathcal{M}$ are presented in the supplemental Fig. E.3.\\

Then, for each patient, by sampling from his posterior distribution using a Monte Carlo method, we propagate the uncertainties from the parameters to the model output and display the hematopoietic dynamics over the treatment. Figure~\ref{fig:ex_dyn} presents three examples when the supplemental Fig.~E.2 presents all dynamics.
Compared to the previous results from Mosca et al., it is interesting to observe that some late data points (at about 2,000 days for patients \#20, 25, and 32) that were previously considered as outliers are now with our improved model the reflection of a decrease of the dosage (or even a treatment interruption) and the sign of a relapse. Such results indicate that the dosage should not be decreased to a too large extent.

\begin{figure}[h]
\centering
    \begin{subfigure}[b]{0.32\textwidth}
        \includegraphics[width=\textwidth]{20}
    \end{subfigure}
    \begin{subfigure}[b]{0.32\textwidth}
        \includegraphics[width=\textwidth]{25}
    \end{subfigure}
    \begin{subfigure}[b]{0.32\textwidth}
        \includegraphics[width=\textwidth]{32}
    \end{subfigure}

    \caption{Examples of inferred dynamics and comparison to the data. Black lines correspond to the inferred VAF in mature cells. Blue and green lines correspond to the inferred CF in homozygous and heterozygous progenitor cells. The dots, triangles, and squares are the experimental data values. The shaded areas represent 95\% credibility intervals. The shaded beige areas
correspond to the dose of IFN$\alpha$ received over time. For patient \#20, the treatment is interrupted after 1218 days of therapy, when, for the two others, the IFN$\alpha$ administration is maintained but with a lower dose after 1280 and 1613 days for patients \#25 and \#32, respectively.
    }
    \label{fig:ex_dyn}
\end{figure}

\FloatBarrier

\subsection{Estimating individual minimal doses}
\label{sec:ind_min_dose}

\begin{figure}[h]
    \centering
    \includegraphics[width=0.8\textwidth]{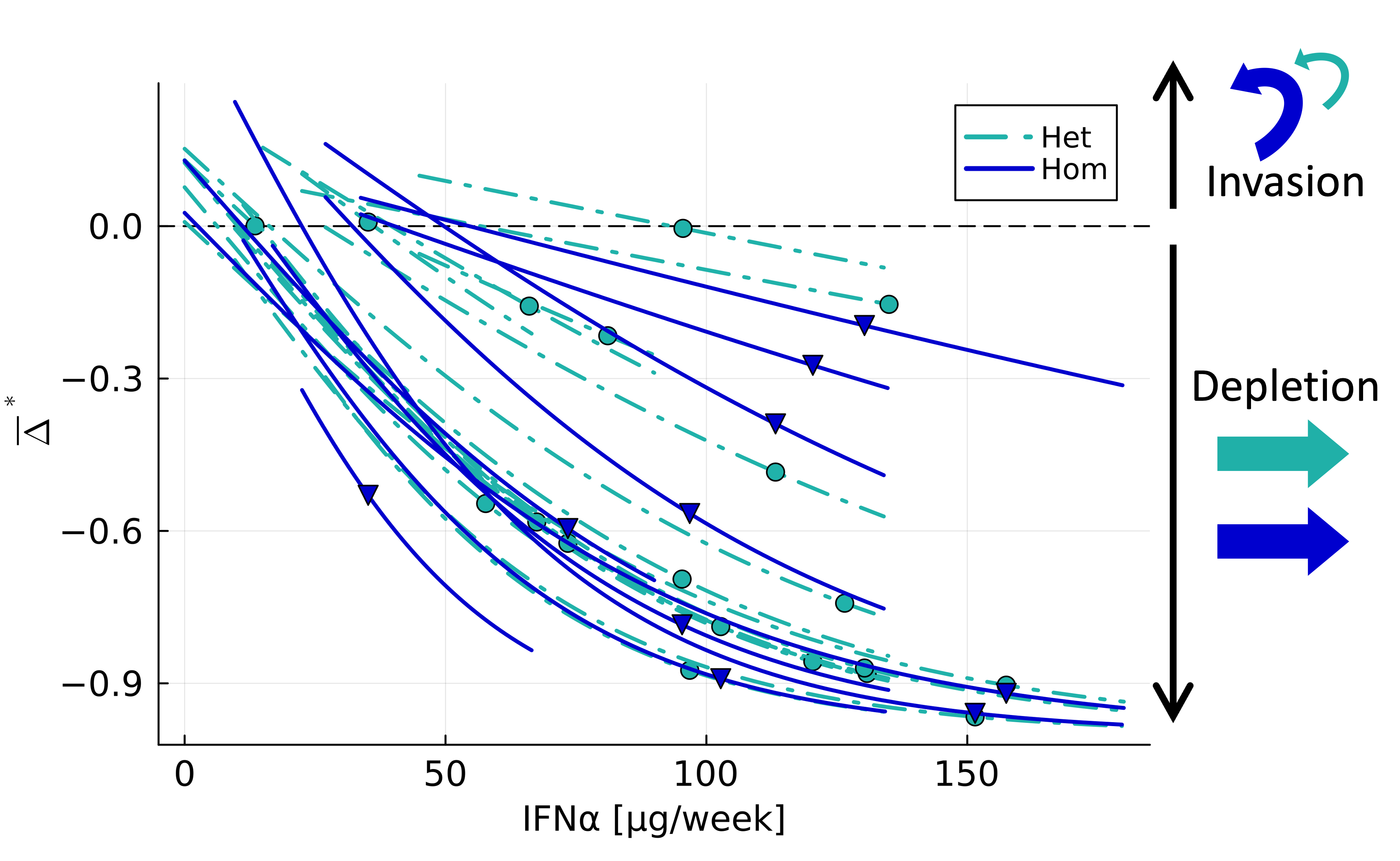}
    \caption{Dose-response relationships for $\bar{\Delta}_{het}^*$ (green) and $\bar{\Delta}^*_{hom}$ (blue) as function of the weekly IFN$\alpha$ dose $d$. Each curve represents a patient. Extremities of the curves correspond to the minimum and maximum dose they received over the treatment. Points and triangles respectively refer to the value of $\bar{\Delta}_{het}^*$ and $\bar{\Delta}_{hom}^*$  at the averaged dosage received over the 450 first days of therapy by the patient.}
    \label{fig:relation_Delta_dose}
\end{figure}

From the estimation results, we see that MPN patients respond differently to the treatment and that there is a risk of relapse if the dose of IFN$\alpha$ is decreased. In the selected model $\mathcal{M}$, the relapse is explained by the fact that decreasing dose $d$ also increases the propensity of mutated HSCs to invade the stem cell pool, i.e., $\bar{\Delta}^*$ is a decreasing function of $d$.
In Fig.~\ref{fig:relation_Delta_dose}, we display for all patients the estimated dose-response relationships $\bar{\Delta}^*_{het}$ and $\bar{\Delta}^*_{hom}$ (for the latter, only if the patient has homozygous subclones) by taking, for the parameters involved in both relations, the means of the posterior distributions for $\Delta_{het}(0), \Delta_{hom}(0), \Delta_{het}^*$ and $\Delta^*_{hom}$.\\

\begin{figure}[h]
\centering
    \begin{subfigure}[b]{0.49\textwidth}
        \includegraphics[width=\textwidth]{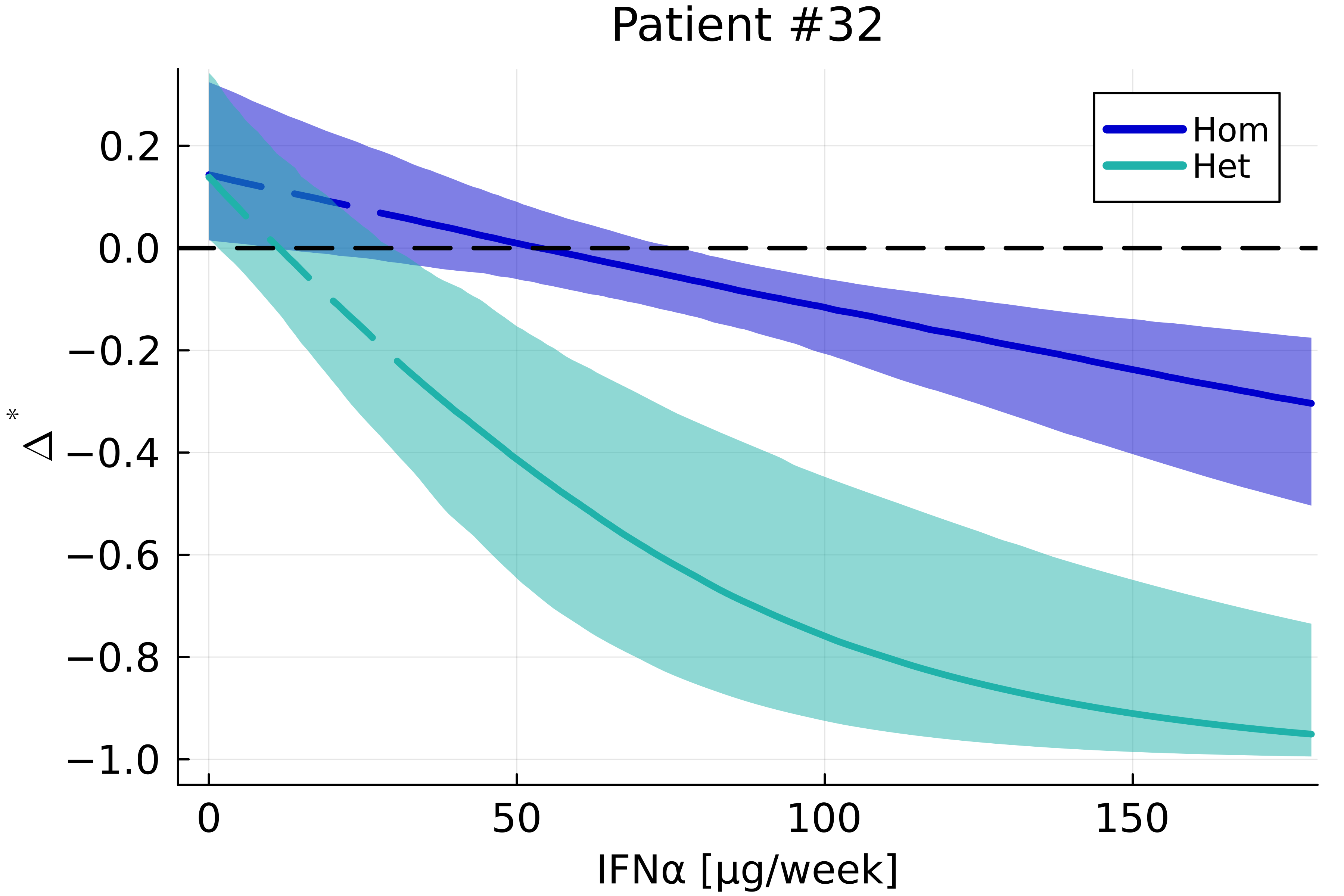}
    \end{subfigure}
    \begin{subfigure}[b]{0.49\textwidth}
        \includegraphics[width=\textwidth]{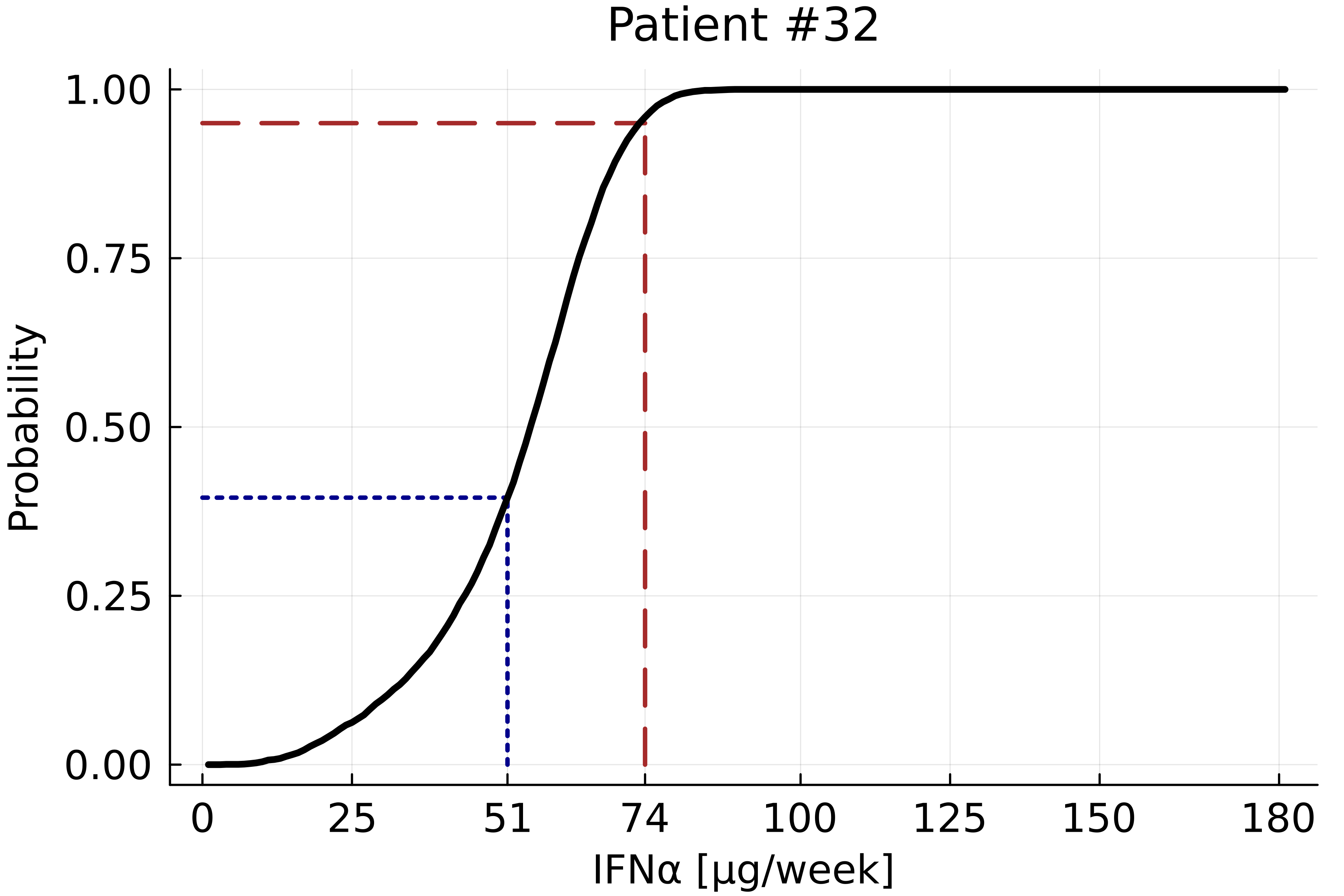}
    \end{subfigure}

    \caption{On the left, dose-response relationships of $\bar{\Delta}^*_{het}$ and $\bar{\Delta}^*_{hom}$ for patient \#32 along with 95\% credibility intervals. On the right, the probability of remission $P_{rem}$ as a function of the dose. The red dashed line indicates the dose above which there is a 97.5\% chance of getting a remission, and the blue dot line corresponds to the estimated (mean \textit{a posteriori}) minimal dose which is $d^{(i)}_{min}=51$~$\mu$g/week for patient $i:=32$.}
    \label{fig:32_example}
\end{figure}

Since $\Delta^*_{het}(0) >0$ and $\bar{\Delta}^*_{het}$ is a decreasing function of $d$, there is, for each heterozygous patient (without homozygous HSCs), a dose $d_{min}^{het}$ such that $\bar{\Delta}_{het}^*(d_{min}^{het}) = 0$ and $\bar{\Delta}^*_{het}(d) > 0$ for $d<d_{min}^{het}$.
The same goes for the homozygous malignant clone and, in the case of patients having simultaneously homozygous and heterozygous HSCs, $d_{min}$ (as introduced in eq.~\eqref{eq:def_dmin}) corresponds to the maximum between $d_{min}^{het}$ and $d_{min}^{hom}$.
Since we obtain the (posterior) distributions of parameters, we can give the credibility intervals of the dose-response relationships to illustrate their uncertainty, as displayed, for example, in Fig.~\ref{fig:32_example} (left) for patient \#32. 
For this patient, his minimal dose $d^{(i)}_{min}$ is estimated (mean \textit{a posteriori}) to be equal to 51~$\mu$g/week (with a 95\% credibility interval: [15, 74]).
In addition, we can estimate for all possible doses $d$ the probability of remission $P_{rem}(d)$ (see eq.~\eqref{eq:P_rem}). Such probability is displayed for patient \#32 in Figure~\ref{fig:32_example} (right side).\\

\begin{figure}[h]
    \centering
    \includegraphics[width=0.95\textwidth]{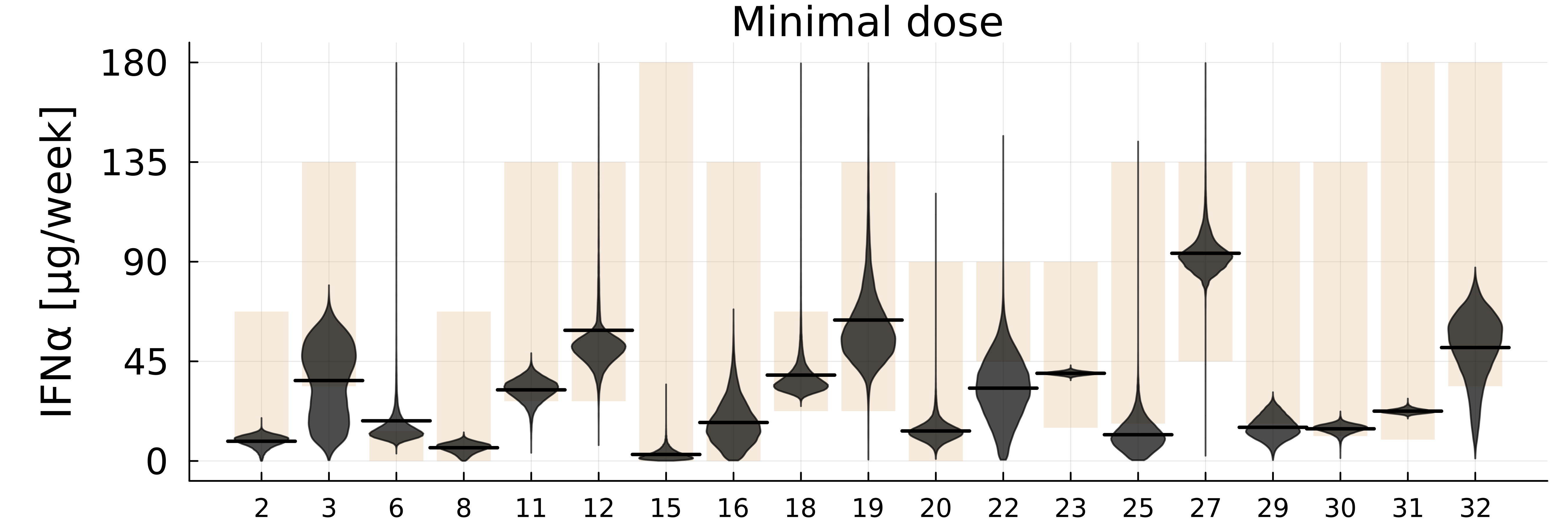}
    \caption{
    Posterior distributions of the minimal dose for each patient (x-axis), where the horizontal black line corresponds to the posterior mean of $d_{min}^{(i)}$. 
The brown-shaded areas represent the ranges of doses administered to each patient.}
    \label{fig:minimal_dose_true_cohort}
\end{figure}

The minimal dose can then be computed for each patient of our cohort (Fig.~\ref{fig:minimal_dose_true_cohort}). 
Our results can then be used as guidelines for clinicians, notably to discourage an abrupt dose de-escalation or a decrease of the IFN$\alpha$ dose for a given patient under the estimated lower limit since it would increase the risk of relapse.\\

\FloatBarrier
\subsection{Determining an initial treatment dose}
\label{sec:starting_dose}

\begin{figure}[h]
\centering
    \begin{subfigure}[b]{0.49\textwidth}
        \includegraphics[width=\textwidth]{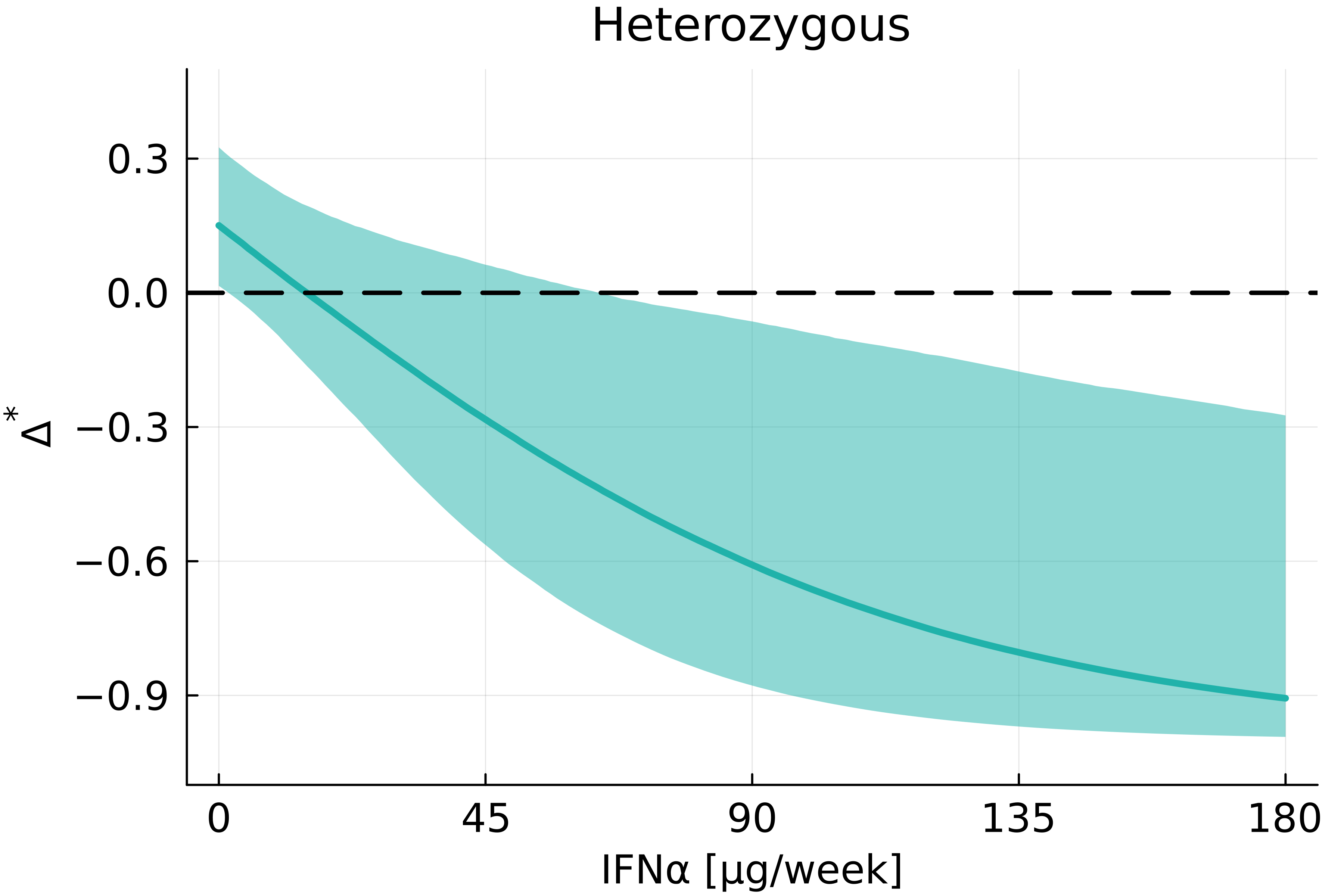}
    \end{subfigure}
    \begin{subfigure}[b]{0.49\textwidth}
        \includegraphics[width=\textwidth]{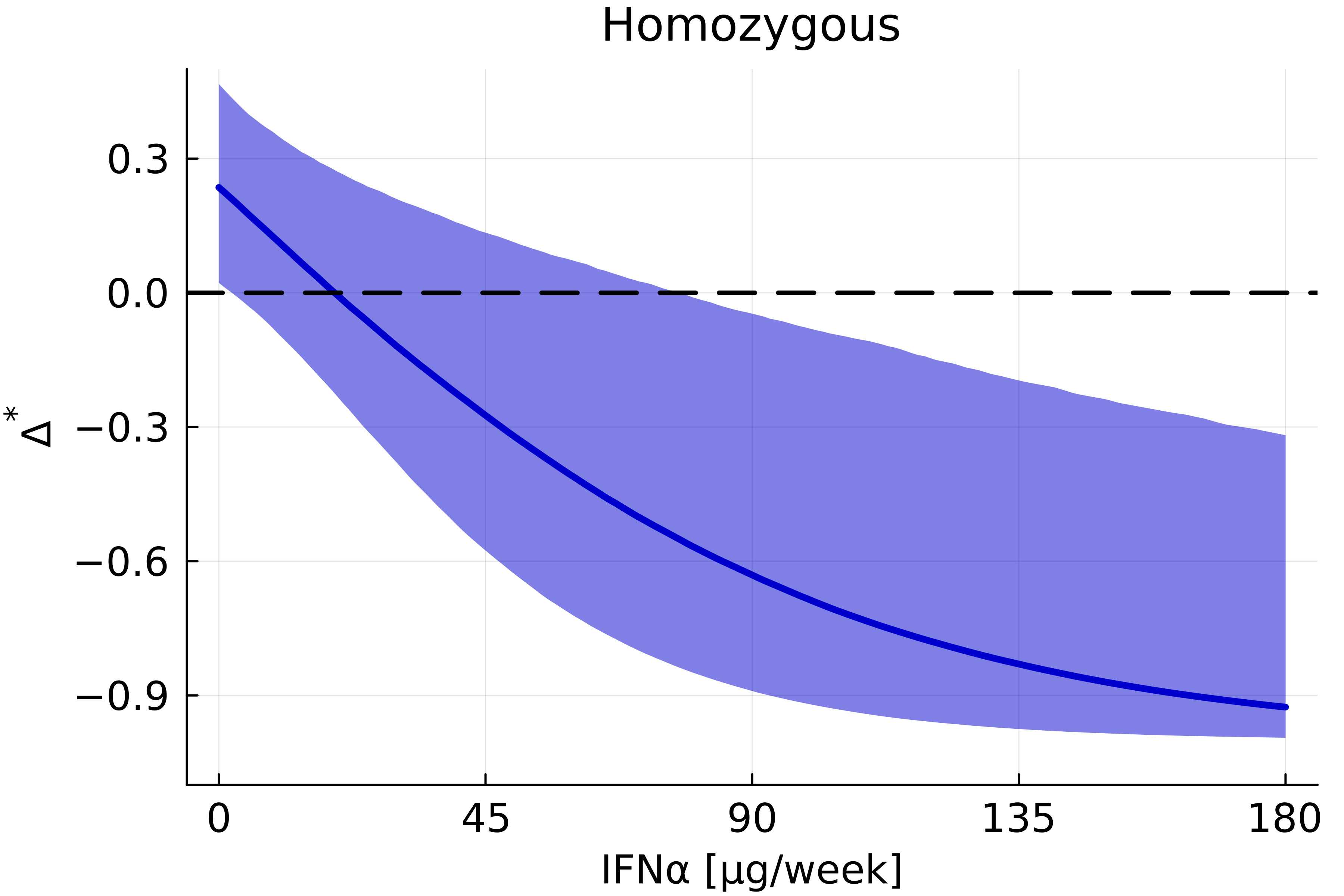}
    \end{subfigure}
    \caption{Prior dose-response relationships of $\bar{\Delta}^*_{het}$ (left) and $\bar{\Delta}^*_{hom}$ (right) for new patients (based on the estimation made from our cohort). Lines indicate the median values and shaded areas indicate 95\% credibility intervals. The black dashed horizontal line indicates $\Delta^*=0$. Below this limit, the treatment induces a long-term remission, according to our model.}
    \label{fig:pop_Delta}
\end{figure}

In the previous section, we computed for all patients of our observational cohort a personalized minimal IFN$\alpha$ dose, both necessary and sufficient to get a long-term remission. Besides, owing to the hierarchical Bayesian framework, the inference is not only at the individual level but also at the population level. \\
For a new patient, we can now consider, as prior distributions for his model parameters, (truncated) Gaussian distributions whose means and variances are the hyper-parameter values estimated on the original cohort of $N=19$ \jakvf MPN patients (see eq.~\eqref{eq:prior_new_patient}).
This new prior can be used \textit{a priori} before having any observation for this patient and can help determine a suitable starting dose. By sampling from these new prior distributions, we can determine the prior behavior of $\bar{\Delta}^*_{het}$ and $\bar{\Delta}^*_{hom}$ according to the dose (Figure~\ref{fig:pop_Delta} left and right respectively).\\

\begin{figure}[h]
\centering
        \includegraphics[width=0.7\textwidth]{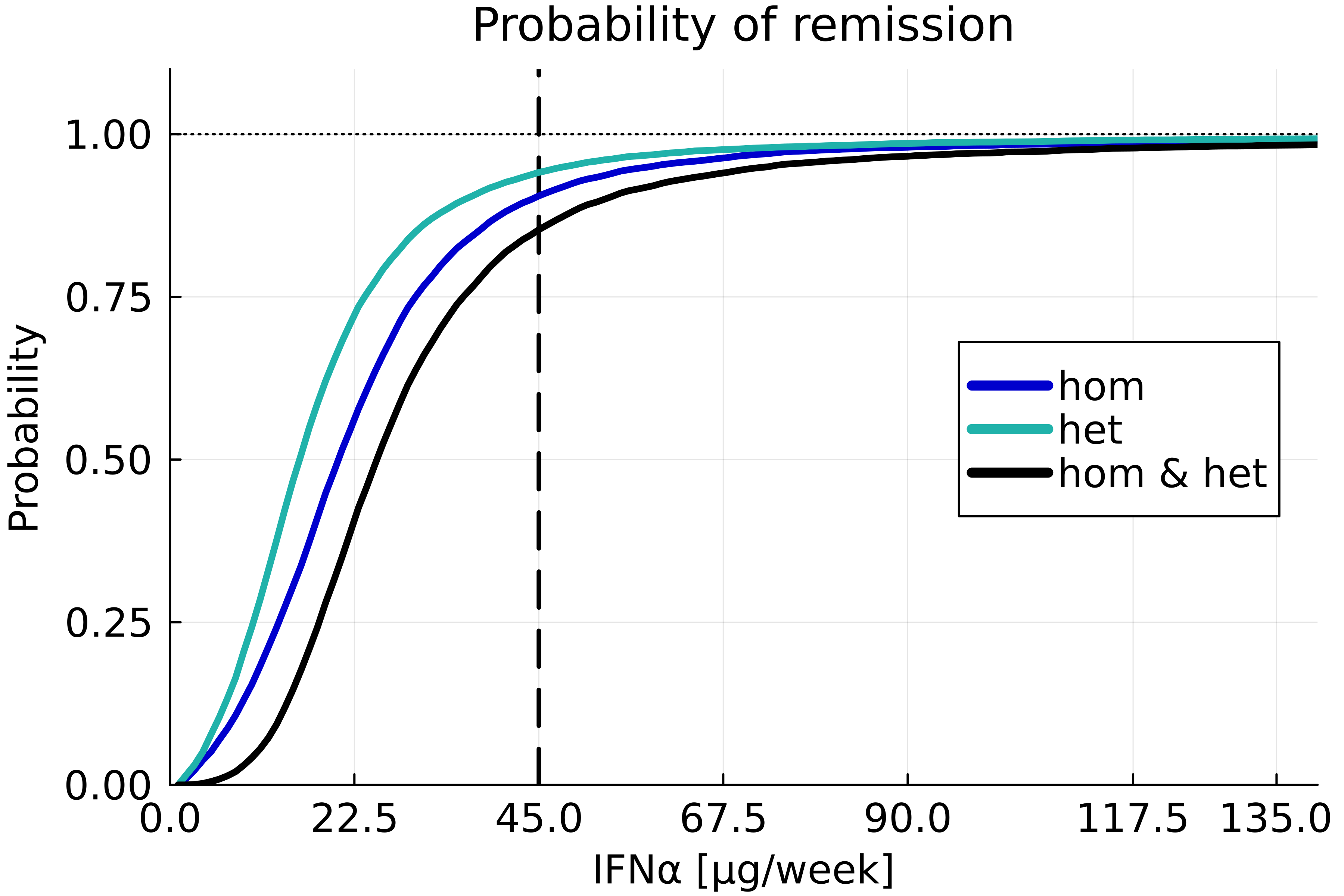}
    \caption{For a new \jakvf MPN patient, probability of having a long-term remission as a function of the initial posology. We distinguish cases where the patients only have homozygous HSCs (blue), only heterozygous HSCs (green), or both (black). The vertical dashed line indicates the initial dose used in clinical trials~\cite{yacoub2019pegylated,knudsen2021genomic}. }
    \label{fig:pop_prob_rem}
\end{figure}

Then, we can compute the probability of long-term remission, as a function of the dose, by considering a patient having only heterozygous, only homozygous, or both malignant subclones (Fig.~\ref{fig:pop_prob_rem}).
For patients with both homozygous and heterozygous \jakvf HSCs, we estimate that a starting dose of 45~$\mu$g/week might lead to long-term remission in 86\% of the cases.
Our results suggest that the minimal initial dose such that 95\% of the patients would reach a remission should equal 71~$\mu$g/week.

\FloatBarrier
\newpage

\section{Discussion}
To determine the minimal dose in IFN$\alpha$ therapy against MPN, we proposed a method combining mathematical modelling, model selection, and hierarchical Bayesian inference. We extended the model of Mosca et al.~\cite{mosca2021} to take into account the variations of posology along treatment.
We proposed several alternative models and used a two-step model selection procedure: first, we discarded most models based on the AIC, then we applied a hierarchical Bayesian inference method to calibrate the remaining models. We selected the best-performing model according to the DIC. 
Finally, we thoroughly analyzed the results obtained for the selected model.\\
We verified to which extent we could draw robust conclusions from a synthetic study by generating 20 datasets, each with 19 virtual patients and observation data comparable to those in Mosca et al. \cite{mosca2021}. We showed that we could select a relevant (if not the true, this latter being found in 50\% of the cases, over the 225 models tested) model from which we could accurately infer the model parameters and the long-term response to the treatment. When it existed, we could also infer the minimal dose and get accurate estimations for most virtual datasets (with a median error of approximately 8~$\mu$g/week). The exceptions were when the selected model involved a constant dose-response relationship, either for $\bar{\Delta}^*_{het}$ or $\bar{\Delta}^*_{hom}$, which is actually not the case of the model we selected based on the true data.\\
The real data used for the calibration came from 19 \jakvf MPN patients.
Our results suggest that IFN$\alpha$ increases the quiescence exit of mutated HSCs, especially the homozygous ones, but we haven't found evidence suggesting that this response may be sensitive to variations of dosage along therapy. A possible explanation could be that homozygous HSCs exit quiescence at the beginning of the therapy and then no longer return to this state. 
The identified major mechanism of action by which the stem cell pool is depleted is HSCs differentiation. IFN$\alpha$ may increase the propensity of mutated HSCs to differentiate into progenitor cells as the dose increases. This mechanism of action would not differ according to the genotype, contrary to the finding of Tong et al.~\cite{tong2021hematopoietic}, who showed using single-cell RNA-seq technique, that homozygous cells are quiescent
while heterozygous cells undergo apoptosis in IFN$\alpha$-treated patients. However, our synthetic study showed that we might not always be able to retrieve the true dose-response relationships concerning $\bar{\Delta}^*_{het}$ and $\bar{\Delta}^*_{hom}$, limiting our capacity to make definitive conclusions on the mechanisms of action of IFN$\alpha$.\\
When a minimal dose is reached, mutated HSCs face more differentiation than self-renewal divisions; long-term remission can be obtained.
For all patients in our cohort, we estimated an individual minimal dose above which both the heterozygous and homozygous malignant clones could ultimately be depleted. 
Our estimated minimal doses range from almost zero (3~$\mu$g/week) to about 90~$\mu$g/week, with a median value equal to 25~$\mu$g/week.
Our findings discourage de-escalating the dose during therapy, below the personalized minimal doses. 
An almost zero dose suggests the patient might be highly susceptible to the therapy. Physicians could use this very low dose to stabilize the patient when a sufficiently low clonal fraction would have been reached and, therefore, avoid a relapse that might occur when definitively interrupting the therapy. However, it is still recommended to use higher doses to target the mutated clones more efficiently (and also to increase the probability of getting remission). Indeed, according to our model, the higher the dose, the better the response.
Our model never recommends a treatment interruption since it would result in a relapse. Indeed, we estimated that mutated HSCs have an initial propensity of invading the stem cell pool, a necessary condition for the expansion of the malignant clone and the appearance of the MPN symptoms, as described in~\cite{hermange2021,van2021reconstructing}. 
Then, in the absence of IFN$\alpha$, even a tiny fraction of mutated cells (that always remain with our deterministic model) would further expand. However, our deterministic model is no longer suitable when the number of mutated HSCs becomes very small since stochastic effects predominate. It could be taken into account by extending the model.  
Other models from the literature could also have been used as baseline models, potentially affecting the results. In particular, models accounting for regulation or interaction between cells could be relevant. Mac Lean et al. have shown, comparing different models, that appropriate modelling of the imatinib treatment against chronic myeloid leukemia (another blood disease affecting the same myeloid lineage as in MPN) had to explicitly account for ecological interaction within the bone-marrow niche~\cite{maclean2014ecology}. Pedersen et al. have analysed a mathematical model of HSC dynamics within the bone-marrow microenvironment~\cite{pedersen2021mathematical}, also in the context of MPN, and their model could be a potential baseline candidate.\\
With our hierarchical Bayesian framework, we inferred a population effect and estimated the minimal dose for all individuals of our cohort, and determined the most suitable initial posology to prescribe to a new patient. 
We estimated that an initial dose of 45~$\mu$g/week, classically used in clinical trials~\cite{yacoub2019pegylated,knudsen2021genomic, Gisslinger2020}, should only induce long-term remission in 86\% of the cases, and we advocate instead to start at about 70~$\mu$g/week. 
A dose escalation remains relevant. Even if we predict a remission with a low dose, it might only be reached in a very long time, and increasing the dose would improve the response to the therapy. Besides, clinicians rarely treat at low doses (e.g., 45~$\mu$g/week) during a long time but rather rapidly increase the posology up to 90 or 135~$\mu$g/week to achieve hematological responses. 
Of course, in clinical routine, physicians also have to consider additional constraints, such as the occurrence of side effects, mainly depression~\cite{lotrich2007depression, trask2000psychiatric}. The occurrence of side effects might force the clinician to decrease the posology or even temporarily interrupt the therapy. To extend this work, we aim to apply optimal control methods to study stop-and-go strategies and determine the period during which the treatment might be interrupted. Since we deal with long-term therapies, it might be relevant to interrupt them regularly when a sufficiently low CF is reached (for which the risk of severe disorders is limited for the patient) and restart them when the CF increases again and before the reappearance of MPN symptoms. Such strategies would avoid treating a patient continuously over many years and might also prevent the development of malignant subclones that could acquire resistance to the treatment. 
Our work could also be extended for prediction purposes, as we proposed in~\cite{hermange2023optimizing}. The posterior distribution of the hyper-parameters (related to the population distributions of the model parameters) could now be used as prior for new individuals. Using data assimilation techniques, we could predict the response to the therapy for new patients and refine our prognostic based on new observations. This work would also entail studying the choice of the population distributions (here independent truncated Gaussian laws) in more detail and also evaluate the relevance of introducing a dependency between hyper-parameters.\\
To conclude, the proposed mathematical approach, used to determine a minimal dose to prescribe, remains general and can be applied to a wide range of problems. Our method illustrates how a model selection procedure can help decipher the mechanism of action of treatment and the potential of hierarchical Bayesian statistics, both for increasing the robustness of the estimations made on an observed cohort of individuals and generalizing results to new patients.

\bibliographystyle{acm}
\bibliography{main}

\newpage
\section*{Acknowledgements} 
P.-H.C. and I.P. are granted by the Prism project, funded by the Agence Nationale de la Recherche under grant number ANR-18-IBHU-0002. This work was also supported
by grants from INCA Plbio2018, 2021 to IP, Ligue Nationale Contre le Cancer ({\'e}quipe
labellis{\'e}e 2019), and ITMO Cancer of Aviesan within the 2021-2030 Cancer Control Strategy framework, on funds administered by Inserm.
This work was performed using HPC resources from the “M{\'e}socentre” computing center of CentraleSup{\'e}lec and {\'e}cole Normale Sup{\'e}rieure Paris-Saclay supported by CNRS and R{\'e}gion Ile-de-France\footnote{http://mesocentre.centralesupelec.fr/}.
We thank A. Della Noce for his proofreading and his help in improving the manuscript.
We thank D. Madhavan for her help in editing the manuscript.\\

\section*{Author contributions} 
G.H., I.P., P.-H.C. conceived the mathematical model, 
G.H., P.-H.C. defined and performed the statistical methods, 
W.V. advised the study, 
G.H., I.P., P.-H.C. wrote the manuscript, 
I.P, P.-H.C. supervised the study, 
all authors revised the manuscript.

\section*{Conflict-of-interest disclosure}
The authors declare no competing financial interests.

\newpage

\appendix
\counterwithin{figure}{section}
\counterwithin{table}{section}
\counterwithin{equation}{section}

\section*{Appendix}

The \rev{baseline} model from Mosca et al.~\cite{mosca2021} that we extend in this work is presented in the Appendix~\ref{sec:details_base}. We derive an analytical solution for this model in §~\ref{sec:analytical_solution} and show its practical identifiability in Appendix~\ref{sec:identif}. The data used for this work are presented in~\ref{sec:observation_model}. Our inference method is detailed in §~\ref{sec:identif_param_estim}. Additional results of our two-step model selection procedure are presented in Appendix~\ref{sec:detail_selection}. Appendix~\ref{sec:detail_model_selected} details the results we get for the selected model (inferred dynamics for each patient, posterior distribution of the parameters, and probability of remission).
In Appendix~\ref{sec:synthetic_study}, we present a synthetic study used to validate our methodology from simulated datasets. In Appendix~\ref{sec:list_models}, we list the models we study and give information about the computation of a minimal dose. In Appendix~\ref{sec:PK}, we present the impact of a pharmacokinetics model on our results.

\newpage
\tableofcontents
\newpage

\section{Details of the \rev{baseline} model}
\label{sec:details_base}

In the \rev{baseline} model, as presented in the main text (§~2.1.1 and Fig.~1) and detailed in Mosca et al.~\cite{mosca2021}, we consider three populations of hematopoietic cells: wild-type (wt), heterozygous (het), or homozygous (hom). The numbers of cells of each type follow the following ODE system (here, we do not make precise the cell genotype):
\begin{equation}
\label{eq:syst_comp}
\left\{
\begin{array}{ll}
  \frac{dN_1(t)}{dt} &= -\gamma N_1(t) + \beta N_2(t)  \\
  \frac{dN_2(t)}{dt} & = \gamma N_1(t) + (\alpha \Delta - \beta) N_2(t) \\
   \frac{dN_i(t)}{dt} &= \alpha ( 1-\Delta) \kappa_i N_2(t) - \delta_i N_i(t) \\
   \frac{dN_{m}(t)}{dt} &= \delta_i \kappa_m N_i(t) - \delta_{m} N_{m}(t) \\		
\end{array}
\right.
\end{equation}
with $\alpha, \beta, \gamma,\kappa_i,\kappa_m,\delta_i,\delta_m>0$ and $\Delta\in[-1,1]$.\\
As presented in the next section, we can derive an analytical solution and thus avoid solving this system numerically. 

\subsection{Analytical solution for the \rev{baseline} model}
\label{sec:analytical_solution}

\subsubsection{Solution for HSC compartments}

The two first equations can be written in a matrix form:
\begin{equation}
\frac{d \vect{N}(t)}{dt} = \matr{A} \vect{N}(t)
\label{eq:syst_matr_HSC}
\end{equation}
with  $\vect{N}(t) := \left( N_1(t),N_2(t) \right)^t$ and:
\begin{equation*}
\matr{A} = \begin{pmatrix}
    - \gamma      & \beta \\
	\gamma & -\beta + \alpha \Delta 
\end{pmatrix}
\end{equation*}
The characteristic polynomial of $\matr{A}$ is:
\begin{align*}
p_{\matr{A}}( X ) & = X^2 - \text{tr}( \matr{A} ) X + \text{det}(\matr{A} ) \\
				& = X^2 + (\gamma + \beta - \alpha \Delta ) X -\gamma \alpha \Delta
\end{align*}
The eigenvectors of $\matr{A}$ are the polynomial roots of $p_{\matr{A}}( X )$. Its discriminant is:
\begin{equation}
D = (\gamma + \beta - \alpha \Delta )^2 + 4 \gamma \alpha \Delta
\label{eq:det_charact_poly}
\end{equation}
Potentially, we should distinguish cases according to the sign of $D$, but it is possible to show that $D>0$ (see section~\ref{sec:proof_det_positif}). Then, the two eigenvalues are distinct real numbers:
\begin{equation*}
\lambda_{\pm} = \frac{-(\gamma + \beta - \alpha \Delta ) \pm \sqrt{D}}{2}
\end{equation*}
with their associated eigenvectors $\vect{P_{\pm}}$:
\begin{equation*}
\vect{P_{\pm}} = \left( 1, \frac{\lambda_{\pm} + \gamma}{\beta } \right)^t
\end{equation*}
Let $\matr{P} = \left( \vect{P_{+}} | \vect{P_{-}} \right)$ and $\matr{D}$ be the diagonal matrix of eigenvalues:
\begin{equation*}
\matr{D} = \begin{pmatrix}
    \lambda_+     & 0 \\
	0 & \lambda_- 
\end{pmatrix}
\end{equation*}
\begin{equation*}
\matr{P} = \begin{pmatrix}
    1     & 1 \\
	 \frac{\lambda_{+} + \gamma}{\beta }  &  \frac{\lambda_{-} + \gamma}{\beta } 
\end{pmatrix}
\end{equation*}
We have:
\begin{equation*}
\matr{P^{-1}} = \frac{- \beta}{\sqrt{D}}\begin{pmatrix}
     \frac{\lambda_{-} + \gamma}{\beta }     & -1 \\
	  -\frac{\lambda_{+} + \gamma}{\beta } &  1 
\end{pmatrix}
\end{equation*}
and $\matr{A} = \matr{P} \matr{D} \matr{P^{-1}}$.
Then, system~\eqref{eq:syst_matr_HSC} is equivalent to:
\begin{align*}
\vect{N}(t) & = \matr{P} e^{\matr{D} t} \matr{P^{-1}} \vect{N}(0) \\
& = \begin{pmatrix}
    1     & 1 \\
	 \frac{\lambda_{+} + \gamma}{\beta }  &  \frac{\lambda_{-} + \gamma}{\beta } 
\end{pmatrix} \begin{pmatrix}
    e^{\lambda_+ t}     & 0 \\
	0 & e^{\lambda_- t}
\end{pmatrix} \frac{- \beta}{\sqrt{D}}\begin{pmatrix}
     \frac{\lambda_{-} + \gamma}{\beta }     & -1 \\
	  -\frac{\lambda_{+} + \gamma}{\beta } &  1 
\end{pmatrix} \vect{N}(0)  \\
& = \frac{- \beta}{\sqrt{D}}\begin{pmatrix}
    1     & 1 \\
	 \frac{\lambda_{+} + \gamma}{\beta }  &  \frac{\lambda_{-} + \gamma}{\beta } 
\end{pmatrix} \begin{pmatrix}
     \frac{\lambda_{-} + \gamma}{\beta } e^{\lambda_+ t}    & -e^{\lambda_+ t} \\
	  -\frac{\lambda_{+} + \gamma}{\beta }e^{\lambda_- t}  & e^{\lambda_- t} 
\end{pmatrix} \vect{N}(0) \\
& = \frac{- \beta}{\sqrt{D}}\begin{pmatrix}
       \frac{\lambda_{-} + \gamma}{\beta } e^{\lambda_+ t}   -\frac{\lambda_{+} + \gamma}{\beta }e^{\lambda_- t}  & e^{\lambda_- t}  -e^{\lambda_+ t}  \\
	\frac{ (\lambda_{-} + \gamma) (\lambda_{+} + \gamma)}{\beta^2 } ( e^{\lambda_+ t}  -  e^{\lambda_- t} ) &   \frac{\lambda_{-} + \gamma}{\beta } e^{\lambda_- t} -  \frac{\lambda_{+} + \gamma}{\beta } e^{\lambda_+ t}
\end{pmatrix} \vect{N}(0) 
\end{align*}
Finally, the solutions of~\eqref{eq:syst_matr_HSC} are:
\begin{align}
N_1(t)= -\frac{\beta}{\sqrt{D}} &\left[  \left(\frac{\lambda_{-} + \gamma}{\beta } N_1(0) - N_2(0)\right)e^{\lambda_+ t} - \left(\frac{\lambda_{+} + \gamma}{\beta } N_1(0) - N_2(0)\right)e^{\lambda_- t} \right] \label{eq:N1}\\
N_2(t) = -\frac{\beta}{\sqrt{D}} &\left[ \left( \frac{ (\lambda_{-} + \gamma) (\lambda_{+} + \gamma)}{\beta^2 } N_1(0) -\frac{\lambda_{+} + \gamma}{\beta } N_2(0)\right)e^{\lambda_+ t} \right. \nonumber\\ 
& \left. - \left(\frac{ (\lambda_{-} + \gamma) (\lambda_{+} + \gamma)}{\beta^2 } N_1(0) -\frac{\lambda_{-} + \gamma}{\beta } N_2(0)\right)e^{\lambda_- t} \right] \label{eq:N2}
\end{align}

\subsubsection{Proof that $D>0$}
\label{sec:proof_det_positif}

Let be $\alpha, \beta$, and $\gamma>0$. Let us consider $D$ as a function of $\Delta$:
\begin{align*}
  D \colon \mathbb{R}&\to \mathbb{R}\\
   \Delta &\mapsto(\gamma + \beta - \alpha \Delta )^2 + 4 \gamma \alpha \Delta
\end{align*}
$D$ is continuous and even $C^\infty$. For $\Delta \in  \mathbb{R}$, we have:
\begin{align*}
    D & = (\gamma + \beta - \alpha \Delta )^2 + 4 \gamma \alpha \Delta \\
    &= (\gamma + \beta)^2 + \alpha^2\Delta^2 -2\alpha\Delta(\gamma + \beta) + 4 \gamma \alpha \Delta \\
    &= \alpha^2\Delta^2 +2\alpha \Delta (\gamma - \beta) + (\gamma+\beta)^2
\end{align*}
The discriminant of the previous polynomial (in $\Delta$) is:
\begin{align*}
    \text{Discriminant} & = \frac{4\alpha^2(\gamma-\beta)^2 - 4\alpha^2(\gamma+\beta)^2}{2\alpha^2}\\
    &=2\left[ (\gamma-\beta)^2 -  (\gamma+\beta)^2\right] \\
    &= 2\left[ (\gamma-\beta-\gamma-\beta) (\gamma-\beta+\gamma+\beta)^2\right] \\
    &= 2\left[ -2\beta \times 2 \gamma \right] \\
    &=-8 \beta \gamma < 0
\end{align*}
The discriminant is strictly negative, meaning that $D$ does not cancel on $\mathbb{R}$. Since $D(\Delta = 0) = (\gamma+\beta)^2>0 $, $D$ is strictly positive $\forall \alpha, \beta, \gamma >0$ and $\forall \Delta  \in \mathbb{R}$.
\subsubsection{Solution for immature cells}
We want to derive the solution $N_i(t)$ of the following equation:
\begin{equation}
\frac{dN_i(t)}{dt} = \alpha ( 1-\Delta) \kappa_i N_2(t) - \delta_i N_i(t) 
\label{eq:Ni}
\end{equation}
$N_2(t)$, expressed in \eqref{eq:N2}, is written more concisely:
\begin{equation*}
N_2(t) = A_{1} e^{\lambda_{1} t} +  A_{2} e^{\lambda_{2} t}
\end{equation*}
Equation~\eqref{eq:Ni} can be easily solved using Duhamel's principle, and we obtain:
\begin{equation}
N_i(t) = K_i e^{-\delta_i t} + \alpha ( 1-\Delta) \kappa_i \sum_{k=1}^{2}\left(
A_{k} \frac{e^{\lambda_{k} t}}{\lambda_{k} + \delta_i} 1_{(\lambda_{k} + \delta_i \neq 0)} +  1_{(\lambda_{k} + \delta_i = 0)}  \cdot A_{k} t e^{-\delta_i t} 
\right)
%\label{eq:Ni}
\end{equation}
with $K_i$ found using initial conditions, that is:
\begin{equation*}
K_i = N_i(0) - \alpha ( 1-\Delta) \kappa_i  \sum_{k=1}^{2}\left(
\frac{A_{k} }{\lambda_{k} + \delta_i} 1_{(\lambda_{k} + \delta_i \neq 0)} \right)
\end{equation*}

\subsubsection{Solution for mature cells}
We derive the solution $N_m(t)$ of the following equation:
\begin{equation}
\frac{dN_m(t)}{dt} = \delta_i \kappa_m N_i(t) - \delta_m N_m(t)
\label{eq:Nm}
\end{equation}
For simplicity, we denote:
\begin{equation*}
N_i(t) = B_{1} e^{\lambda_{1} t} +  B_{2} e^{\lambda_{2} t} + B_{3} e^{\lambda_{3} t}+ C t e^{-\delta_i t}
\end{equation*}
with at least one of the constants $B_1, B_2, B_3, C$, equal to 0 (and the only possibility for $C$ to be different from zero would be to have $\delta_i=-\lambda_1$ or $\delta_i=-\lambda_2$; which would be highly improbable when sampling the parameters from their prior distribution). Remember that $\lambda_+ \neq \lambda_-$ because $D>0$.\\
As for the previous case, we use Duhamel's method, and we get:
\begin{align}
N_m(t) = & K_m e^{-\delta_m t} \nonumber\\
& +  \delta_i \kappa_m \sum_{k=1}^{3}\left(
B_{k} \frac{e^{\lambda_{k} t}}{\lambda_{k} + \delta_m} 1_{(\lambda_{k} + \delta_m \neq 0)} +  1_{(\lambda_{k} + \delta_m = 0)}  \cdot B_{k} t e^{-\delta_m t} 
\right) \\
& + C \delta_i \kappa_m \left( \frac{1}{\delta_m - \delta_i}\left(t-\frac{1}{\delta_m - \delta_i}\right) 1_{( \delta_m \neq \delta_i)} + \frac{t^2}{2} 1_{( \delta_m =\delta_i)} 
\right) e^{-\delta_i t} \nonumber
\end{align}
with $K_m$ found thanks to the initial conditions:
\begin{equation*}
K_m = N_m(0) -\delta_i \kappa_m  \sum_{k=1}^{3}\left(
\frac{B_{k} }{\lambda_{k} + \delta_m} 1_{(\lambda_{k} + \delta_m \neq 0)} \right) + C \delta_i \kappa_m \frac{1}{(\delta_m - \delta_i)^2} 1_{\delta_m \neq \delta_i}
\end{equation*}

\subsubsection{Summary of the analytical solution}
\label{sec:as_sol}

For a given set of parameters, the solution of the system~\eqref{eq:syst_comp} is:
\begin{align*}
N_1(t) &= -\frac{\beta}{\sqrt{D}} \left[  \left(\frac{\lambda_{-} + \gamma}{\beta } N_1(0) - N_2(0)\right)e^{\lambda_+ t} - \left(\frac{\lambda_{+} + \gamma}{\beta } N_1(0) - N_2(0)\right)e^{\lambda_- t} \right] \\
N_2(t) &= A_{+} e^{\lambda_{+} t} +  A_{-} e^{\lambda_{-} t}
\end{align*}
with:
\begin{align*}
A_+ &= -\frac{\beta}{\sqrt{D}} \left[ \left( \frac{ (\lambda_{-} + \gamma) (\lambda_{+} + \gamma)}{\beta^2 } N_1(0) -\frac{\lambda_{+} + \gamma}{\beta } N_2(0)\right) \right] \\
A_- &= +\frac{\beta}{\sqrt{D}} \left[  \left(\frac{ (\lambda_{-} + \gamma) (\lambda_{+} + \gamma)}{\beta^2 } N_1(0) -\frac{\lambda_{-} + \gamma}{\beta } N_2(0)\right) \right]
\end{align*}
and:
\begin{equation*}
N_i(t) = K_i e^{-\delta_i t} + \alpha ( 1-\Delta) \kappa_i \left(
A_{+} \frac{e^{\lambda_{+} t}}{\lambda_{+} + \delta_i}  + A_{-} \frac{e^{\lambda_{-} t}}{\lambda_{-} + \delta_i}  
\right)
\end{equation*}
with:
\begin{equation*}
K_i = N_i(0) - \alpha ( 1-\Delta) \kappa_i  \left(
\frac{A_{+} }{\lambda_{+} + \delta_i}  + \frac{A_{-} }{\lambda_{-} + \delta_i} \right)
\end{equation*}
and finally:
\begin{align*}
N_m(t) =  K_m e^{-\delta_m t}  +  \delta_i \kappa_m \left[ K_i \frac{e^{-\delta_i t}}{\delta_m-\delta_i} +   \alpha ( 1-\Delta) \kappa_i\left(
   \frac{A_{+}}{\lambda_+ + \delta_i} \frac{e^{\lambda_{+} t}}{\lambda_{+} + \delta_m}   + \frac{A_{-}}{\lambda_- + \delta_i} \frac{e^{\lambda_{-} t}}{\lambda_{-} + \delta_m}  
\right) \right]
\end{align*}
with:
\begin{equation*}
K_m = N_m(0) -\delta_i \kappa_m \left[\frac{ K_i }{\delta_m-\delta_i} +   \alpha ( 1-\Delta) \kappa_i\left(
   \frac{A_{+}}{ (\lambda_+ + \delta_i)(\lambda_{+} + \delta_m)}   + \frac{A_{-}}{(\lambda_- + \delta_i)(\lambda_{-} + \delta_m)}  
\right) \right]
\end{equation*}

\subsection{Normalization, initial conditions, and simplification}

Data from Mosca et al.~\cite{mosca2021} do not provide information about absolute values for quantities of wt, het, or hom cells separately (that are solutions of the ODE systems), but rather their relative proportions (for more details, see Appendix~\ref{sec:observation_model}).
To take this into account, we consider as outputs of our model no longer the numbers of cells but the proportions of immature heterozygous cells (and similarly for hom cells):
\begin{equation*}
    z_{het}(t)=\frac{N_{i,het}(t)}{N_i(t)+N_{i,het}(t)+N_{i,hom}(t)}
\end{equation*}
 as well as the mature variant allele frequency (VAF) among granulocytes:
 \begin{equation*}
y(t) =\frac{0.5 \cdot N_{m,het}(t) + N_{m,hom}(t)}{N_{m}(t)+N_{m,het}(t)+N_{m,hom}(t)}     
 \end{equation*}

Following the idea of Michor et al.~\cite{michor2005dynamics}, it is considered that IFN$\alpha$ acts by modifying the values of some parameters in the model.
Time $t=0$ corresponds to the beginning of the treatment. Before that time, equations~\eqref{eq:syst_comp} are still valid, but homeostatic conditions are assumed to be satisfied, i.e., the system is in a quasi-stationary state. Homeostatic conditions are, of course,  verified for wt cells as soon as $\Delta = 0$, leading to the following initial conditions: 
\begin{align*}
    N_{1}(0) &=\frac{\beta}{\beta + \gamma}N_{HSC} \\
    N_{2}(0) & =  \frac{\gamma}{\beta + \gamma}N_{HSC} \\
    N_{i}(0) & =\frac{\kappa_i \alpha}{\delta_i} N_{2}(0) \\
    N_{m}(0) & = \frac{\kappa_m \delta_i}{\delta_{m}} N_{i}(0)
\end{align*}
where $N_{HSC}$, the total wild-type HSC number, is considered constant. 
For mutated cells, the \rev{baseline} model assumes that $\Delta_{het}  \approx \Delta_{hom} \approx 0^+$. We also introduce $\eta_{het} = \frac{N_{1,het}(0) + N_{2,het}(0)}{N_{HSC} }$ and $\chi_{het} = \frac{N_{2,het}(0)}{N_{1,het}(0) + N_{2,het}(0) }$ for expressing the initial conditions for het and hom cells.
From $t=0$, patients are under treatment. IFN$\alpha$ is assumed to modify the values of some parameters, potentially in different ways depending on the cell type. In terms of notation, the superscript $^*$ is added to  parameters impacted by the drug. 
From $t\geq0$, eq.~\eqref{eq:syst_comp} remains valid with new parameters, and an equilibrium shift induces new dynamics.\\

Given the solutions of the ODE systems and the way we normalize them to obtain ratios, we get some simplifications. Parameter $N_{HSC}$ is not relevant anymore: no matter its value, it will not change the final output of the model (when considering the ratios $z_{het}, z_{hom}$, and $y$).
By introducing $k_{i,het}$ and $k_{i,hom}$ such that $k_{i,het} = \kappa_{i,het} / \kappa_i$ and $k_{i,hom} = \kappa_{i,hom} / \kappa_i$, parameter $\kappa_i$ no longer needs to be estimated. The same goes for $\kappa_m$.\\

Based on prior biological knowledge, many other assumptions are detailed and justified by Mosca et al.~\cite{mosca2021} to decrease the number of parameters to estimate for each patient. These assumptions are summarized in table \rev{1 (main text)}. For each patient, we end up with 7 parameters to estimate for the \rev{baseline} model. To verify if these assumptions are sufficient to ensure the identifiability of the model, and before further extending the model with potentially more parameters to estimate, we generated virtual data (§~\ref{sec:virtual_data}) and tried to retrieve the model parameters that generated them (§~\ref{sec:res_identif}).

\FloatBarrier
\newpage

\newpage
\section{Data and observation model}
\label{sec:observation_model}

Data $\mathcal{D}=\{\mathcal{D}_i\}_{i \in \{1,\cdots, N\}}$ come from Mosca et al.~\cite{mosca2021}. From the cohort of patients of Mosca et al., we consider $N=19$ MPN patients: those for which there are enough observations, and only those having the mutation \jakvf. Identical patient IDs than in~\cite{mosca2021} are used (when displaying the results). 
For a patient $i$, the data consist of observations at different times $t_k^{(i)}$, from the beginning of the therapy ($t=0$), where Variant Allele Frequencies (VAF) $\hat{y}_k^{(i)}$ among mature cells and the clonal architecture among progenitor cells $(\hat{n}_{k}^{(i)},\hat{n}_{k,het}^{(i)},\hat{n}_{k,hom}^{(i)})$ were measured. $N_k^{(i)} = \hat{n}_{k}^{(i)} + \hat{n}_{k,het}^{(i)} + \hat{n}_{k,hom}^{(i)}$ corresponds to the number of progenitor cells sampled and, consequently, the number of colonies genotyped at time $t_k^{(i)}$.\\

In Tab.~\ref{tab:info_patients}, we provide some summary statistics concerning the observations and doses of IFN$\alpha$ that have been administered.\\

\begin{table}[b]
    \centering
    \footnotesize 
    \begin{tabular}{|c|c|c|c|c|c|c|c|c|c|c|c|c|c|}
    \hline
    \multicolumn{3}{|c|}{ID} & \multicolumn{5}{c|}{Observation times}  & Mean &\multicolumn{5}{c|}{Information about the posology (over 3,000 days)} \\ \cline{1-8}\cline{9-14}
    %%%%%
         \multirow{2}{*}{$i$}	&	\multirow{2}{*}{label} &	\rev{MPN} 	&	Nb. of 		&	First \rev{obs.} & \multicolumn{3}{c|}{\rev{Sampling period [days]}}	&	dose over & \rev{Max.}	&	\rev{Min.}	&	Mean	&	Standard	&	\multirow{2}{*}{COV}		\\  \cline{6-8}
         %%%
         & & \rev{type}&\rev{obs.} & time [days] & \rev{Mean} & \rev{Min} & \rev{Max} & 450 days & dose  & dose 	&	dose 	&	deviation	&	 
         \\
         \hline
         %%%%
1	&	\#2	& \rev{PV} &	13	&	0	& \rev{163} & \rev{104} & \rev{473}&	67.5	&	67.5	&	9.64	&	20.71	&	22.14	&	1.07	\\ \hline
2	&	\#3	&\rev{ET} &	8	&	0	& \rev{163} & \rev{81} & \rev{313}&	120.4	&	135.0	&	33.75	&	58.93	&	39.74	&	0.67	\\ \hline
3	&	\#6	&\rev{ET} &	13	&	134 & \rev{169} & \rev{84} & \rev{314}	&	13.5	&	13.5	&	0.0	&	9.78	&	6.01	&	0.61	\\ \hline
4	&	\#8	&\rev{PV} &	13	&	40 & \rev{166} & \rev{64 } & \rev{517}	&	57.7	&	67.5	&	0.0	&	14.76	&	21.15	&	1.43	\\ \hline
5	&	\#11	&\rev{PV} &	12	&	1	& \rev{174 } & \rev{90 } & \rev{413} &	96.8	&	135.0	&	27.0	&	48.29	&	40.26	&	0.83	\\ \hline
6	&	\#12	&\rev{PV} &	10	&	0 & \rev{102 } & \rev{63 } & \rev{161}	&	113.25	&	135.0	&	27.0	&	45.4	&	32.84	&	0.72	\\ \hline
7	&	\#15	&\rev{PV} &	14	&	349 & \rev{106 } & \rev{44 } & \rev{172}	&	157.47	&	180.0	&	0.0	&	38.35	&	65.13	&	1.7	\\ \hline
8	&	\#16	&	\rev{PV} &12	&	0	& \rev{132 } & \rev{91 } & \rev{194} &	126.5	&	135.0	&	0.0	&	50.22	&	53.73	&	1.07	\\ \hline
9	&	\#18	&\rev{PMF} &	14	&	0	& \rev{98 } & \rev{28 } & \rev{165} &	35.17	&	67.5	&	22.5	&	24.85	&	5.48	&	0.22	\\ \hline
10	&	\#19	&\rev{PV} &	12	&	390 & \rev{171 } & \rev{89 } & \rev{422}	&	135.0	&	135.0	&	22.5	&	58.44	&	49.45	&	0.85	\\ \hline
11	&	\#20	&\rev{PV} &	12	&	0 & \rev{168 } & \rev{78 } & \rev{342}	&	73.45	&	90.0	&	0.0	&	21.43	&	29.43	&	1.37	\\ \hline
12	&	\#22	&\rev{PV} &	7	&	0	& \rev{187 } & \rev{50 } & \rev{546} &	81.1	&	90.0	&	45.0	&	88.66	&	7.59	&	0.09	\\ \hline
13	&	\#23	&\rev{PV} &	14	&	0 & \rev{187 } & \rev{63 } & \rev{505}	&	66.05	&	90.0	&	15.0	&	47.61	&	18.7	&	0.39	\\ \hline
14	&	\#25	&\rev{PV} &	15	&	0 & \rev{139 } & \rev{77 } & \rev{377}	&	130.8	&	135.0	&	16.87	&	49.67	&	45.9	&	0.92	\\ \hline
15	&	\#27	&\rev{ET} &	12	&	273	 & \rev{157 } & \rev{105 } & \rev{454} &	95.51	&	135.0	&	45.0	&	108.89	&	27.69	&	0.25	\\ \hline
16	&	\#29	&\rev{ET} &	15	&	26 & \rev{149 } & \rev{83 } & \rev{427}	&	95.34	&	135.0	&	16.87	&	33.42	&	30.22	&	0.9	\\ \hline
17	&	\#30	&\rev{ET} &	13	&	184 & \rev{149 } & \rev{103 } & \rev{362}	&	102.75	&	135.0	&	11.25	&	29.51	&	36.32	&	1.23	\\ \hline
18	&	\#31	&\rev{PV} &	10	&	0	 & \rev{234 } & \rev{126 } & \rev{603} &	151.5	&	180.0	&	9.64	&	50.19	&	51.86	&	1.03	\\ \hline
19	&	\#32	&\rev{PV} &	13	&	1	& \rev{172 } & \rev{77 } & \rev{620} &	130.3	&	180.0	&	33.75	&	88.77	&	59.58	&	0.67	\\ \hline
    \end{tabular}
    
    \caption{Some characteristics associated to the patients from Mosca et al.~\cite{mosca2021}. For each of these patients (the $i^{th}$ patient is assigned to a label corresponding to the one used in the original paper), we give some information concerning \rev{the MPN type (ET for Essential Thrombocythemia, PV for Polycythemia Vera, and PMF for Primary Myelofibrosis),} the observations, and the posology. The number of observations indicated in the table corresponds to the number of VAF measurements. The first observation time corresponds to the time - from the start of the IFN$\alpha$ therapy - when the first measurement (either a VAF measurement or a clonal architecture) has been made. The value 0 indicates that initial observations have been made at or before the start of the therapy. 
    \rev{The sampling period refers to the time between two VAF measurements. We give the mean, minimal and maximal time between two measurements for each patient.}
    We indicate the mean dose administered over the first 450 days of the therapy (in $\mu$g/week). Concerning the other information about the posology, we give some summary statistics concerning the IFN$\alpha$ doses received over the first 3,000 days of therapy: the maximal, minimal, mean dose, the standard deviation (each expressed in $\mu$g/week), as well as the coefficient of variation (COV) corresponding to the standard deviation divided by the mean.
    }
    \label{tab:info_patients}
\end{table}

\begin{figure}[h]
\centering
\includegraphics[width=0.8\textwidth]{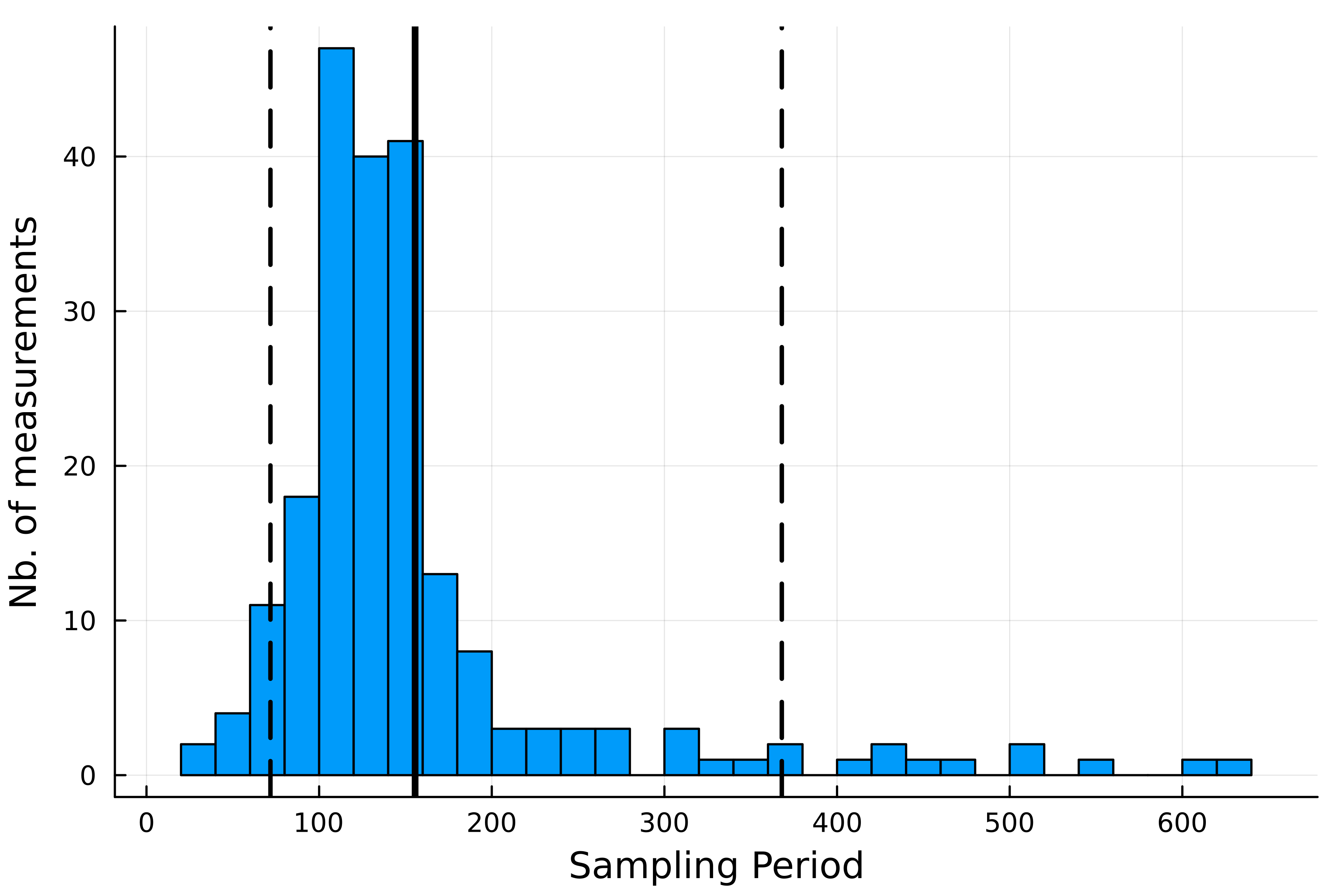}
\caption{\rev{Distribution of the sampling periods (time between two VAF measurements) for the whole cohort, in days. The vertical line corresponds to the mean (156 days) and the dashed lines to the 5 and 95 percentiles.}}
\label{fig:hist_sampling}
\end{figure}

\begin{figure}[h]
\centering
\includegraphics[width=0.9\textwidth]{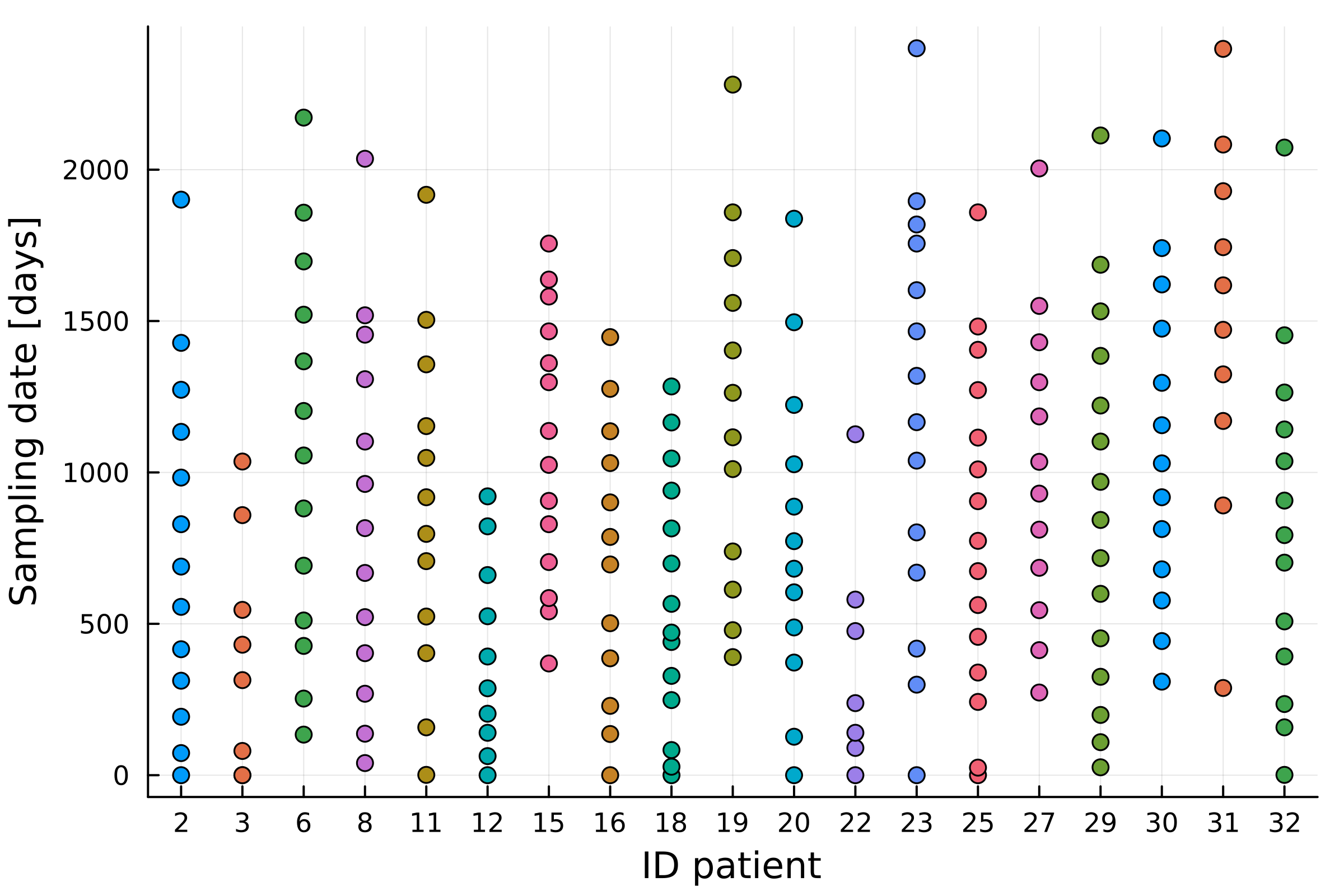}
\caption{\rev{For each patient (x-axis), we indicate all the sampling times (for a VAF measurement) in days from the beginning of the therapy ($t=0$ days).}}
\label{fig:sampling_dates_per_patient}
\end{figure}

\rev{In Fig.~\ref{fig:hist_sampling} and~\ref{fig:sampling_dates_per_patient}, we show some information regarding the sampling period or the dates at which we have observations.}\\


In the following, we will omit the patient index $i$ for clarity.\\
We use the same observation model as Mosca et al.~\cite{mosca2021}. For mature cells, it is assumed a Gaussian noise with a variance that depends on the true VAF $y_k \in [0,1]$ at time $t_k$:
\begin{equation}
\label{eq:noise_VAF}
\hat{y}_{k} \mid y_{k} \sim \mathcal{N}\left(y_{k}, y_{k}\left(1-y_{k}\right) \sigma_{m}^{2}
\right)
\end{equation}
with $\sigma_m$ to be estimated.\\
Concerning the observations among immature cells, we observe some counts of wt, het, and hom progenitor cells at time $t_k$, respectively denoted by $\hat{n}_{k}$, $\hat{n}_{k,het}$ and $\hat{n}_{k,hom}$. These cells are assumed to be randomly sampled from an unknown but very large set of immature cells, so that a multinomial distribution might model the uncertainty, as it was also done in~\cite{catlin2001statistical}:
\begin{equation}
\label{eq:noise_CF}
\mathbb{P}\left[\hat{n}_{k}=n_{1}, \hat{n}_{k,  het }=n_{2}, \hat{n}_{k,  hom }=n_{3} \mid z_{k,  het}, z_{k, hom}\right]=\frac{\left(n_{1}+n_{2}+n_{3}\right) !}{n_{1} ! n_{2} ! n_{3} !} z_{k}^{n_{1}} z_{k, het}^{n_{2}} z_{k,hom}^{n_{\mathrm{3}}}
\end{equation}
where $z_k$, $z_{k, het}$, and $z_{k,hom}$ are the true CF for wt, het, and hom progenitor cells respectively (with $z_k = 1 - z_{k, het} - z_{k,hom}$).

\FloatBarrier
\newpage
\section{Practical Identifiability of the \rev{baseline} model}
\label{sec:identif}

\subsection{Generating virtual data}
\label{sec:virtual_data}

To verify the identifiability of the \rev{baseline} model, we generate 30 virtual patients. We sample 30 parameter vectors (from a chosen population distribution), leading to 30 different dynamics. For each of them, we sample observation times and add some noise to reproduce the data from the cohort of Mosca et al.~\cite{mosca2021}. \\
We only consider 19 patients in this work, yet the number of \jakvf \rev{MPN} patients in the original cohort was higher, but many patients had only a few data points. We also consider such patients with little information in the virtual data generated here (see fig.~\ref{fig:distr_data}).\\

Five parameter values and two initial conditions are required to generate one patient's virtual dynamics (see tab.\rev{ 1 in the main text}).
The population distributions used to generate the parameters are the following:
\begin{itemize}
\item $\eta_{het} \sim \mathcal{U}([0,1])$ (uniform)
\item $\eta_{hom} \sim \mathcal{U}([0,2.5])$ (uniform)
\item $\Delta^*_{het}  \sim \mathcal{N}(-0.1,0.1^2)$ (truncated over $[-1,1]$)
\item $\Delta_{hom}^* \sim \mathcal{N}(-0.3,0.1^2)$ (truncated over $[-1,1]$)
\item $\gamma_{het}^* \sim \mathcal{N}(1/40,0.01^2)$  (truncated over $[1/300, 1/10]$)
\item $\gamma_{hom}^* \sim \mathcal{N}(1/120,0.01^2)$  (truncated over $[1/300, 1/10]$)
\item $k_{m,het} \sim \mathcal{N}(10,1)$  (truncated over $[1, 20]$)\\
\end{itemize}

\begin{figure}[p]
\centering
\includegraphics[width=1\textwidth]{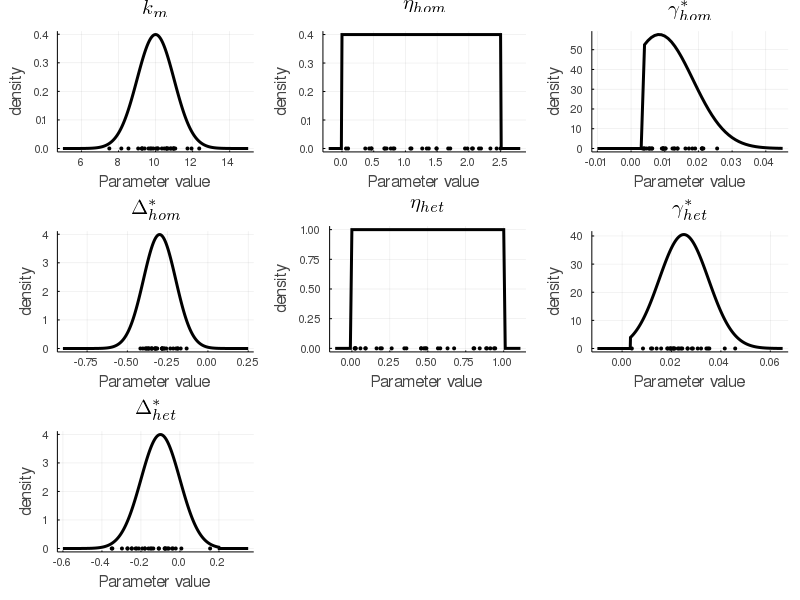}
\caption{Generated parameters for 30 virtual patients. Points are sampled parameters. The line represents the actual population density, from which we have sampled values.}
\label{fig:generation_params}
\end{figure}

\begin{figure}[p]
\centering
\includegraphics[width=0.5\textwidth]{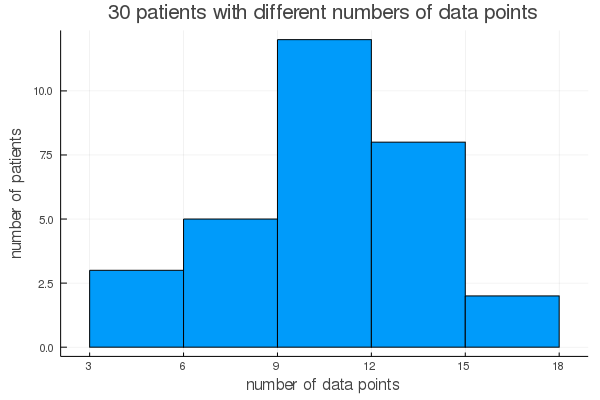}
\caption{Distribution of the number of data points $n$ sampled among our 30 virtual patients, from 3 to 15 observation times, with a mean value of 10, consistent with the cohort from Mosca et al.~\cite{mosca2021}.}
\label{fig:distr_data}
\end{figure}

In figure~\ref{fig:generation_params}, we show all parameters that have been sampled and the corresponding population distributions.\\

Now that we have the true dynamics for 30 virtual patients (since our model is deterministic), we want to generate noisy data for each of them. Not all patients will have the same amount of data $n_i$. Indeed, to be consistent with the dataset from Mosca et al.~\cite{mosca2021}, we generate a sparse data set, as depicted in figure~\ref{fig:distr_data}. This figure shows the distribution of the number of observation times for the virtual cohort. The mean number of observations is equal to 10 (the same as in the cohort of Mosca et al.), with some virtual patients with only 3 data points while others have 15 data points.\\

Then, we can generate a number $n_i$ of noisy observations from each virtual patient's actual dynamics $i$. First, we generate the times (in days) when observations are made. We always consider that the first measure is made at the initial time. Then, we generate the others randomly according to the following method:
\begin{verbatim}
function sample_time(n)
    t = zeros(n)
    t[1] = 0.0
    for i in 2:n
        t[i] = t[i-1] + rand(truncated(Normal(100, 10),30, 300))
    end
    return t
end
\end{verbatim}
to consider observations about each trimester.\\
Then, from the actual ratios for immature and mature cells at this time, we add some noise. We use the same noise model as described in Appendix~\ref{sec:observation_model}.\\

At this stage, we have 30 datasets  comparable to those from the cohort of Mosca et al.\\
The objective is thus to retrieve the actual patient's dynamics using our inference method.

\subsection{Model calibration}
\label{sec:identif_param_estim}

\subsubsection{Posterior distribution}
The model identification method used to retrieve the parameters of all patients and the population parameters (called hyper-parameters) is the same as the one used by Mosca et al.~\cite{mosca2021}.\\
 $\vect{\theta}=\left\{\vect{\theta}^{(i)}\right\}_{i \in \{1,\cdots,N\}}$ denotes the set of all patient-related model parameters with:
\begin{equation*}
\begin{array}{c}
\vect{\theta}^{(1)}=\left(\theta_{1}^{(1)}, \cdots, \theta_{P}^{(1)}\right) \\
\vdots \\
\vect{\theta}^{(N)}=\left(\theta_{1}^{(N)}, \cdots, \theta_{P}^{(N)}\right)
\end{array}
\end{equation*}
where $P=7$ is the number of parameters to estimate. With the hierarchical inference method, instead of estimating each $\vect{\theta}^{(i)}$ independently, we introduce some hyper-parameters (HP) $\vect{\tau}=\left(\tau_{1}, \cdots, \tau_{P}\right)$ and $\vect{\sigma^2}=\left(\sigma_{1}^{2}, \cdots, \sigma_{P}^{2}\right)$ so that, \textit{a priori}:
\begin{equation*}
\forall i \in \{1,\cdots,N\}, \forall k \in\{1, \cdots, P\}, \theta_{k}^{(i)} \mid \tau_{k}, \sigma_{k}^{2} \sim \mathcal{N}_{c, k}\left(\tau_{k}, \sigma_{k}^{2}\right)
\end{equation*}
where $\mathcal{N}_{c, k}$ is defined as a truncated (over a range that depends on the considered parameter $k$, as described in section~\ref{sec:virtual_data}) Gaussian distribution. 
Note that, for parameters $\eta_{het}$ and $\eta_{hom}$, we do not consider hyper-parameters since there is no reason to have initial conditions sampled from population distributions.
Now, we can estimate the joint posterior distributions of $\vect{\theta}^{(1)}, \cdots, \vect{\theta}^{(N)}$ and hyperparameters $\vect{\tau}$ and $\vect{\sigma^2}$:
\begin{align}
\label{eq:suppl_posterior}
\mathbb{P}\left[\vect{\theta}^{(1)}, \cdots, \vect {\theta}^{(N)}, \vect{\tau}, \vect {\sigma^{2}} \mid \mathcal{D}\right] & \propto \mathbb{P}\left[\mathcal{D} \mid \vect{\theta}^{(1)}, \cdots, \vect {\theta}^{(N)}, \vect{\tau}, \vect{\sigma^{2}}\right]\mathbb{P}\left[\vect{\theta}^{(1)}, \cdots, \vect {\theta}^{(N)}, \vect{\tau}, \vect{\sigma^{2}} \right] \nonumber  \\
& \propto \mathbb{P}\left[\mathcal{D} \mid \vect{\theta}^{(1)}, \cdots, \vect {\theta}^{(N)}\right]\mathbb{P}\left[\vect{\theta}^{(1)}, \cdots, \vect {\theta}^{(N)}, \vect{\tau}, \vect{\sigma^{2}} \right] \nonumber  \\
&\propto \prod_{i \in \{1, \cdots, N\}}\left(\mathbb{P}\left[\mathcal{D}_{i} \mid \vect{\theta}^{(i)}\right] \right) \mathbb{P}\left[\vect{\theta}^{(1)}, \cdots, \vect {\theta}^{(N)} \mid \vect{\tau}, \vect{\sigma^{2}} \right] \mathbb{P}\left[ \vect{\tau}, \vect{\sigma^{2}} \right]\nonumber \\
& \propto \prod_{i \in \{1, \cdots, N\}}\left(\mathbb{P}\left[\mathcal{D}_{i} \mid \vect{\theta}^{(i)}\right]\mathbb{P}\left[\vect{\theta}^{(i)} \mid \vect{\tau}, \vect{\sigma^{2}}\right]\right)\mathbb{P}[\vect{\tau}]\mathbb{P}\left[\vect{\sigma^{2}}\right] 
\end{align}
The previous relation is obtained by considering independence between patients, conditionally on the hyper-parameter values. 
In addition, we further simplify the relation above by assuming that the components of our parameter and hyper-parameter vectors are independent. We get for patient $i$:
\begin{equation*}
\mathbb{P}\left[\vect{\theta}^{(i)} \mid \tau, \sigma^{2}\right]=\prod_{k \in\{1, \cdots, P\}}\mathbb{P}\left[\theta_{k}^{(i)} \mid \tau_{k}, \sigma_{k}^{2}\right]
\end{equation*}
and:
\begin{equation*}
\mathbb{P}[\vect{\tau}]=\prod_{k \in\{1, \cdots, P\}}\mathbb{P}\left[\tau_{k}\right]
\end{equation*}
\begin{equation*}
\mathbb{P}\left[\vect{\sigma^{2}}\right]=\prod_{k \in\{1, \cdots, P\}}\mathbb{P}\left[\sigma_{k}^{2}\right]
\end{equation*}
Likelihood $\mathbb{P}\left[\mathcal{D}_{i} \mid \vect{\theta}^{(i)}\right]$ is expressed according to the observation model described in Appendix~\ref{sec:observation_model}.

\subsubsection{Conditional laws of the hyper-parameters}
The resulting posterior distribution is approximated numerically using the Metropolis-Hasting within Gibbs algorithm. Before using this computational method, it is useful to express the HP conditional posterior distributions.\\
Concerning hyper-parameter $\vect{\tau}$, for $k \in \{1, \cdots, P\}$, by keeping in~\eqref{eq:suppl_posterior} only the terms involving $\tau_i$ (since we are here only interested in its distribution ignoring a multiplication factor), we get:
\begin{equation*}
\mathbb{P}\left[\tau_k | \vect{\theta^{(1)}}, \cdots, \vect{\theta^{(N)}}, \vect{\sigma^2}, \mathcal{D} \right] \propto \mathbb{P}\left[ \theta_k^{(1)} | \tau_k, \sigma_k^2 \right]\cdots \mathbb{P}\left[ \theta_k^{(N)} | \tau_k, \sigma_k^2 \right] \mathbb{P}\left[ \tau_k \right]
\end{equation*}
Prior distributions for $\tau_k$ are chosen uniform over intervals $[a_k, b_k]$ with the same upper and lower bounds that the ones used to truncate the Gaussian law for $\theta_k | \tau_k, \sigma^2_k$.
Thus, we derive the (conditional posterior) probability density function $f_{\tau_k}$ for $t\in\mathbb{R}$:
\begin{align*}
f_{\tau_k}(t) & \propto\frac{1}{\sqrt{2\pi \sigma^2_k}} \exp\left( -\frac{1}{2\sigma_k^2 }( \theta_k^{(1)} - t)^2 \right)\cdots \frac{1}{\sqrt{2\pi \sigma^2_k}} \exp\left( -\frac{1}{2\sigma_k^2 }( \theta_k^{(N)} - t)^2 \right) 1_{[a_k, b_k]}(t)\\
&\propto \exp\left( -\frac{1}{2\sigma_k^2}\left((\theta_k^{(1)} - t)^2 + \cdots + ( \theta_k^{(N)} - t)^2  \right) \right)1_{[a_k, b_k]}(t)
\end{align*}
Since we have: 
\begin{align*}
 (\theta_k^{(1)} - t)^2 + \cdots + ( \theta_k^{(N)} - t)^2 & = \sum_{i \in \{1,\cdots,N\}} (\theta_k^{(i)})^2 + N t^2 - 2 t \sum_{i \in \{1,\cdots,N\}} \theta_k^{(i)}\\
 &= N(t -  \frac{1}{N}\sum_{i \in \{1,\cdots,N\}} \theta_k^{(i)} )^2 + \cdots
\end{align*}
where symbol $\cdots$ indicate terms that do not involve $t$, we finally get that:
\begin{equation}
f_{\tau_k}(t) \propto\exp\left( -\frac{N}{2\sigma_k^2 }  \left(t - \frac{1}{N}\sum_{i \in \{1,\cdots,N\}} \theta_k^{(i)}\right)^2 \right) 1_{[a_k, b_k]}(t)
\end{equation}
Thus, we deduce that $\tau_k$ (more precisely, its conditional posterior distribution) follows a Gaussian law that is truncated over interval $[a_k, b_k]$: 
\begin{equation}
\tau_k \sim \mathcal{N}_c(\frac{1}{N}\sum_{i \in \{1,\cdots,N\}} \theta_k^{(i)}, \frac{\sigma_k^2}{N})
\end{equation}

Concerning hyper-parameter $\sigma^2_k$ (for $k\in\{1,\cdots,P\}$), we do the same as for hyper-parameter $\tau_k$ to express $\sigma^2_k |  \vect{\theta^{(1)}}, \cdots, \vect{\theta^{(N)}}, \vect{\tau},  \mathcal{D} $ and find for its probability density function, for $t\in\mathbb{R}$:
\begin{equation*}
f_{\sigma_k^2}(t) \propto \left(\frac{1}{t}\right)^{N/2+1} \exp\left(-\frac{1}{2t} \sum_{i \in \{1,\cdots,N\}} \left( \theta_k^{(i)} - \tau_k \right)^2 \right) 1_{[a'_k, b'_k]}(t)
\end{equation*}
where we considered that $\sigma^2_k$ followed an improper prior distribution - namely an inverse-gamma (0,0) law - truncated over $[a'_k, b'_k]$ (we choose $a'_k=0$ and $b'_k=2$).
We recognise the expression of a (truncated) inverse-gamma law. As a reminder, a random variable $X$ that follows an inverse-gamma law of parameters $\alpha, \beta$ has for its density, for $x\in \mathbb{R}$:
\begin{equation*}
f_X(x)=\frac{\beta^{\alpha}}{\Gamma(\alpha)}(1 / x)^{\alpha+1} \exp (-\beta / x)
\end{equation*}
Then, $\sigma^2_k |  \vect{\theta^{(1)}}, \cdots, \vect{\theta^{(N)}}, \vect{\tau},  \mathcal{D} $ follows a (truncated) inverse-gamma law with parameters $\alpha_k = \alpha = N/2$ and $\beta_k = 1/2 \sum_{i\in\{1,\cdots,N\}}\left(\theta_k^{(i)} - \tau_k\right)^2$.\\

\subsubsection{Metropolis within Gibbs algorithm}

The previous joint posterior distribution~\eqref{eq:suppl_posterior} is estimated by generating a Markov-Chain Monte-Carlo (MCMC). We use the Metropolis within Gibbs algorithm, which allows us to iteratively sample values for parameters $\vect{\theta}^{(1)}, \cdots, \vect{\theta}^{(N)} $ and hyper-parameters $\vect{\tau}$ and $\vect{\sigma^2}$.
After initializing the MCMC chain, we successively sample HP values and then parameter values for each patient (independently, conditionally on the HP values).\\

To initialize the parameter values for each patient, we first run an optimization algorithm - namely the CMA-ES algorithm - to find the parameter vector that maximizes the likelihood. The CMA-ES algorithm also provides a covariance matrix that we will use for the proposal in the Metropolis-Hasting scheme.
HPs are initialized randomly.\\

We use the Gibbs method for sampling values for hyper-parameters (component-wise). The proposal of the Metropolis algorithm is the conditional posterior law (that we made explicit in the previous section). With this choice, the new sample is always accepted. \\

Then, conditionally on the HP values previously sampled, we sample parameter values for each patient, one by one.
This is achieved using a standard Metropolis-Hasting scheme, with a Multivariate Gaussian law for the proposal. As a setting of the algorithm, a proper choice for the covariance matrix has to be found. 
We choose to set the proposal's covariance matrix to the one estimated by the CMA-ES algorithm (adjusted by a multiplication factor tuned to get suitable acceptance rates).\\
The algorithm pursues a huge number of iterations until achieving convergence (when the ergodic means do not vary anymore).\\

Model calibration was achieved by implementing the previous methods in the Julia programming language.  \\
The codes used for this work are available.
The framework used for parameter estimation (and that can be used for a wide range of problems) is available at: 
\begin{verbatim}
https://gitlab-research.centralesupelec.fr/2012hermangeg/bayesian-inference
\end{verbatim}
The implementation of the \rev{baseline} model and the settings to use the previous inference framework is available at:
\begin{verbatim}
https://gitlab-research.centralesupelec.fr/2012hermangeg/identifiability-base-model
\end{verbatim}

\FloatBarrier

\subsection{Results of the identifiability study}
\label{sec:res_identif}

\begin{figure}
\centering
\includegraphics[width=0.7\textwidth]{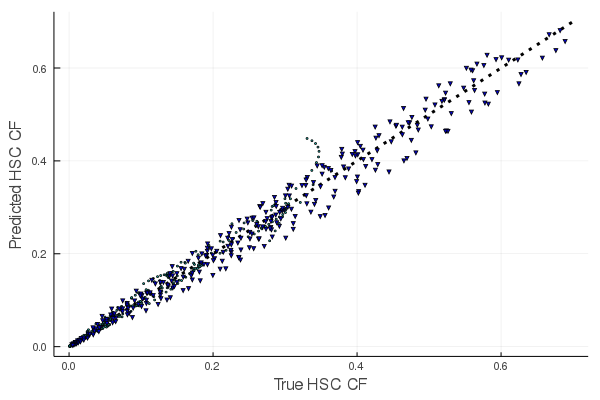}
\caption{Comparison of the true HSC clonal fractions and the inferred ones. The values are taken at the observation time from our virtual data set. Blue triangles refer to homozygous HSCs, and green circles to heterozygous HSCs.}
\label{fig:fits_HSC}
\end{figure}

\begin{figure}[]
    \centering
    \begin{subfigure}[b]{0.19\textwidth}
        \includegraphics[width=\textwidth]{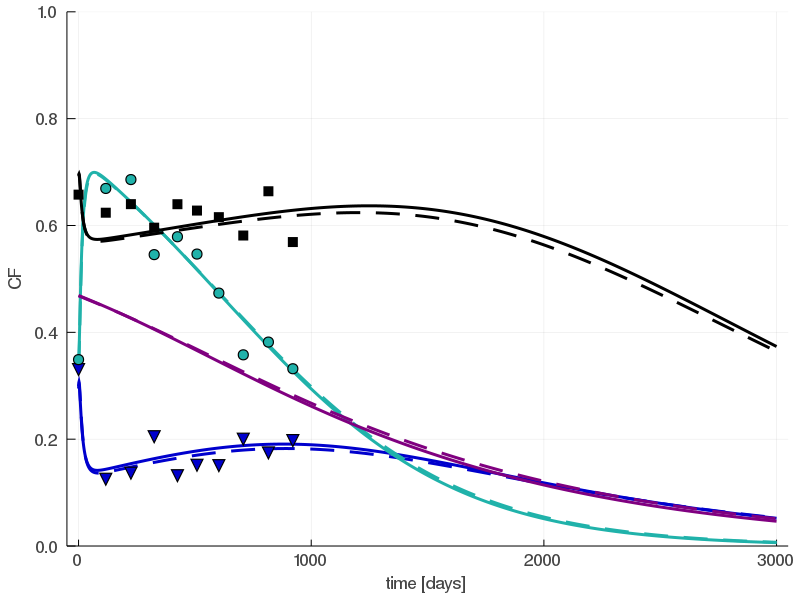}
    \end{subfigure}
	\begin{subfigure}[b]{0.19\textwidth}
        \includegraphics[width=\textwidth]{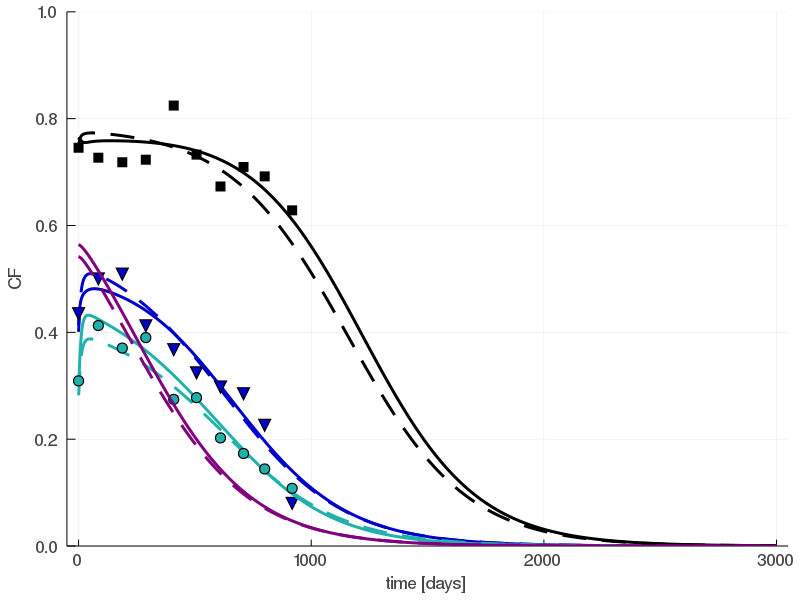}
    \end{subfigure}
	 \begin{subfigure}[b]{0.19\textwidth}
        \includegraphics[width=\textwidth]{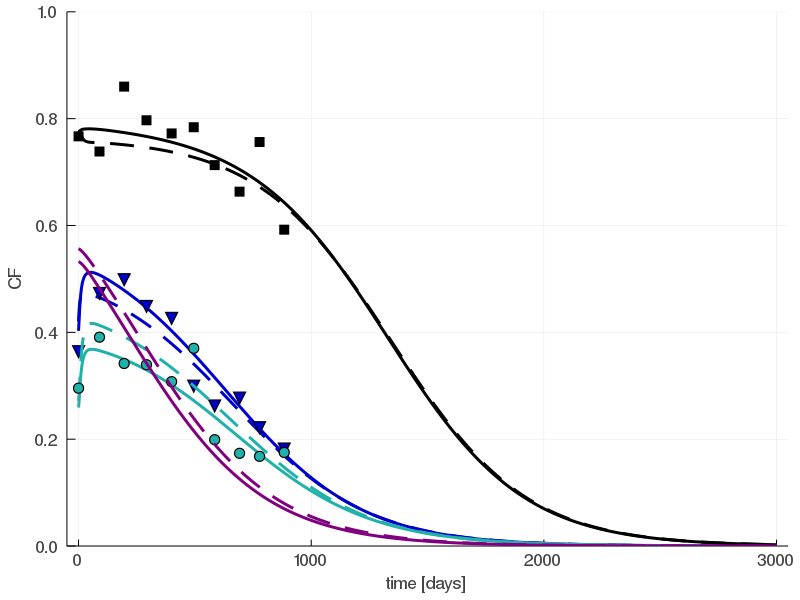}
    \end{subfigure}
	\begin{subfigure}[b]{0.19\textwidth}
        \includegraphics[width=\textwidth]{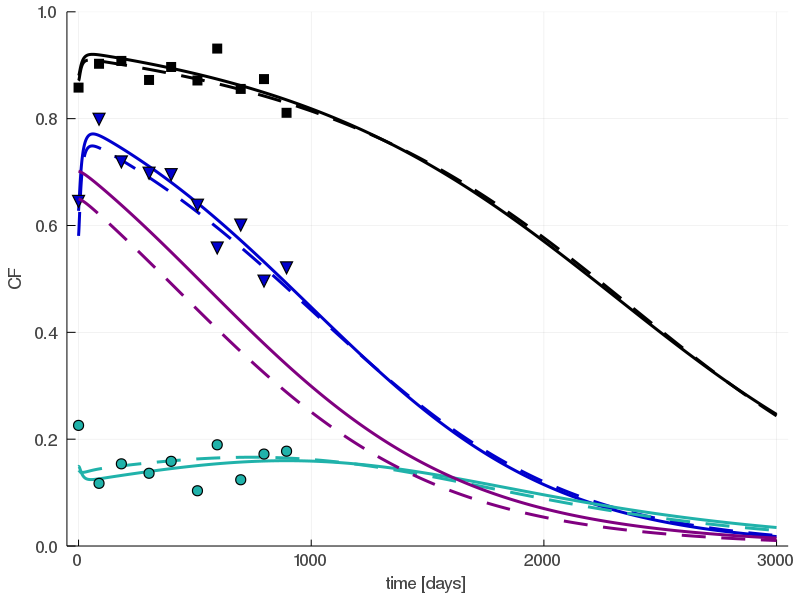}
    \end{subfigure}
	 \begin{subfigure}[b]{0.19\textwidth}
        \includegraphics[width=\textwidth]{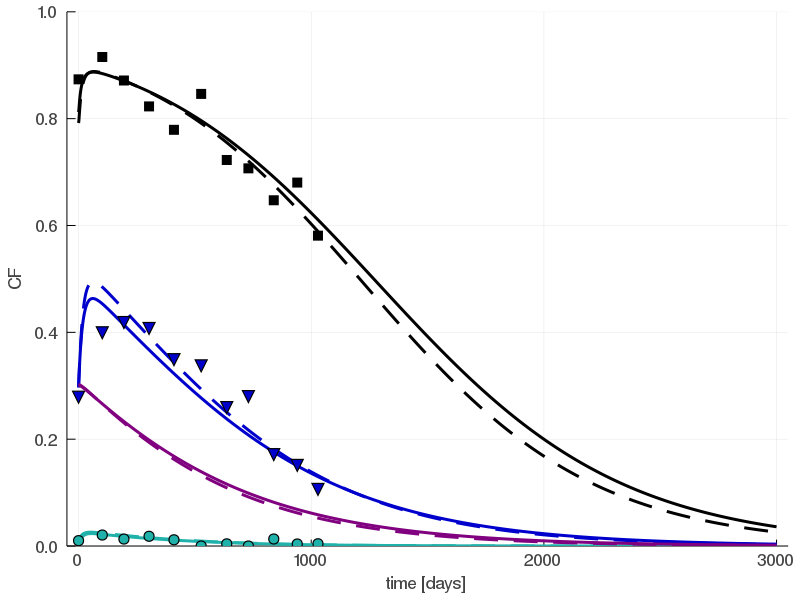}
    \end{subfigure}
	\begin{subfigure}[b]{0.19\textwidth}
        \includegraphics[width=\textwidth]{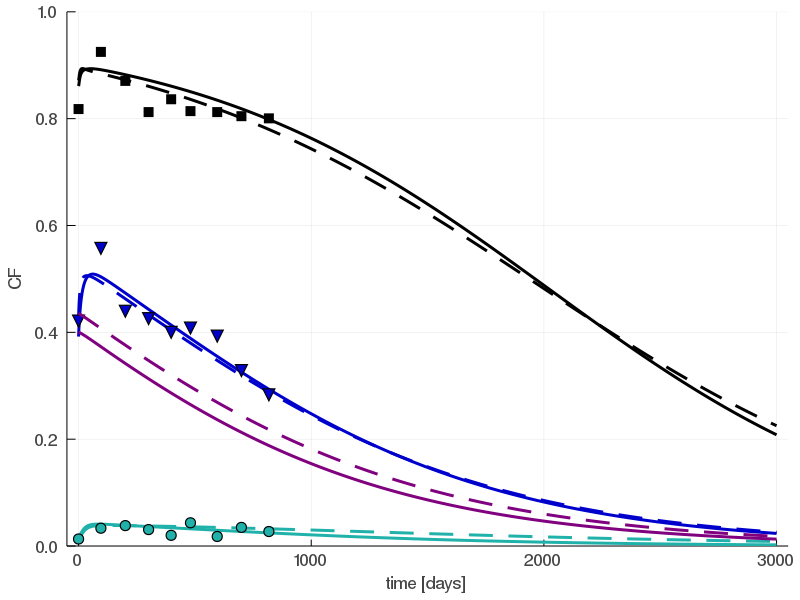}
    \end{subfigure}
	\begin{subfigure}[b]{0.19\textwidth}
        \includegraphics[width=\textwidth]{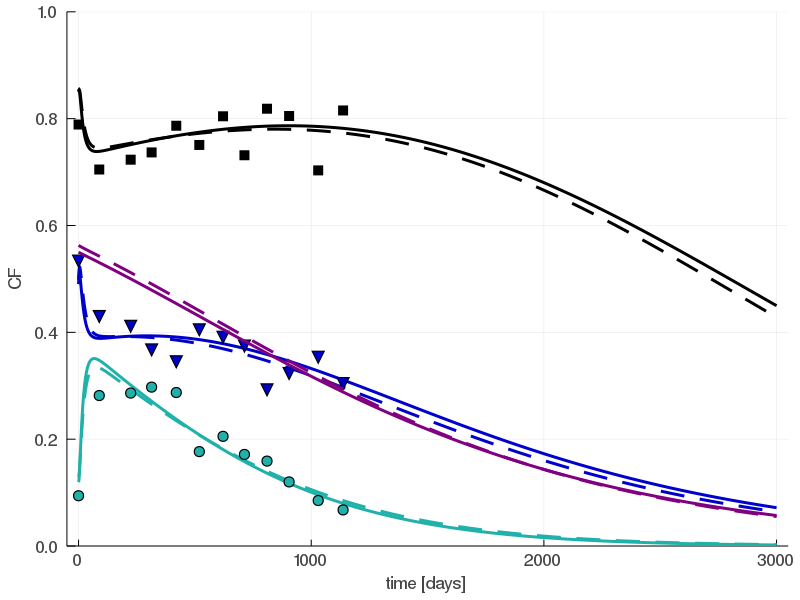}
    \end{subfigure}
	 \begin{subfigure}[b]{0.19\textwidth}
        \includegraphics[width=\textwidth]{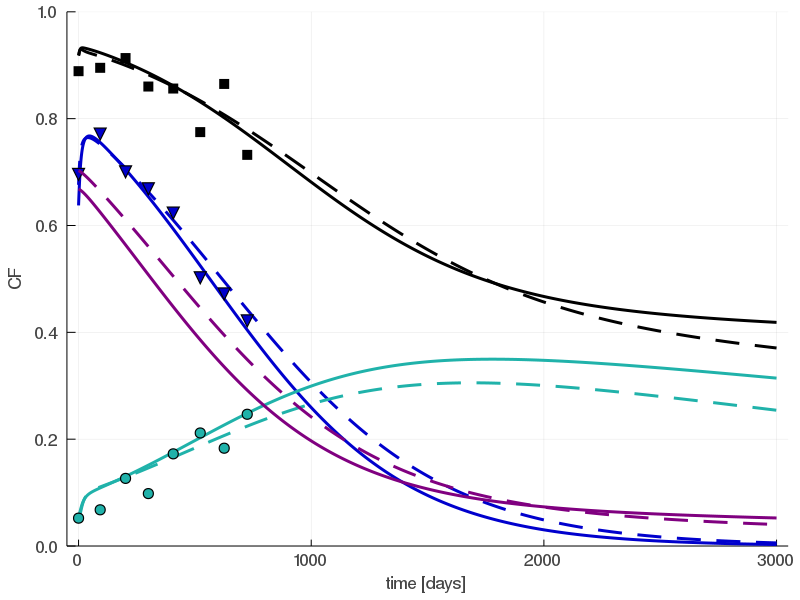}
    \end{subfigure}
	\begin{subfigure}[b]{0.19\textwidth}
        \includegraphics[width=\textwidth]{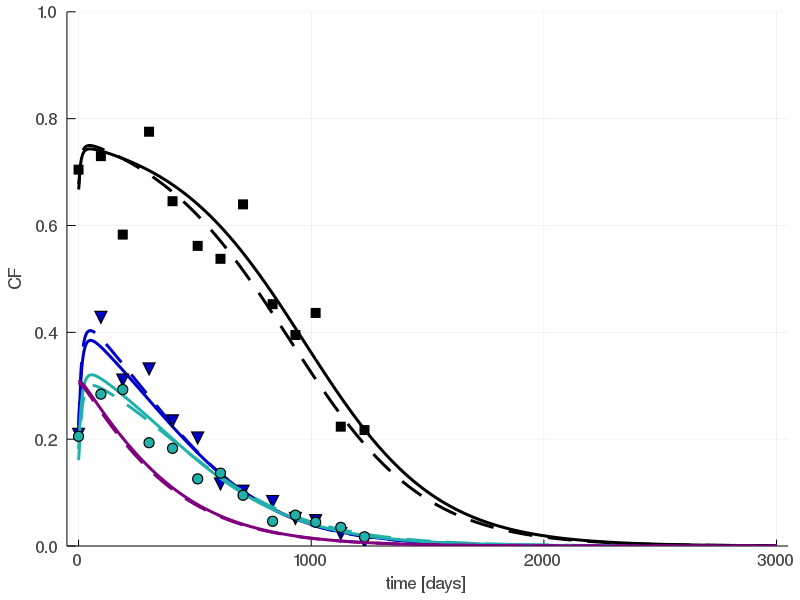}
    \end{subfigure}
	 \begin{subfigure}[b]{0.19\textwidth}
        \includegraphics[width=\textwidth]{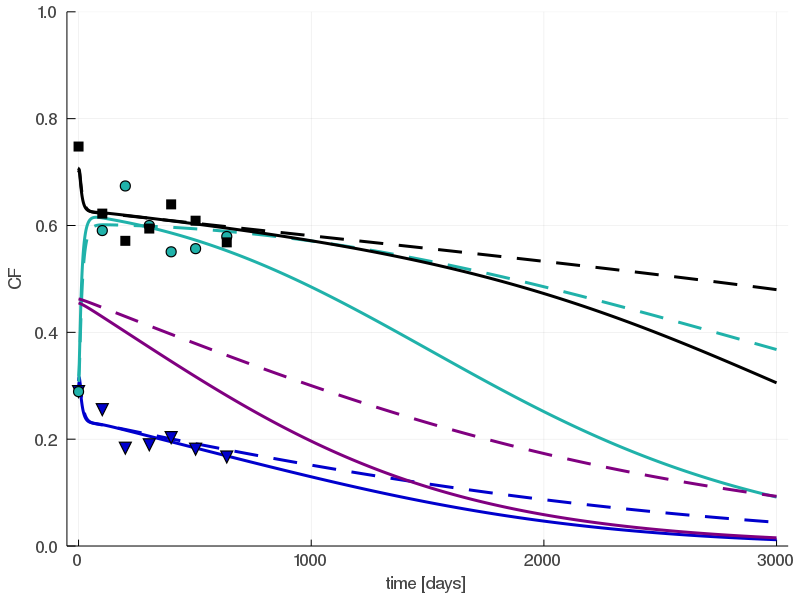}
    \end{subfigure}
\begin{subfigure}[b]{0.19\textwidth}
        \includegraphics[width=\textwidth]{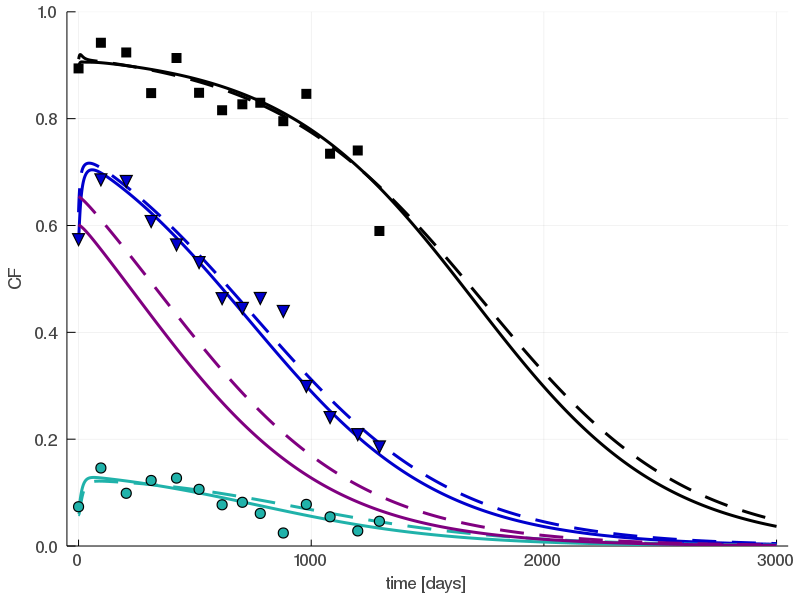}
    \end{subfigure}
	\begin{subfigure}[b]{0.19\textwidth}
        \includegraphics[width=\textwidth]{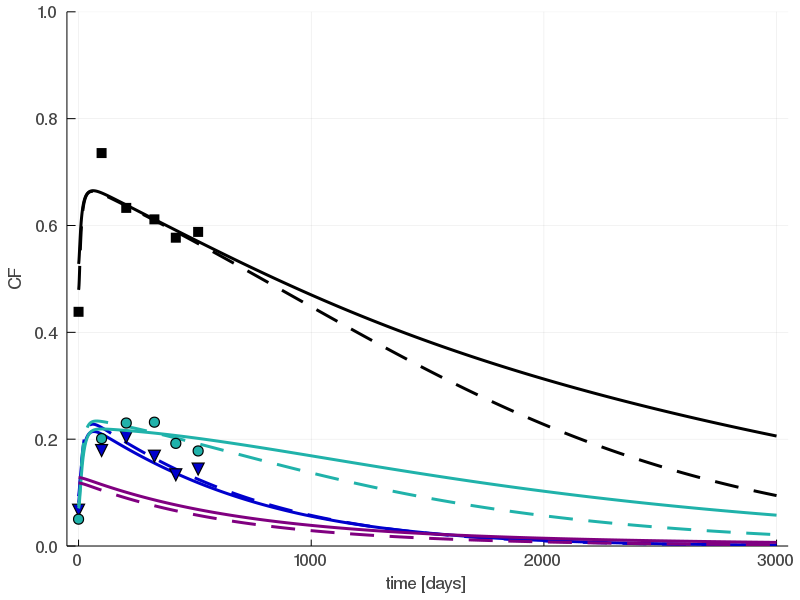}
    \end{subfigure}
	 \begin{subfigure}[b]{0.19\textwidth}
        \includegraphics[width=\textwidth]{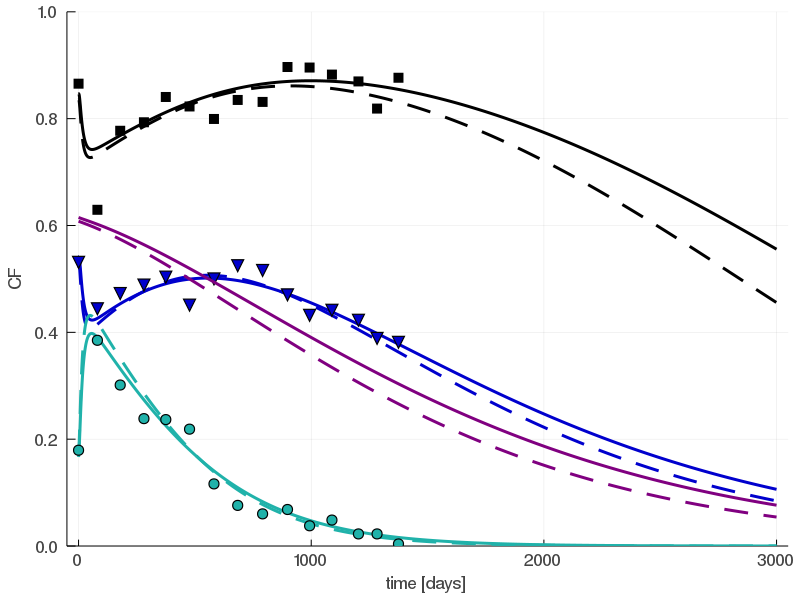}
    \end{subfigure}
	\begin{subfigure}[b]{0.19\textwidth}
        \includegraphics[width=\textwidth]{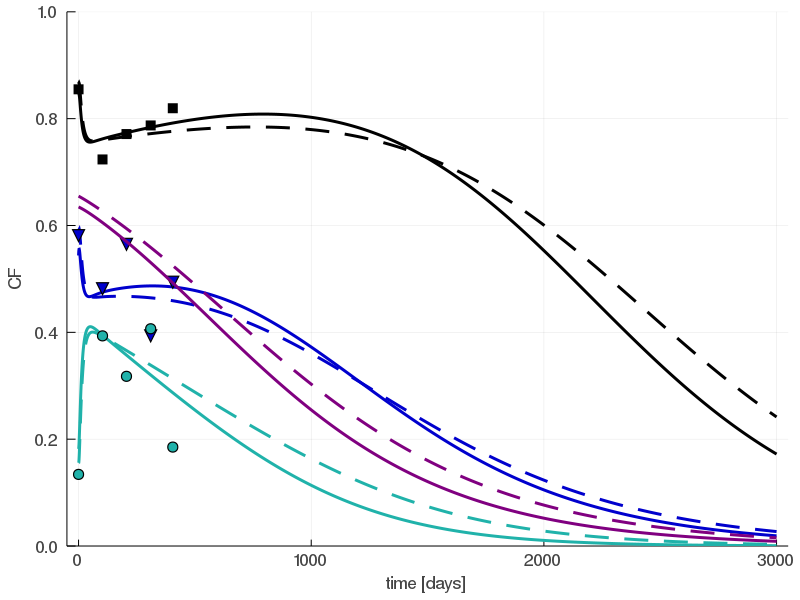}
    \end{subfigure}
	 \begin{subfigure}[b]{0.19\textwidth}
        \includegraphics[width=\textwidth]{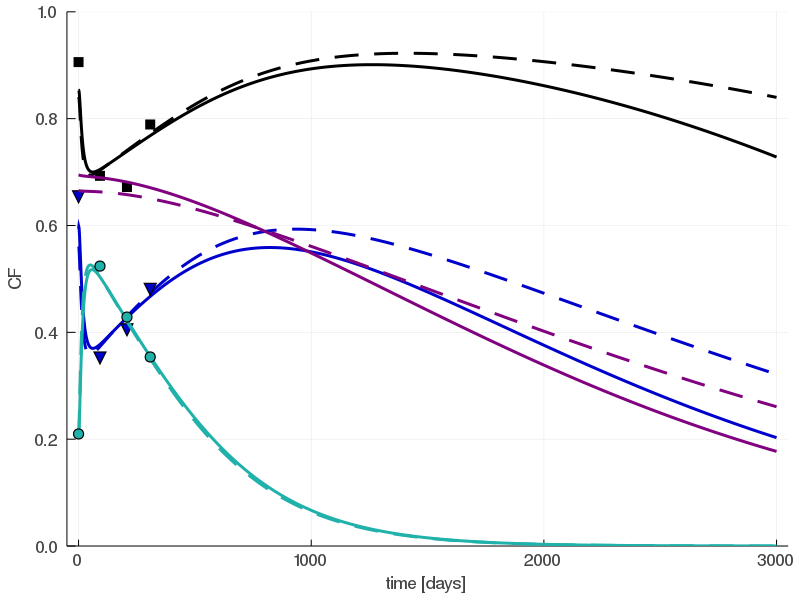}
    \end{subfigure}
    \begin{subfigure}[b]{0.19\textwidth}
        \includegraphics[width=\textwidth]{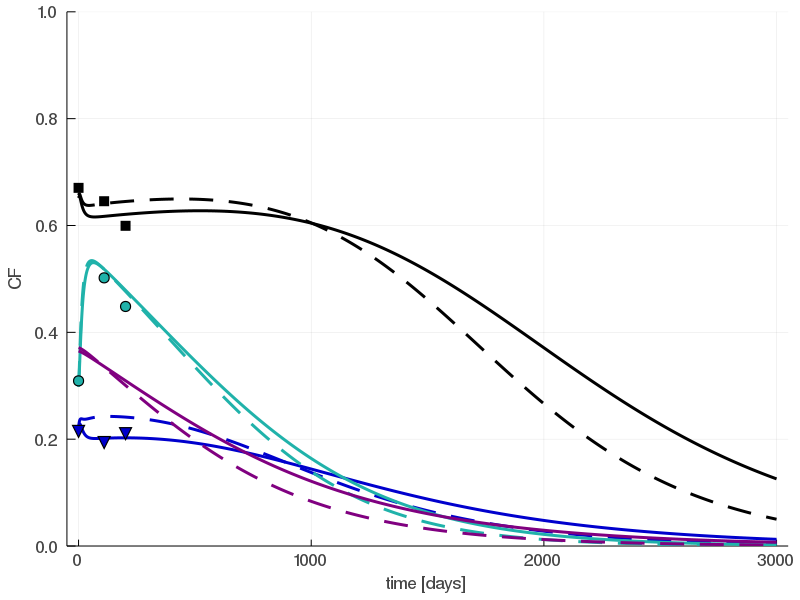}
    \end{subfigure}
	\begin{subfigure}[b]{0.19\textwidth}
        \includegraphics[width=\textwidth]{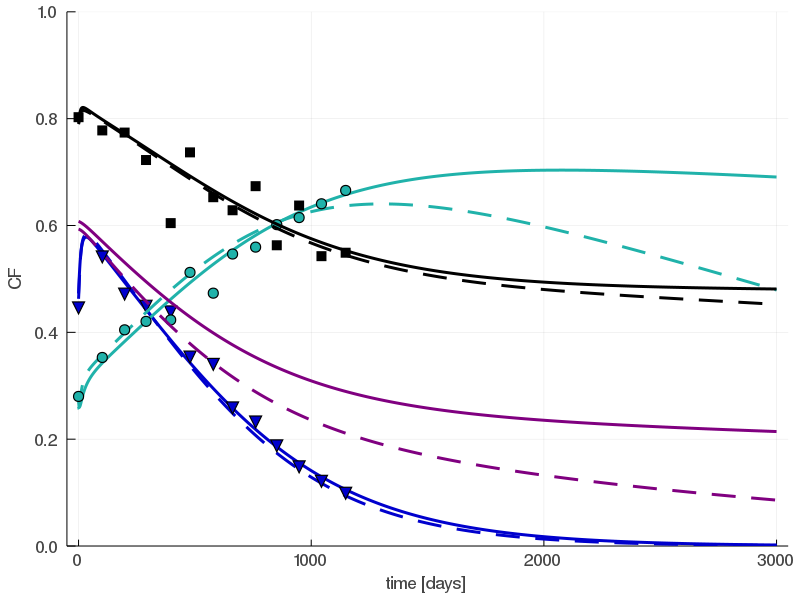}
    \end{subfigure}
	 \begin{subfigure}[b]{0.19\textwidth}
        \includegraphics[width=\textwidth]{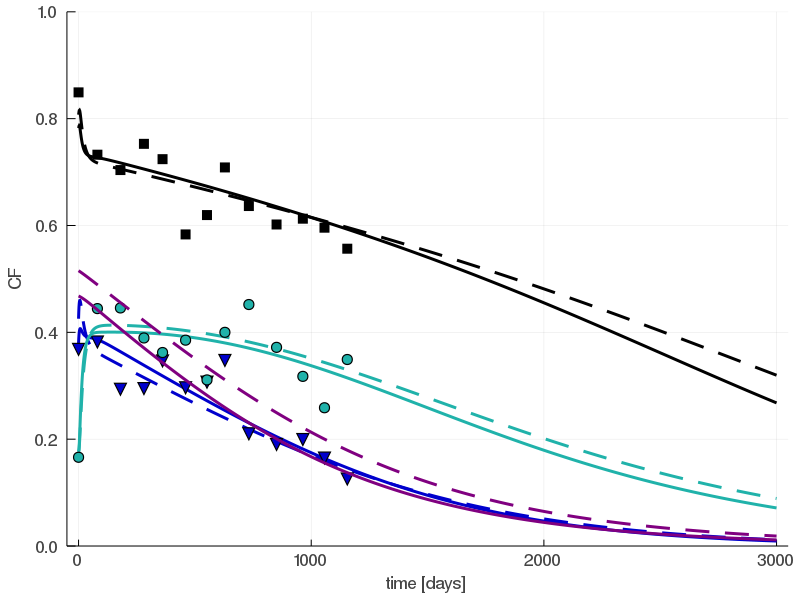}
    \end{subfigure}
	\begin{subfigure}[b]{0.19\textwidth}
        \includegraphics[width=\textwidth]{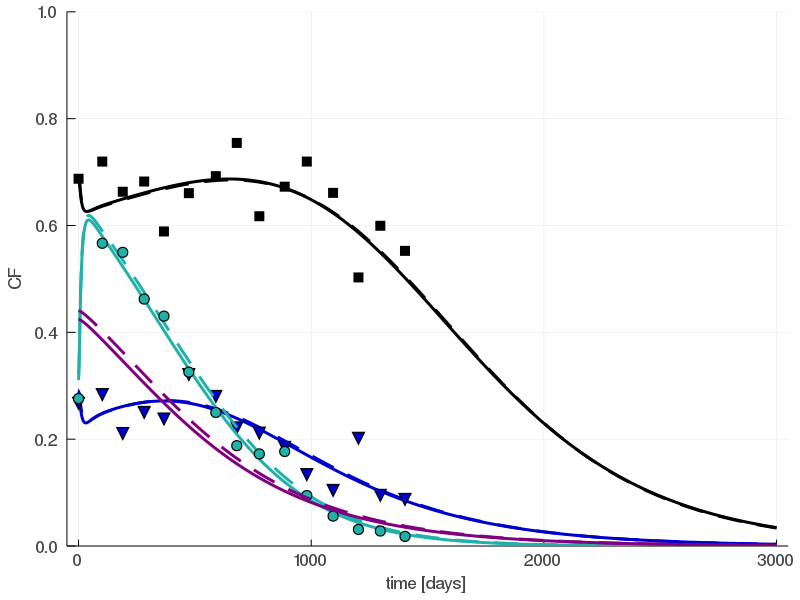}
    \end{subfigure}
	 \begin{subfigure}[b]{0.19\textwidth}
        \includegraphics[width=\textwidth]{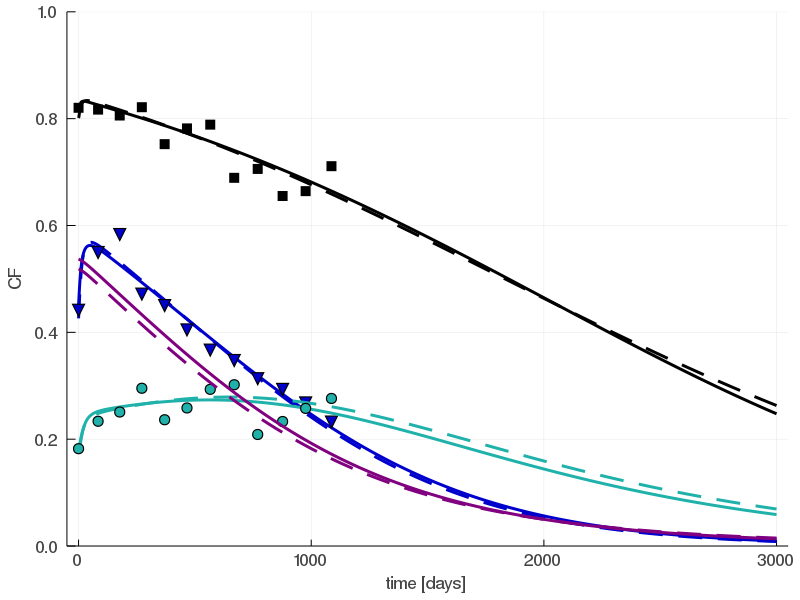}
    \end{subfigure}
    \begin{subfigure}[b]{0.19\textwidth}
        \includegraphics[width=\textwidth]{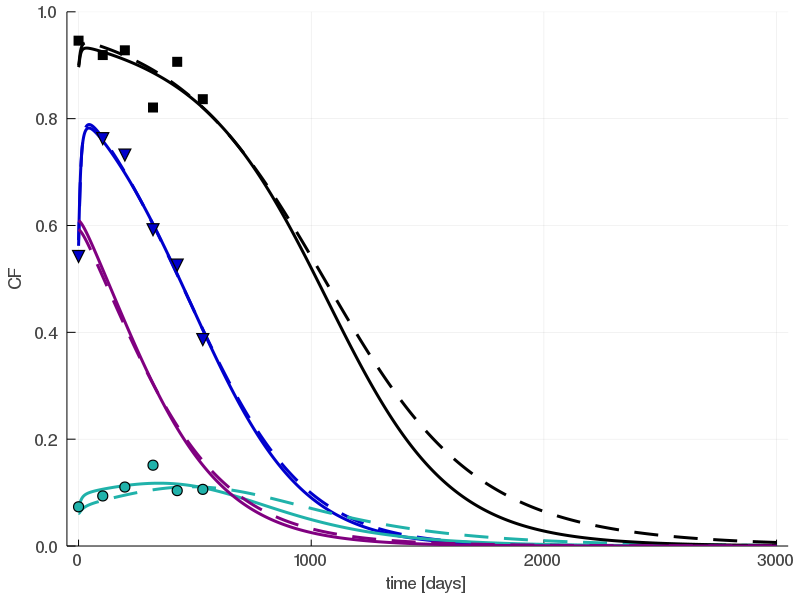}
    \end{subfigure}
	\begin{subfigure}[b]{0.19\textwidth}
        \includegraphics[width=\textwidth]{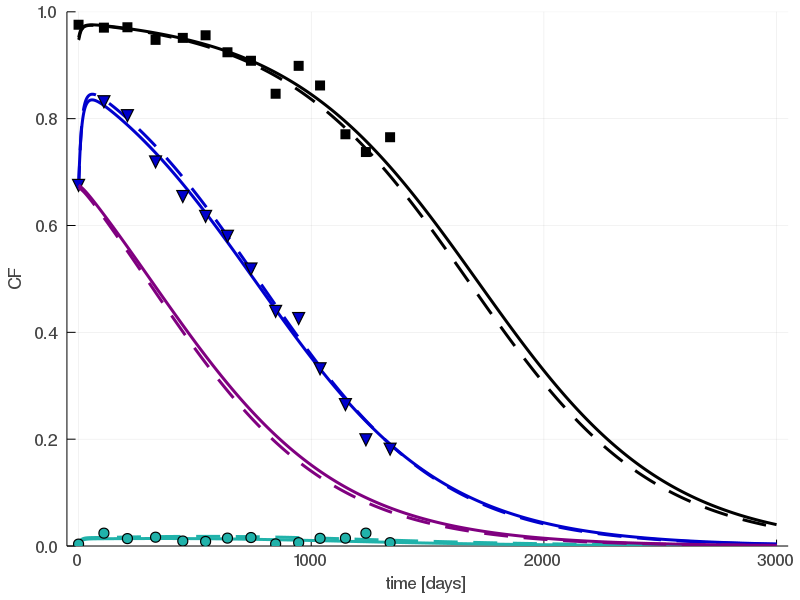}
    \end{subfigure}
	 \begin{subfigure}[b]{0.19\textwidth}
        \includegraphics[width=\textwidth]{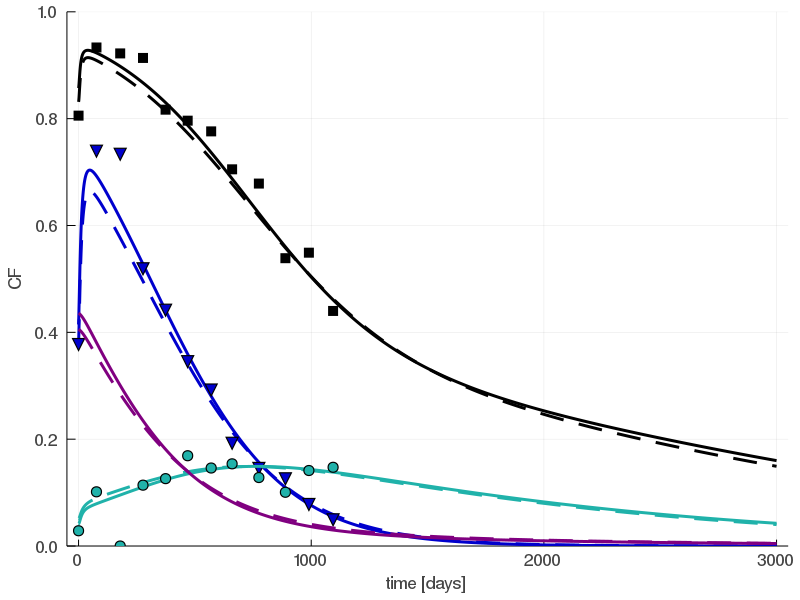}
    \end{subfigure}
	\begin{subfigure}[b]{0.19\textwidth}
        \includegraphics[width=\textwidth]{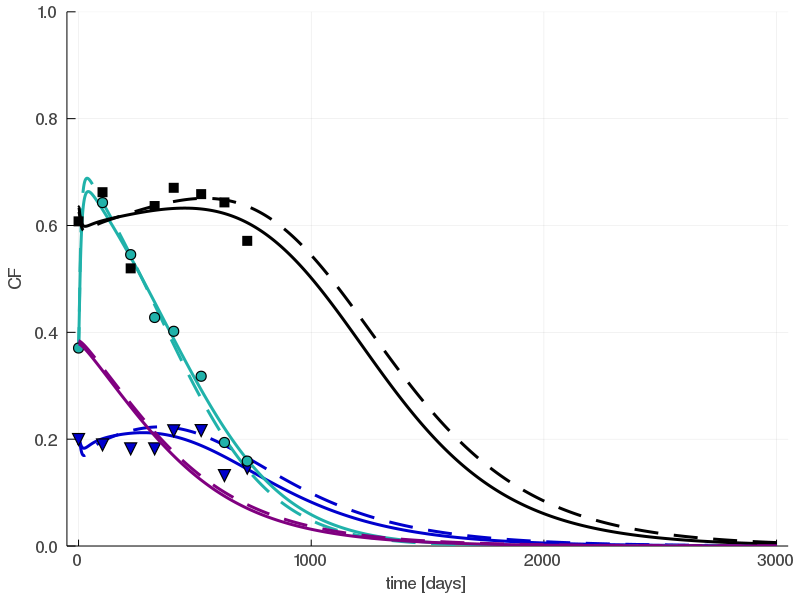}
    \end{subfigure}
	 \begin{subfigure}[b]{0.19\textwidth}
        \includegraphics[width=\textwidth]{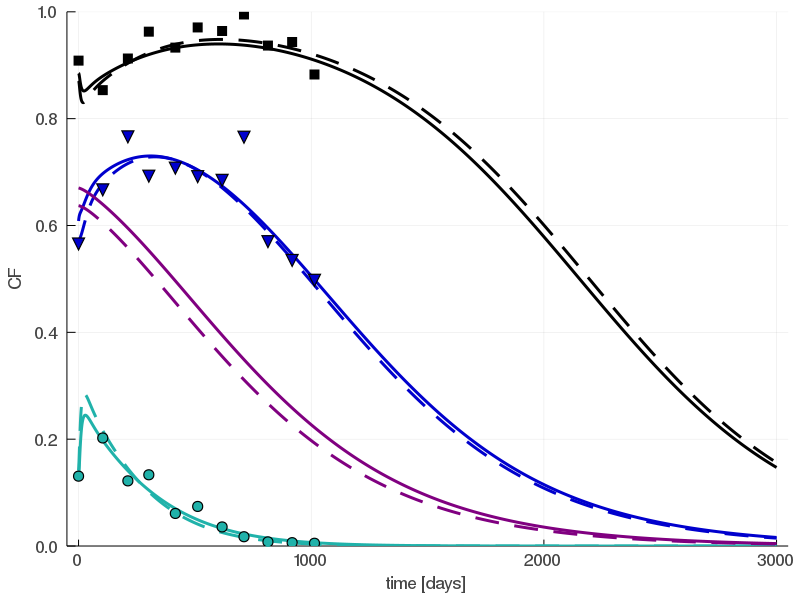}
    \end{subfigure}
     \begin{subfigure}[b]{0.19\textwidth}
        \includegraphics[width=\textwidth]{patient_21.png}
    \end{subfigure}
    \begin{subfigure}[b]{0.19\textwidth}
        \includegraphics[width=\textwidth]{patient_22.png}
    \end{subfigure}
	\begin{subfigure}[b]{0.19\textwidth}
        \includegraphics[width=\textwidth]{patient_23.png}
    \end{subfigure}
	 \begin{subfigure}[b]{0.19\textwidth}
        \includegraphics[width=\textwidth]{patient_24.png}
    \end{subfigure}
	\begin{subfigure}[b]{0.19\textwidth}
        \includegraphics[width=\textwidth]{patient_25.png}
    \end{subfigure}
	 \begin{subfigure}[b]{0.19\textwidth}
        \includegraphics[width=\textwidth]{patient_26.png}
    \end{subfigure}
     \begin{subfigure}[b]{0.19\textwidth}
        \includegraphics[width=\textwidth]{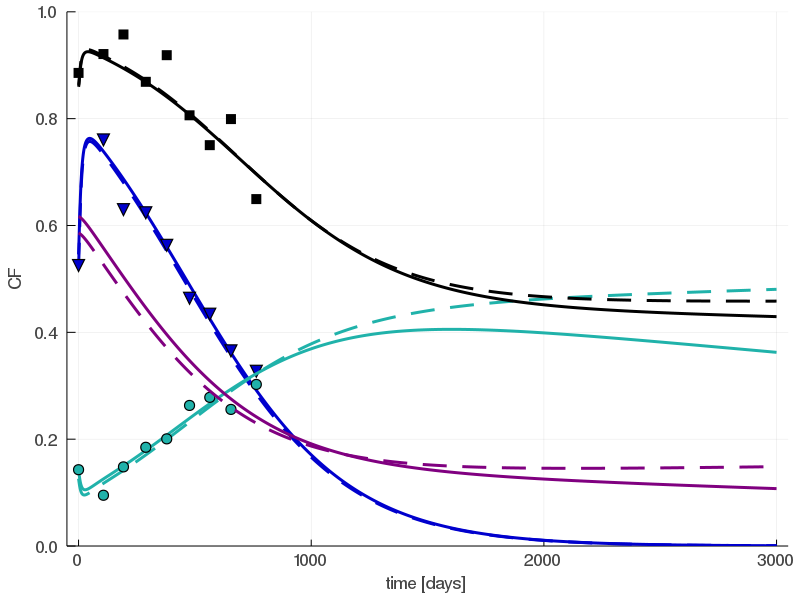}
    \end{subfigure}
    \begin{subfigure}[b]{0.19\textwidth}
        \includegraphics[width=\textwidth]{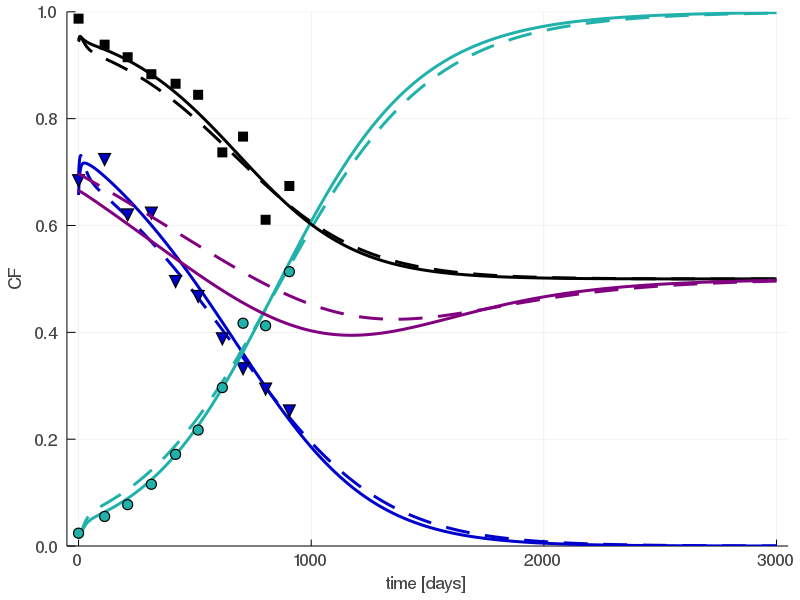}
    \end{subfigure}
	\begin{subfigure}[b]{0.19\textwidth}
        \includegraphics[width=\textwidth]{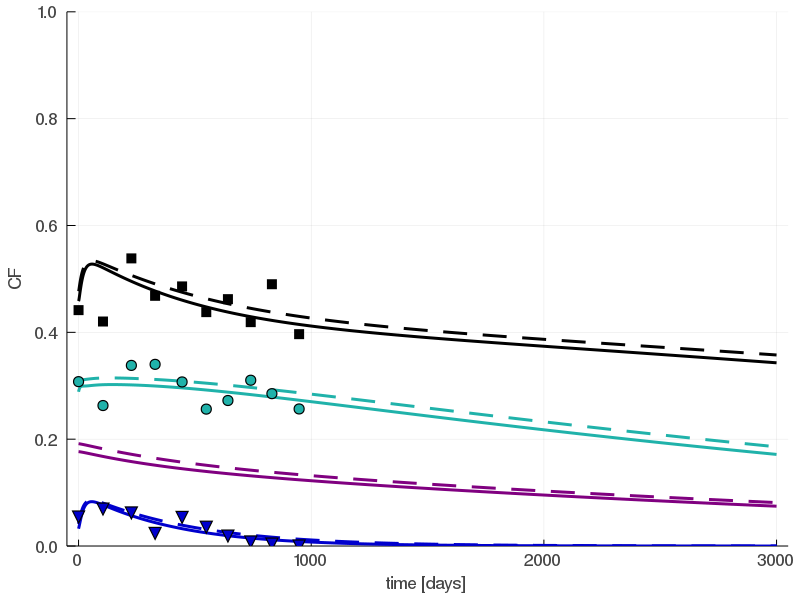}
    \end{subfigure}
	 \begin{subfigure}[b]{0.19\textwidth}
        \includegraphics[width=\textwidth]{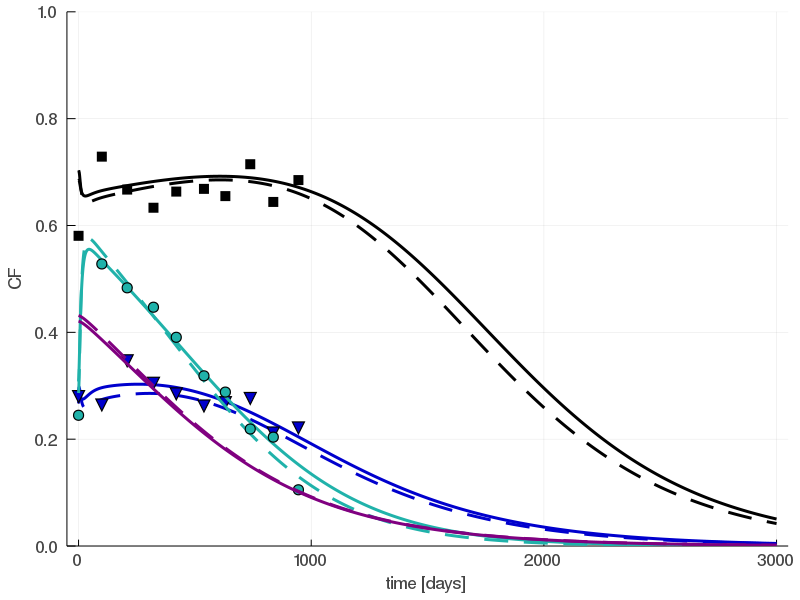}
    \end{subfigure}
    \caption{Comparison of true and estimated dynamics for our 30 virtual patients. Dash lines are the true dynamics. Solid lines are the estimated dynamics (based on the median estimated parameter vector). The black squares, green circles, and blue triangles are our noisy data for mature VAF, immature heterozygous, and homozygous clonal fractions respectively. The purple line refers to the HSC-mutated VAF.}
    \label{fig:fits30patients}
\end{figure}

We ran the previous estimation method over 13 million iterations, with a burn-in length of 2 million. \\
All the predicted dynamics are presented in Figure~\ref{fig:fits30patients} for our 30 virtual patients. We also compare the predicted dynamics (solid lines) with the actual dynamics (dash lines). There is good agreement between both. \\
Figure~\ref{fig:fits_HSC} compares predicted \textit{vs.} actual HSC Clonal Fractions (CF) for heterozygous and homozygous cells. Values are taken at the same times as the ones in our dataset. Even if we do not have data for HSCs, the estimation method can find the true values with accuracy.\\

\begin{figure}[]
    \centering
    \begin{subfigure}[b]{0.49\textwidth}
        \includegraphics[width=\textwidth]{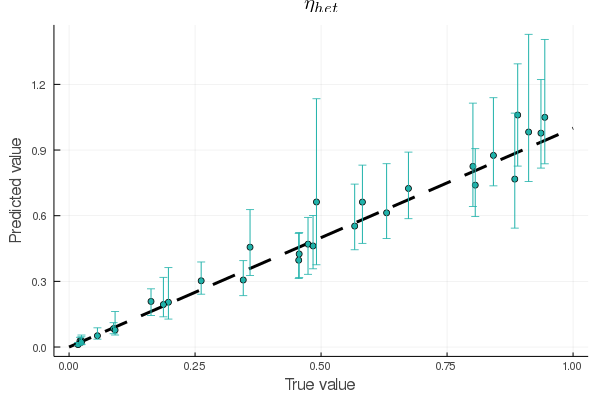}
    \end{subfigure}
     \begin{subfigure}[b]{0.49\textwidth}
        \includegraphics[width=\textwidth]{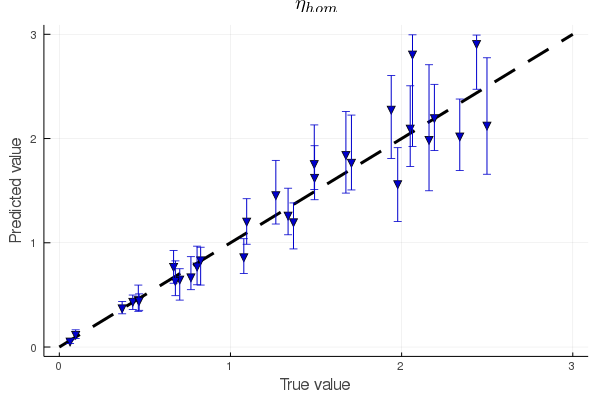}
    \end{subfigure}
    \begin{subfigure}[b]{0.49\textwidth}
        \includegraphics[width=\textwidth]{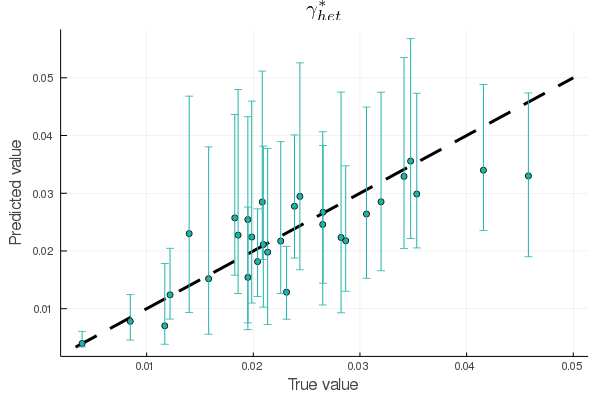}
    \end{subfigure}
     \begin{subfigure}[b]{0.49\textwidth}
        \includegraphics[width=\textwidth]{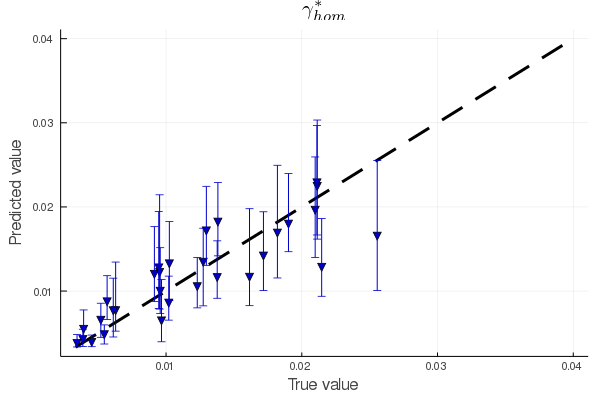}
    \end{subfigure}
    \begin{subfigure}[b]{0.49\textwidth}
        \includegraphics[width=\textwidth]{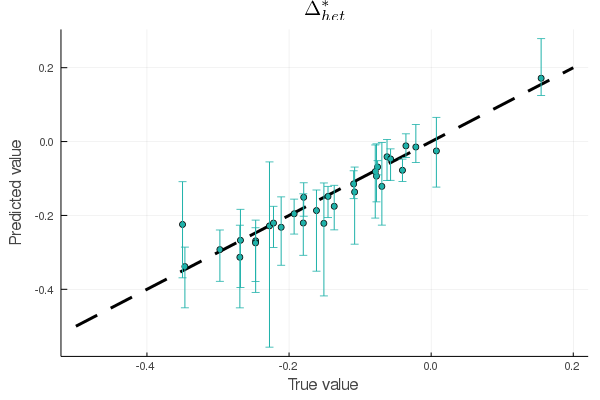}
    \end{subfigure}
     \begin{subfigure}[b]{0.49\textwidth}
        \includegraphics[width=\textwidth]{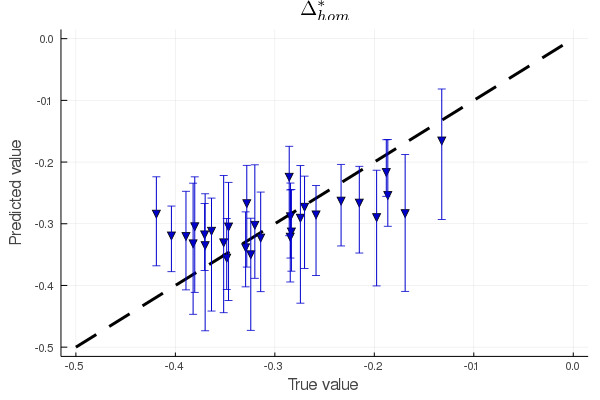}
    \end{subfigure}
    \begin{subfigure}[b]{0.49\textwidth}
        \includegraphics[width=\textwidth]{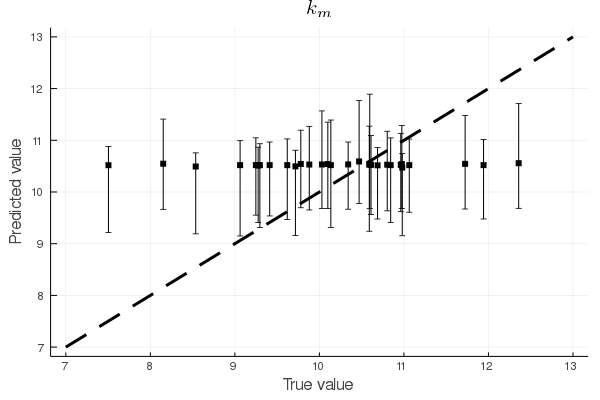}
    \end{subfigure}
    \caption{Comparison between true and inferred parameter values for our 30 virtual patients. Inferred values are presented with their 95\% credibility intervals.}
    \label{fig:params_estimated}
\end{figure}

At the individual level, the parameter estimation results are shown in Fig.~\ref{fig:params_estimated}.
$\eta_{het}$, $\eta_{hom}$, $\gamma^*_{het}$, $\gamma^*_{hom}$ and $\Delta^*_{het}$ are quite good inferred, with 95\% credibility intervals that always contain the true value. 
$k_m$ is poorly estimated: all values concentrate around 10.5 so that the population's effect seems to predominate over inter-individual variability, suggesting that the model output might not be so highly sensitive to the value $k_m$.
Unlike $\Delta^*_{het}$, $\Delta^*_{hom}$ is not predicted with accuracy, and we see a substantial population effect that was not seen for heterozygous HSCs. This might be because true values for $\Delta^*_{hom}$ are lower than those for $\Delta^*_{het}$. Low values might induce a stiffer increase within the first days of therapy, making it more tedious to estimate the value with accuracy since we do not have data for the very first days of therapy (except at the initial time).\\

\begin{figure}[]
    \centering
    \begin{subfigure}[b]{0.49\textwidth}
        \includegraphics[width=\textwidth]{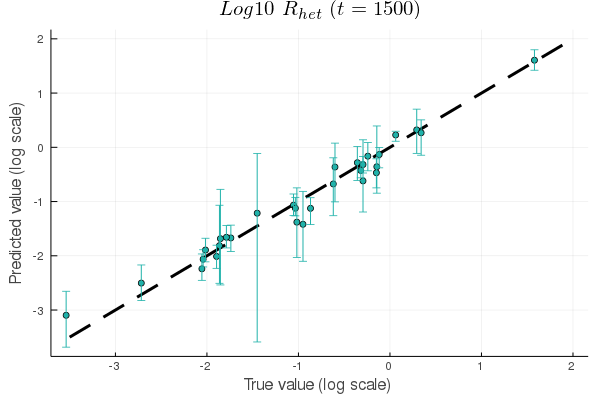}
    \end{subfigure}
     \begin{subfigure}[b]{0.49\textwidth}
        \includegraphics[width=\textwidth]{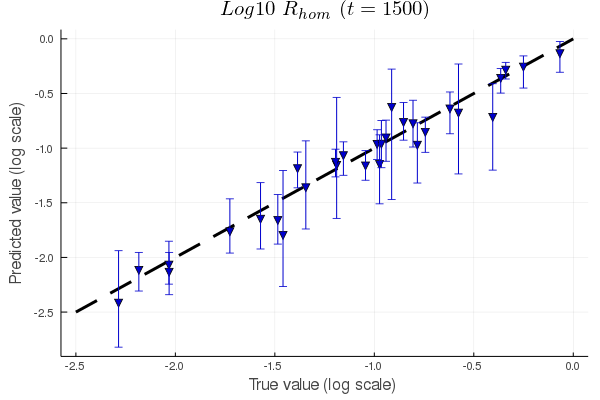}
    \end{subfigure}
    \caption{Comparison between the true R-factor (x-axis) and inferred R-factor (y-axis). The response factor is calculated here at $t=1,500$ days. Axes are in logarithmic scale. On the left we have the heterozygous R-factor, and on the right the one calculated on homozygous HSCs. The error bar corresponds to a 95\% credibility interval.}
    \label{fig:R_factor}
\end{figure}

%R-factor
In addition, we can evaluate the capacity of the model to accurately infer not only the model parameters, but also some key quantities related to the molecular response to the treatment. For that purpose, Mosca et al. introduced and analyzed in detail a variable they called R-factor ($R$). It is defined as the proportion of the inferred mutated CF (either heterozygous or homozygous) among HSCs at a given time \textit{vs} at the initial time. For heterozygous mutated HSCs, the heterozygous $R$-factor is then expressed by:
\begin{equation}
R_{het}(t) = \frac{N_{1,het}(t)+N_{2,het}(t)}{N_{1,het}(0)+N_{2,het}(0)}
\label{eq:R_factor_het}
\end{equation}
and similarly for $R_{hom}(t)$. 
Here, we will focus on the response at 1,500 days. We will note $R := R(t=1,500)$ (and not specify anymore that this quantity is evaluated at 1,5000 days). The lower the value, the better the response. 
Since the R-factor quantifies the long-term molecular response, it is important to verify if this quantity can be accurately estimated. In Fig.~\ref{fig:R_factor}, we show that both the heterozygous and homozygous R-factors could be indeed accurately estimated.\\

%%%

\begin{figure}[]
    \centering
    \begin{subfigure}[b]{0.49\textwidth}
        \includegraphics[width=\textwidth]{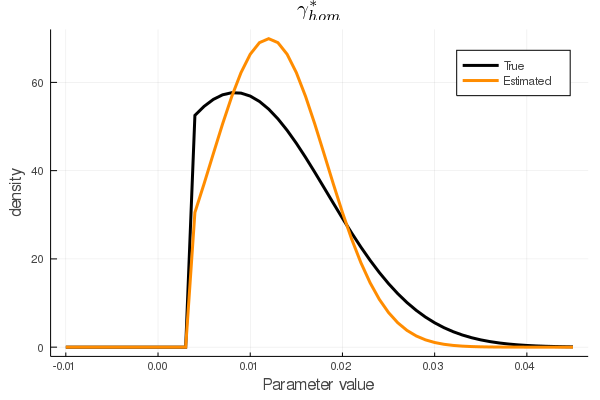}
    \end{subfigure}
     \begin{subfigure}[b]{0.49\textwidth}
        \includegraphics[width=\textwidth]{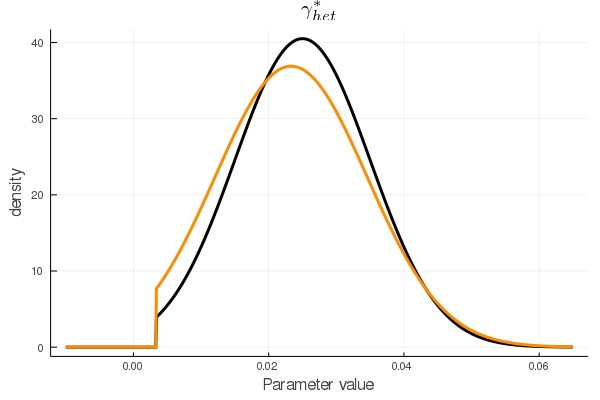}
    \end{subfigure}
    \begin{subfigure}[b]{0.49\textwidth}
        \includegraphics[width=\textwidth]{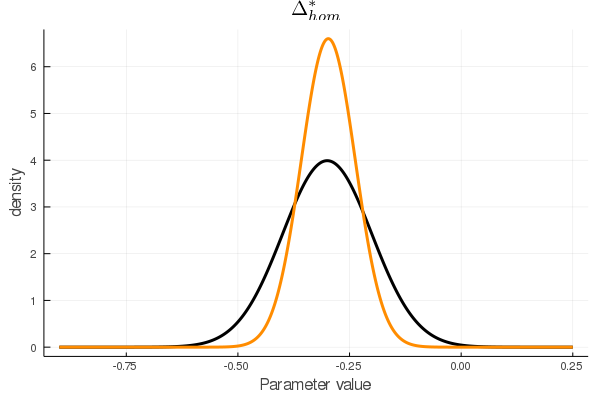}
    \end{subfigure}
     \begin{subfigure}[b]{0.49\textwidth}
        \includegraphics[width=\textwidth]{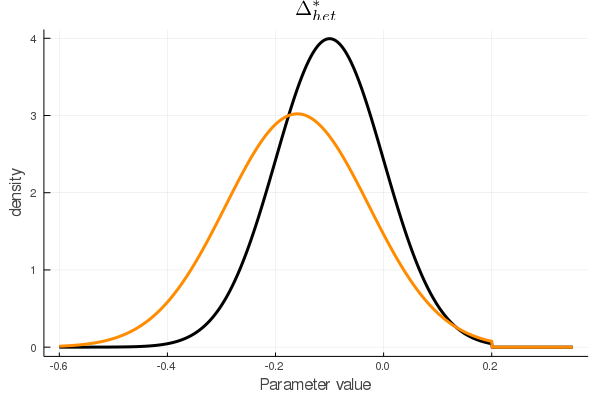}
    \end{subfigure}
    \begin{subfigure}[b]{0.49\textwidth}
        \includegraphics[width=\textwidth]{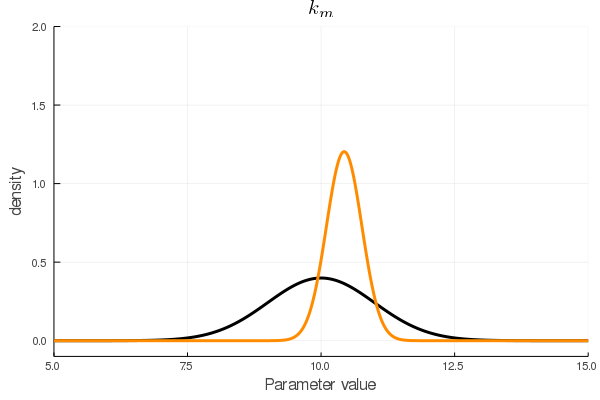}
    \end{subfigure}
    \caption{Comparison of the true population's density (black) used to generate the parameter values and the ones estimated (orange) through our hierarchical Bayesian framework. The density in orange is the estimated population density. That is, a truncated Gaussian law with mean and variance that are estimated hyper-parameters (mean value of their posterior distributions).}
    \label{fig:HP_estimated}
\end{figure}

We now analyze the previous results at the population level: we compare the estimated population's density to the actual one. 
Results are shown in figure~\ref{fig:HP_estimated}. In orange, we have the results of the estimation. 
We see from this figure how we can infer not only the individual values but also the population's density. It works very well for $\gamma^*_{het}$.
For $\Delta^*_{hom}$ we also have good results, even if we predict less variance than in reality, but the mean of the population's density is very well estimated. It is interesting to note because the estimations at the individual level were not so good.
For $k_m$, the mean of the population's density is well estimated, but the variance is lower than in reality. \\

Finally, this practical identifiability study demonstrates the ability of our estimation method to infer with accuracy the patient's dynamics. We also see that most of the parameters are well estimated at the individual level and that the hierarchical framework can retrieve the population distributions that generate the individual parameters.

\FloatBarrier
\newpage
\section{Detailed results of the model selection procedure}
\label{sec:detail_selection}

\begin{figure}[h]
    \centering
    \includegraphics[width=\textwidth]{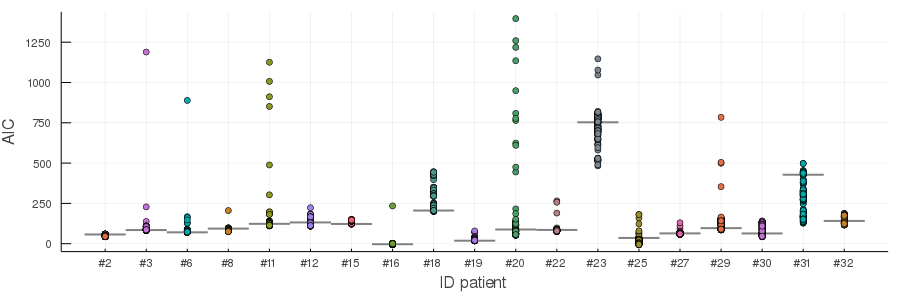}
    \caption{AIC$_{i,j}$ for each model $j$ and each patient $i$.}
    \label{fig:AIC}
\end{figure}

The first step in our model selection procedure is to calibrate all 225 models for 19 MPN patients. This results in a vast amount of models to calibrate, and there is thus a need for an efficient inference method. For this purpose, we use the CMA-ES algorithm to compute the maximum likelihood $\mathcal{L}_{i,j}$ for each model $\mathcal{M}_j$, given data from patient $i$.\\
For all models, the likelihood is expressed using the same observation model as described in Appendix~\ref{sec:observation_model}.\\
Then, we compute the AIC (see eq.~(8) in the main text) for each combination: patient $\times$ model. Results are displayed in Fig.~\ref{fig:AIC} (and Fig.~3 in the main text to compare the models that perform the best). Each dot corresponds to a model; the horizontal line corresponds to the AIC obtained with the \rev{baseline} model. This latter was already a good model, with very good fits for patients \#6, 15, 16, 19. However, the \rev{baseline} model had the inconvenience of resulting in very poor fits for other patients, such as patients \#20, 23, 31. Thus, there was a need for a model performing better for all patients in the cohort.

\begin{figure}[h]
    \centering
    \includegraphics[width=0.8\textwidth]{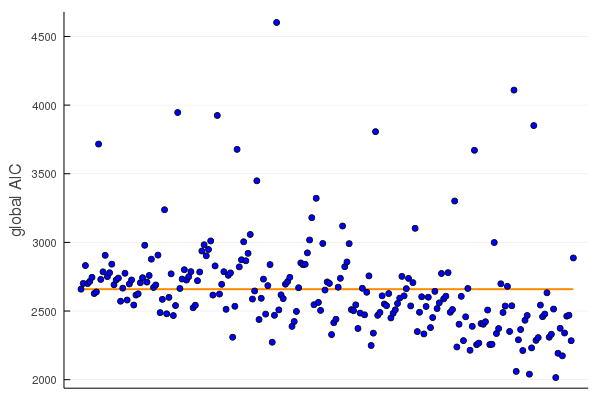}
    \caption{Global AIC$_j$ for each model $j$. The orange horizontal line represents the global AIC of the \rev{baseline} model.}
    \label{fig:global_AIC}
\end{figure}

\begin{figure}[h]
    \centering
    \includegraphics[width=\textwidth]{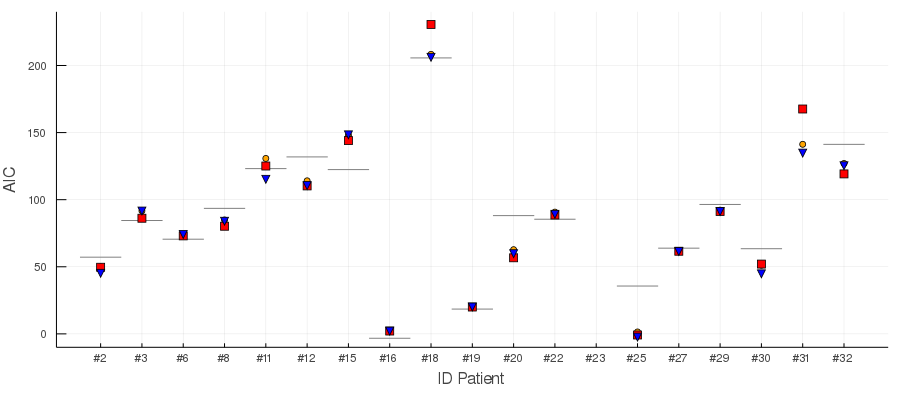}
    \caption{Focus on the 3 best model according to their global AIC. The orange circle corresponds to the model with a constant dose-response relationship for $\bar{\gamma}^*_{het}$ and $\bar{\gamma}^*_{hom}$, a sigmoid affine one for $\bar{\Delta}^*_{hom}$, and an affine one for $\bar{\Delta}^*_{het}$. 
    The red square corresponds to the model with a constant dose-response relationship for $\bar{\gamma}^*_{het}$, an affine one for $\bar{\gamma}^*_{hom}$, a sigmoid affine one for $\bar{\Delta}^*_{hom}$, and an affine one for $\bar{\Delta}^*_{het}$. 
    The blue triangle corresponds to the model with a constant dose-response relationship for $\bar{\gamma}^*_{het}$ and $\bar{\gamma}^*_{hom}$, and a sigmoid affine one for $\bar{\Delta}^*_{hom}$ and $\bar{\Delta}^*_{het}$. Horizontal black line corresponds to the \rev{baseline} model. y-axis is truncated for clarity. }
    \label{fig:AIC_3best}
\end{figure}

Nevertheless, the best model for a given patient is not necessarily the best for another one. 
To compare the models, not at the individual level but for the population (of our 19 \jakvf \rev{MPN} patients) level, we compute a global AIC (see eq.~(9) in the main text) that sums the contribution of all individual AIC. In Fig.~\ref{fig:global_AIC}, we display the global AIC of all models and compare them to the global AIC of the \rev{baseline} model (horizontal orange line). Many models improve the \rev{baseline} model on the whole cohort. In particular, three models stand out, with a global AIC of around 2,000. 
They are presented in Tab.~2 in the main text, and we display how they perform for each patient in Fig.~\ref{fig:AIC_3best}. These models are the following:
\begin{itemize}
    \item Model "orange" (or $m=199$, see section~\ref{sec:list_models}): a constant dose-response relationship for $\bar{\gamma}^*_{het}$ and $\bar{\gamma}^*_{hom}$, an affine sigmoid  one for $\bar{\Delta}^*_{hom}$, and an affine one for $\bar{\Delta}^*_{het}$
    \item Model "red" ($m=205$): constant dose-response relationship for $\bar{\gamma}^*_{het}$, an affine one for $\bar{\gamma}^*_{hom}$, an affine sigmoid one for $\bar{\Delta}^*_{hom}$, and an affine one for $\bar{\Delta}^*_{het}$
    \item Model "blue" ($m=217$): constant dose-response relationships for $\bar{\gamma}^*_{het}$ and $\bar{\gamma}^*_{hom}$, and an affine sigmoid one for $\bar{\Delta}^*_{hom}$ and $\bar{\Delta}^*_{het}$
\end{itemize}
Models "orange" and "red" are better than the \rev{baseline} model for 11 (over 19) patients, and model "blue" improve the fits of 12 patients.\\ 

Note that, instead of the AIC, we could have used the Bayesian Information Criterion (BIC), defined by:
\begin{equation*}
B I C_{i, j}=-2 \log \left(\mathcal{L}_{i, j}\right)+k_{j} \cdot \log \left(N_{i}\right)
\end{equation*}
with $N_i$ the number of data points (both for mature and progenitor cells) for patient $i$. The same three best models were selected when applying this criterion.\\

In a second step, we only consider the three models that stand out and further compare them using a more rigorous hierarchical Bayesian inference method. \\
The same estimation method used for our virtual dataset (Appendix~\ref{sec:identif}) is used. Results are presented in Tab.~2 in the main text.
The selected model is model "blue" (model $m=217$), with constant dose-response relationships for $\bar{\gamma}^*_{het}$ and $\bar{\gamma}^*_{hom}$ and a sigmoid affine one for $\bar{\Delta}^*_{hom}$ and $\bar{\Delta}^*_{het}$. It is interesting to see how the results are consistent across the several selection criteria. The selected model is the model that has the best AIC, DIC, and that also improves the results for 12 patients instead of 11 for the two others.

\FloatBarrier
\newpage
\section{Detailed analysis of the selected model}
\label{sec:detail_model_selected}

\begin{figure}[h]
    \centering
    \includegraphics[width=0.7\textwidth]{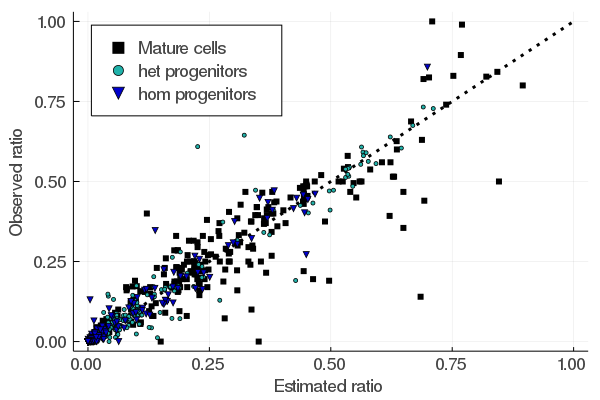}
    \caption{Comparison between the observed ratios (CF for progenitor cells, VAF for mature cells) and the estimated ones (based on the mean posterior parameter vector). With our 19 patients, we have 232 data points for mature cells and 255 for heterozygous and homozygous progenitors. The linear dotted line represents exact fit.}
    \label{fig:pred_vs_obs}
\end{figure}

The best model, selected after our two-steps model selection procedure, is the one with a constant dose-response relationship for $\bar{\gamma}^*_{het}$ and $\bar{\gamma}^*_{hom}$, and a sigmoid affine relation for $\bar{\Delta}^*_{hom}$ and $\bar{\Delta}^*_{het}$. \\
To visualize to which extent this model fits well the cohort data, we compare in Fig.~\ref{fig:pred_vs_obs} the observed and inferred values, both for mature cells (VAF) and for het and hom progenitor cells (CF). We observe overall a good agreement between observations and inferred values.\\

\begin{figure}[h]
    \centering
    \begin{subfigure}[b]{0.26\textwidth}
        \includegraphics[width=\textwidth]{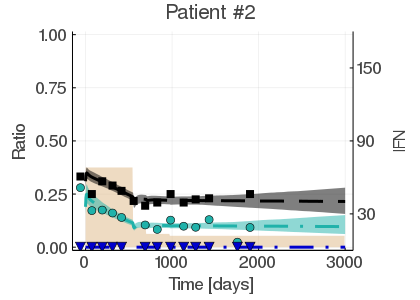}
    \end{subfigure}
    \begin{subfigure}[b]{0.26\textwidth}
        \includegraphics[width=\textwidth]{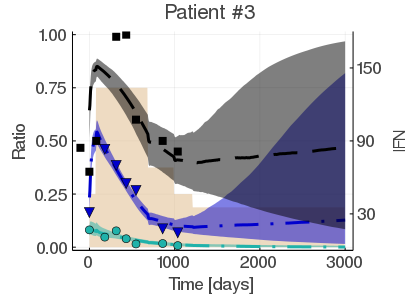}
    \end{subfigure}
    \begin{subfigure}[b]{0.26\textwidth}
        \includegraphics[width=\textwidth]{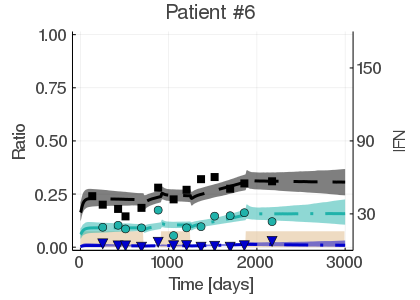}
    \end{subfigure}
    \begin{subfigure}[b]{0.26\textwidth}
        \includegraphics[width=\textwidth]{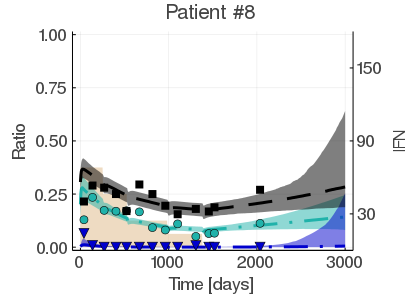}
    \end{subfigure}
    \begin{subfigure}[b]{0.26\textwidth}
        \includegraphics[width=\textwidth]{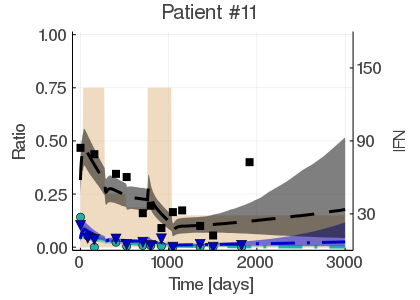}
    \end{subfigure}
    \begin{subfigure}[b]{0.26\textwidth}
        \includegraphics[width=\textwidth]{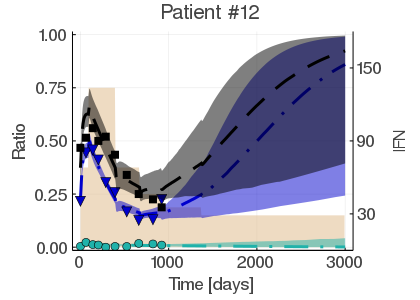}
    \end{subfigure}
    \begin{subfigure}[b]{0.26\textwidth}
        \includegraphics[width=\textwidth]{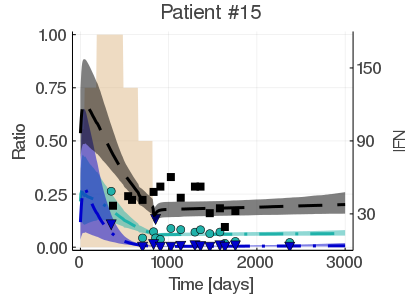}
    \end{subfigure}
    \begin{subfigure}[b]{0.26\textwidth}
        \includegraphics[width=\textwidth]{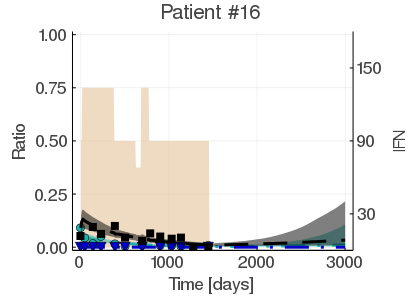}
    \end{subfigure}
    \begin{subfigure}[b]{0.26\textwidth}
        \includegraphics[width=\textwidth]{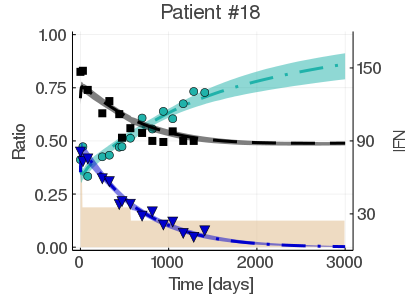}
    \end{subfigure}
    \begin{subfigure}[b]{0.26\textwidth}
        \includegraphics[width=\textwidth]{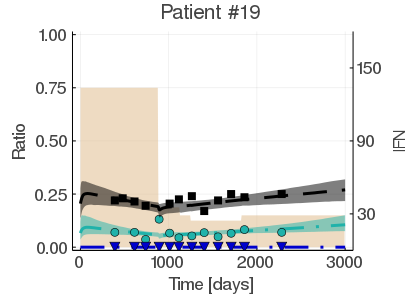}
    \end{subfigure}
    \begin{subfigure}[b]{0.26\textwidth}
        \includegraphics[width=\textwidth]{20}
    \end{subfigure}
    \begin{subfigure}[b]{0.26\textwidth}
        \includegraphics[width=\textwidth]{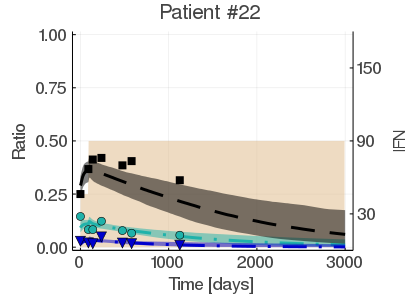}
    \end{subfigure}
    \begin{subfigure}[b]{0.26\textwidth}
        \includegraphics[width=\textwidth]{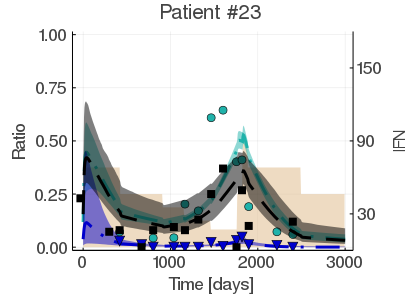}
    \end{subfigure}
    \begin{subfigure}[b]{0.26\textwidth}
        \includegraphics[width=\textwidth]{25}
    \end{subfigure}
    \begin{subfigure}[b]{0.26\textwidth}
        \includegraphics[width=\textwidth]{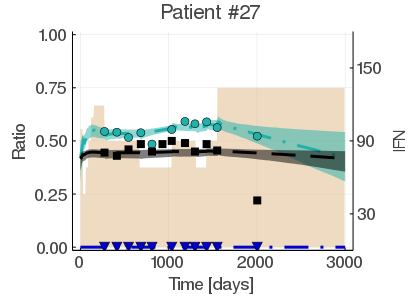}
    \end{subfigure}
    \begin{subfigure}[b]{0.26\textwidth}
        \includegraphics[width=\textwidth]{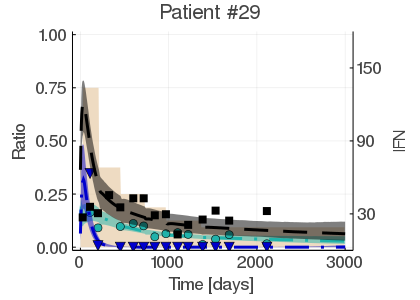}
    \end{subfigure}
    \begin{subfigure}[b]{0.26\textwidth}
        \includegraphics[width=\textwidth]{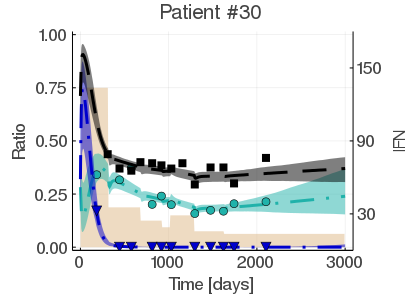}
    \end{subfigure}
    \begin{subfigure}[b]{0.26\textwidth}
        \includegraphics[width=\textwidth]{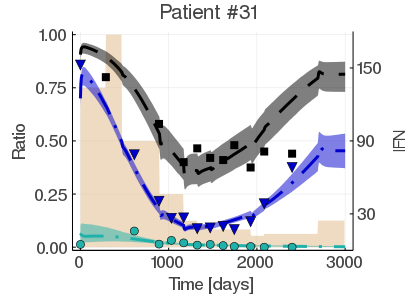}
    \end{subfigure}
    \begin{subfigure}[b]{0.26\textwidth}
        \includegraphics[width=\textwidth]{32}
    \end{subfigure}
    
    \caption{Dynamics of inferred homozygous (blue) and heterozygous (green) mutated progenitors (CF), and mature cells (VAF, in black) are presented for 19 \jakvf \rev{MPN} patients for whom we made the parameter estimation using our hierarchical Bayesian framework and the selected model. Triangles, dots, and squares are experimental data values. The curves were determined with the model (median values). The shaded areas represent 95\% credibility intervals. The shaded beige areas correspond to the dose of IFN$\alpha$ received overtime.}
    \label{fig:Dyn_estimated}
\end{figure}

Fig.~\ref{fig:Dyn_estimated} displays the inferred dynamics with 95\% credibility intervals for the mature VAF and progenitor CF. There is a good agreement between observed and inferred values for most patients. We also display on this figure how the IFN$\alpha$ doses vary over therapy for each patient. It is interesting to observe how the proportion of mutated cells increases when the dose decreases to a too large extent. This model outperforms the \rev{baseline} model of Mosca et al.~\cite{mosca2021}, particularly for patients \#23, 31, 20, and 25.\\

\begin{figure}[h]
    \centering
    \begin{subfigure}[b]{0.49\textwidth}
        \includegraphics[width=\textwidth]{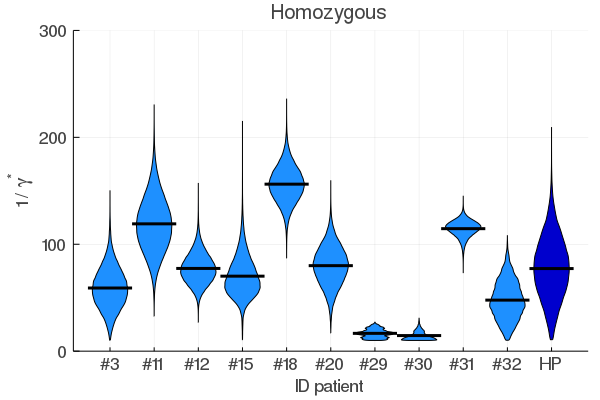}
    \end{subfigure}
     \begin{subfigure}[b]{0.49\textwidth}
        \includegraphics[width=\textwidth]{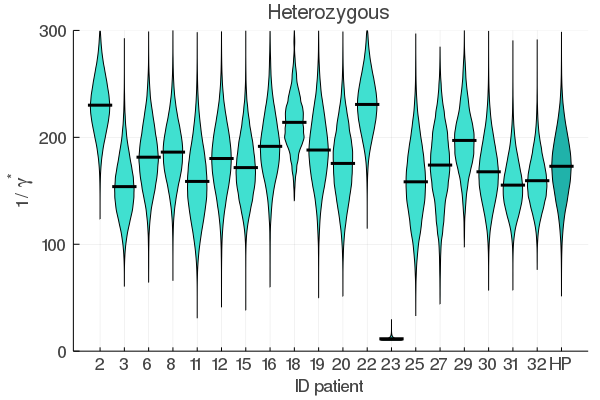}
    \end{subfigure}
    \begin{subfigure}[b]{0.49\textwidth}
        \includegraphics[width=\textwidth]{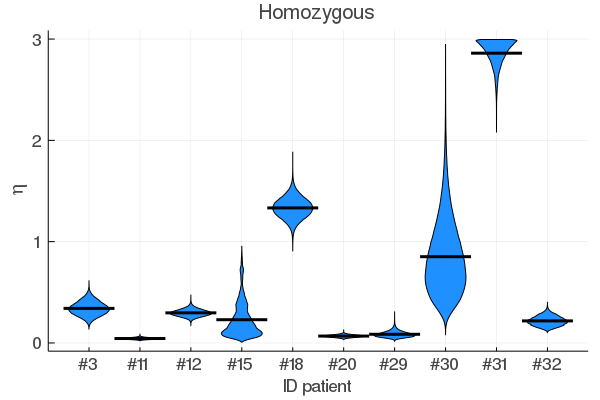}
    \end{subfigure}
     \begin{subfigure}[b]{0.49\textwidth}
        \includegraphics[width=\textwidth]{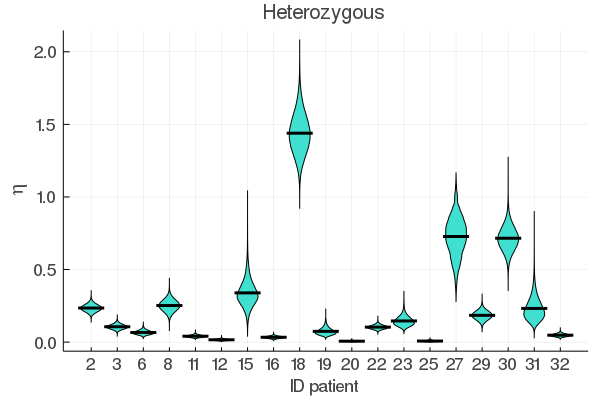}
    \end{subfigure}
    \begin{subfigure}[b]{0.49\textwidth}
        \includegraphics[width=\textwidth]{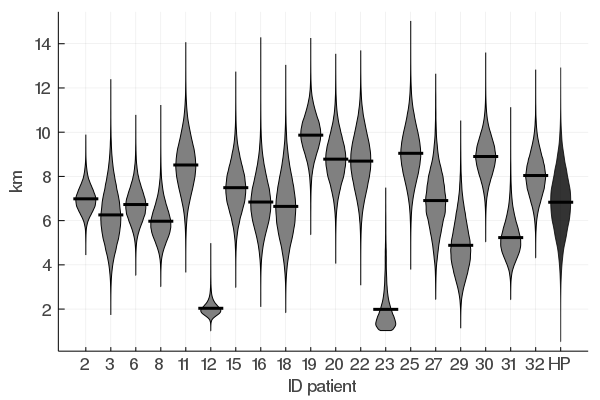}
    \end{subfigure}
    \caption{Posterior distributions of the parameters.
    HP indicates the population distribution, described by a (truncated) Gaussian distribution with mean $\mathbb{E}[\vect{\tau} |\mathcal{D}]$ and variance $\mathbb{E}[\vect{\sigma^2} |\mathcal{D}]$. 
     For parameters related to homozygous cells (blue), only patients that exhibit homozygous clones are presented. Horizontal lines indicate mean values.}
    \label{fig:suppl_distributions_param}
\end{figure}

\begin{figure}[h]
    \centering
    \includegraphics[width=0.6\textwidth]{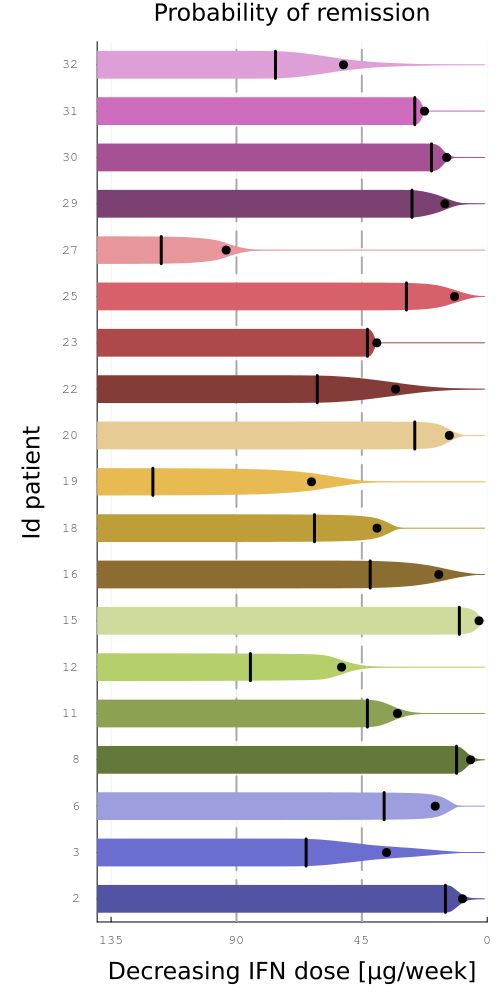}
    \caption{
    Estimation of the minimal dose that should be given to each
    patient. The x-axis indicates a dose de-escalation, from 135~$\mu$g/week (left) to 0~$\mu$g/week (right). Decreasing the dose also decreases the probability of remission. The black dots indicate the minimal estimated IFN$\alpha$ dose  $d^{(i)}_{min}$, and the vertical black lines the dose that should be given to patient $i$ such that there is a 97.5\% chance of getting a long-term molecular remission.}
    \label{fig:decreasing_dose_cohort}
\end{figure}

We also display in Fig.~\ref{fig:suppl_distributions_param} the posterior distributions of the model parameters (in complement to those presented in Fig. 5 in the main text) and in Fig.~\ref{fig:decreasing_dose_cohort} the probability of remission for each patient.

\FloatBarrier
\newpage

\section{Robustness of the post-selection inference based on a synthetic study}
\label{sec:synthetic_study}

We conduct a synthetic study to study to which extent we can draw robust conclusions from our model selection procedure and the subsequent post-selection inference. 
In this study, we simulate several datasets. 
Each dataset consists in generating - from one of the 225 models considered - the dynamics of 19 virtual patients.
Then, we try to retrieve, using our two-step model selection procedure, the model that, among the 225 potential models, is the most susceptible to having generated the dataset. %

    \subsection{Simulating datasets}

\subsubsection{Overview of the simulation process}

We aim to simulate synthetic observations in the style of the Mosca et al. data, from a given model for which the true parameters would be known. 
Let $\mathcal{M}_m$ be the model we consider, with $m\in \{1, \cdots, 225\}$. For this model, we will construct the synthetic dataset $\mathcal{D}^{(m)} = \{\mathcal{D}^{(m)}_i \}_{i \in \{1, \cdots, N\}}$ consisting of synthetic observations of $N=19$ virtual patients. The set of the 19 virtual patients generated from model $\mathcal{M}_m$ is defined in the following as the (virtual) cohort $m$.
The $i^{th}$ virtual patient of (virtual) cohort $m$ should be consistent with the $i^{th}$ patient from the (real) cohort of Mosca et al.~\cite{mosca2021}, that is:
\begin{itemize}
    \item Same variations of doses over time: $d^{(i)}: t \mapsto d^{(i)}(t)$
    \item Same observation times $t_k^{(i)}$, with $k$ indexing the observations
    \item Same numbers of colonies genotyped at time $t_k^{(i)}$: $N_{k}^{(i)}$\\
\end{itemize}
The parameter vector $\vect{\theta}^{(i,m)}$ of the $i^{th}$ virtual patient of the cohort $m$ is known. \\
Given the known parameter vector $\vect{\theta}^{(i,m)}$ and the dose input $d^{(i)}$, the cell populations dynamics over time can be calculated from the model $\mathcal{M}_m$. From this latter, we compute:
\begin{itemize}
    \item The VAF among mature cells at time $t_k^{(i)}$: $y_{\vect{\theta}^{(i,m)}}(t_k^{(i)})$
    \item The heterozygous and homozygous CF among progenitors at time $t_k^{(i)}$: $z_{\vect{\theta}^{(i,m)}, het}(t_k^{(i)})$ and $z_{\vect{\theta}^{(i,m)}, hom}(t_k^{(i)})$
    \item The minimal dose: $d_{min}^{(i,m)}$ (if applicable, see section~\ref{sec:list_models})
    \item The heterozygous and homozygous response factor, $R_{het}$ and $R_{hom}$ after 1,500 days from the beginning of the therapy (see section~\ref{sec:res_identif})
\end{itemize}
At each observation time $t_k^{(i)}$, we generate noisy VAF (pseudo)-measurements $\hat{y}_k^{(i,m)}$ using the observation model~\eqref{eq:noise_VAF}, and sample $\hat{n}_{k,het}^{(i,m)}$ and $\hat{n}_{k,hom}^{(i,m)}$ heterozygous and homozygous mutated progenitors, respectively, using the observation model described in eq.~\eqref{eq:noise_CF}. %
This results in a virtual dataset $\mathcal{D}^{(m)}_i$ for the $i^{th}$ patient of cohort $m$ and ultimately in the synthetic dataset $\mathcal{D}^{(m)}$.
Concerning the parameter values of each virtual patient $i$ from cohort $m$, they are sampled from a given probability distribution to mimic a population effect (see~\ref{sec:synth_pop_distrib} for more details).

\subsubsection{Considered models}
\label{sec:considered_models_for_selection}

We aim to study the complete model-selection process not only for one particular model (since we do not know \textit{a priori} which choice would be the most suitable) but for many of them. However, it is not conceivable to generate 225 virtual cohorts - one for each potential model - for which we would apply the model selection process (and, therefore, select among 225 models). \\
We thus have to choose a restricted number of models from which we will generate virtual cohorts. We choose to consider 20 models.
Our 225 models differ according to the relation linking the four parameters $\bar{\Delta}^*_{het}$, $\bar{\Delta}^*_{hom}$, $\bar{\gamma}^*_{het}$, and $\bar{\gamma}^*_{hom}$ to the dose $d$. Therefore, some models might be more similar than others. For example, models 10 and 28 (see Tab.~\ref{tab:list_models_1}) differ only according to how the dose would affect $\bar{\Delta}^*_{het}$, but all three other parameters would be affected by the dose in the same manner. 
On the contrary, we could expect models 10 and 90 (see Tab.\ref{tab:list_models_2}) to have different behaviours since they have no dose-response relationships in common. 
Based on that observation, among the 225 models studied, we will choose 20 models that differ - a priori - as much as possible. 
To do that, we sample 20 quadruplets $(r_1, r_2, r_3, r_4)$ in $\{0, 1, 2, 3, 4\}^2 \times \{0, 1, 2\}^2$ using a Latin Hypercube sampler (algorithm from Urquhart et al.~\cite{urquhart2020surrogate} based on the work from Bates et al.~\cite{bates2004formulation}). The value for $r_1$ will correspond to the index of the relation linking $\bar{\Delta}_{hom}^*$ to the dose $d$ ($r_1=0$ for constant, 1 for linear, 2 for affine, 3 for sigmoid, and 4 for affine sigmoid), $r_2$ will be associated to $\bar{\Delta}_{het}^*$, $r_3$ to $\bar{\gamma}_{hom}^*$ ($r_3 = 0$ for constant, 1 for inverse, and 2 for affine), and $r_4$ to $\bar{\gamma}_{het}^*$, such that the model index corresponds to $m =1+ r_4 + 3\times r_3 + 3^2 \times  r_2 + 5\times 3^2 \times r_1$ (see section~\ref{sec:list_models}).\\
Finally, the 20 models (each associated with a virtual cohort of 19 patients) are those presented in Tab.~\ref{tab:synthetic_cohort}. \\
Note that we ensure to have model 217 in the set of the 20 randomly selected models since model 217 corresponds to the one selected when running the two-step model selection procedure on the real dataset of Mosca et al.~\cite{mosca2021}.

\begin{table}[ht]
\centering
   \begin{tabular}{ | c | c | c | c | c |c | }
     \hline
      \rule{0pt}{10pt}$m$ & $\bar{\Delta}^*_{hom}$ & $\bar{\Delta}^*_{het}$ & $\bar{\gamma}^*_{hom}$ & $\bar{\gamma}^*_{het}$ \\ \hline
	2 & constant & 	constant &	constant &	inverse \\ \hline
21 &constant	&affine&	constant&	affine \\ \hline	
25 &constant	&affine	&affine&	constant \\ \hline
45 &constant&	affine sigmoid	&affine	&affine \\ \hline
52 &linear	&constant	&affine	&constant \\ \hline
60 &linear	&linear	&inverse	&affine \\ \hline
76 &linear	&sigmoid	&inverse	&constant \\ \hline
83 &linear	&affine sigmoid	&constant	&inverse \\ \hline
101 &affine	&linear	&constant	&inverse \\ \hline
106 &affine	&linear	&affine	&constant \\ \hline
113 &affine	&affine	&inverse	&inverse \\ \hline
120 &affine	&sigmoid	&constant	&affine \\ \hline
141 &sigmoid	&constant	&inverse	&affine \\ \hline
152 &sigmoid	&linear	&affine	&inverse \\ \hline
163 &sigmoid	&sigmoid	&constant	&constant \\ \hline
180 &sigmoid	&affine sigmoid	&affine	&affine \\ \hline
184 &affine sigmoid	&constant	&inverse	&constant \\ \hline
204 &affine sigmoid	&affine	&inverse	&affine \\ \hline
215 &affine sigmoid	&sigmoid	&affine	&inverse \\ \hline
217 &affine sigmoid	&affine sigmoid	&constant	&constant \\ 
     \hline
   \end{tabular}
   
\caption{Models considered for the synthetic study.}
\label{tab:synthetic_cohort}
 \end{table}

\subsubsection{Population distributions}
\label{sec:synth_pop_distrib}

Within a virtual cohort, the patients are not independent: their parameter values are sampled from population distributions that we choose beforehand, as follows:
\begin{itemize}
    \item $k_{m,het} \sim \mathcal{N}(8, 1)$ (truncated over $[1,20]$)
    \item $\eta_{hom} \sim \mathcal{U}([0, 2.5])$ (uniform)
    \item $\eta_{het} \sim \mathcal{U}([0, 1.0])$ (uniform)
    \item Concerning $\bar{\Delta}^*_{hom}$, if the relationship is:
\begin{itemize}
    \item[$\bullet$] constant: $\Delta_{hom}^* \sim \mathcal{N}(-0.3, 0.1)$ (truncated over $[-1,1]$)
    \item[$\bullet$] linear: $\Delta_{hom}^* \sim \mathcal{N}(-0.5, 0.1)$ (truncated over $[-1,0]$)
    \item[$\bullet$] affine: $\Delta_{hom}\sim \mathcal{N}(0.2, 0.1)$ (truncated over $[0,0.5]$) and $\Delta_{hom}^* \sim \mathcal{N}(-0.6, 0.1)$ (truncated over $[-1,0]$)
    \item[$\bullet$] sigmoid: $\Delta_{hom}^* \sim \mathcal{N}(7.0, 0.2)$ (truncated over $[0,10.0]$)
    \item[$\bullet$] affine sigmoid: $\Delta_{hom} \sim \mathcal{N}(0.2, 0.1)$ (truncated over $[0,0.5]$) and $\Delta_{hom}^* \sim \mathcal{N}(7.0, 0.2)$ (truncated over $[0,10]$)
\end{itemize}
\item Concerning $\bar{\Delta}^*_{het}$, if the relationship is:
\begin{itemize}
    \item[$\bullet$] constant: $\Delta_{het}^* \sim \mathcal{N}(-0.1, 0.1)$ (truncated over $[-1,1]$)
    \item[$\bullet$] linear: $\Delta_{het}^* \sim \mathcal{N}(-0.3, 0.1)$ (truncated over $[-1,0]$)
    \item[$\bullet$] affine: $\Delta_{het}\sim \mathcal{N}(0.1, 0.1)$ (truncated over $[0,0.5]$) and $\Delta_{het}^* \sim \mathcal{N}(-0.3, 0.1)$ (truncated over $[-1,0]$)
    \item[$\bullet$] sigmoid: $\Delta_{het}^* \sim \mathcal{N}(4.0, 0.2)$ (truncated over $[0,10.0]$)
    \item[$\bullet$] affine sigmoid: $\Delta_{het} \sim \mathcal{N}(0.1, 0.1)$ (truncated over $[0,0.5]$) and $\Delta_{het}^* \sim \mathcal{N}(4.0, 0.2)$ (truncated over $[0,10]$)
\end{itemize}
\item Concerning $\bar{\gamma}^*_{hom}$, if the relationship is:
\begin{itemize}
    \item[$\bullet$] constant: $\tau_{hom}^* = 1/ \gamma_{hom}^* \sim \mathcal{N}(50,2)$ (truncated over $[10, 300]$)
    \item[$\bullet$] inverse: $\tau_{hom}^*\sim \mathcal{N}(-230,2)$ (truncated over $[-290, 0]$)
    \item[$\bullet$] affine: $\gamma_{hom}^*\sim\mathcal{N}(22/300, 0.01)$ (truncated over $[0, 29/300]$)
\end{itemize}
\item Concerning $\bar{\gamma}^*_{het}$, if the relationship is:
\begin{itemize}
    \item[$\bullet$] constant: $\tau_{het}^* = 1/ \gamma_{het}^* \sim \mathcal{N}(200,2)$ (truncated over $[10, 300]$)
    \item[$\bullet$] inverse: $\tau_{het}^*\sim \mathcal{N}(-100,2)$ (truncated over $[-290, 0]$)
    \item[$\bullet$] affine: $\gamma_{het}^*\sim\mathcal{N}(10/300, 0.01)$ (truncated over $[0, 29/300]$) \\
\end{itemize}
\end{itemize}
All previous population distributions involve (truncated) Gaussian distributions $\mathcal{N}(\mu, \sigma)$ of mean $\mu$ and variance $\sigma^2$.
When running the hierarchical Bayesian inference procedure, the mean and variance of these laws have to be estimated as hyperparameters.

\subsubsection{Virtual cohorts}

\begin{figure}[h]
    \centering

    \begin{subfigure}[b]{0.26\textwidth}
        \includegraphics[width=\textwidth]{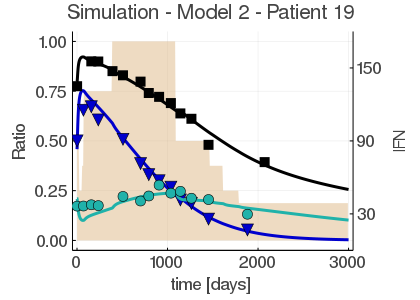}
    \end{subfigure}
    \begin{subfigure}[b]{0.26\textwidth}
        \includegraphics[width=\textwidth]{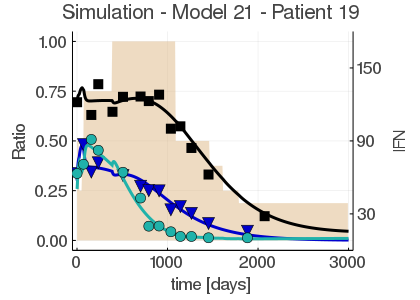}
    \end{subfigure}
    \begin{subfigure}[b]{0.26\textwidth}
        \includegraphics[width=\textwidth]{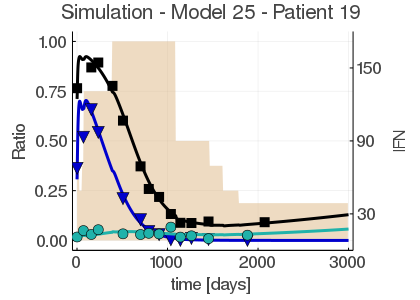}
    \end{subfigure}
    \begin{subfigure}[b]{0.26\textwidth}
        \includegraphics[width=\textwidth]{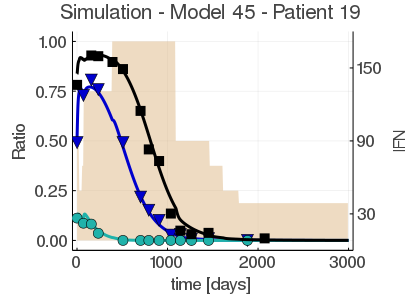}
    \end{subfigure}
    \begin{subfigure}[b]{0.26\textwidth}
        \includegraphics[width=\textwidth]{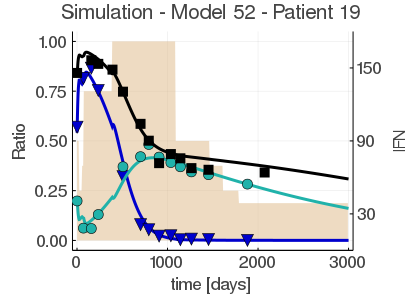}
    \end{subfigure}
    \begin{subfigure}[b]{0.26\textwidth}
        \includegraphics[width=\textwidth]{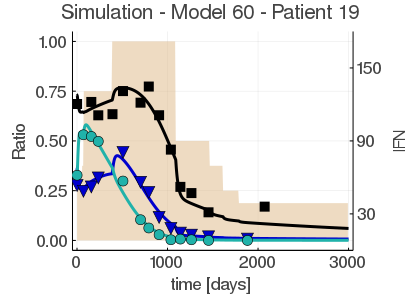}
    \end{subfigure}
    \begin{subfigure}[b]{0.26\textwidth}
        \includegraphics[width=\textwidth]{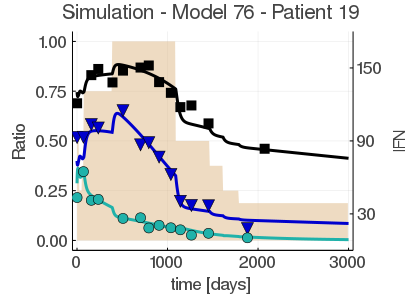}
    \end{subfigure}
    \begin{subfigure}[b]{0.26\textwidth}
        \includegraphics[width=\textwidth]{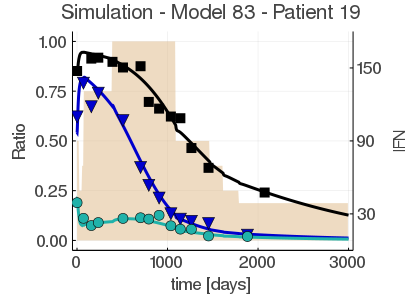}
    \end{subfigure}
    \begin{subfigure}[b]{0.26\textwidth}
        \includegraphics[width=\textwidth]{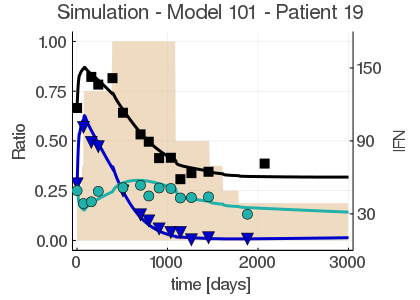}
    \end{subfigure}
    \begin{subfigure}[b]{0.26\textwidth}
        \includegraphics[width=\textwidth]{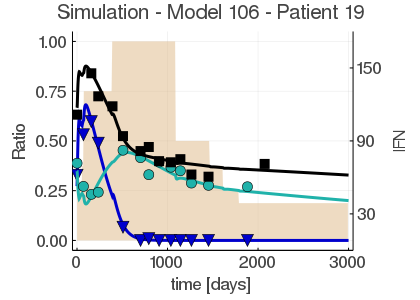}
    \end{subfigure}
    \begin{subfigure}[b]{0.26\textwidth}
        \includegraphics[width=\textwidth]{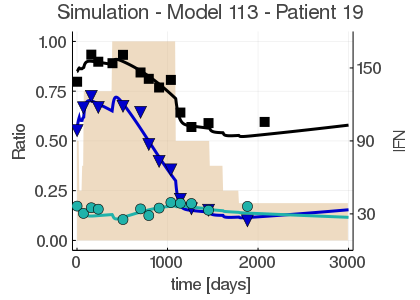}
    \end{subfigure}
    \begin{subfigure}[b]{0.26\textwidth}
        \includegraphics[width=\textwidth]{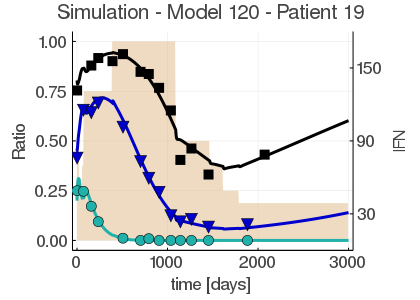}
    \end{subfigure}
    \begin{subfigure}[b]{0.26\textwidth}
        \includegraphics[width=\textwidth]{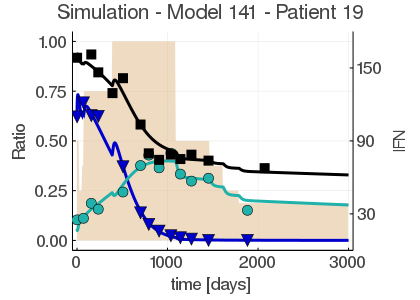}
    \end{subfigure}
    \begin{subfigure}[b]{0.26\textwidth}
        \includegraphics[width=\textwidth]{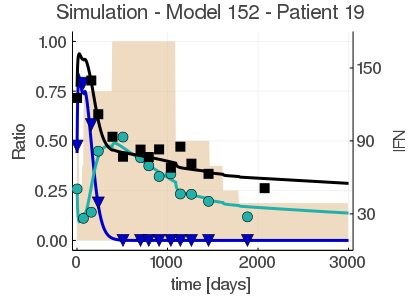}
    \end{subfigure}
    \begin{subfigure}[b]{0.26\textwidth}
        \includegraphics[width=\textwidth]{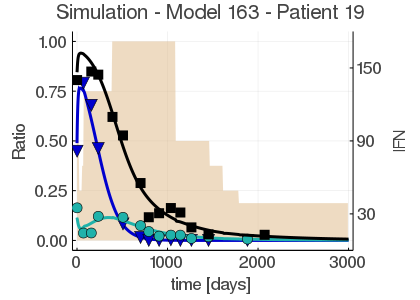}
    \end{subfigure}
    \begin{subfigure}[b]{0.26\textwidth}
        \includegraphics[width=\textwidth]{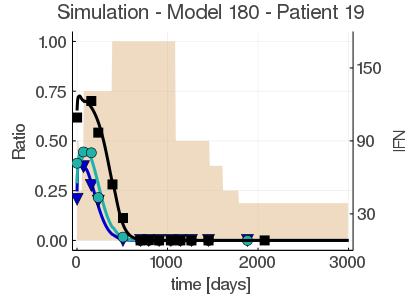}
    \end{subfigure}
    \begin{subfigure}[b]{0.26\textwidth}
        \includegraphics[width=\textwidth]{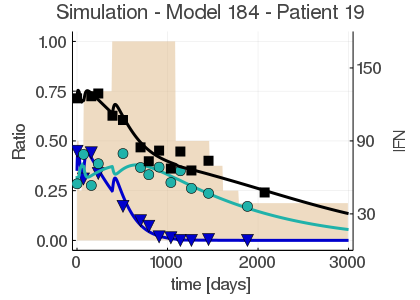}
    \end{subfigure}
    \begin{subfigure}[b]{0.26\textwidth}
        \includegraphics[width=\textwidth]{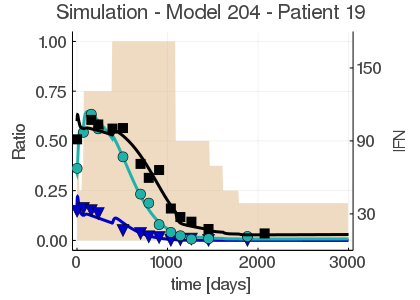}
    \end{subfigure}
    \begin{subfigure}[b]{0.26\textwidth}
        \includegraphics[width=\textwidth]{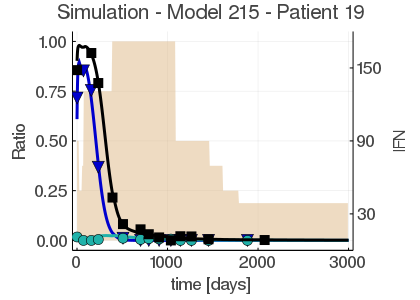}
    \end{subfigure}
    \begin{subfigure}[b]{0.26\textwidth}
        \includegraphics[width=\textwidth]{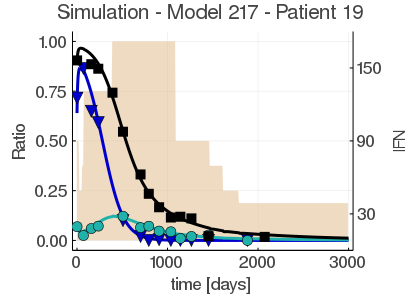}
    \end{subfigure}
    \caption{Simulated cell population dynamics of the $19^{th}$ virtual patient for each of the 20 cohorts considered. Virtual patients $i=19$ correspond to the true patient \#32 from Mosca et al.~\cite{mosca2021} (see Fig.~\ref{fig:Dyn_estimated}). The black line corresponds to the dynamics of the VAF in mature cells, and the blue and green lines to the dynamics of the homozygous and heterozygous CF in progenitors, respectively. Each $i^{th}$ virtual patient will have the same dose variations (in brown) and observation times (the observations correspond to the black squares, blue triangles, and green circles) as the associated $i^{th}$ patient from Mosca et al. }
    \label{fig:virtual_19_all_cohorts}
\end{figure}

\begin{figure}[h]
    \centering
    \begin{subfigure}[b]{0.26\textwidth}
        \includegraphics[width=\textwidth]{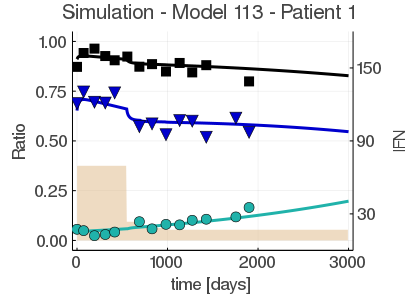}
    \end{subfigure}
    \begin{subfigure}[b]{0.26\textwidth}
        \includegraphics[width=\textwidth]{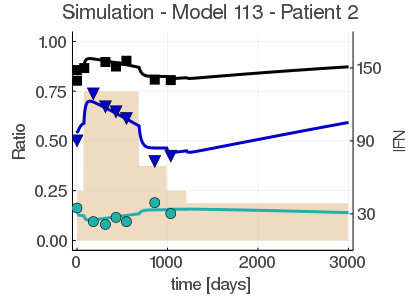}
    \end{subfigure}
    \begin{subfigure}[b]{0.26\textwidth}
        \includegraphics[width=\textwidth]{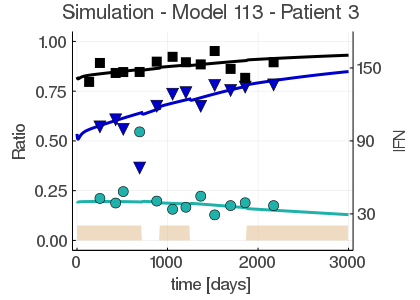}
    \end{subfigure}
    \begin{subfigure}[b]{0.26\textwidth}
        \includegraphics[width=\textwidth]{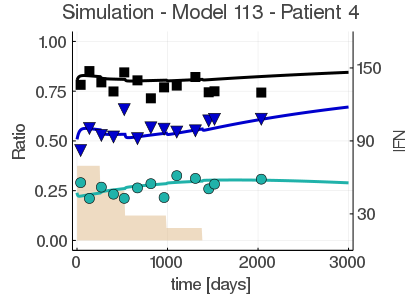}
    \end{subfigure}
    \begin{subfigure}[b]{0.26\textwidth}
        \includegraphics[width=\textwidth]{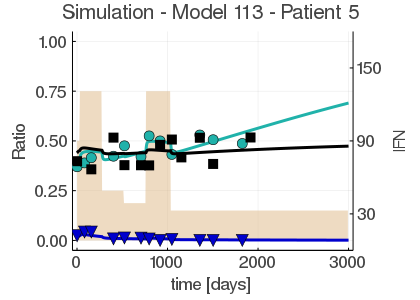}
    \end{subfigure}
    \begin{subfigure}[b]{0.26\textwidth}
        \includegraphics[width=\textwidth]{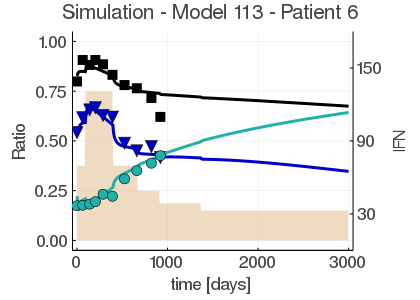}
    \end{subfigure}
    \begin{subfigure}[b]{0.26\textwidth}
        \includegraphics[width=\textwidth]{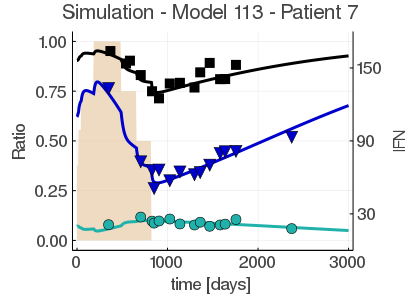}
    \end{subfigure}
    \begin{subfigure}[b]{0.26\textwidth}
        \includegraphics[width=\textwidth]{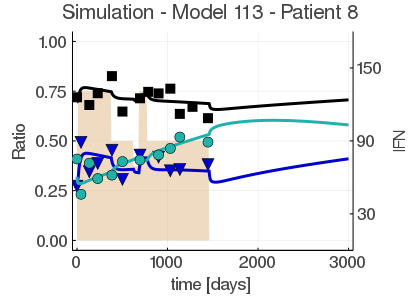}
    \end{subfigure}
    \begin{subfigure}[b]{0.26\textwidth}
        \includegraphics[width=\textwidth]{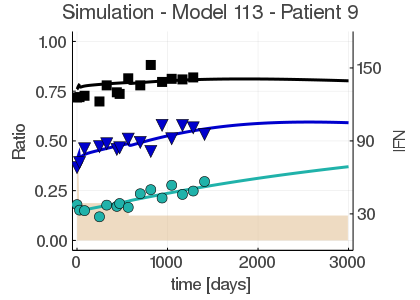}
    \end{subfigure}
    \begin{subfigure}[b]{0.26\textwidth}
        \includegraphics[width=\textwidth]{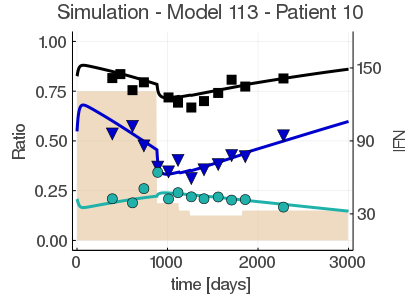}
    \end{subfigure}
    \begin{subfigure}[b]{0.26\textwidth}
        \includegraphics[width=\textwidth]{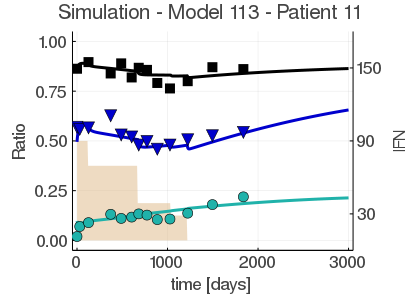}
    \end{subfigure}
    \begin{subfigure}[b]{0.26\textwidth}
        \includegraphics[width=\textwidth]{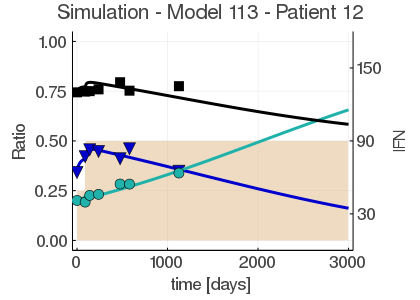}
    \end{subfigure}
    \begin{subfigure}[b]{0.26\textwidth}
        \includegraphics[width=\textwidth]{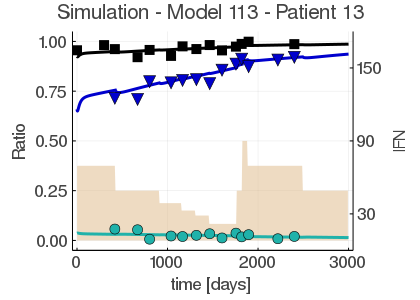}
    \end{subfigure}
    \begin{subfigure}[b]{0.26\textwidth}
        \includegraphics[width=\textwidth]{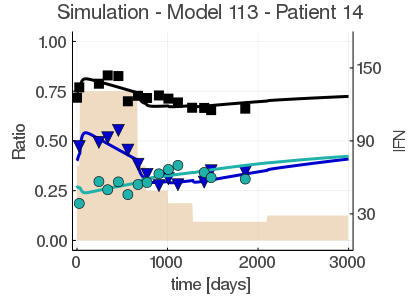}
    \end{subfigure}
    \begin{subfigure}[b]{0.26\textwidth}
        \includegraphics[width=\textwidth]{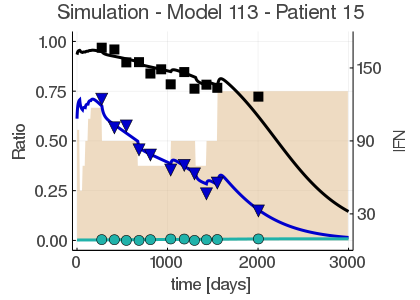}
    \end{subfigure}
    \begin{subfigure}[b]{0.26\textwidth}
        \includegraphics[width=\textwidth]{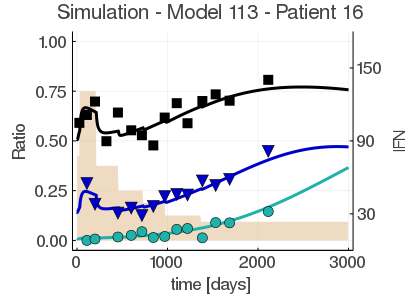}
    \end{subfigure}
    \begin{subfigure}[b]{0.26\textwidth}
        \includegraphics[width=\textwidth]{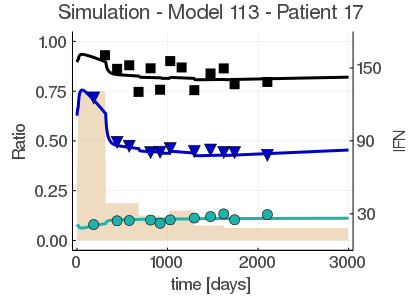}
    \end{subfigure}
    \begin{subfigure}[b]{0.26\textwidth}
        \includegraphics[width=\textwidth]{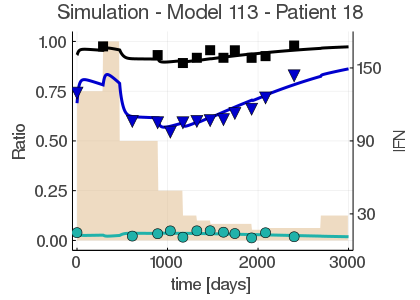}
    \end{subfigure}
    \begin{subfigure}[b]{0.26\textwidth}
        \includegraphics[width=\textwidth]{model113_patient19}
    \end{subfigure}
    
    \caption{Simulated dynamics and simulated observations for each virtual patient from cohort 113 (that is, the data are generated from the model $m=113$).}
    \label{fig:virtual_cohort_113}
\end{figure}

Ultimately, we simulate 20 cohorts of 19 virtual patients, that is, 380 virtual patients. 
In Fig.~\ref{fig:virtual_19_all_cohorts}, we display the true dynamics and the noisy pseudo-observations of the 19$^{th}$ patient in each virtual cohort. These virtual patients are associated with patient \#32 from the real cohort of Mosca et al.~\cite{mosca2021} (same dose variations and observation times as patient \#32). 
In Fig.~\ref{fig:virtual_cohort_113}, we display the dynamics of each virtual patient from the cohort $m=113$.\\

For some of the studied models, depending on the parameter values, there exists a range of doses for which there is no remission (see section~\ref{sec:list_models}). When the model at the origin of a given virtual cohort $m$ falls in that case, there exists for each virtual patient $i$ a minimal IFN$\alpha$ dose $d_{min}^{(i,m)}$ required to reach remission, which we indicate in Tab.~\ref{tab:synthetic_minimal_ind_dose}.
This minimal dose is a quantity of interest that we aim to estimate accurately to provide relevant guidelines to the clinicians.

\begin{table}[]
    \centering
    \footnotesize\
    \begin{tabular}{|c|c|c|c|c|c|c|c|c|c|c|c|c|c|c|c|c|c|c|c|c|}
    \hline
\diagbox[width=1cm]{$i$}{$m$}		&	21	&	25	&	45	&	83	&	101	&	106	&	113	&	120	&	180	&	184	&	204	&	215	&	217	\\ \hline
1		&	0.099	&	0.294	&	0.077	&	0.127	&	0.336	&	0.229	&	0.54	&	0.26	&	0.071	&	0.101	&	0.123	&	0.063	&	0.123	\\ \hline
2		&	0.885	&	0.508	&	0.097	&	0.092	&	0.284	&	0.545	&	0.443	&	0.238	&	0.093	&	Inf	&	$> 1$	&	0.021	&	0.037	\\ \hline
3		&	$> 1$	&	$> 1$	&	0.068	&	0.039	&	0.426	&	0.515	&	0.638	&	0.143	&	0.063	&	0.062	&	0.429	&	0.035	&	0.066	\\ \hline
4		&	$> 1$	&	0.064	&	0.047	&	0.065	&	0.124	&	0.257	&	0.711	&	0.53	&	0.0	&	0.026	&	0.06	&	0.065	&	0.03	\\ \hline
5	&	0.371	&	$> 1$	&	0.007	&	0.056	&	0.2	&	0.043	&	0.591	&	0.325	&	0.04	&	0.03	&	0.474	&	0.054	&	0.024	\\ \hline
6	&	$> 1$	&	0.525	&	0.079	&	0.019	&	0.488	&	0.393	&	$> 1$	&	0.517	&	0.051	&	0.039	&	0.222	&	0.026	&	0.042	\\ \hline
7		&	0.313	&	0.288	&	0.054	&	0.016	&	0.623	&	0.398	&	0.343	&	0.031	&	0.109	&	0.047	&	0.172	&	0.006	&	0.098	\\ \hline
8	&	0.432	&	0.087	&	0.078	&	0.035	&	0.416	&	0.319	&	$> 1$	&	0.374	&	0.11	&	Inf	&	0.659	&	0.052	&	0.052	\\ \hline
9	&	0.313	&	0.194	&	0.008	&	0.063	&	0.398	&	0.361	&	0.862	&	0.338	&	0.058	&	0.052	&	0.479	&	0.022	&	0.041	\\ \hline
10		&	0.551	&	0.165	&	0.041	&	0.03	&	0.644	&	0.246	&	0.362	&	0.285	&	0.01	&	0.035	&	0.523	&	0.085	&	0.077	\\ \hline
11		&	0.419	&	0.097	&	0.038	&	0.04	&	0.618	&	0.219	&	0.781	&	0.291	&	0.06	&	0.042	&	0.052	&	0.025	&	0.102	\\ \hline
12		&	0.524	&	0.296	&	0.111	&	0.029	&	0.083	&	0.197	&	$> 1$	&	0.133	&	0.076	&	0.048	&	0.476	&	0.024	&	0.032	\\ \hline
13		&	0.425	&	0.134	&	0.003	&	0.12	&	0.34	&	0.051	&	0.638	&	0.028	&	0.1	&	0.033	&	0.481	&	0.02	&	0.062	\\ \hline
14		&	0.351	&	0.024	&	0.07	&	0.067	&	0.157	&	0.476	&	0.548	&	0.21	&	0.015	&	0.042	&	0.494	&	0.061	&	0.127	\\ \hline
15	&	0.628	&	0.449	&	0.101	&	0.088	&	0.29	&	0.507	&	0.581	&	0.354	&	0.081	&	0.055	&	0.415	&	0.04	&	0.067	\\ \hline
16		&	0.399	&	0.127	&	0.046	&	0.026	&	0.466	&	0.434	&	$> 1$	&	0.312	&	0.058	&	0.083	&	$> 1$	&	0.058	&	0.098	\\ \hline
17		&	$> 1$	&	0.687	&	0.043	&	0.019	&	0.544	&	0.487	&	0.137	&	0.418	&	0.018	&	0.083	&	0.045	&	0.065	&	0.053	\\ \hline
18		&	0.344	&	0.456	&	0.052	&	0.018	&	0.305	&	0.415	&	0.465	&	0.347	&	0.092	&	Inf	&	0.74	&	0.076	&	0.072	\\ \hline
19		&	0.251	&	0.416	&	0.12	&	0.017	&	0.246	&	0.303	&	0.322	&	0.297	&	0.056	&	0.043	&	0.217	&	0.064	&	0.055	\\ \hline
    \end{tabular}

    \normalsize
    \caption{For some of the models used to generate the virtual cohort (index $m$), according to the individual (index $i$) parameter values, there is a minimal dose $d_{min}^{(i,m)}$ under which one malignant clone (either heterozygous or homozygous) would theoretically continue to expand. The table lists these (normalized, that is, divided by 180 $\mu$g/week) minimal doses. When the minimal dose would be equal to a value higher than the maximal dose  of 180 $\mu$g/week, we indicate $>1$. When a model has one of its dose-relationship which is constant (e.g. model 184), if the individual value $\Delta^*>0$, then no remission would be possible. In that case, we indicate Inf.}
    \label{tab:synthetic_minimal_ind_dose}
\end{table}

\FloatBarrier
\subsection{First step of the model selection (AIC-based)}

\subsubsection{Capacity to (pre)select the true model }

\begin{figure}[h]
    \centering
    \begin{subfigure}[b]{0.44\textwidth}
        \includegraphics[width=\textwidth]{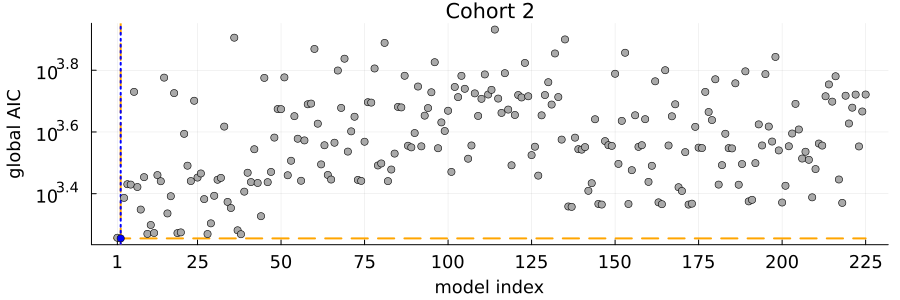}
    \end{subfigure}
    \begin{subfigure}[b]{0.44\textwidth}
        \includegraphics[width=\textwidth]{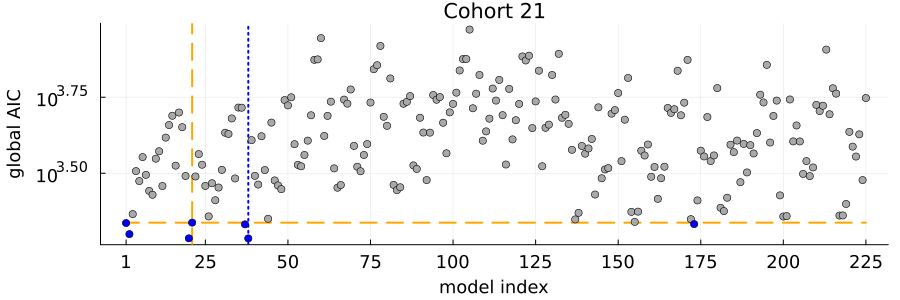}
    \end{subfigure}
\begin{subfigure}[b]{0.44\textwidth}
        \includegraphics[width=\textwidth]{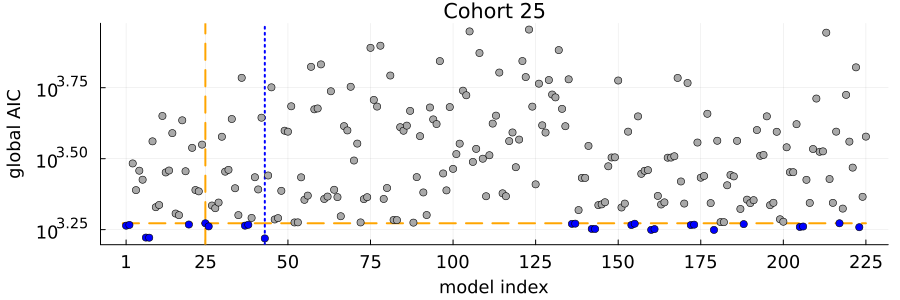}
    \end{subfigure}
    \begin{subfigure}[b]{0.44\textwidth}
        \includegraphics[width=\textwidth]{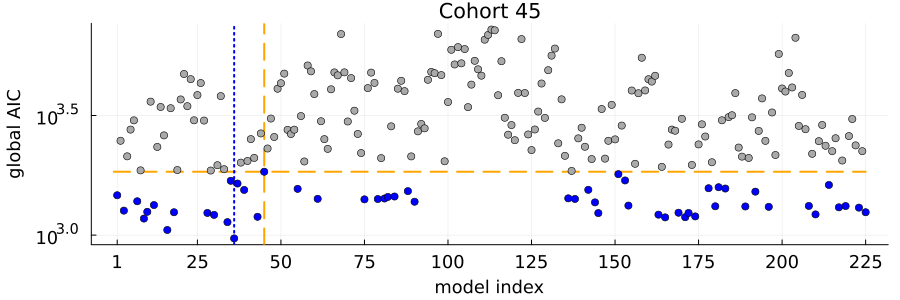}
    \end{subfigure}
    \begin{subfigure}[b]{0.44\textwidth}
        \includegraphics[width=\textwidth]{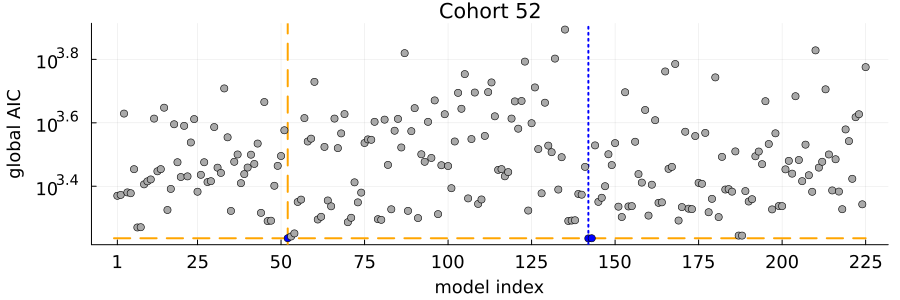}
    \end{subfigure}
    \begin{subfigure}[b]{0.44\textwidth}
        \includegraphics[width=\textwidth]{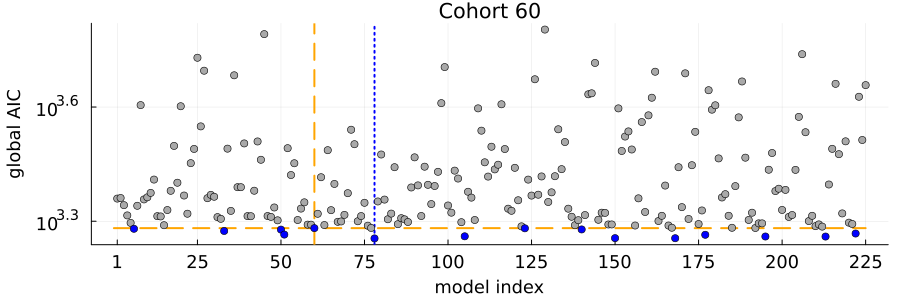}
    \end{subfigure}
    \begin{subfigure}[b]{0.44\textwidth}
        \includegraphics[width=\textwidth]{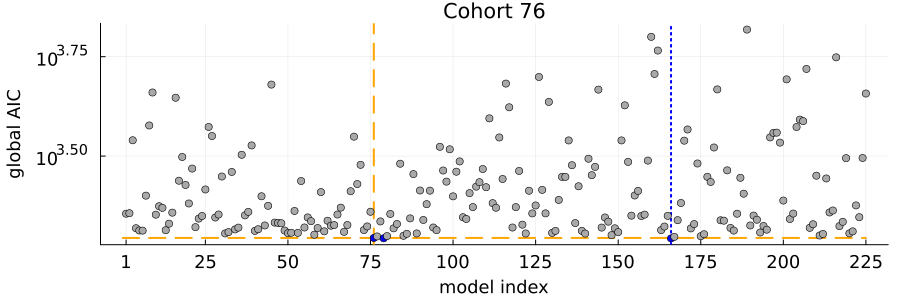}
    \end{subfigure}
    \begin{subfigure}[b]{0.44\textwidth}
        \includegraphics[width=\textwidth]{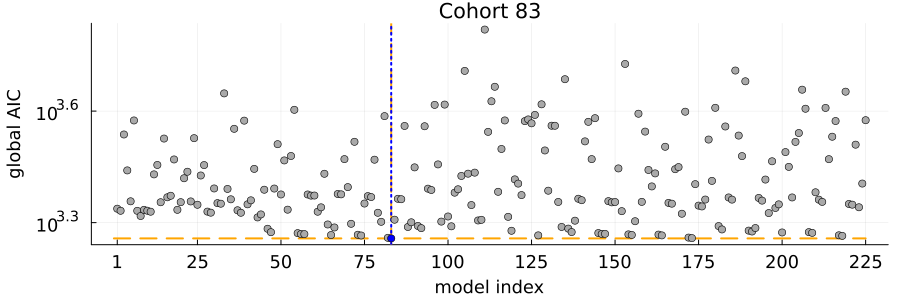}
    \end{subfigure}
    \begin{subfigure}[b]{0.44\textwidth}
        \includegraphics[width=\textwidth]{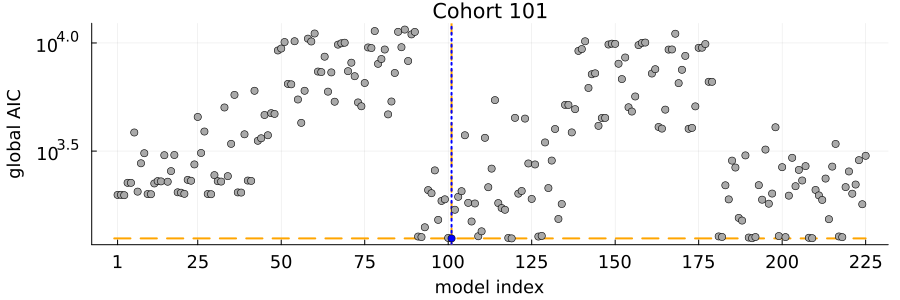}
    \end{subfigure}
    \begin{subfigure}[b]{0.44\textwidth}
        \includegraphics[width=\textwidth]{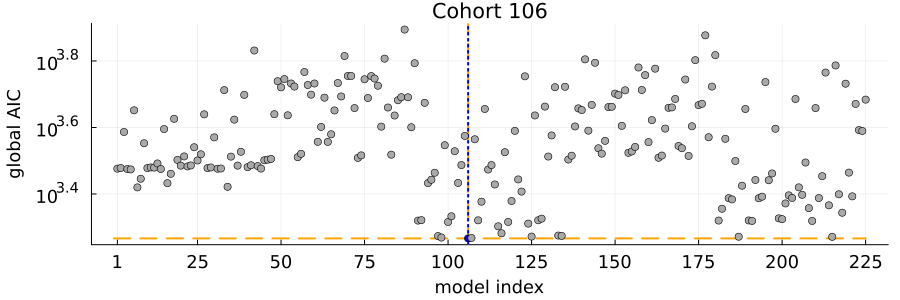}
    \end{subfigure}
    \begin{subfigure}[b]{0.44\textwidth}
        \includegraphics[width=\textwidth]{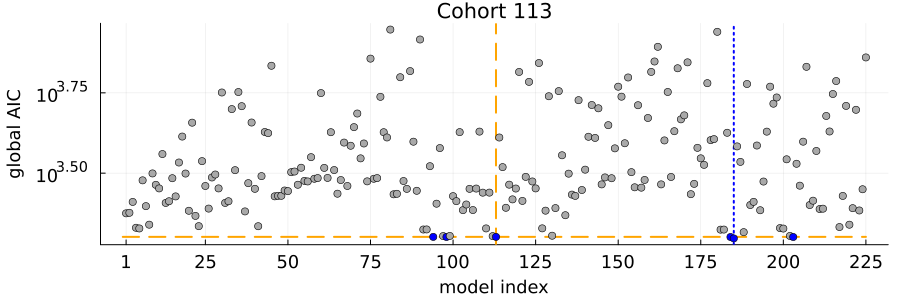}
    \end{subfigure}
    \begin{subfigure}[b]{0.44\textwidth}
        \includegraphics[width=\textwidth]{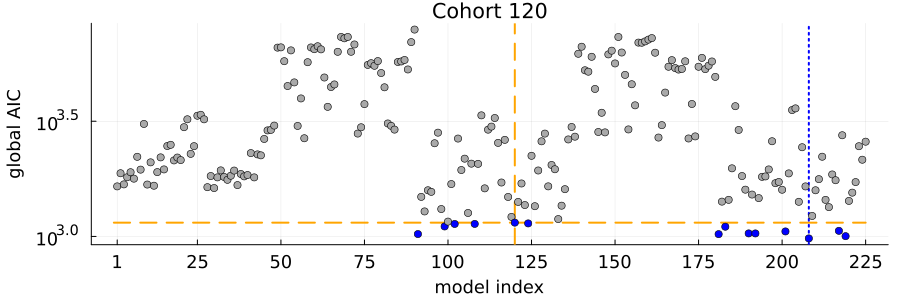}
    \end{subfigure}
\begin{subfigure}[b]{0.44\textwidth}
        \includegraphics[width=\textwidth]{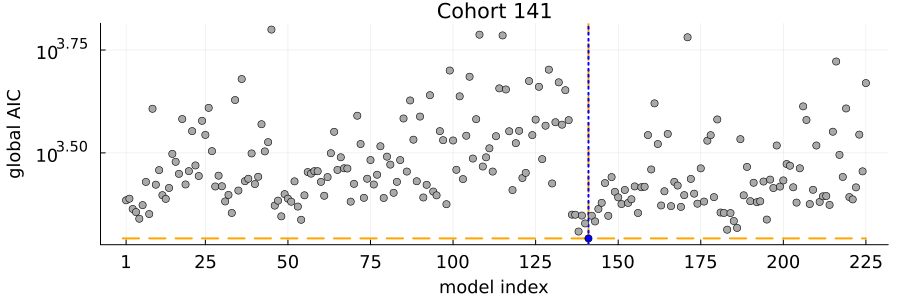}
    \end{subfigure}
    \begin{subfigure}[b]{0.44\textwidth}
        \includegraphics[width=\textwidth]{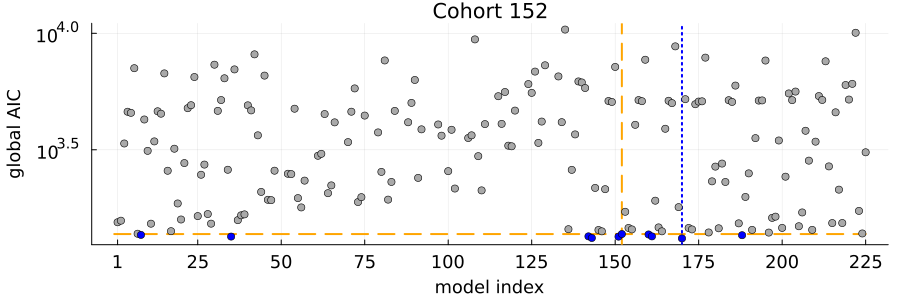}
    \end{subfigure}
    \begin{subfigure}[b]{0.44\textwidth}
        \includegraphics[width=\textwidth]{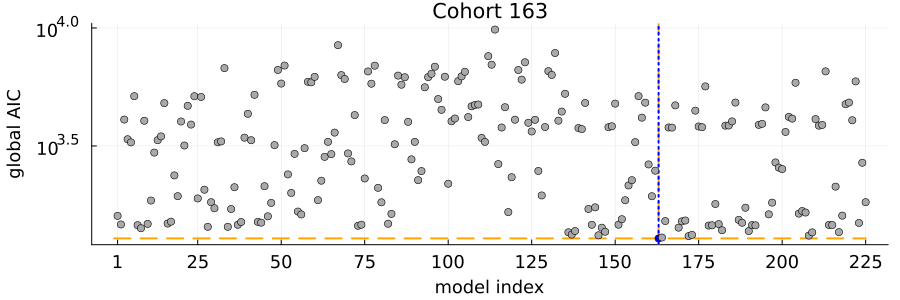}
    \end{subfigure}
    \begin{subfigure}[b]{0.44\textwidth}
        \includegraphics[width=\textwidth]{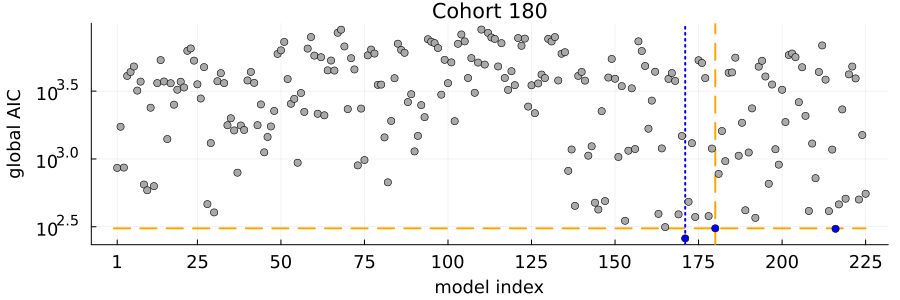}
    \end{subfigure}
    \begin{subfigure}[b]{0.44\textwidth}
        \includegraphics[width=\textwidth]{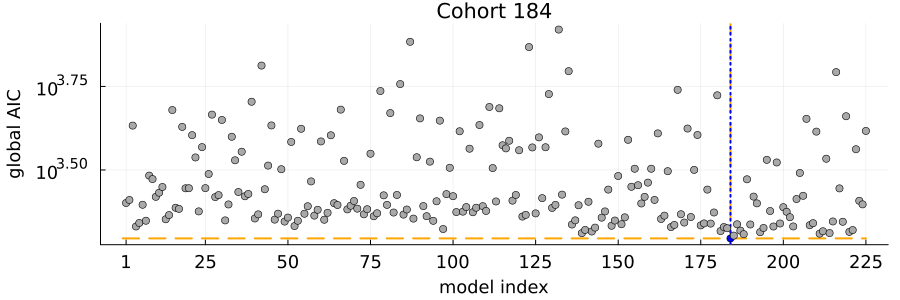}
    \end{subfigure}
    \begin{subfigure}[b]{0.44\textwidth}
        \includegraphics[width=\textwidth]{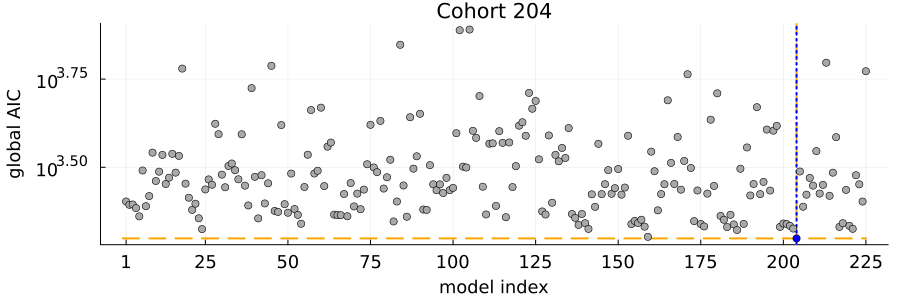}
    \end{subfigure}
    \begin{subfigure}[b]{0.44\textwidth}
        \includegraphics[width=\textwidth]{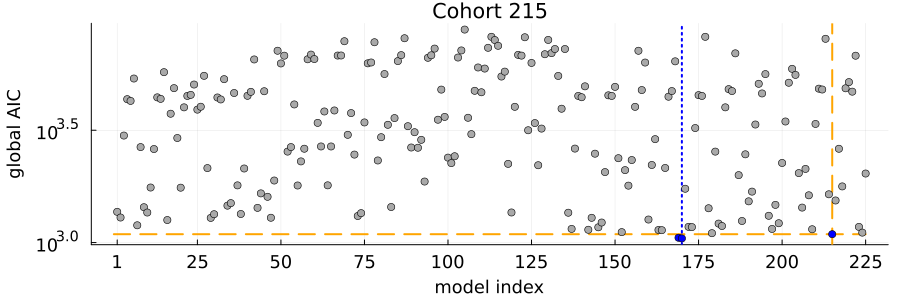}
    \end{subfigure}
    \begin{subfigure}[b]{0.44\textwidth}
        \includegraphics[width=\textwidth]{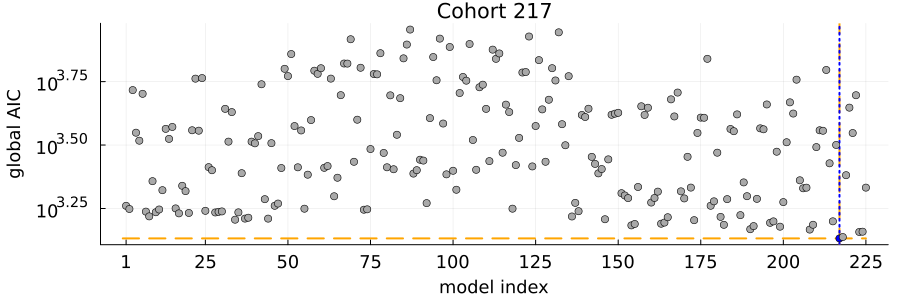}
    \end{subfigure}
    \caption{Global AIC value (y-axis) of each model $j$ (x-axis) for each virtual cohort $m$. The orange dashed line corresponds to the model $m$ used to generate the data. The vertical dot blue line indicates the best model. The blue points correspond to the models which perform better than the true one.}
    \label{fig:synthetic_global_AIC}
\end{figure}

\begin{figure}
    \centering
    \includegraphics[width=0.7\textwidth]{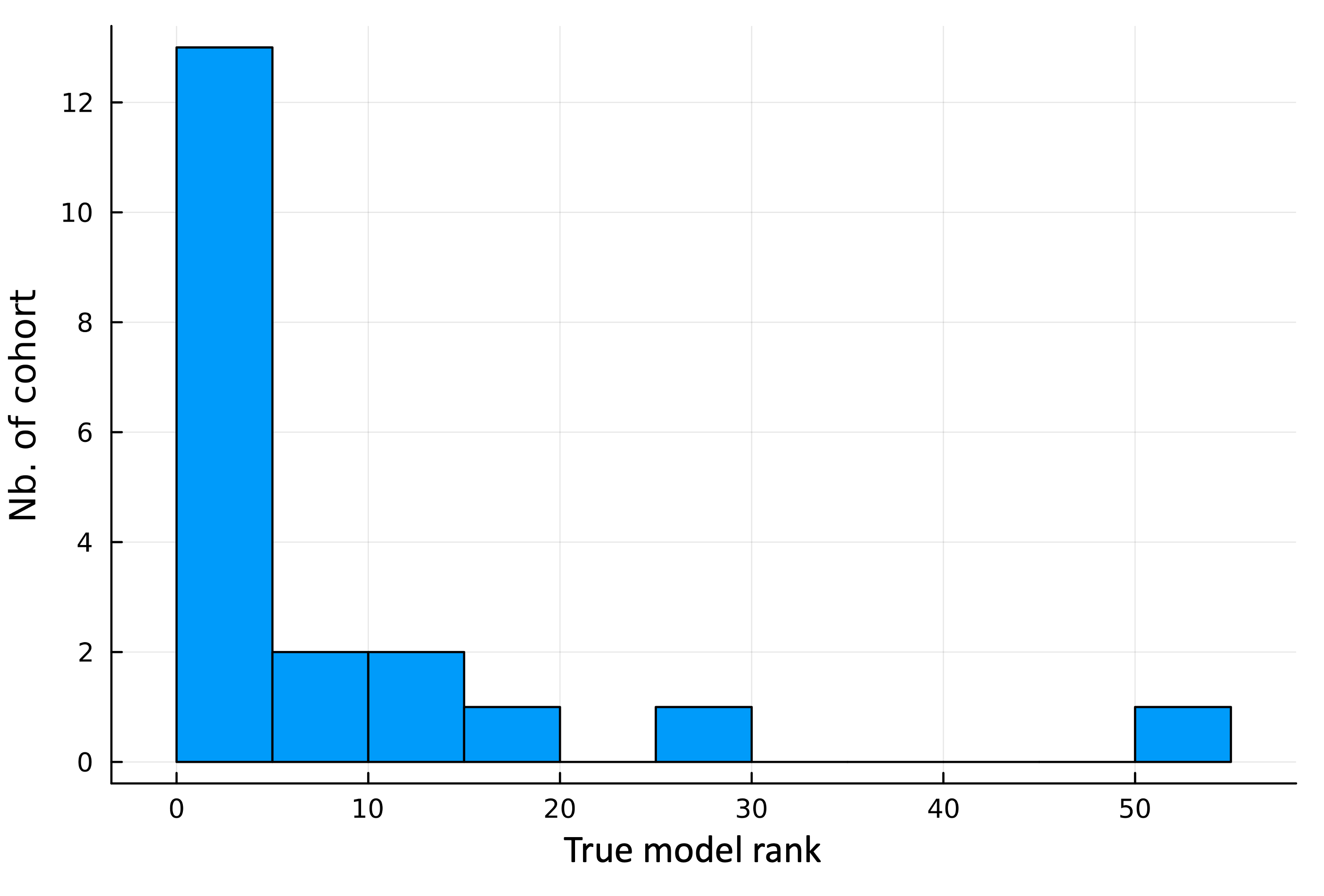}
    \caption{Distribution of the true model ranks over the 225 models studied, sorted according to the global AIC value.}
    \label{fig:synthetic_range_AIC}
\end{figure}

For each of the 20 virtual cohorts (for which we know the true generative model), we first apply the model selection based on AIC as described in §2.2.2 in the main text. This first step can be considered a pre-selection step before running the hierarchical Bayesian inference procedure.\\
First, we have to compute the $AIC_{i, m,j}$ of model $j$ for the $i$th patient of cohort $m$ (Fig.~\ref{fig:synthetic_ind_AIC}). For that purpose, we have to numerically run the CMA-ES algorithm at least $19\times 20\times 225 = 85,500$ times. 
Yet, even if efficient, there is still a risk that the CMA-ES procedure does not converge to the maximum likelihood. To minimize that risk, we execute this procedure five times, randomly changing the setting of the algorithm (starting point and initial covariance matrix). In the end, we have run the CMA-ES procedure 427,500 times. \\
Secondly, for each model $j$ and each virtual cohort $m$, we compute a global $AIC_{m,j}$ value, as displayed in Fig.~\ref{fig:synthetic_global_AIC}.
Contrary to what we obtained when analyzing the true data, for which three models stood out, we get a higher number of potential candidates overall. We should run the hierarchical estimation procedure for each to further discriminate between these good models. However, it would result in too many models. Therefore, we consider only the three best models for each virtual cohort, the same number as what we did for the true cohort.
For each virtual cohort, the three selected models are presented in Tab.~\ref{tab:synthetic_AIC_selection}.
We see that, with this first selection based on the AIC, the true model is the best one (over 225 potential candidates) in 45\% of the cases (9 over 20 virtual cohorts) and within the best three models in 65\% of the cases (13 over 20 virtual cohorts). We also display in Fig.~\ref{fig:synthetic_range_AIC} the distribution of the true model ranks (over the 20 studied virtual cohorts). There are only two cohorts ($m=45$ and $m=25$) for which the true model is not among the 10\% best models. \\
Even if the true model is among the first three models in 65\% of the cases - and therefore included in the set of models for which the parameters will be estimated within the hierarchical Bayesian framework - this still lets seven cohorts for which we will not be able to select the true model. 
However, as we introduced in~\ref{sec:considered_models_for_selection} and detailed in section~\ref{sec:list_models}, some models may be more similar than others based on the dose-response relationships they have in common. In other words, the selected subsets of the top three models might not contain the true models but still be close to it, that is, have most of the true dose-response relationships in common with the true model. We explore this point in the next paragraph.

\begin{table}[]
    \centering
    
    \begin{tabular}{|c|c|c|c|c|c|c|}
    \hline
\multirow{2}{*}{$m$} & \multicolumn{3}{c|}{Selected models} & True model & \multicolumn{2}{c|}{Nb. of actual relations found}  \\ 
\cline{2-4} \cline{6-7}
& First & Second & Third & rank & in the first & for the top three \\ \hline
2	&	2	&	1	&	\textbf{38}	&	1	&	4	&	4	\\ \hline
\textbf{21}	&	\textbf{38}	&	\textbf{20}	&	2	&	7	&	2	&	3	\\ \hline
\textbf{25}	&	\textbf{43}	&	8	&	7	&	26	&	3	&	3	\\ \hline
\textbf{45}	&	36	&	16	&	34	&	54	&	3	&	3	\\ \hline
52	&	142	&	143	&	52	&	3	&	3	&	4	\\ \hline
60	&	78	&	168	&	150	&	15	&	3	&	4	\\ \hline
76	&	166	&	79	&	76	&	3	&	3	&	4	\\ \hline
\textbf{83}	&	\textbf{83}	&	\textbf{173}	&	\textbf{82}	&	1	&	4	&	4	\\ \hline
\textbf{101}	&	\textbf{101}	&	\textbf{119}	&	\textbf{191}	&	1	&	4	&	4	\\ \hline
\textbf{106}	&	\textbf{106}	&	\textbf{107}	&	\textbf{98}	&	1	&	4	&	4	\\ \hline
\textbf{113}	&	\textbf{185}	&	\textbf{203}	&	\textbf{184}	&	6	&	2	&	3	\\ \hline
\textbf{120}	&	\textbf{208}	&	\textbf{219}	&	\textbf{181}	&	14	&	2	&	3	\\ \hline
141	&	141	&	138	&	\textbf{183}	&	1	&	4	&	4	\\ \hline
152	&	170	&	143	&	151	&	10	&	3	&	4	\\ \hline
163	&	163	&	164	&	\textbf{172}	&	1	&	4	&	4	\\ \hline
\textbf{180}	&	171	&	\textbf{216}	&	\textbf{180}	&	3	&	3	&	4	\\ \hline
\textbf{184}	&	\textbf{184}	&	\textbf{185}	&	\textbf{188}	&	1	&	4	&	4	\\ \hline
\textbf{204}	&	\textbf{204}	&	\textbf{159}	&	\textbf{186}	&	1	&	4	&	4	\\ \hline
\textbf{215}	&	170	&	169	&	\textbf{215}	&	3	&	3	&	4	\\ \hline
\textbf{217}	&	\textbf{217}	&	\textbf{218}	&	\textbf{223}	&	1	&	4	&	4	\\ \hline
    \end{tabular}
    
    \caption{Results of the first step model selection procedure. For each virtual cohort $m$ - where $m$ corresponds to the index of the model used to simulate the cohort - we indicate the first three best models based on the global AIC value. In addition, among the 225 models studied, we indicate the rank of the true model (when ranking the models according to their global AIC value). The 225 models studied differ regarding the dose-response relationship linking the dose $d$ to the four parameters $\bar{\Delta}^*_{het}$, $\bar{\Delta}^*_{hom}$, $\bar{\gamma}^*_{het}$ and $\bar{\gamma}^*_{hom}$. We indicate the number of relations found to be true in the best model (4 means that the best model is the true one) and within the three best models. \\
Some models allow the existence of a minimal IFN$\alpha$ dose $d_{min}$; they are highlighted in bold. 
}
   \label{tab:synthetic_AIC_selection}
\end{table}

\FloatBarrier
\subsubsection{Capacity to (pre)select the true dose-response relationships}

Here, we explore to which extent the best pre-selected models, if not the true ones, are close to it. To evaluate this point, we consider for each virtual cohort the best model (based on the global AIC). Tab.~\ref{tab:synthetic_AIC_selection} indicates the number of true dose-response relationships correctly retrieved. For 17 virtual cohorts over 20, at least 3 of the relations found are the good ones. To investigate that point in more detail, we study  each of the four quantities $\bar{\Delta}^*_{het}$, $\bar{\Delta}^*_{hom}$, $\bar{\gamma}^*_{het}$ and $\bar{\gamma}^*_{hom}$ separately and study if some particular dose-response relationships are less retrieved than others. We present in Tab.~\ref{tab:confusion_matrix_bestAIC_Delta_het}, \ref{tab:confusion_matrix_bestAIC_Delta_hom}, \ref{tab:confusion_matrix_bestAIC_gamma_het}, and~\ref{tab:confusion_matrix_bestAIC_gamma_hom}  the confusion matrices associated to $\bar{\Delta}^*_{het}$, $\bar{\Delta}^*_{hom}$, $\bar{\gamma}^*_{het}$ and $\bar{\gamma}^*_{hom}$, respectively. 
First, we observe that the true dose-response relationships associated with $\bar{\gamma}^*_{hom}$ are correctly retrieved for each virtual cohort. For $\bar{\gamma}^*_{het}$, one affine relation has been classified as inverse, and another as constant, when the relations in all other 18 cohorts have been correctly classified. By "classified", we mean "found in the selected model (based on the global AIC value)". 
These results suggest that the first-step model selection procedure based on AIC is in capacity to correctly  retrieve the correct dose-response relationships, among the ones that are studied, concerning $\bar{\gamma}^*_{hom}$, and $\bar{\gamma}^*_{het}$, which are the quantities associated with the quiescence exit of mutated HSCs. 
The reason might be that the three relations studied for $\bar{\gamma}^*$ (but also the choice of the parameter population distributions) might induce different behaviours that can be discriminated against with the available observations. Note that the three relations studied (constant, affine, and inverse) all involve one parameter to estimate.\\
It is with $\bar{\Delta}^*_{het}$ and $\bar{\Delta}^*_{hom}$ that we have more misclassifications, which could be expected since we also study more relations for them (5 relations) compared to $\bar{\gamma}^*$  (3 relations), with two of them having one additional degree of freedom and, therefore, being penalized when applying the AIC criterion.
Concerning $\bar{\Delta}^*_{hom}$:
\begin{itemize}
    \item When the true dose-response relation is the constant or the sigmoid one, it is correctly retrieved.
    \item In half of the cases, the linear relation might be confused with the sigmoid one or the affine relation can be confused with the affine sigmoid. The reason is that the sigmoid relation approaches the linear one for low doses, such that both relations might model the same low-dose behaviour.
    \item One (over four) affine sigmoid relation has been classified as sigmoid, which is justified by the fact that this latter has a degree of freedom less, and thus is less penalized with the AIC criterion.
\end{itemize}
We find analogous results for $\bar{\Delta}^*_{het}$ with two more misclassifications compared to $\bar{\Delta}^*_{hom}$: an affine dose-response relation being classified as constant, and an additional affine-sigmoid relation classified as sigmoid. In both cases, the selected relation is more parsimonious than the true one. The fact that we have more misclassifications for $\bar{\Delta}^*_{het}$ than $\bar{\Delta}^*_{hom}$ can be explained by the choice of the parameter population distributions, with $\Delta_{het}$ (that corresponds to $\bar{\Delta}_{het}^*(d=0)$) taking (individual) values around 0.1 when $\Delta_{hom}$ takes values around 0.2 (see §~\ref{sec:synth_pop_distrib}). Therefore, the individual dose-response relationships for $\bar{\Delta}^*_{het}$ are more likely to be close to a linear (or sigmoid) one, for which $\bar{\Delta}_{het}^*(d=0) = 0$.\\
Overall, these results indicate that the selected models (based on the global AIC value), if not the true ones, still remain close to it.\\
In the next paragraph, we explore in more detail how the virtual patient's contributions to the global AIC can be responsible for not selecting the true model.

\begin{table}[]
    \centering
    
    \begin{tabular}{|c|c|c|c|c|c|c|c|}
\hline
        \diagbox[]{True}{Best} & constant & linear & affine & sigmoid & affine sigmoid & $\sum$ \\ \hline
constant & 4 & 0 & 0 & 0  &0& 4\\ \hline
linear& 0 & 2 & 0 & 2 & 0 & 4\\ \hline
affine& 0 & 0 & 2 & 0 & 2& 4\\ \hline
sigmoid& 0 & 0 & 0 & 4 & 0& 4\\ \hline
affine sigmoid& 0 & 0 & 0 & 1 & 3& 4\\ \hline
$\sum$ &4 & 2 & 2& 7 & 5 & 20 \\\hline
    \end{tabular}

    \caption{Confusion matrix associated with the dose-response relationship of $\bar{\Delta}^*_{hom}$, where - for a given cohort - the best model is based on the global AIC value.   }
    \label{tab:confusion_matrix_bestAIC_Delta_hom}
\end{table}

\begin{table}[]
    \centering
    
    \begin{tabular}{|c|c|c|c|c|c|c|c|}
\hline
        \diagbox[]{True}{Best} & constant & linear & affine & sigmoid & affine sigmoid & $\sum$ \\ \hline
constant & 4 & 0 & 0 & 0  &0& 4\\ \hline
linear& 0 & 2 & 0 & 2 & 0 & 4\\ \hline
affine& 1 & 0 & 1 & 0 & 2& 4\\ \hline
sigmoid& 0 & 0 & 0 & 4 & 0& 4\\ \hline
affine sigmoid& 0 & 0 & 0 & 2 & 2& 4\\ \hline
$\sum$ &5 & 2 & 1& 8 & 4 & 20 \\\hline
    \end{tabular}

    \caption{
    Confusion matrix associated with the dose-response relationship of $\bar{\Delta}^*_{het}$, where - for a given cohort - the best model is based on the global AIC value. }
    \label{tab:confusion_matrix_bestAIC_Delta_het}
\end{table}

\begin{table}[]
    \centering
    
    \begin{tabular}{|c|c|c|c|c|c|}
\hline
        \diagbox[]{True}{Best} & constant & inverse & affine & $\sum$ \\ \hline
constant &  7 & 0 & 0 & 7\\\hline
inverse & 0 & 6 & 0 & 6\\\hline
affine & 0 & 0 & 7 & 7\\\hline
$\sum$ & 7 & 6 & 7 & 20\\\hline
    \end{tabular}

    \caption{
    Confusion matrix associated with the dose-response relationship of $\bar{\gamma}^*_{hom}$, where - for a given cohort - the best model is based on the global AIC value.}
    \label{tab:confusion_matrix_bestAIC_gamma_hom}
\end{table}

\begin{table}[]
    \centering
    
    \begin{tabular}{|c|c|c|c|c|c|}
\hline
        \diagbox[]{True}{Best} & constant & inverse & affine & $\sum$ \\ \hline
constant &  7 & 0 & 0 & 7\\\hline
inverse & 0 & 6 & 0 & 6\\\hline
affine & 1 & 1 & 5 & 7\\\hline
$\sum$ & 8 & 7 & 5 & 20\\\hline
    \end{tabular}

    \caption{
    Confusion matrix associated with the dose-response relationship of $\bar{\gamma}^*_{het}$, where - for a given cohort - the best model is based on the global AIC value. }
    \label{tab:confusion_matrix_bestAIC_gamma_het}
\end{table}

\FloatBarrier

\subsubsection{Investigating the influence of the posology and the observation times on the results of the first-step selection procedure}

\begin{figure}[h]
    \centering
    \begin{subfigure}[b]{0.44\textwidth}
        \includegraphics[width=\textwidth]{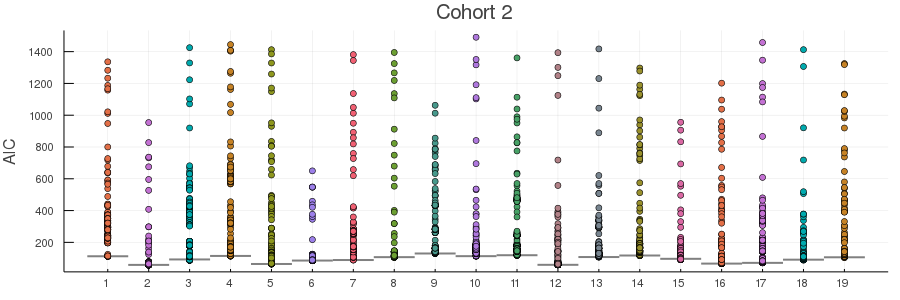}
    \end{subfigure}
    \begin{subfigure}[b]{0.44\textwidth}
        \includegraphics[width=\textwidth]{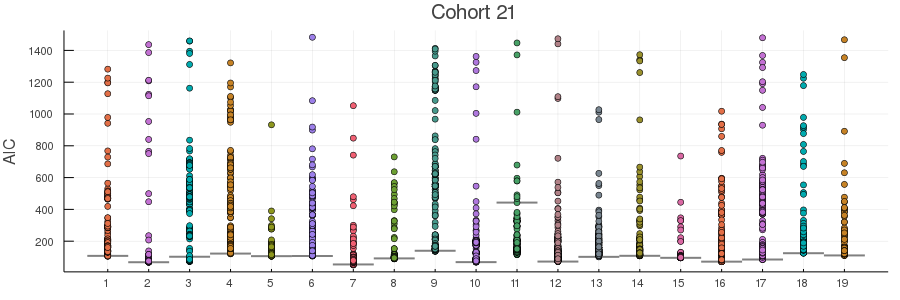}
    \end{subfigure}
\begin{subfigure}[b]{0.44\textwidth}
        \includegraphics[width=\textwidth]{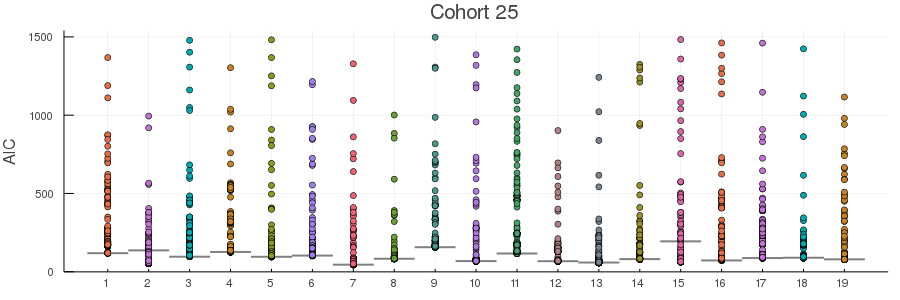}
    \end{subfigure}
    \begin{subfigure}[b]{0.44\textwidth}
        \includegraphics[width=\textwidth]{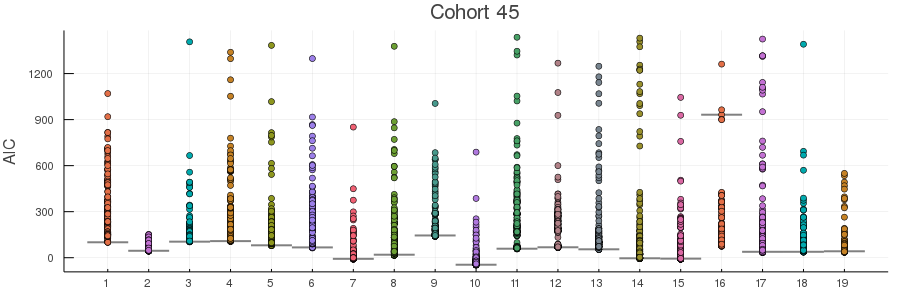}
    \end{subfigure}
    \begin{subfigure}[b]{0.44\textwidth}
        \includegraphics[width=\textwidth]{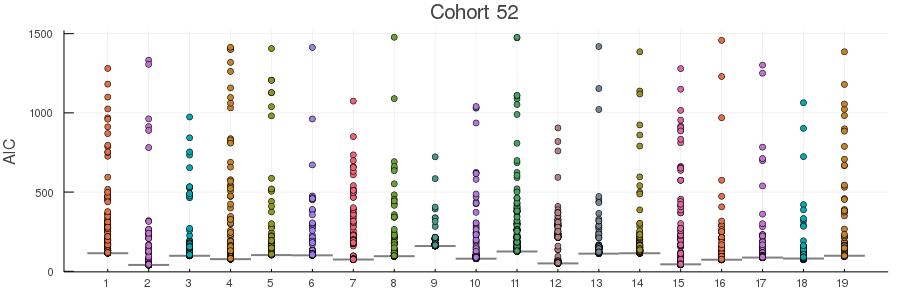}
    \end{subfigure}
    \begin{subfigure}[b]{0.44\textwidth}
        \includegraphics[width=\textwidth]{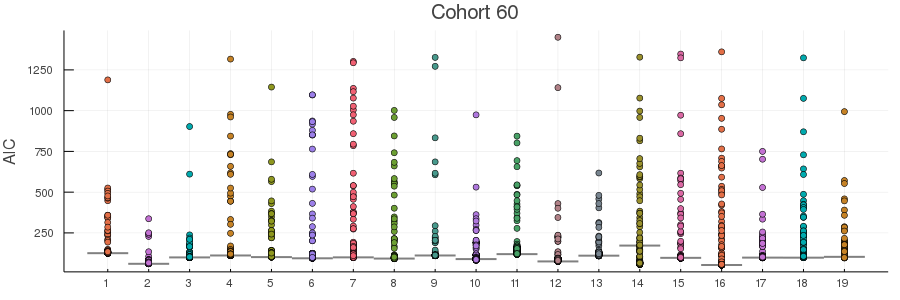}
    \end{subfigure}
    \begin{subfigure}[b]{0.44\textwidth}
        \includegraphics[width=\textwidth]{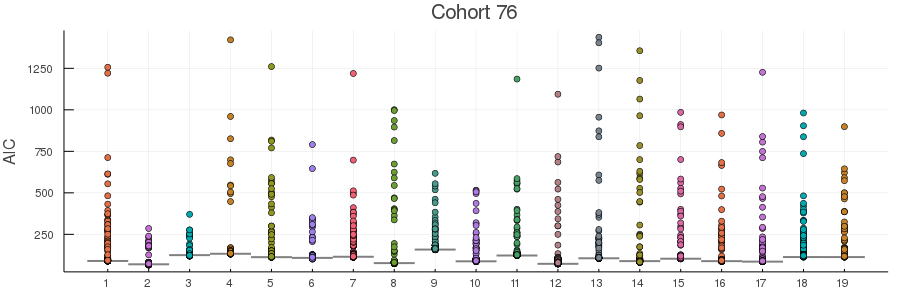}
    \end{subfigure}
    \begin{subfigure}[b]{0.44\textwidth}
        \includegraphics[width=\textwidth]{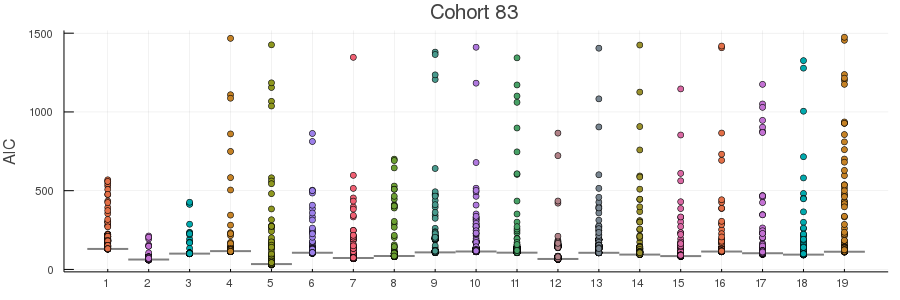}
    \end{subfigure}
    \begin{subfigure}[b]{0.44\textwidth}
        \includegraphics[width=\textwidth]{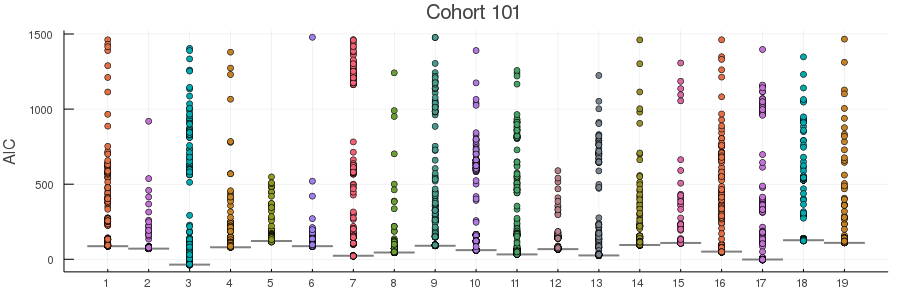}
    \end{subfigure}
    \begin{subfigure}[b]{0.44\textwidth}
        \includegraphics[width=\textwidth]{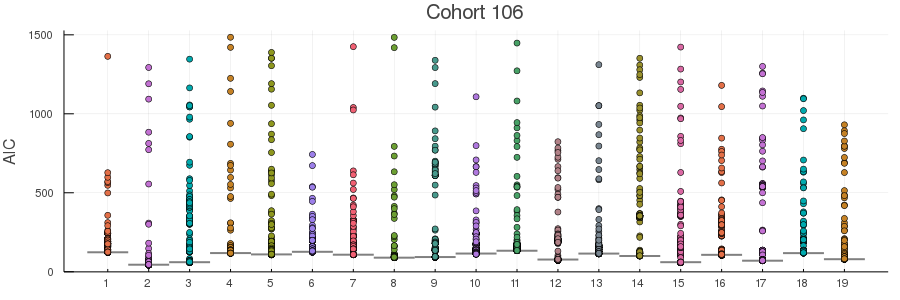}
    \end{subfigure}
    \begin{subfigure}[b]{0.44\textwidth}
        \includegraphics[width=\textwidth]{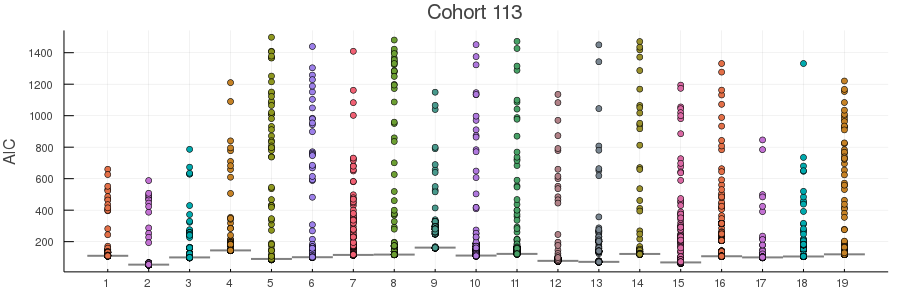}
    \end{subfigure}
    \begin{subfigure}[b]{0.44\textwidth}
        \includegraphics[width=\textwidth]{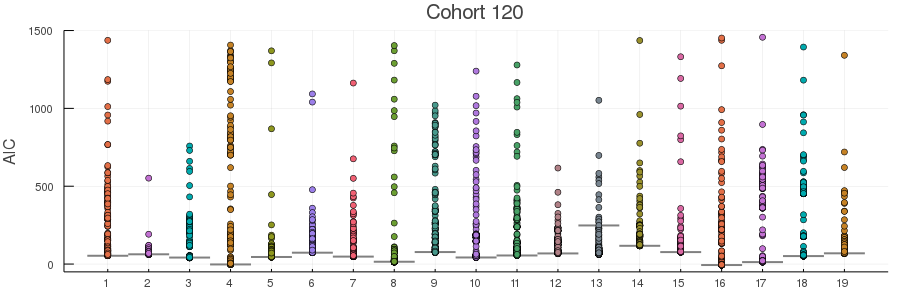}
    \end{subfigure}
\begin{subfigure}[b]{0.44\textwidth}
        \includegraphics[width=\textwidth]{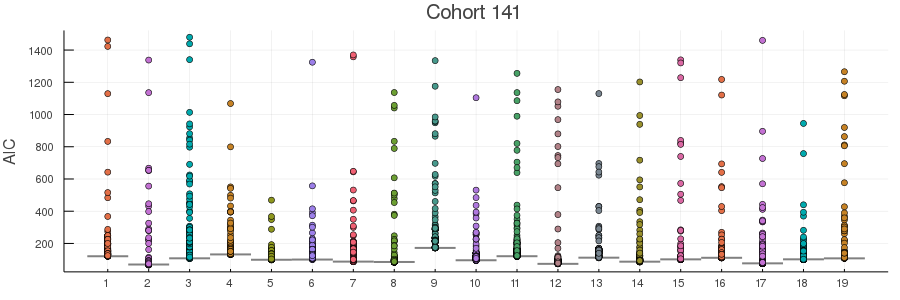}
    \end{subfigure}
    \begin{subfigure}[b]{0.44\textwidth}
        \includegraphics[width=\textwidth]{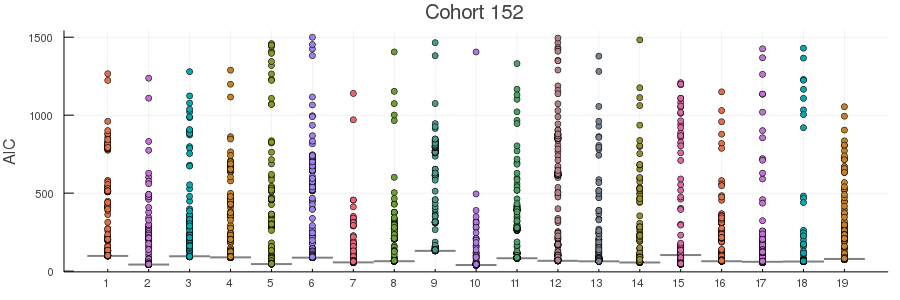}
    \end{subfigure}
    \begin{subfigure}[b]{0.44\textwidth}
        \includegraphics[width=\textwidth]{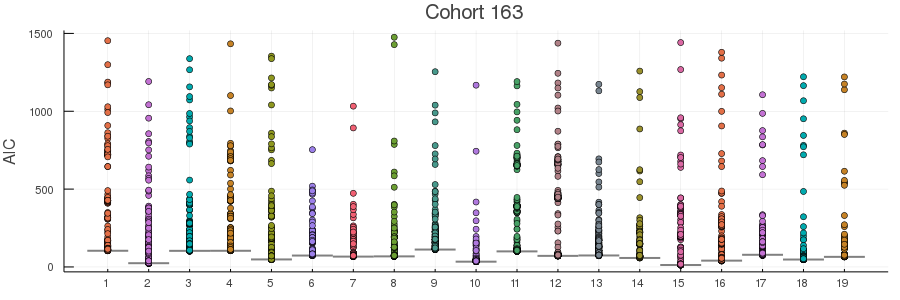}
    \end{subfigure}
    \begin{subfigure}[b]{0.44\textwidth}
        \includegraphics[width=\textwidth]{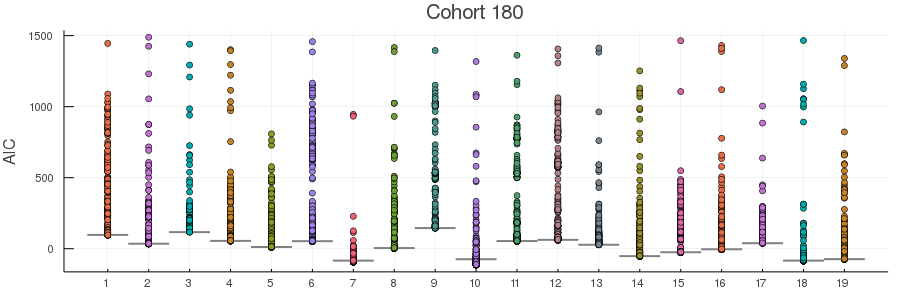}
    \end{subfigure}
    \begin{subfigure}[b]{0.44\textwidth}
        \includegraphics[width=\textwidth]{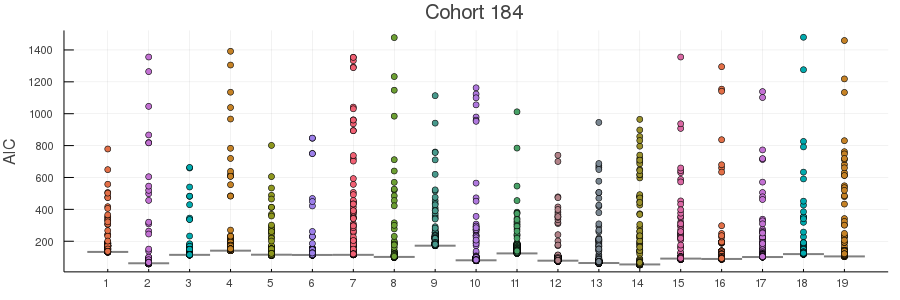}
    \end{subfigure}
    \begin{subfigure}[b]{0.44\textwidth}
        \includegraphics[width=\textwidth]{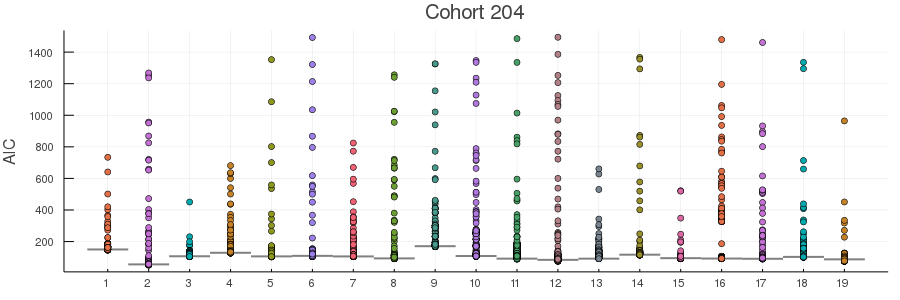}
    \end{subfigure}
    \begin{subfigure}[b]{0.44\textwidth}
        \includegraphics[width=\textwidth]{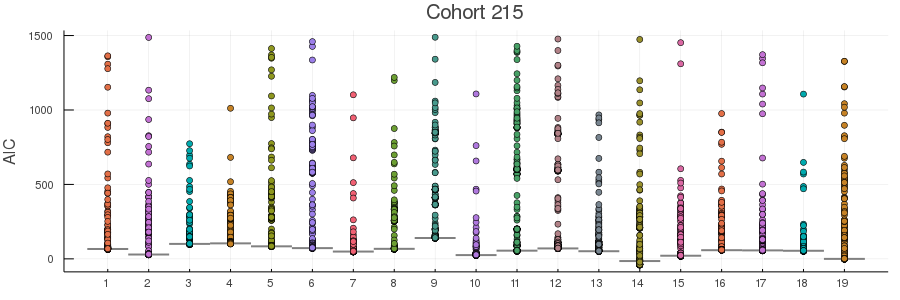}
    \end{subfigure}
    \begin{subfigure}[b]{0.44\textwidth}
        \includegraphics[width=\textwidth]{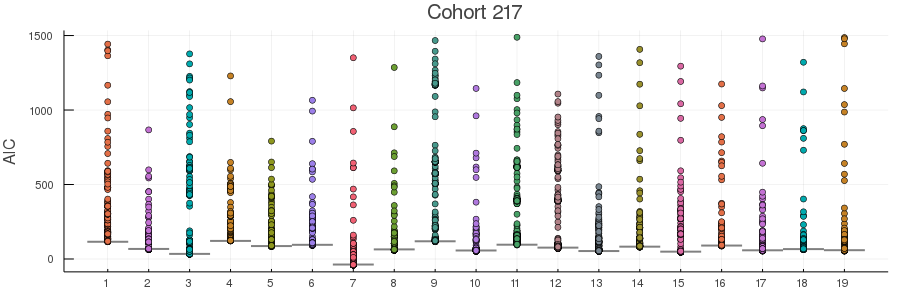}
    \end{subfigure}
    \caption{AIC$_{i,j, m}$ (y-axis) for each model $j$ (colored circle) and each  $i^{th}$ virtual patient (x-axis) from the synthetic cohort $m$ (subfigure). Horizontal lines correspond to the AIC value of the true model.}
    \label{fig:synthetic_ind_AIC}
\end{figure}

To compute the global AIC value of each studied model and each virtual cohort, we first have to compute the individual AIC values. These are displayed in Fig.~\ref{fig:synthetic_ind_AIC}. On this figure, for each virtual patient, we indicate the AIC value obtained with the model used to generate the data. For most virtual patients, we observe that the true model is among the best ones. There are a few exceptions: for the $16^{th}$ patient from cohort 45, the AIC of the true model is nearly the worst (223/225). The selection ranks of the true models (based on the AIC value) for each patient $i$ from virtual cohort $m$ are given in Tab.~\ref{tab:synth_ind_AIC}. 
We also indicate that for each value of $i$ (corresponding to the $i^{th}$ patient of a given cohort), the median true model rank is computed over all cohorts. Since all patients with the same index $i$ also have their posology and observations times in common, this median value indicates whether these shared characteristics might adversely affect the first-step selection procedure. 
In particular, we observe that the true model has a median AIC value of 40 when computed over all the $15^{th}$ patients when this median value is below 25 in all other cases. When looking in more detail at the characteristics of this patient (which corresponds to patient \#27 in the real cohort from Mosca et al.~\cite{mosca2021}), two observations emerge:
\begin{itemize}
    \item First, no observations are made at the initial time, with the first measurement being made 273 days after the start of the therapy.
    \item Second, most of the observations are obtained when the dose administered is a medium dose, about 67.5-90 $\mu$g/week, when the dose is finally increased up to 135 $\mu$g/week 1,550 days after the start of the therapy.
\end{itemize}
Patient \#27 is not the only one with no observations at the initial time; patients \#6, \#15, \#19, and \#30 also fall in that case (corresponding respectively to $i=$3, 7, 10, and 17).
However, patient \#27 is the only one to encounter such dosage dynamics, with a dose escalation rather than a de-escalation.\\
We further analyze if some patient characteristics, as described in Tab.~\ref{tab:info_patients}, could be correlated to the median true model range computed over all cohorts. There is a correlation only with the first observation time: when testing the nullity of the coefficient in the linear regression (Student Test), the null hypothesis is rejected with a $p$-value equal to 0.0291. However, the conditions for applying this statistical test are not totally respected; the first observation time is not a continuous variable since it takes most of its value at 0 (at the start of the therapy).

\begin{table}[]
    \centering
    
    \footnotesize\begin{tabular}
    {|c||c|c|c|c|c|c|c|c|c|c|c|c|c|c|c|c|c|c|c|}
    \hline
    \diagbox[]{$m$}{$i$} & 1 & 2 & 3 & 4 & 5 & 6 & 7 & 8 & 9 & 10 & 11 & 12 & 13 & 14 & 15 & 16 & 17 & 18 & 19 \\ \hline\hline
2	&	6	&	20	&	13	&	4	&	1	&	19	&	1	&	7	&	1	&	1	&	4	&	8	&	2	&	4	&	53	&	1	&	1	&	7	&	3	\\ \hline
21	&	5	&	2	&	45	&	3	&	4	&	2	&	4	&	16	&	7	&	1	&	215	&	1	&	2	&	3	&	1	&	5	&	2	&	2	&	3	\\ \hline
25	&	4	&	180	&	1	&	15	&	6	&	8	&	16	&	24	&	5	&	43	&	7	&	31	&	61	&	37	&	185	&	3	&	6	&	48	&	24	\\ \hline
45	&	2	&	20	&	2	&	7	&	6	&	4	&	9	&	15	&	41	&	14	&	3	&	13	&	5	&	20	&	60	&	223	&	19	&	9	&	36	\\ \hline
52	&	1	&	17	&	2	&	3	&	1	&	4	&	6	&	16	&	1	&	8	&	2	&	1	&	5	&	4	&	49	&	10	&	11	&	36	&	59	\\ \hline
60	&	11	&	1	&	32	&	6	&	2	&	3	&	8	&	9	&	72	&	111	&	11	&	5	&	20	&	189	&	38	&	11	&	15	&	1	&	20	\\ \hline
76	&	2	&	39	&	115	&	26	&	5	&	34	&	45	&	52	&	14	&	24	&	4	&	10	&	30	&	103	&	5	&	28	&	16	&	2	&	2	\\ \hline
83	&	8	&	29	&	33	&	49	&	33	&	54	&	97	&	93	&	22	&	38	&	9	&	30	&	8	&	1	&	20	&	17	&	75	&	17	&	11	\\ \hline
101	&	8	&	32	&	32	&	2	&	13	&	19	&	16	&	29	&	7	&	26	&	2	&	31	&	18	&	18	&	16	&	38	&	59	&	20	&	2	\\ \hline
106	&	13	&	56	&	27	&	18	&	26	&	16	&	15	&	4	&	24	&	22	&	6	&	122	&	27	&	1	&	43	&	51	&	13	&	3	&	9	\\ \hline
113	&	66	&	116	&	76	&	29	&	117	&	21	&	37	&	30	&	66	&	29	&	25	&	42	&	44	&	43	&	68	&	24	&	59	&	35	&	33	\\ \hline
120	&	2	&	9	&	17	&	8	&	5	&	3	&	4	&	57	&	19	&	6	&	2	&	11	&	211	&	9	&	7	&	4	&	3	&	6	&	7	\\ \hline
141	&	2	&	10	&	5	&	2	&	3	&	2	&	3	&	26	&	3	&	7	&	10	&	2	&	3	&	10	&	18	&	8	&	27	&	25	&	1	\\ \hline
152	&	9	&	5	&	24	&	12	&	3	&	3	&	8	&	18	&	5	&	32	&	27	&	8	&	8	&	11	&	177	&	1	&	53	&	7	&	16	\\ \hline
163	&	3	&	12	&	7	&	1	&	8	&	2	&	1	&	2	&	4	&	4	&	3	&	7	&	10	&	1	&	4	&	11	&	17	&	7	&	4	\\ \hline
180	&	7	&	12	&	1	&	10	&	20	&	15	&	152	&	33	&	4	&	127	&	12	&	15	&	3	&	18	&	28	&	16	&	5	&	66	&	10	\\ \hline
184	&	7	&	33	&	14	&	2	&	29	&	14	&	1	&	41	&	3	&	42	&	2	&	59	&	24	&	29	&	58	&	12	&	3	&	8	&	3	\\ \hline
204	&	30	&	42	&	22	&	11	&	17	&	28	&	20	&	22	&	42	&	20	&	10	&	78	&	4	&	18	&	105	&	16	&	8	&	19	&	84	\\ \hline
215	&	32	&	27	&	40	&	11	&	7	&	9	&	95	&	29	&	12	&	56	&	41	&	19	&	20	&	88	&	34	&	3	&	39	&	47	&	15	\\ \hline
217	&	2	&	87	&	14	&	1	&	15	&	10	&	89	&	103	&	11	&	101	&	3	&	40	&	19	&	16	&	72	&	11	&	26	&	23	&	37	\\ \hline\hline
median	&	6	&	23	&	19	&	7	&	6	&	9	&	12	&	25	&	9	&	25	&	6	&	14	&	14	&	17	&	40	&	11	&	15	&	13	&	10	\\ \hline
    \end{tabular}
    
    \normalsize
    \caption{Ranks of the true model $m$ (among the 225, based on the computation of an individual AIC value) when considering each patient $i$ (from each virtual cohort $m$) independent. The last line indicates the median true model rank computed over all $i^{th}$ patients. }
    \label{tab:synth_ind_AIC}
\end{table}

\FloatBarrier

\subsection{Second step combining hierarchical modelling and DIC}

After having preselected three models based on the computation of a global AIC value for each virtual cohort, we continue with our second (and last) step of the model selection procedure. It consists in running a hierarchical Bayesian inference method for each of the three potential models and then computing a DIC value. The selected model is then the one with the lowest DIC value. In this second step, the individuals are no longer assumed to be independent since we introduce a population effect through population distributions (see~\ref{sec:identif_param_estim}).
The results of this second step are presented in Tab.~\ref{tab:results_DIC_synth}. 
The true model is the best one, after that second step, in 50\% of the cases. We get slightly better results than if we had limited ourselves to the selection based on the AIC (where the true models were found to be the bests in 45\% of the cases).
In particular, we find that:
\begin{itemize}
    \item A dose-response relationship which is constant is always correctly retrieved;
    \item For the cohorts $m=215$ and $52$, the true models were the third ones based on the global AIC and are the best ones based on the DIC;
    \item For cohort $m=180$, the true model moves up from third to second place after this second step, with relatively minor differences between the DIC values. 
    \item For the cohort $m=163$, for which the true model was found to be the best based on the global AIC value, it is now in the second place.
    \item Overall, when the true model allows for a minimal dose, it will be similar for the selected model. The only exception are for cohorts $45$ and 180 (what we justify later in~\ref{sec:min_dose_synth_res} and Fig.~\ref{fig:min_dose_inferred_vs_true} given that the true individual minimal doses were actually low for the virtual patients from these two synthetic cohorts).
\end{itemize}

\begin{table}[h]
    \centering
    
    \begin{tabular}{|c|cc|cc|cc|}
    \hline
    Cohort & \multicolumn{2}{c|}{First}  & \multicolumn{2}{c|}{Second}  & \multicolumn{2}{c|}{Third} 	\\ \cline{2-7}
         $m$	&	model	&	DIC	&	model	&	DIC	&	Model	&	DIC	\\ \hline
2	&	2	&	1791.75	&	1	&	1793.56	&	\textbf{38}	&	1805.66	\\ \hline
\textbf{21}	&	\textbf{38}	&	1771.89	&	\textbf{20}	&	1908.01	&	2	&	1960.64	\\ \hline
\textbf{25}	&	\textbf{43}	&	1508.89	&	7	&	1522.49	&	8	&	1688.87	\\ \hline
\textbf{45}	&	36	&	968.18	&	34	&	1022.02	&	16	&	 1058.43	\\ \hline
52	&	52	&	1613.59	&	143	&	1657.40	&	142	&	1674.0\\ \hline
60	&	168	&	1702.54	&	150	&	1862.72	&	78	&	1868.85	\\ \hline
76	&	79	&	1905.0	&	166	&	1920.87	&	76	&	1925.25	\\ \hline
\textbf{83}	&	\textbf{83}	&	1638.62	&	\textbf{173}	&	1641.70	&	\textbf{82}	&	1682.44	\\ \hline
\textbf{101}	&	\textbf{101}	&	1118.15	&	\textbf{191}	&	1170.86	&	\textbf{119}	&	1206.29	\\ \hline
\textbf{106}	&	\textbf{106}	&	1693.92	&	\textbf{107}	&	1862.57&	\textbf{98}	&	1909.55	\\ \hline
\textbf{113}	&	\textbf{185}	&	1980.25	&	\textbf{184}	&	 2004.38	&	\textbf{203}	&	7667.08	\\ \hline
\textbf{120}	&	\textbf{181}	&	935.60	&	\textbf{208}	&	987.14	&	\textbf{219}	&	1051.0	\\ \hline
141	&	141	&	1880.43	&	138	&	1978.44	&	\textbf{183}	&	2123.43	\\ \hline
152	&	143	&	1199.1	&	170	&	1212.90	&	151	&	1250.42	\\ \hline
163	&	\textbf{172}	&	1150.18	&	163	&	1340.62	&	164	&	1342.63	\\ \hline
\textbf{180}	&	171	&	241.67	&	\textbf{180}	&	244.16	&	\textbf{216}	&	248.34	\\ \hline
\textbf{184}	&	\textbf{184}	&	1854.29	&	\textbf{185}	&	1946.20	&	\textbf{188}	&	1977.91	\\ \hline
\textbf{204}	&	\textbf{204}	&	2038.09	&	\textbf{159}	&	2075.77	&	\textbf{186}	&	2204.90	\\ \hline
\textbf{215}	&	\textbf{215}	&	1027.38	&	170	&	1068.34	&	169	&	1093.89	\\ \hline
\textbf{217}	&	\textbf{217}	&	1230.08	&	\textbf{218}	&	 1246.21	&	\textbf{223}	&	1344.06\\ \hline 
    \end{tabular}
    
    \caption{Results of the second-step model selection procedure. 
    For each virtual cohort $m$ - where $m$ corresponds to the index of the model used to simulate the observations - we indicate the first three best models based on the DIC value. Some models allow the existence of a minimal IFN$\alpha$ dose $d_{min}$; they are highlighted in bold. }
    \label{tab:results_DIC_synth}
\end{table}

According to our results, in 50\% of the cases, the selected model (among 225 studied models) is different from the true one but might still be quite similar to it, as we already mentioned. We explore in the next paragraph if we can still make accurate inferences, especially regarding the long-term response of the treatment and the inferred minimal dose (when applicable).
\FloatBarrier
\newpage

\subsection{Post-selection inference}

For each virtual cohort $m$, we now consider the model previously selected (based on the DIC) for which we inferred the posterior distributions of the model parameters of each virtual individual (using a hierarchical Bayesian inference method).

\subsubsection{Estimated individual parameters}

\begin{figure}[h]
    \centering

    \begin{subfigure}[b]{0.22\textwidth}
        \includegraphics[width=\textwidth]{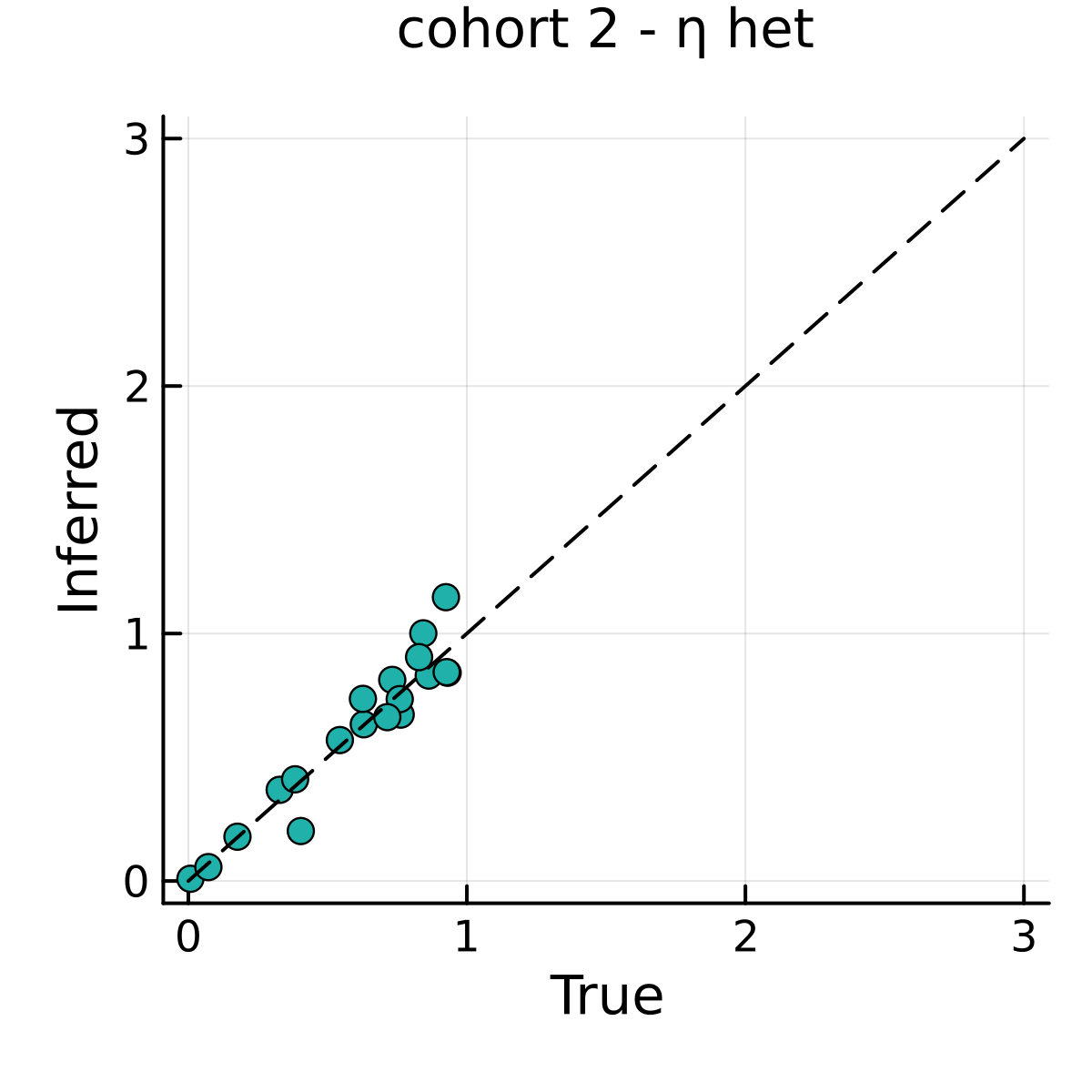}
    \end{subfigure}
     \begin{subfigure}[b]{0.22\textwidth}
        \includegraphics[width=\textwidth]{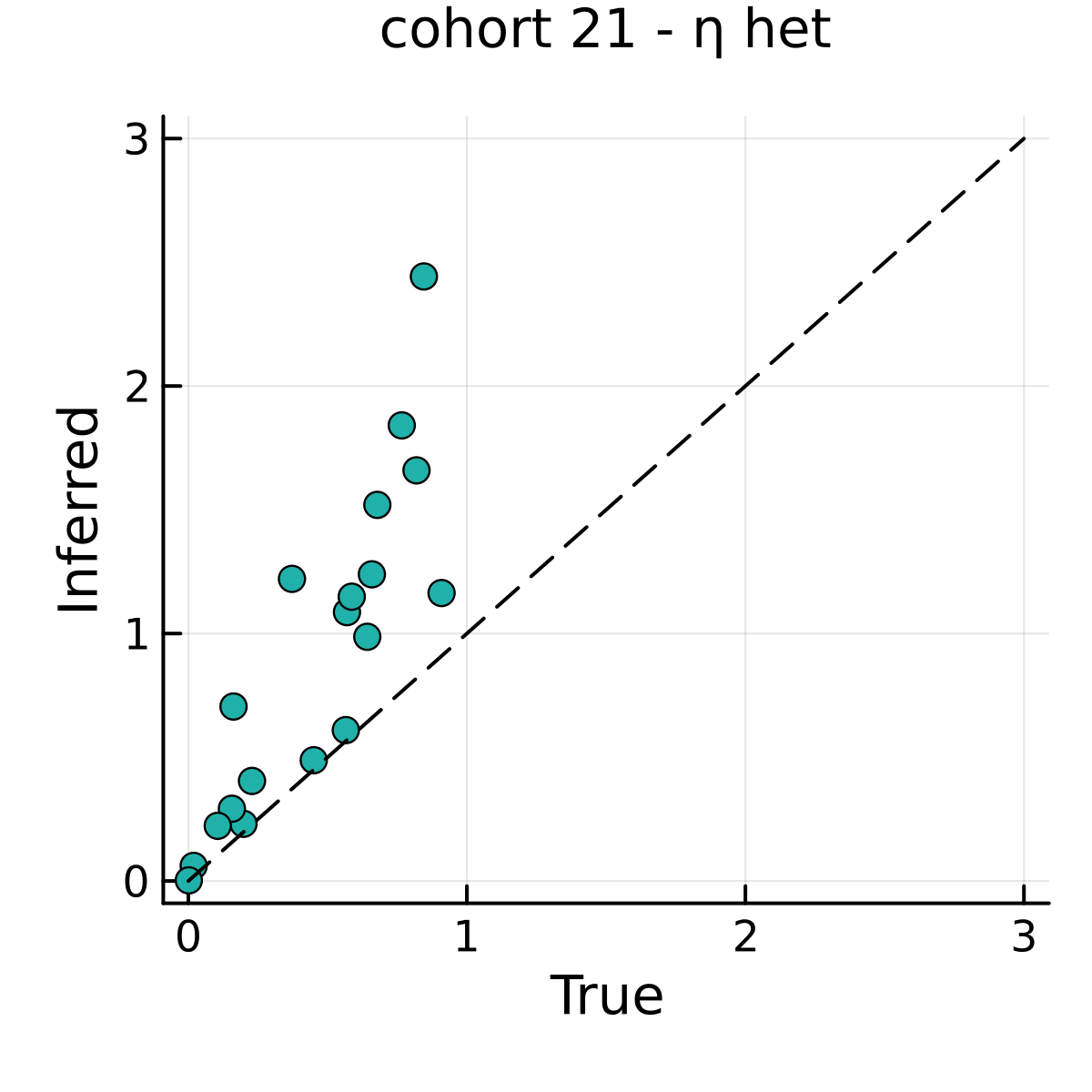}
    \end{subfigure}
     \begin{subfigure}[b]{0.22\textwidth}
        \includegraphics[width=\textwidth]{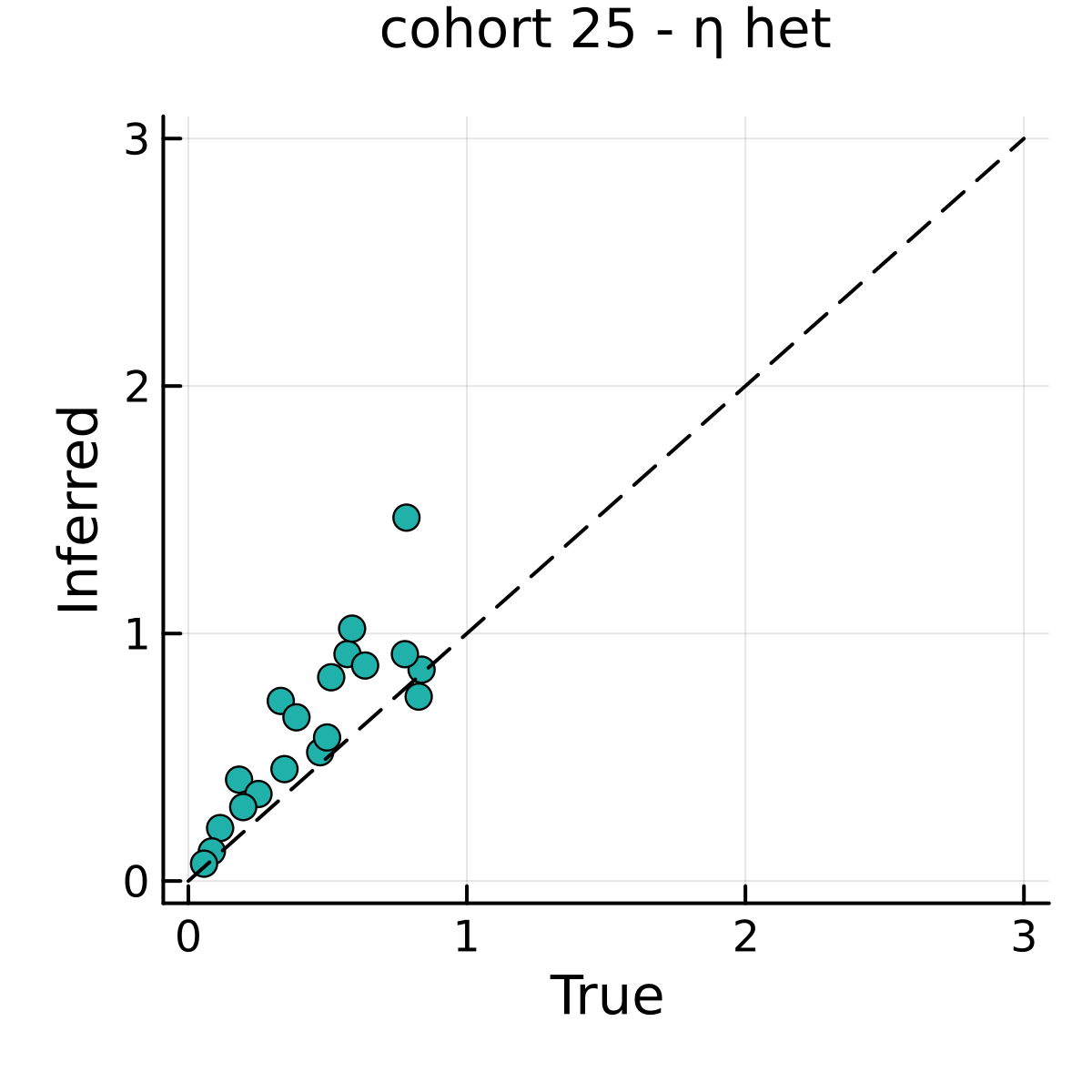}
    \end{subfigure}
     \begin{subfigure}[b]{0.22\textwidth}
        \includegraphics[width=\textwidth]{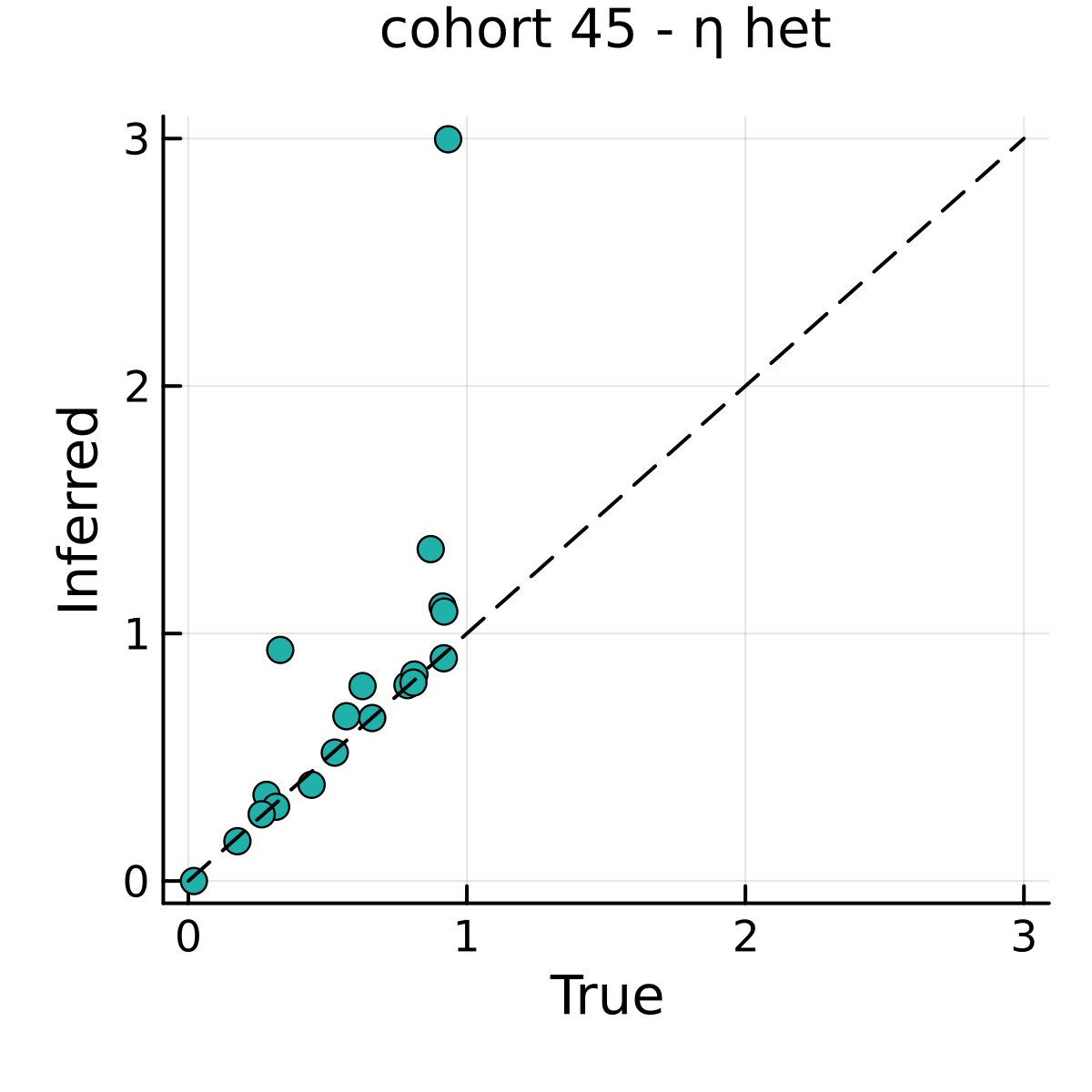}
    \end{subfigure}
     \begin{subfigure}[b]{0.22\textwidth}
        \includegraphics[width=\textwidth]{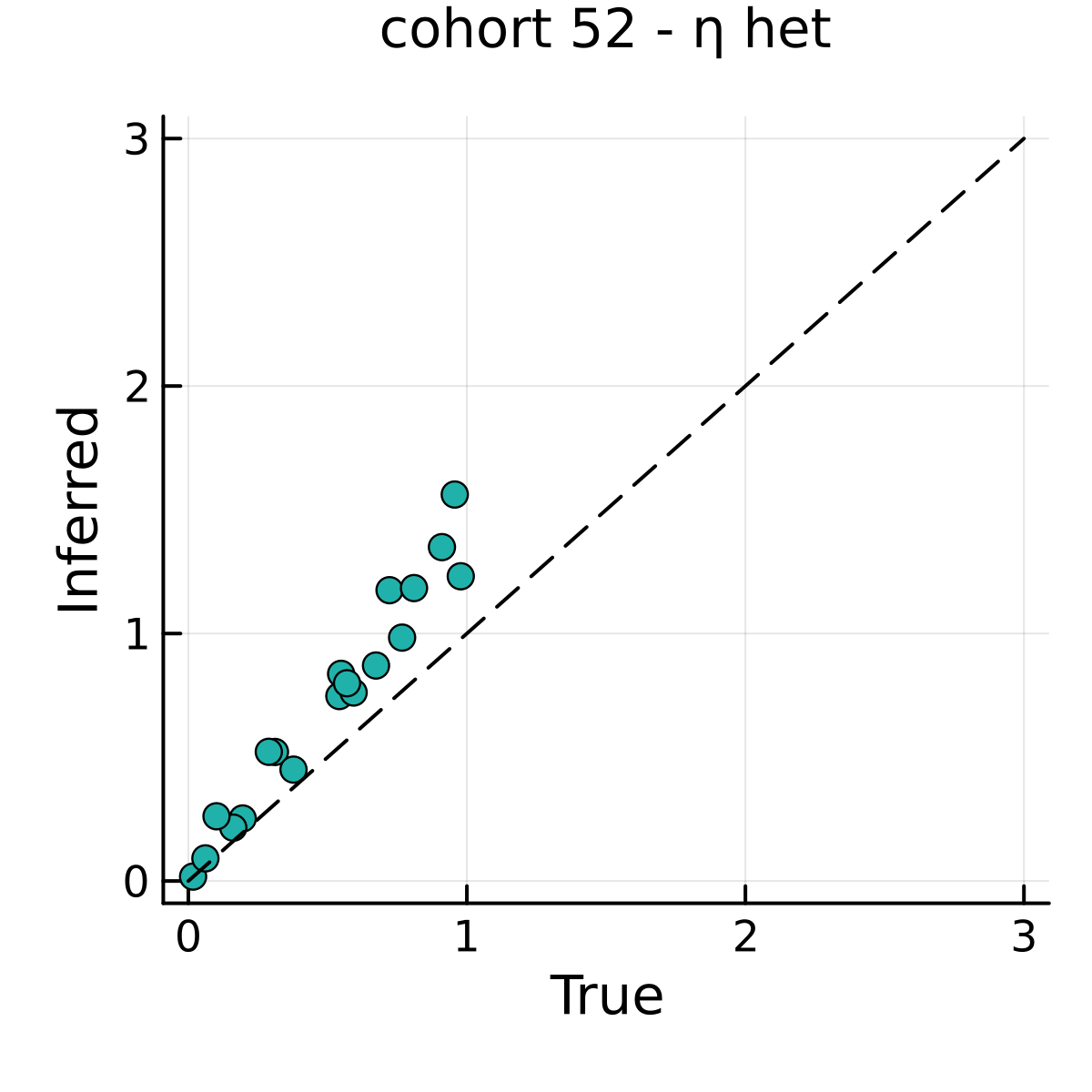}
    \end{subfigure}
     \begin{subfigure}[b]{0.22\textwidth}
        \includegraphics[width=\textwidth]{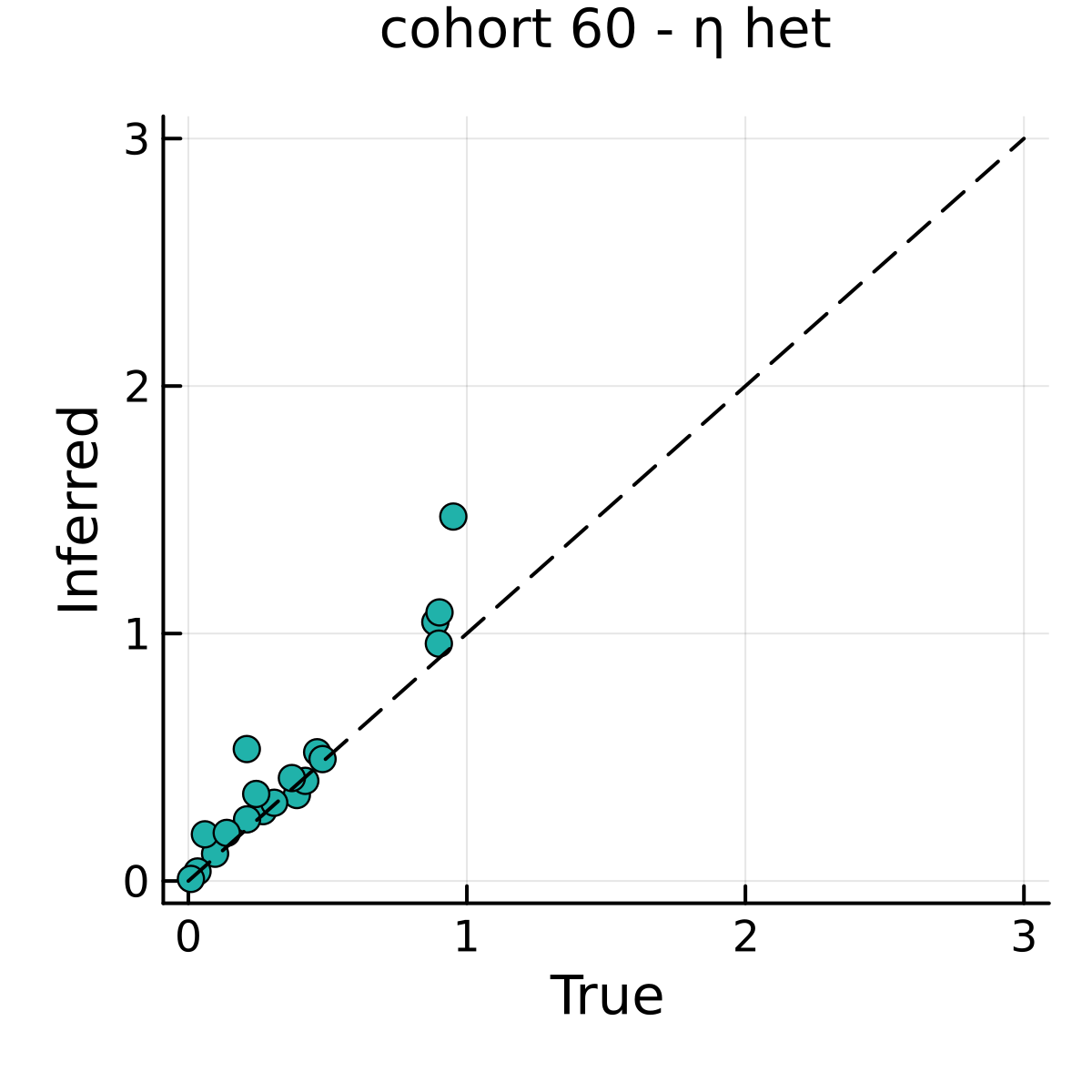}
    \end{subfigure}
     \begin{subfigure}[b]{0.22\textwidth}
        \includegraphics[width=\textwidth]{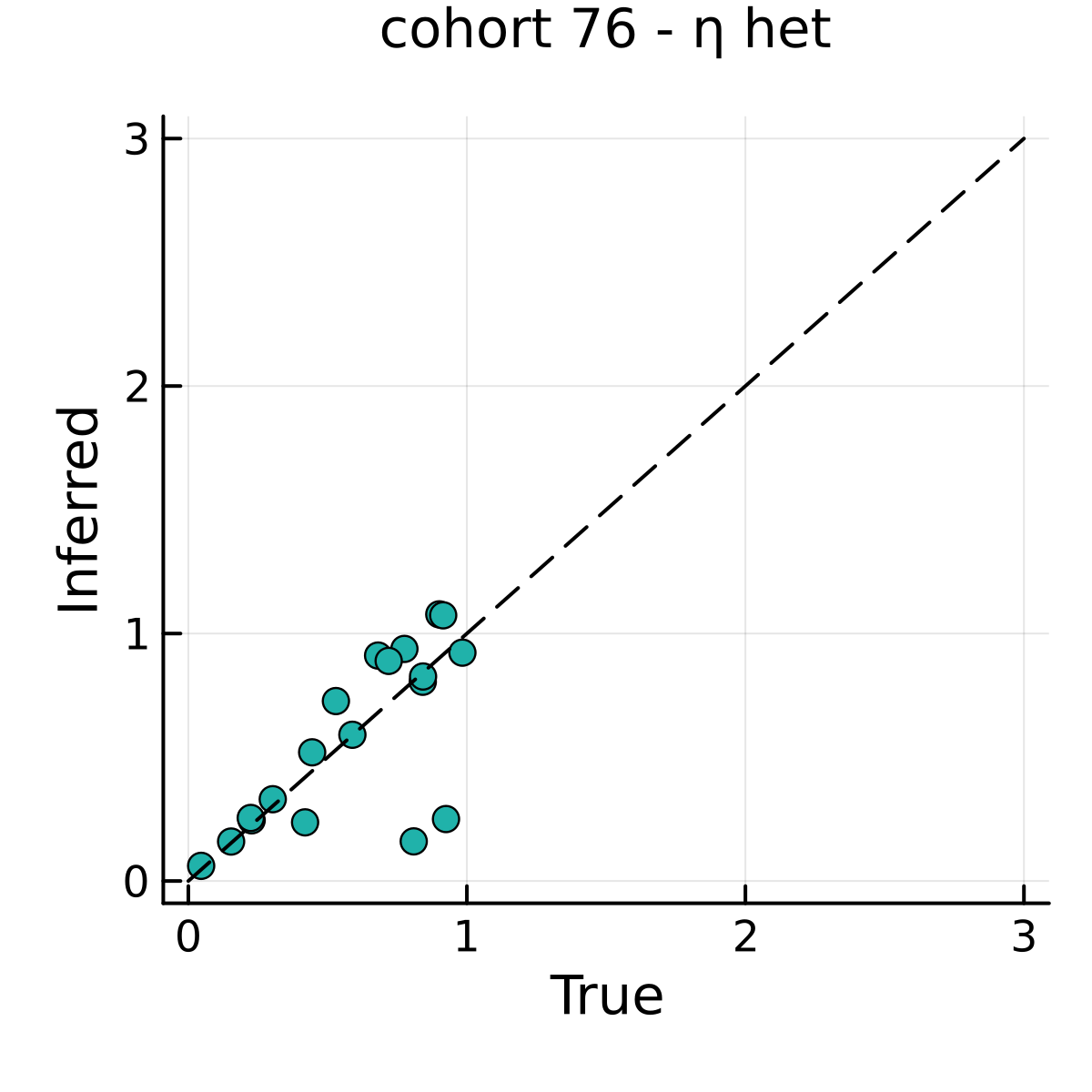}
    \end{subfigure}
     \begin{subfigure}[b]{0.22\textwidth}
        \includegraphics[width=\textwidth]{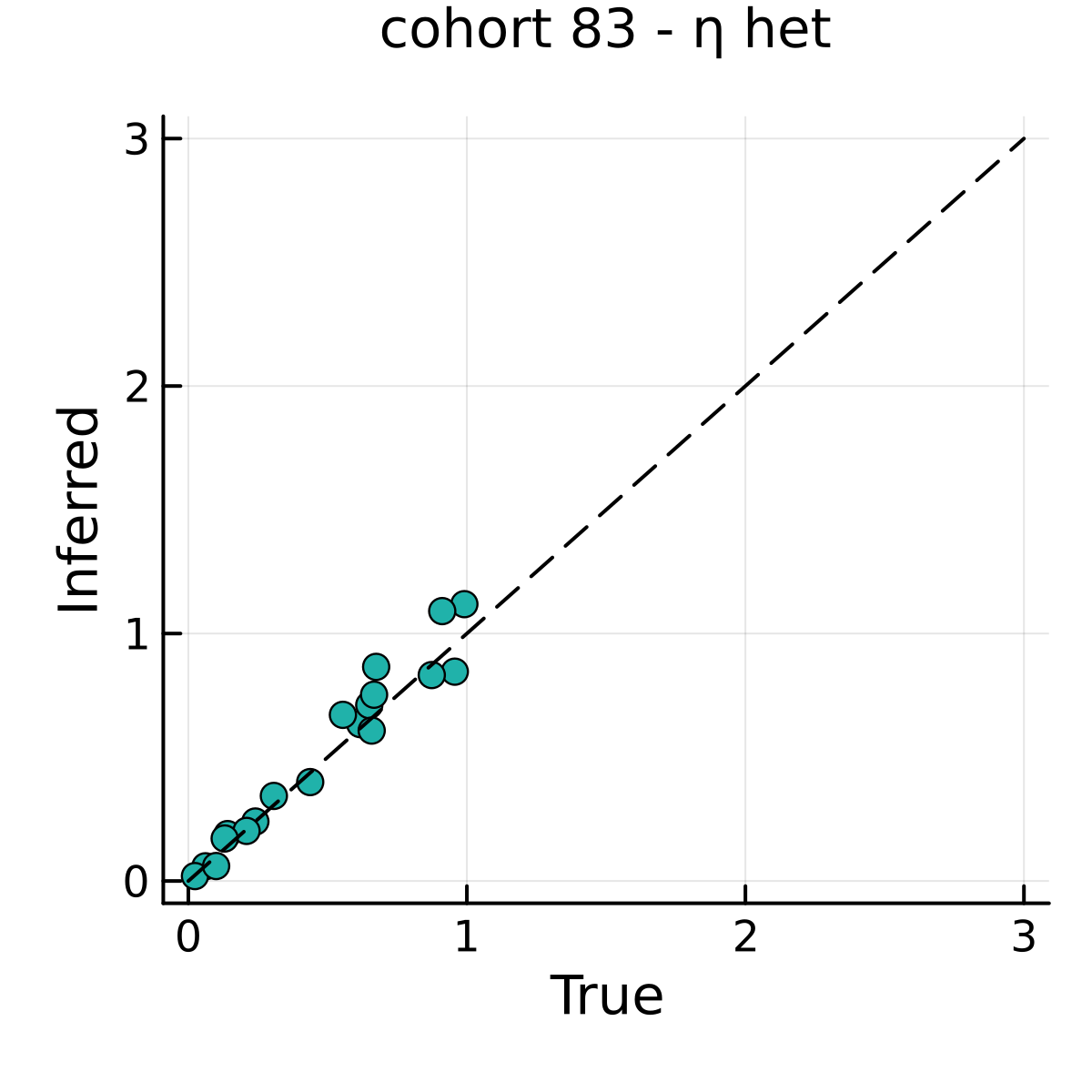}
    \end{subfigure}
     \begin{subfigure}[b]{0.22\textwidth}
        \includegraphics[width=\textwidth]{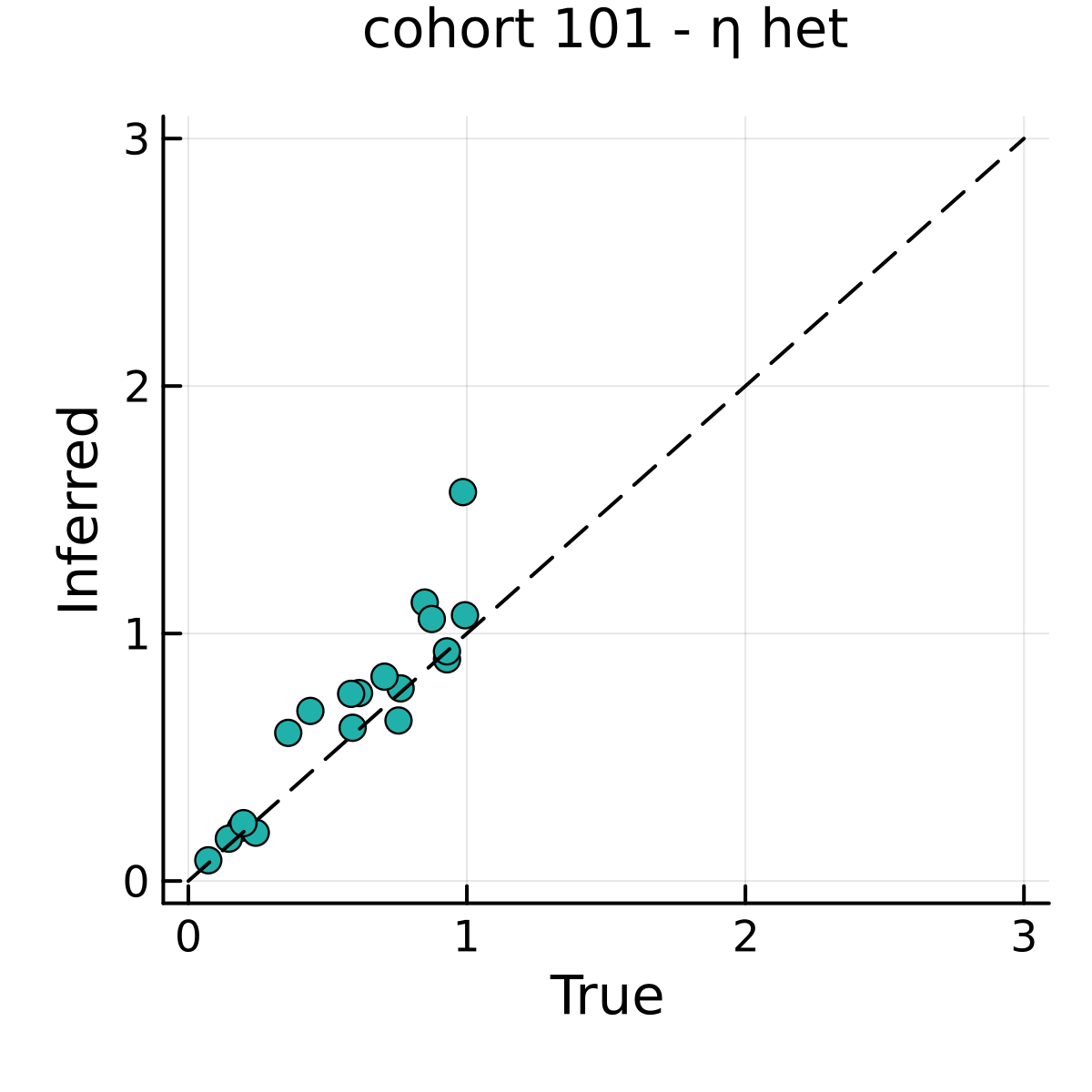}
    \end{subfigure}
     \begin{subfigure}[b]{0.22\textwidth}
        \includegraphics[width=\textwidth]{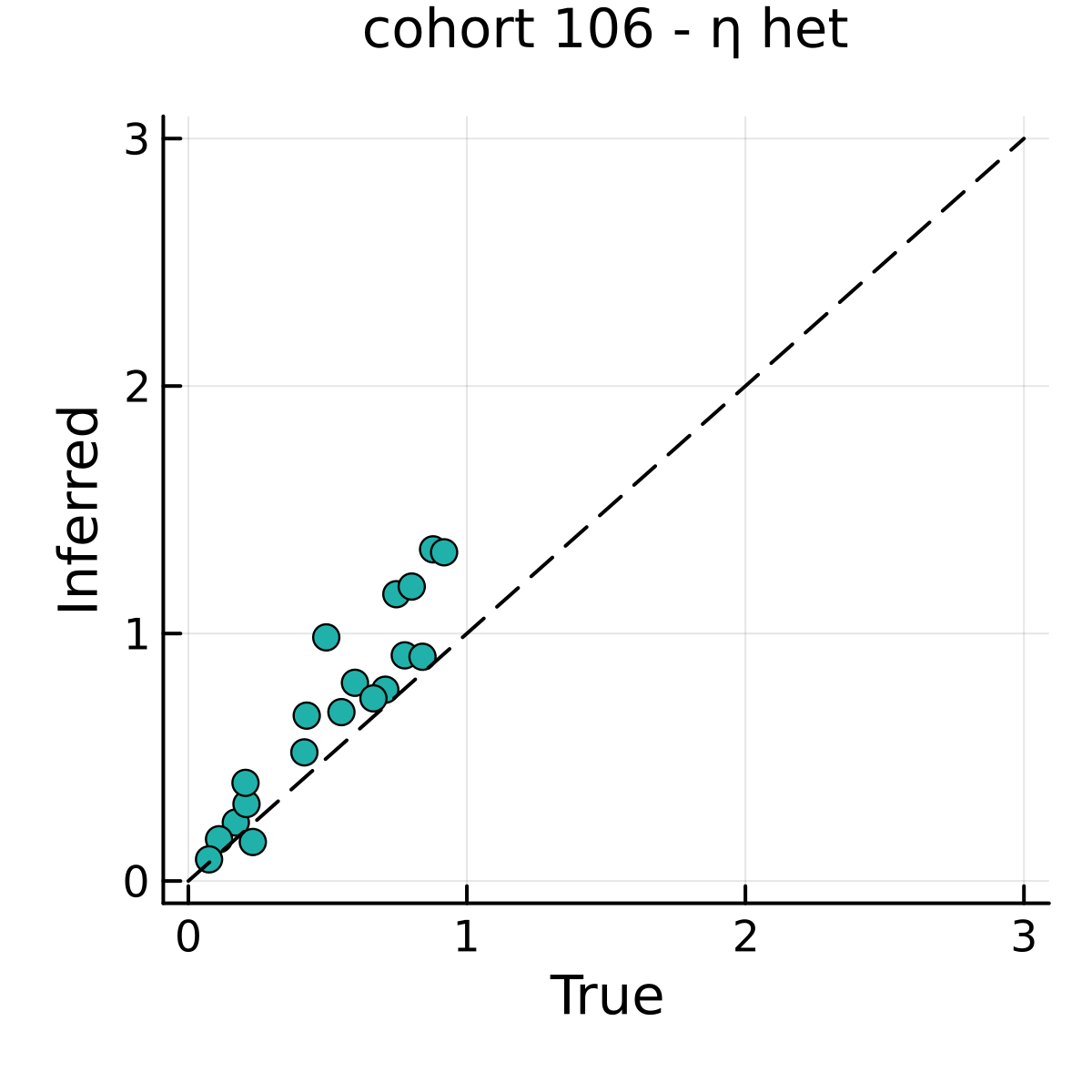}
    \end{subfigure}
     \begin{subfigure}[b]{0.22\textwidth}
        \includegraphics[width=\textwidth]{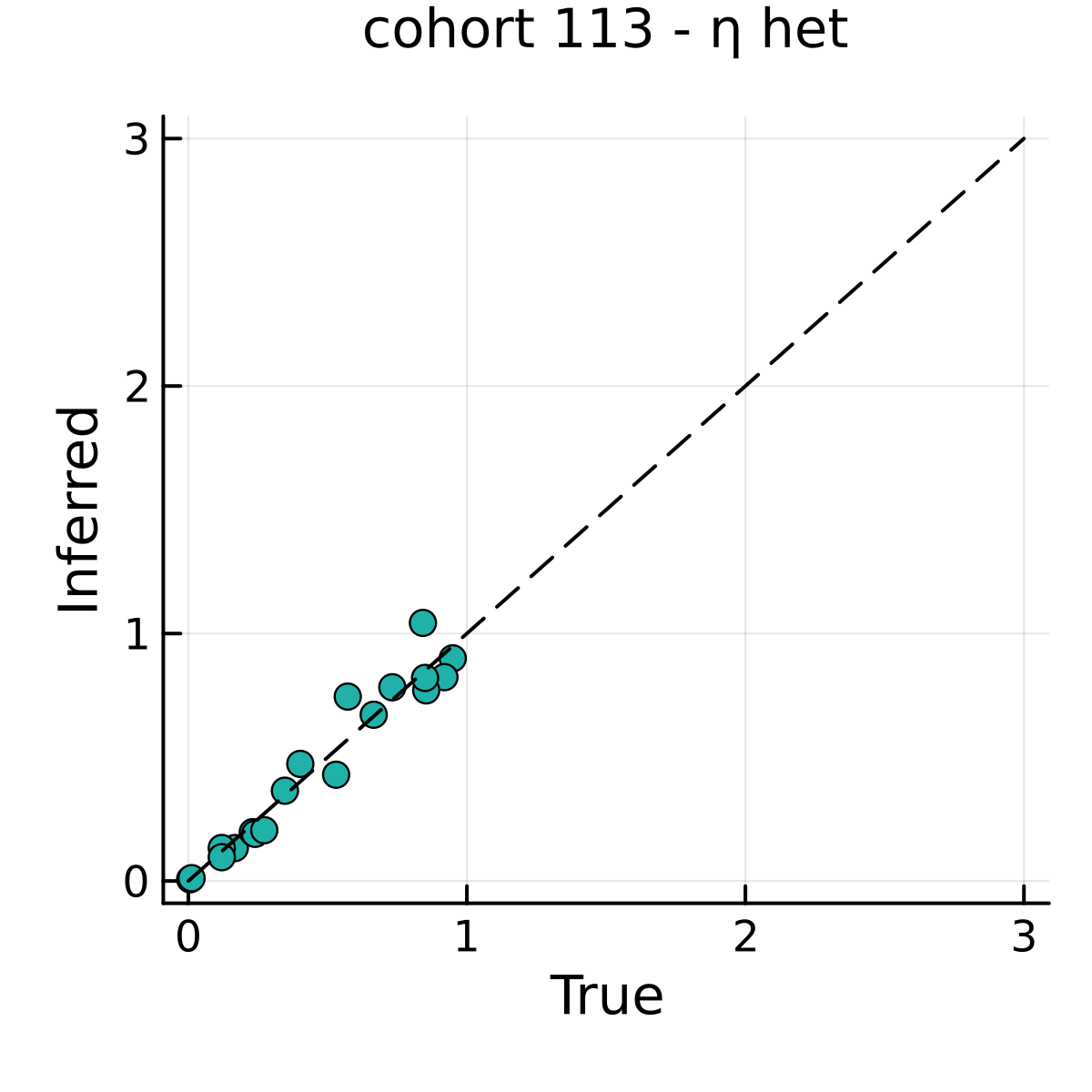}
    \end{subfigure}
     \begin{subfigure}[b]{0.22\textwidth}
        \includegraphics[width=\textwidth]{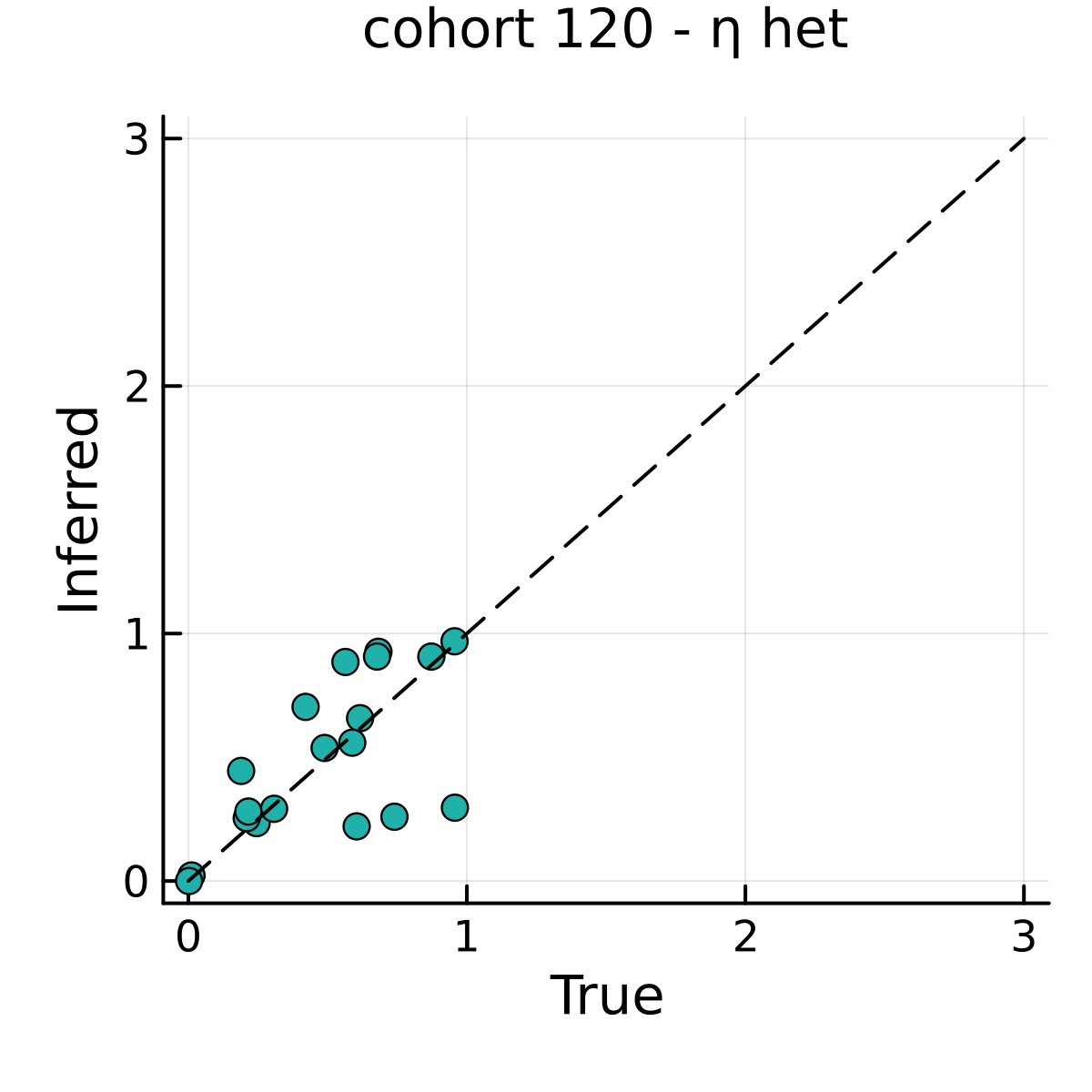}
    \end{subfigure}
     \begin{subfigure}[b]{0.22\textwidth}
        \includegraphics[width=\textwidth]{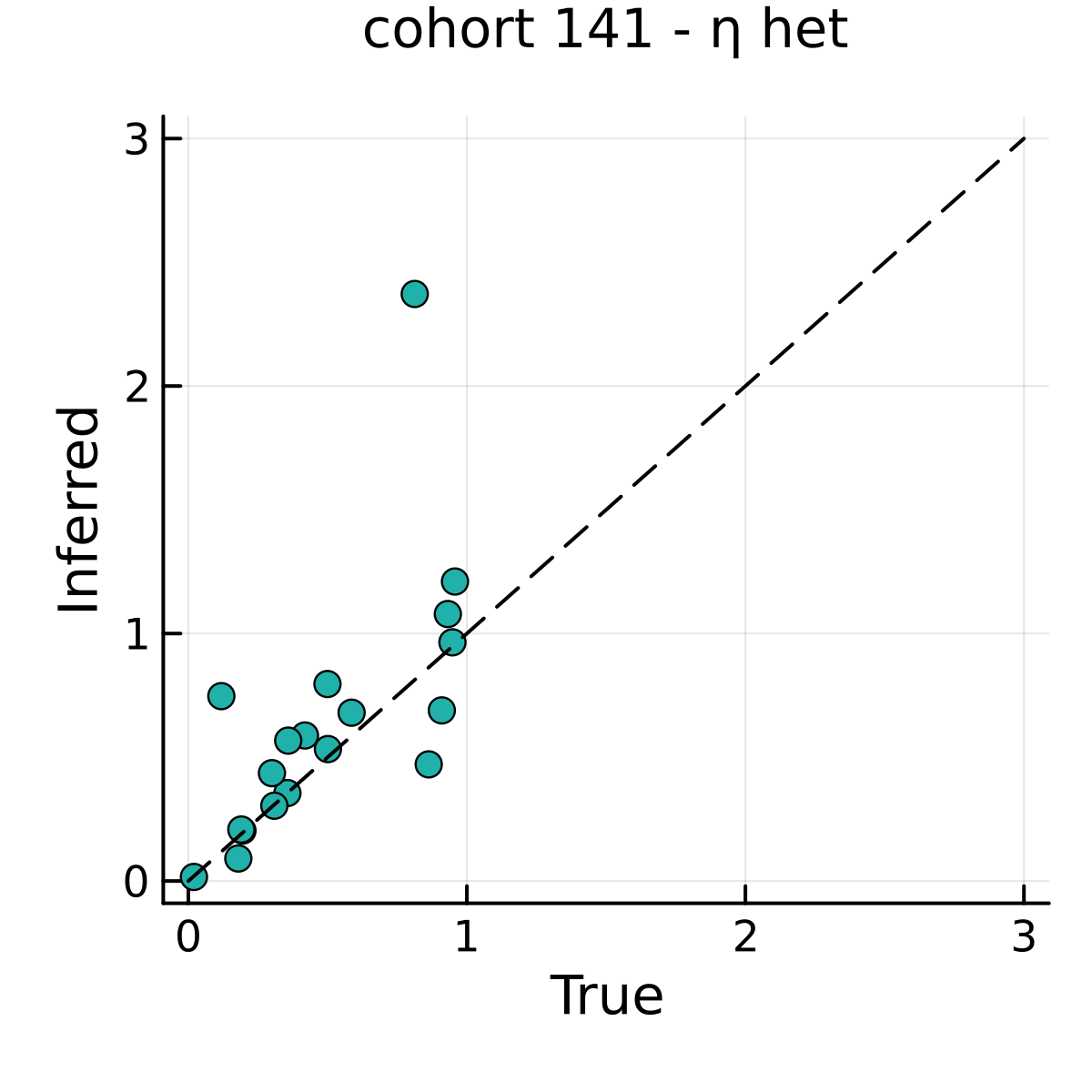}
    \end{subfigure}
     \begin{subfigure}[b]{0.22\textwidth}
        \includegraphics[width=\textwidth]{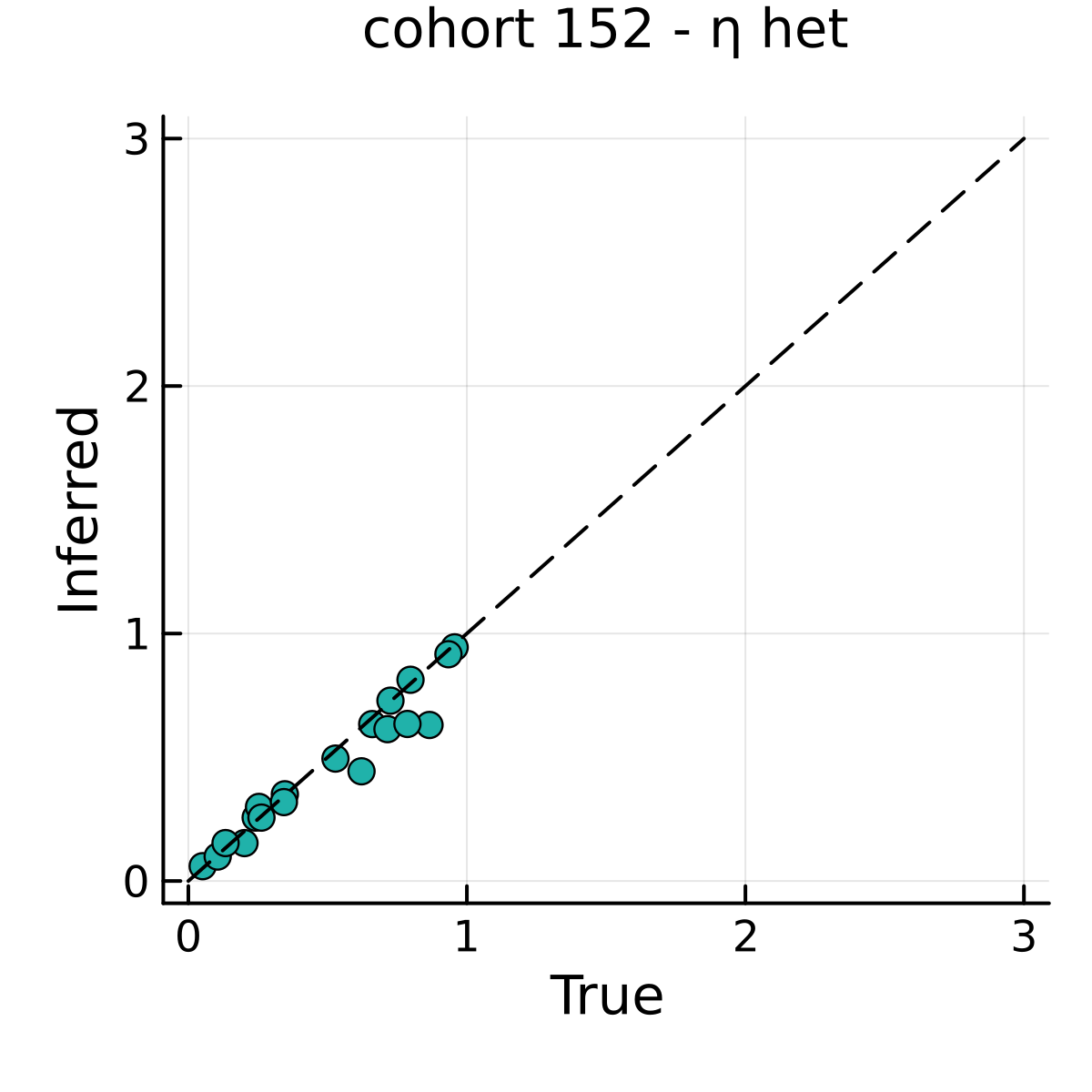}
    \end{subfigure}
     \begin{subfigure}[b]{0.22\textwidth}
        \includegraphics[width=\textwidth]{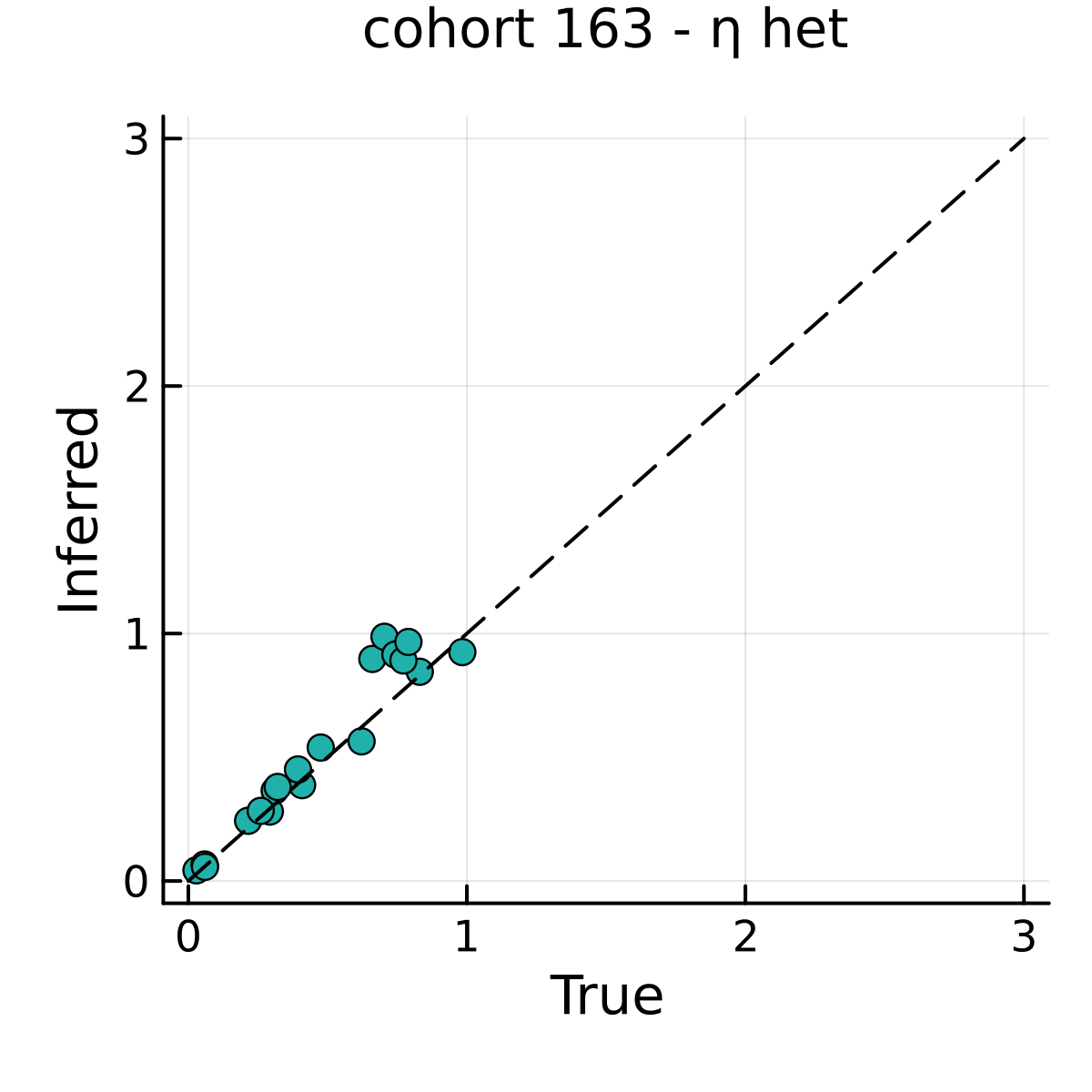}
    \end{subfigure}
     \begin{subfigure}[b]{0.22\textwidth}
        \includegraphics[width=\textwidth]{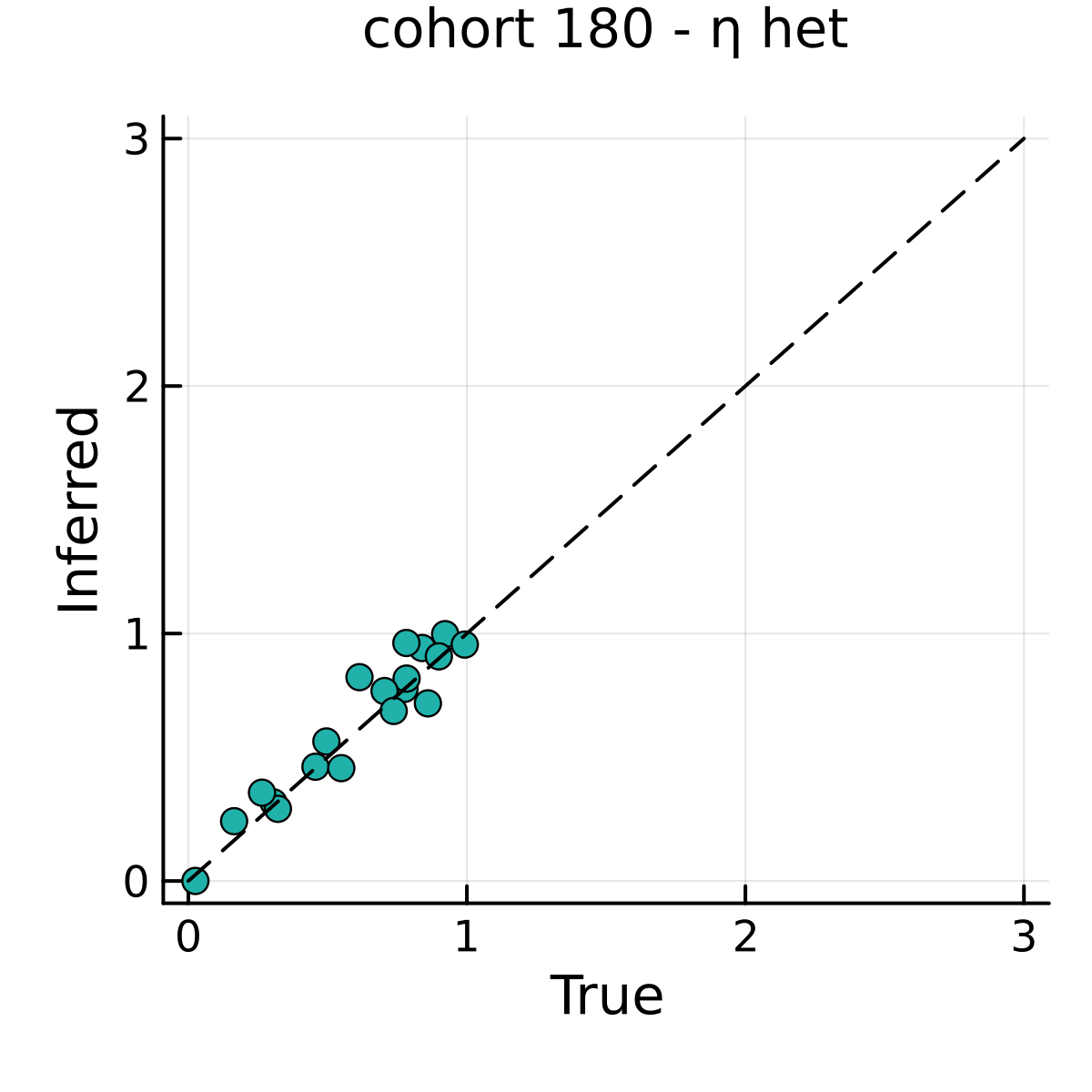}
    \end{subfigure}
    \begin{subfigure}[b]{0.22\textwidth}
        \includegraphics[width=\textwidth]{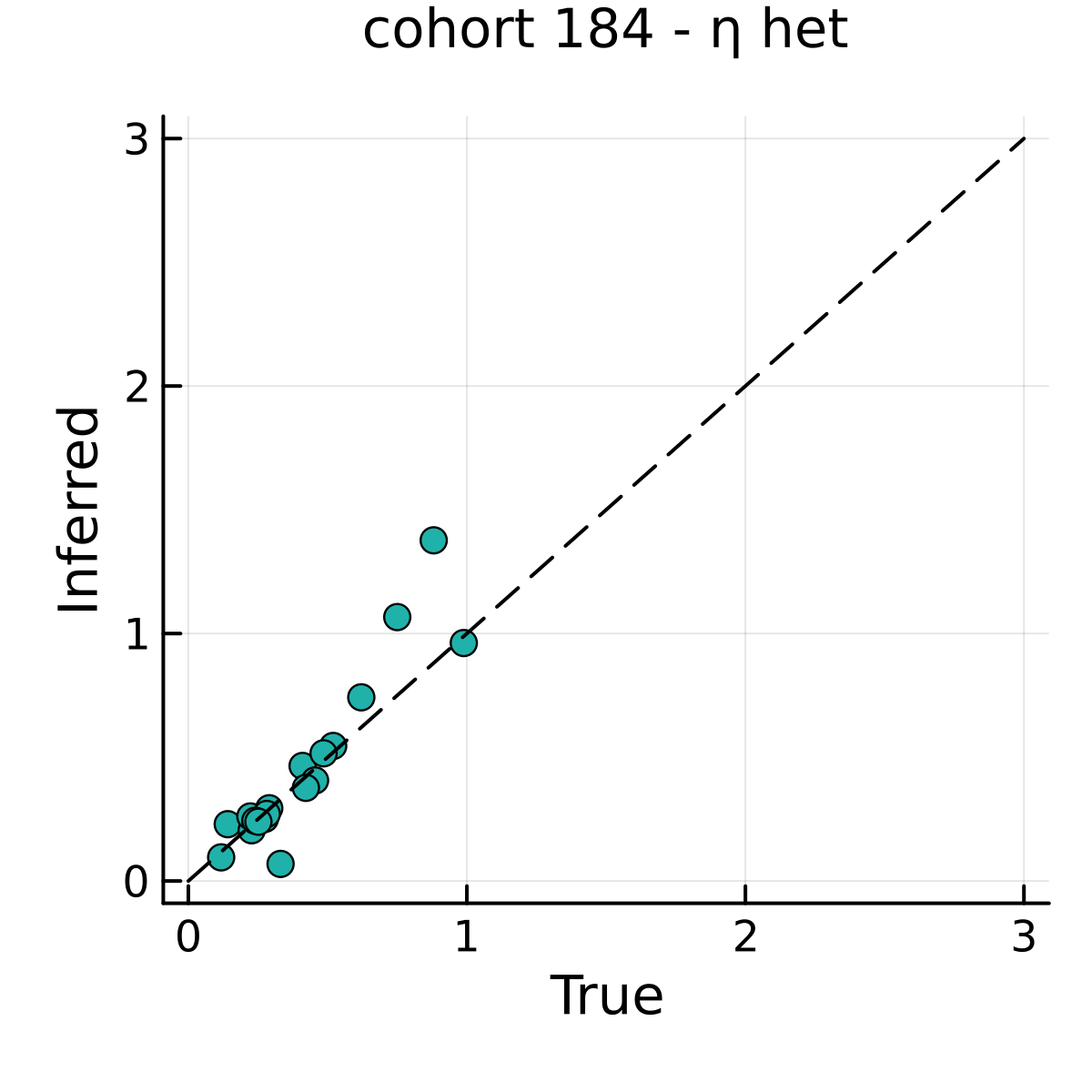}
    \end{subfigure}
     \begin{subfigure}[b]{0.22\textwidth}
        \includegraphics[width=\textwidth]{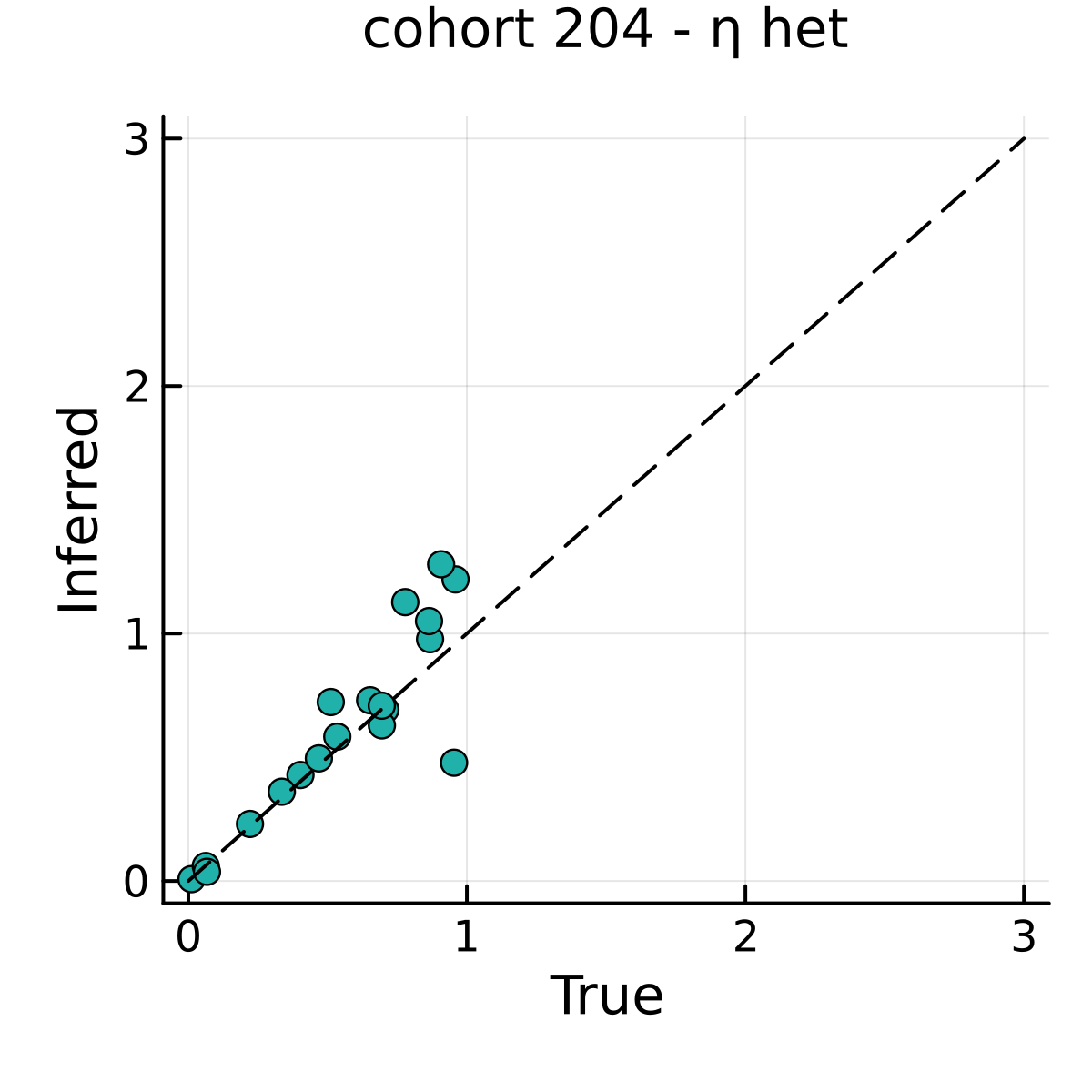}
    \end{subfigure} 
    \begin{subfigure}[b]{0.22\textwidth}
        \includegraphics[width=\textwidth]{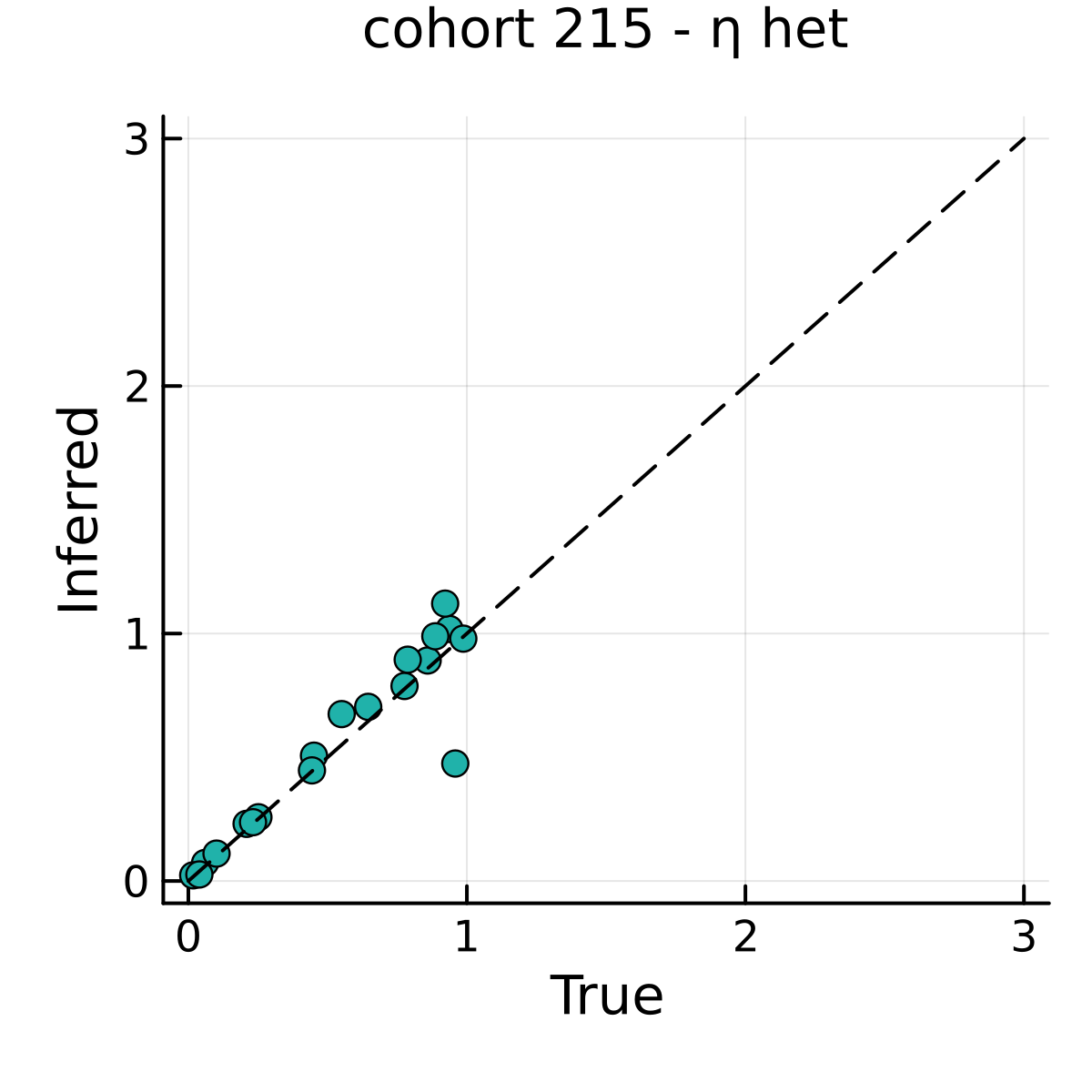}
    \end{subfigure}
     \begin{subfigure}[b]{0.22\textwidth}
        \includegraphics[width=\textwidth]{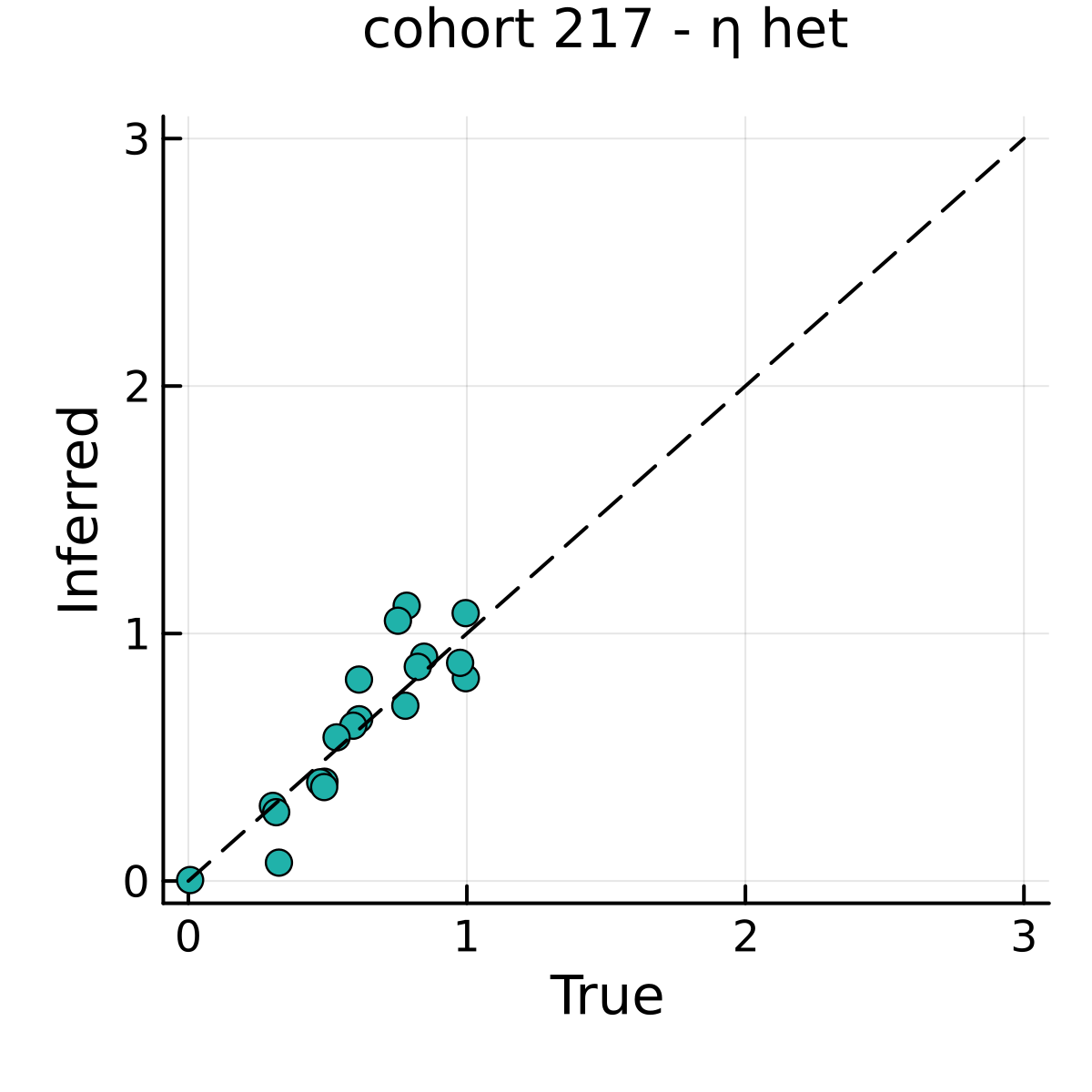}
    \end{subfigure}
    
    \caption{Comparison between the inferred (posterior mean, y-axis) value of parameter $\eta_{het}$ and the true one (x-axis) for each virtual cohort (subfigures). }
    \label{fig:synth_eta_het}
\end{figure}

\begin{figure}[h]
    \centering
    \begin{subfigure}[b]{0.22\textwidth}
        \includegraphics[width=\textwidth]{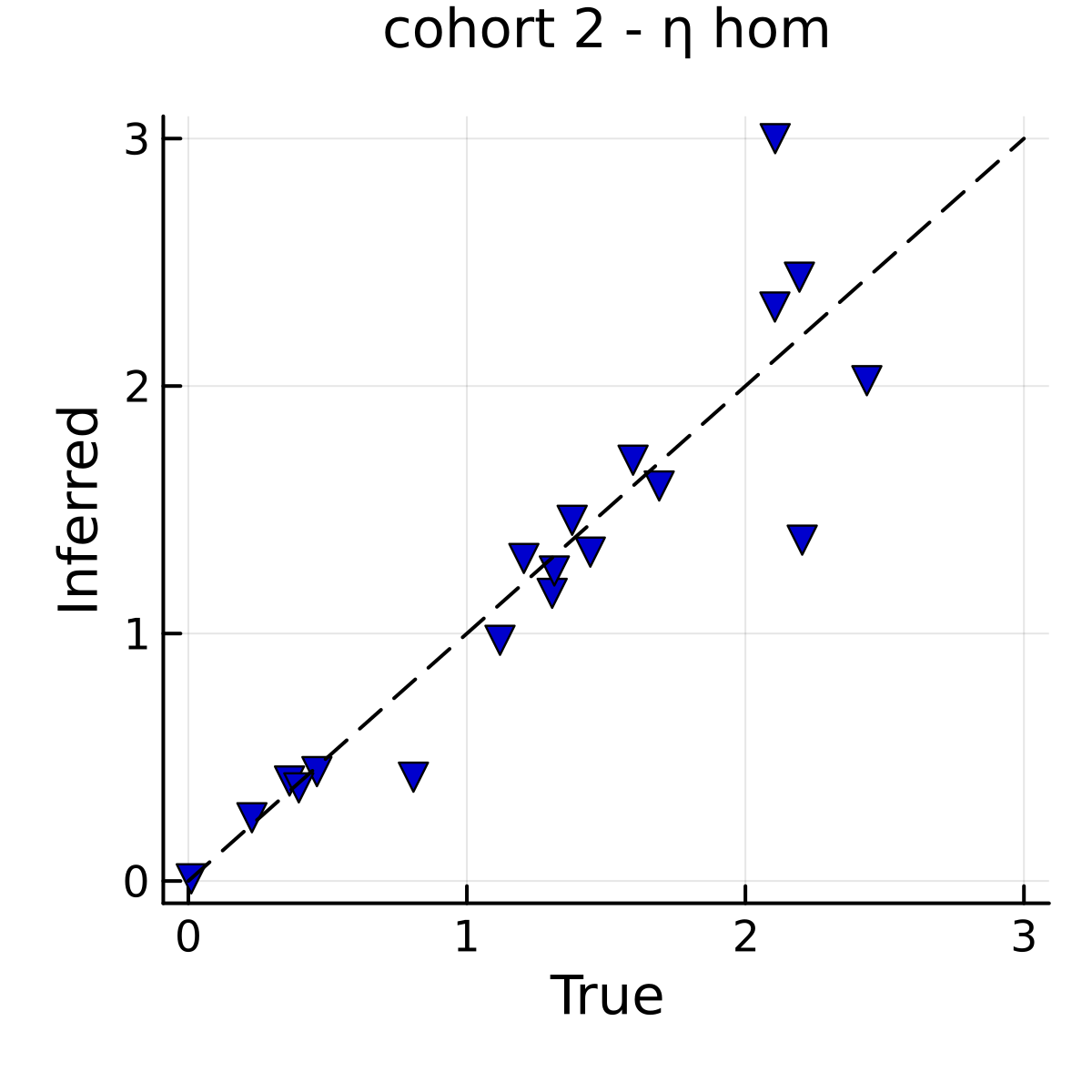}
    \end{subfigure}
     \begin{subfigure}[b]{0.22\textwidth}
        \includegraphics[width=\textwidth]{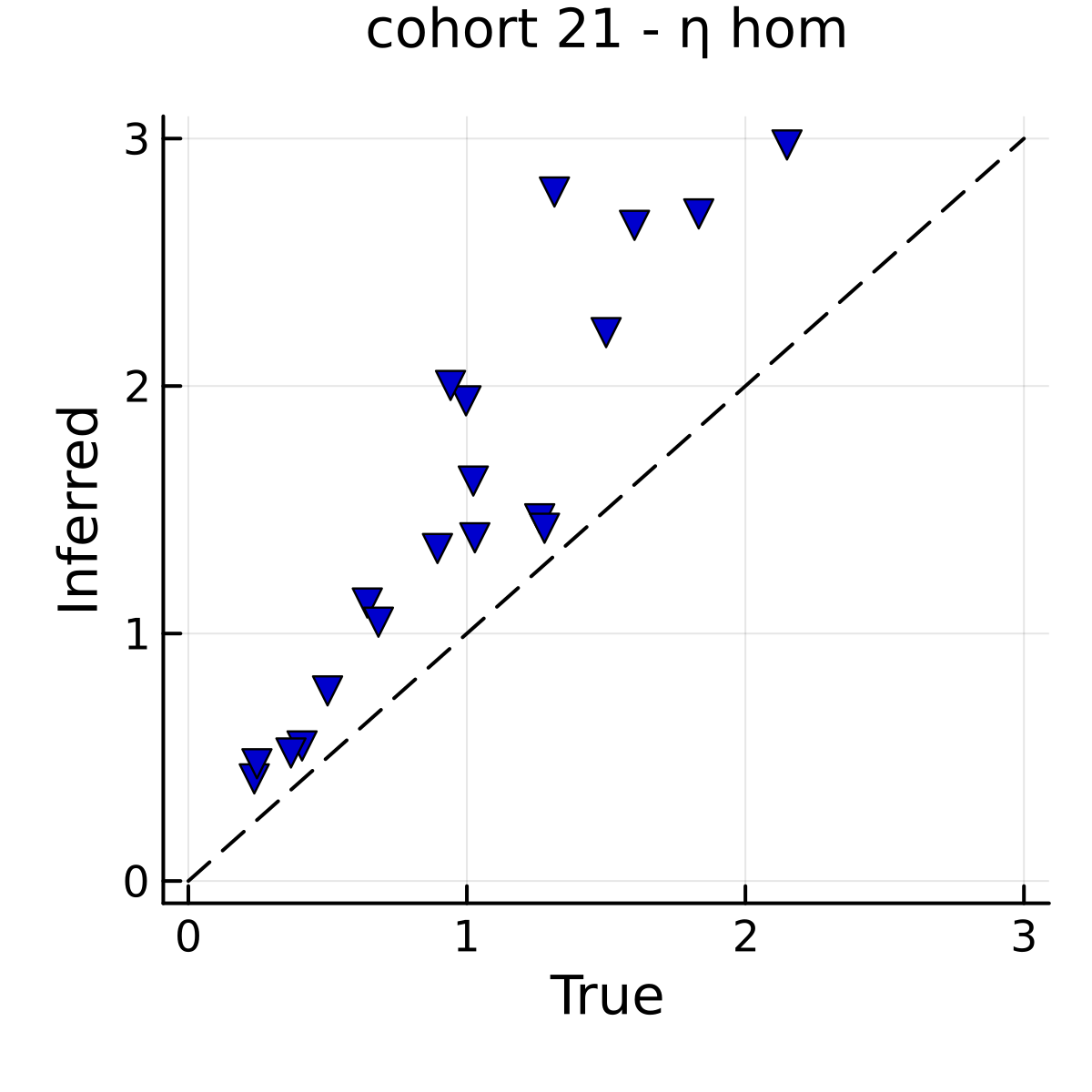}
    \end{subfigure}
     \begin{subfigure}[b]{0.22\textwidth}
        \includegraphics[width=\textwidth]{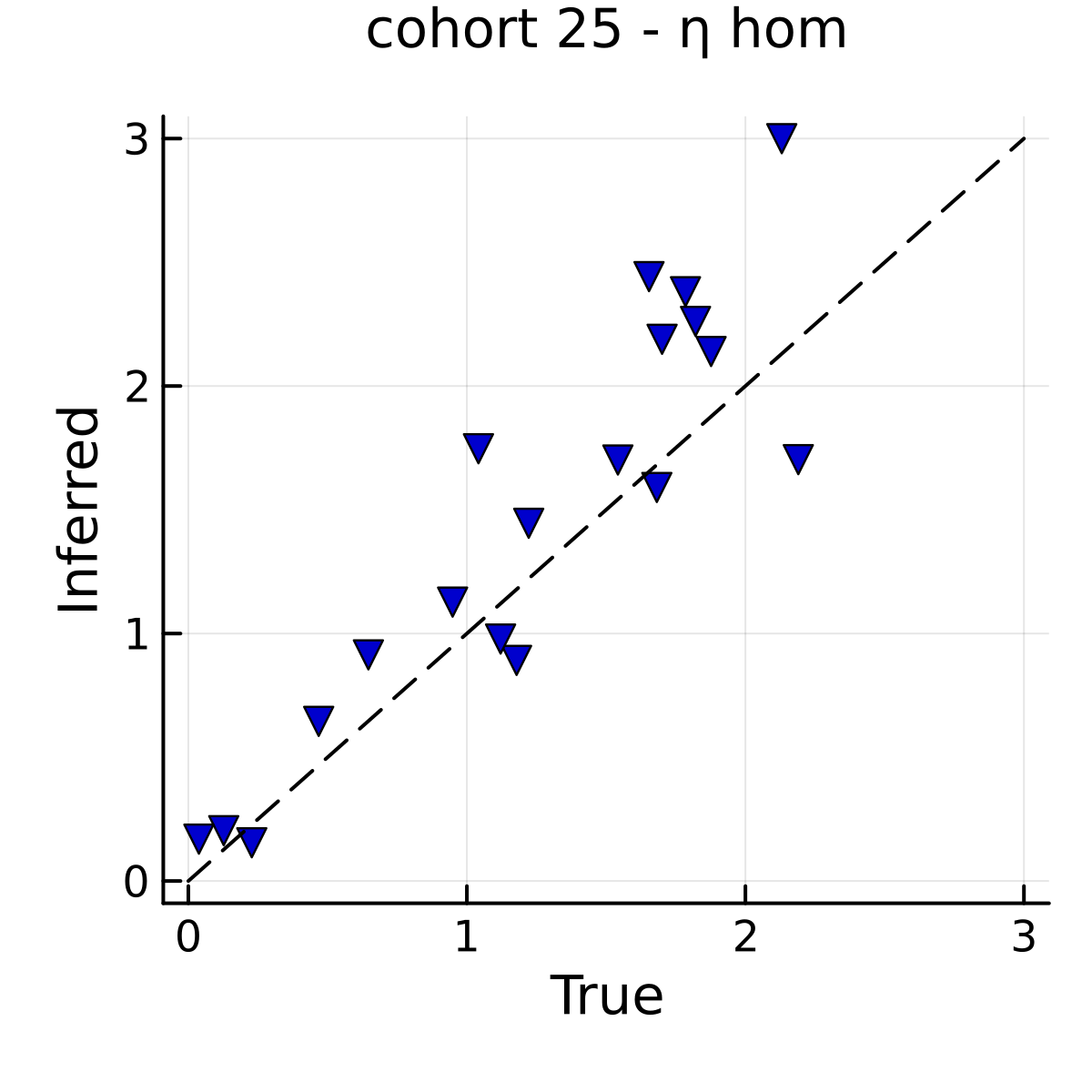}
    \end{subfigure}
     \begin{subfigure}[b]{0.22\textwidth}
        \includegraphics[width=\textwidth]{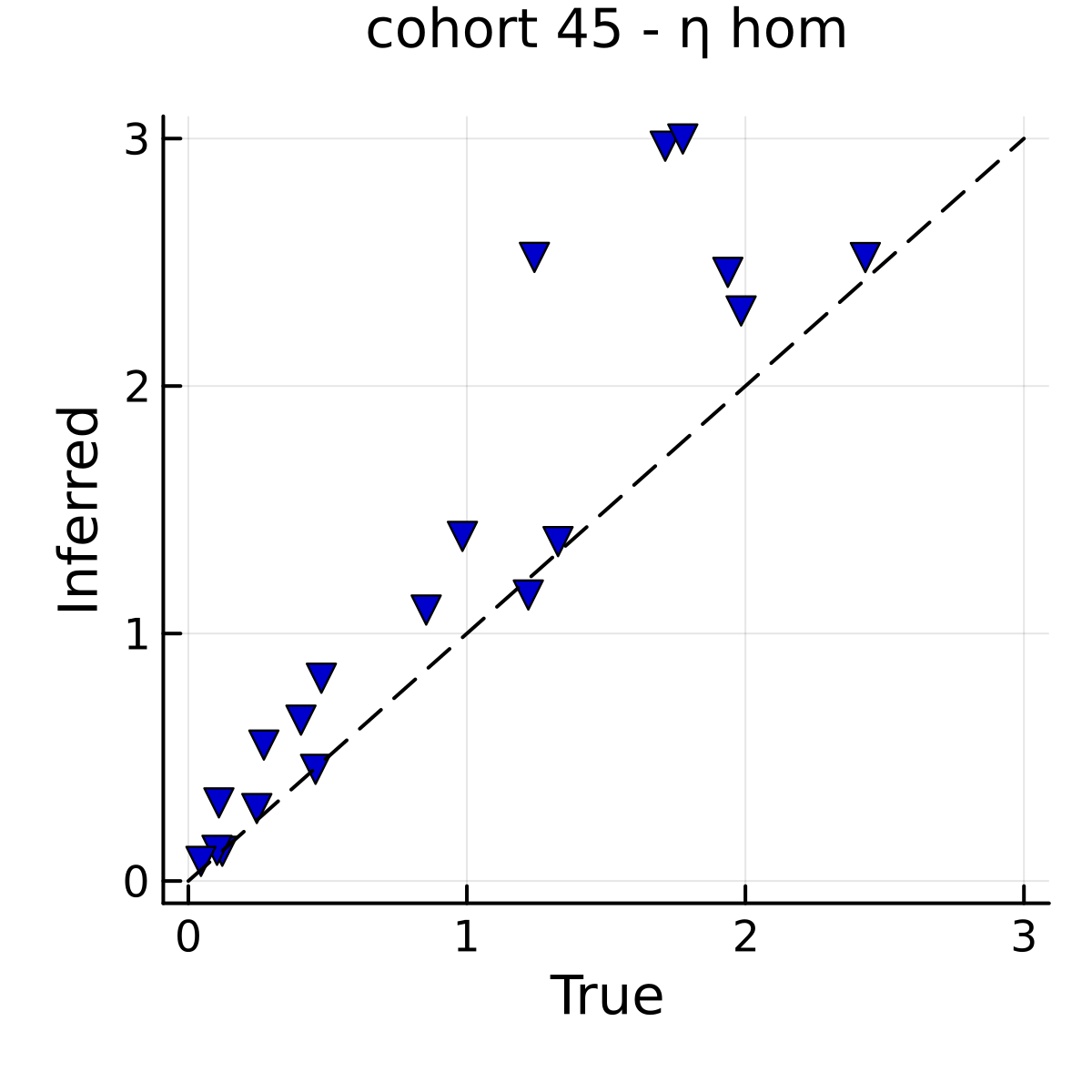}
    \end{subfigure}
     \begin{subfigure}[b]{0.22\textwidth}
        \includegraphics[width=\textwidth]{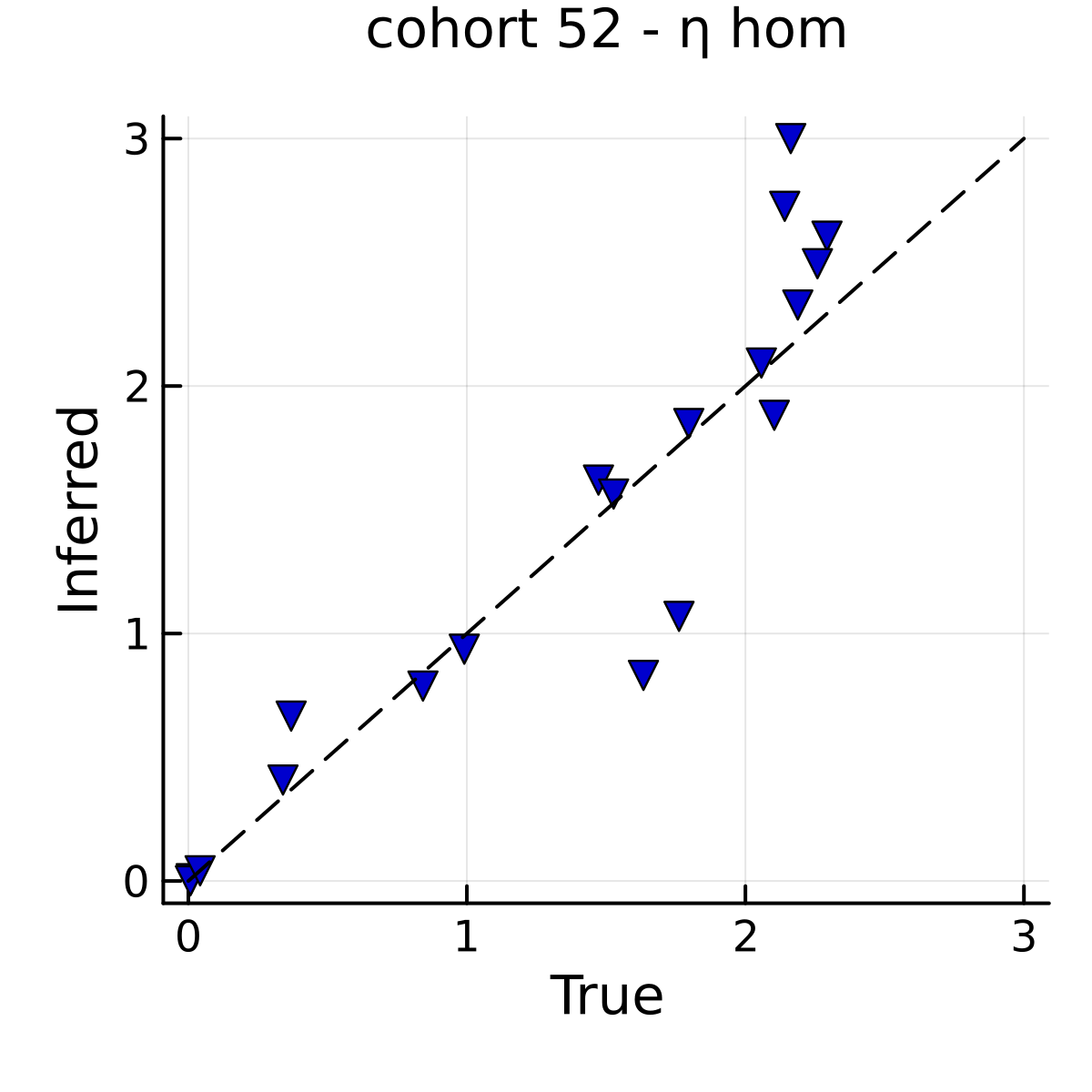}
    \end{subfigure}
     \begin{subfigure}[b]{0.22\textwidth}
        \includegraphics[width=\textwidth]{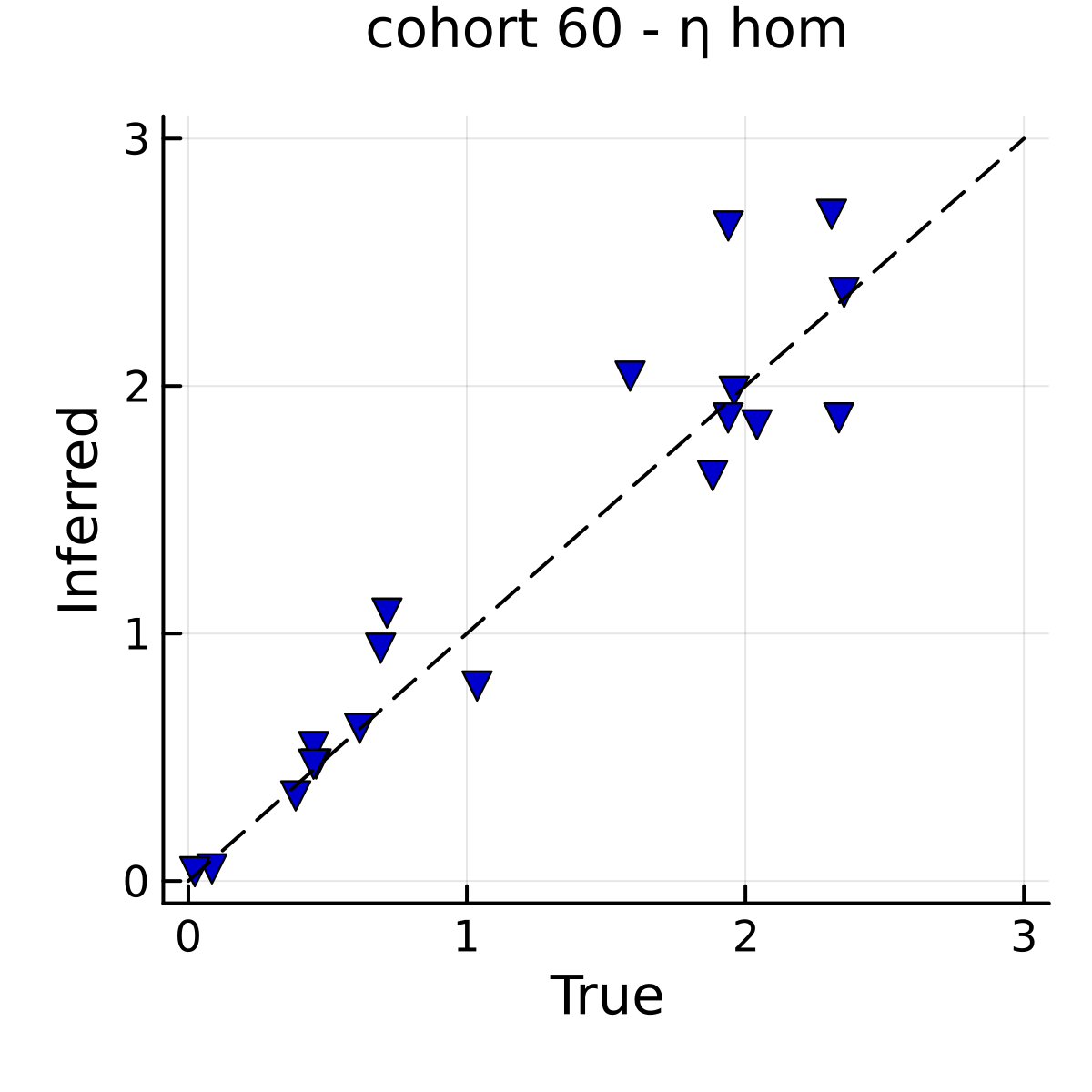}
    \end{subfigure}
     \begin{subfigure}[b]{0.22\textwidth}
        \includegraphics[width=\textwidth]{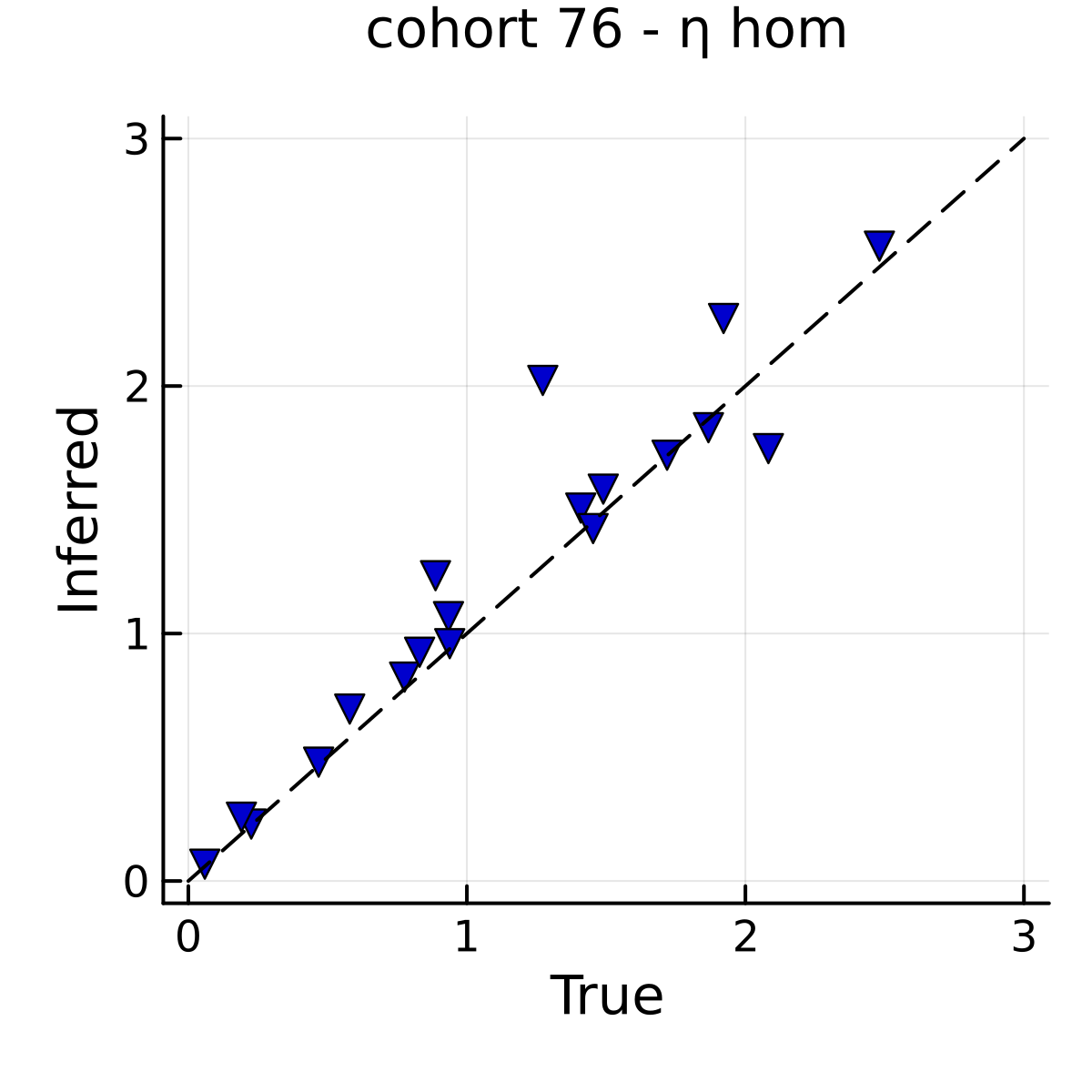}
    \end{subfigure}
     \begin{subfigure}[b]{0.22\textwidth}
        \includegraphics[width=\textwidth]{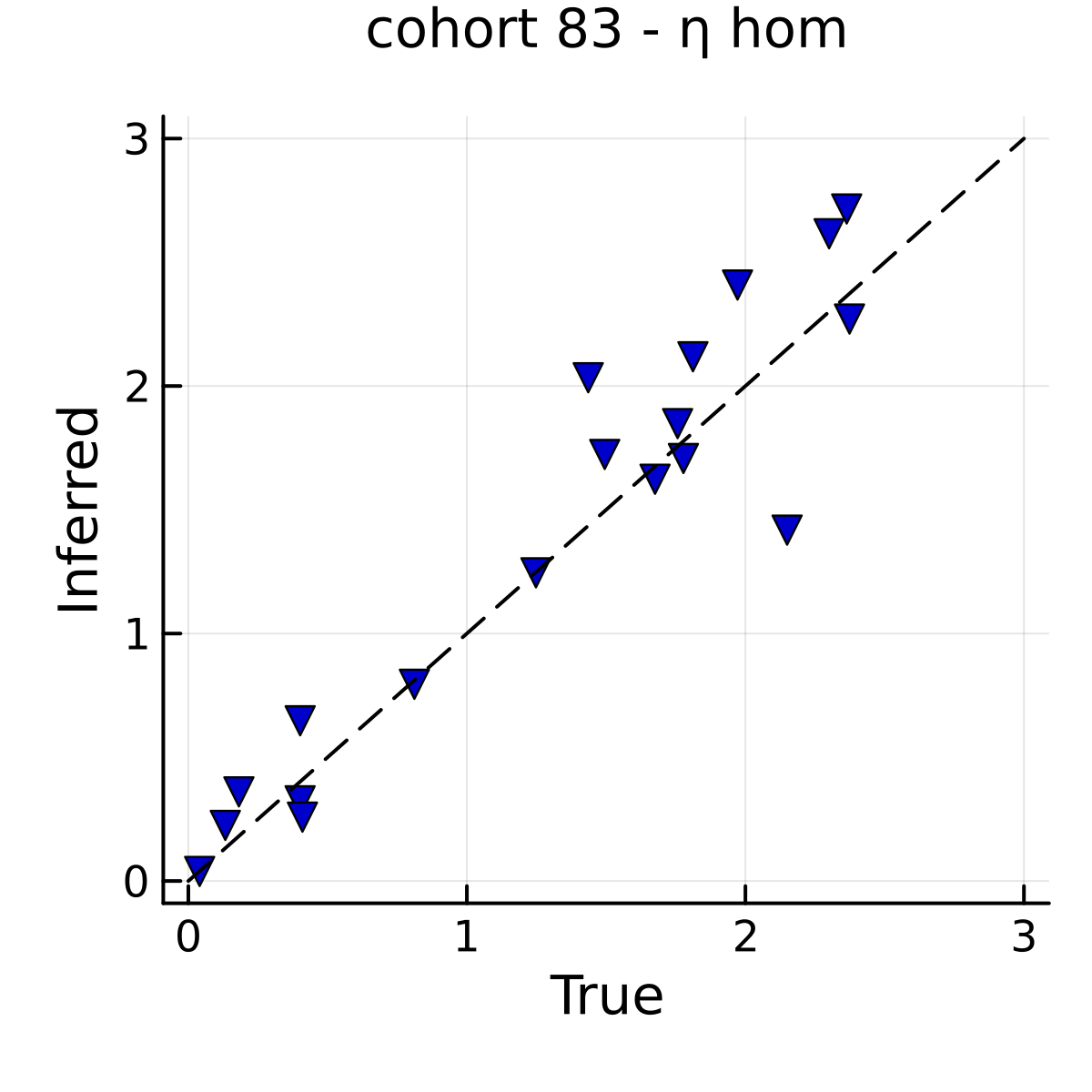}
    \end{subfigure}
     \begin{subfigure}[b]{0.22\textwidth}
        \includegraphics[width=\textwidth]{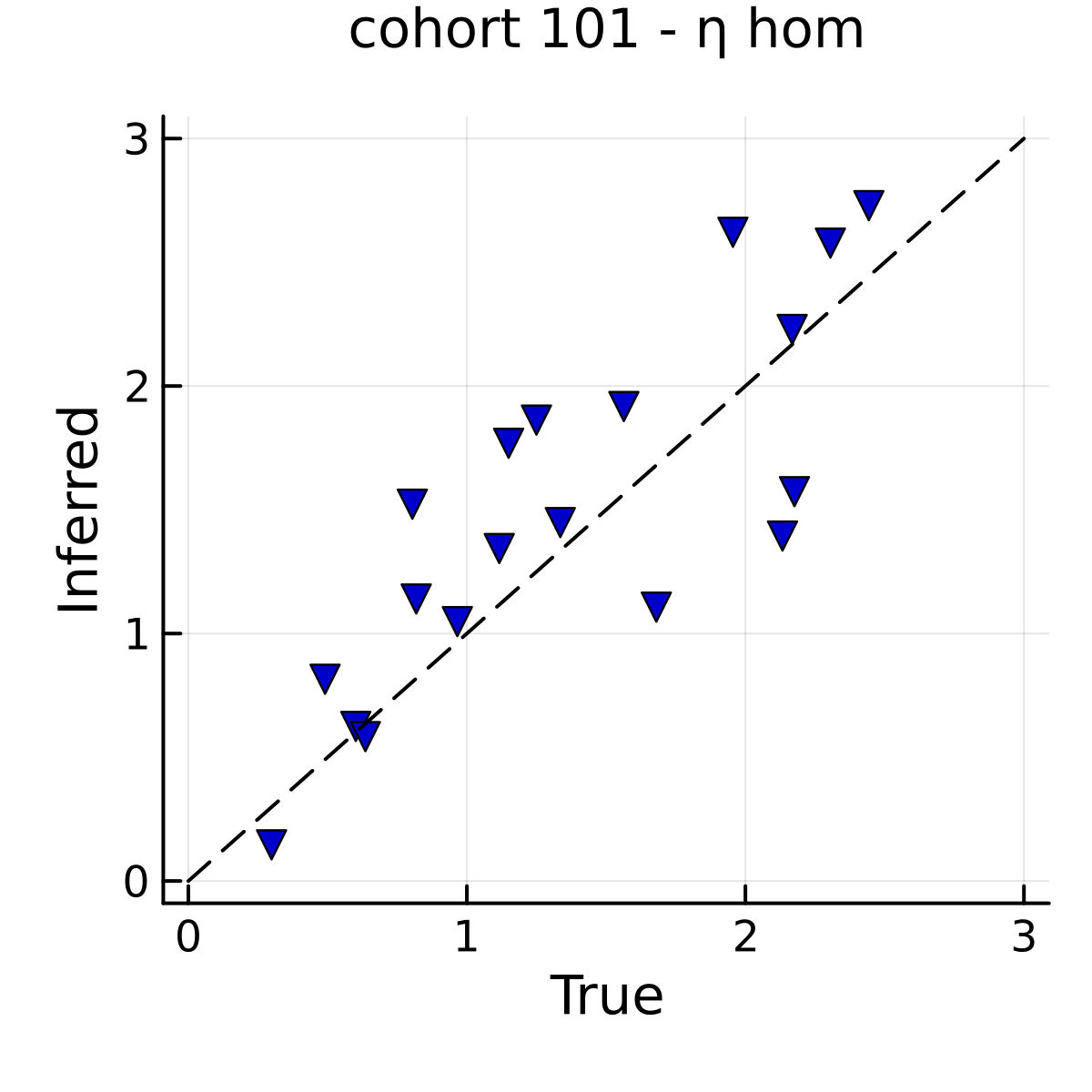}
    \end{subfigure}
     \begin{subfigure}[b]{0.22\textwidth}
        \includegraphics[width=\textwidth]{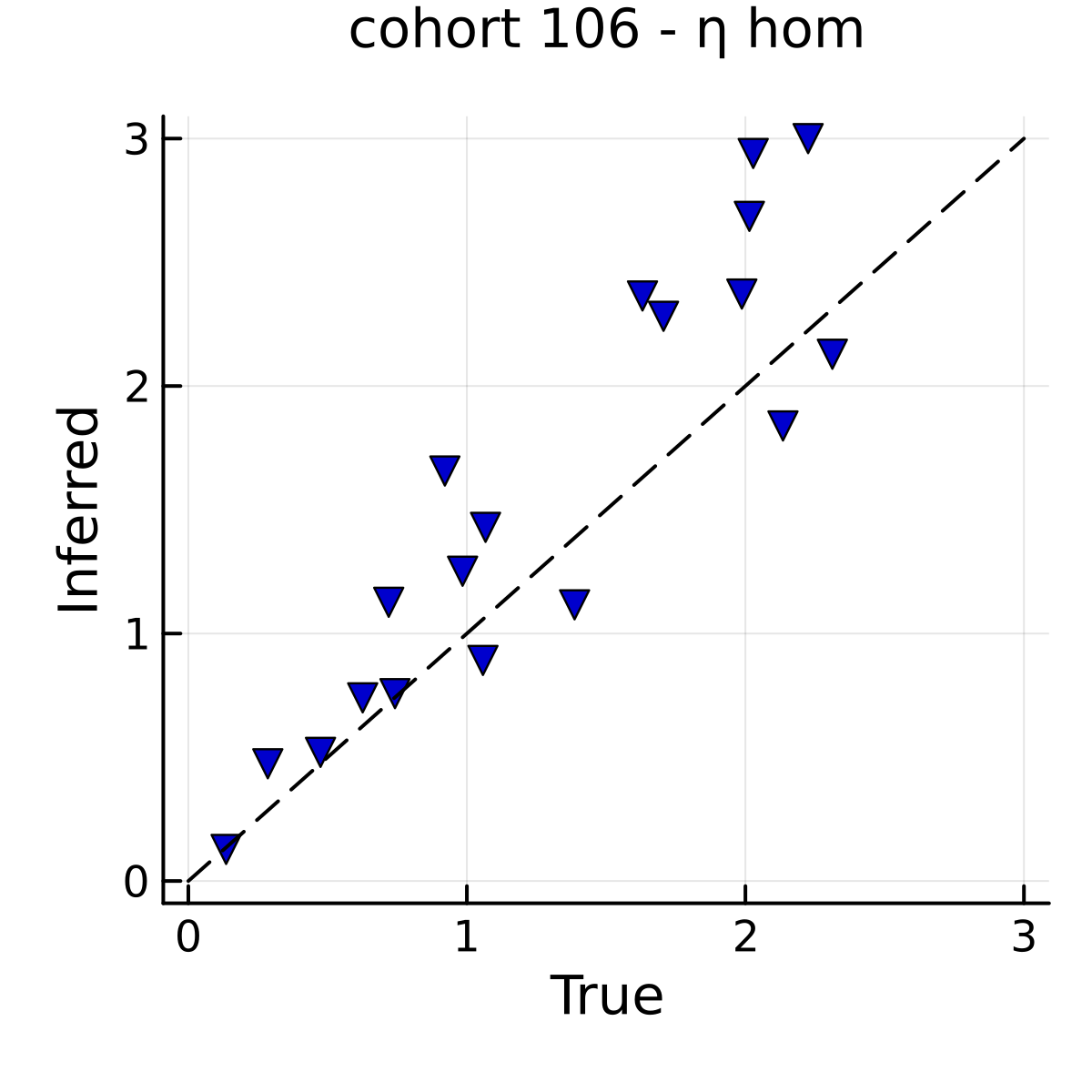}
    \end{subfigure}
     \begin{subfigure}[b]{0.22\textwidth}
        \includegraphics[width=\textwidth]{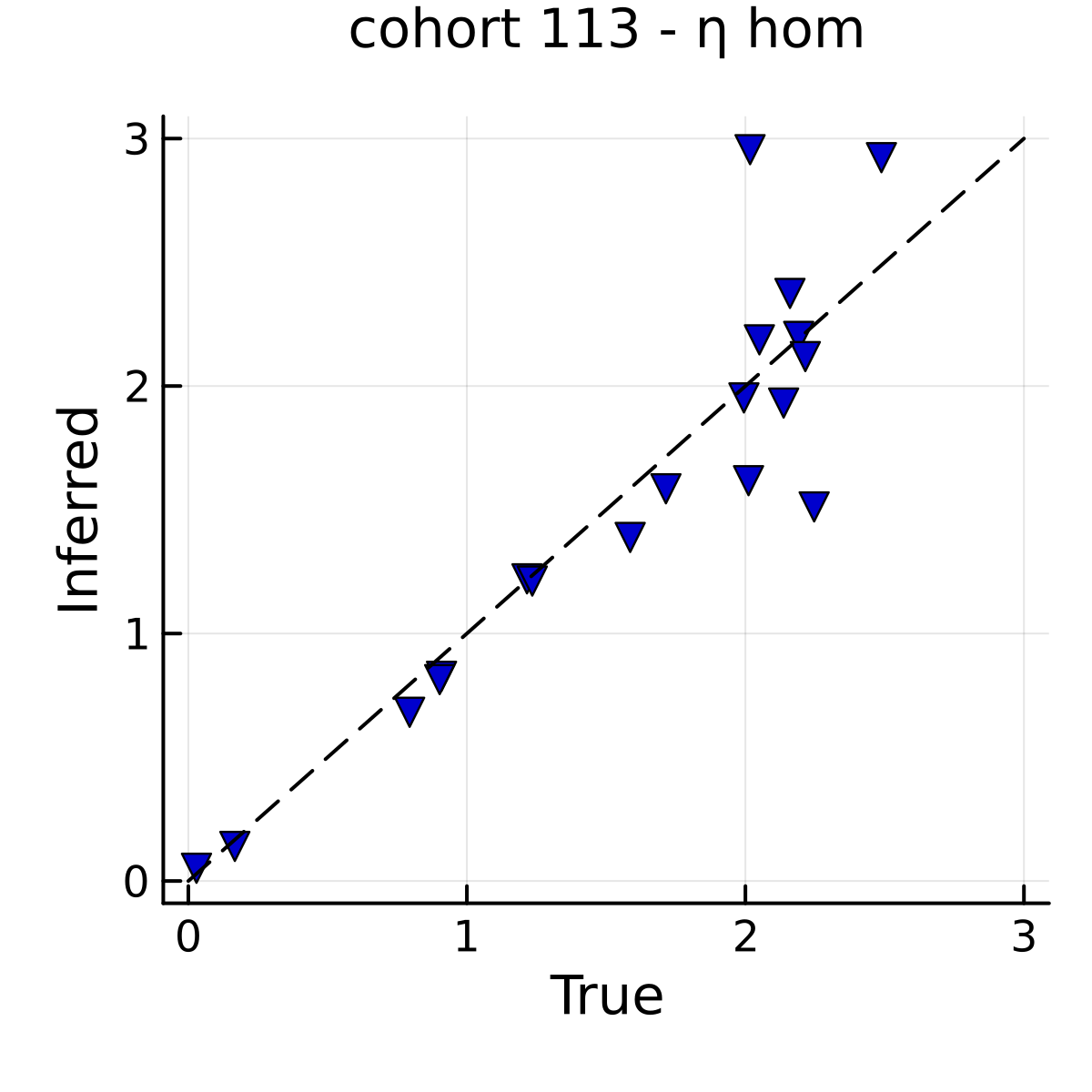}
    \end{subfigure}
     \begin{subfigure}[b]{0.22\textwidth}
        \includegraphics[width=\textwidth]{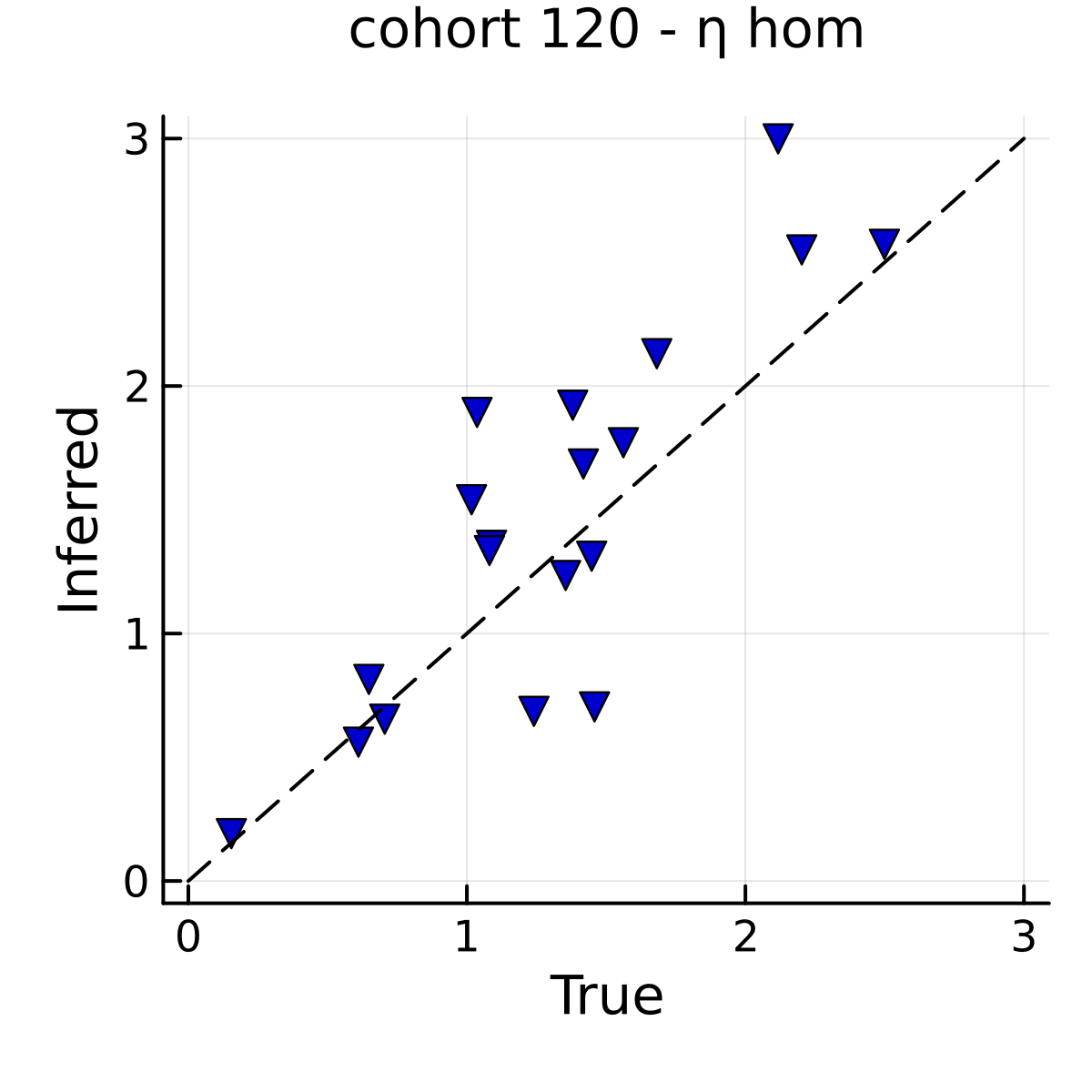}
    \end{subfigure}
     \begin{subfigure}[b]{0.22\textwidth}
        \includegraphics[width=\textwidth]{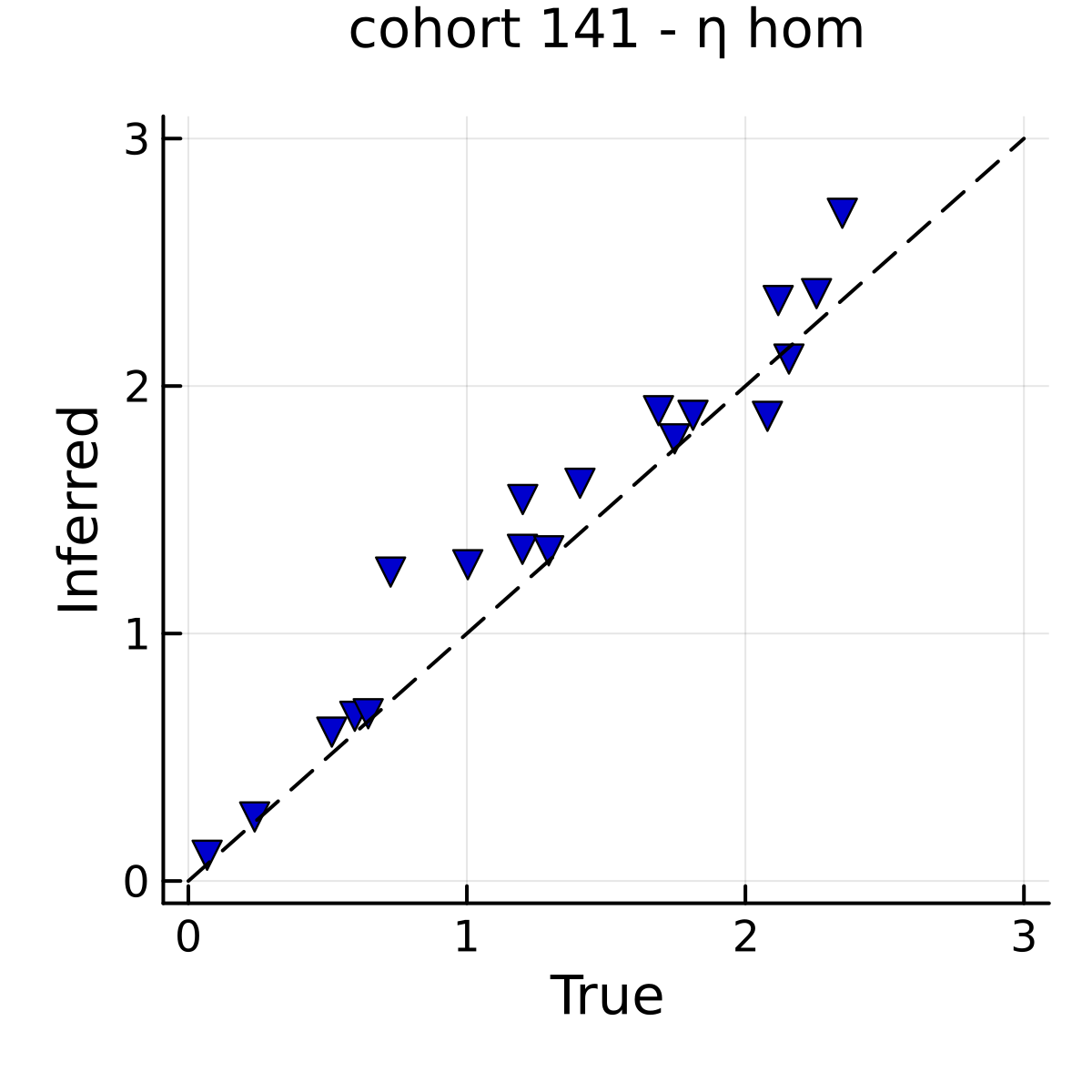}
    \end{subfigure}
     \begin{subfigure}[b]{0.22\textwidth}
        \includegraphics[width=\textwidth]{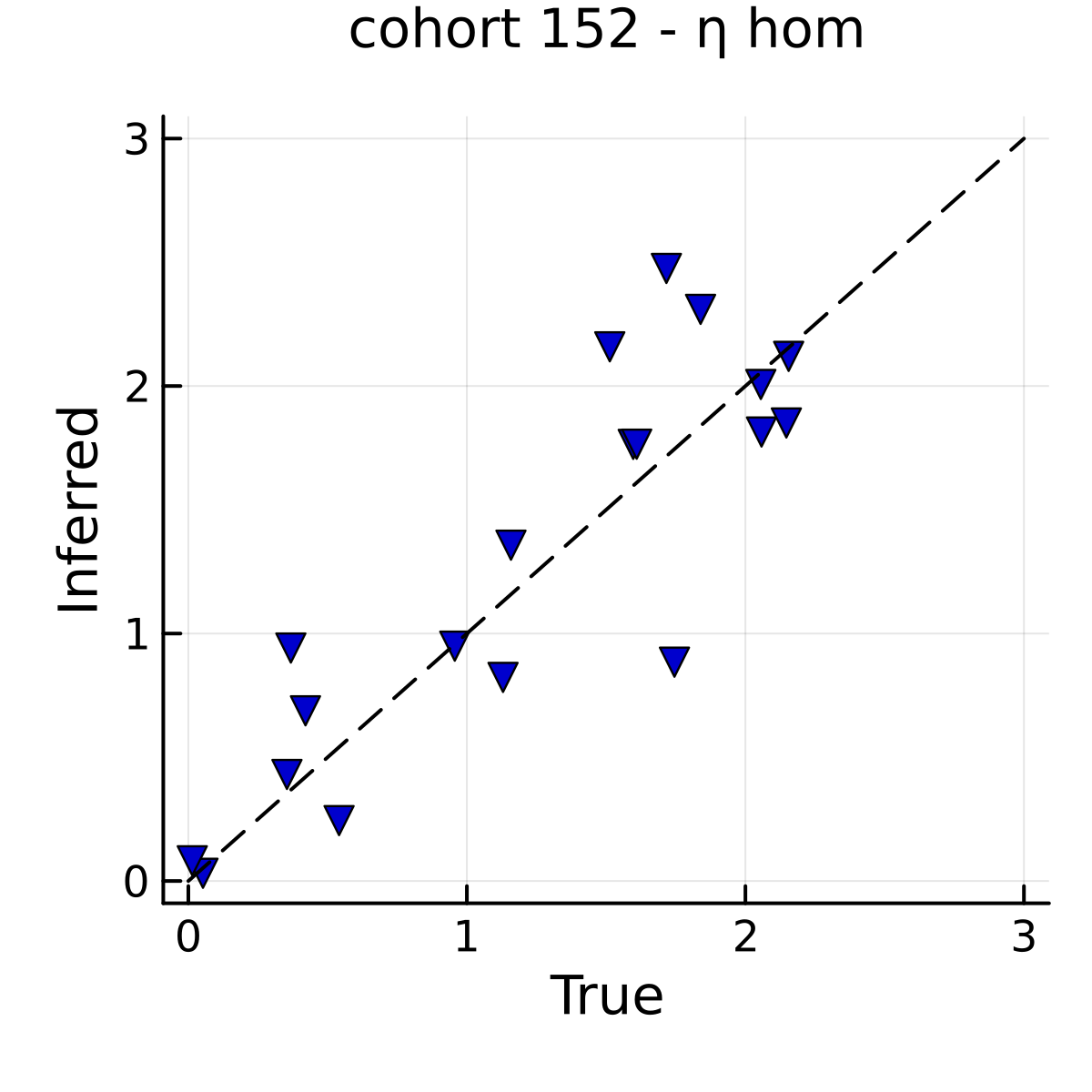}
    \end{subfigure}
     \begin{subfigure}[b]{0.22\textwidth}
        \includegraphics[width=\textwidth]{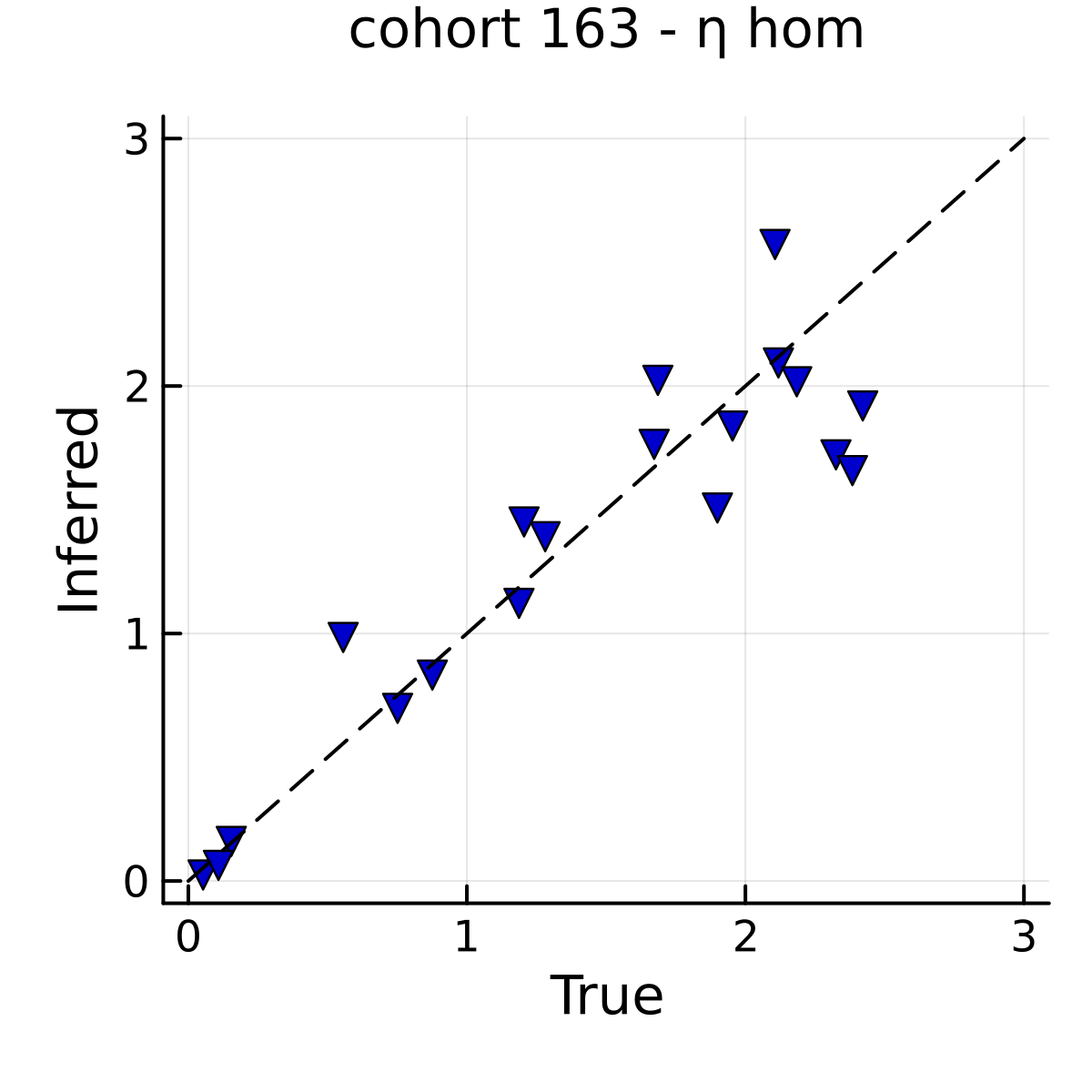}
    \end{subfigure}
     \begin{subfigure}[b]{0.22\textwidth}
        \includegraphics[width=\textwidth]{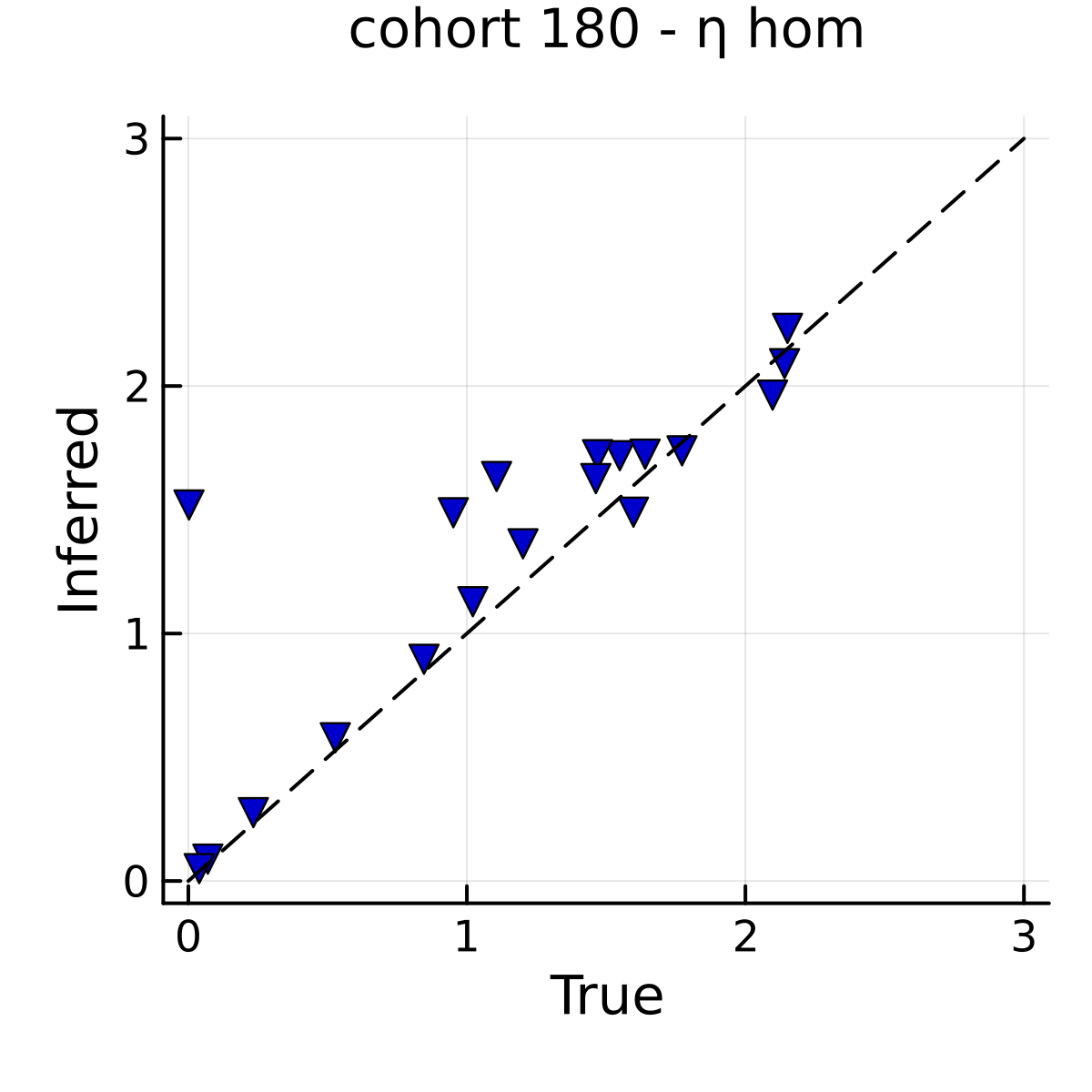}
    \end{subfigure}
    \begin{subfigure}[b]{0.22\textwidth}
        \includegraphics[width=\textwidth]{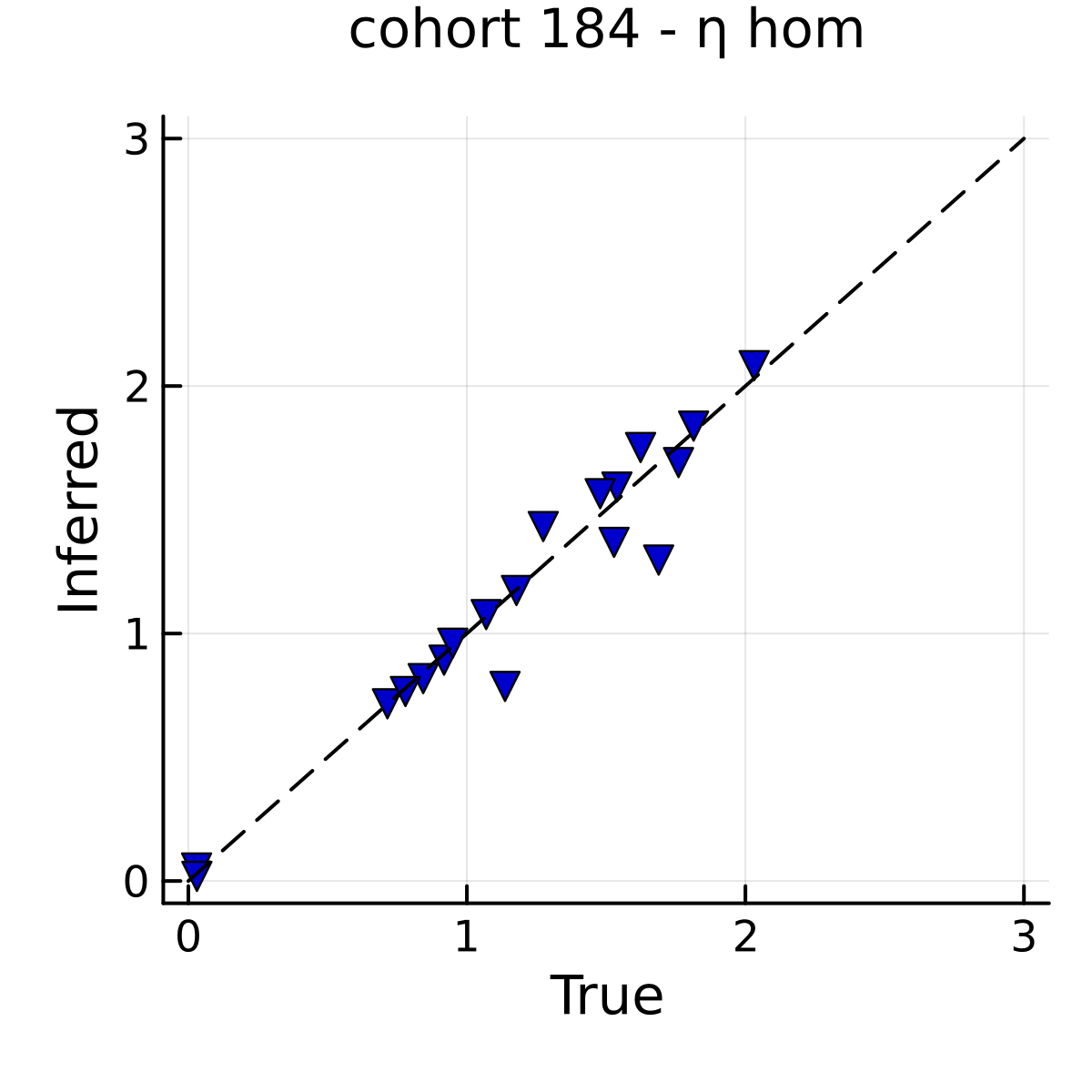}
    \end{subfigure}
     \begin{subfigure}[b]{0.22\textwidth}
        \includegraphics[width=\textwidth]{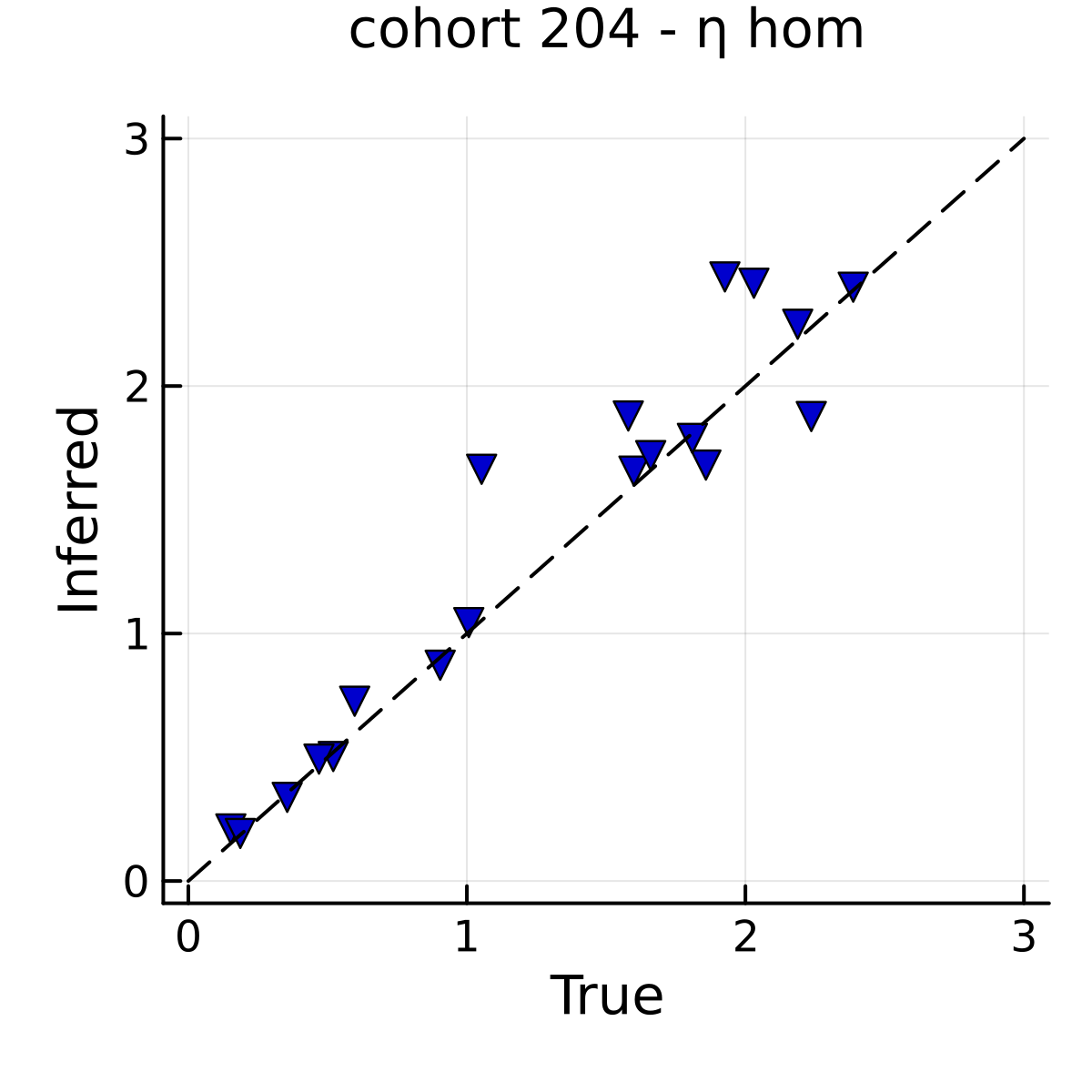}
    \end{subfigure} 
    \begin{subfigure}[b]{0.22\textwidth}
        \includegraphics[width=\textwidth]{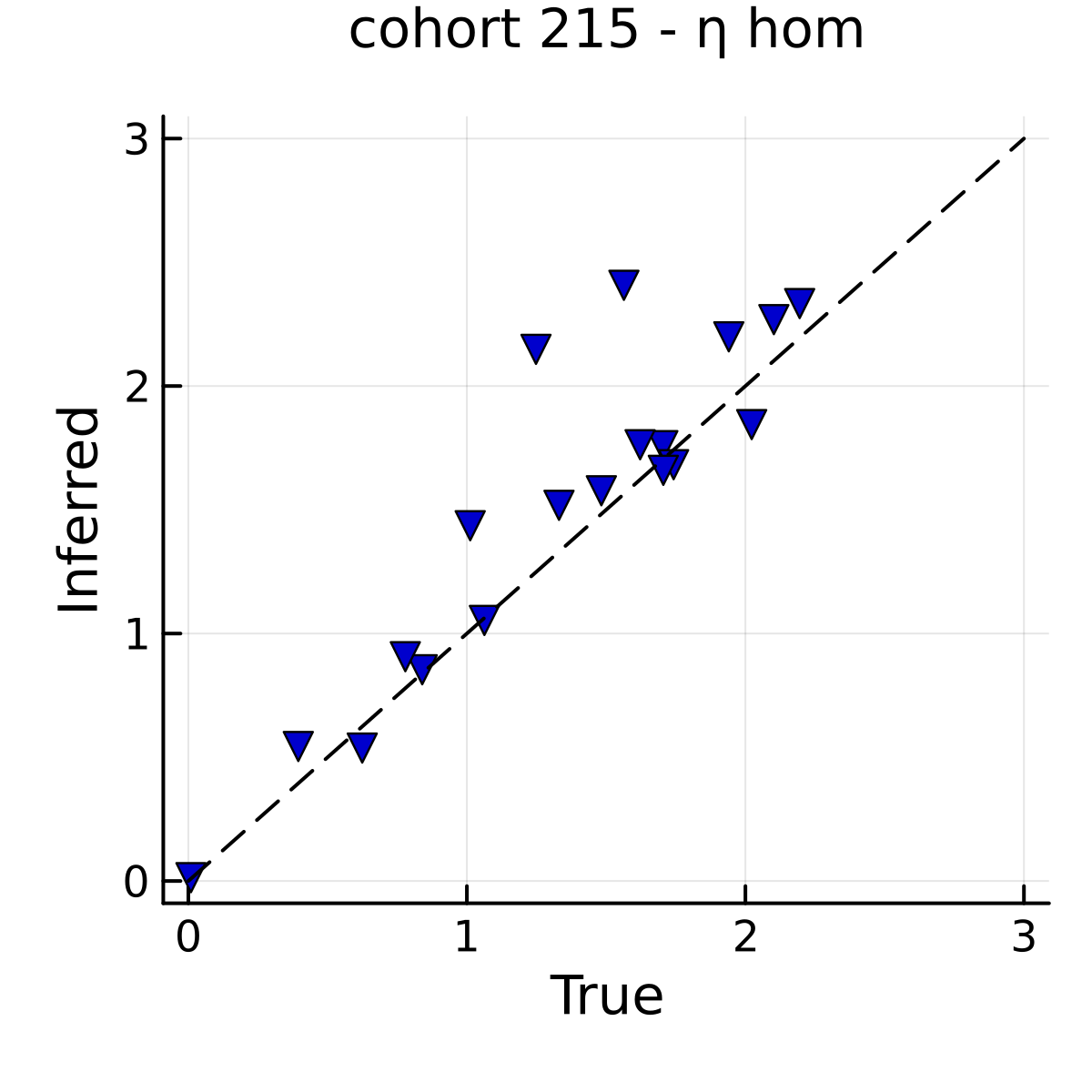}
    \end{subfigure}
     \begin{subfigure}[b]{0.22\textwidth}
        \includegraphics[width=\textwidth]{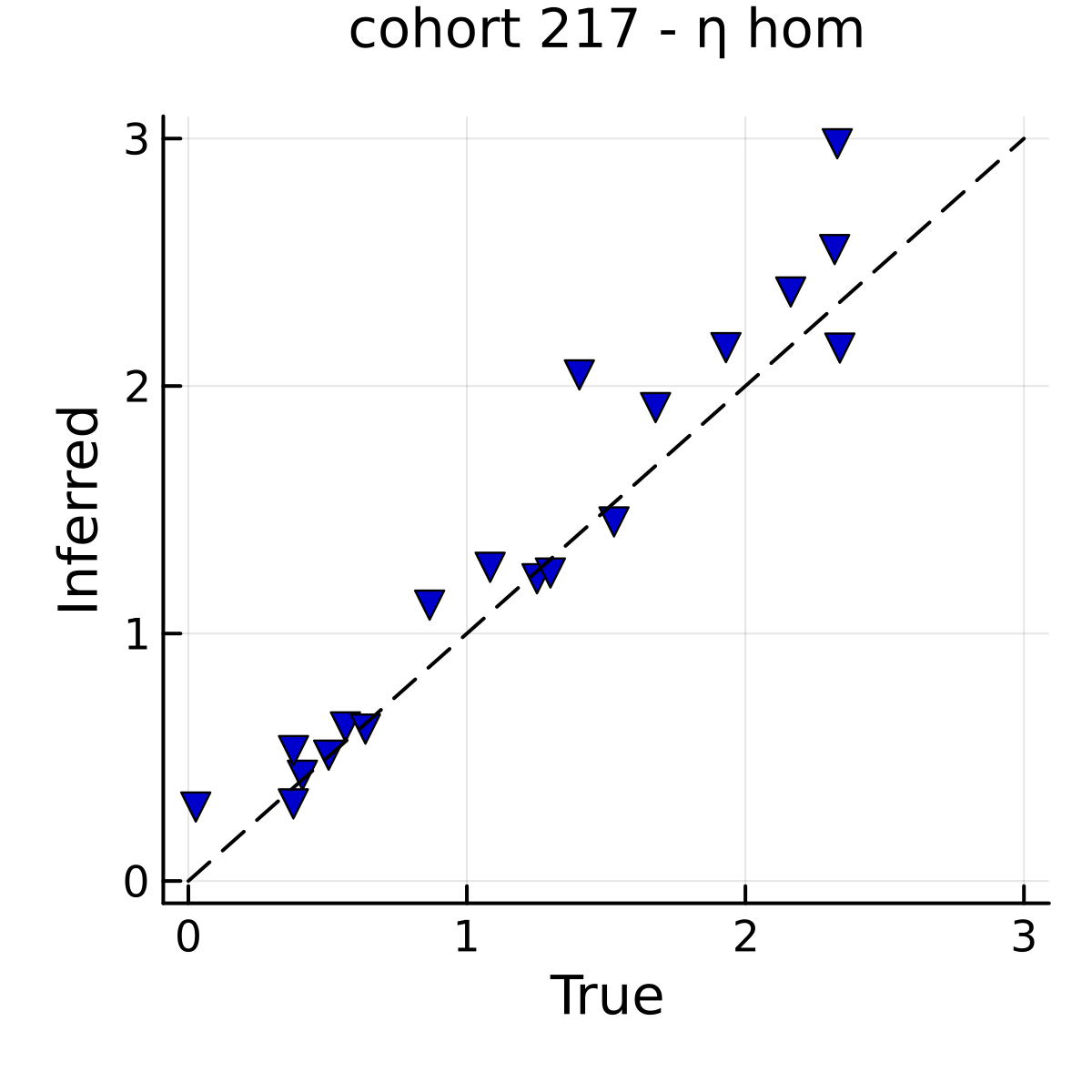}
    \end{subfigure}
    \caption{Comparison between the inferred (posterior mean, y-axis) value of parameter $\eta_{hom}$ and the true one (x-axis) for each virtual cohort. }
    \label{fig:synth_eta_hom}
\end{figure}

\begin{figure}[h]
    \centering
    \begin{subfigure}[b]{0.22\textwidth}
        \includegraphics[width=\textwidth]{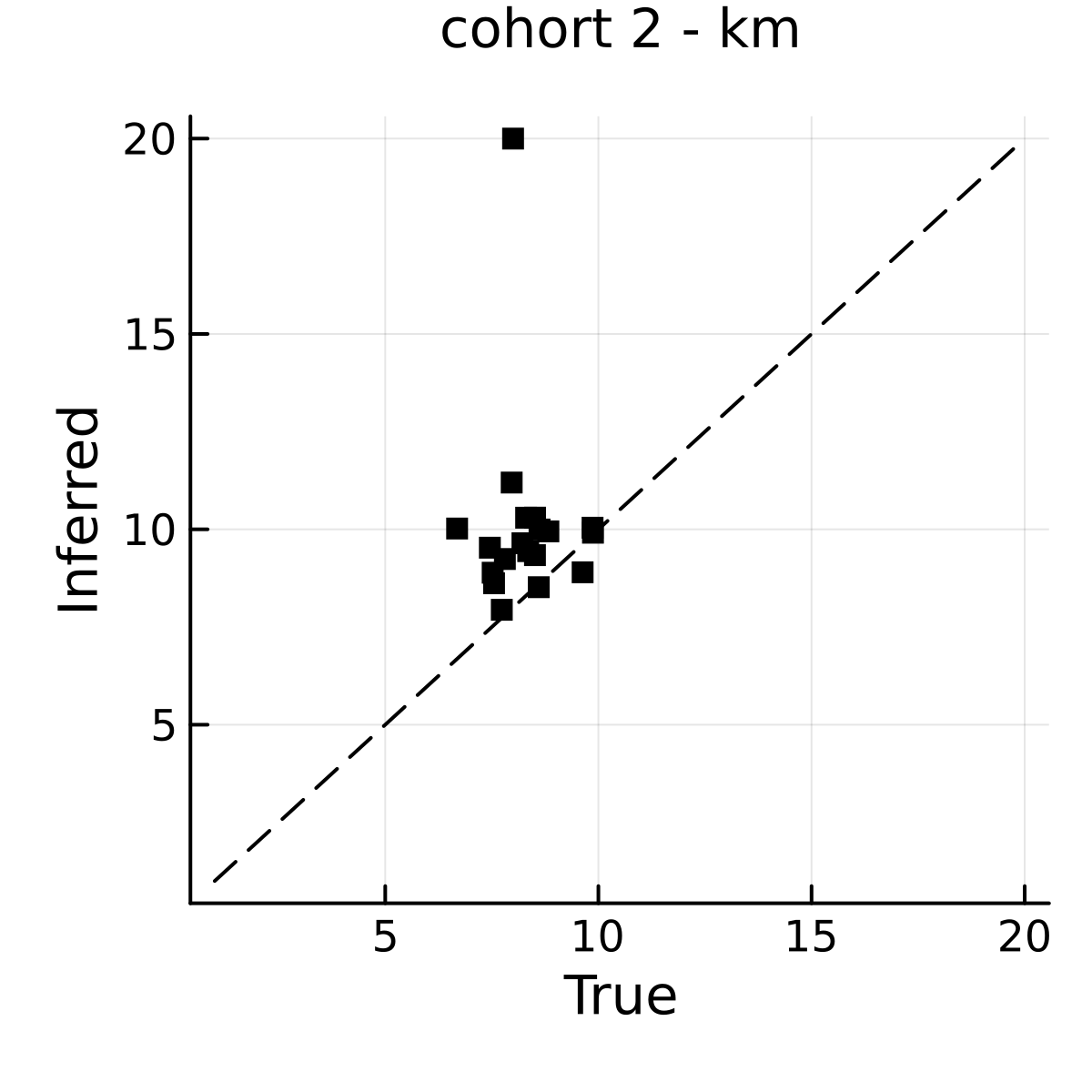}
    \end{subfigure}
    \begin{subfigure}[b]{0.22\textwidth}
        \includegraphics[width=\textwidth]{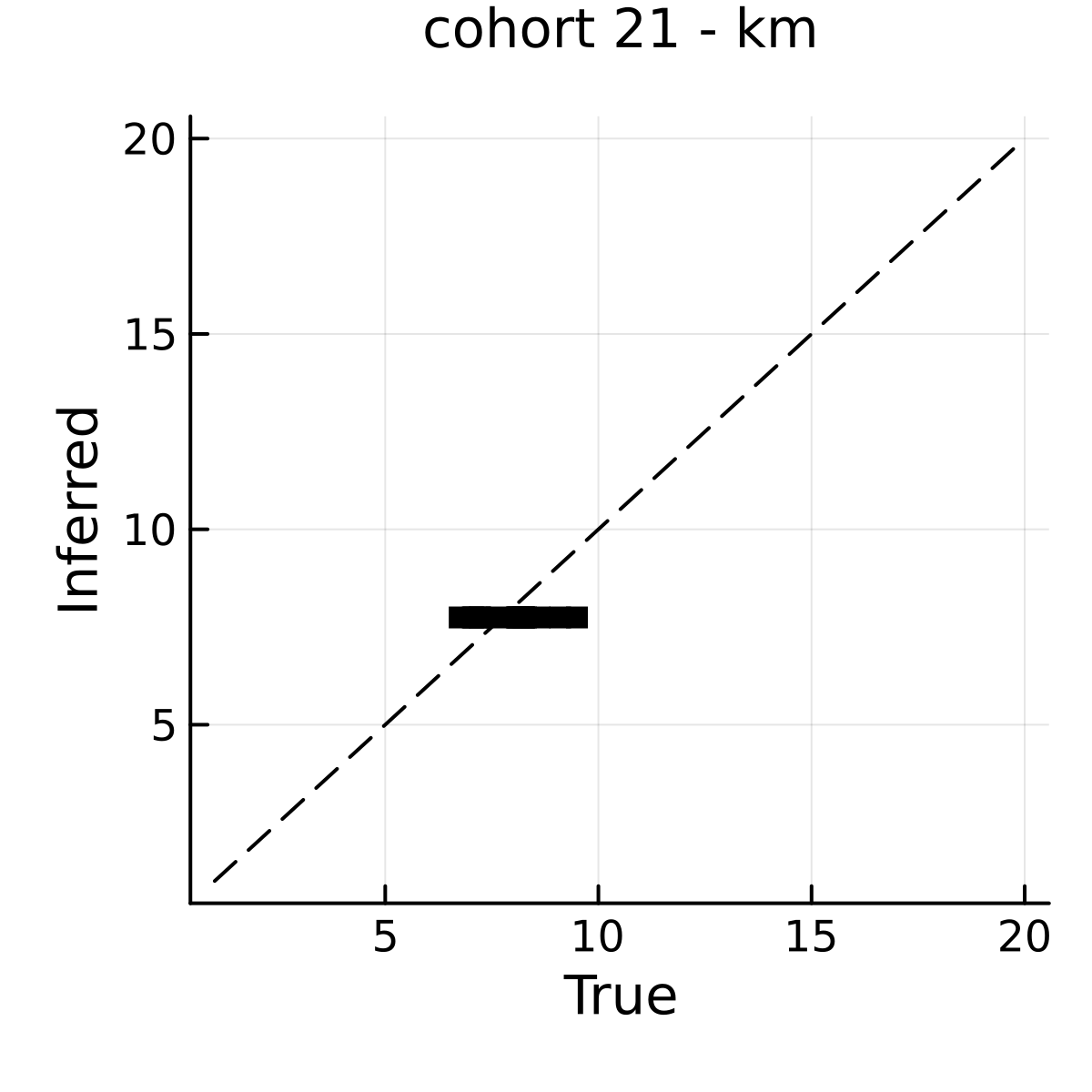}
    \end{subfigure}
    \begin{subfigure}[b]{0.22\textwidth}
        \includegraphics[width=\textwidth]{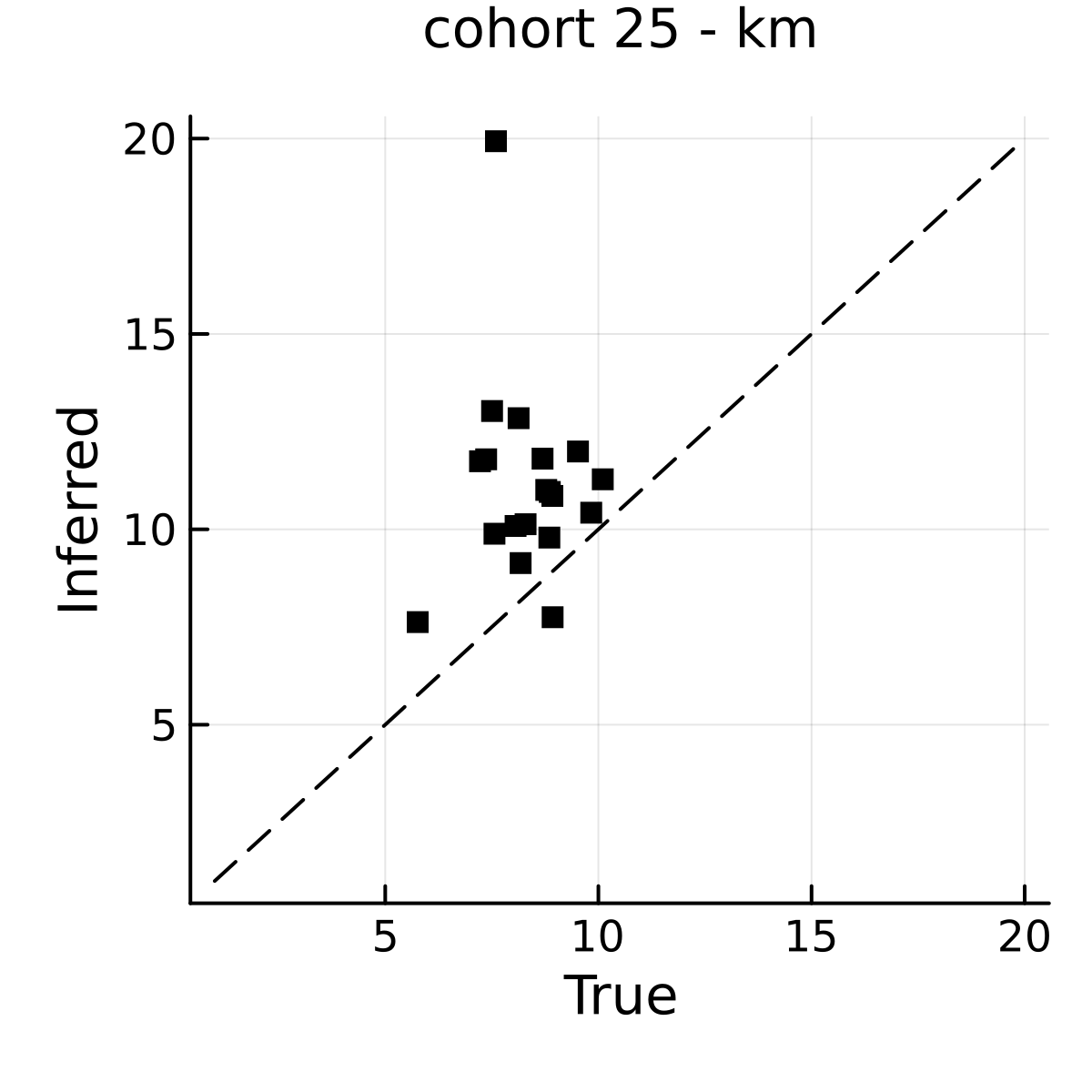}
    \end{subfigure}
    \begin{subfigure}[b]{0.22\textwidth}
        \includegraphics[width=\textwidth]{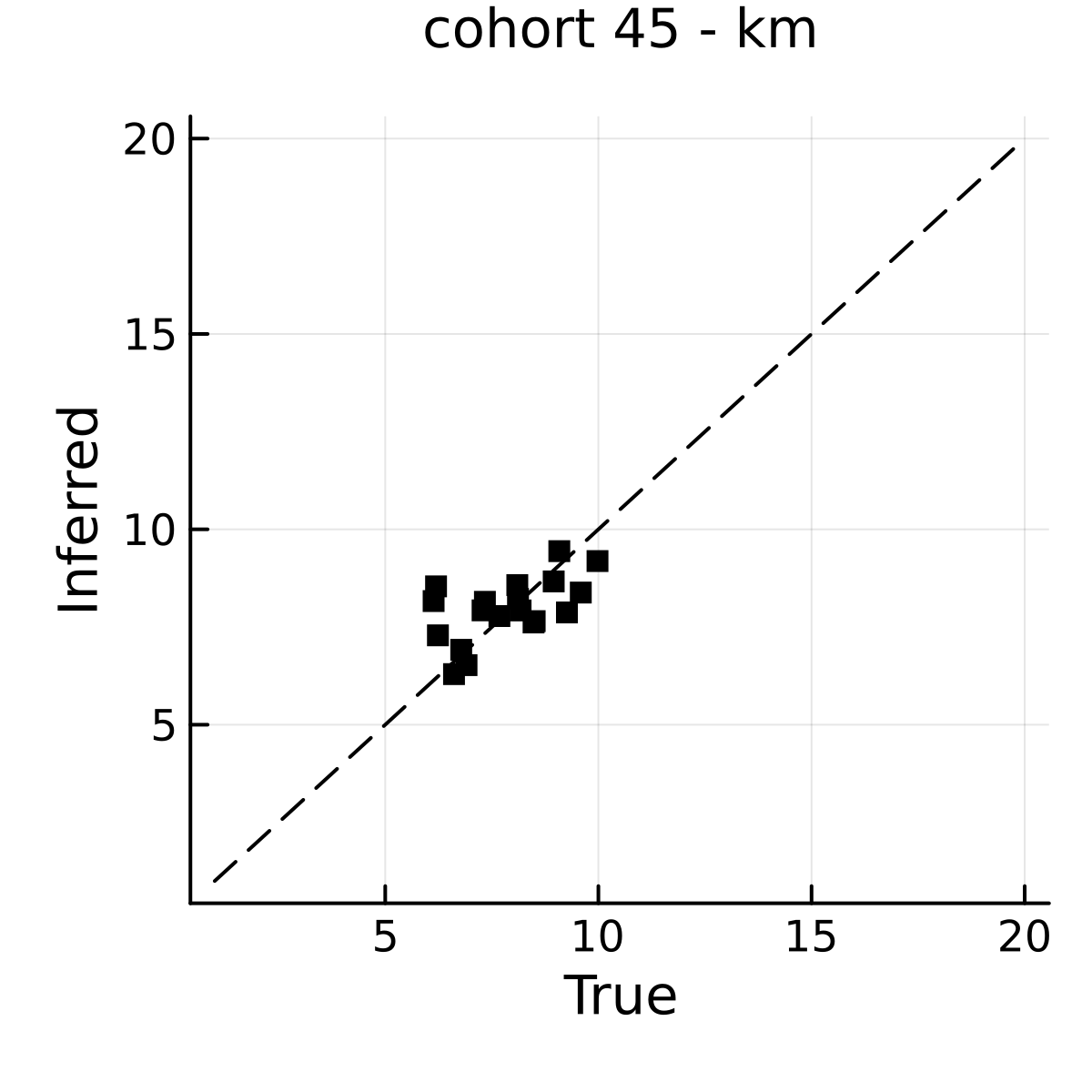}
    \end{subfigure}
    \begin{subfigure}[b]{0.22\textwidth}
        \includegraphics[width=\textwidth]{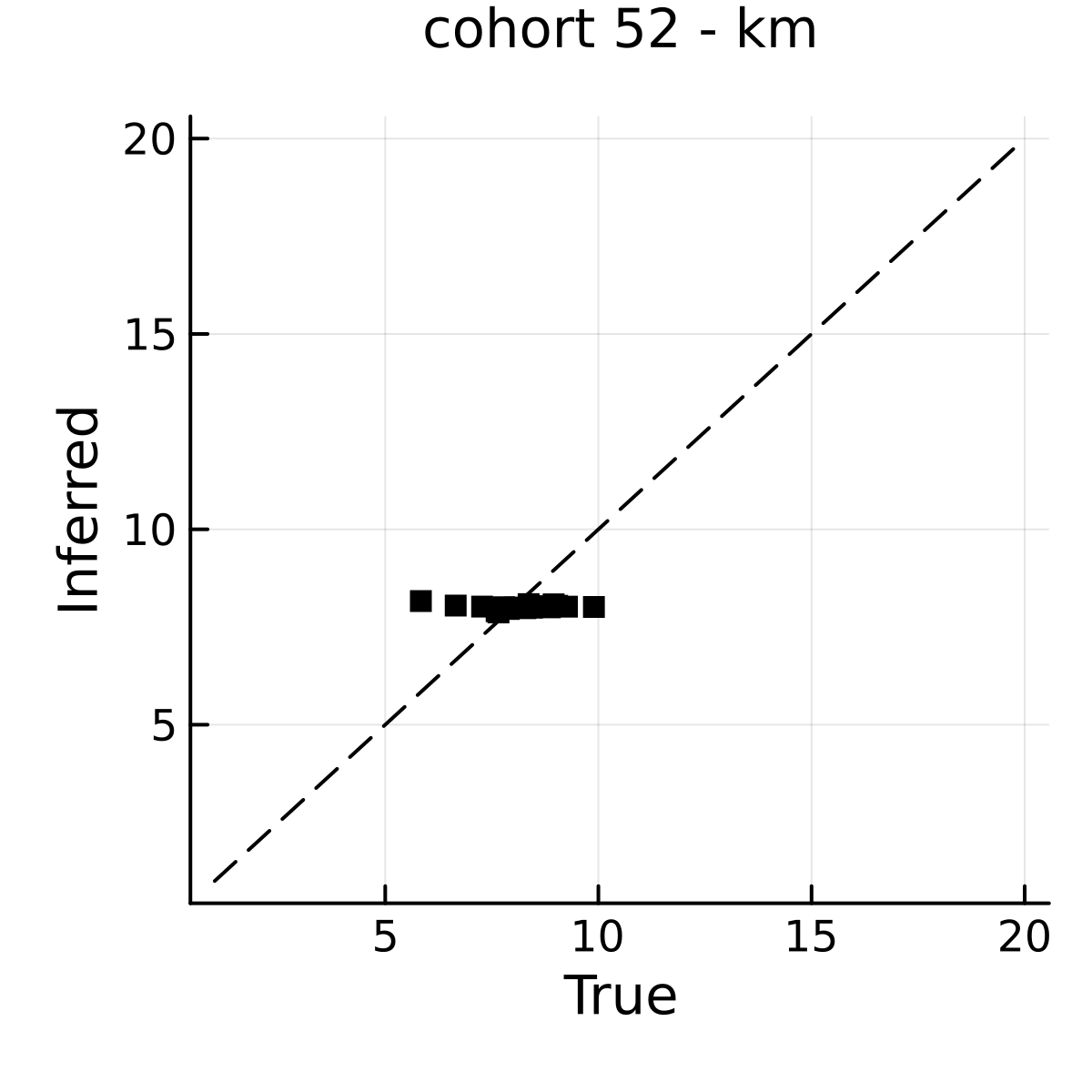}
    \end{subfigure}
    \begin{subfigure}[b]{0.22\textwidth}
        \includegraphics[width=\textwidth]{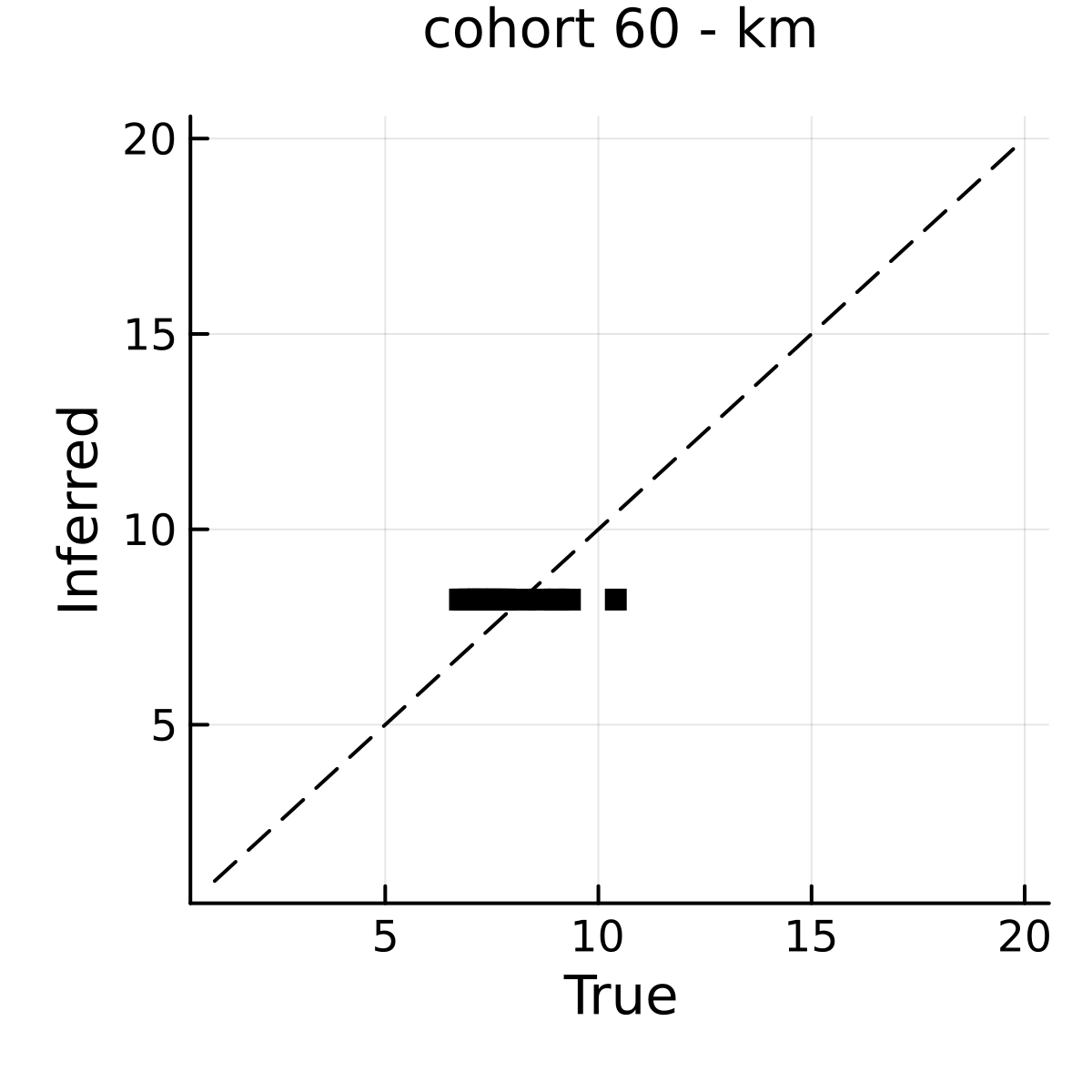}
    \end{subfigure}
    \begin{subfigure}[b]{0.22\textwidth}
        \includegraphics[width=\textwidth]{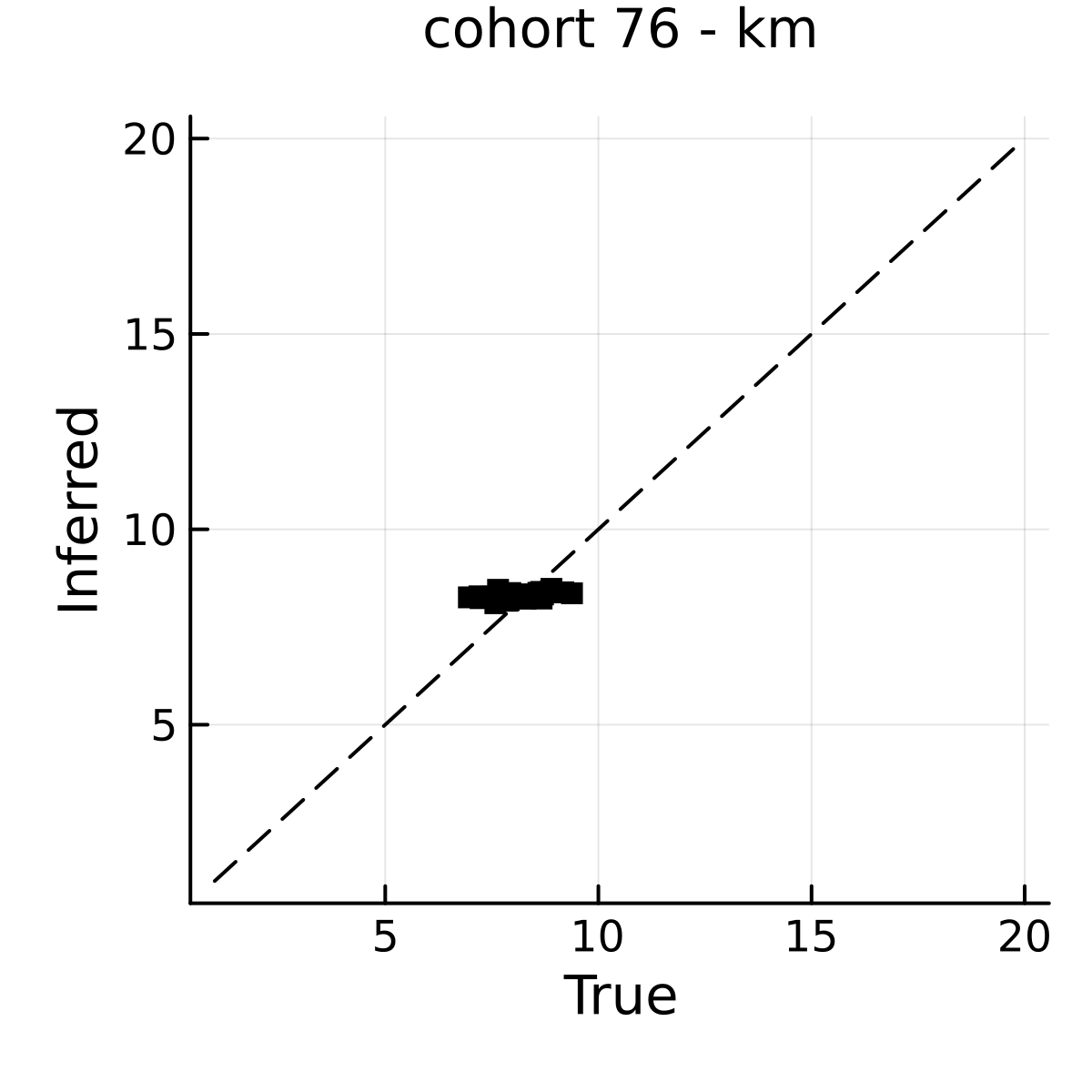}
    \end{subfigure}
    \begin{subfigure}[b]{0.22\textwidth}
        \includegraphics[width=\textwidth]{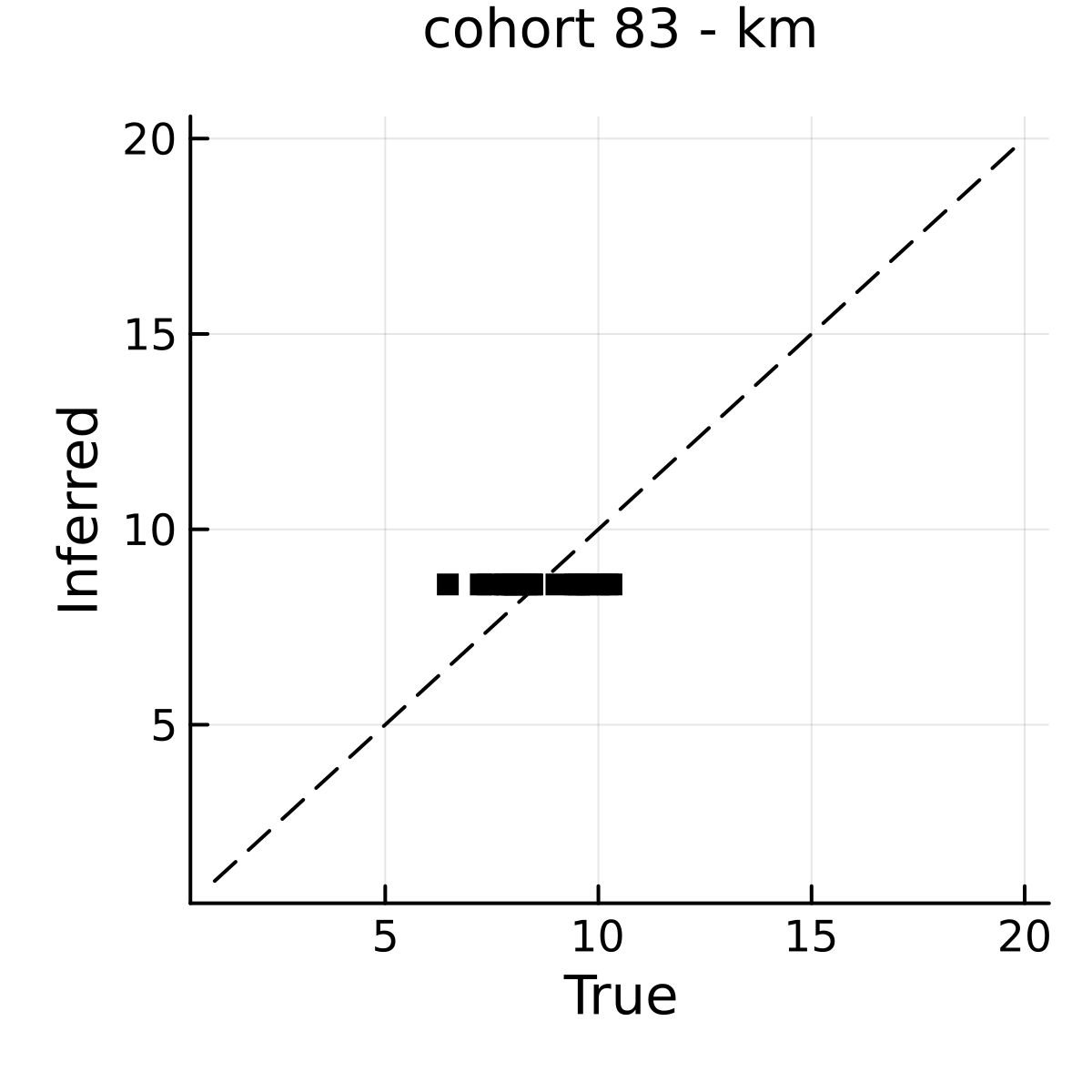}
    \end{subfigure}
    \begin{subfigure}[b]{0.22\textwidth}
        \includegraphics[width=\textwidth]{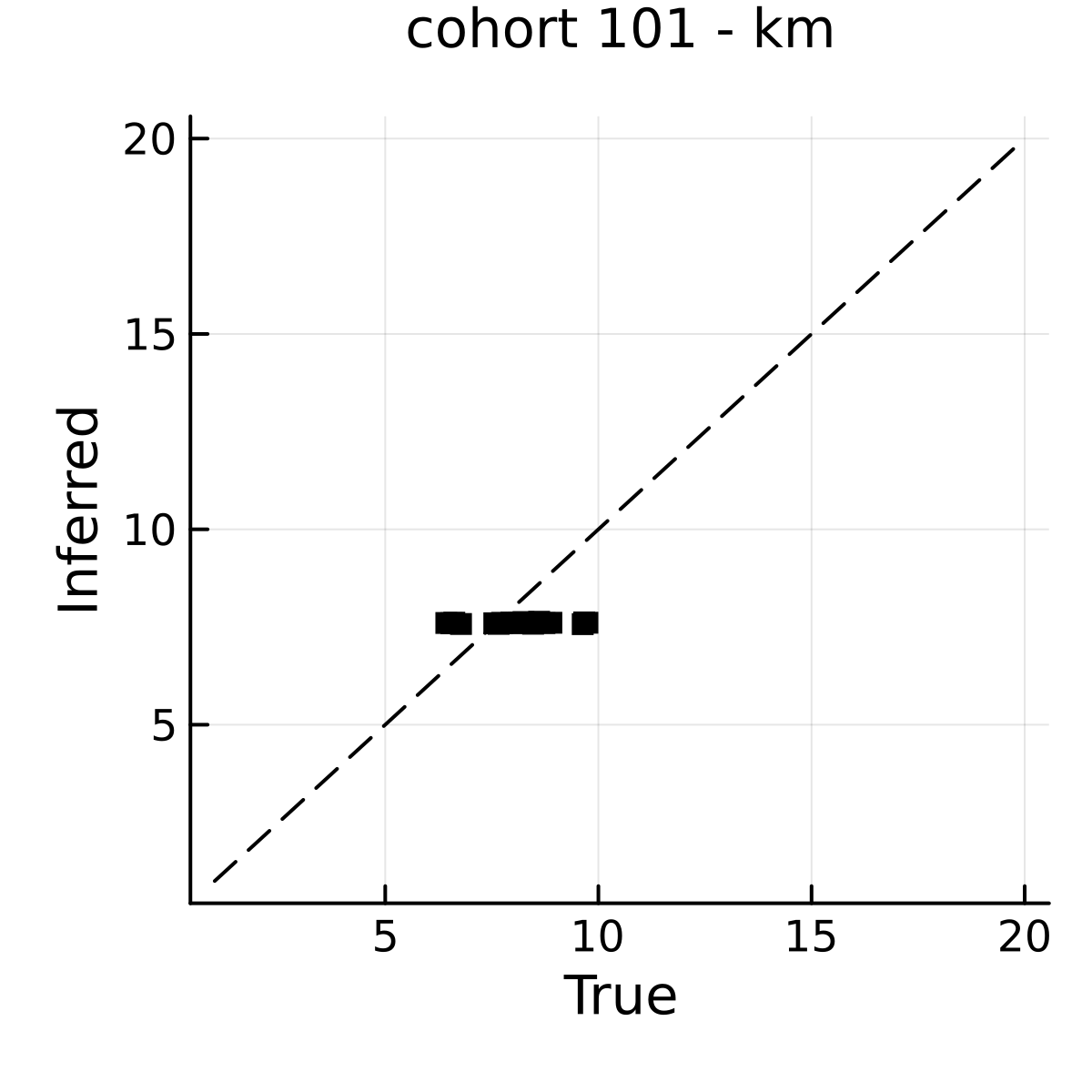}
    \end{subfigure}
    \begin{subfigure}[b]{0.22\textwidth}
        \includegraphics[width=\textwidth]{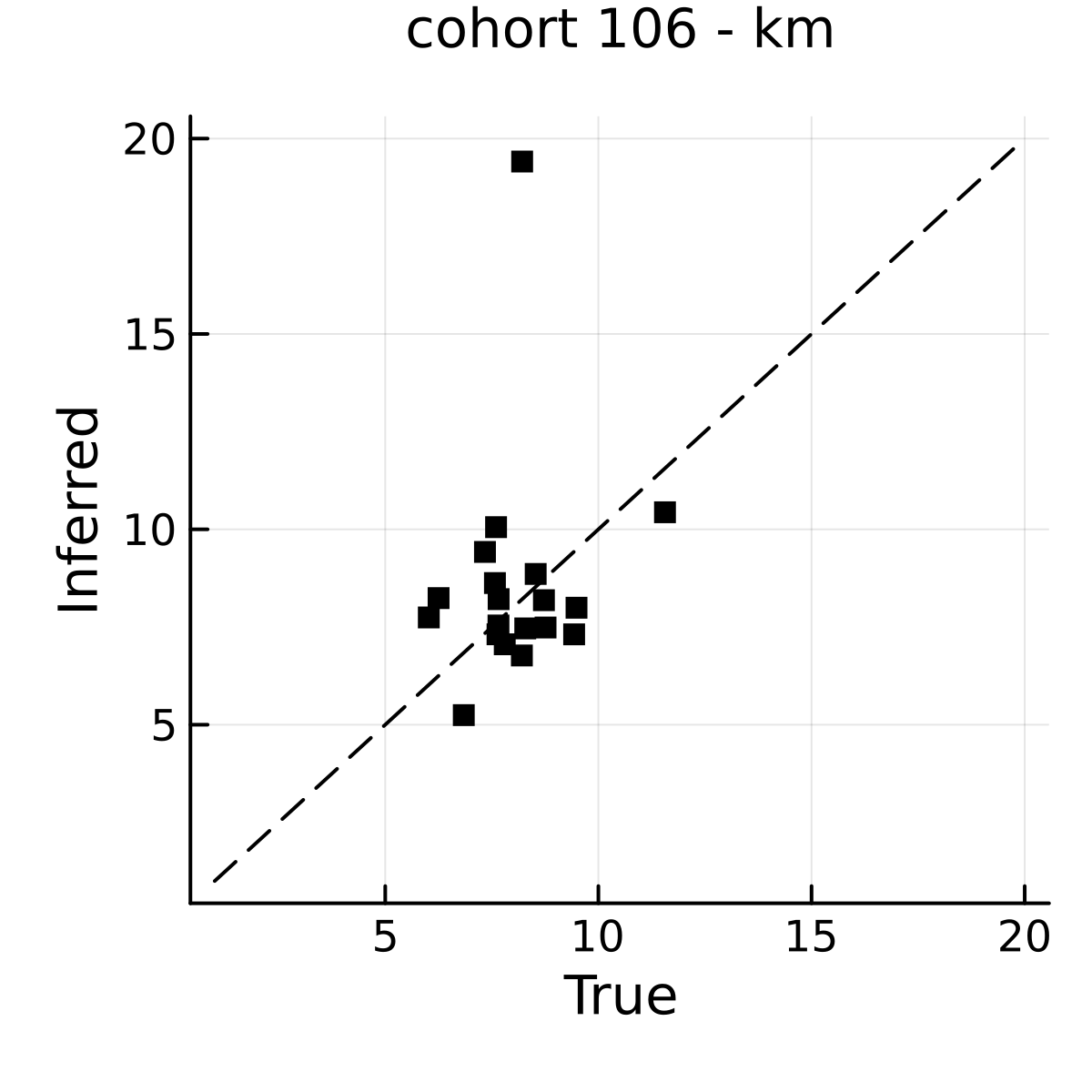}
    \end{subfigure}
    \begin{subfigure}[b]{0.22\textwidth}
        \includegraphics[width=\textwidth]{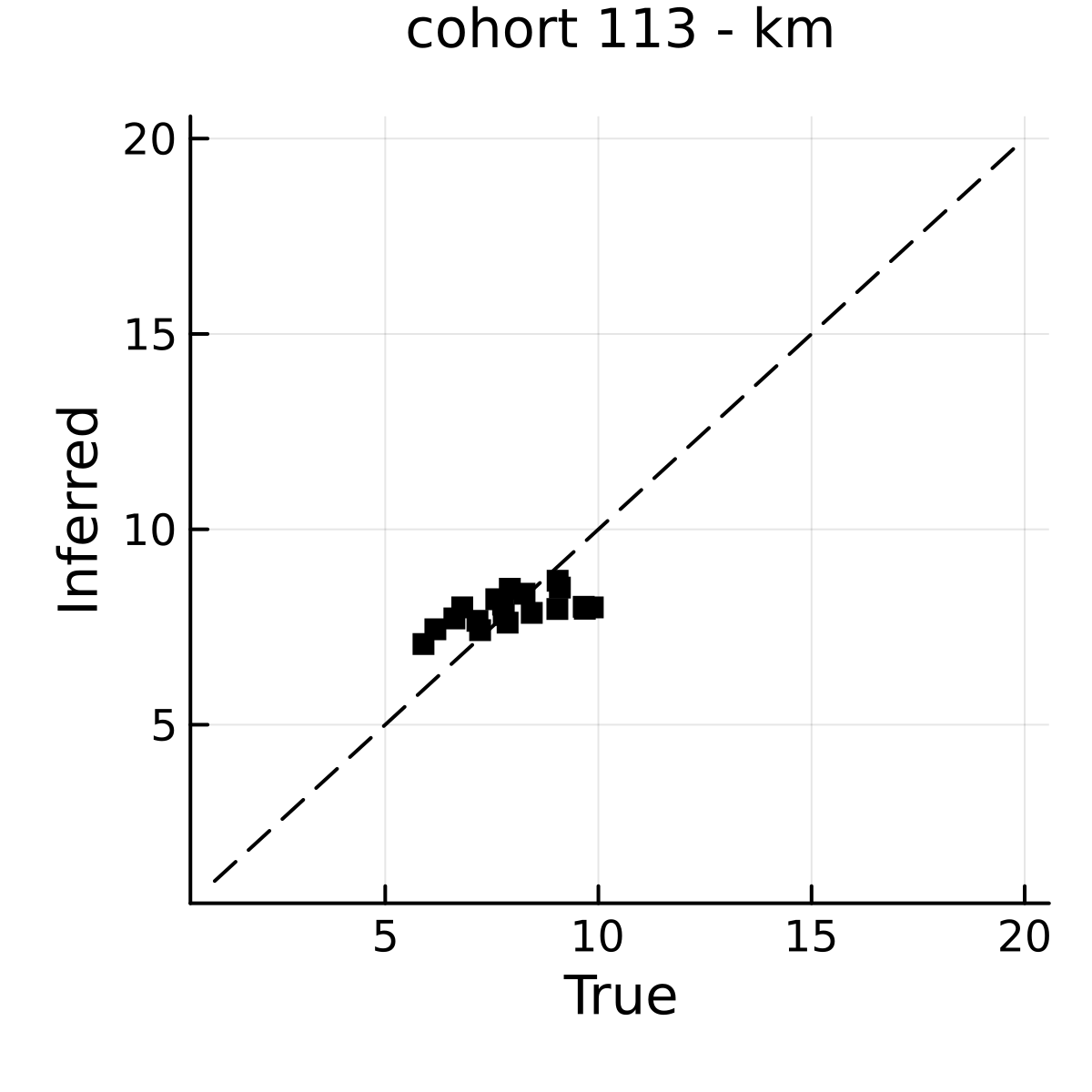}
    \end{subfigure}
    \begin{subfigure}[b]{0.22\textwidth}
        \includegraphics[width=\textwidth]{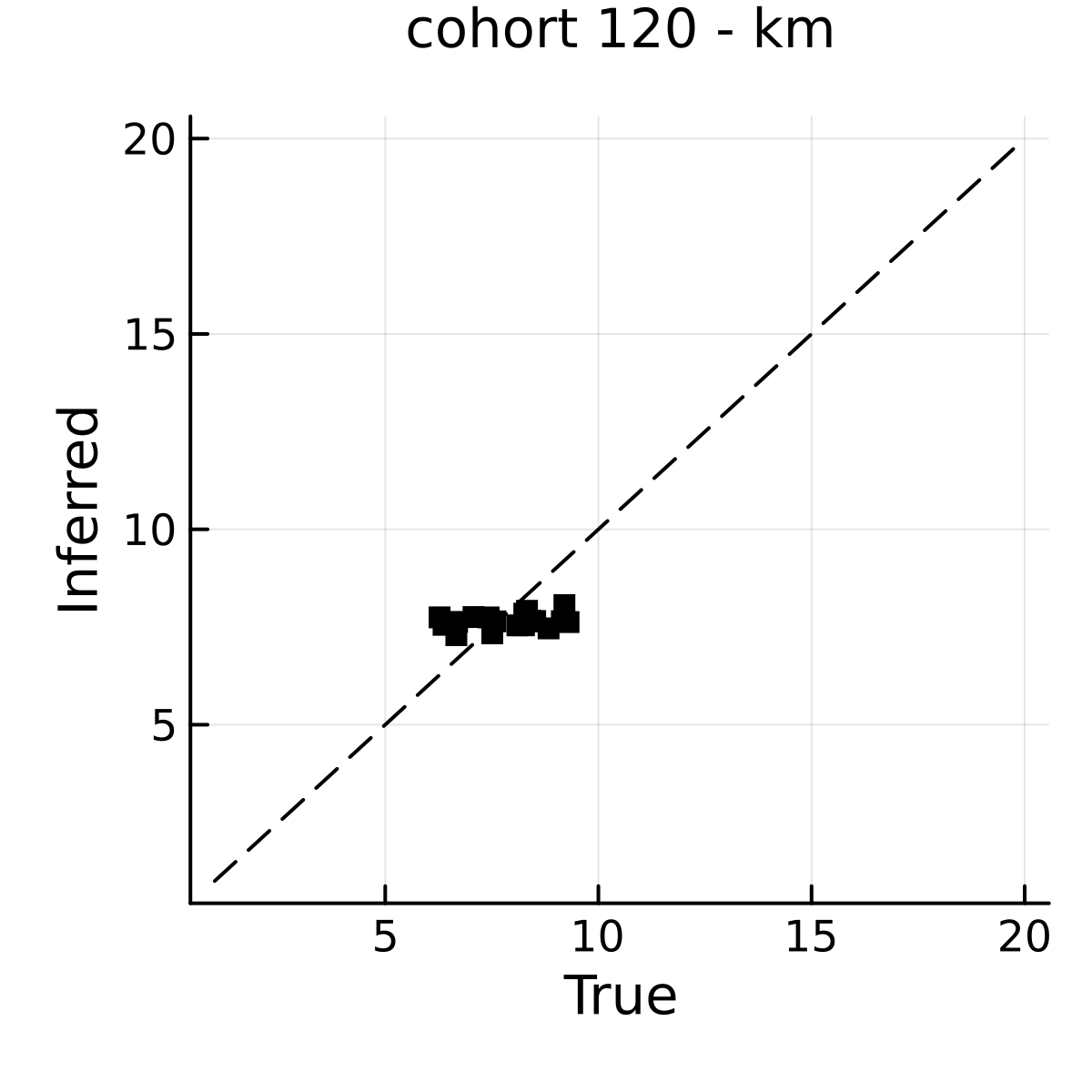}
    \end{subfigure}
    \begin{subfigure}[b]{0.22\textwidth}
        \includegraphics[width=\textwidth]{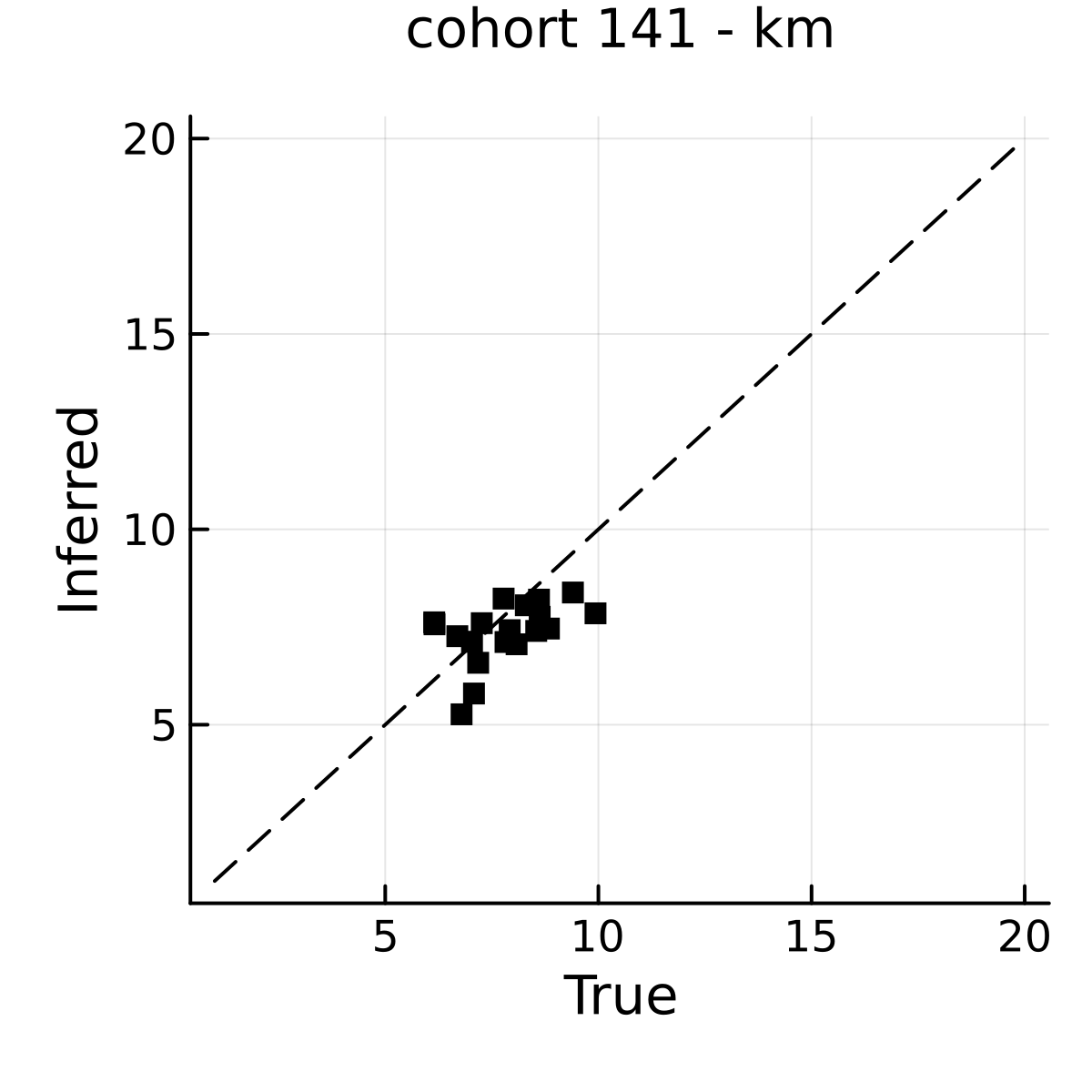}
    \end{subfigure}
    \begin{subfigure}[b]{0.22\textwidth}
        \includegraphics[width=\textwidth]{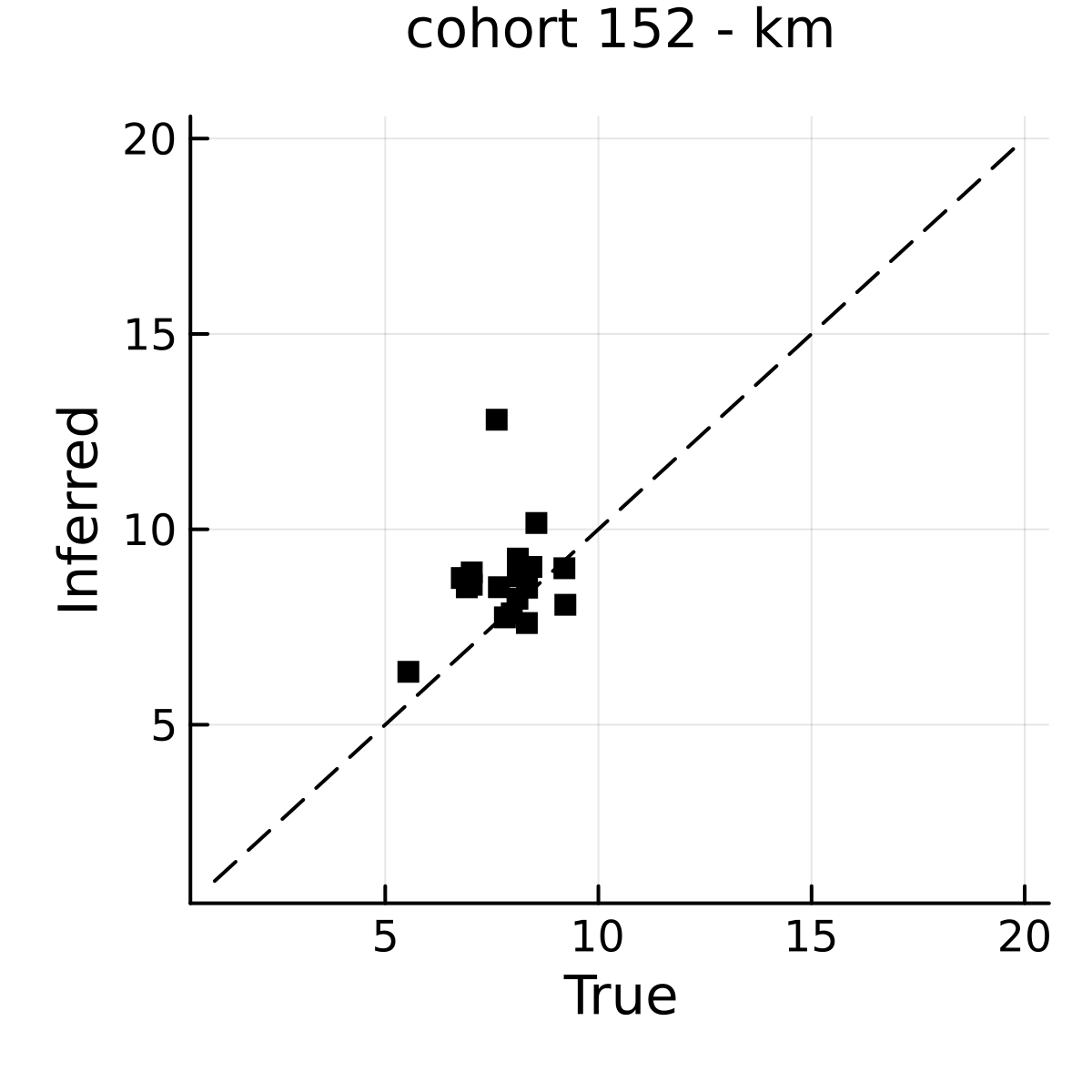}
    \end{subfigure}
    \begin{subfigure}[b]{0.22\textwidth}
        \includegraphics[width=\textwidth]{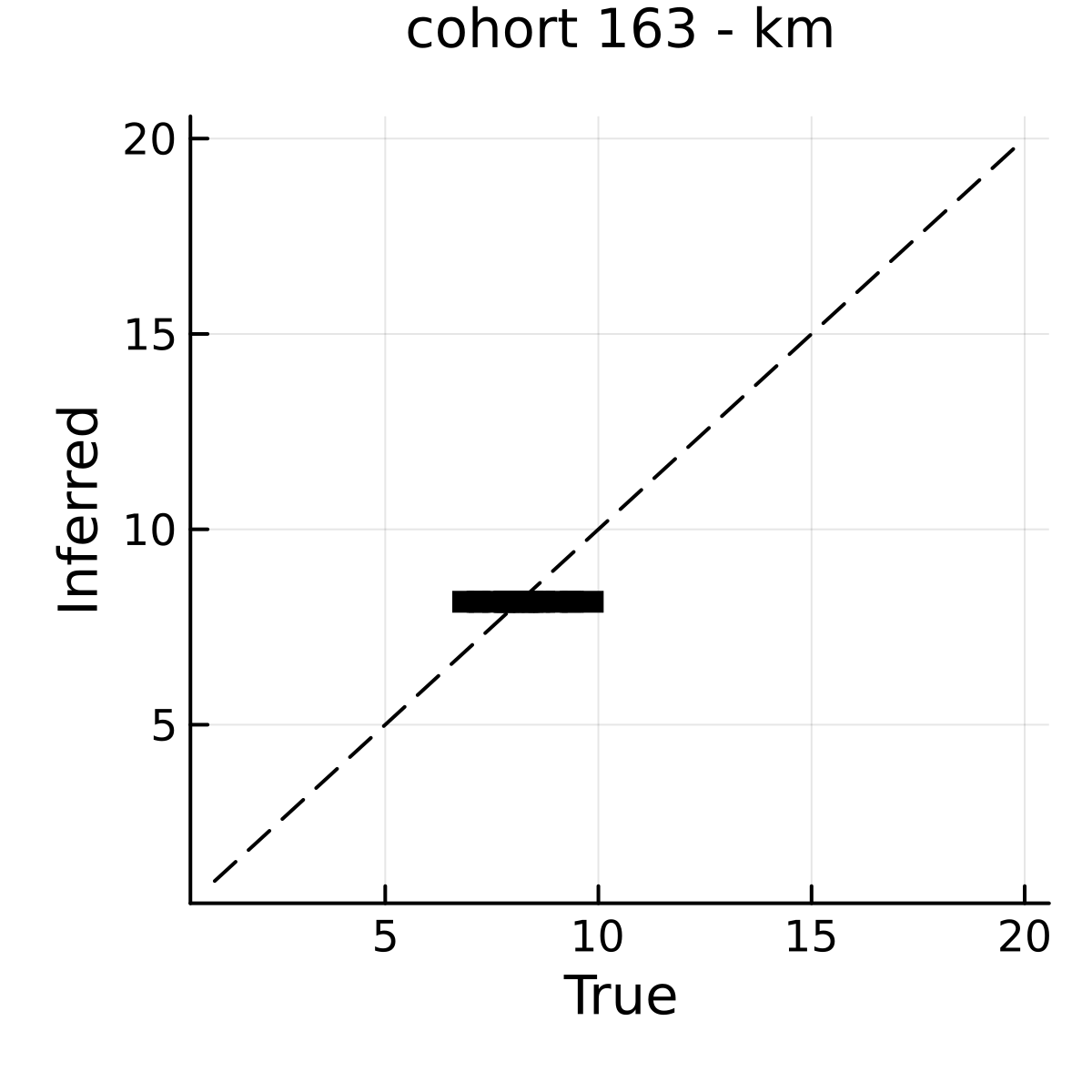}
    \end{subfigure}
    \begin{subfigure}[b]{0.22\textwidth}
        \includegraphics[width=\textwidth]{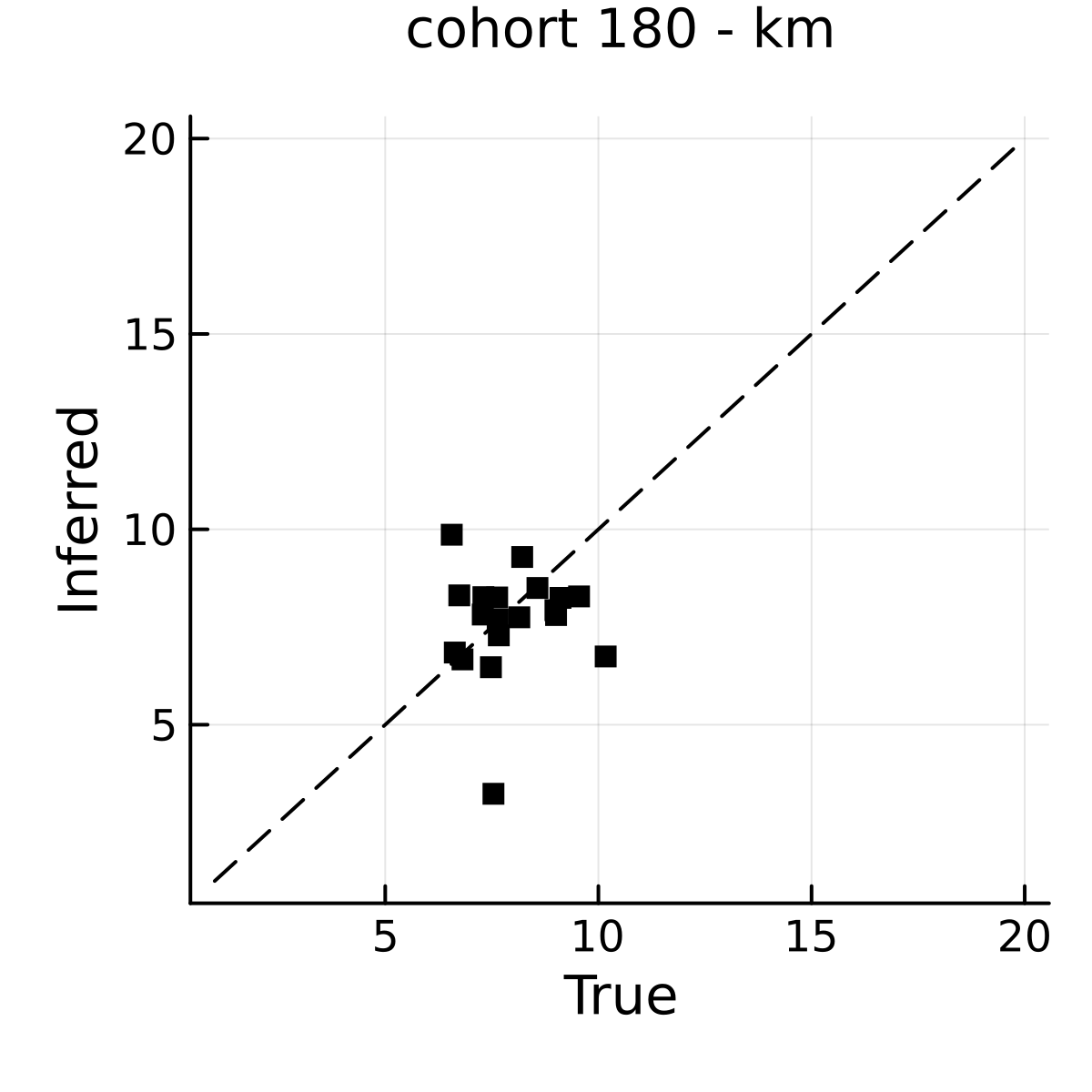}
    \end{subfigure}
    \begin{subfigure}[b]{0.22\textwidth}
        \includegraphics[width=\textwidth]{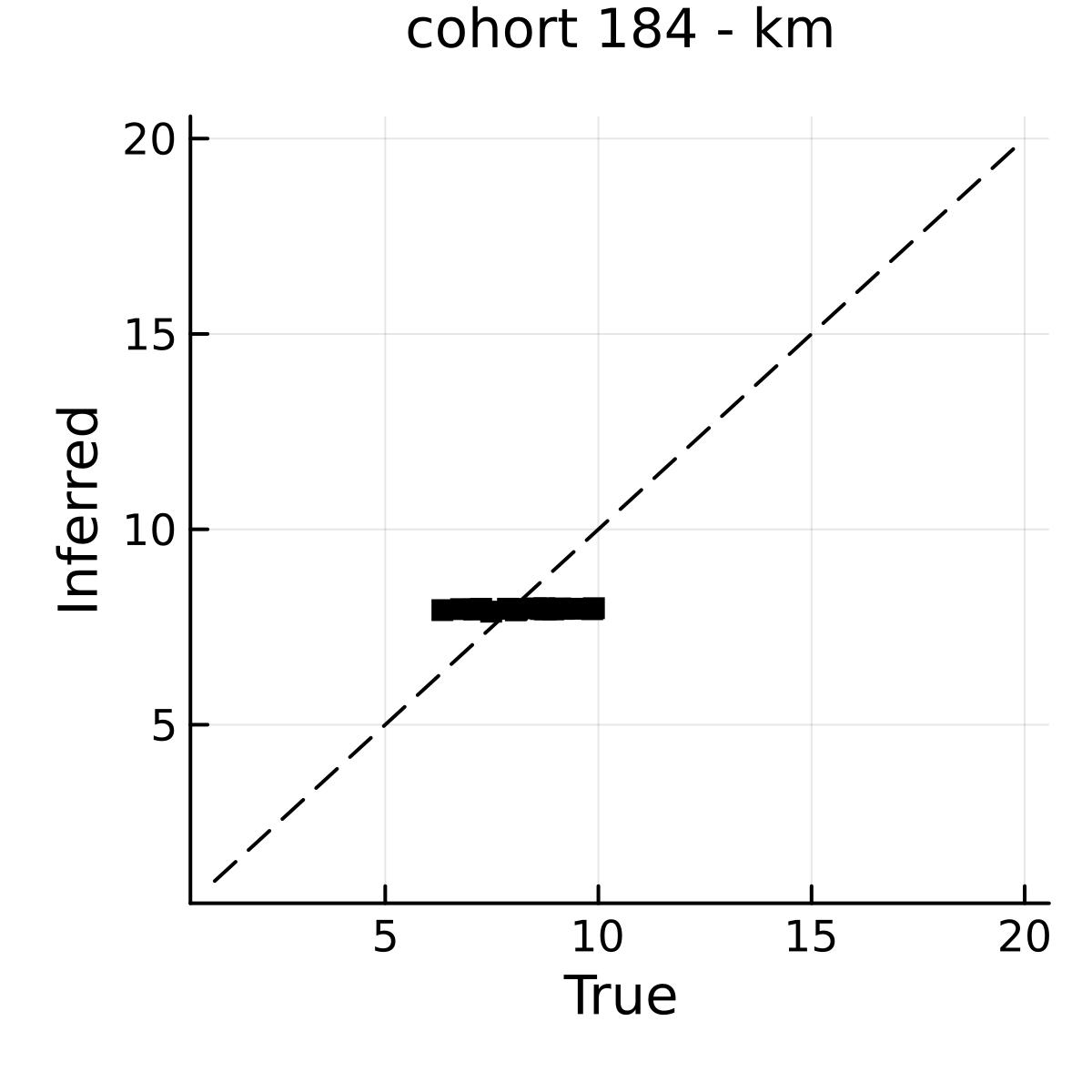}
    \end{subfigure}
    \begin{subfigure}[b]{0.22\textwidth}
        \includegraphics[width=\textwidth]{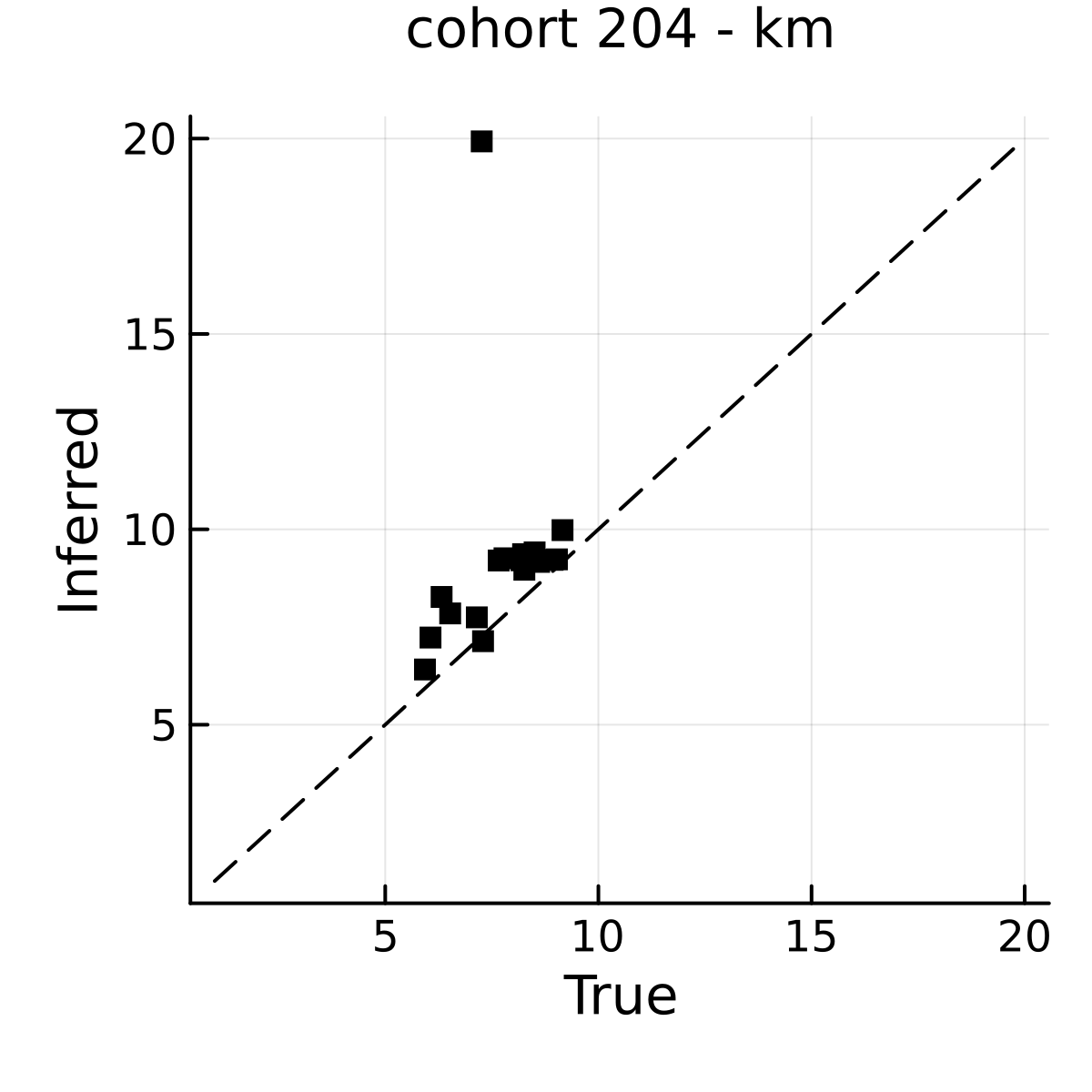}
    \end{subfigure}
    \begin{subfigure}[b]{0.22\textwidth}
        \includegraphics[width=\textwidth]{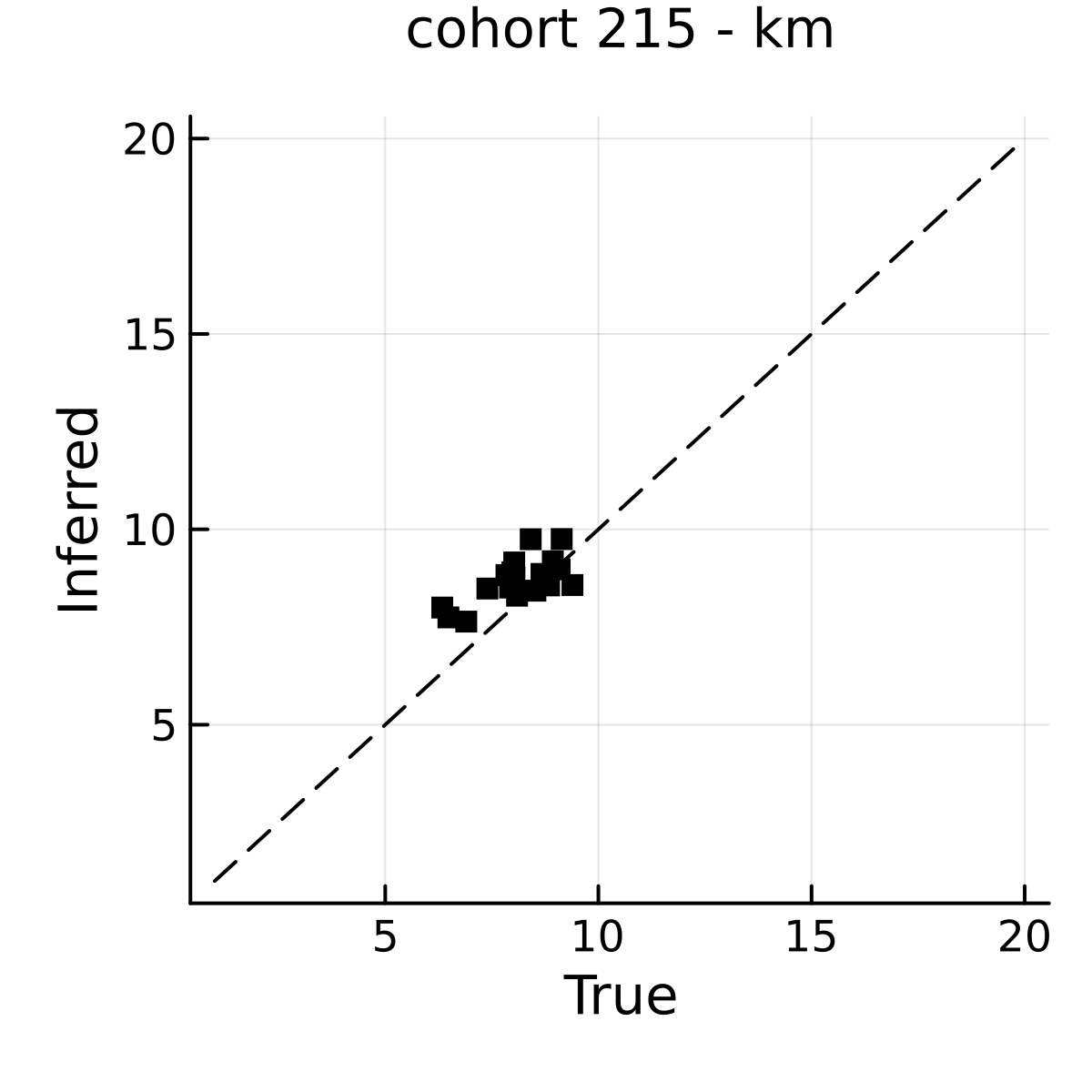}
    \end{subfigure}
    \begin{subfigure}[b]{0.22\textwidth}
        \includegraphics[width=\textwidth]{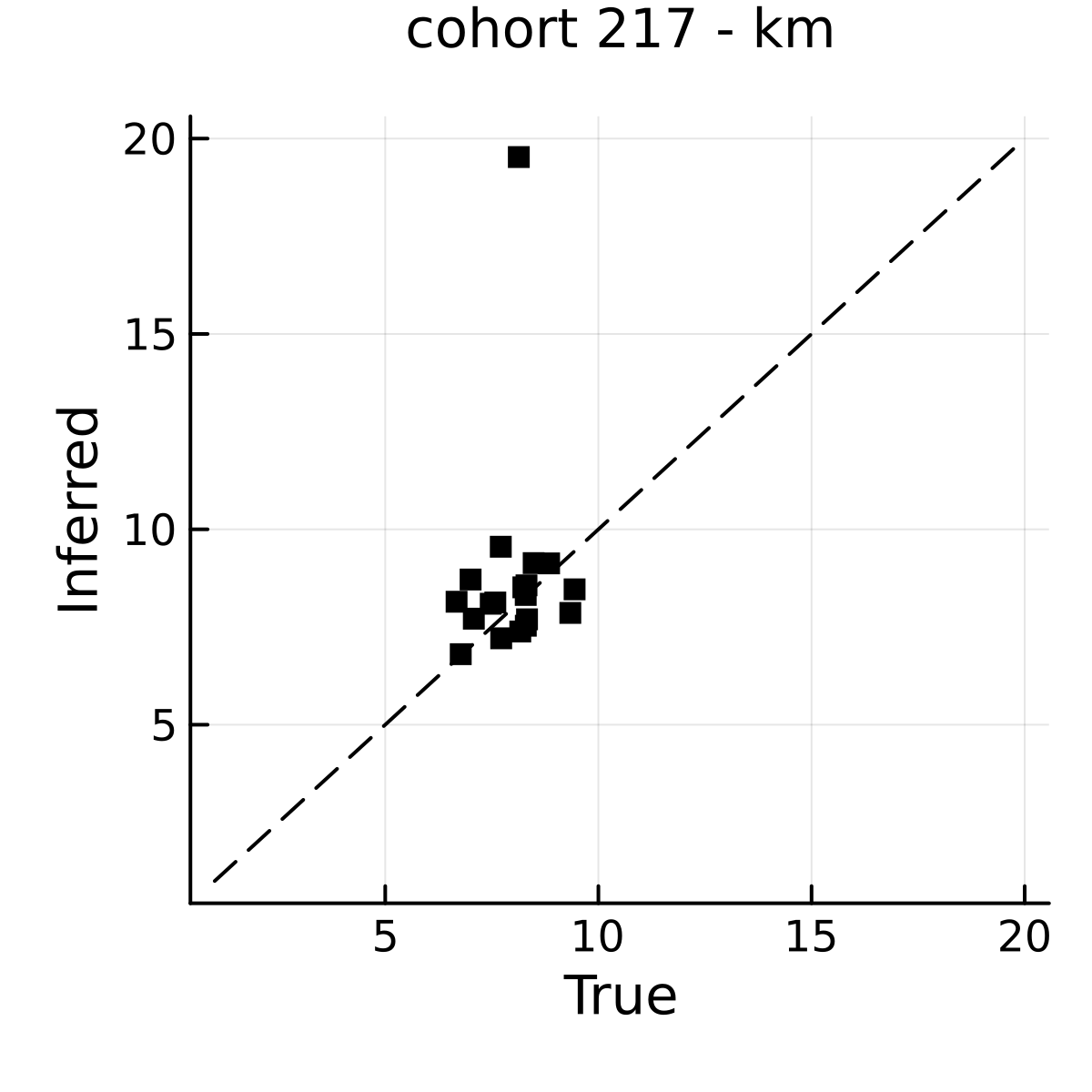}
    \end{subfigure}
    \caption{Comparison between the inferred (posterior mean, y-axis) value of parameter $k_m$ and the true one (x-axis) for each virtual cohort.}
    \label{fig:synth_km}
\end{figure}

\begin{figure}[h]
    \centering
    \begin{subfigure}[b]{0.22\textwidth}
        \includegraphics[width=\textwidth]{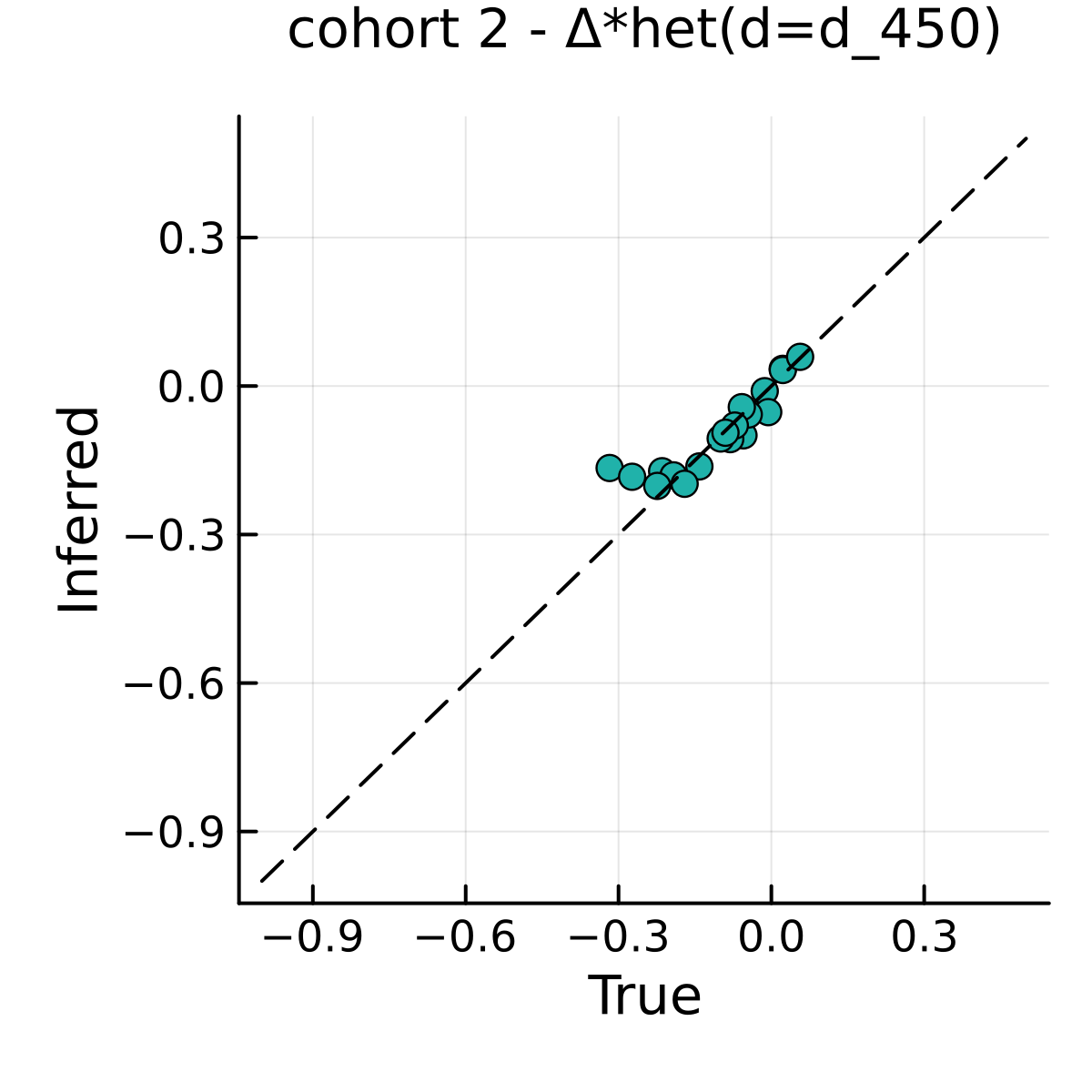}
    \end{subfigure}
    \begin{subfigure}[b]{0.22\textwidth}
        \includegraphics[width=\textwidth]{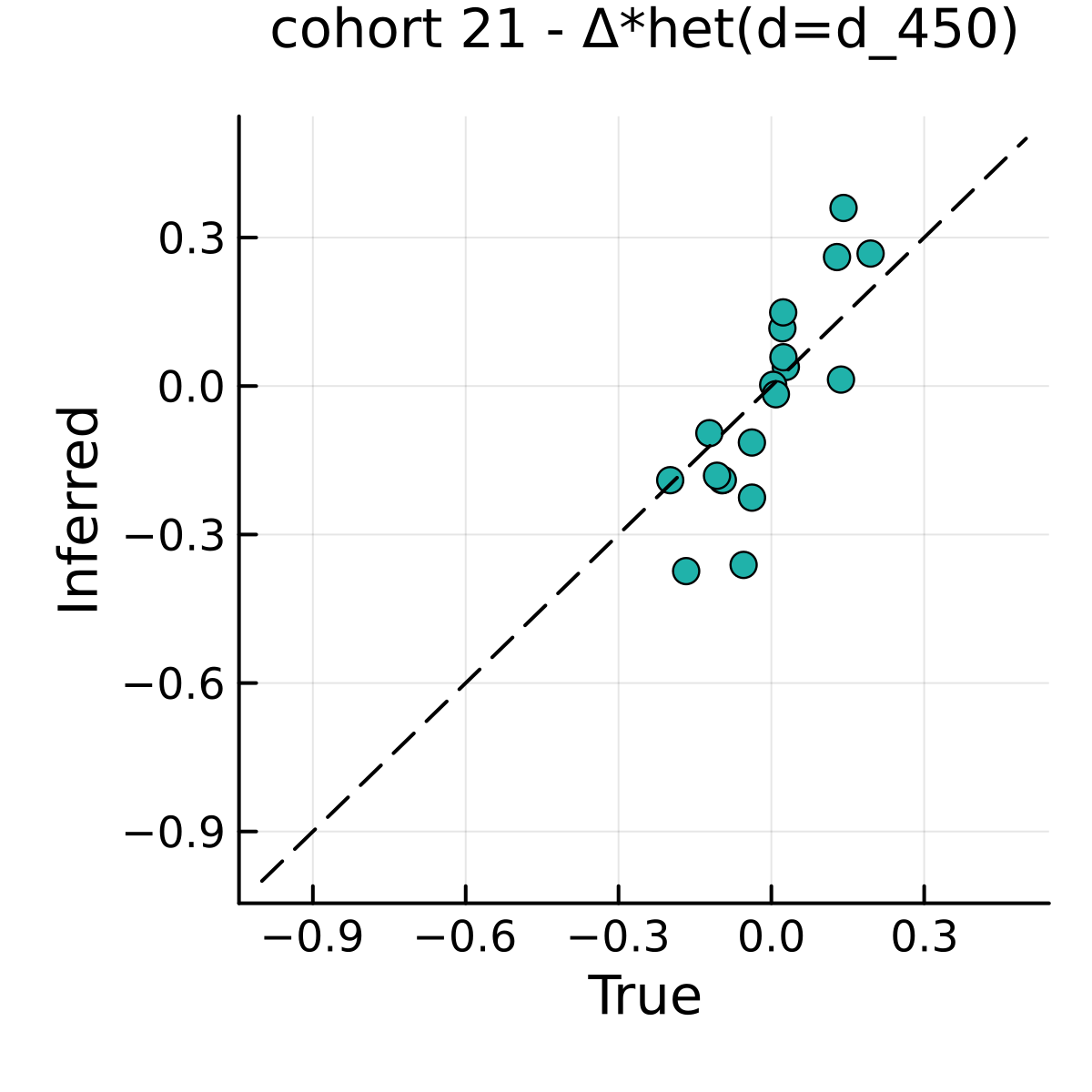}
    \end{subfigure}
    \begin{subfigure}[b]{0.22\textwidth}
        \includegraphics[width=\textwidth]{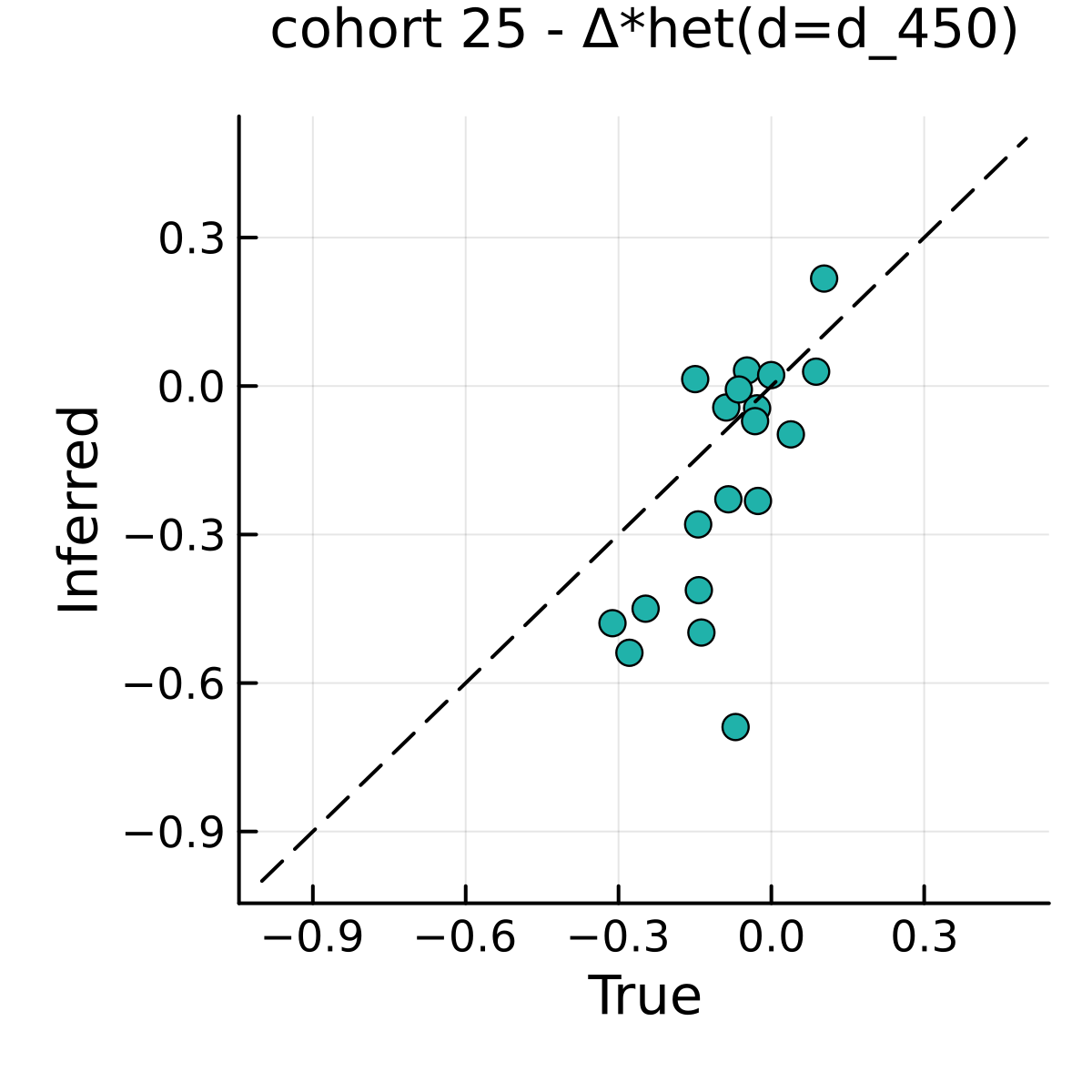}
    \end{subfigure}
    \begin{subfigure}[b]{0.22\textwidth}
        \includegraphics[width=\textwidth]{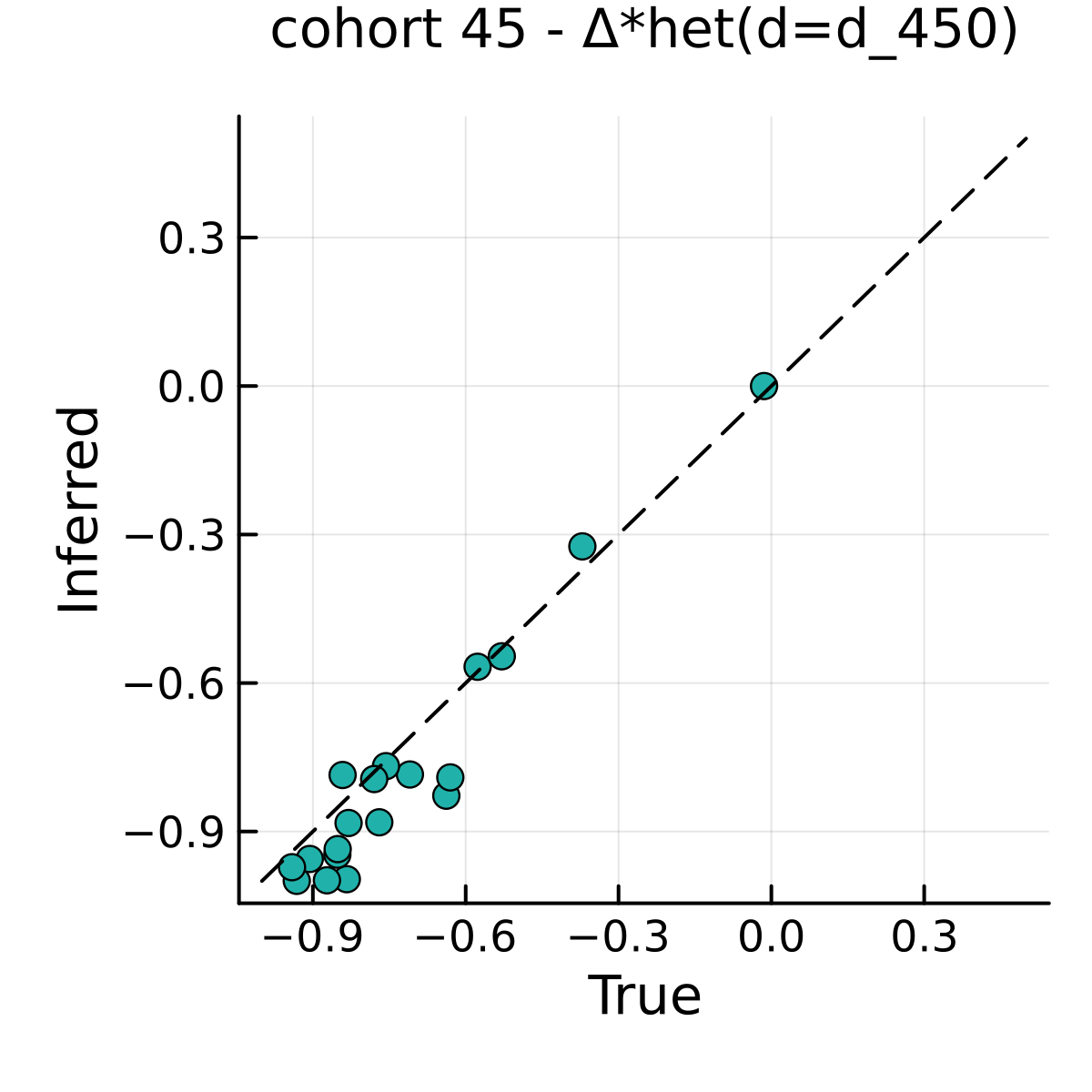}
    \end{subfigure}
    \begin{subfigure}[b]{0.22\textwidth}
        \includegraphics[width=\textwidth]{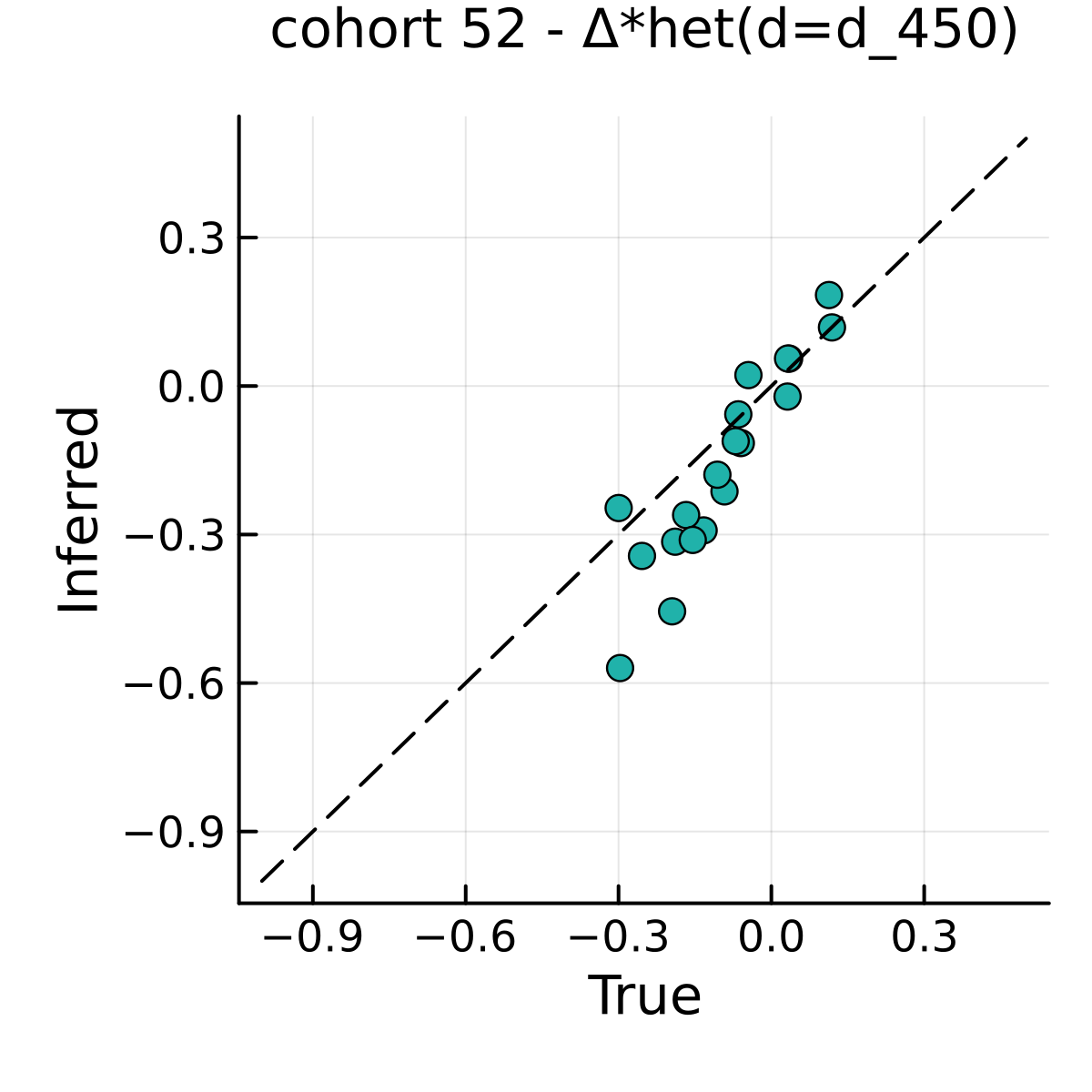}
    \end{subfigure}
    \begin{subfigure}[b]{0.22\textwidth}
        \includegraphics[width=\textwidth]{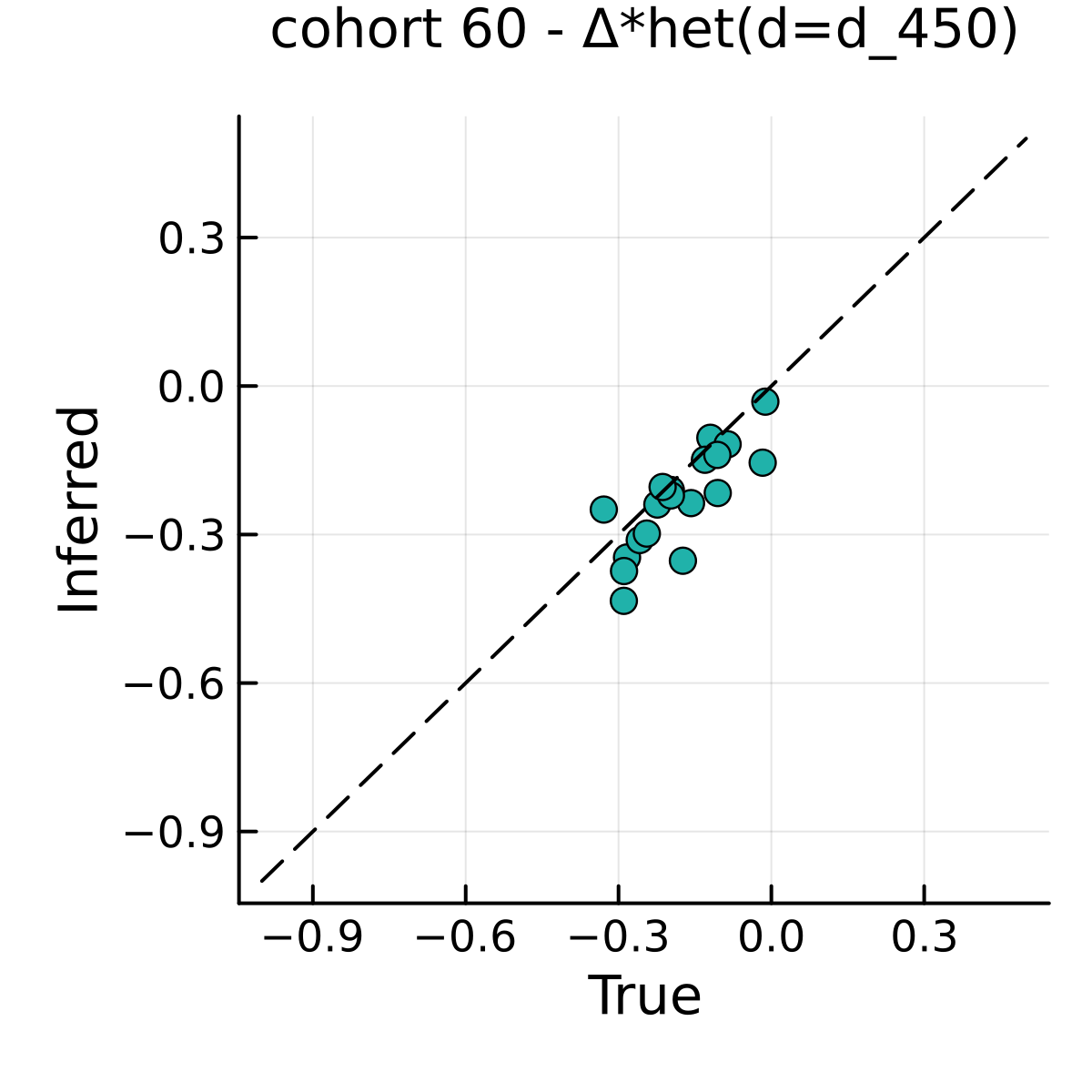}
    \end{subfigure}
    \begin{subfigure}[b]{0.22\textwidth}
        \includegraphics[width=\textwidth]{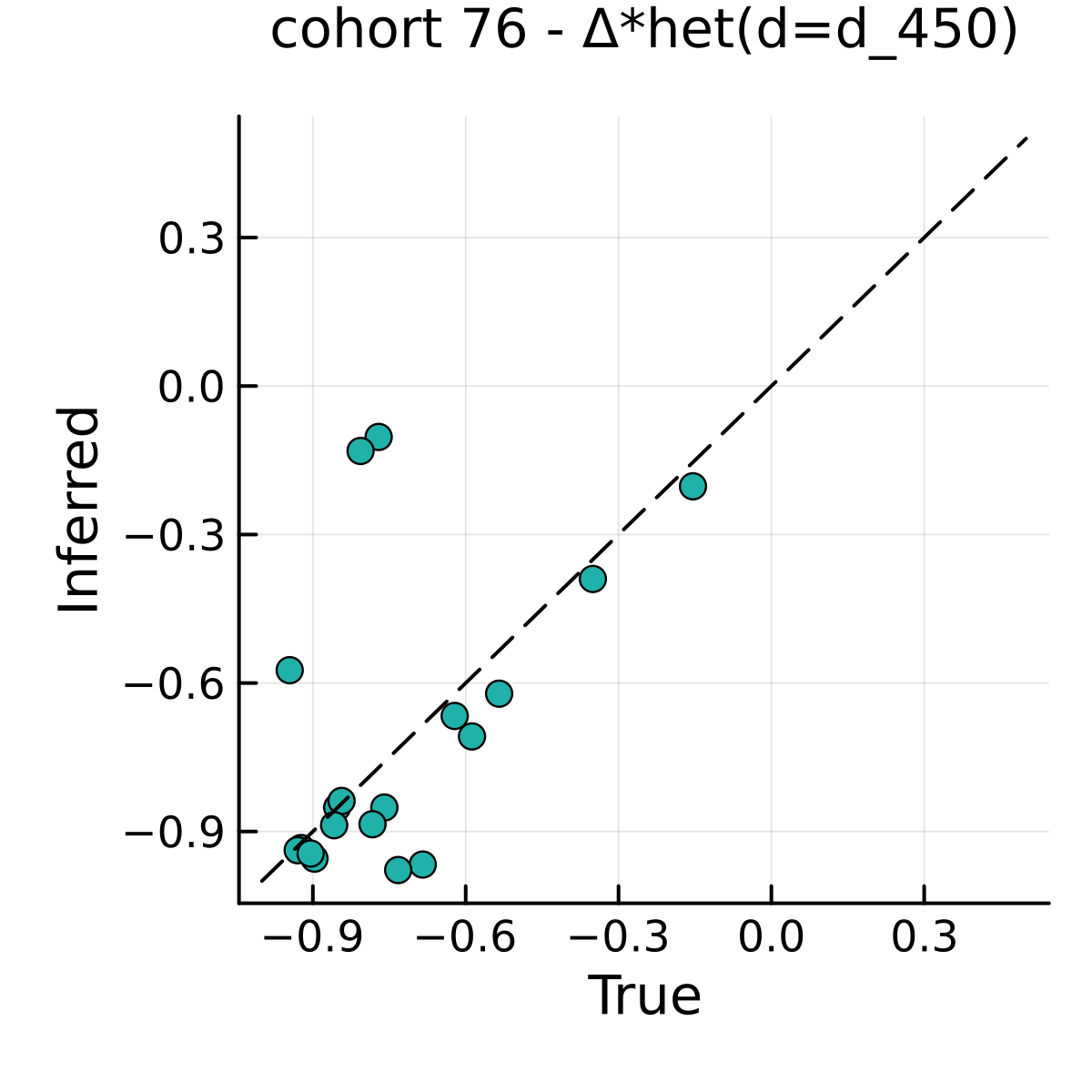}
    \end{subfigure}
    \begin{subfigure}[b]{0.22\textwidth}
        \includegraphics[width=\textwidth]{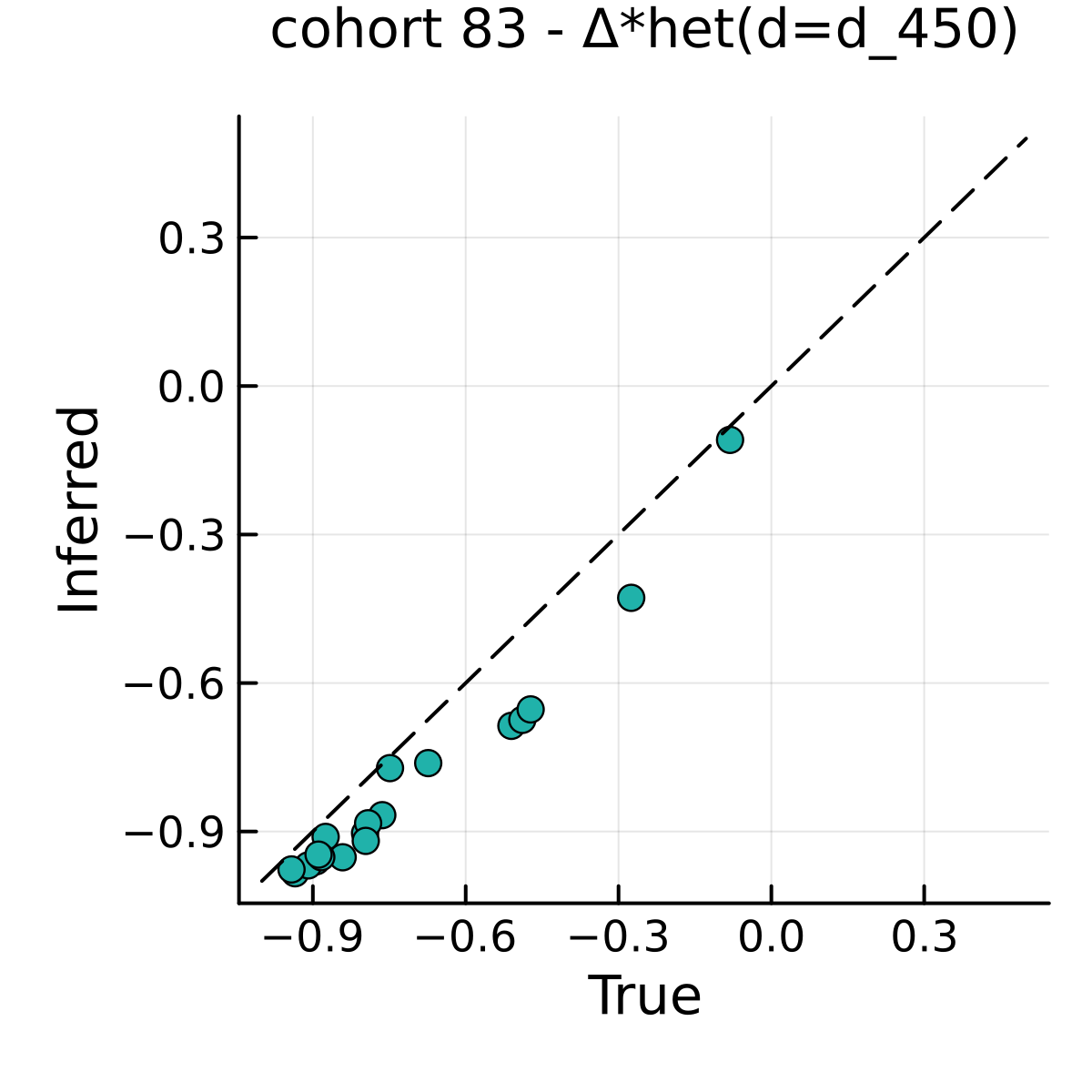}
    \end{subfigure}
    \begin{subfigure}[b]{0.22\textwidth}
        \includegraphics[width=\textwidth]{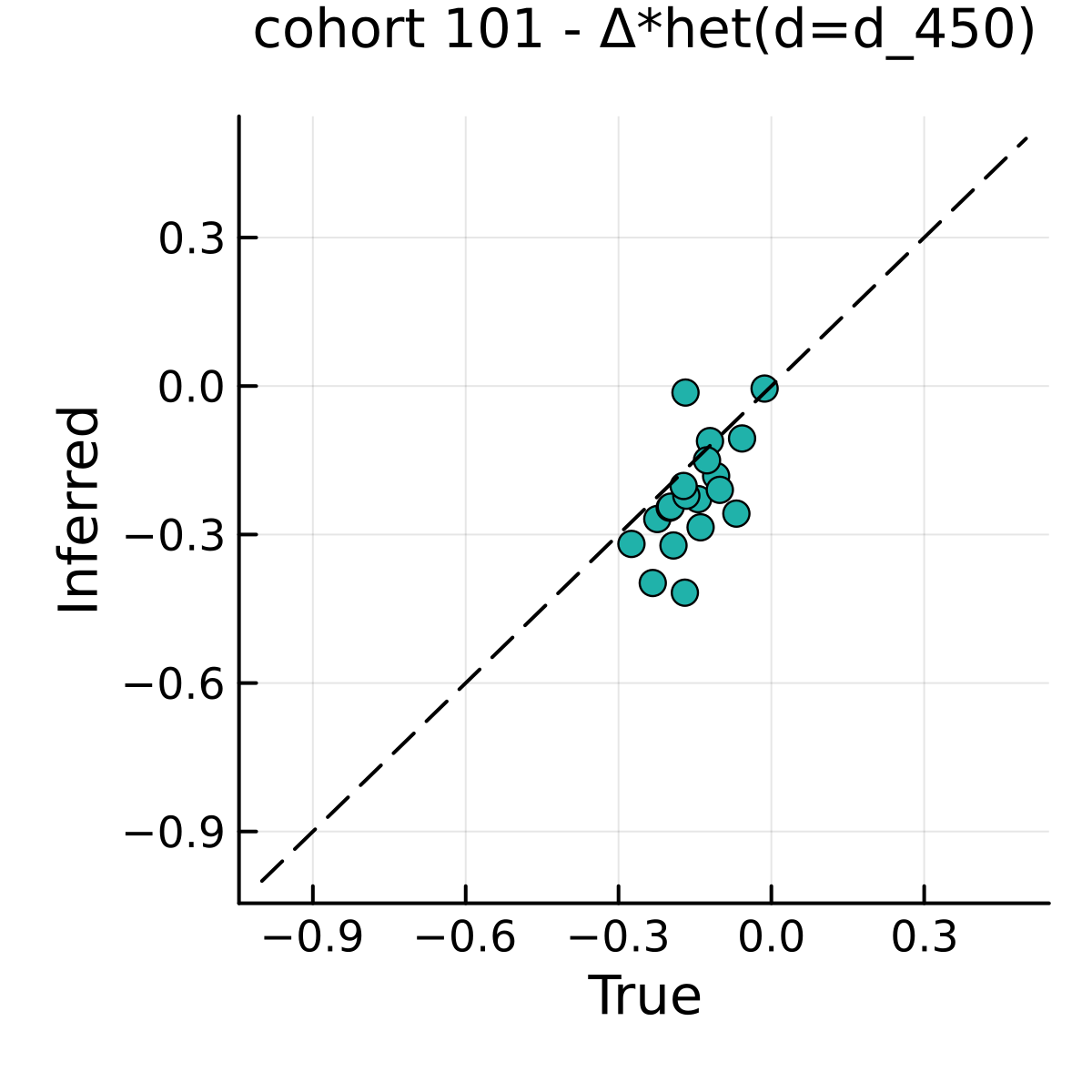}
    \end{subfigure}
    \begin{subfigure}[b]{0.22\textwidth}
        \includegraphics[width=\textwidth]{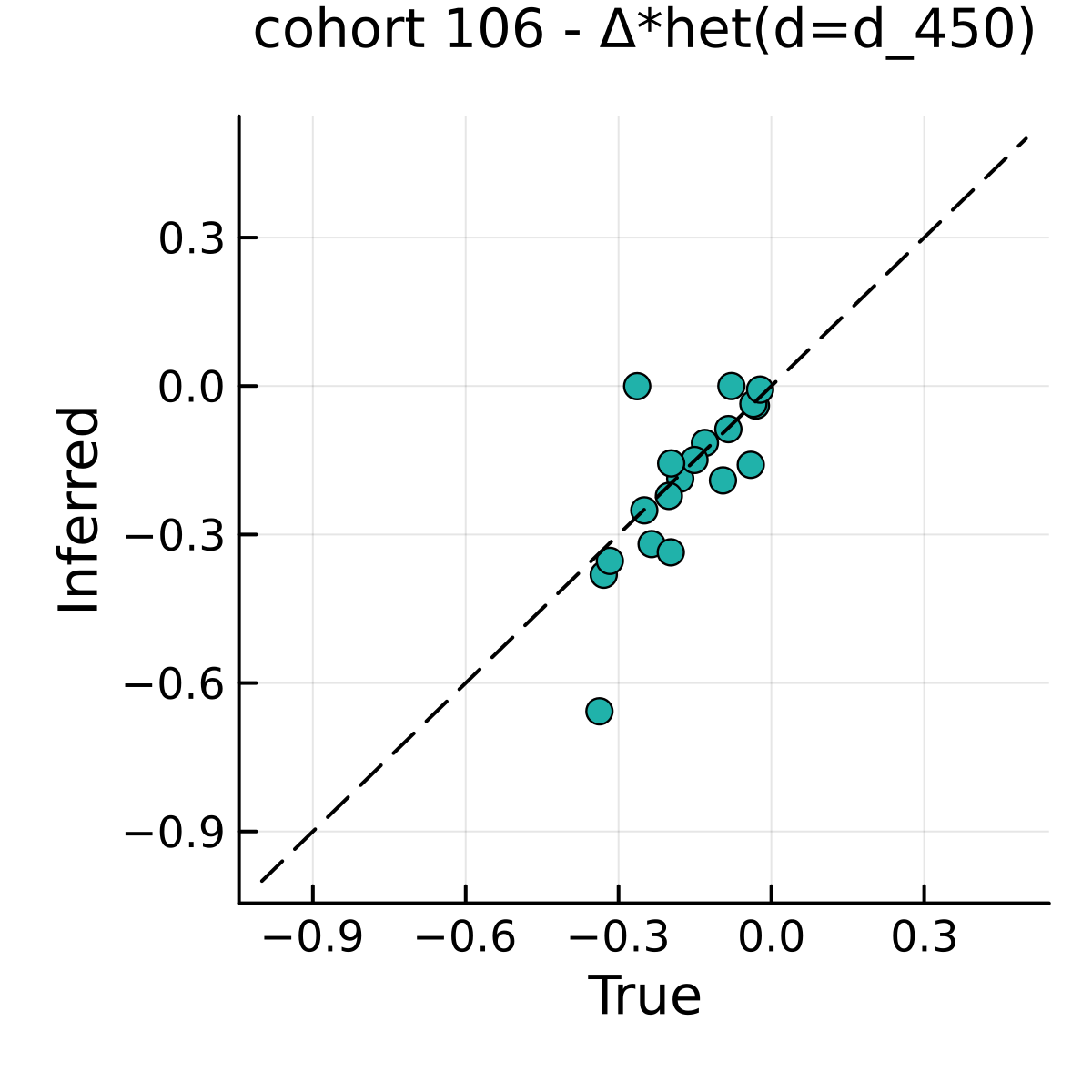}
    \end{subfigure}
    \begin{subfigure}[b]{0.22\textwidth}
        \includegraphics[width=\textwidth]{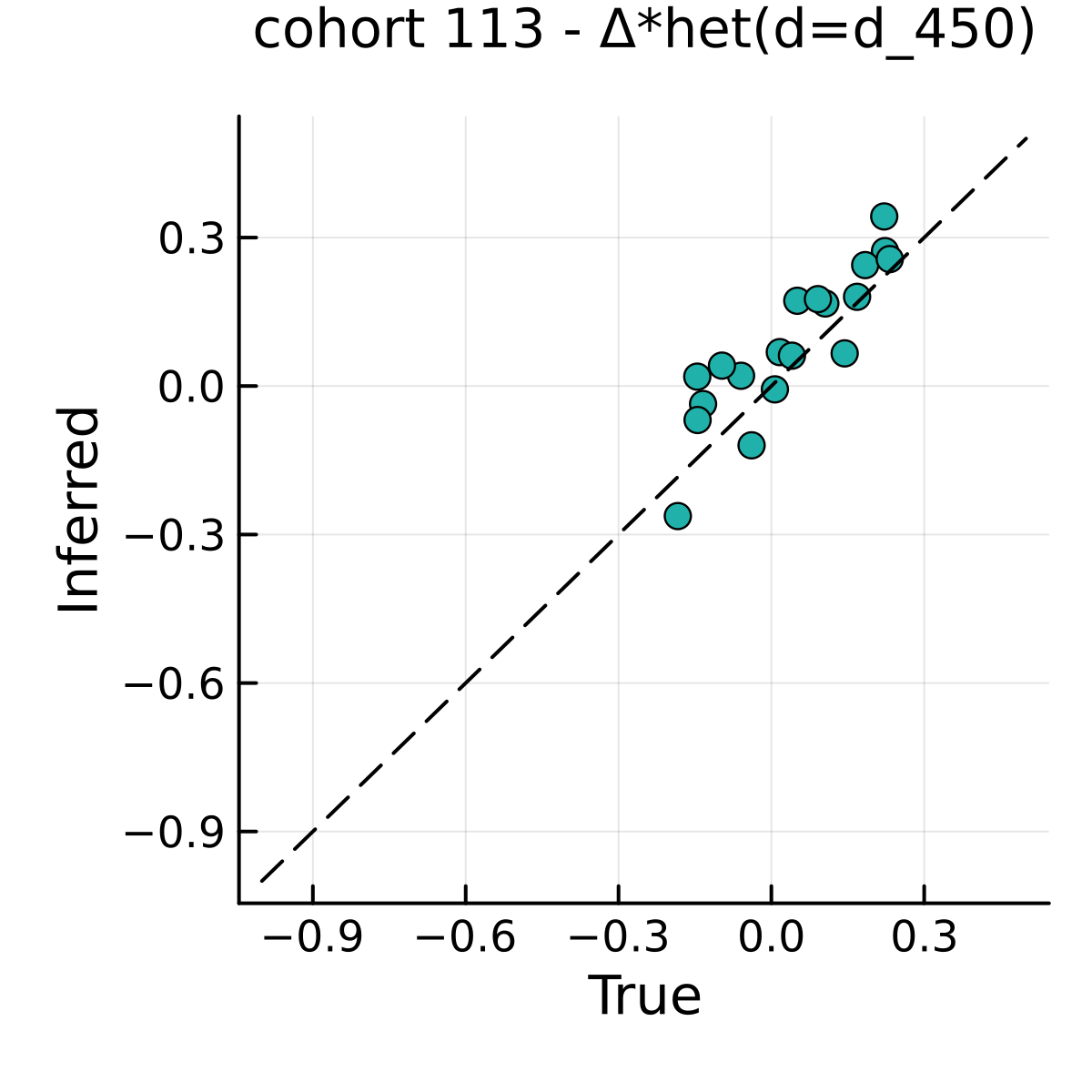}
    \end{subfigure}
    \begin{subfigure}[b]{0.22\textwidth}
        \includegraphics[width=\textwidth]{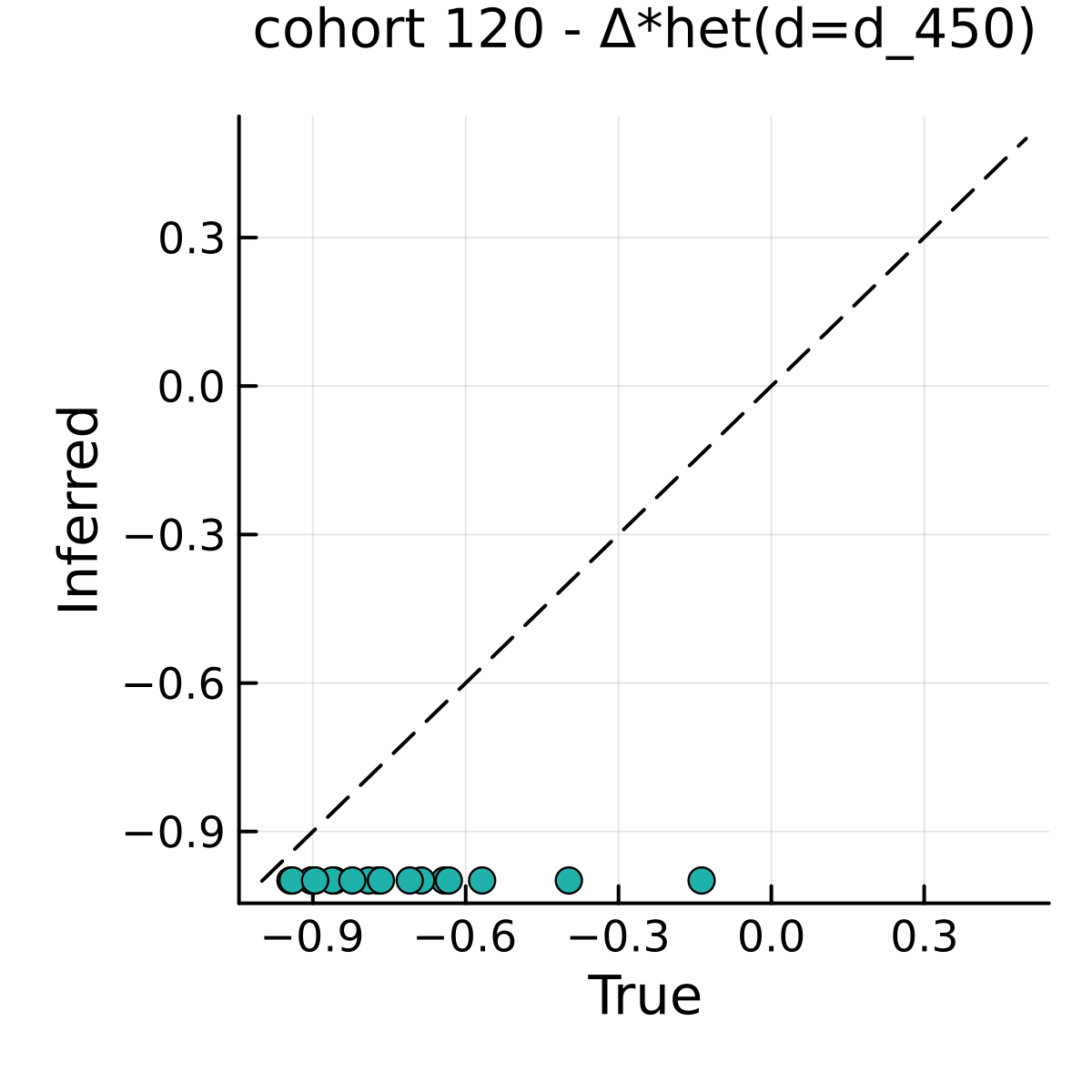}
    \end{subfigure}
    \begin{subfigure}[b]{0.22\textwidth}
        \includegraphics[width=\textwidth]{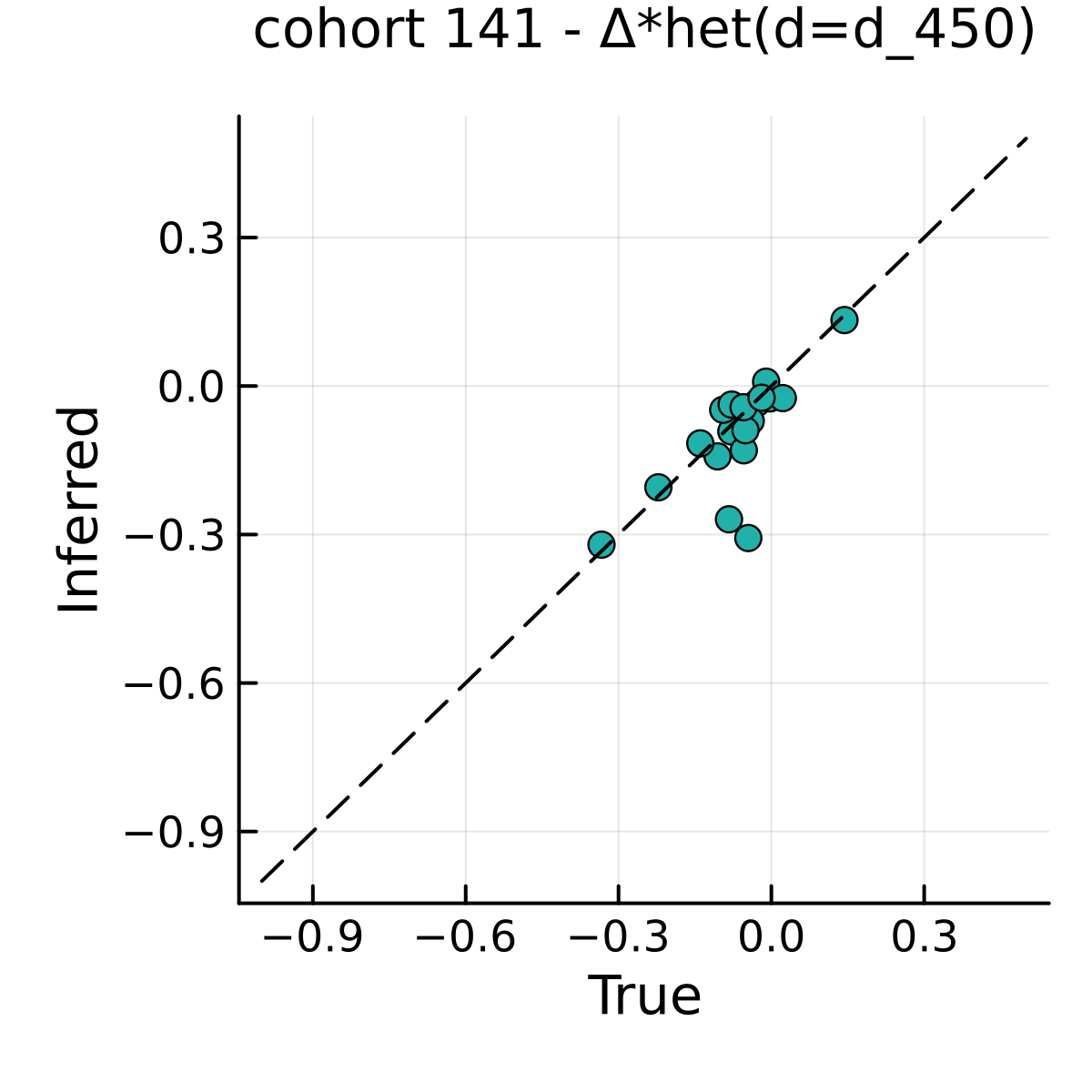}
    \end{subfigure}
    \begin{subfigure}[b]{0.22\textwidth}
        \includegraphics[width=\textwidth]{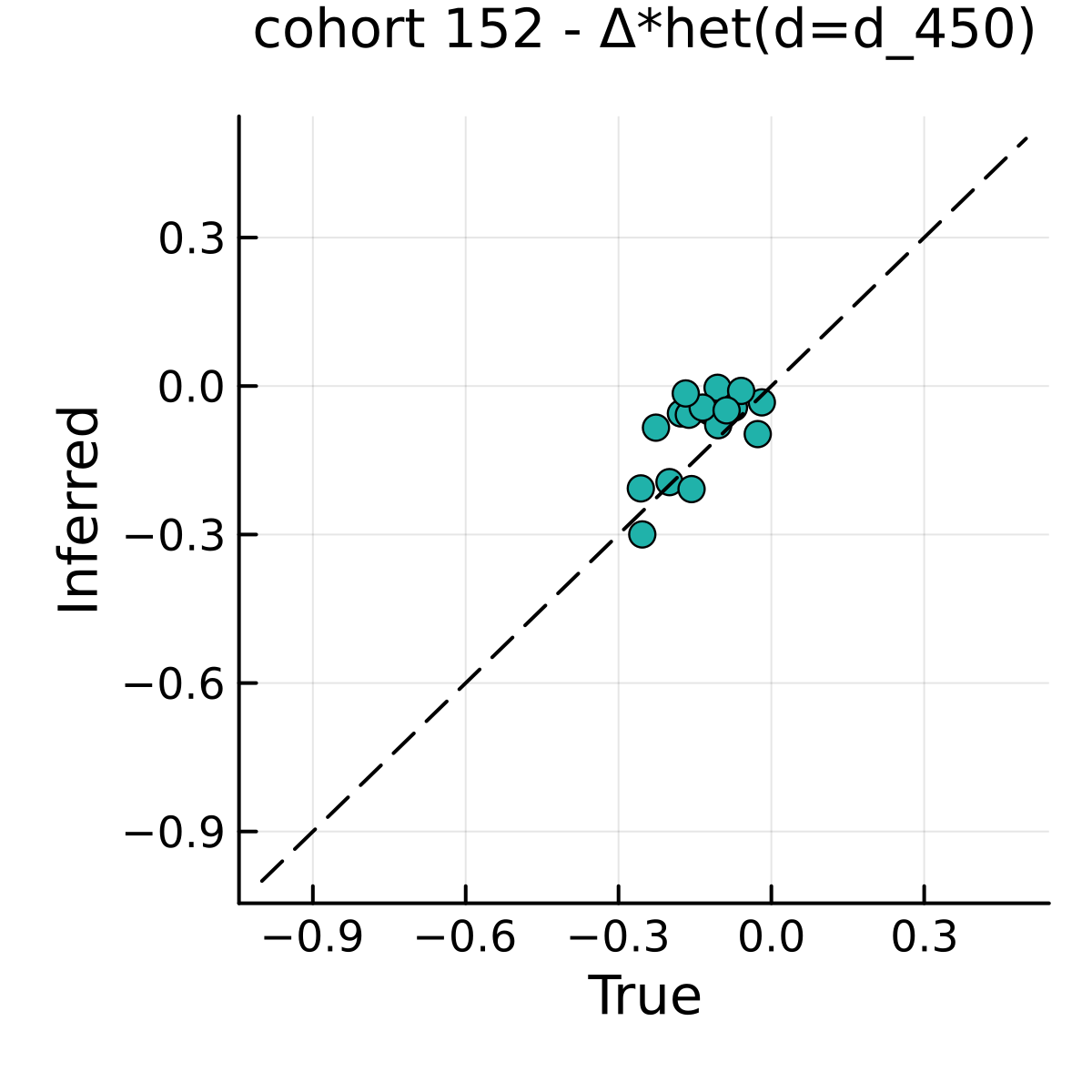}
    \end{subfigure}
    \begin{subfigure}[b]{0.22\textwidth}
        \includegraphics[width=\textwidth]{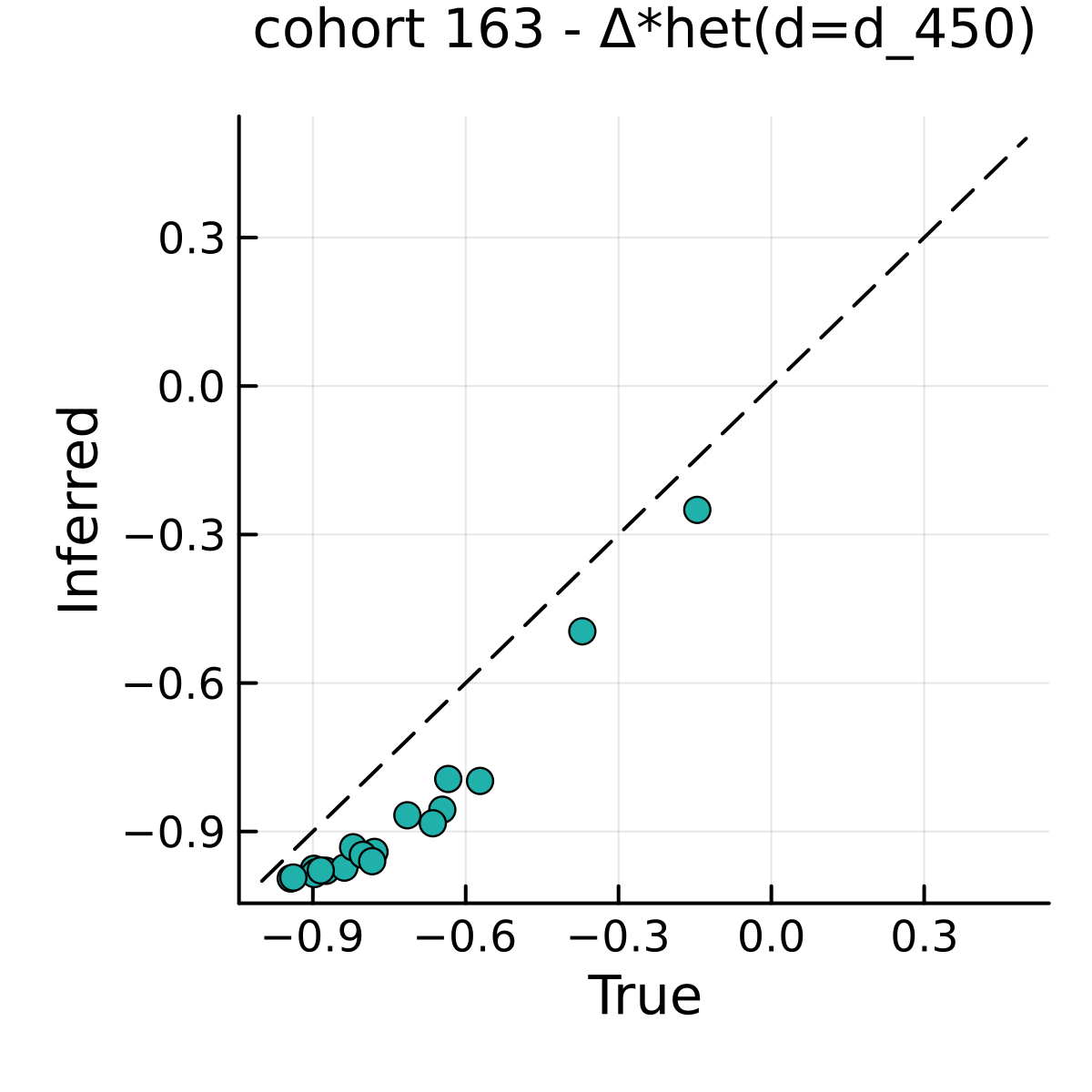}
    \end{subfigure}
    \begin{subfigure}[b]{0.22\textwidth}
        \includegraphics[width=\textwidth]{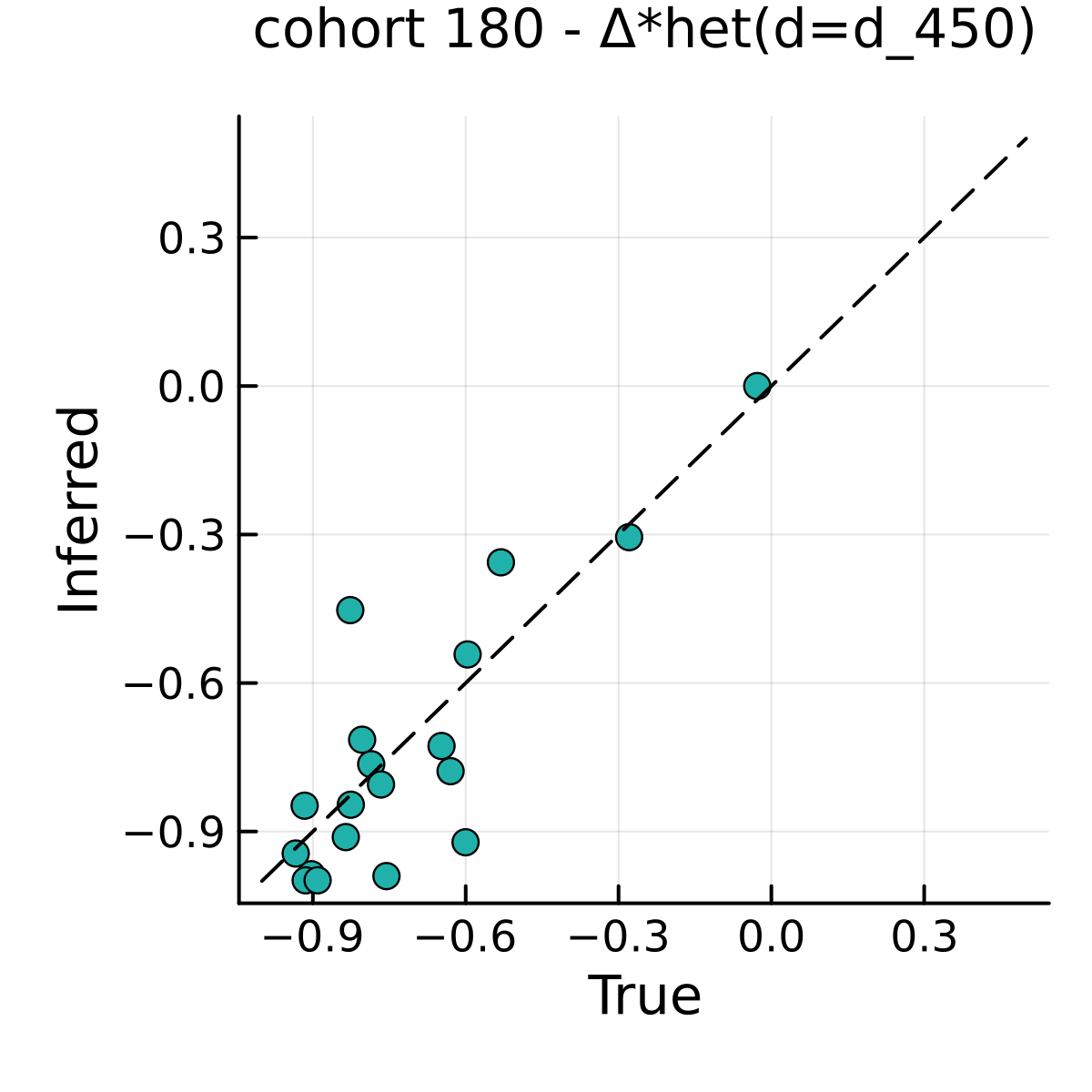}
    \end{subfigure}
    \begin{subfigure}[b]{0.22\textwidth}
        \includegraphics[width=\textwidth]{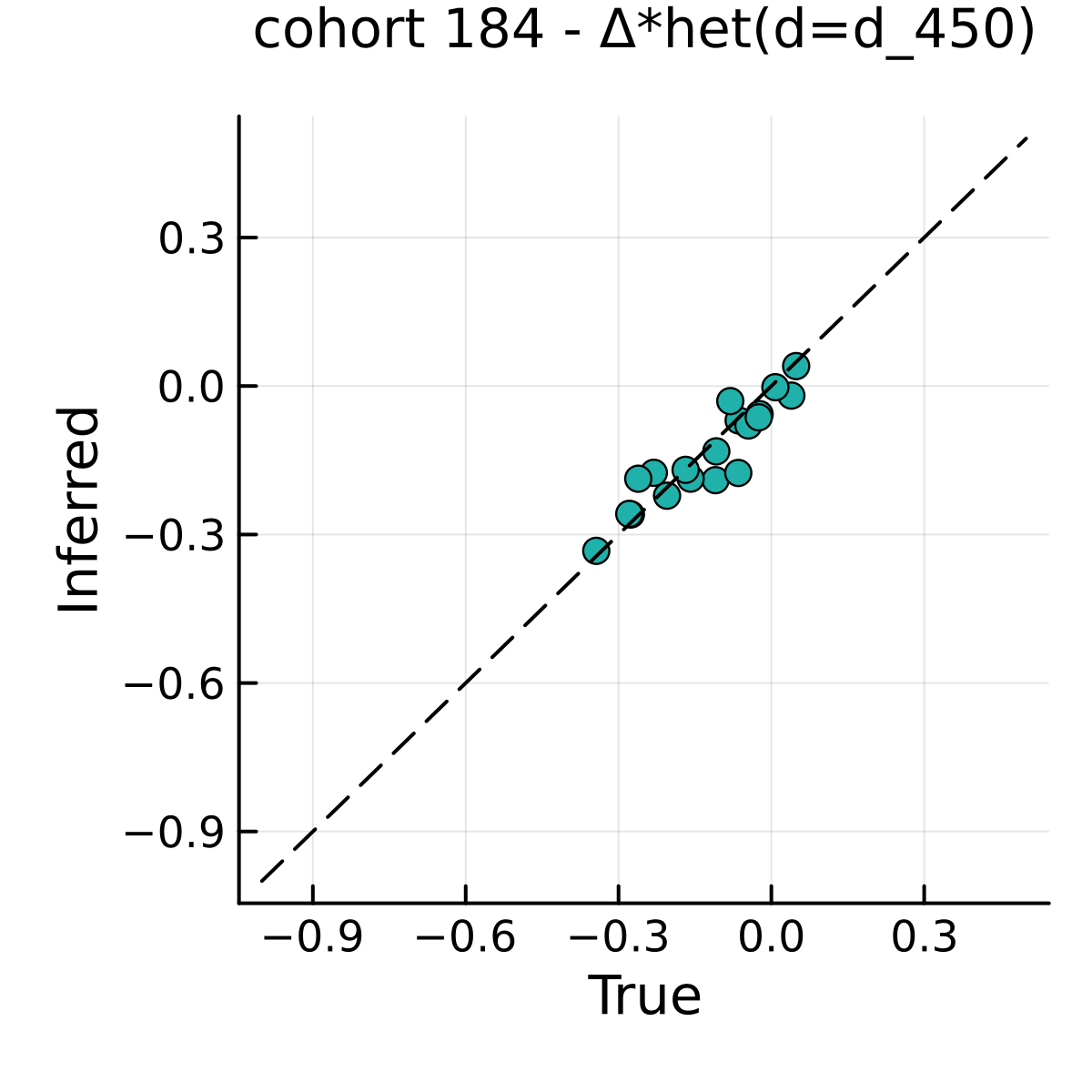}
    \end{subfigure}
    \begin{subfigure}[b]{0.22\textwidth}
        \includegraphics[width=\textwidth]{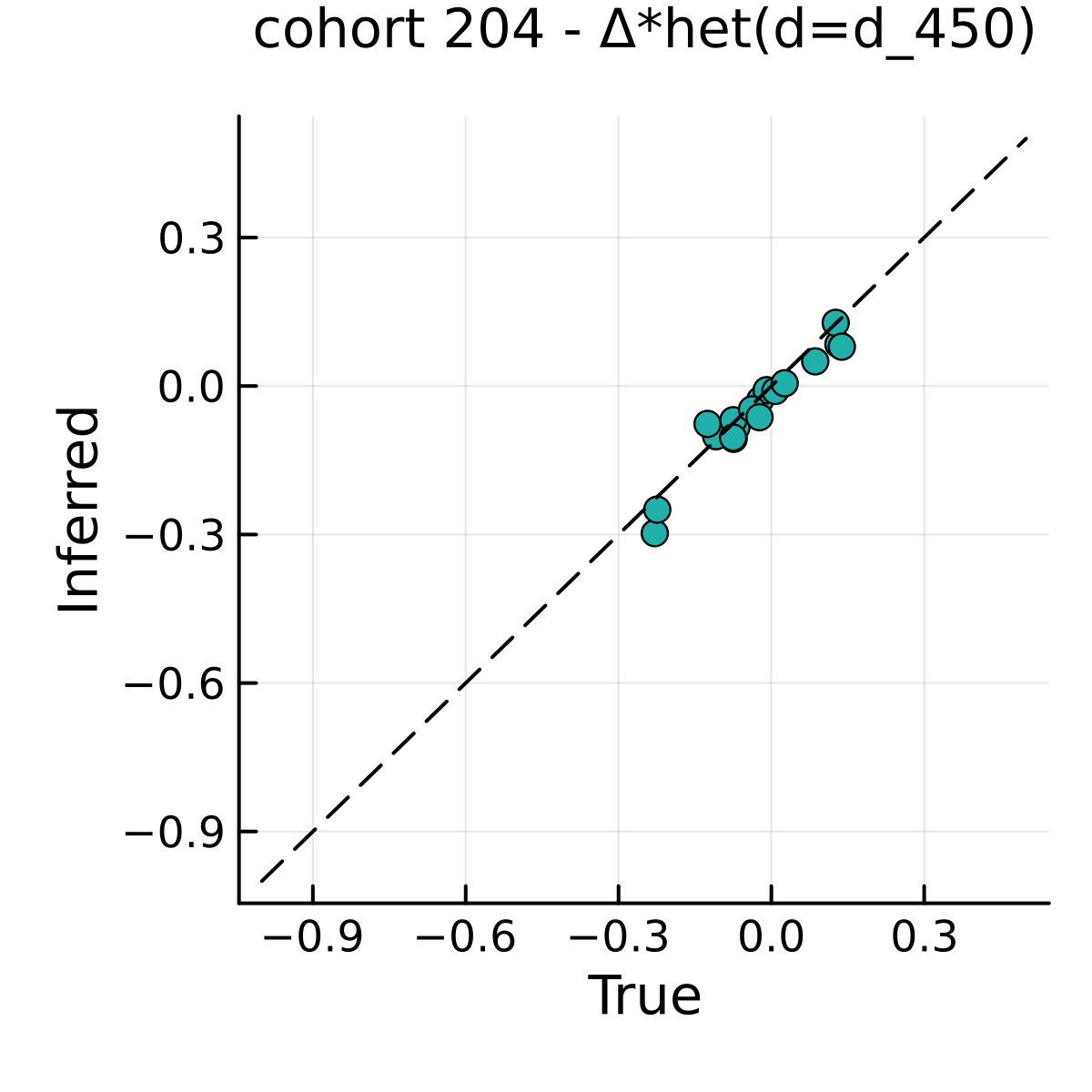}
    \end{subfigure}
    \begin{subfigure}[b]{0.22\textwidth}
        \includegraphics[width=\textwidth]{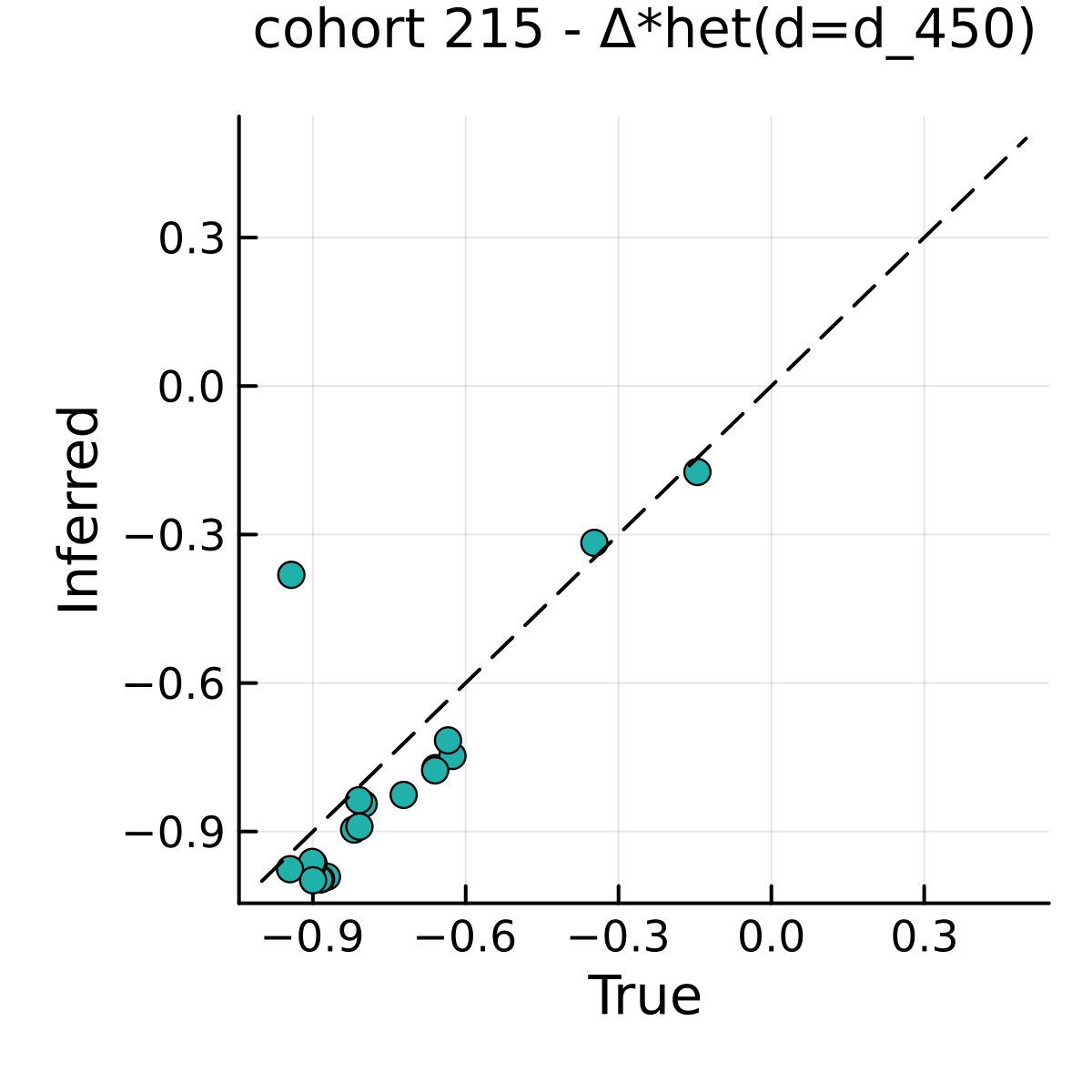}
    \end{subfigure}
    \begin{subfigure}[b]{0.22\textwidth}
        \includegraphics[width=\textwidth]{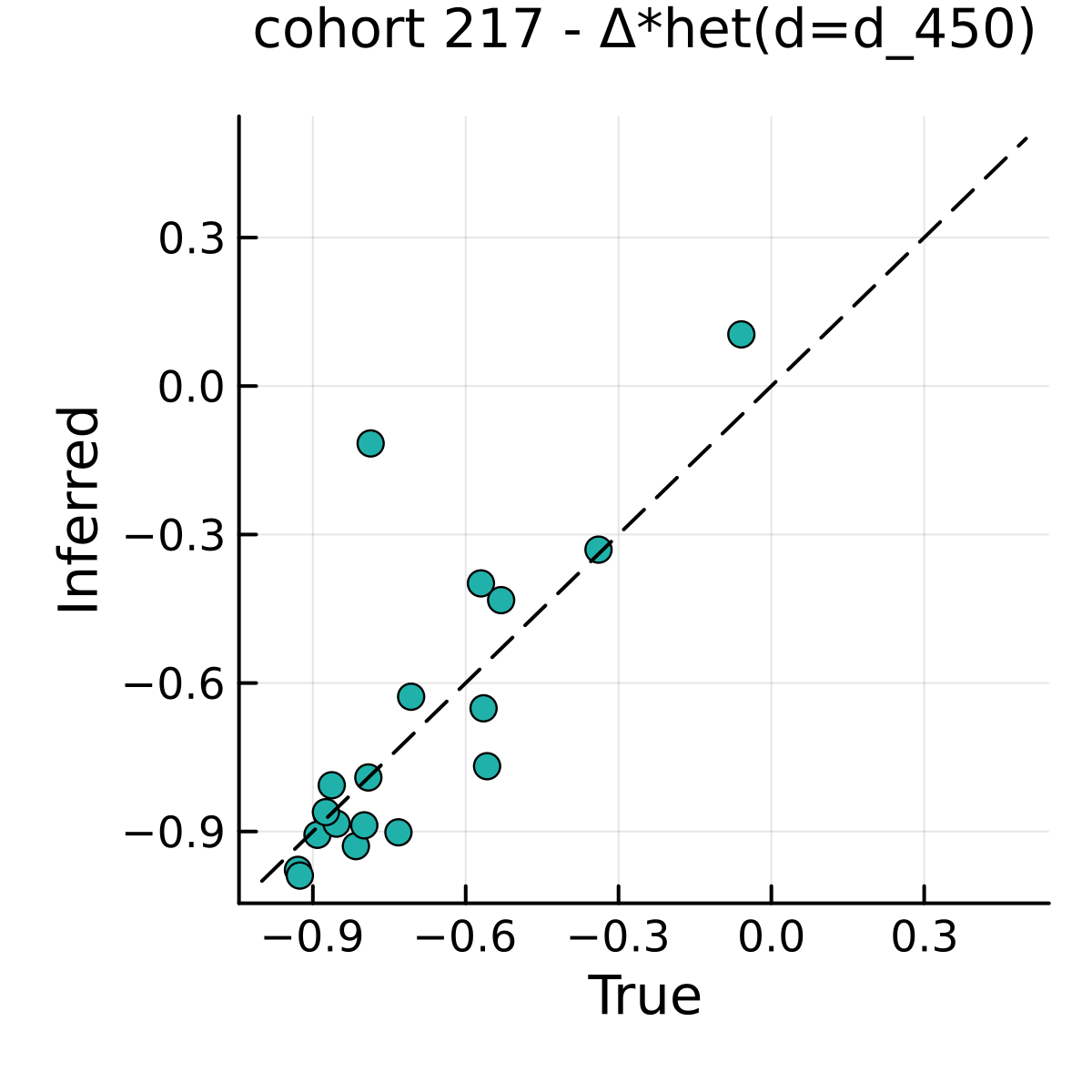}
    \end{subfigure}
    \caption{Comparison between the mean posterior value of $\bar{\Delta}^*_{het}(d_{450}^{(i)})$ (y-axis) for each virtual patient $i$ and each synthetic cohort $m$, and the true one (x-axis). $d_{450}^{(i)}$ corresponds to the mean dose received by the $i^{th}$ patient over 450 days of therapy.}
    \label{fig:synth_Delta_het}
\end{figure}

\begin{figure}[h]
    \centering
    \begin{subfigure}[b]{0.22\textwidth}
        \includegraphics[width=\textwidth]{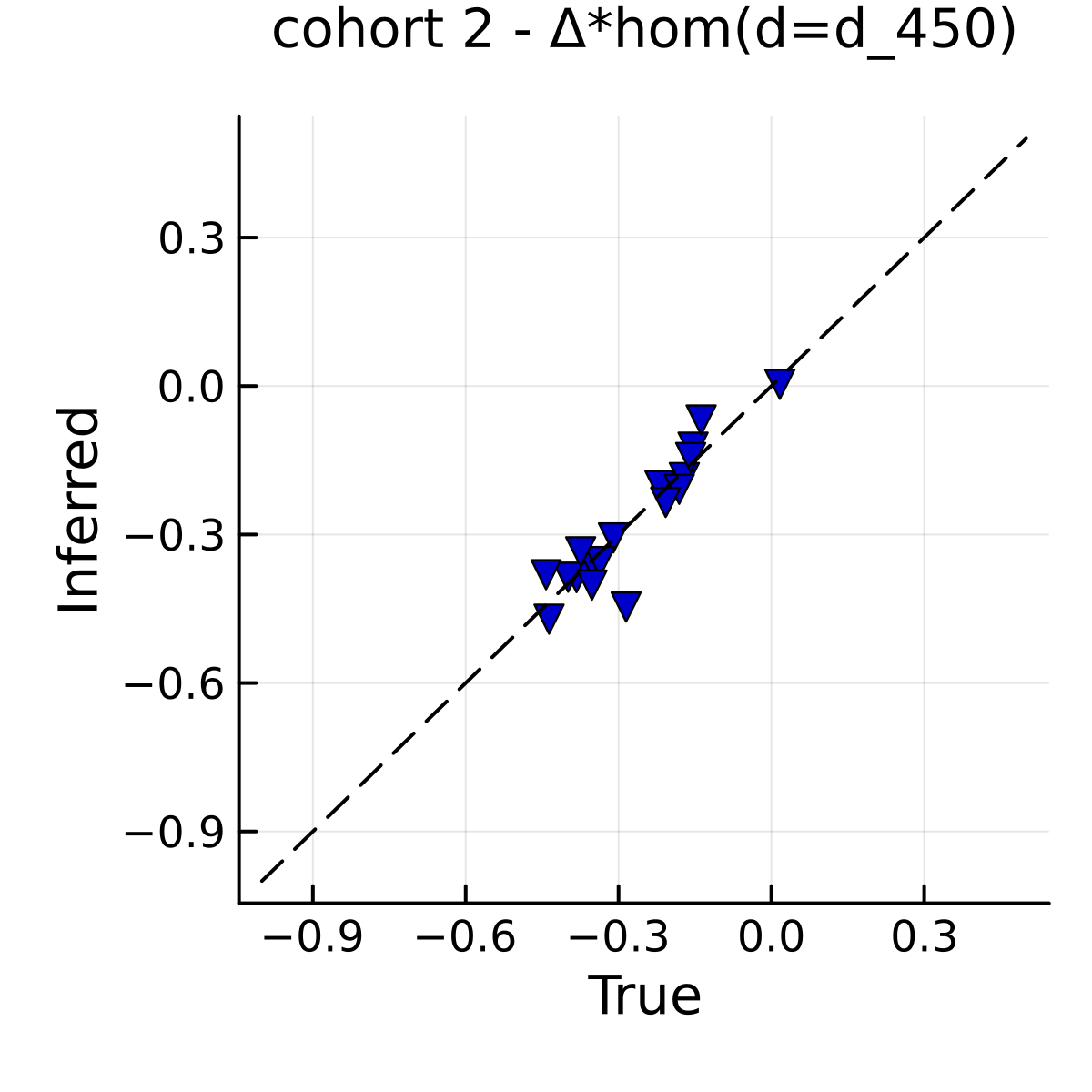}
    \end{subfigure}
    \begin{subfigure}[b]{0.22\textwidth}
        \includegraphics[width=\textwidth]{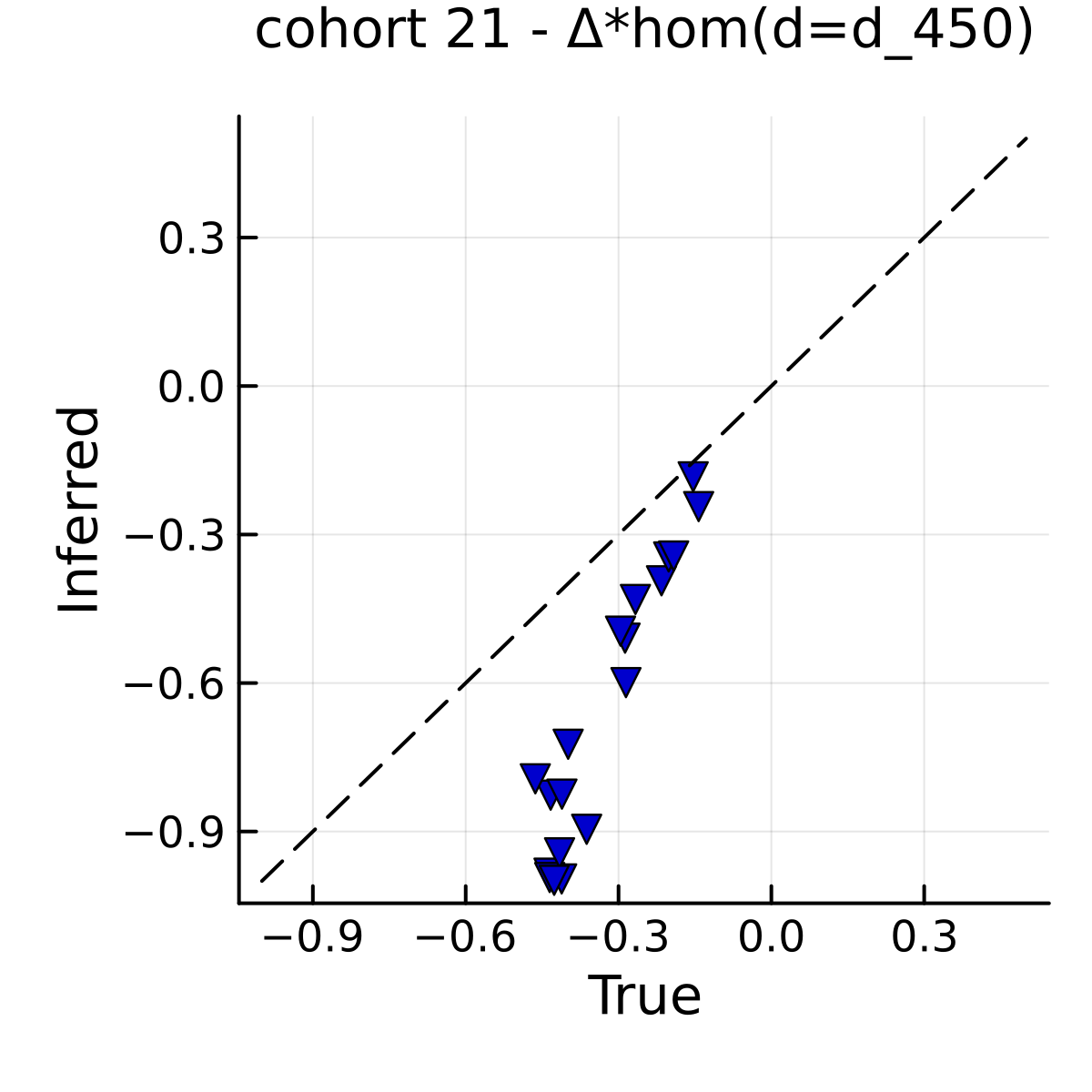}
    \end{subfigure}
    \begin{subfigure}[b]{0.22\textwidth}
        \includegraphics[width=\textwidth]{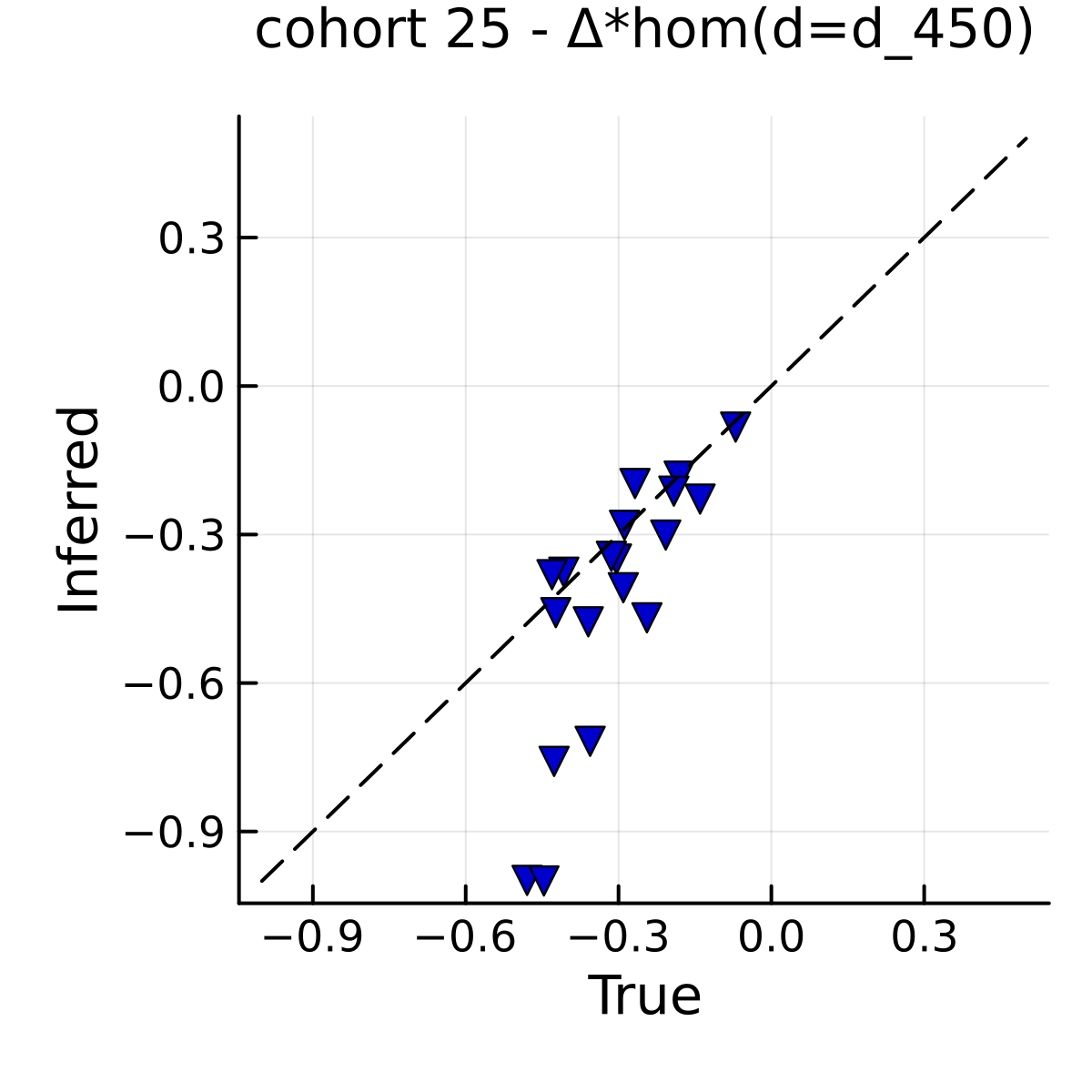}
    \end{subfigure}
    \begin{subfigure}[b]{0.22\textwidth}
        \includegraphics[width=\textwidth]{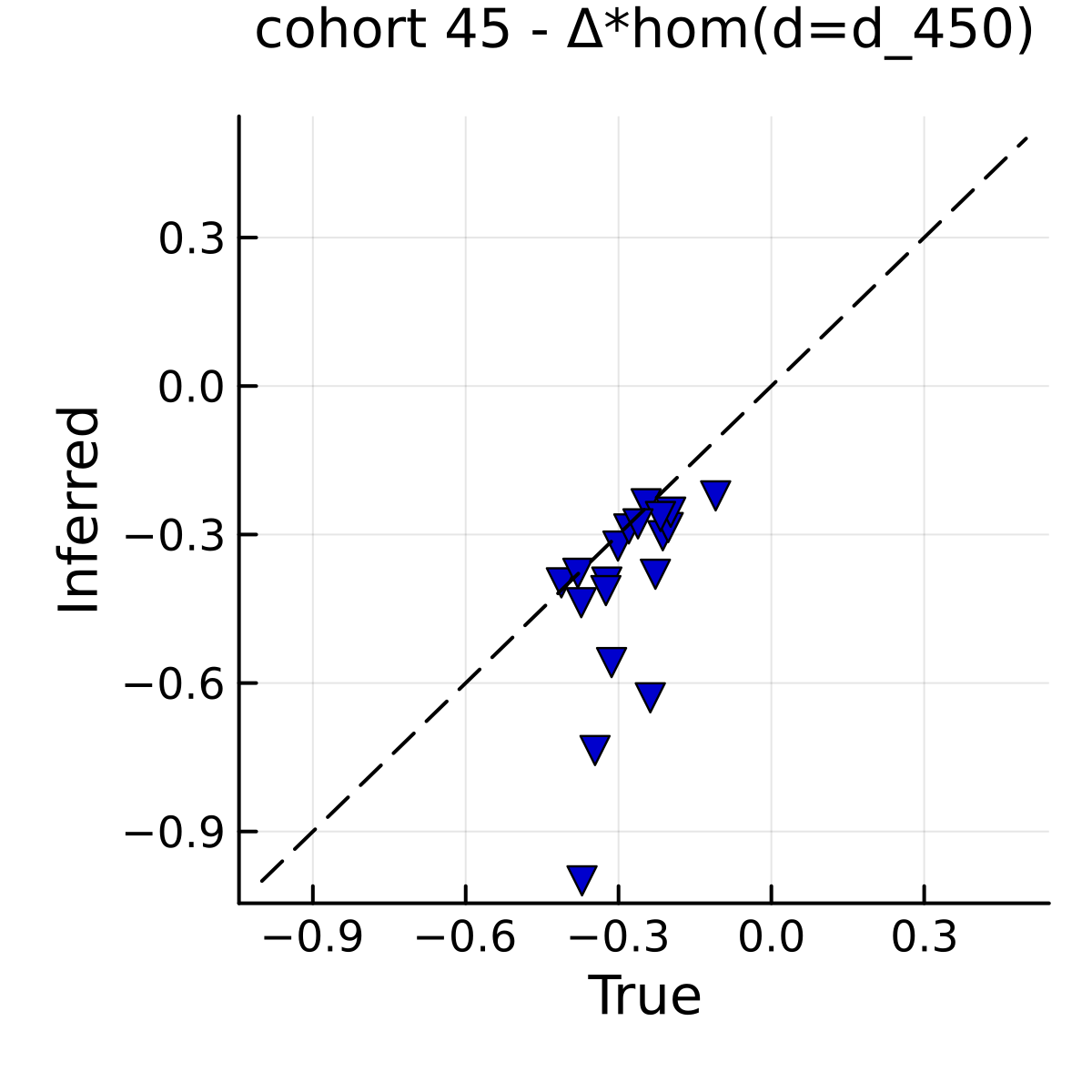}
    \end{subfigure}
    \begin{subfigure}[b]{0.22\textwidth}
        \includegraphics[width=\textwidth]{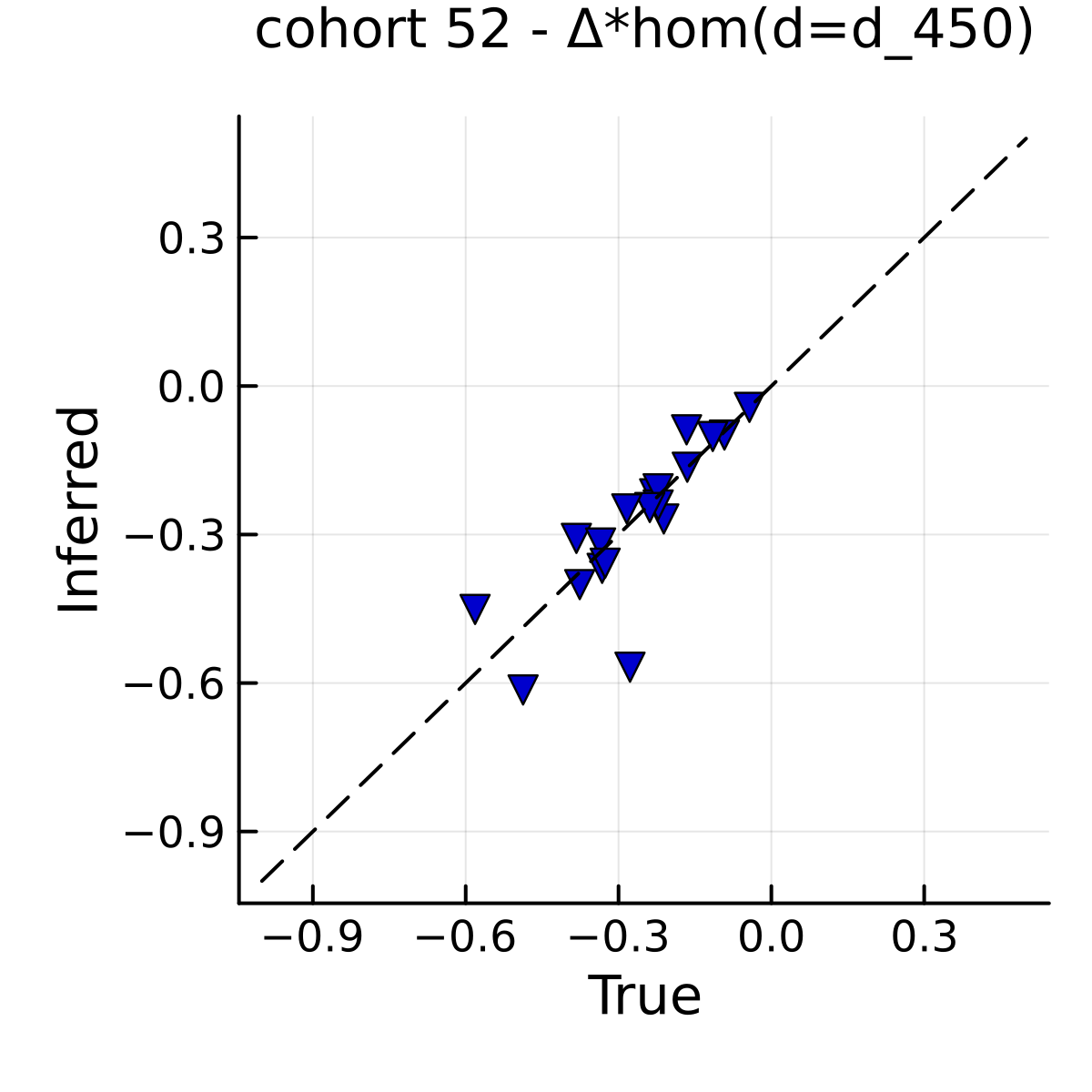}
    \end{subfigure}
    \begin{subfigure}[b]{0.22\textwidth}
        \includegraphics[width=\textwidth]{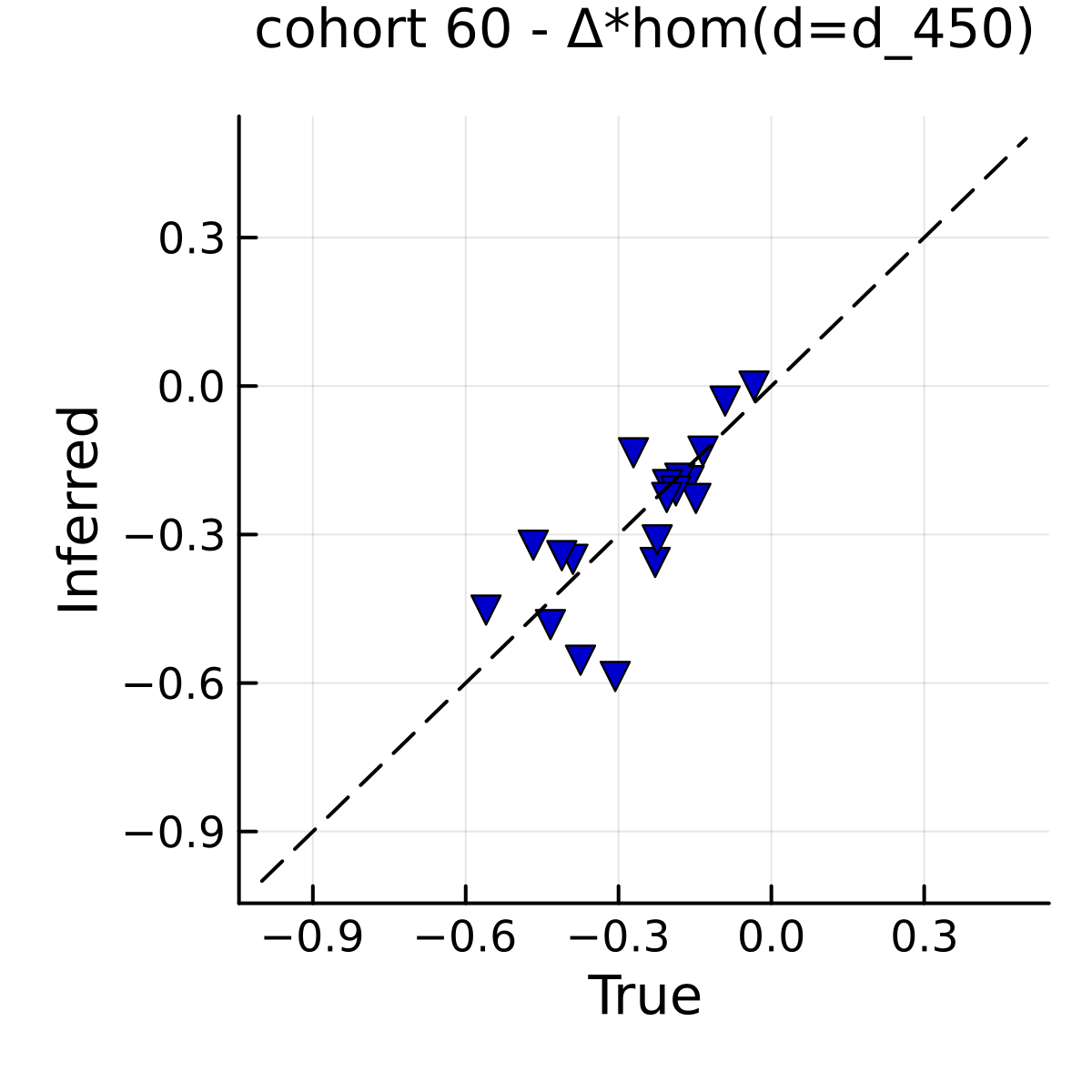}
    \end{subfigure}
    \begin{subfigure}[b]{0.22\textwidth}
        \includegraphics[width=\textwidth]{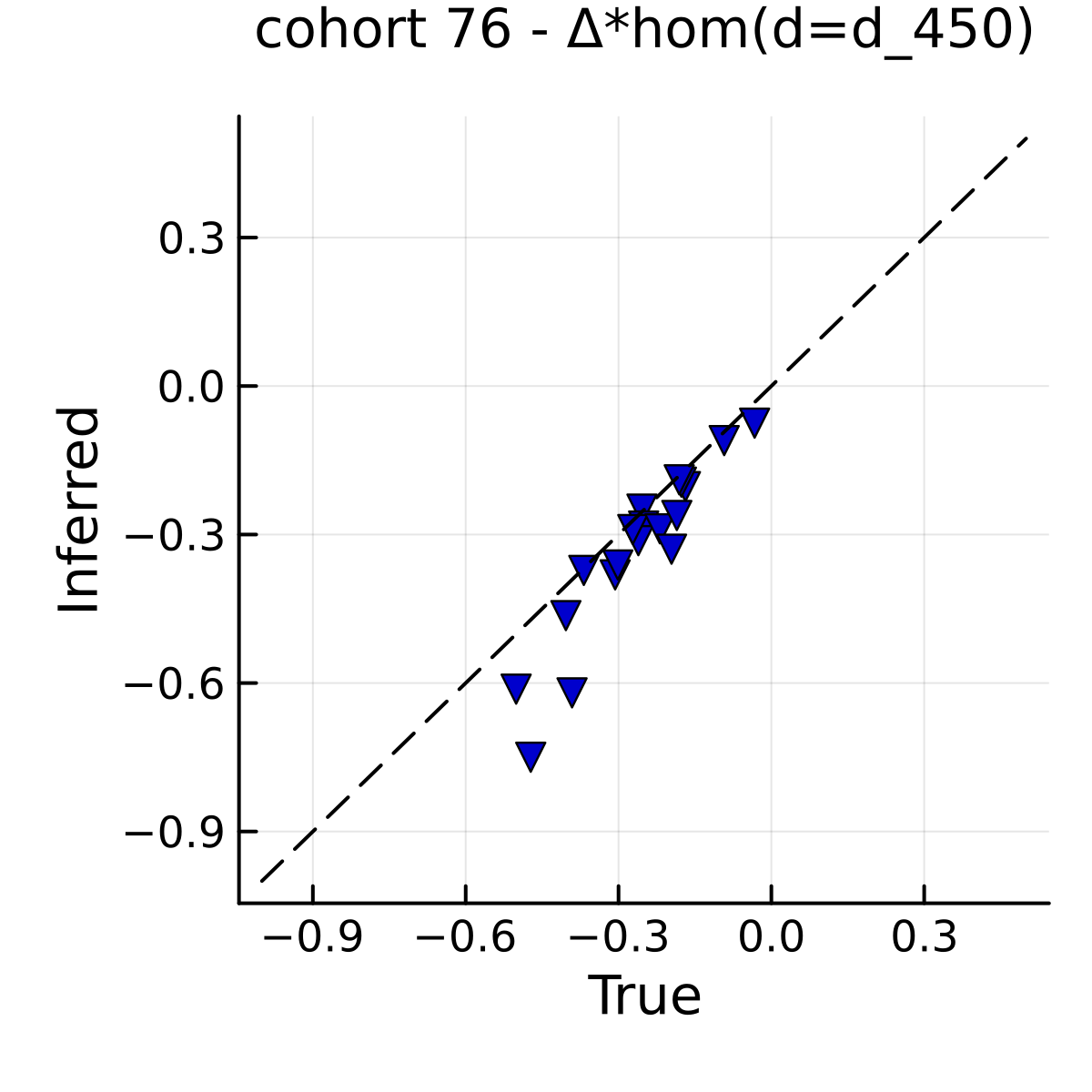}
    \end{subfigure}
    \begin{subfigure}[b]{0.22\textwidth}
        \includegraphics[width=\textwidth]{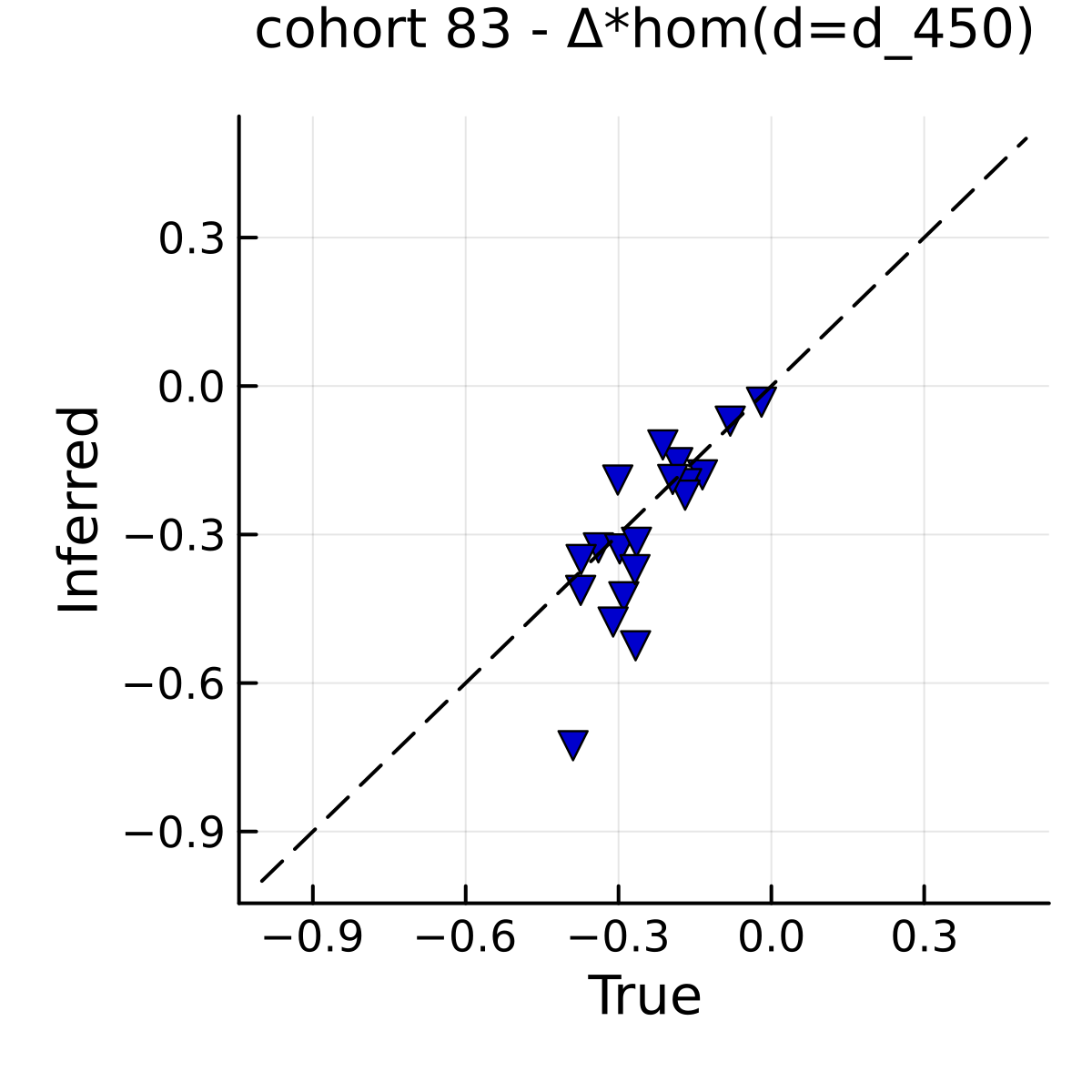}
    \end{subfigure}
    \begin{subfigure}[b]{0.22\textwidth}
        \includegraphics[width=\textwidth]{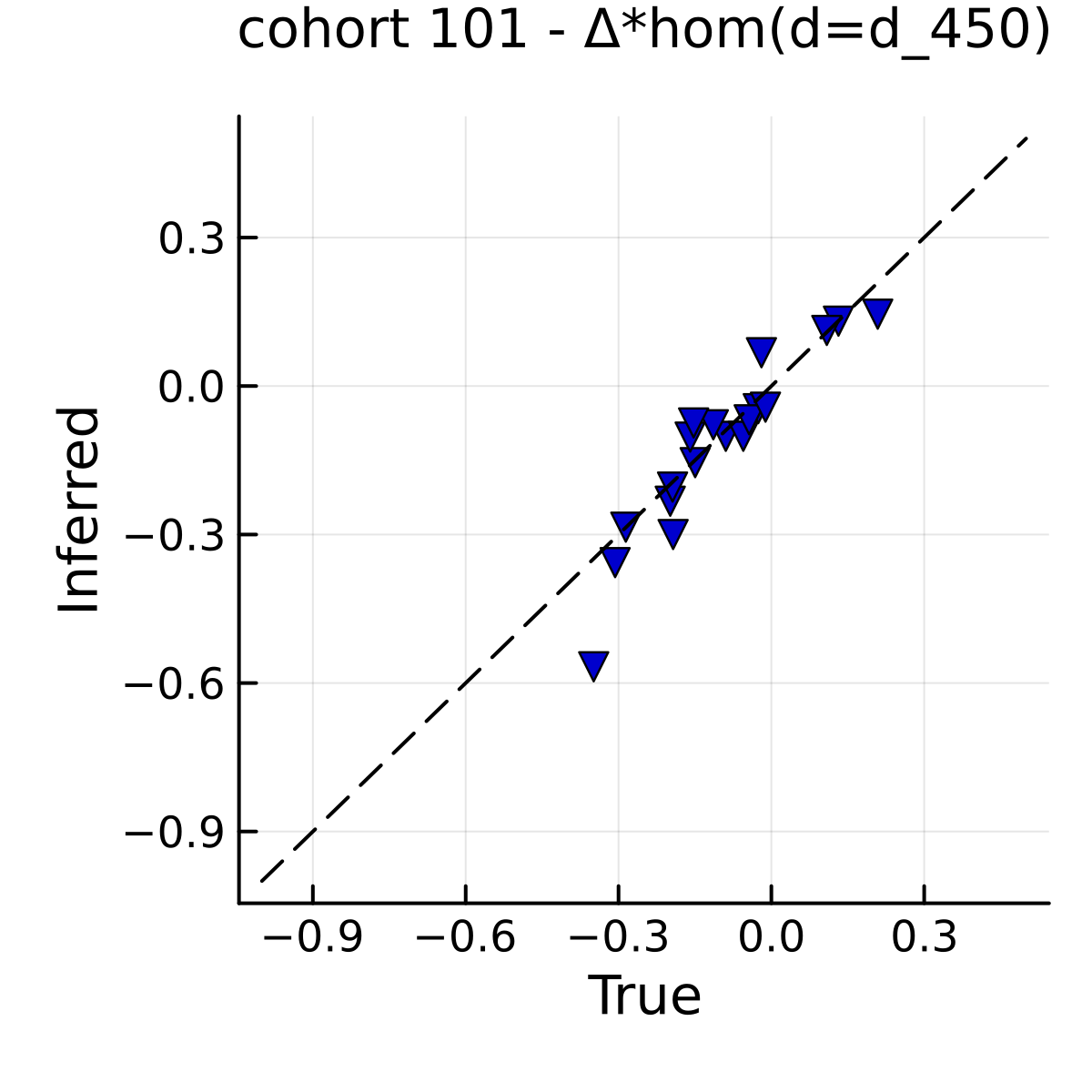}
    \end{subfigure}
    \begin{subfigure}[b]{0.22\textwidth}
        \includegraphics[width=\textwidth]{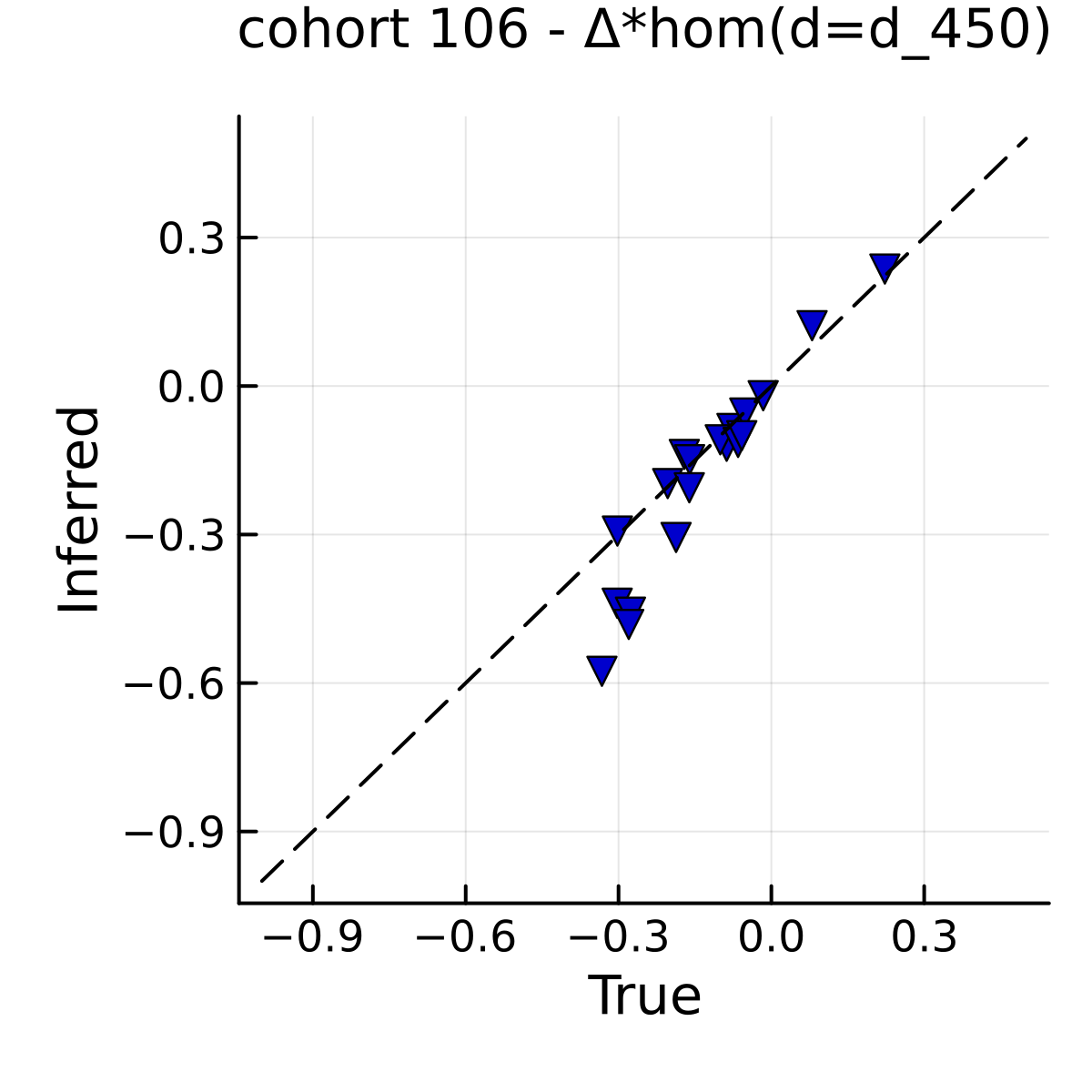}
    \end{subfigure}
    \begin{subfigure}[b]{0.22\textwidth}
        \includegraphics[width=\textwidth]{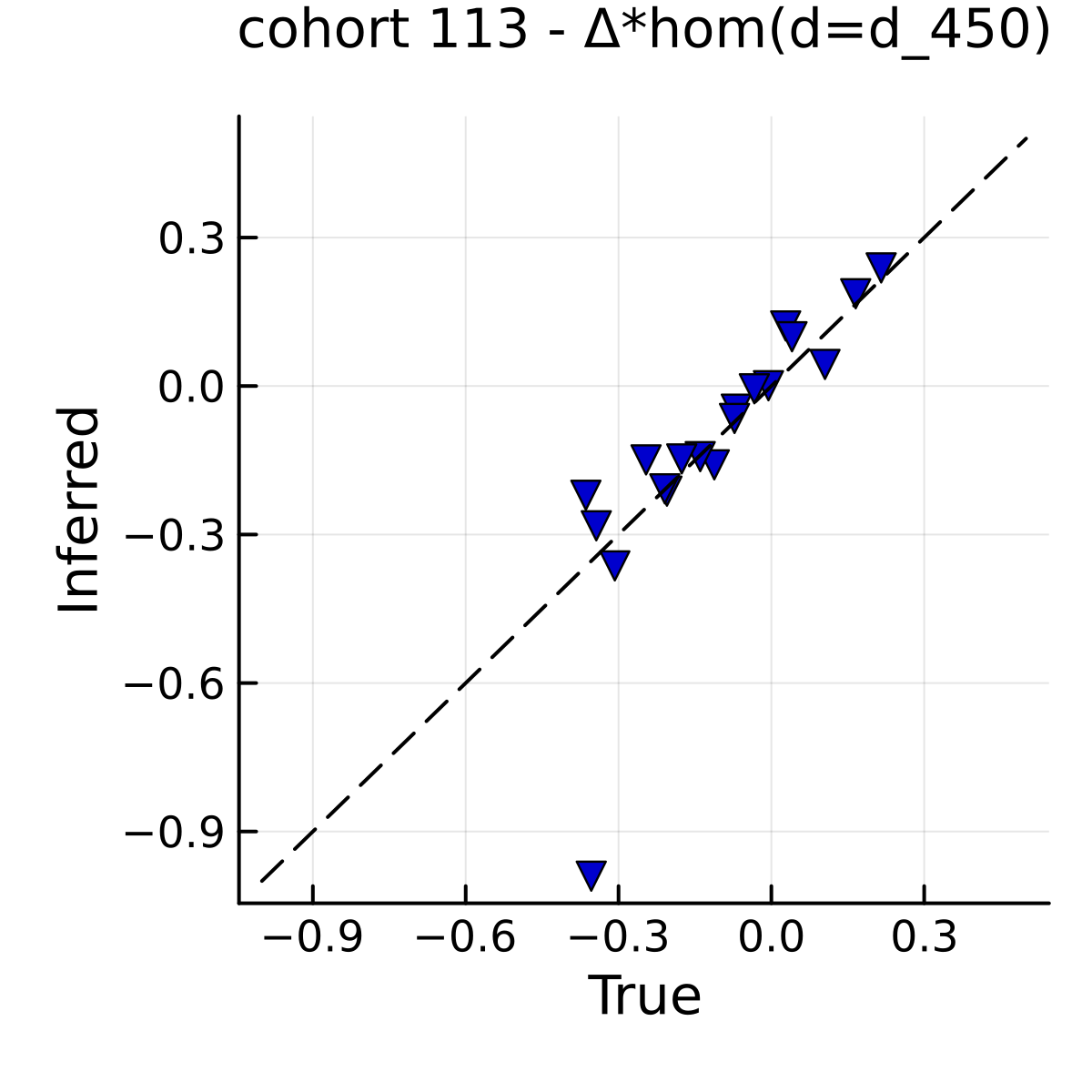}
    \end{subfigure}
    \begin{subfigure}[b]{0.22\textwidth}
        \includegraphics[width=\textwidth]{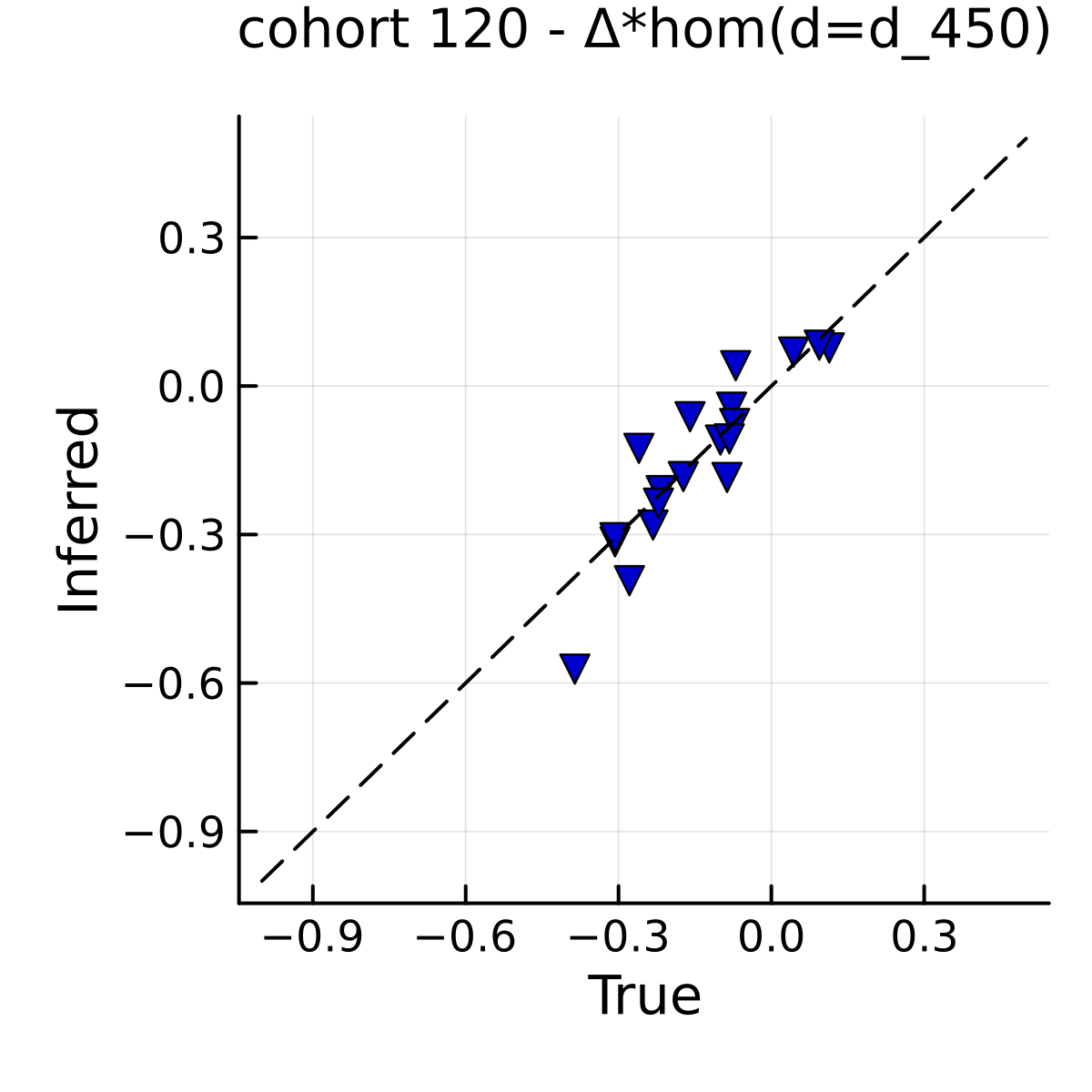}
    \end{subfigure}
    \begin{subfigure}[b]{0.22\textwidth}
        \includegraphics[width=\textwidth]{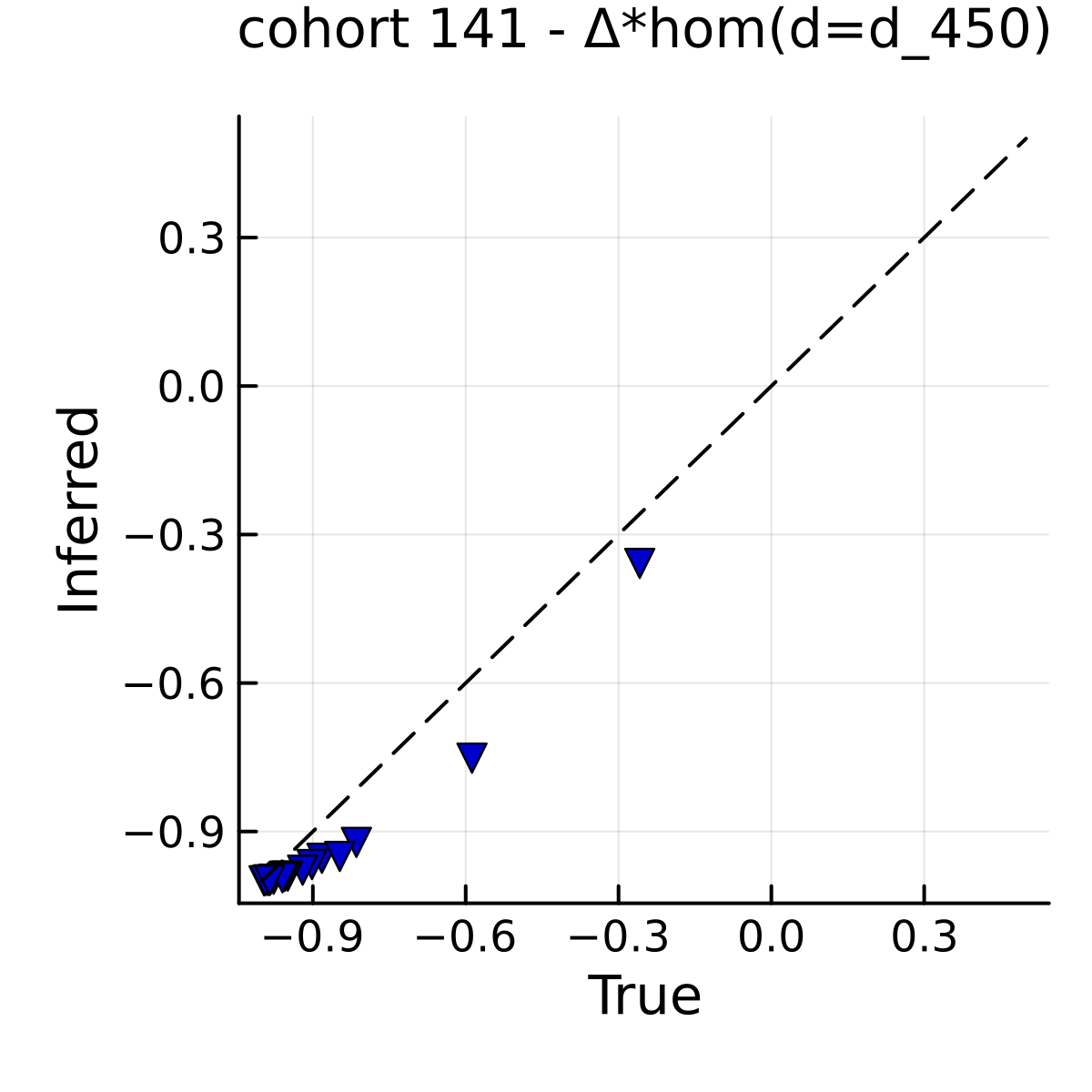}
    \end{subfigure}
    \begin{subfigure}[b]{0.22\textwidth}
        \includegraphics[width=\textwidth]{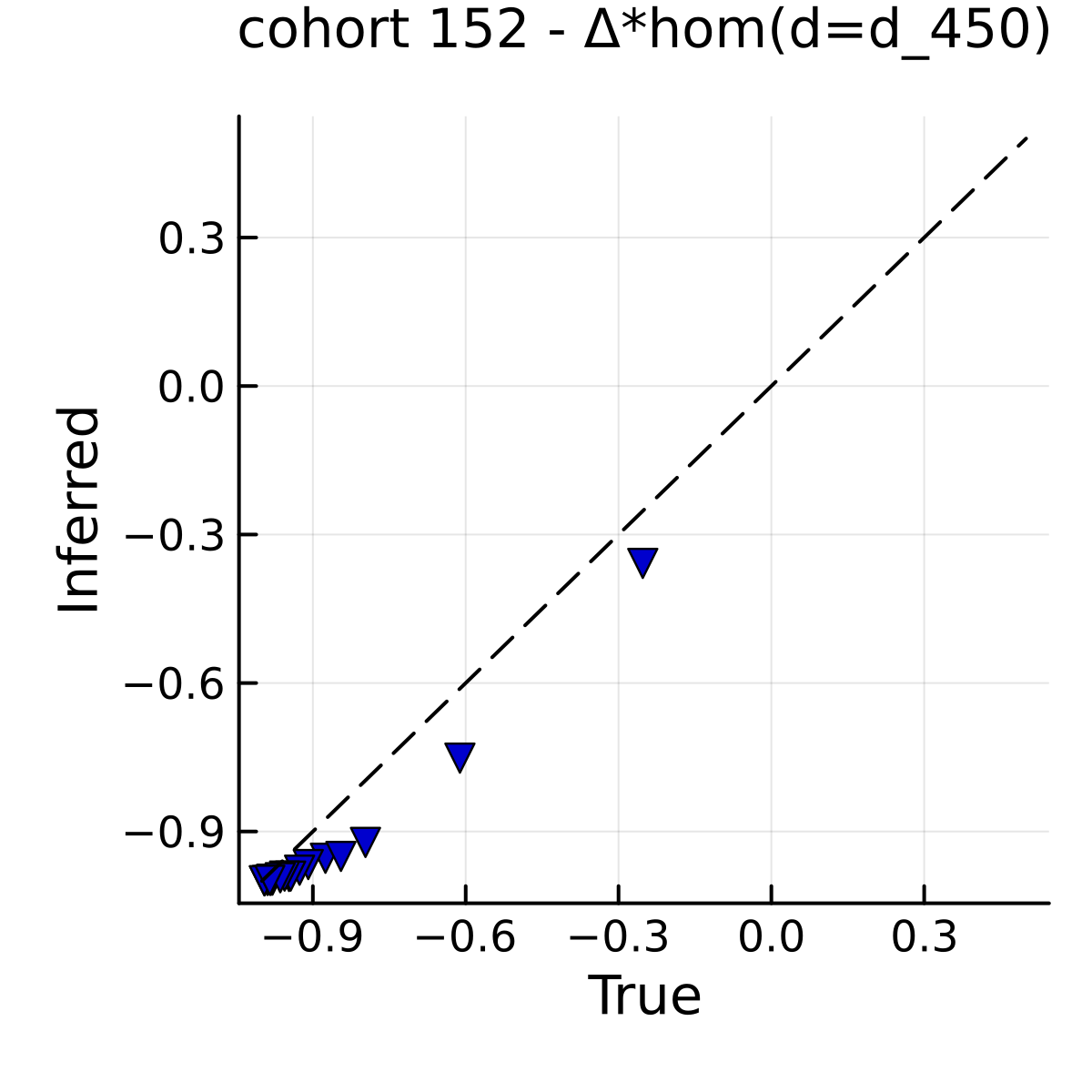}
    \end{subfigure}
    \begin{subfigure}[b]{0.22\textwidth}
        \includegraphics[width=\textwidth]{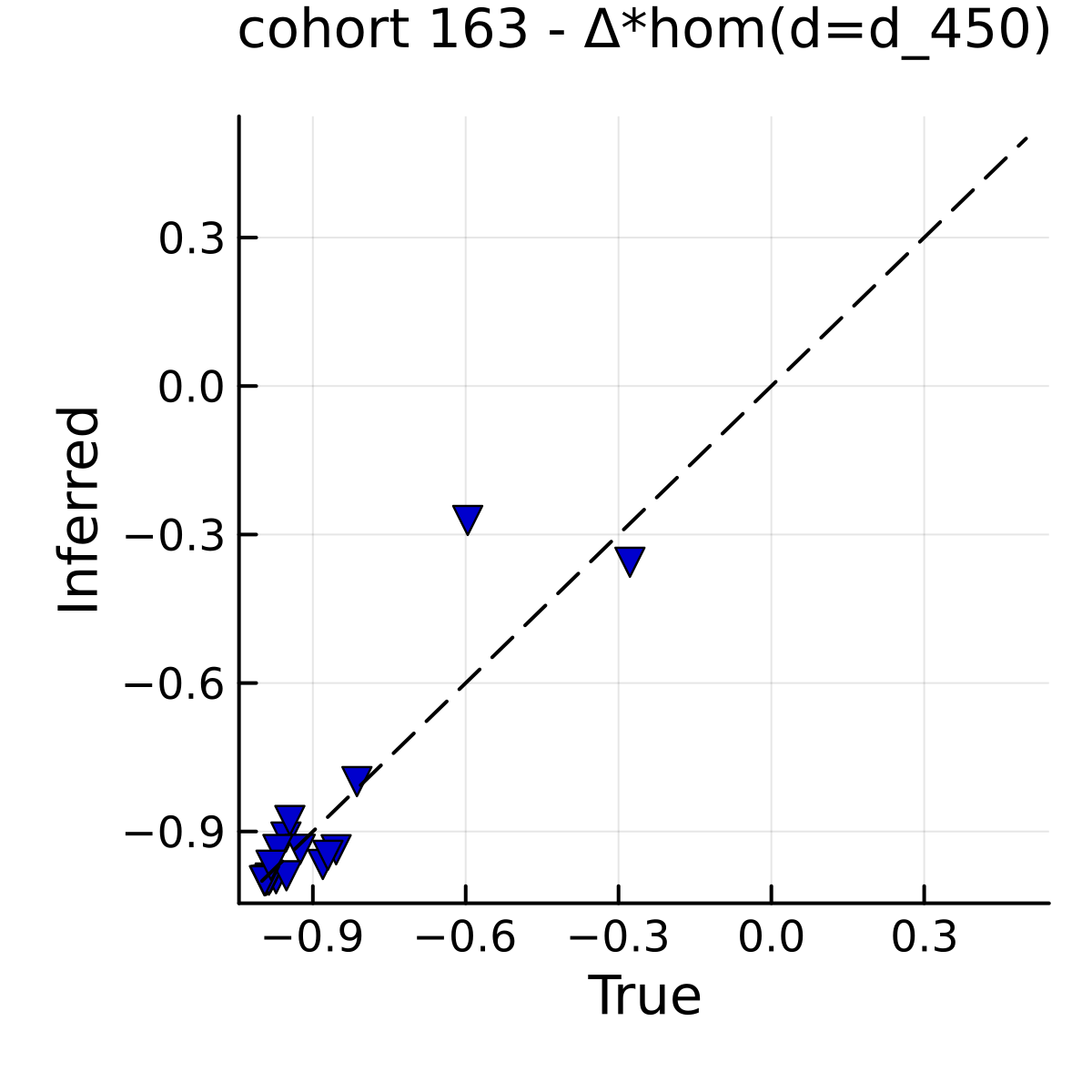}
    \end{subfigure}
    \begin{subfigure}[b]{0.22\textwidth}
        \includegraphics[width=\textwidth]{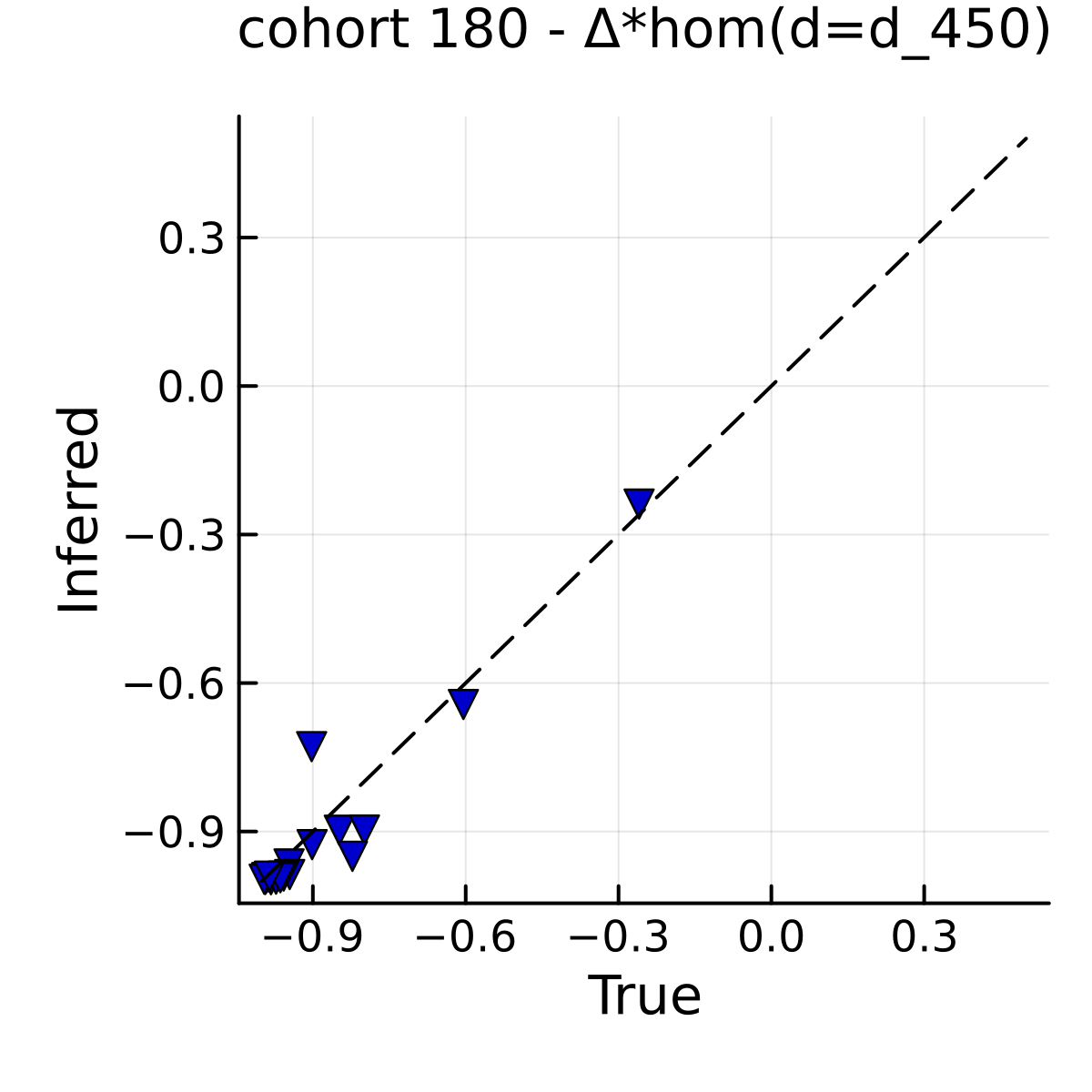}
    \end{subfigure}
    \begin{subfigure}[b]{0.22\textwidth}
        \includegraphics[width=\textwidth]{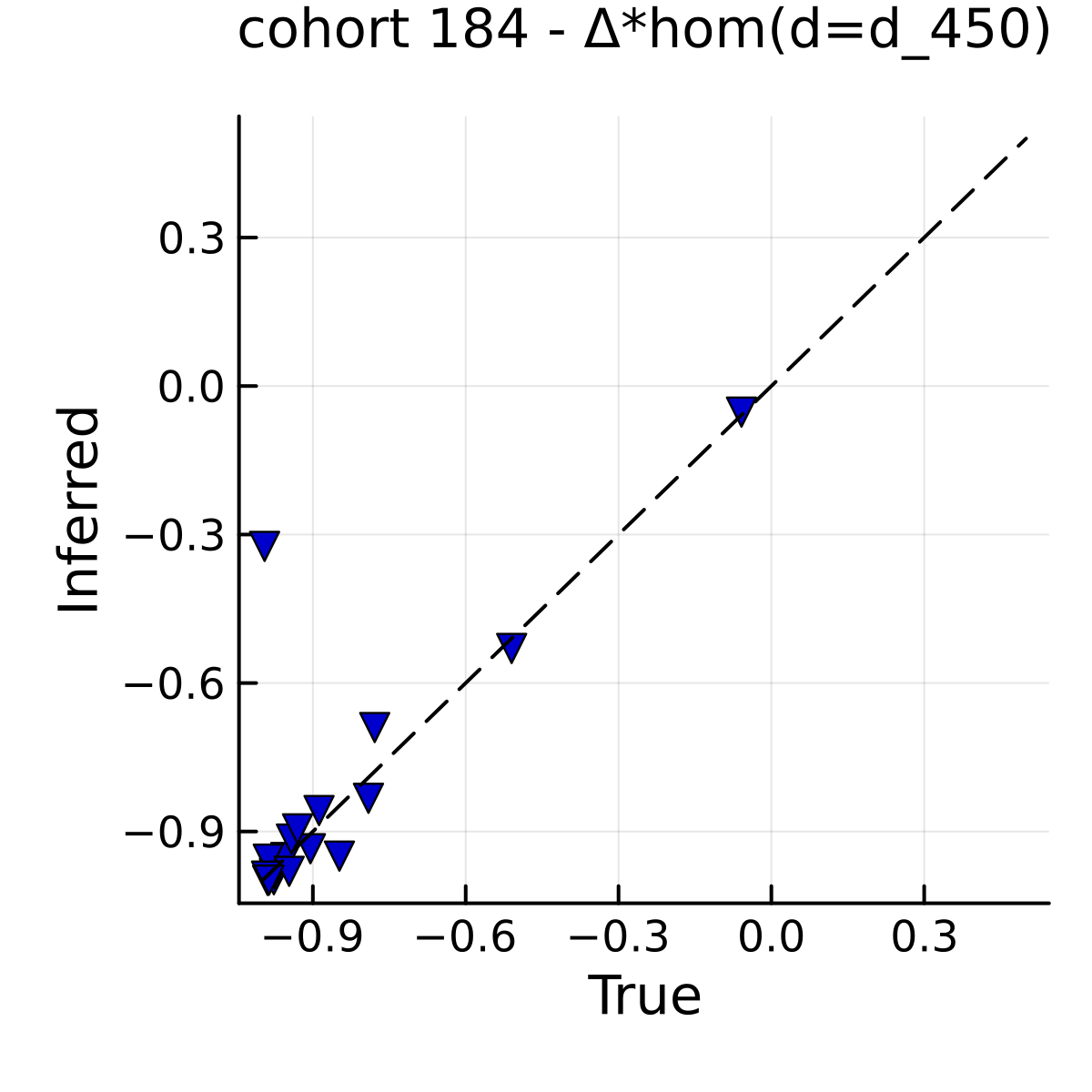}
    \end{subfigure}
    \begin{subfigure}[b]{0.22\textwidth}
        \includegraphics[width=\textwidth]{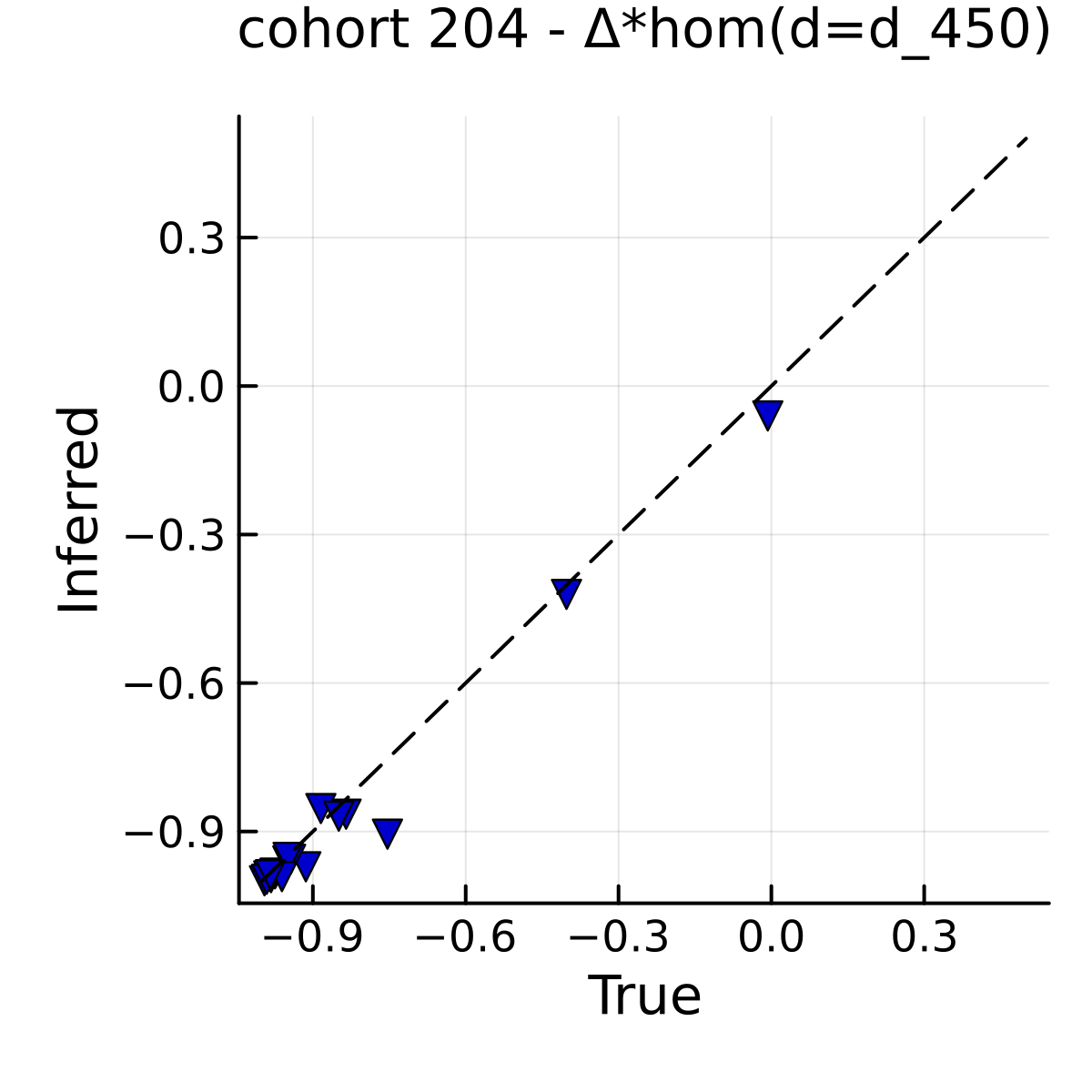}
    \end{subfigure}
    \begin{subfigure}[b]{0.22\textwidth}
        \includegraphics[width=\textwidth]{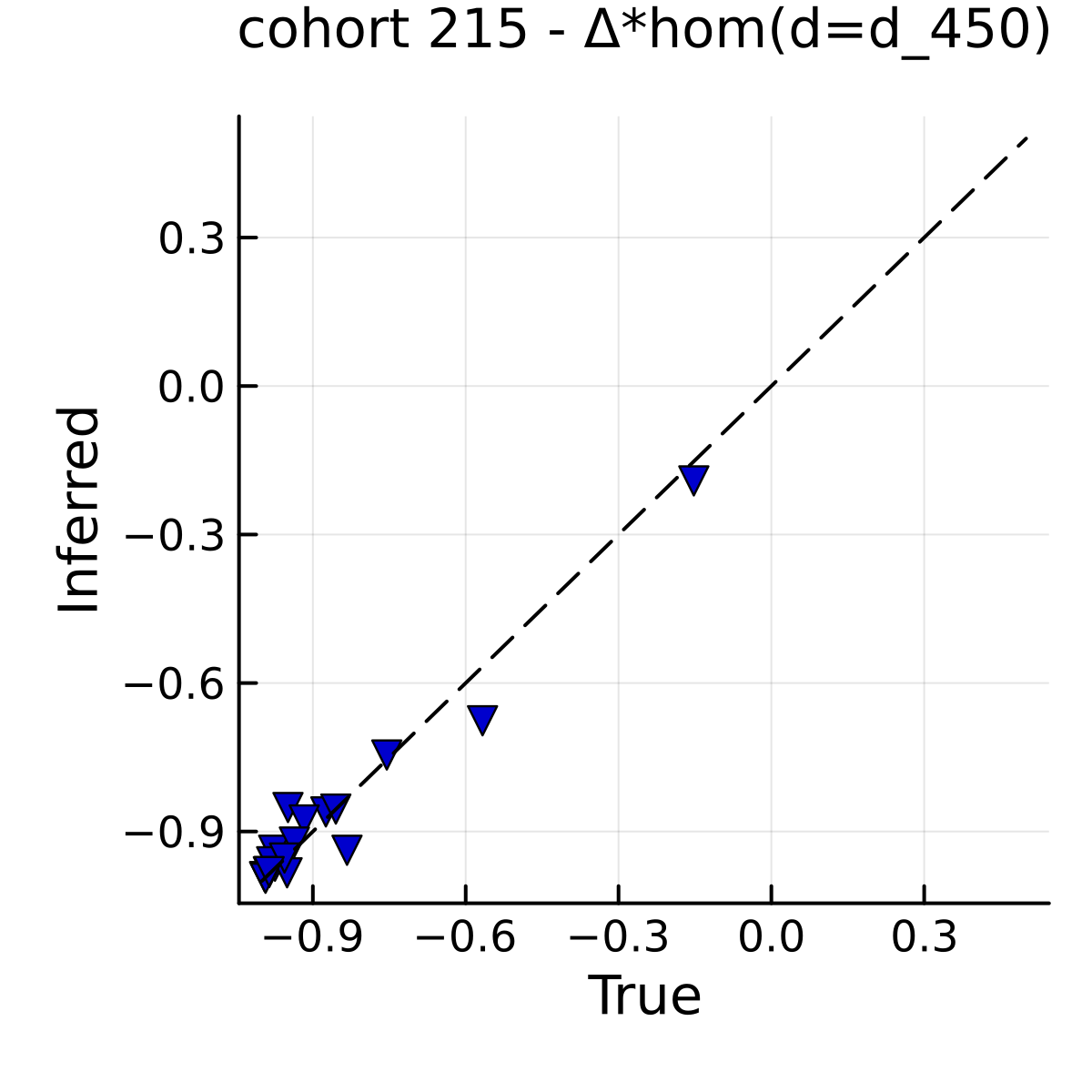}
    \end{subfigure}
    \begin{subfigure}[b]{0.22\textwidth}
        \includegraphics[width=\textwidth]{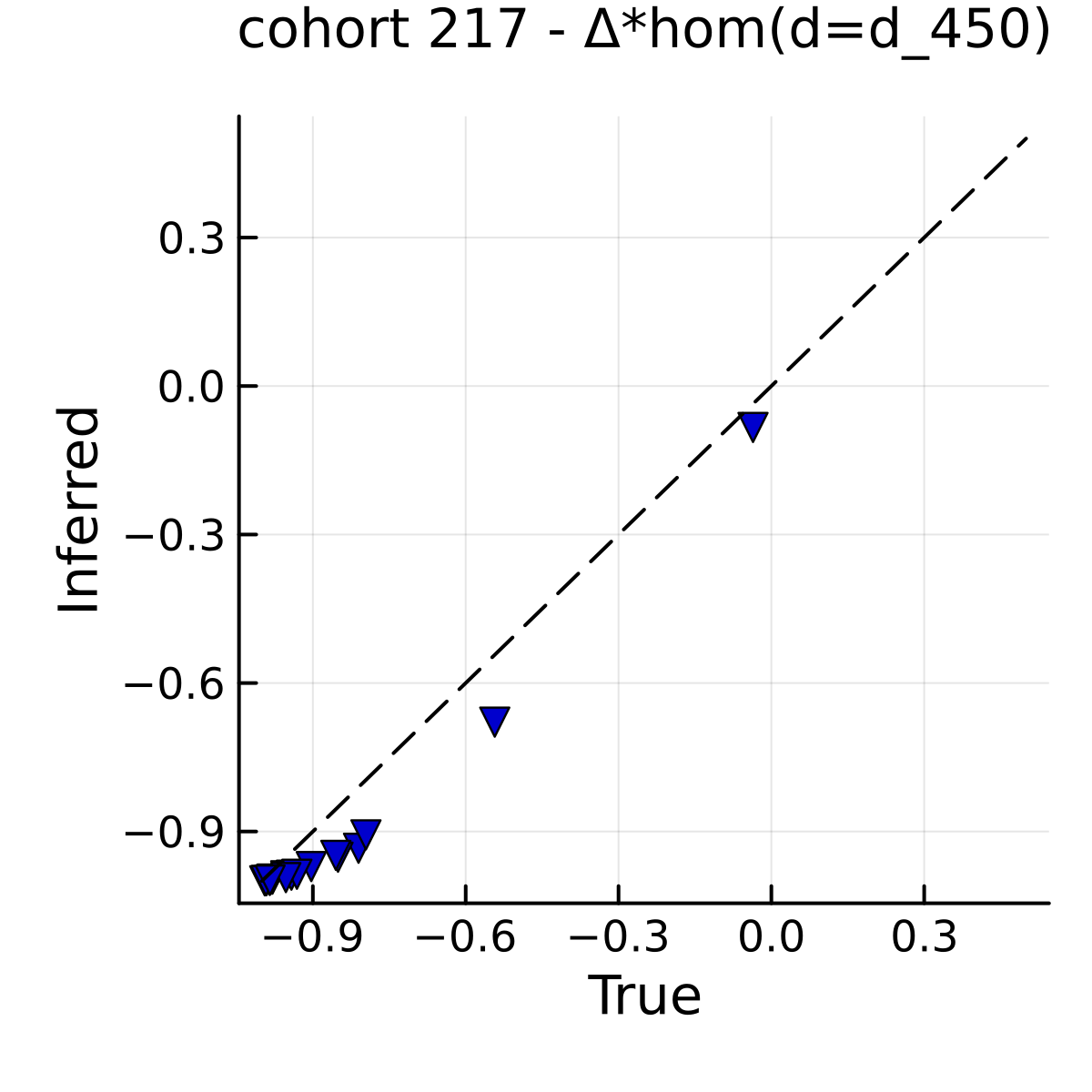}
    \end{subfigure}
    \caption{Comparison between the mean posterior value of $\bar{\Delta}^*_{hom}(d_{450}^{(i)})$ (y-axis) for each virtual patient $i$ and each synthetic cohort $m$, and the true one (x-axis). $d_{450}^{(i)}$ corresponds to the mean dose received by patient $i$ over 450 days of therapy.}
    \label{fig:synth_Delta_hom}
\end{figure}

\begin{figure}[h]
    \centering
    \begin{subfigure}[b]{0.22\textwidth}
        \includegraphics[width=\textwidth]{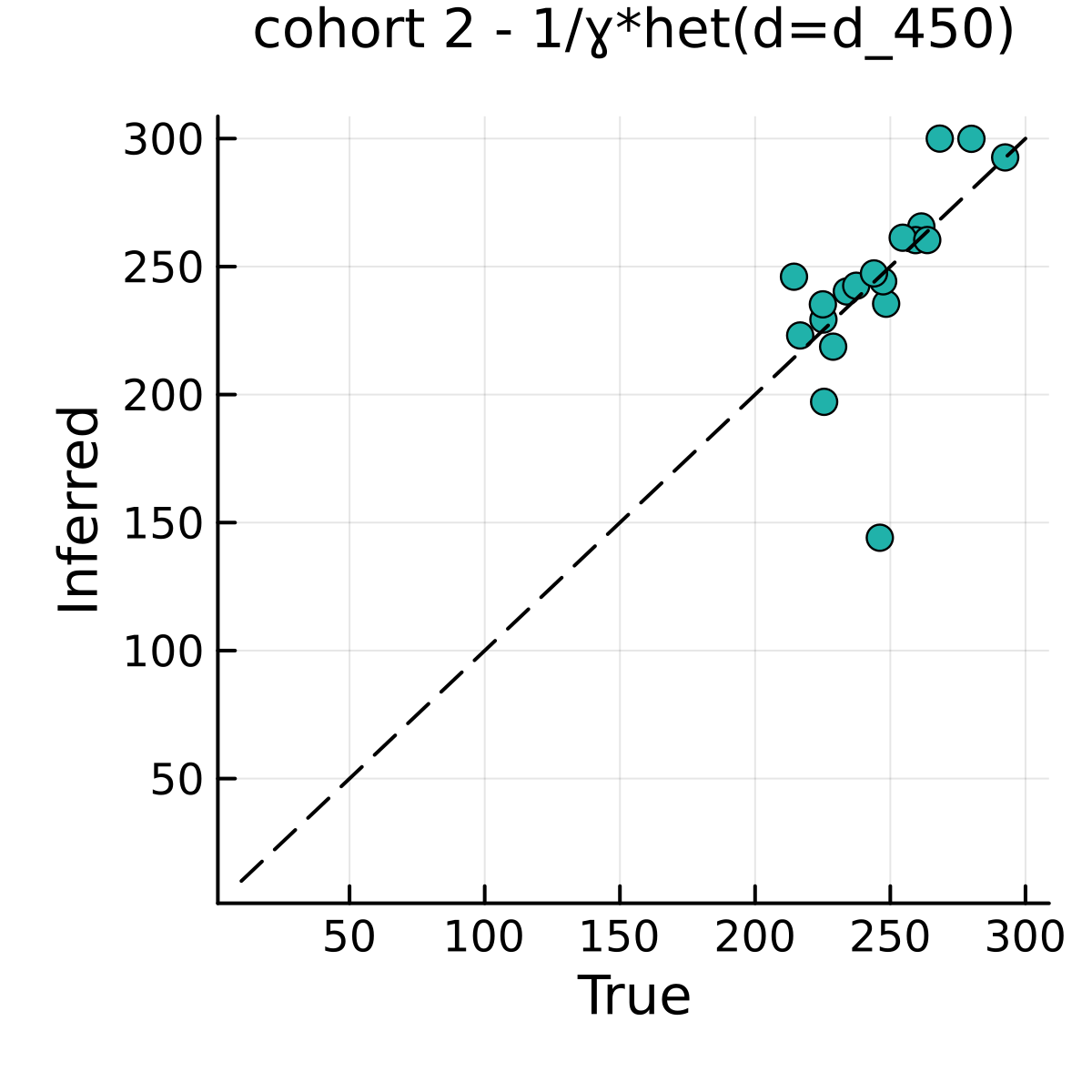}
    \end{subfigure}
    \begin{subfigure}[b]{0.22\textwidth}
        \includegraphics[width=\textwidth]{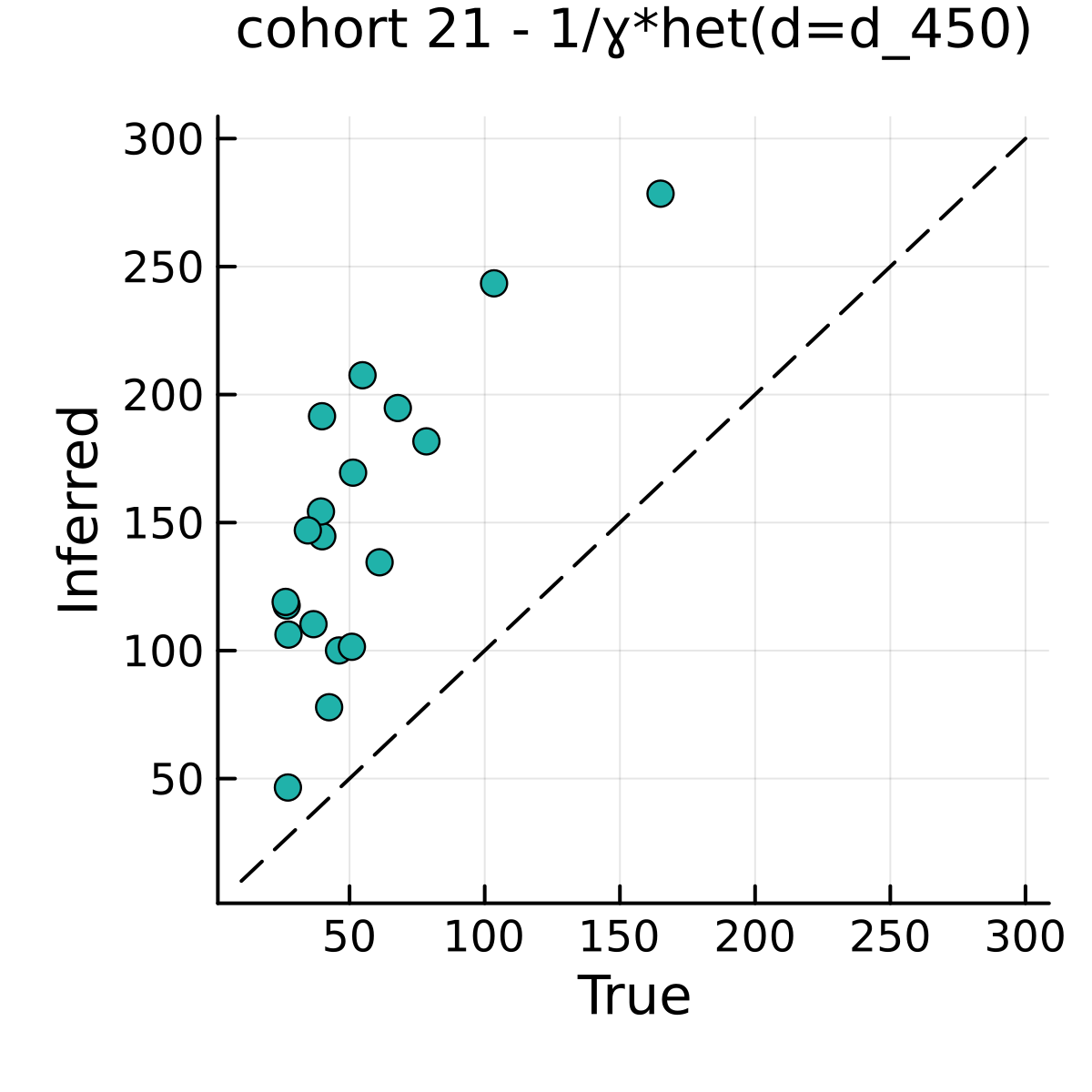}
    \end{subfigure}
    \begin{subfigure}[b]{0.22\textwidth}
        \includegraphics[width=\textwidth]{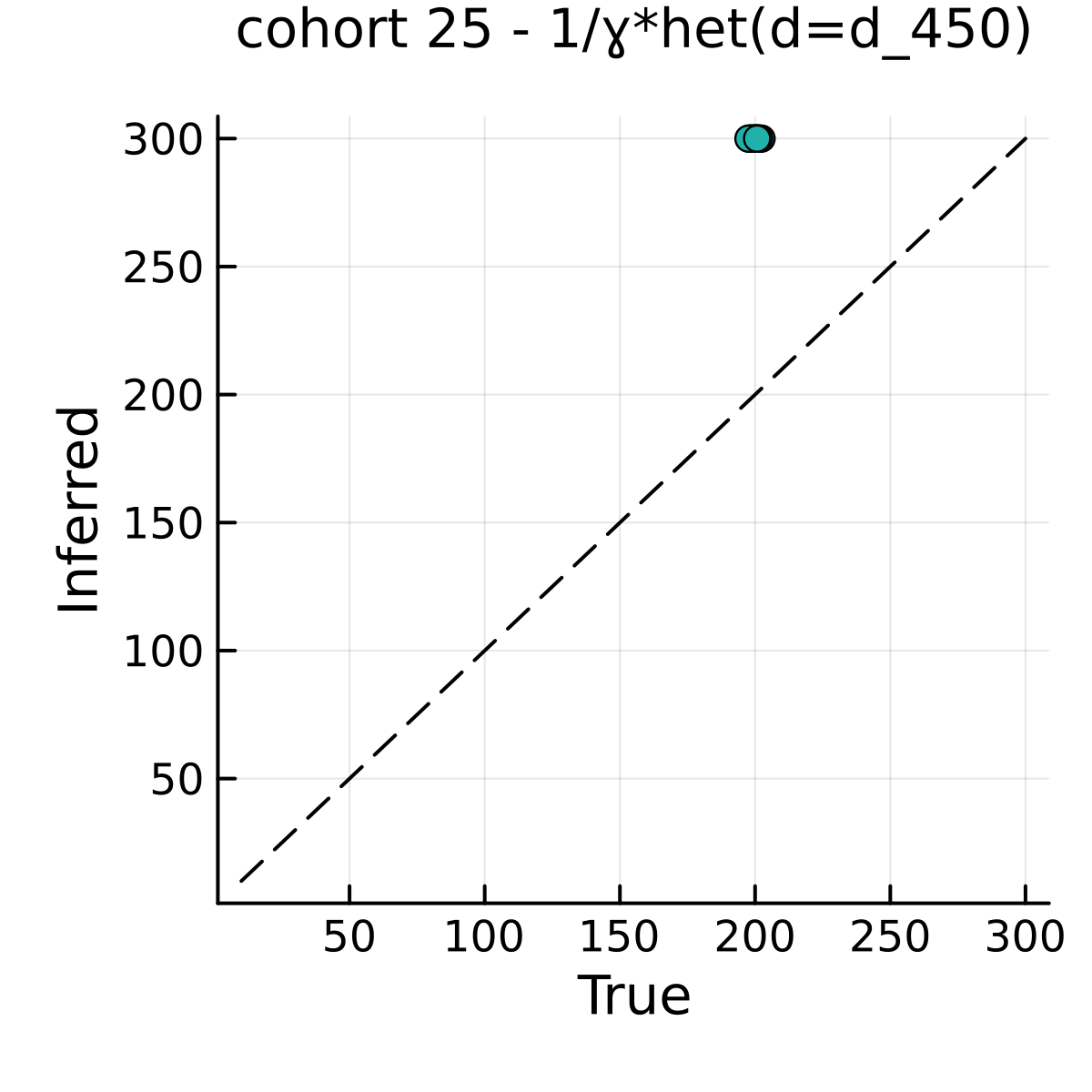}
    \end{subfigure}
    \begin{subfigure}[b]{0.22\textwidth}
        \includegraphics[width=\textwidth]{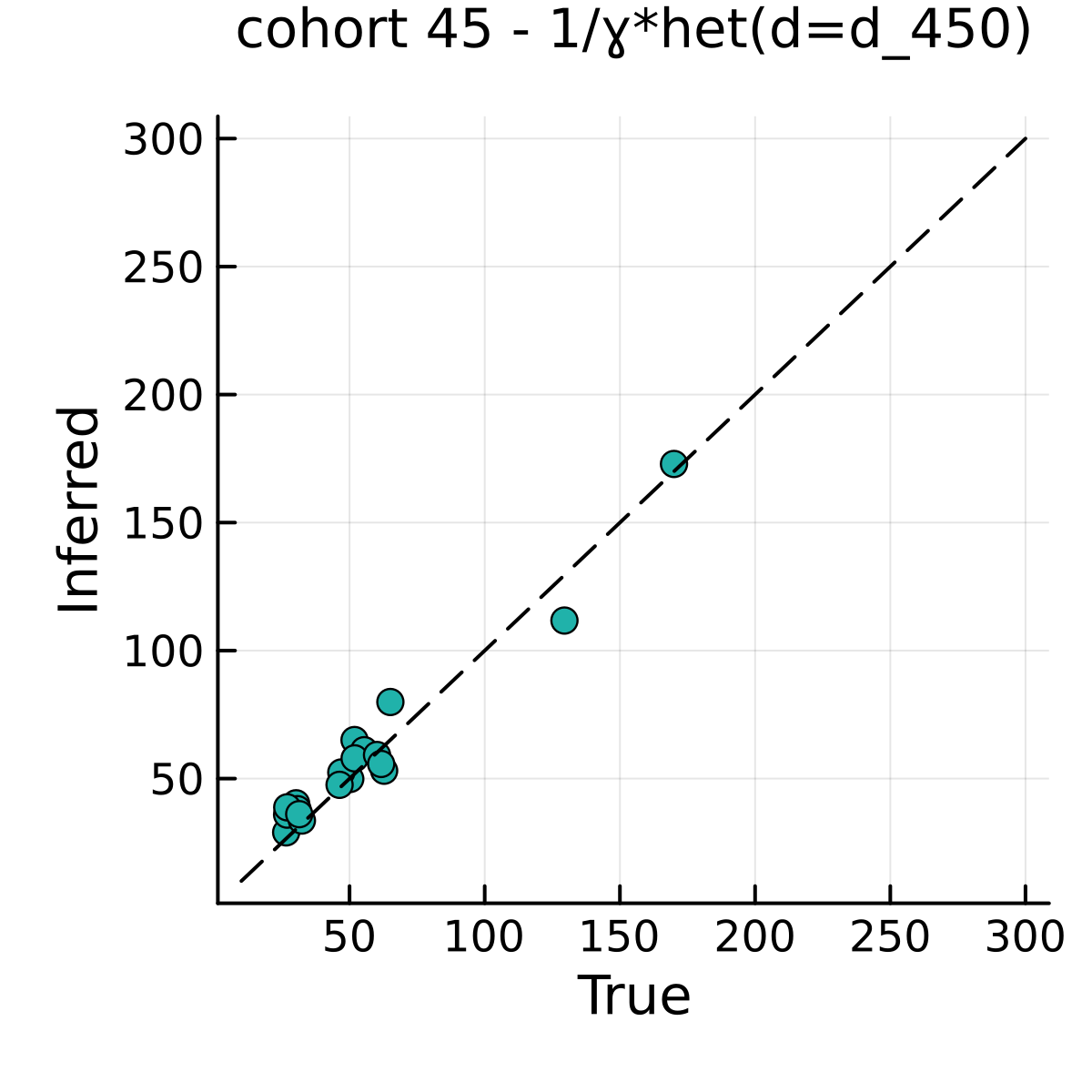}
    \end{subfigure}
    \begin{subfigure}[b]{0.22\textwidth}
        \includegraphics[width=\textwidth]{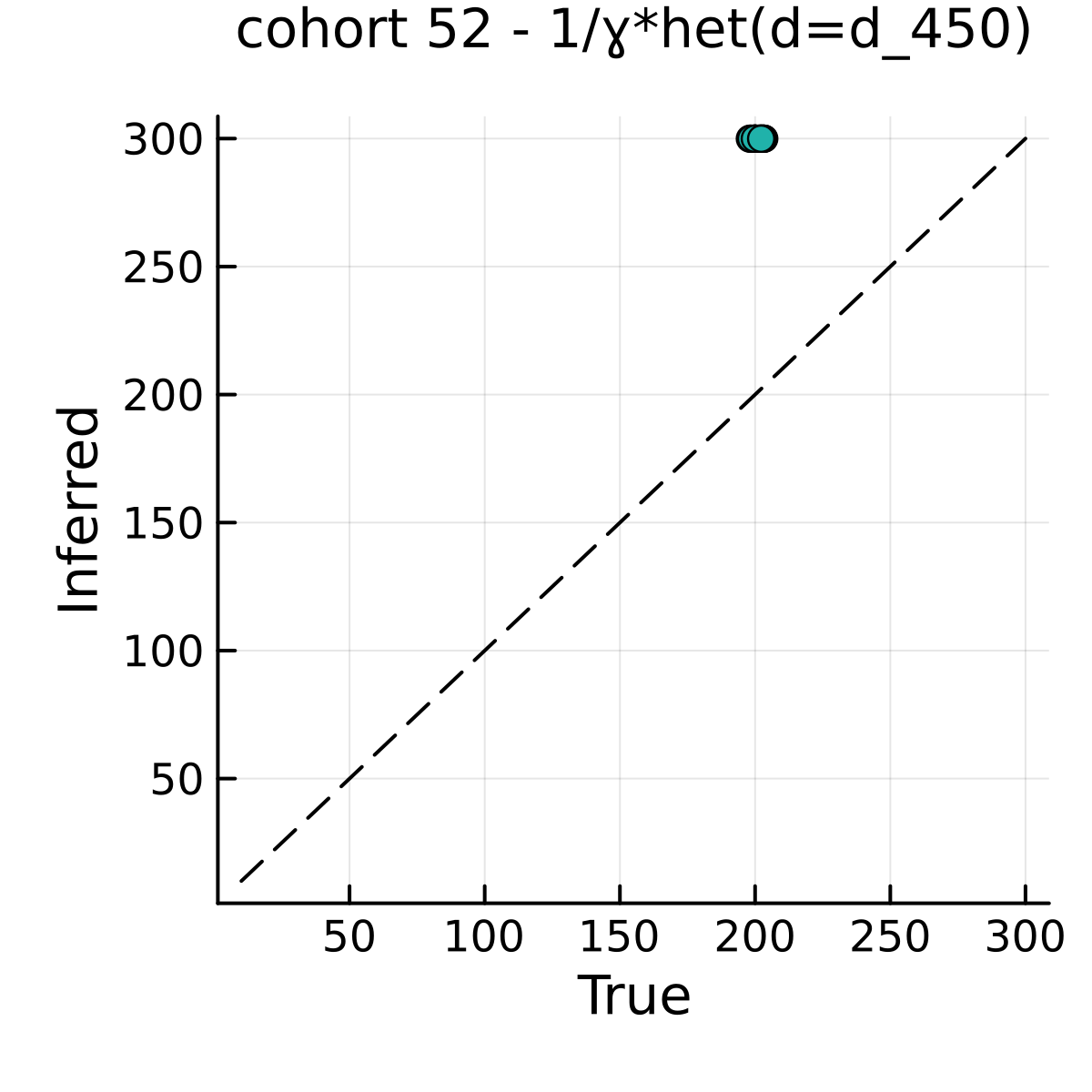}
    \end{subfigure}
    \begin{subfigure}[b]{0.22\textwidth}
        \includegraphics[width=\textwidth]{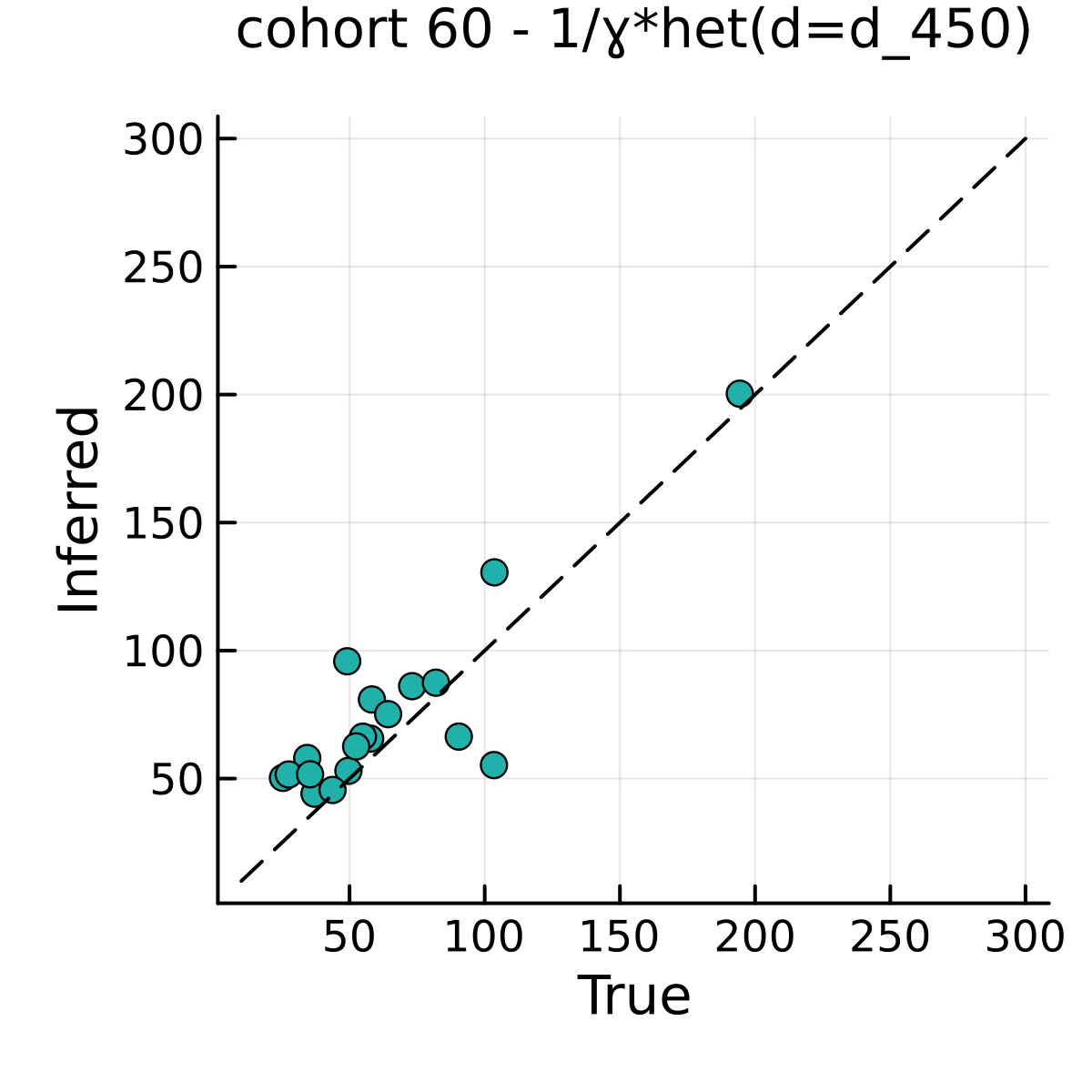}
    \end{subfigure}
    \begin{subfigure}[b]{0.22\textwidth}
        \includegraphics[width=\textwidth]{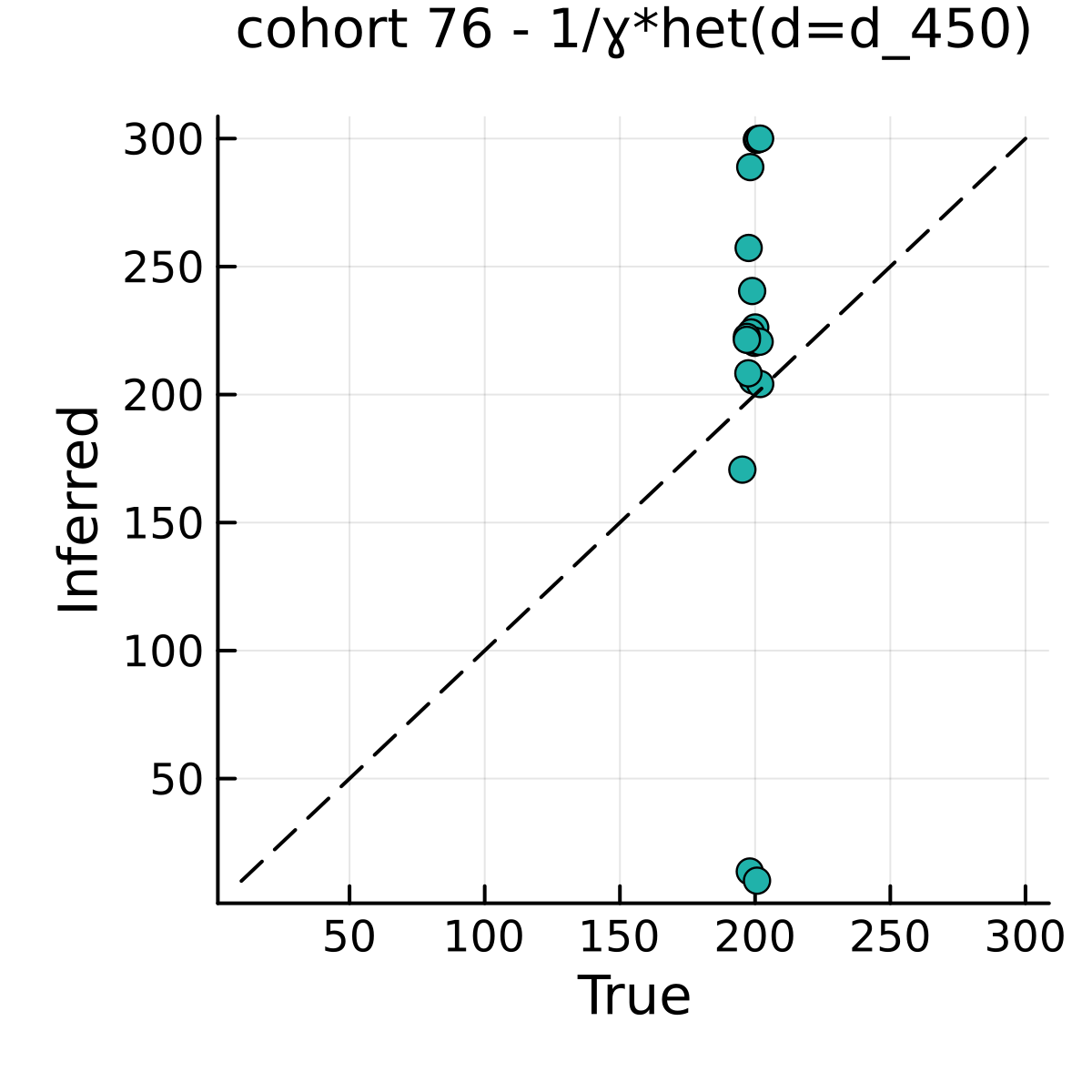}
    \end{subfigure}
    \begin{subfigure}[b]{0.22\textwidth}
        \includegraphics[width=\textwidth]{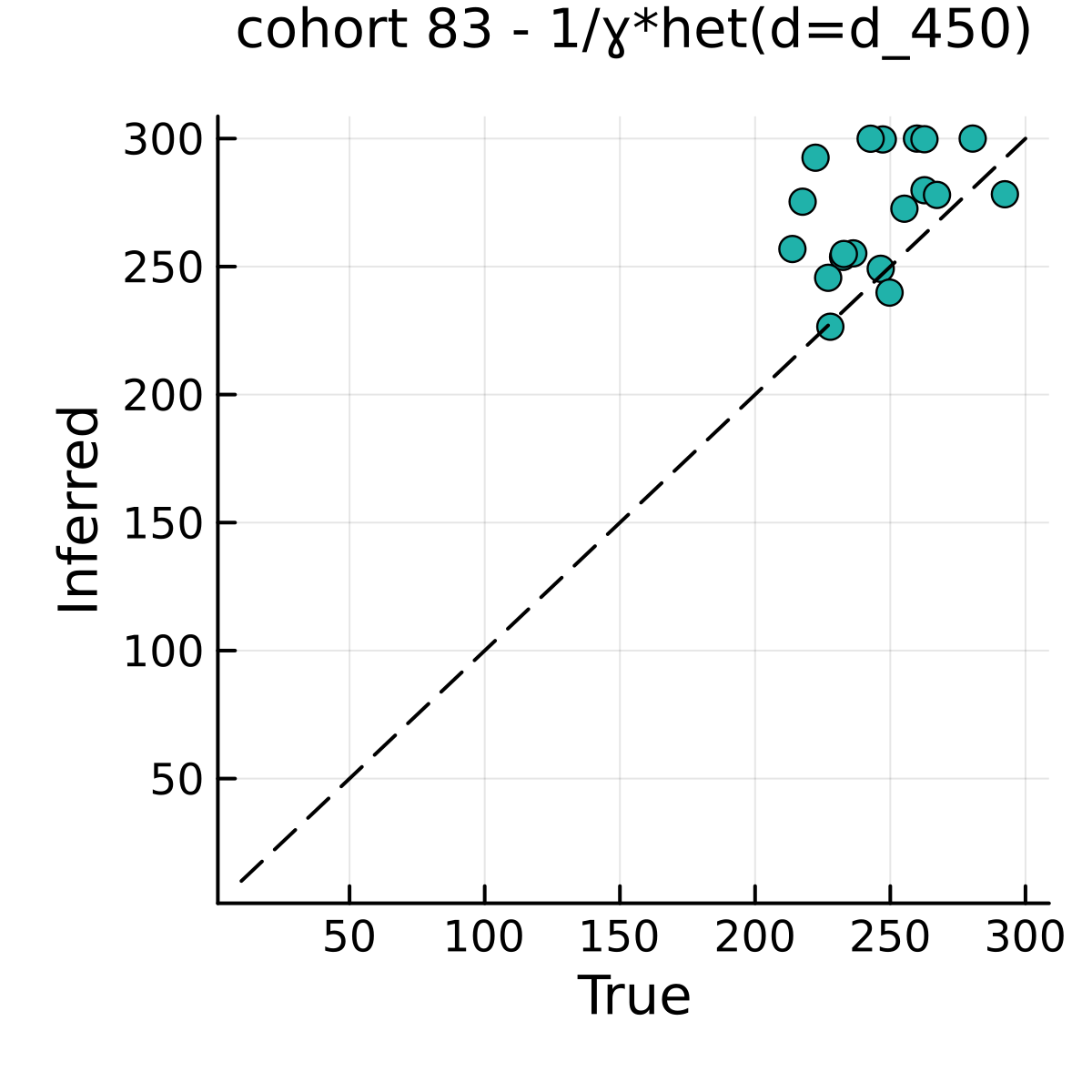}
    \end{subfigure}
    \begin{subfigure}[b]{0.22\textwidth}
        \includegraphics[width=\textwidth]{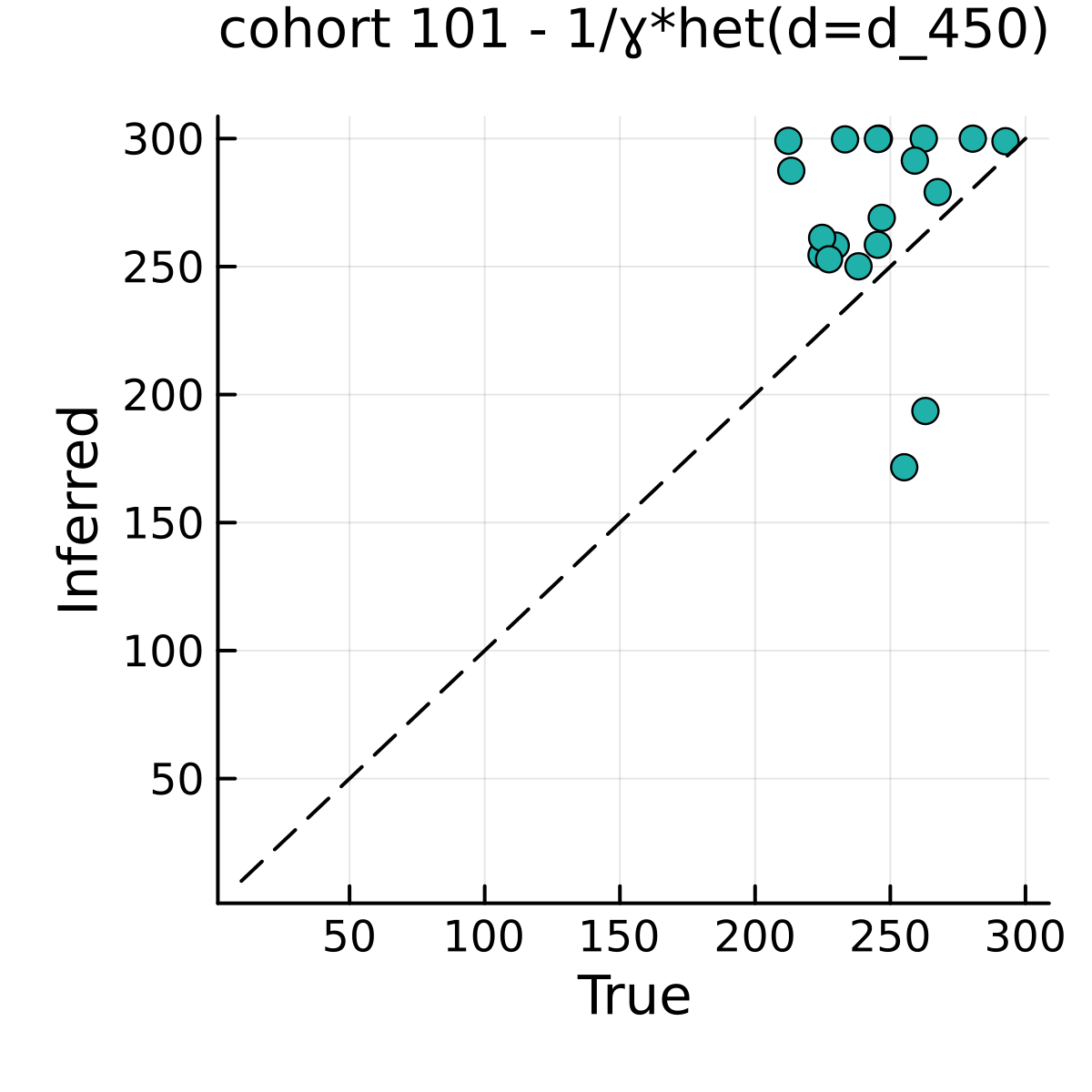}
    \end{subfigure}
    \begin{subfigure}[b]{0.22\textwidth}
        \includegraphics[width=\textwidth]{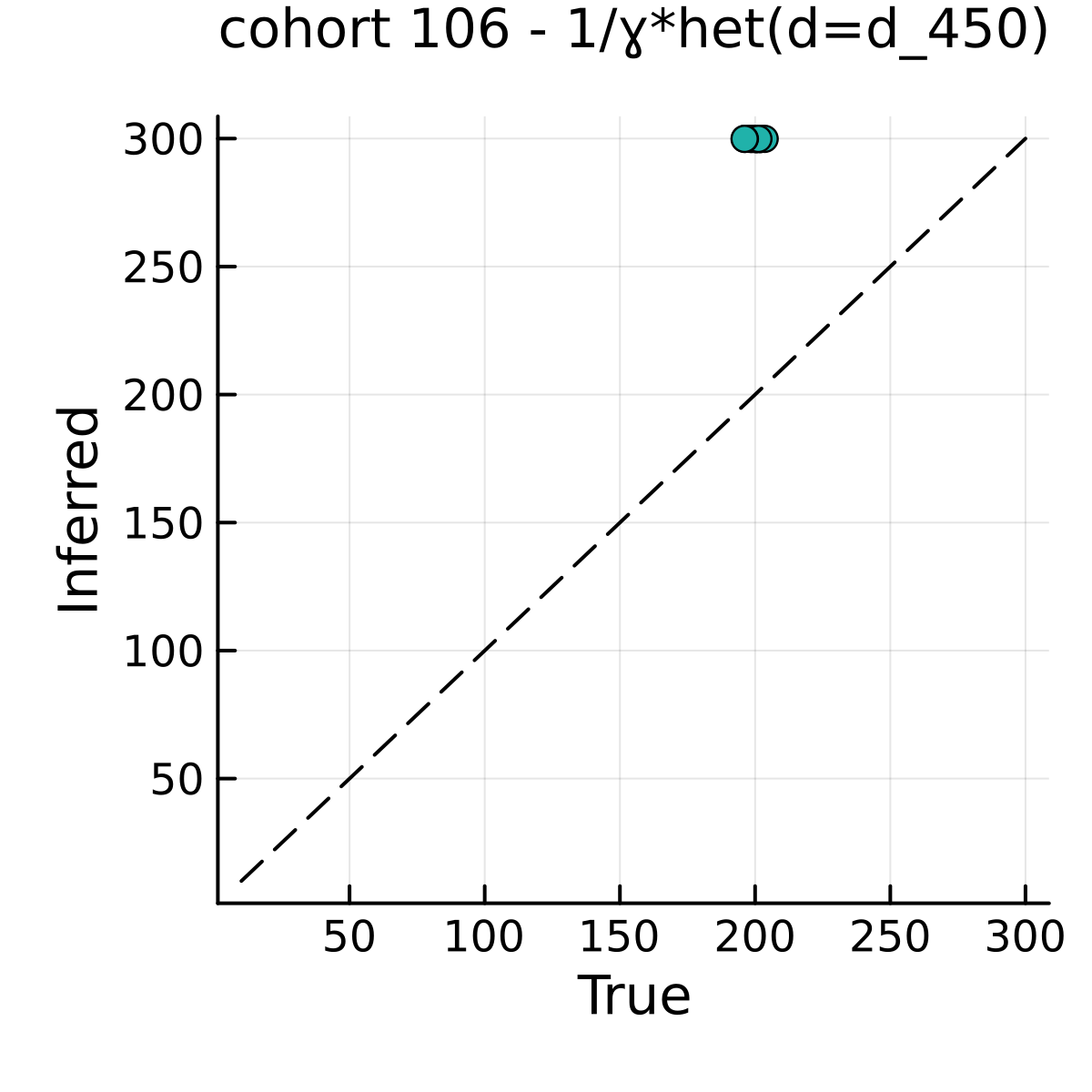}
    \end{subfigure}
    \begin{subfigure}[b]{0.22\textwidth}
        \includegraphics[width=\textwidth]{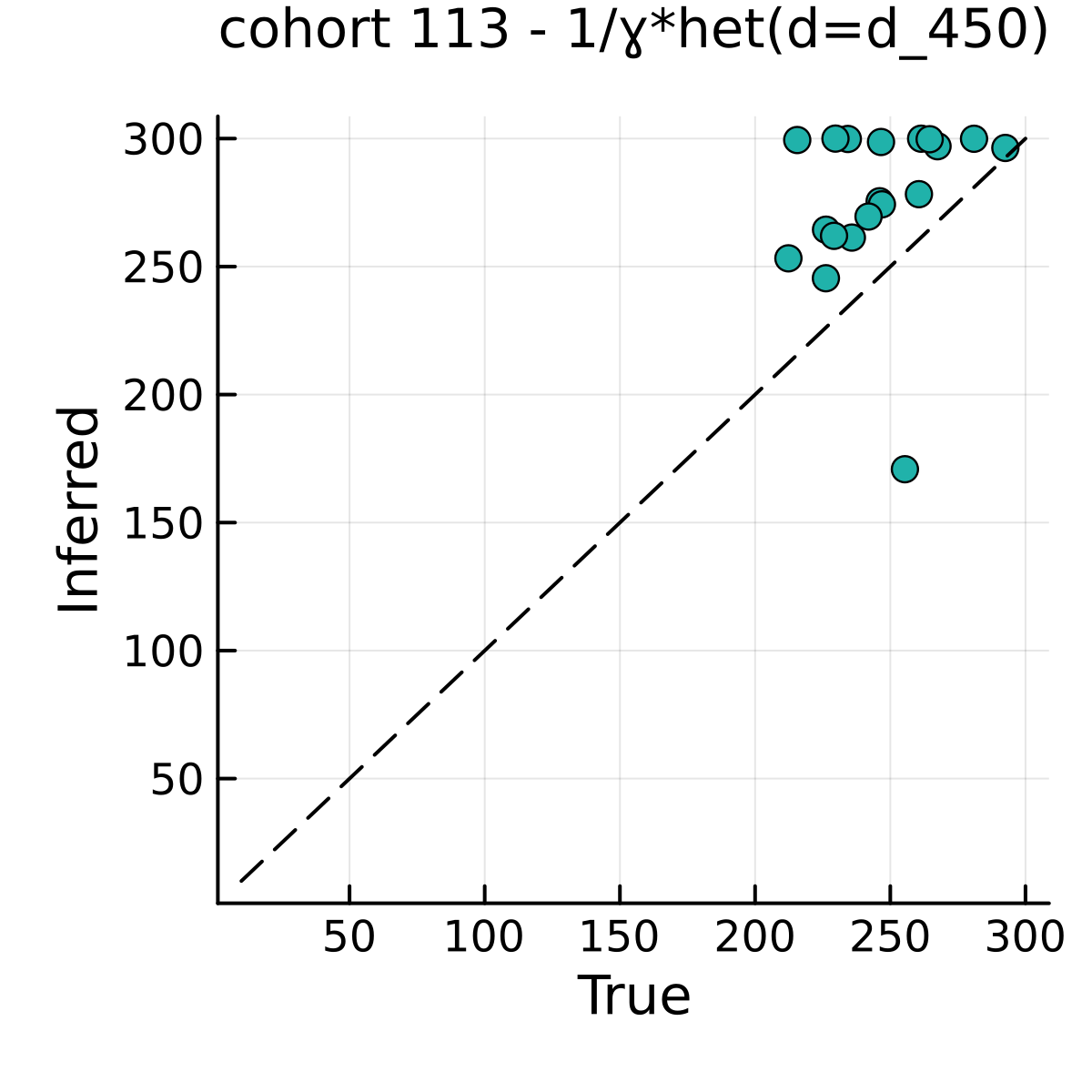}
    \end{subfigure}
    \begin{subfigure}[b]{0.22\textwidth}
        \includegraphics[width=\textwidth]{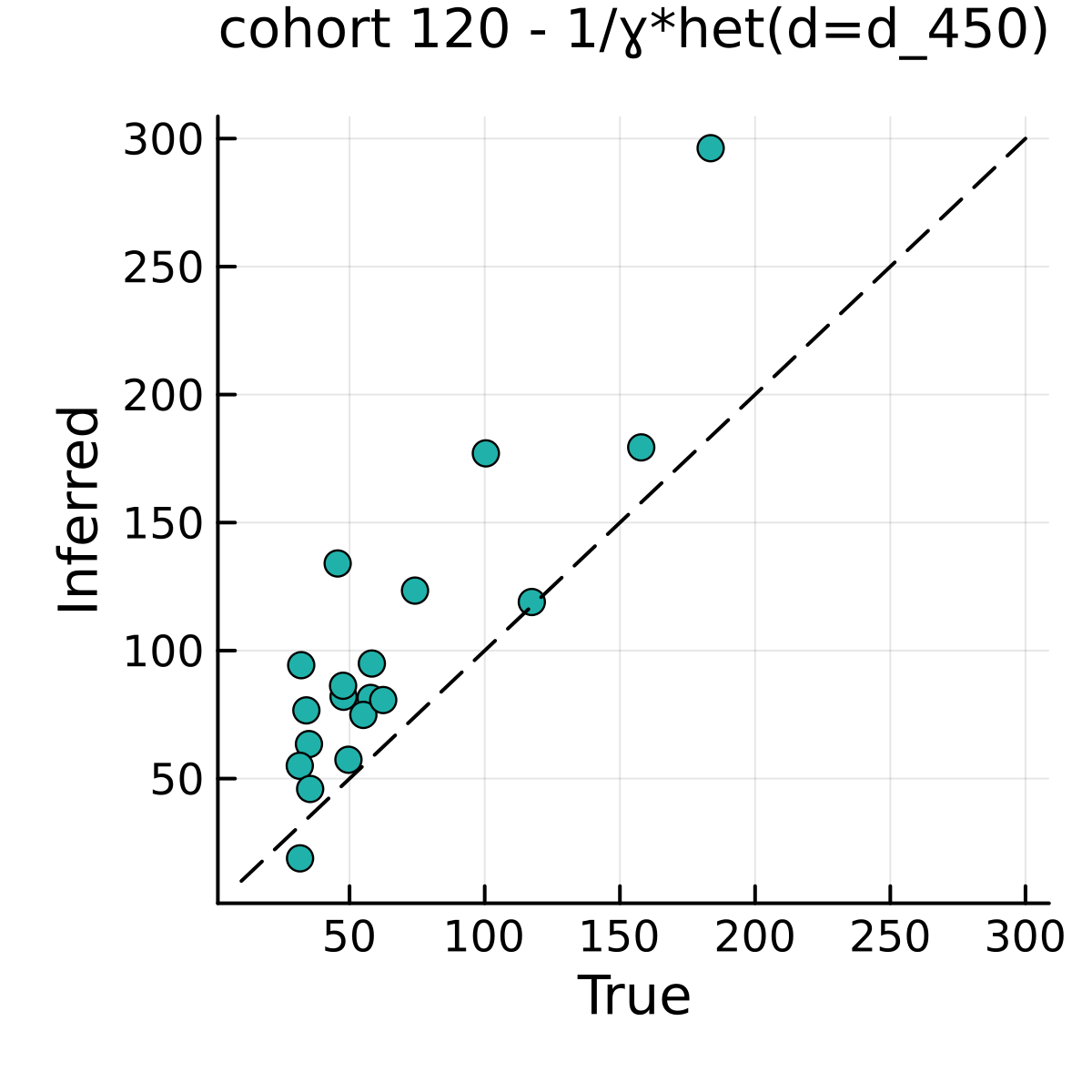}
    \end{subfigure}
    \begin{subfigure}[b]{0.22\textwidth}
        \includegraphics[width=\textwidth]{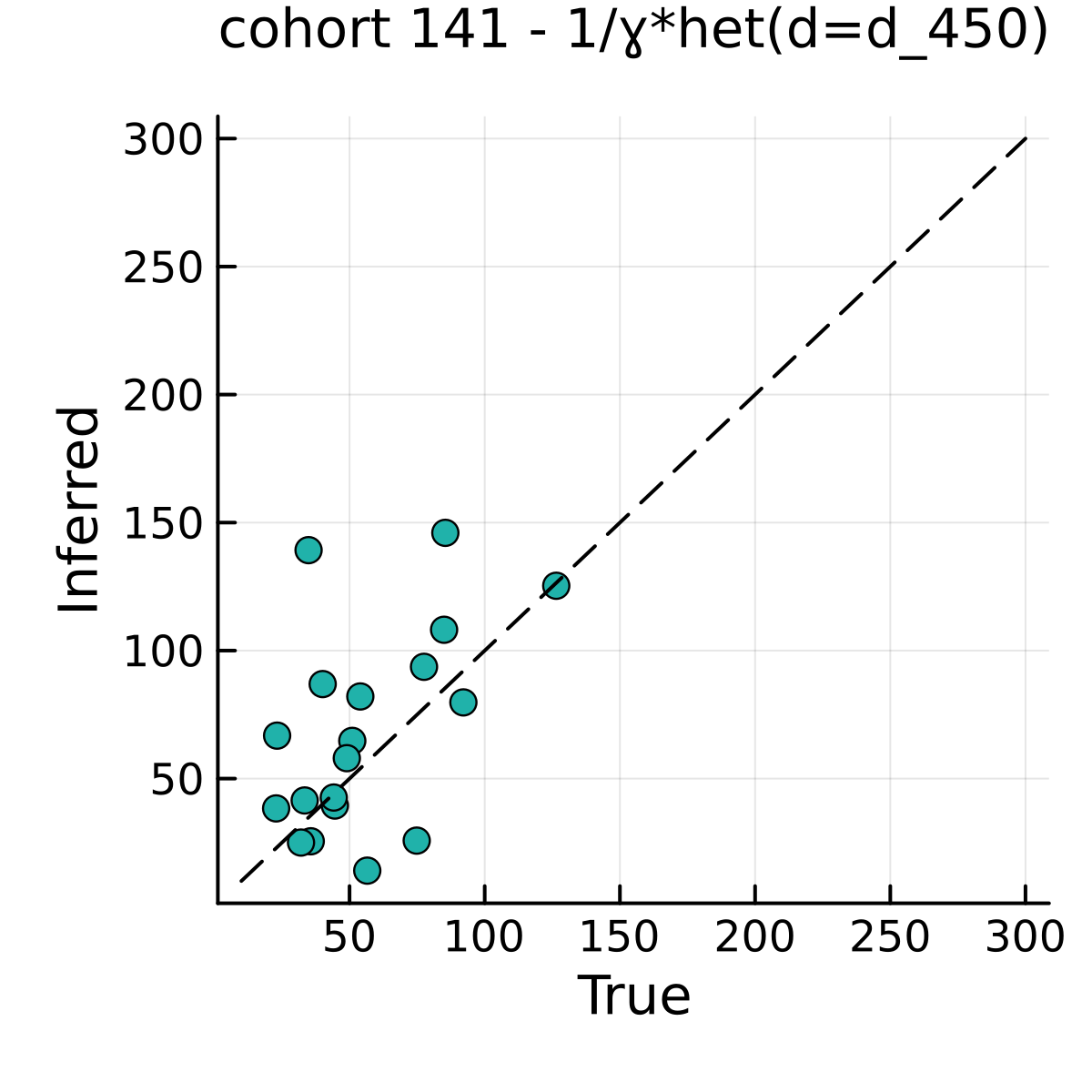}
    \end{subfigure}
    \begin{subfigure}[b]{0.22\textwidth}
        \includegraphics[width=\textwidth]{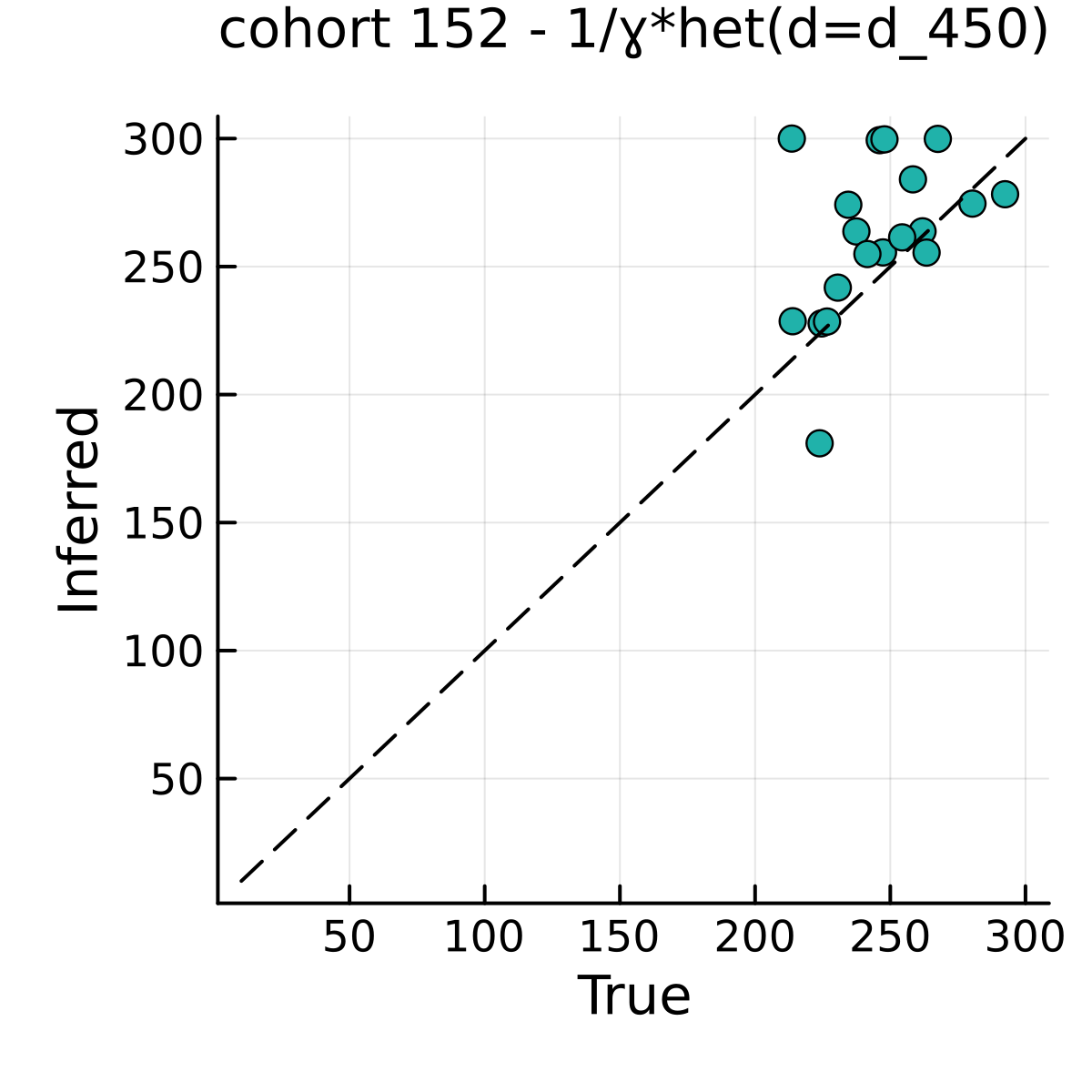}
    \end{subfigure}
    \begin{subfigure}[b]{0.22\textwidth}
        \includegraphics[width=\textwidth]{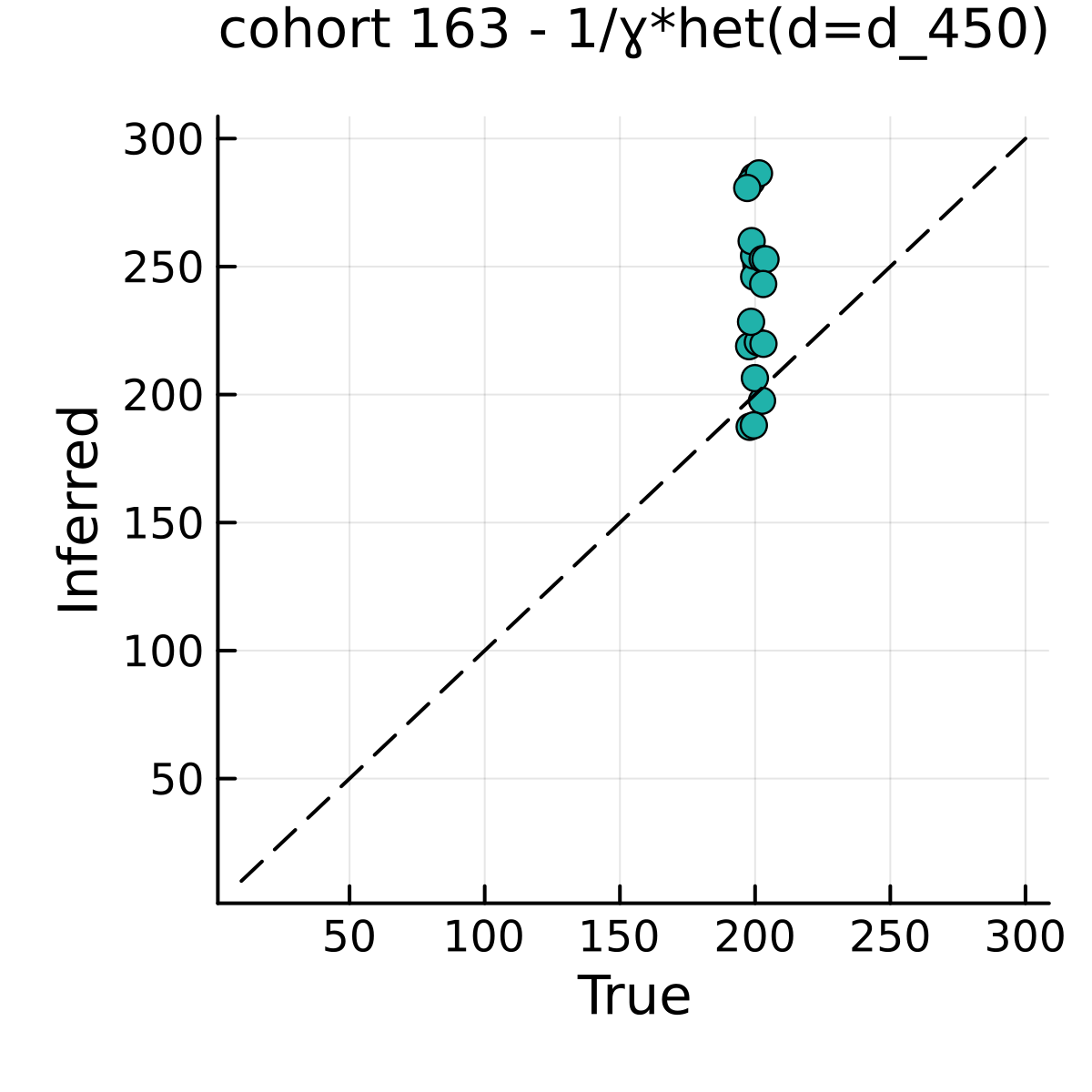}
    \end{subfigure}
    \begin{subfigure}[b]{0.22\textwidth}
        \includegraphics[width=\textwidth]{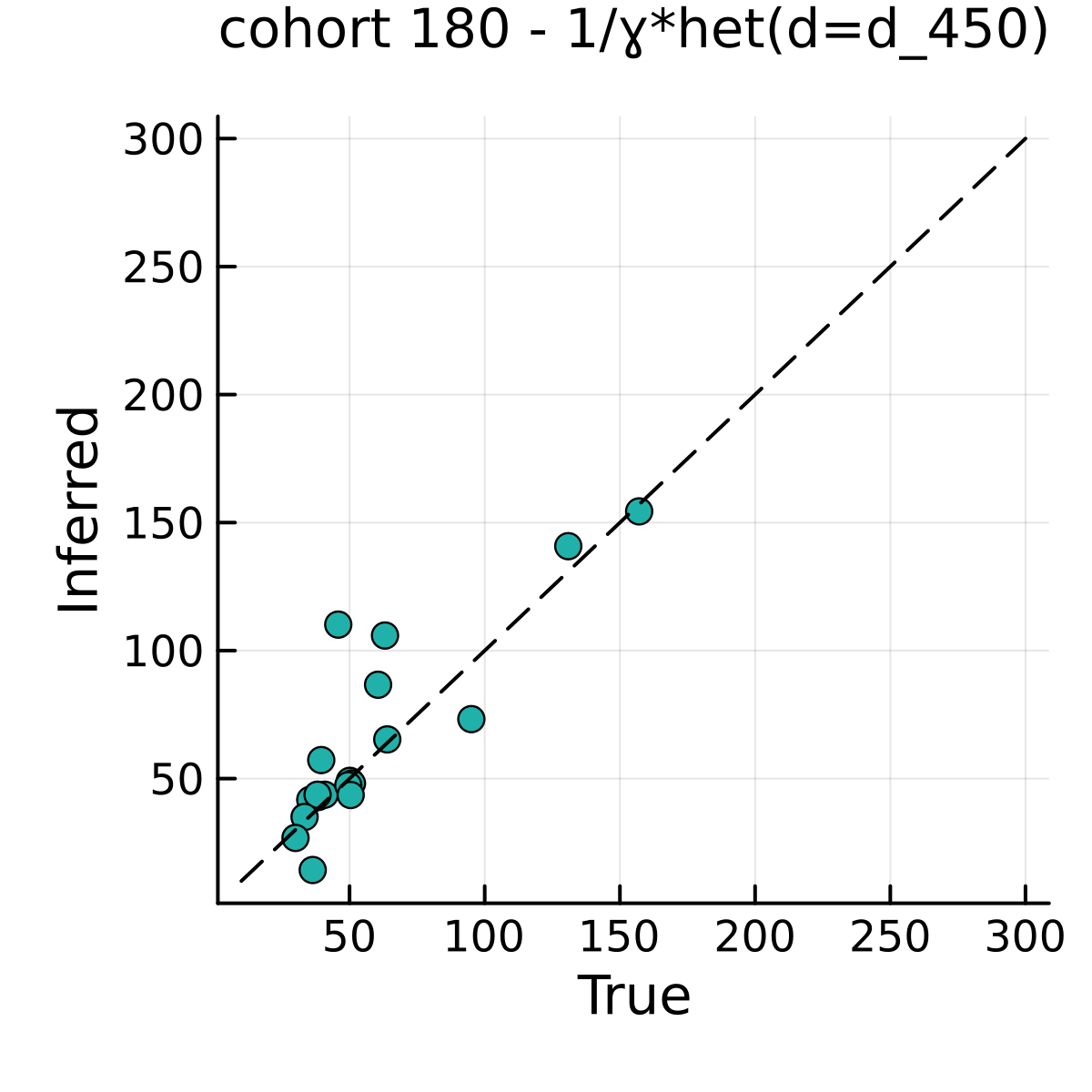}
    \end{subfigure}
    \begin{subfigure}[b]{0.22\textwidth}
        \includegraphics[width=\textwidth]{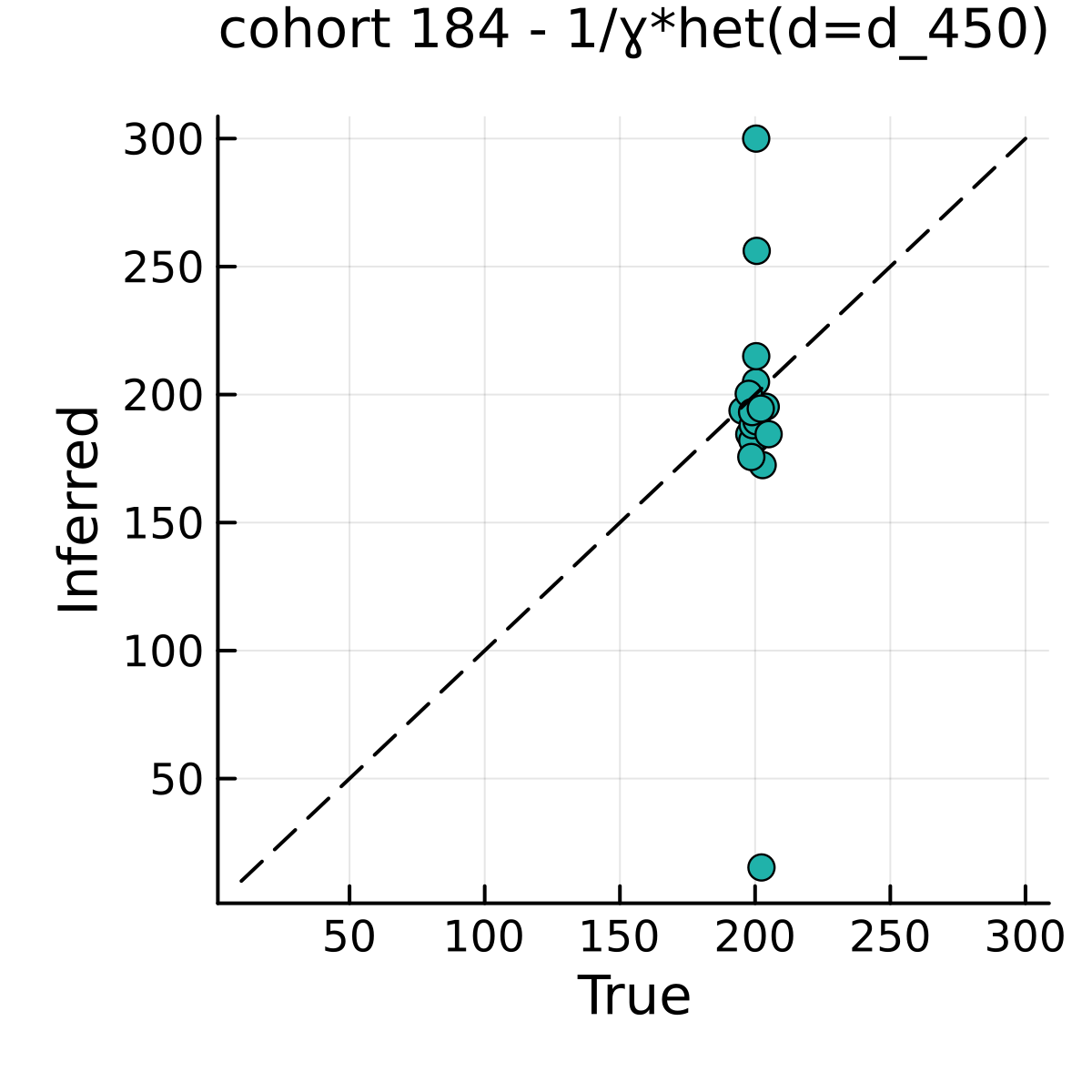}
    \end{subfigure}
    \begin{subfigure}[b]{0.22\textwidth}
        \includegraphics[width=\textwidth]{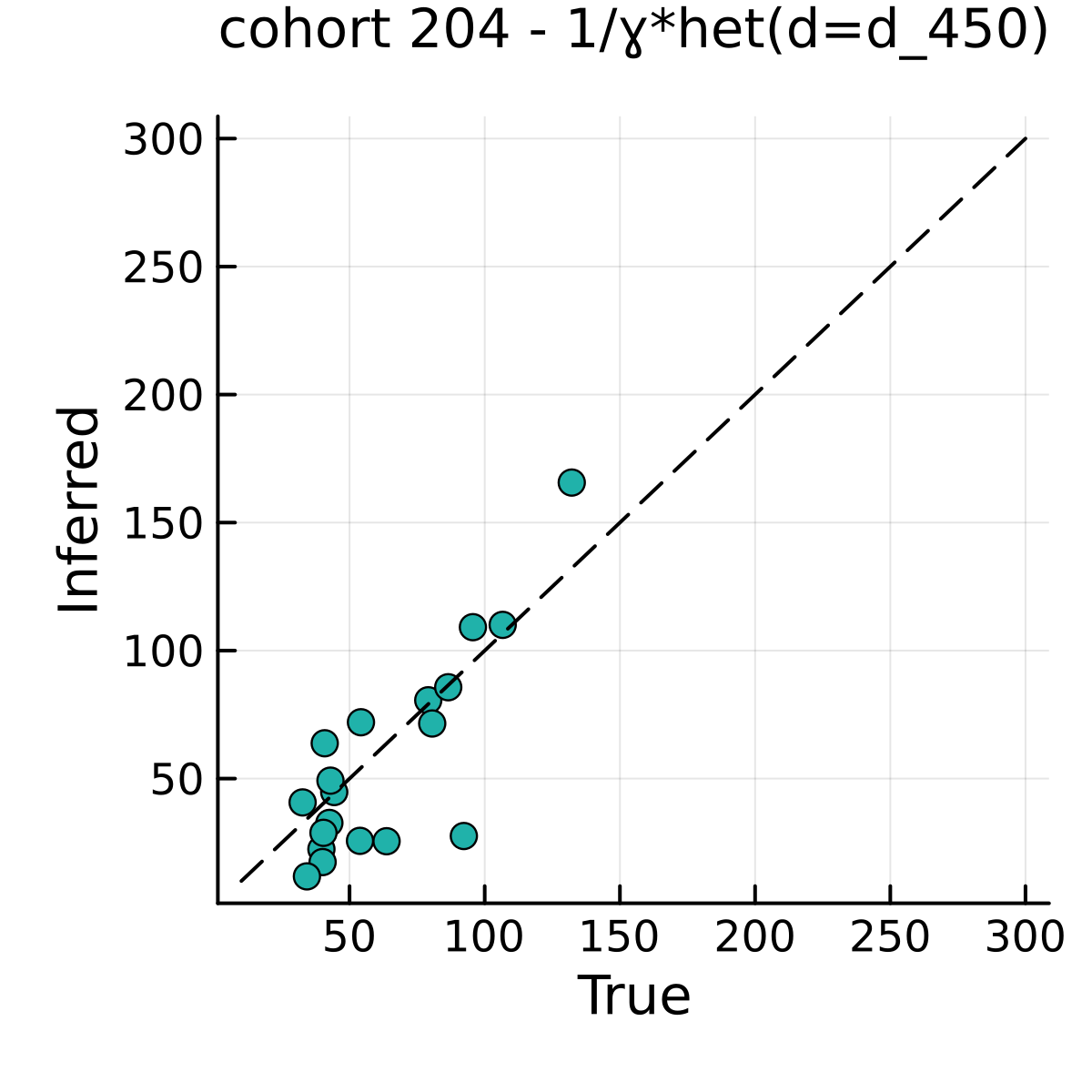}
    \end{subfigure}
    \begin{subfigure}[b]{0.22\textwidth}
        \includegraphics[width=\textwidth]{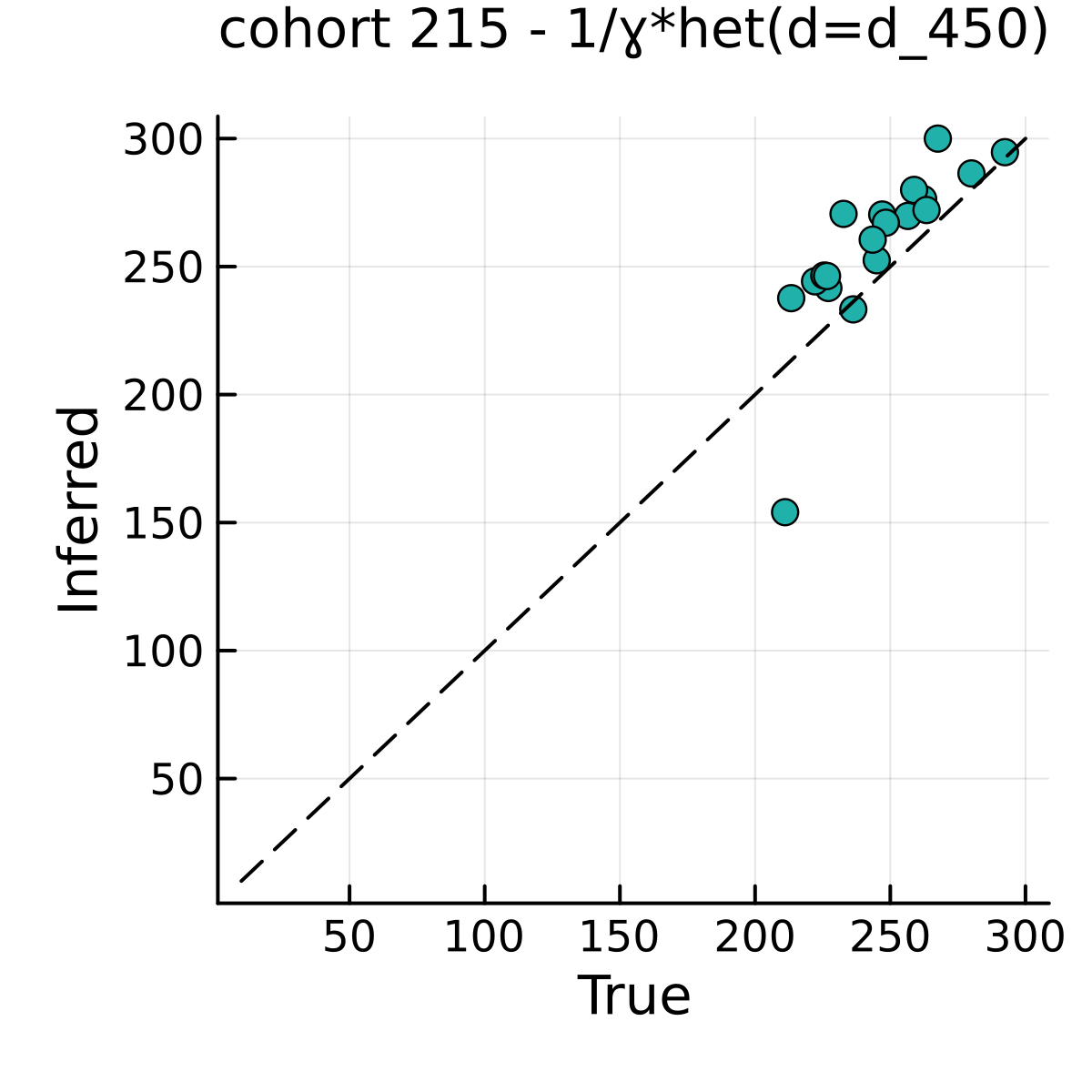}
    \end{subfigure}
    \begin{subfigure}[b]{0.22\textwidth}
        \includegraphics[width=\textwidth]{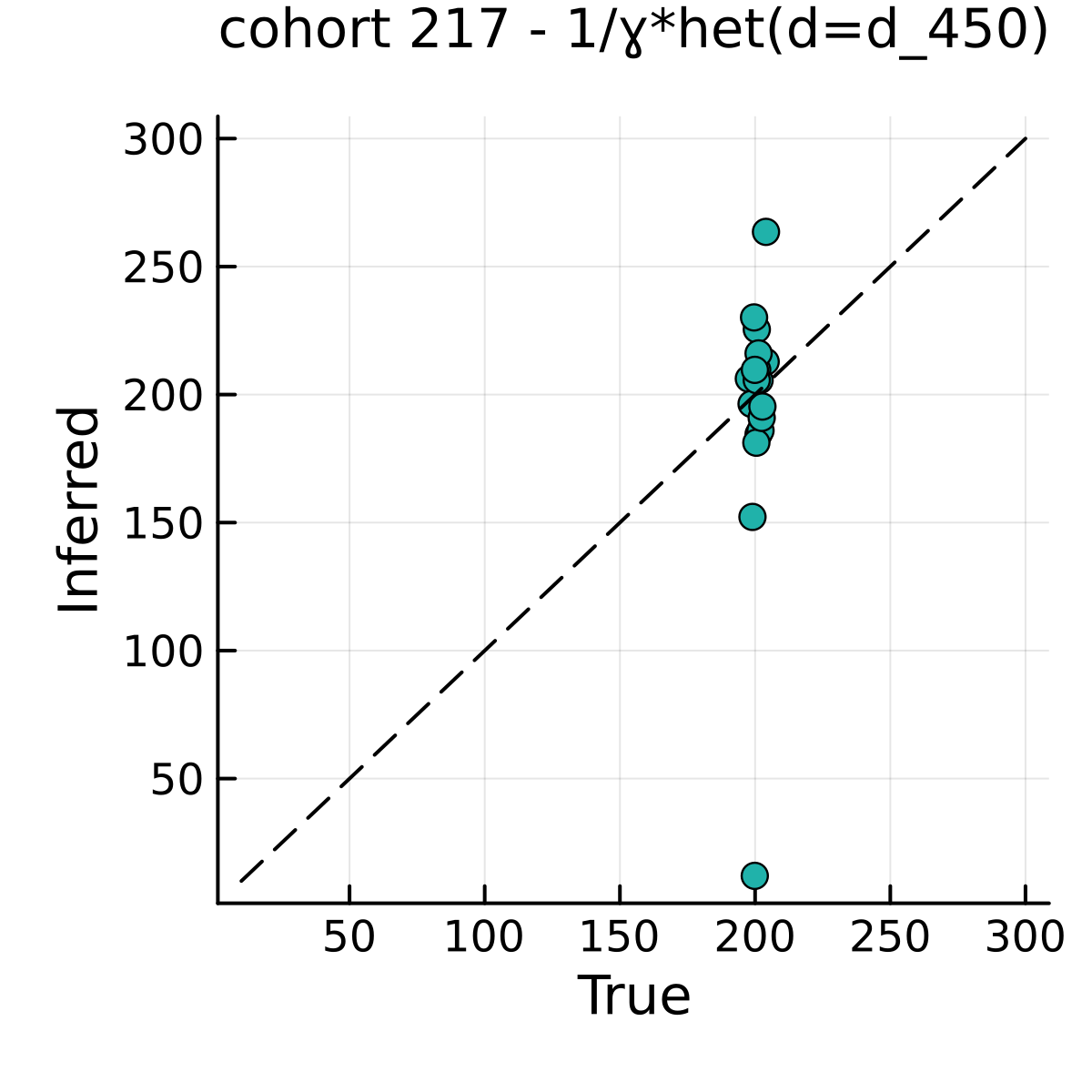}
    \end{subfigure}
    \caption{Comparison between the mean posterior value of $1/\bar{\gamma}^*_{het}(d_{450}^{(i)})$ (y-axis) for each virtual patient $i$ and each synthetic cohort $m$, and the true one (x-axis). $d_{450}^{(i)}$ corresponds to the mean dose received by patient $i$ over 450 days of therapy.}
    \label{fig:synth_gamma_het}
\end{figure}

\begin{figure}[h]
    \centering
    \begin{subfigure}[b]{0.22\textwidth}
        \includegraphics[width=\textwidth]{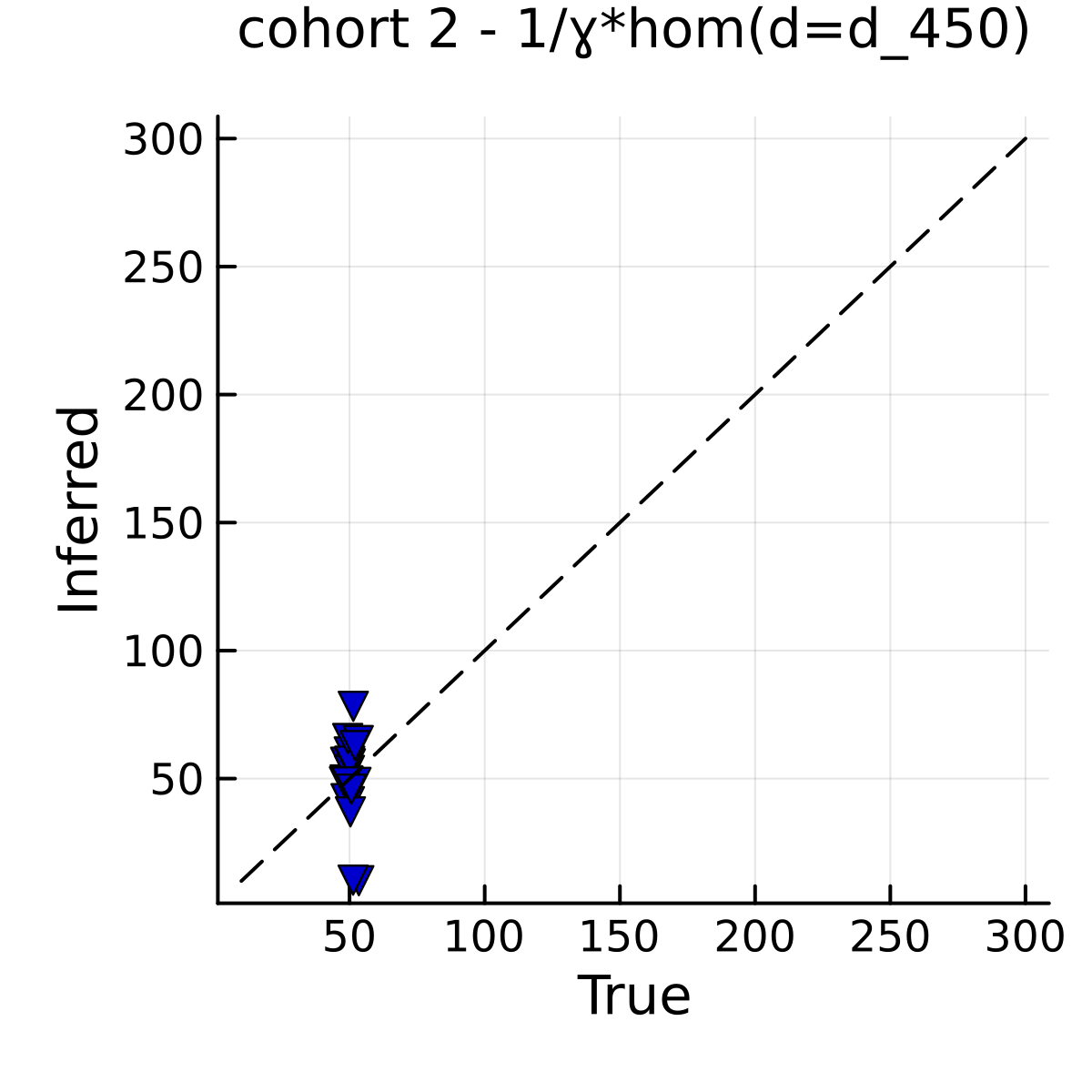}
    \end{subfigure}
    \begin{subfigure}[b]{0.22\textwidth}
        \includegraphics[width=\textwidth]{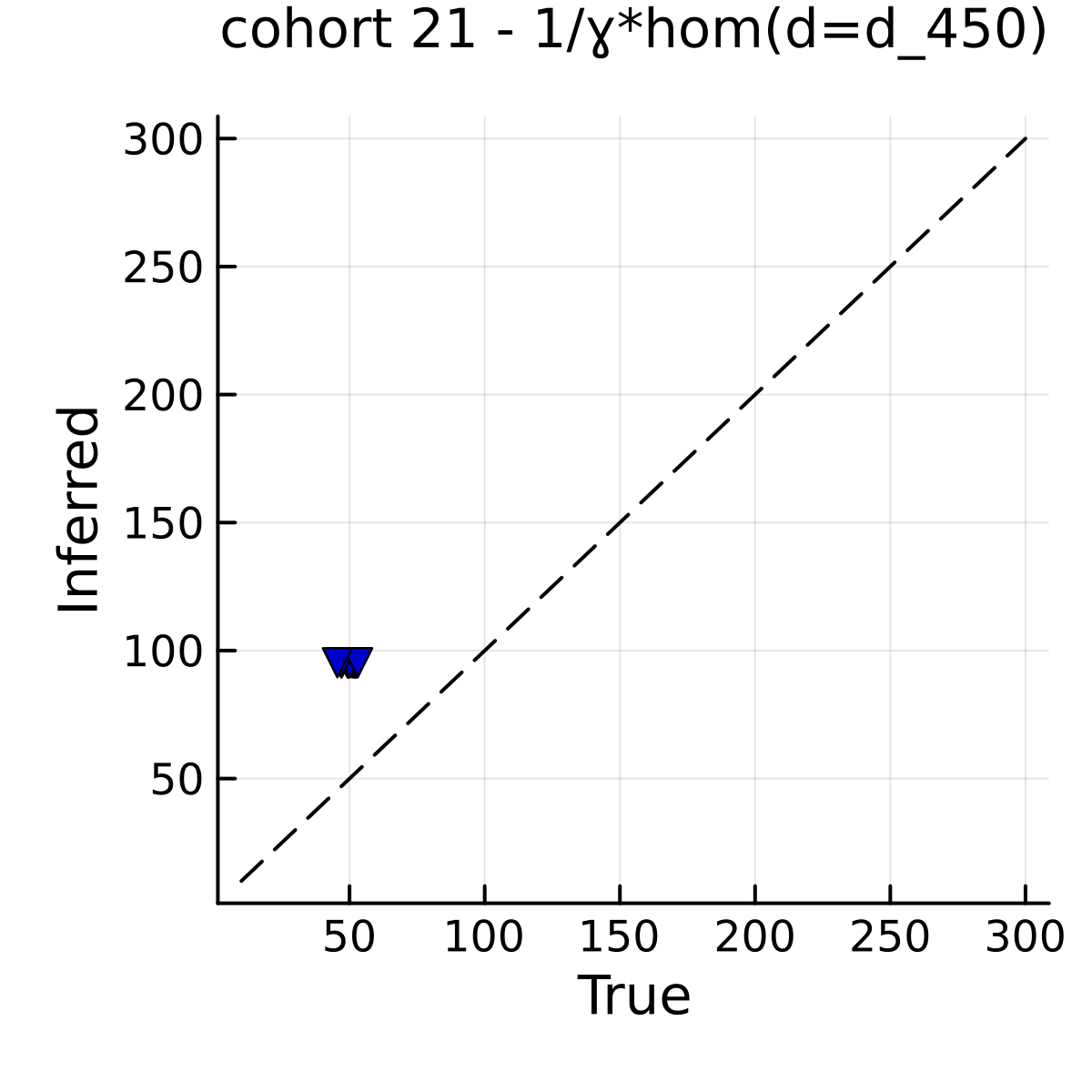}
    \end{subfigure}
    \begin{subfigure}[b]{0.22\textwidth}
        \includegraphics[width=\textwidth]{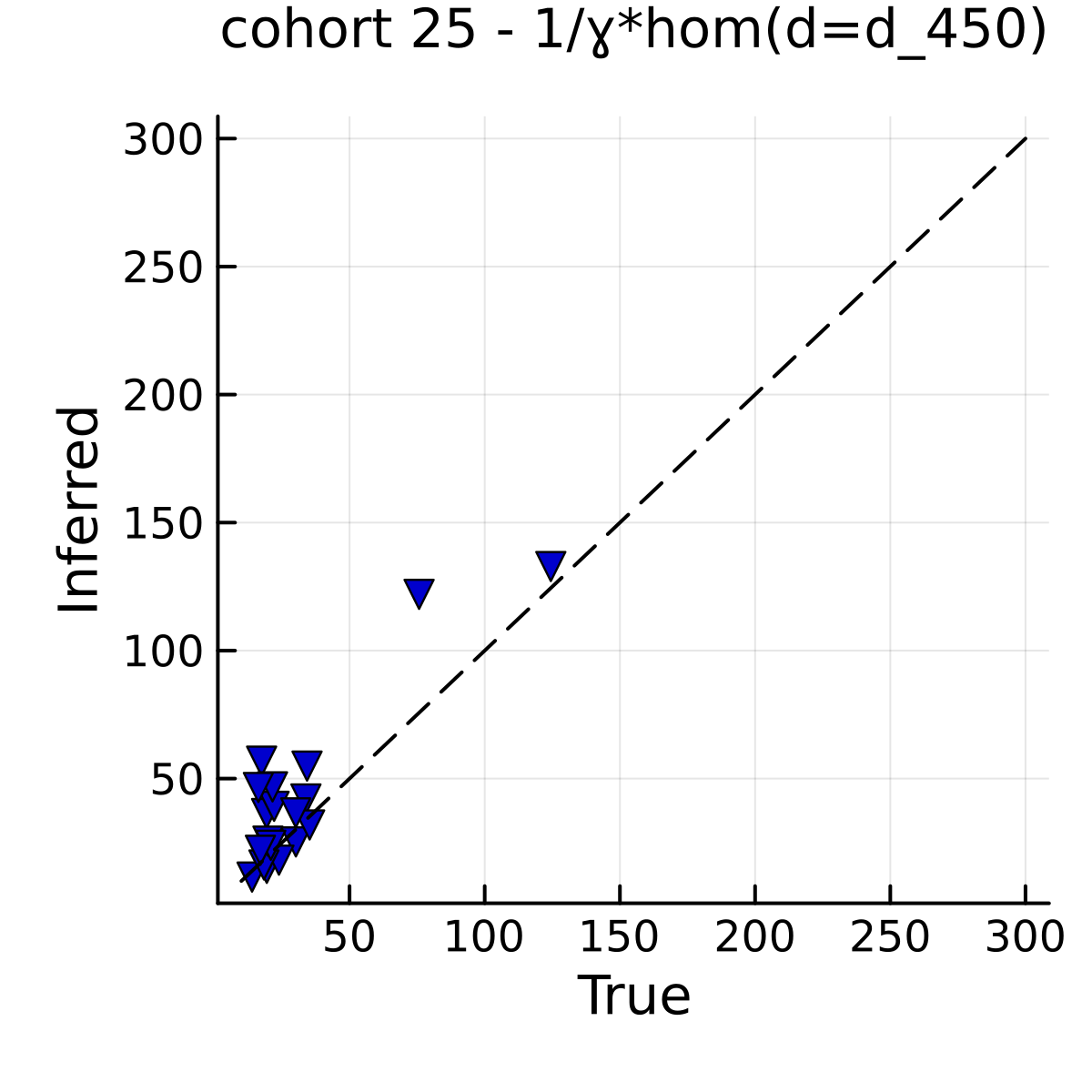}
    \end{subfigure}
    \begin{subfigure}[b]{0.22\textwidth}
        \includegraphics[width=\textwidth]{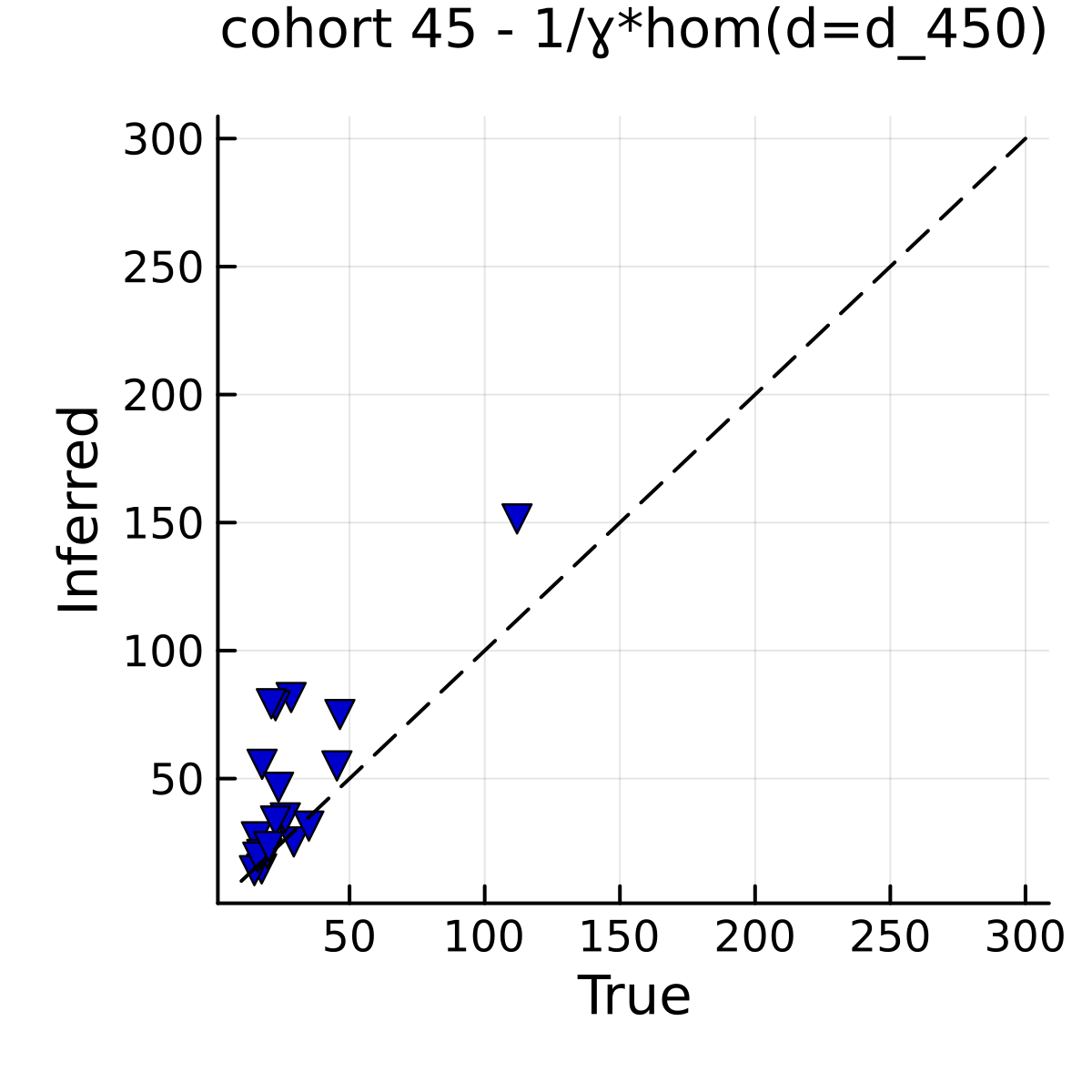}
    \end{subfigure}
    \begin{subfigure}[b]{0.22\textwidth}
        \includegraphics[width=\textwidth]{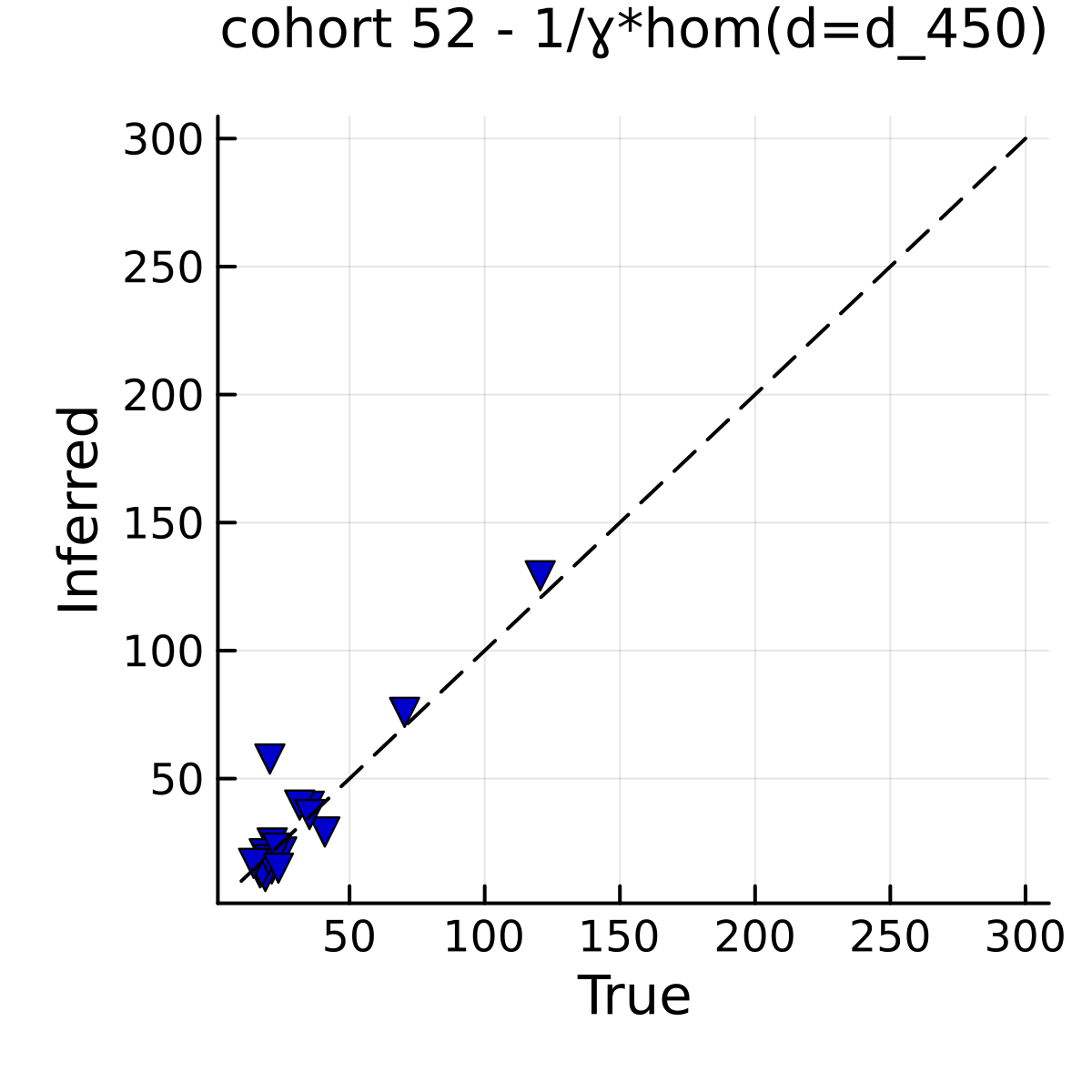}
    \end{subfigure}
    \begin{subfigure}[b]{0.22\textwidth}
        \includegraphics[width=\textwidth]{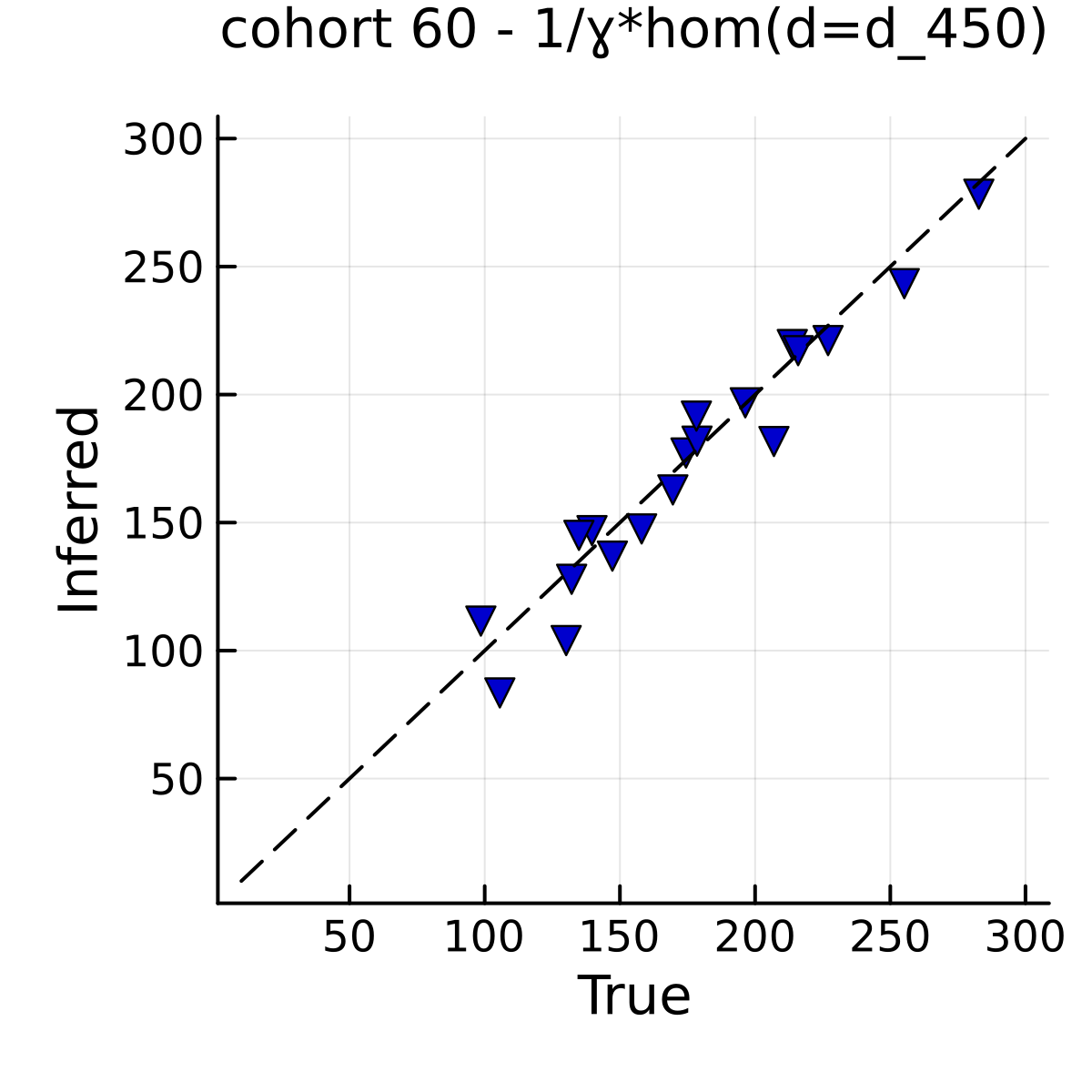}
    \end{subfigure}
    \begin{subfigure}[b]{0.22\textwidth}
        \includegraphics[width=\textwidth]{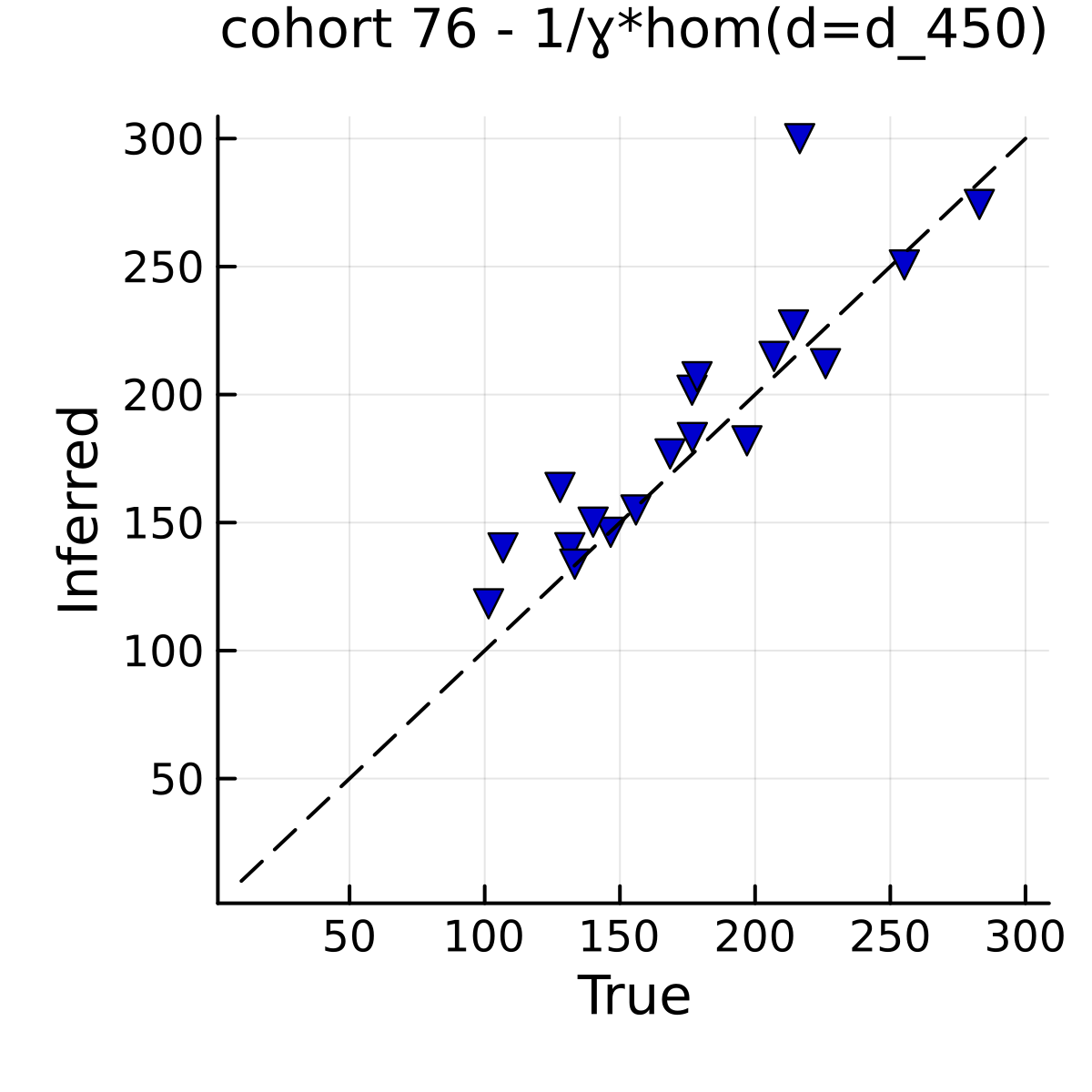}
    \end{subfigure}
    \begin{subfigure}[b]{0.22\textwidth}
        \includegraphics[width=\textwidth]{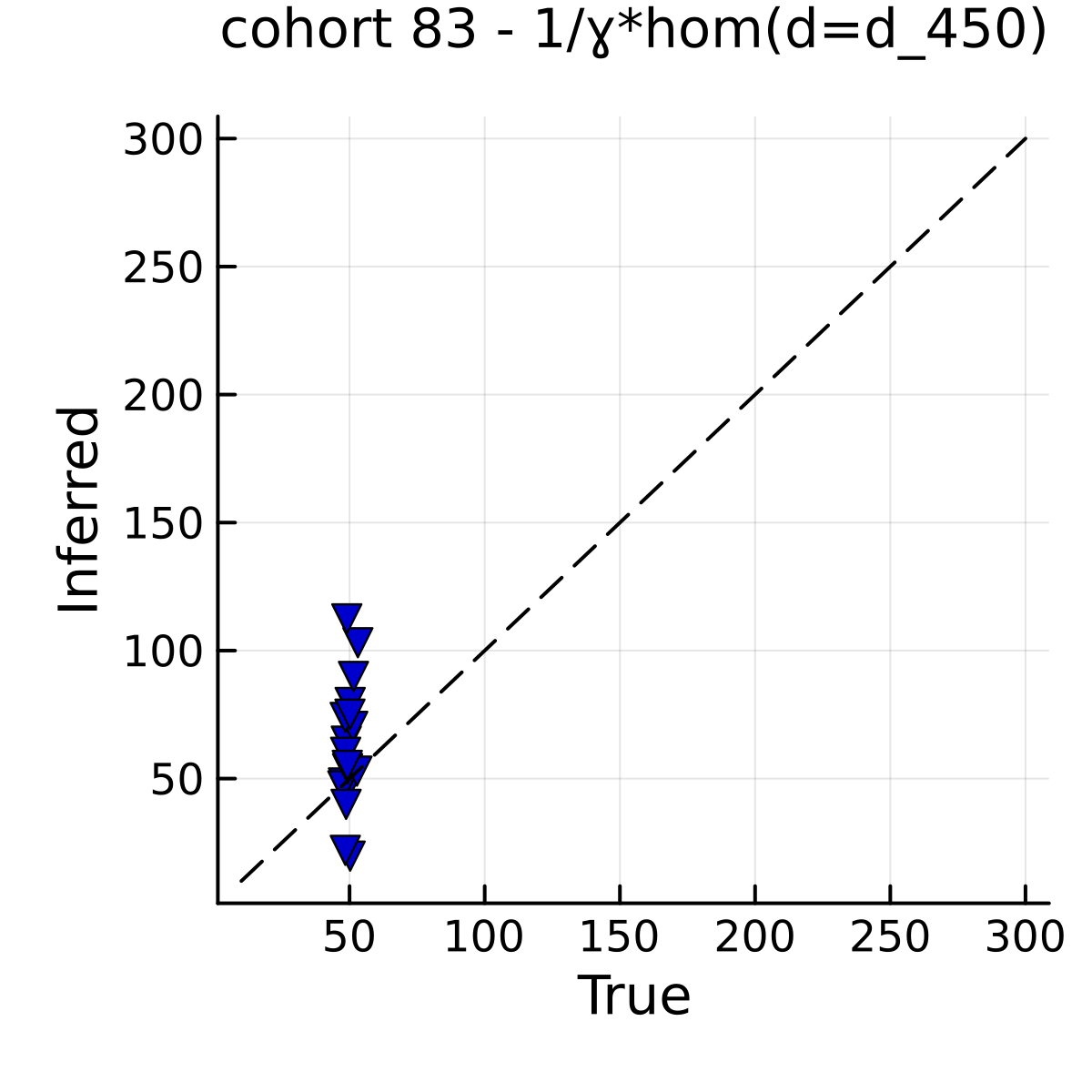}
    \end{subfigure}
    \begin{subfigure}[b]{0.22\textwidth}
        \includegraphics[width=\textwidth]{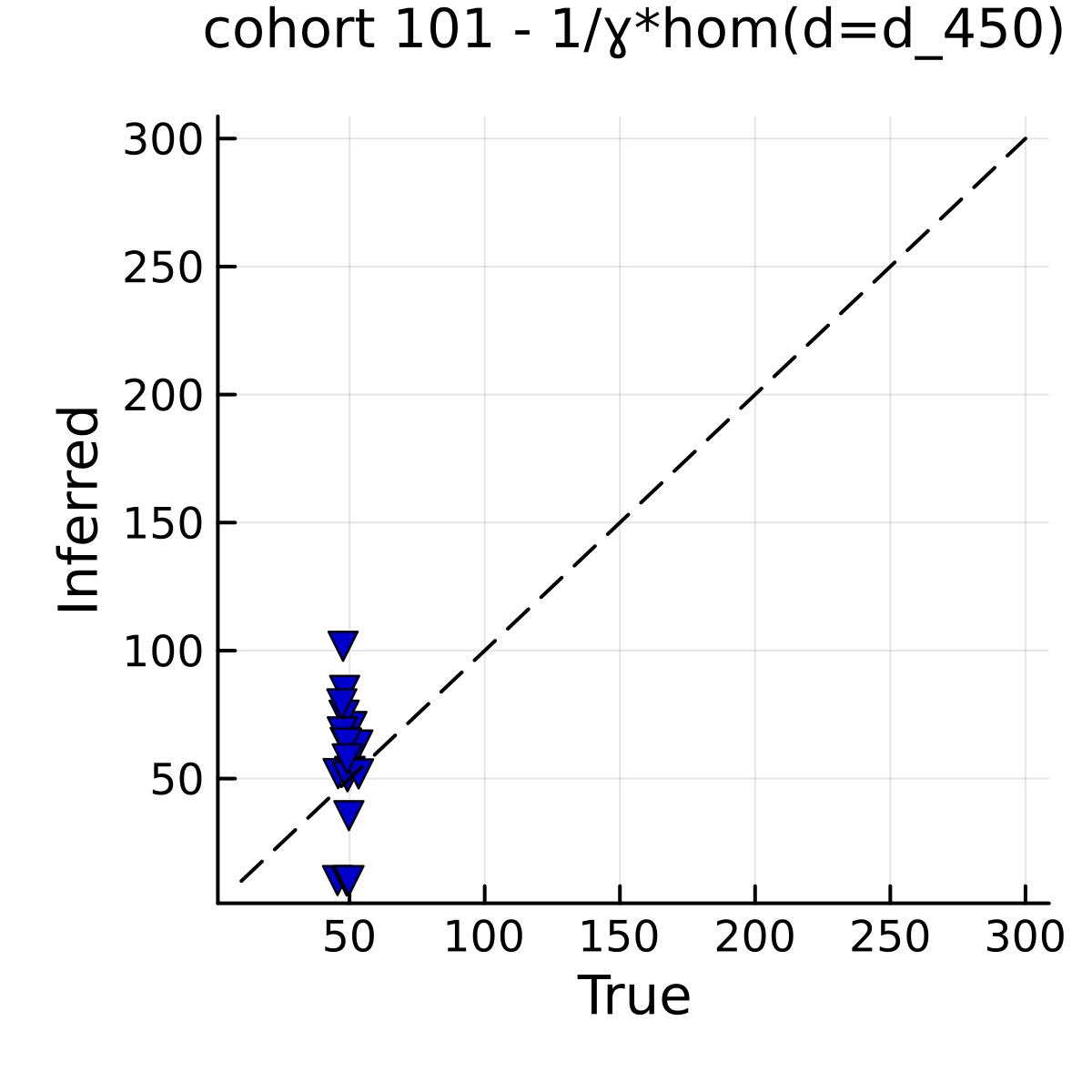}
    \end{subfigure}
    \begin{subfigure}[b]{0.22\textwidth}
        \includegraphics[width=\textwidth]{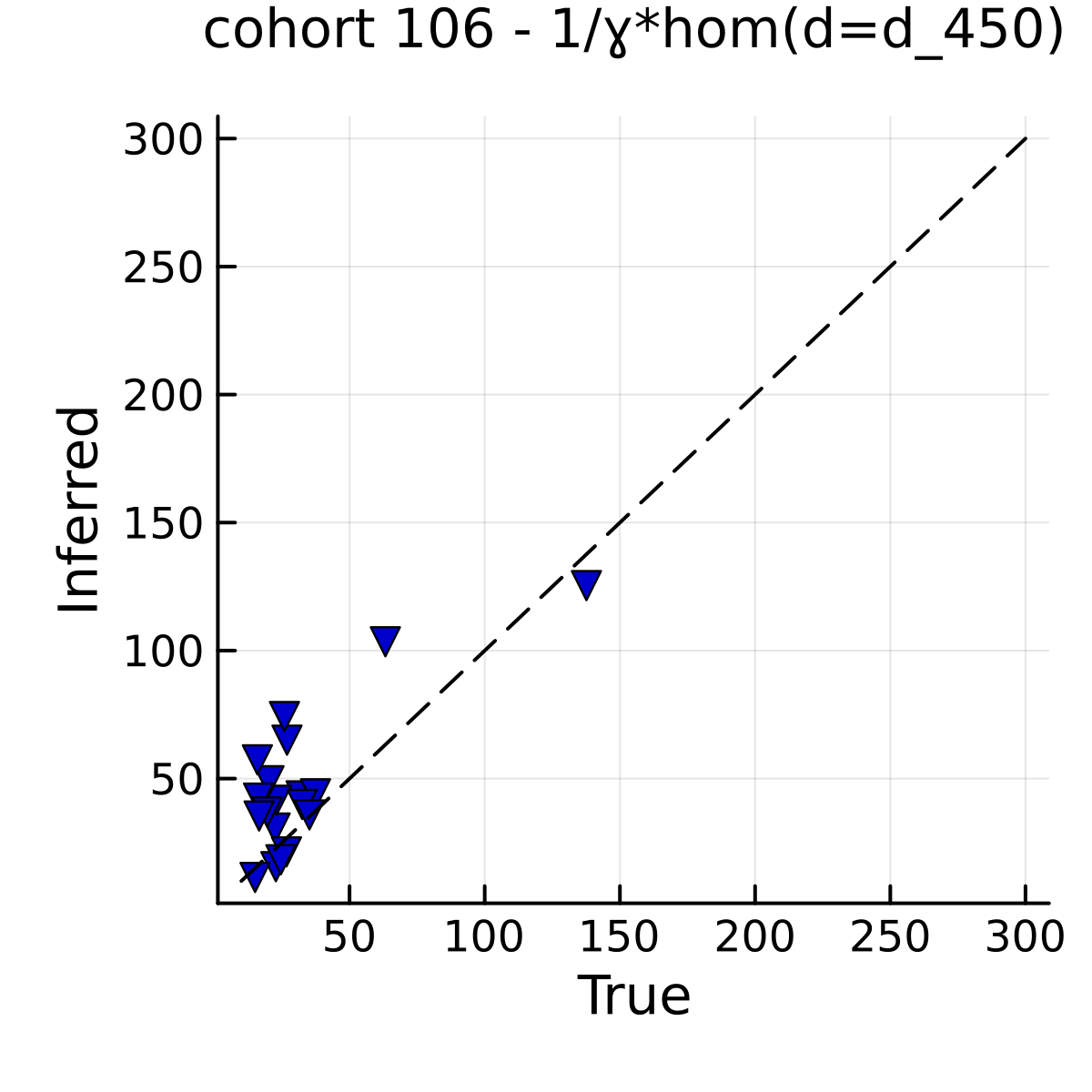}
    \end{subfigure}
    \begin{subfigure}[b]{0.22\textwidth}
        \includegraphics[width=\textwidth]{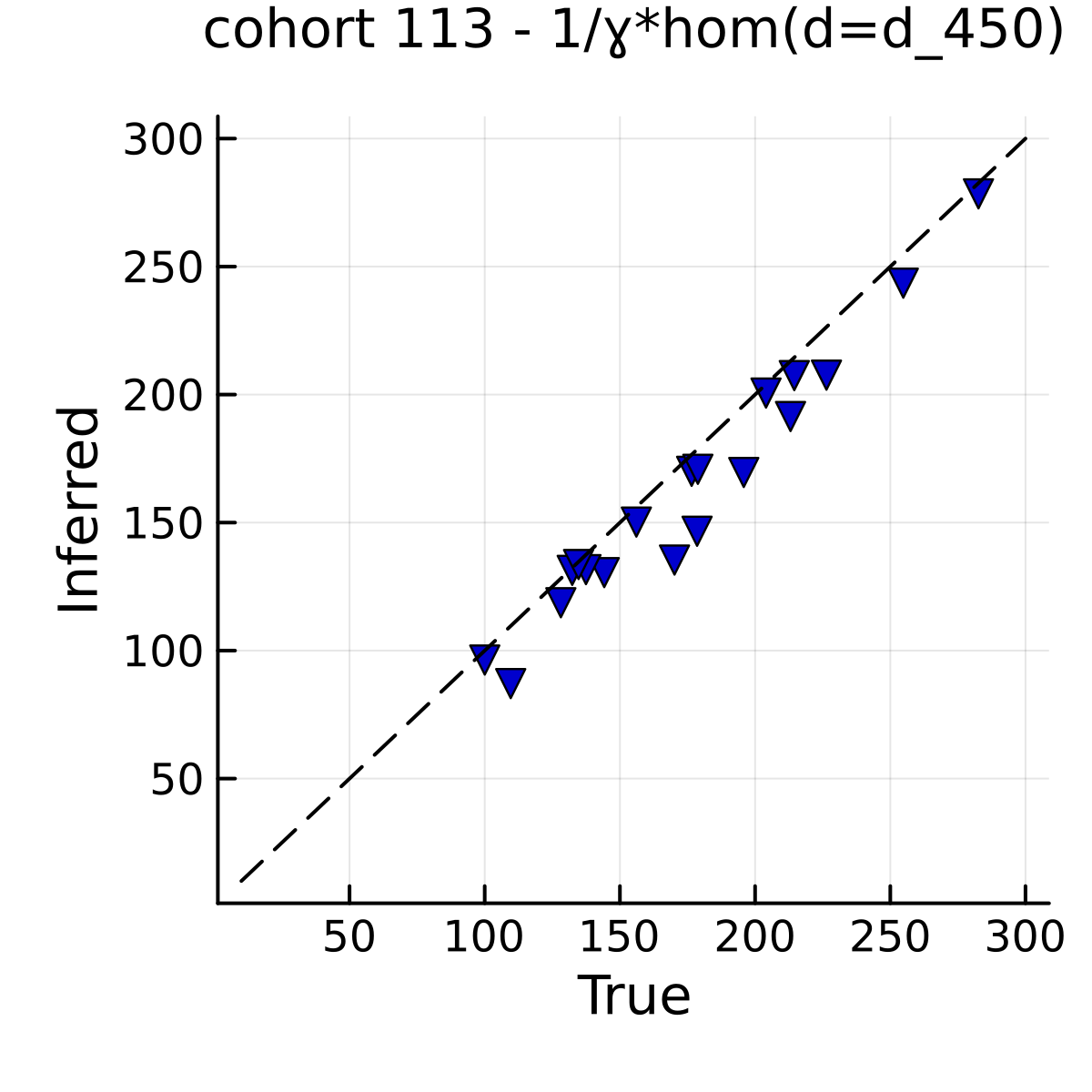}
    \end{subfigure}
    \begin{subfigure}[b]{0.22\textwidth}
        \includegraphics[width=\textwidth]{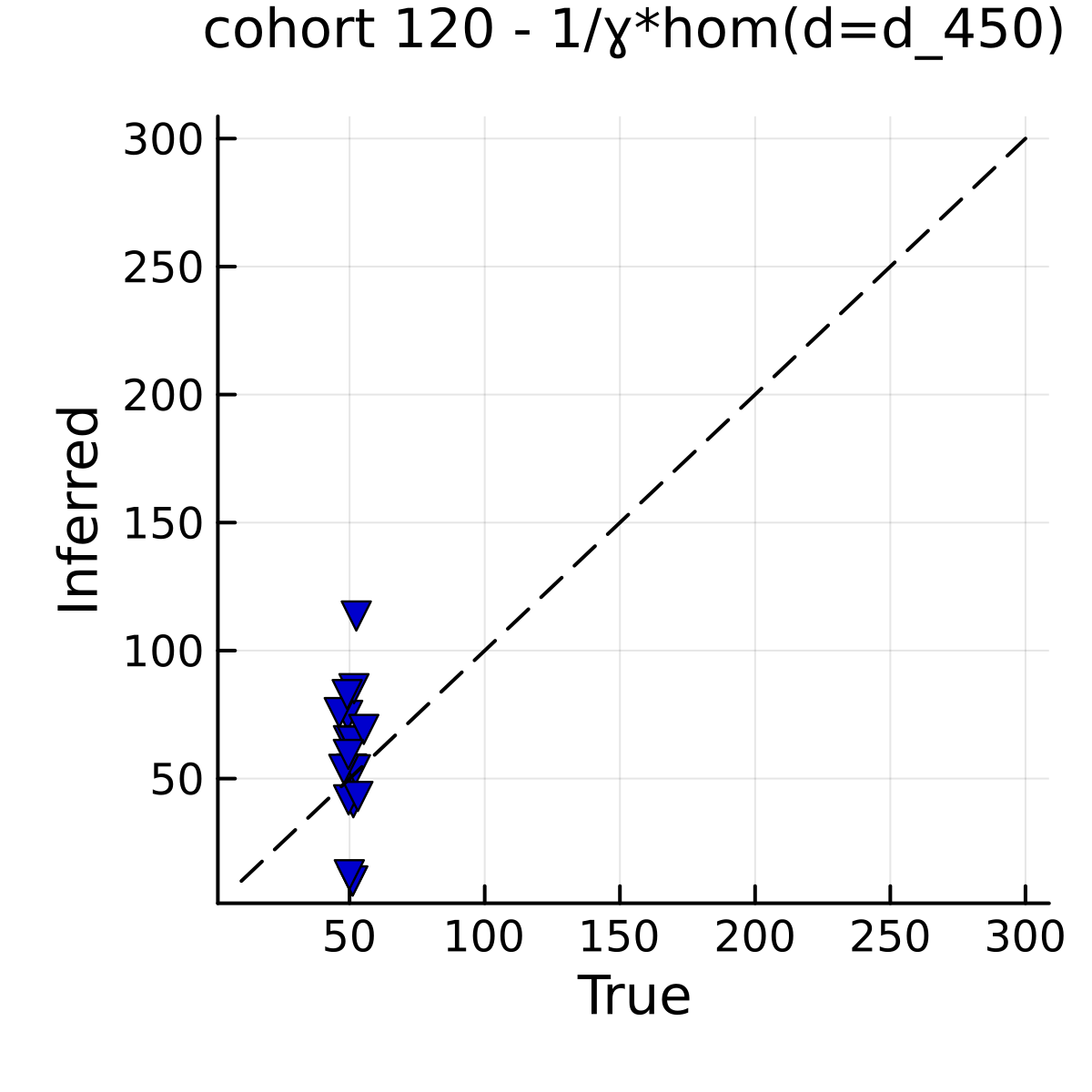}
    \end{subfigure}
    \begin{subfigure}[b]{0.22\textwidth}
        \includegraphics[width=\textwidth]{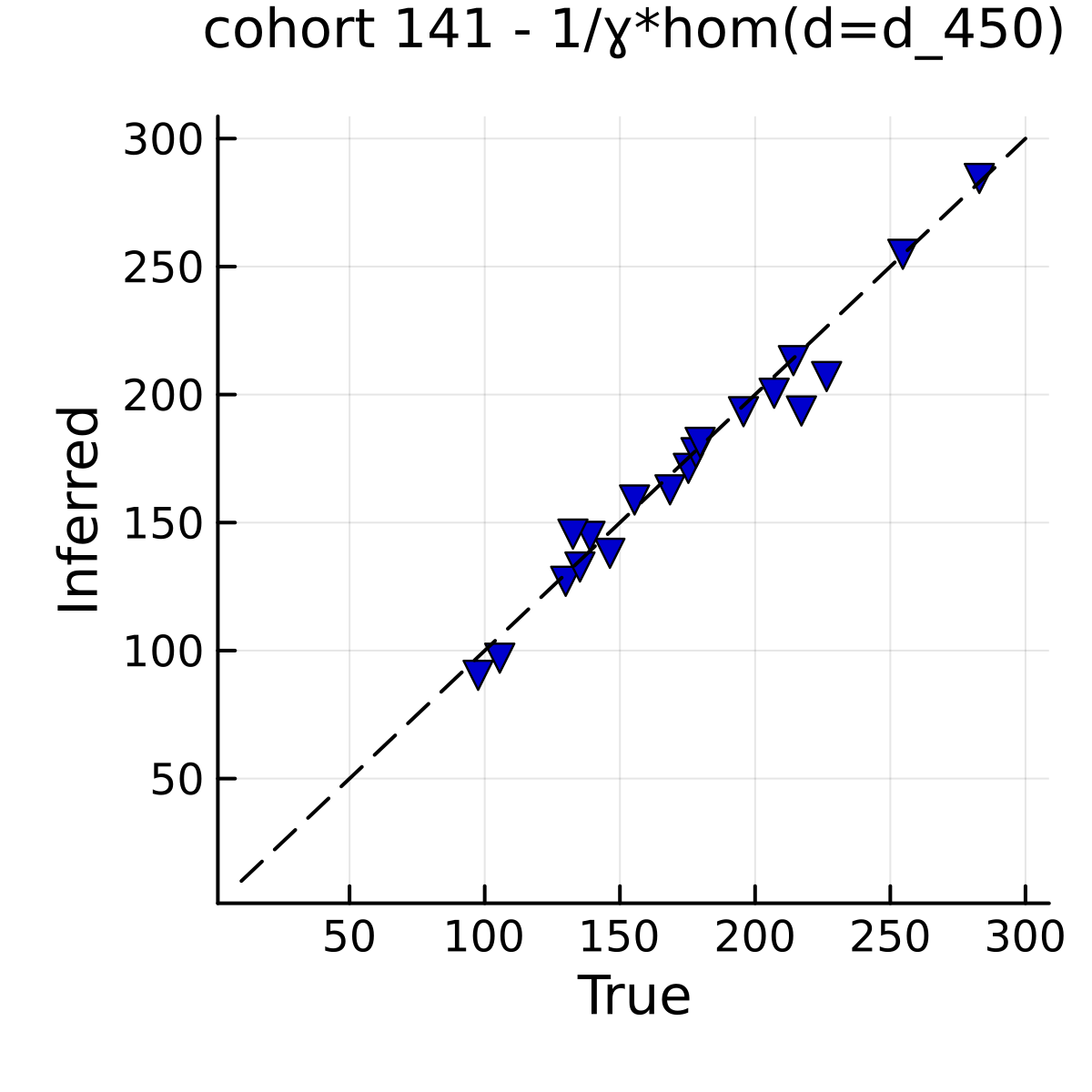}
    \end{subfigure}
    \begin{subfigure}[b]{0.22\textwidth}
        \includegraphics[width=\textwidth]{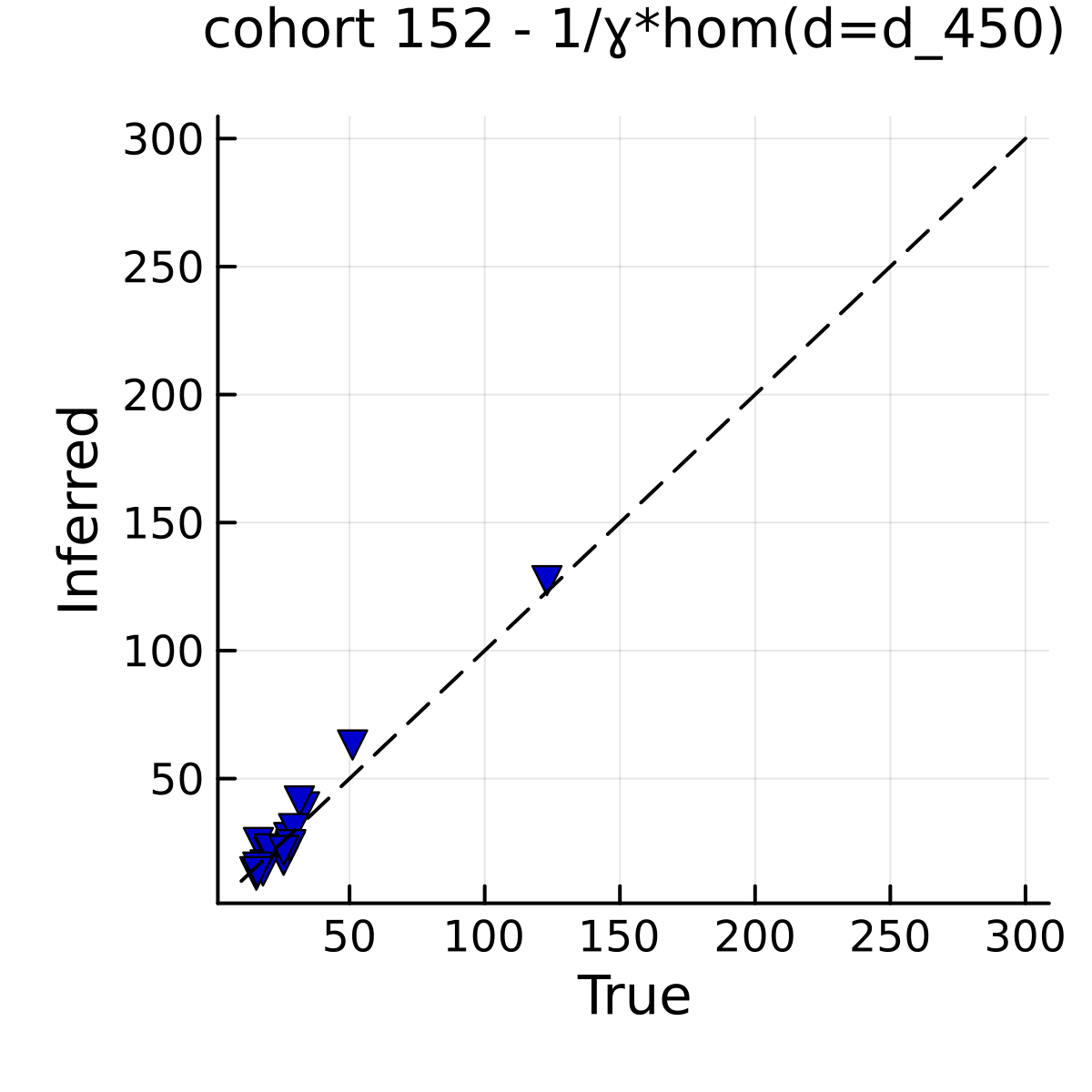}
    \end{subfigure}
    \begin{subfigure}[b]{0.22\textwidth}
        \includegraphics[width=\textwidth]{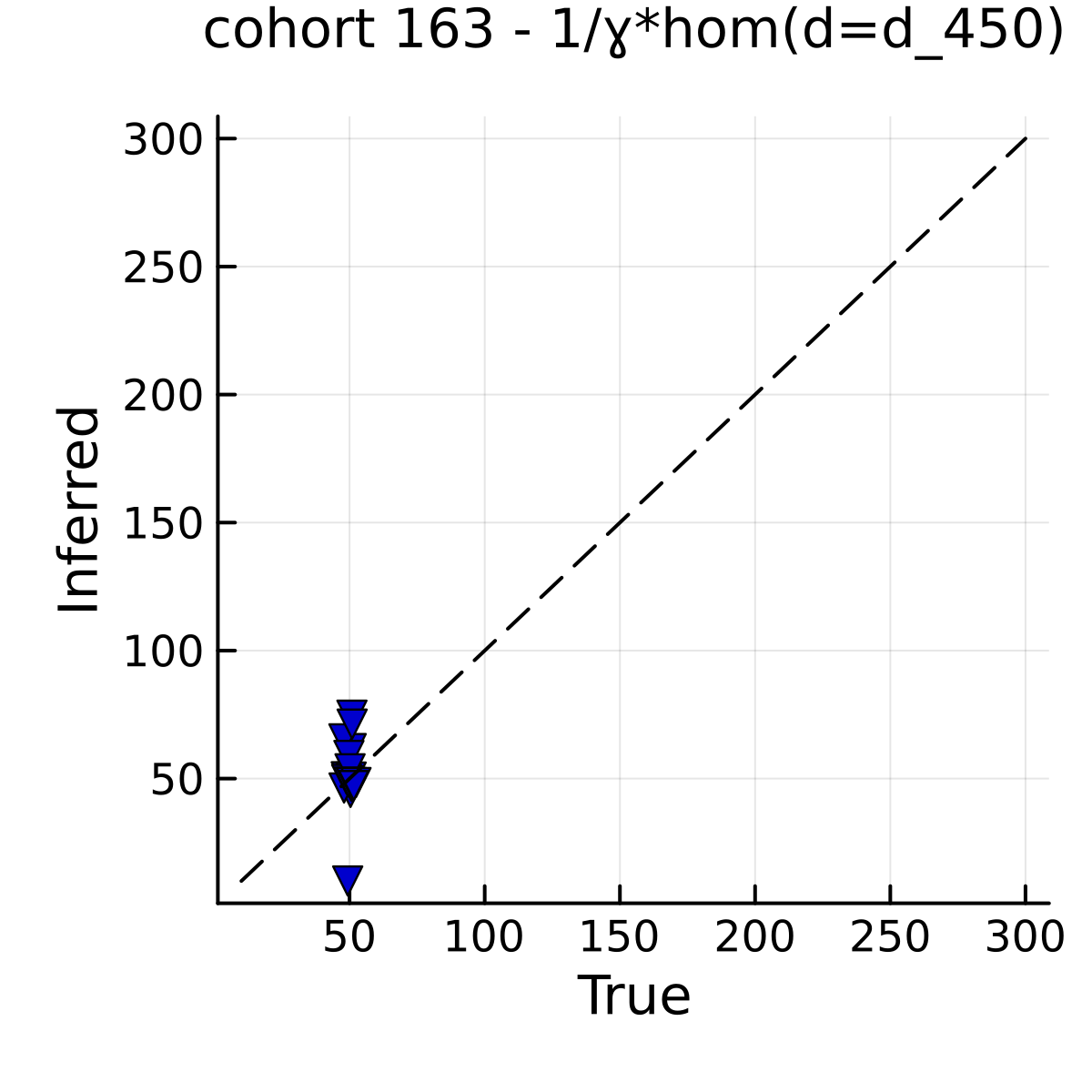}
    \end{subfigure}
    \begin{subfigure}[b]{0.22\textwidth}
        \includegraphics[width=\textwidth]{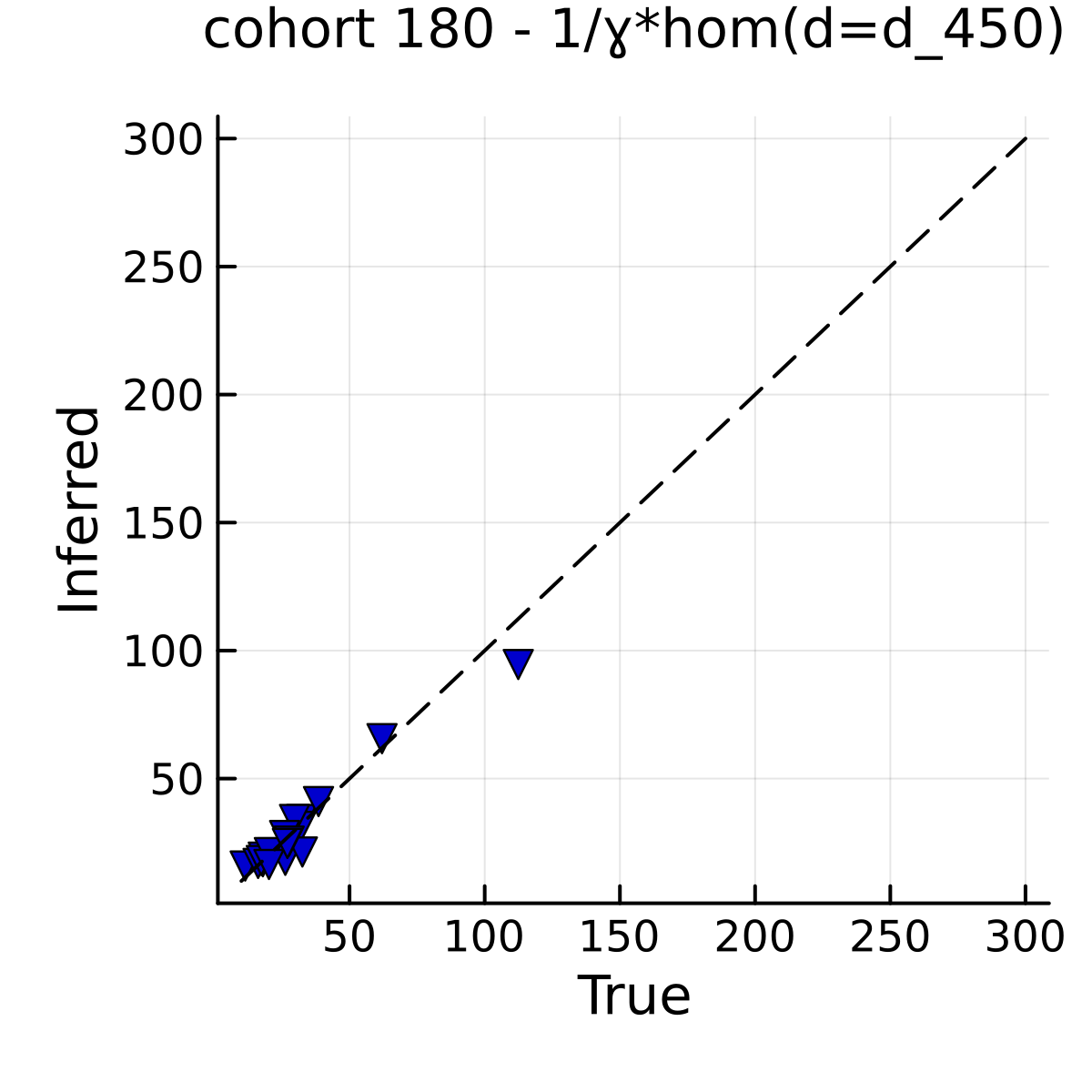}
    \end{subfigure}
    \begin{subfigure}[b]{0.22\textwidth}
        \includegraphics[width=\textwidth]{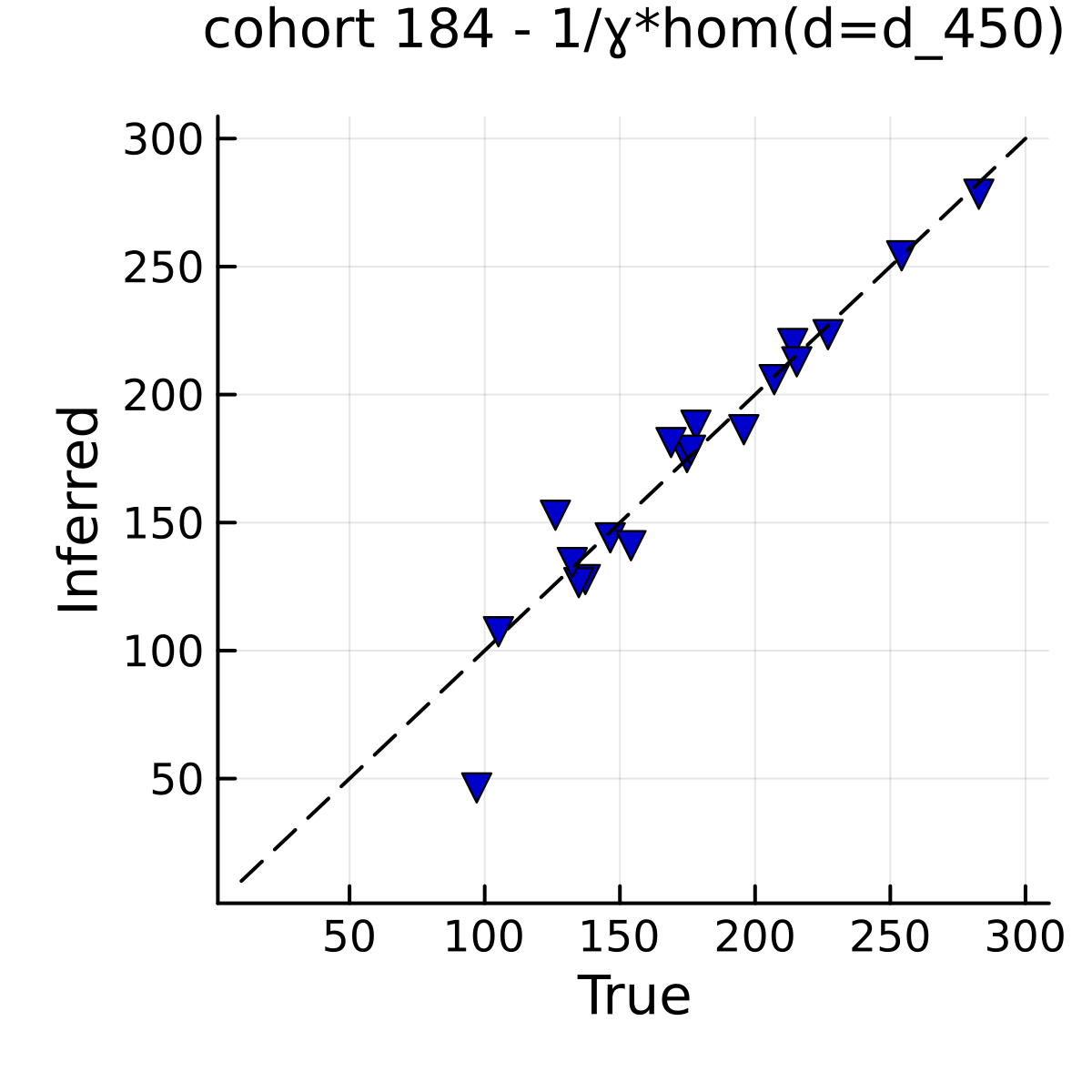}
    \end{subfigure}
    \begin{subfigure}[b]{0.22\textwidth}
        \includegraphics[width=\textwidth]{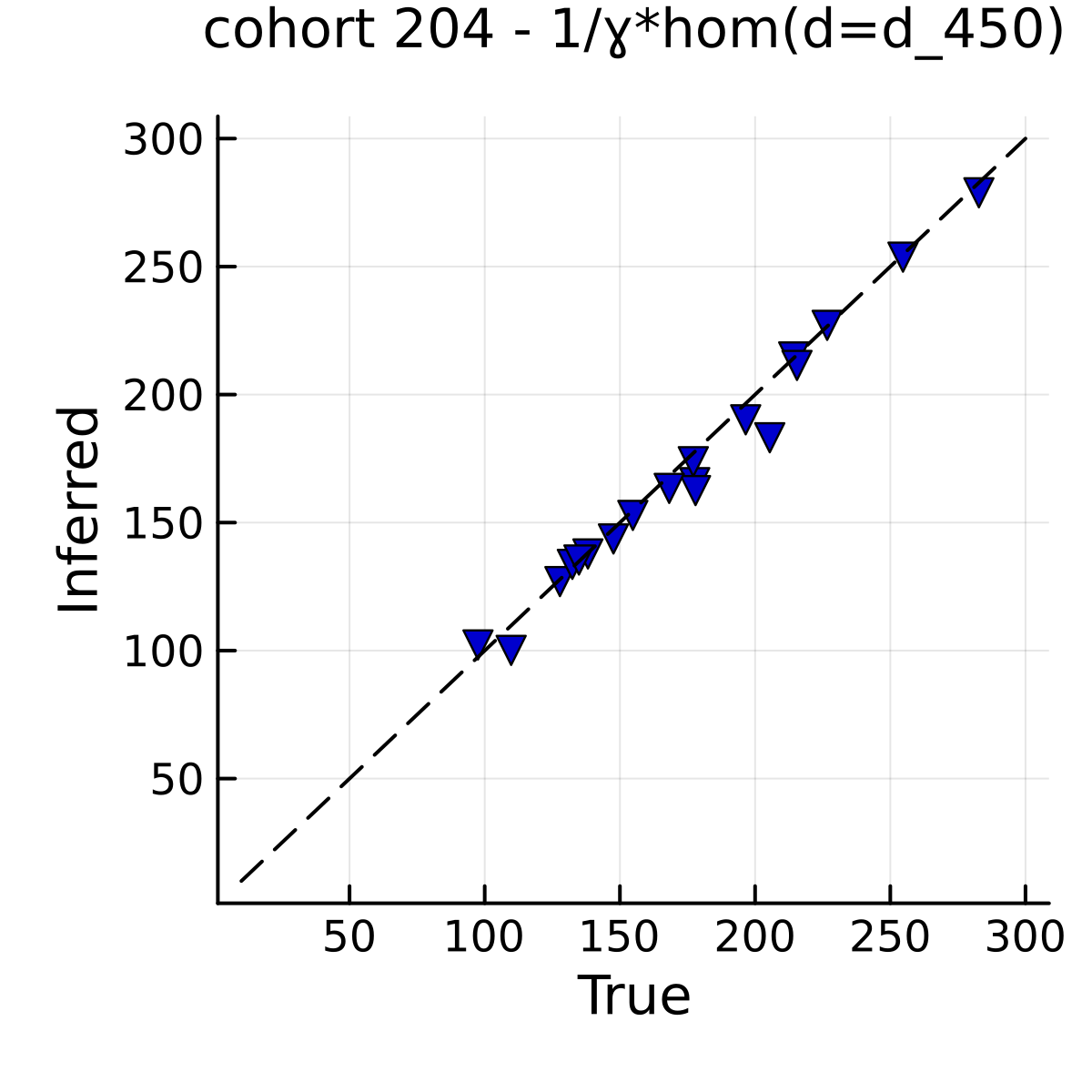}
    \end{subfigure}
    \begin{subfigure}[b]{0.22\textwidth}
        \includegraphics[width=\textwidth]{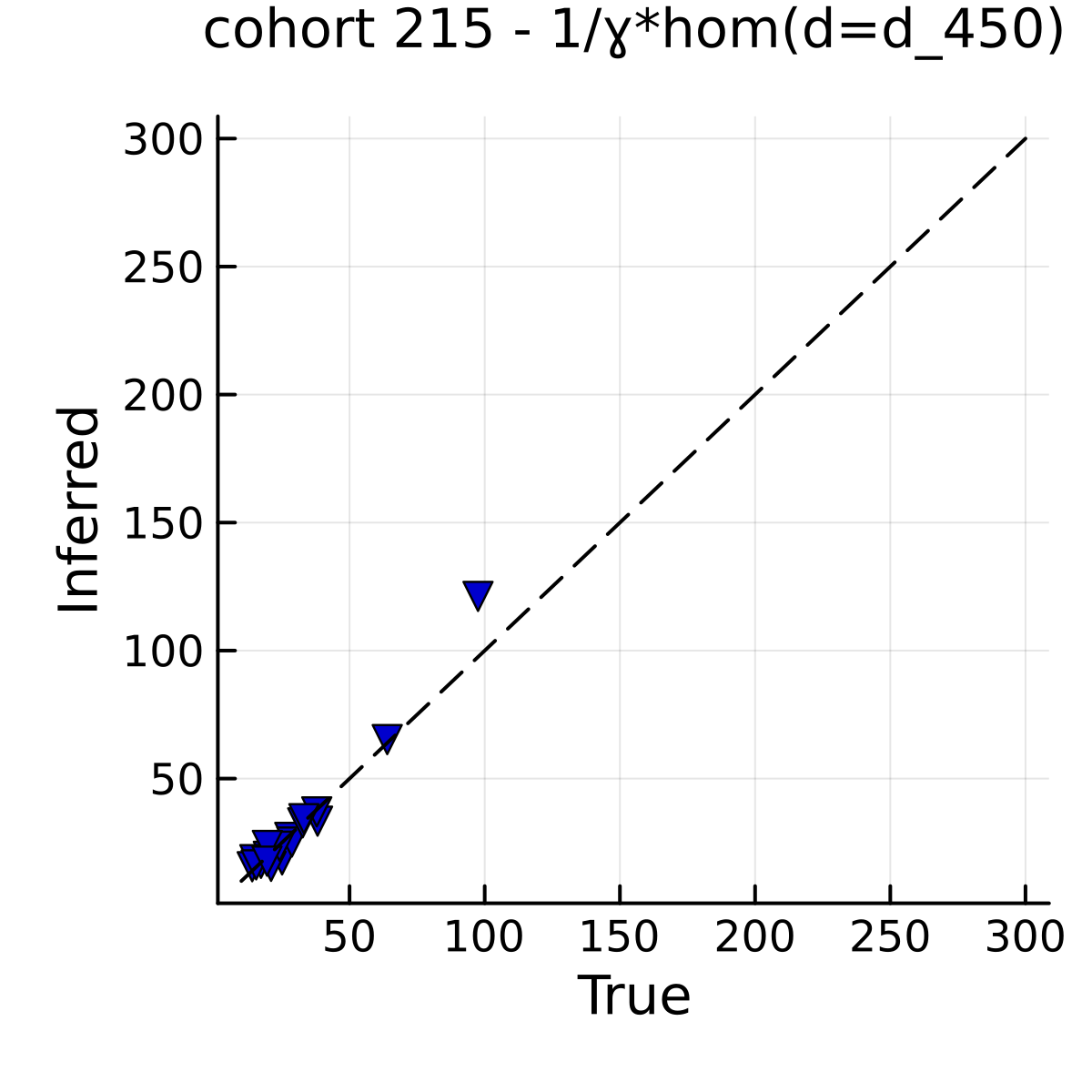}
    \end{subfigure}
    \begin{subfigure}[b]{0.22\textwidth}
        \includegraphics[width=\textwidth]{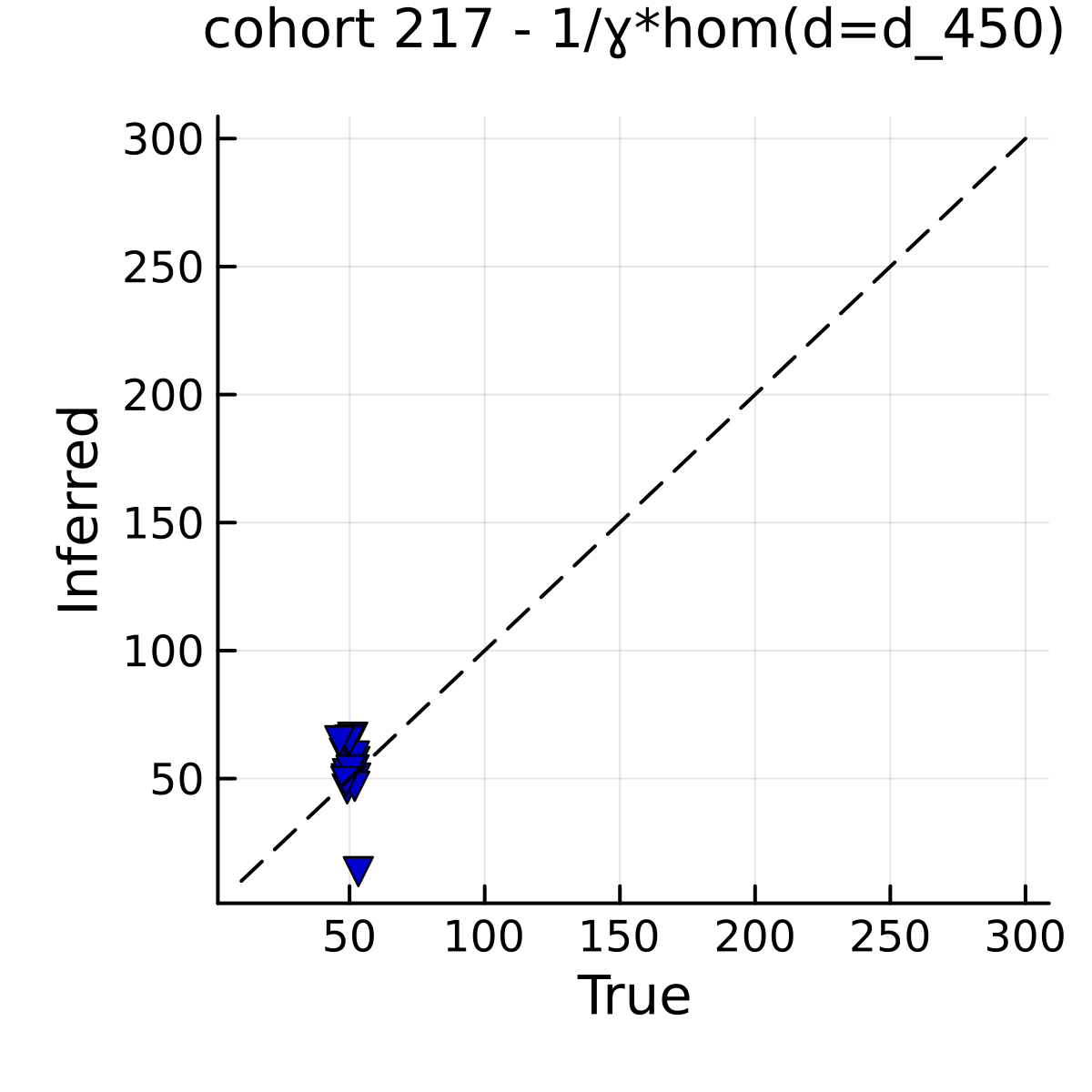}
    \end{subfigure}
    \caption{Comparison between the mean posterior value of $1/\bar{\gamma}^*_{hom}(d_{450}^{(i)})$ (y-axis) for each virtual patient $i$ and each synthetic cohort $m$, and the true one (x-axis). $d_{450}^{(i)}$ corresponds to the mean dose received by the $i^{th}$ patient over 450 days of therapy.}
    \label{fig:synth_gamma_hom}
\end{figure}

In this section, we confront the inferred parameter values (mean \textit{a posteriori}) to the true ones for each individual $i$ from each virtual cohort $m$.
For some cohorts, the selected model differs from the true one and, consequently, the definition of the model parameters too.
Parameters $\eta_{het}$ (Fig.~\ref{fig:synth_eta_het}), $\eta_{hom}$ (Fig.~\ref{fig:synth_eta_hom}), and $k_m$ (Fig.~\ref{fig:synth_km}) are common across all the 225 considered models. We can directly compare the inferred values to the true ones for them.\\
The models differ according to the dose-response relationships of $\bar{\Delta}^*_{het}$, $\bar{\Delta}^*_{hom}$, $\bar{\gamma}^*_{het}$, and $\bar{\gamma}^*_{hom}$ and, therefore, to the parameters involved in these relationships. Then, if we cannot directly compare the parameters related to these relationships, we can still compare the estimated values of $\bar{\Delta}^{* (i,m)}_{het}(d)$ to the true ones, for a given value of the dose $d$ (and similarly for the three other dose-response relationships).
For patient $i$, we define $d_{450}^{(i)}$ as the mean dose he receives over the 450 days of therapy (see Tab.~\ref{tab:info_patients}). 
Then, we compare for each individual $i$ of each cohort $m$ the true value $\bar{\Delta}^{*, (i, m)}_{het}(d_{450}^{(i)})$ to the inferred one (mean \textit{a posteriori}). The results are presented in Fig.~\ref{fig:synth_Delta_het}.
In a similar manner, we present the results concerning $\bar{\Delta}^{*, (i, m)}_{hom}(d_{450}^{(i)})$, $\bar{\gamma}^{*, (i, m)}_{het}(d_{450}^{(i)})$, and $\bar{\gamma}^{*, (i, m)}_{hom}(d_{450}^{(i)})$ in Fig.~\ref{fig:synth_Delta_hom}, \ref{fig:synth_gamma_het}, and~\ref{fig:synth_gamma_hom}, respectively.\\

Overall, parameters $\eta_{het}$ and $\eta_{hom}$ are correctly estimated. Let us recall that, for these two parameters only, no population effect is considered.\\

Concerning $k_m$, we retrieve similar results as in section~\ref{sec:identif} (practical identifiability of the \rev{baseline} model). For most cohorts, the population effect seems to predominate over inter-individual variability, with the mean of the population distribution being correctly retrieved.\\

Concerning the values of $\bar{\Delta}^*_{het}$ (Fig.~\ref{fig:synth_Delta_het}) and  $\bar{\Delta}^*_{hom}$ (Fig.~\ref{fig:synth_Delta_hom}) taken at the (individual) mean dose computed over the 450 days of therapy, we also get overall accurate estimations. 
Yet, concerning $\bar{\Delta}^*_{hom}(d_{450}^{(i)})$, we observe that we underestimate this value for some individuals from the cohorts $m=21$, 25, 45 (and, to a lesser extent, for the cohorts 76, 83, 106). \\

Concerning the values of $\bar{\gamma}^*_{het}$ and  $\bar{\gamma}^*_{hom}$ (Fig.~\ref{fig:synth_gamma_het} and~\ref{fig:synth_gamma_hom}, where we represent the inverse of these values, corresponding biologically to the mean time of quiescence exit) taken at the (individual) mean dose computed over the 450 days of therapy, we also get pretty good results overall, yet with some cases deserving more attention. \\
For some cohorts (more precisely, for some dose-response relationships), the population distributions used to sample individual parameters related to $1/\bar{\gamma}^*$ were chosen with very low variance. So, the individual true parameter values were very close, almost as if the parameter was constant for the population.
Concerning $\bar{\gamma}^*_{het}$, cohorts $m=25$, 52, 76, 106, 163, 184, and 217 fall into that case.
For three of them, namely cohorts 25, 52, and 106, we infer, indeed, a substantial population effect. However, the values for $1/\bar{\gamma}^*_{het}(d_{450}^{(i)})$ are overestimated, concentrating in the upper bound of the prior range of values (that is, near the value of $300$ days associated with the value of the quiescence exit of WT cells). For cohorts 76 and 163, we infer an inter-individual variability (when it was not the case in reality) while overestimating the values $1/\bar{\gamma}^*_{het}(d_{450}^{(i)})$.  For cohorts 184 and 217, we infer inter-individual variability but accurately estimate the mean value associated with the population distribution. \\
For cohort 21, we get poor estimations of $1/\bar{\gamma}^*_{het}(d_{450}^{(i)})$. \\
We observe better results for $1/\bar{\gamma}^*_{hom}$.

\FloatBarrier

\subsubsection{Response at 1,500 days}

\begin{figure}[h]
    \centering

    \begin{subfigure}[b]{0.22\textwidth}
        \includegraphics[width=\textwidth]{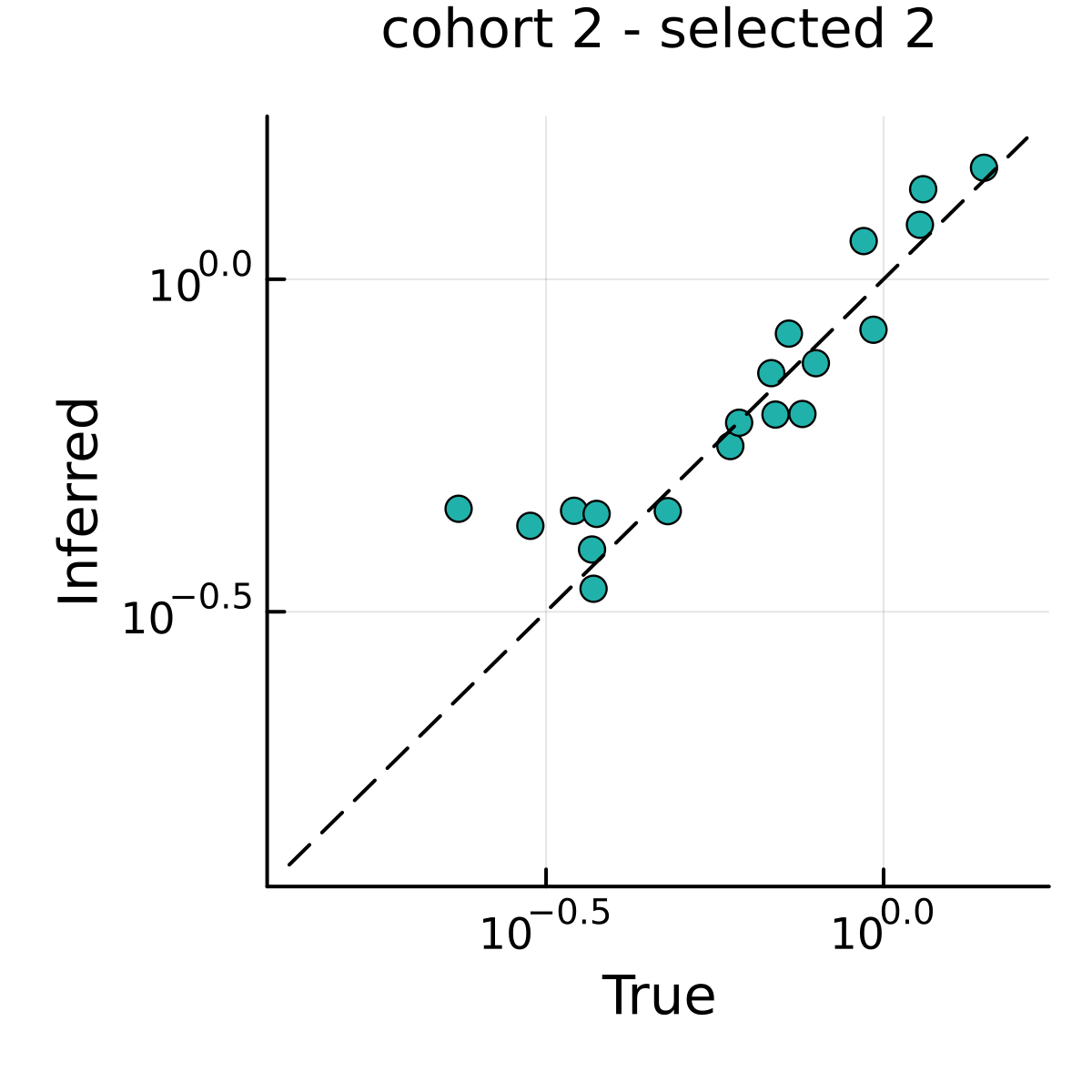}
    \end{subfigure}
     \begin{subfigure}[b]{0.22\textwidth}
        \includegraphics[width=\textwidth]{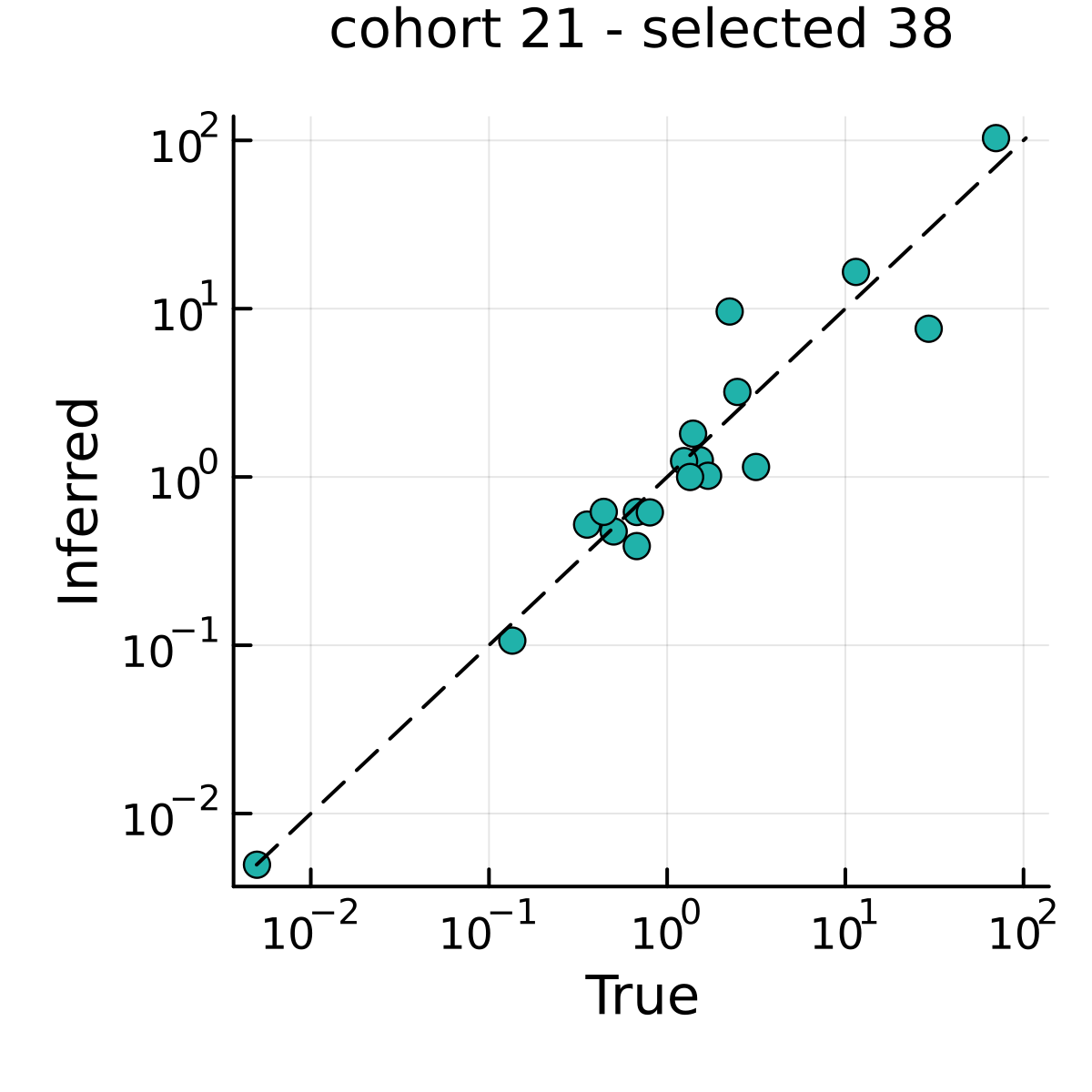}
    \end{subfigure}
     \begin{subfigure}[b]{0.22\textwidth}
        \includegraphics[width=\textwidth]{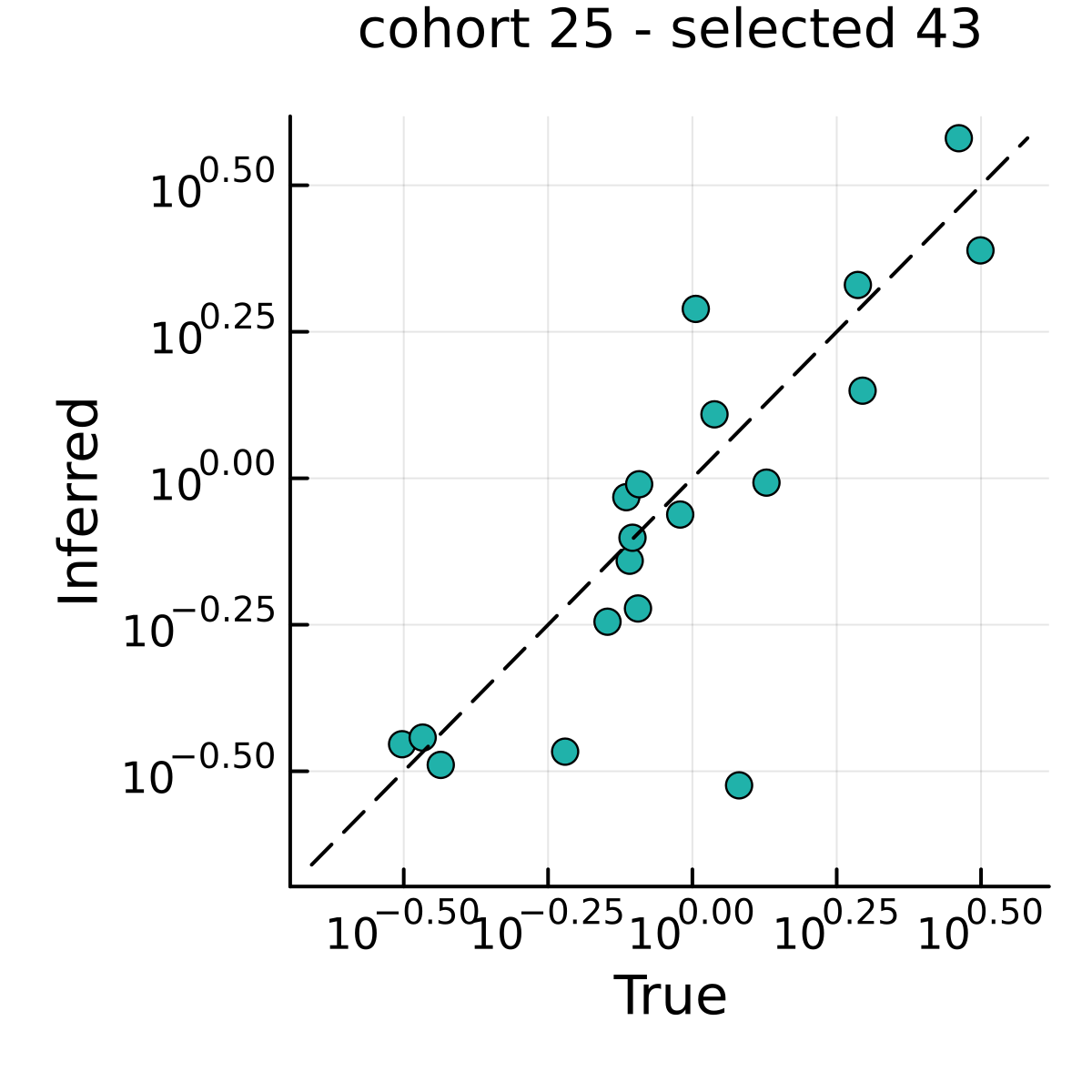}
    \end{subfigure}
     \begin{subfigure}[b]{0.22\textwidth}
        \includegraphics[width=\textwidth]{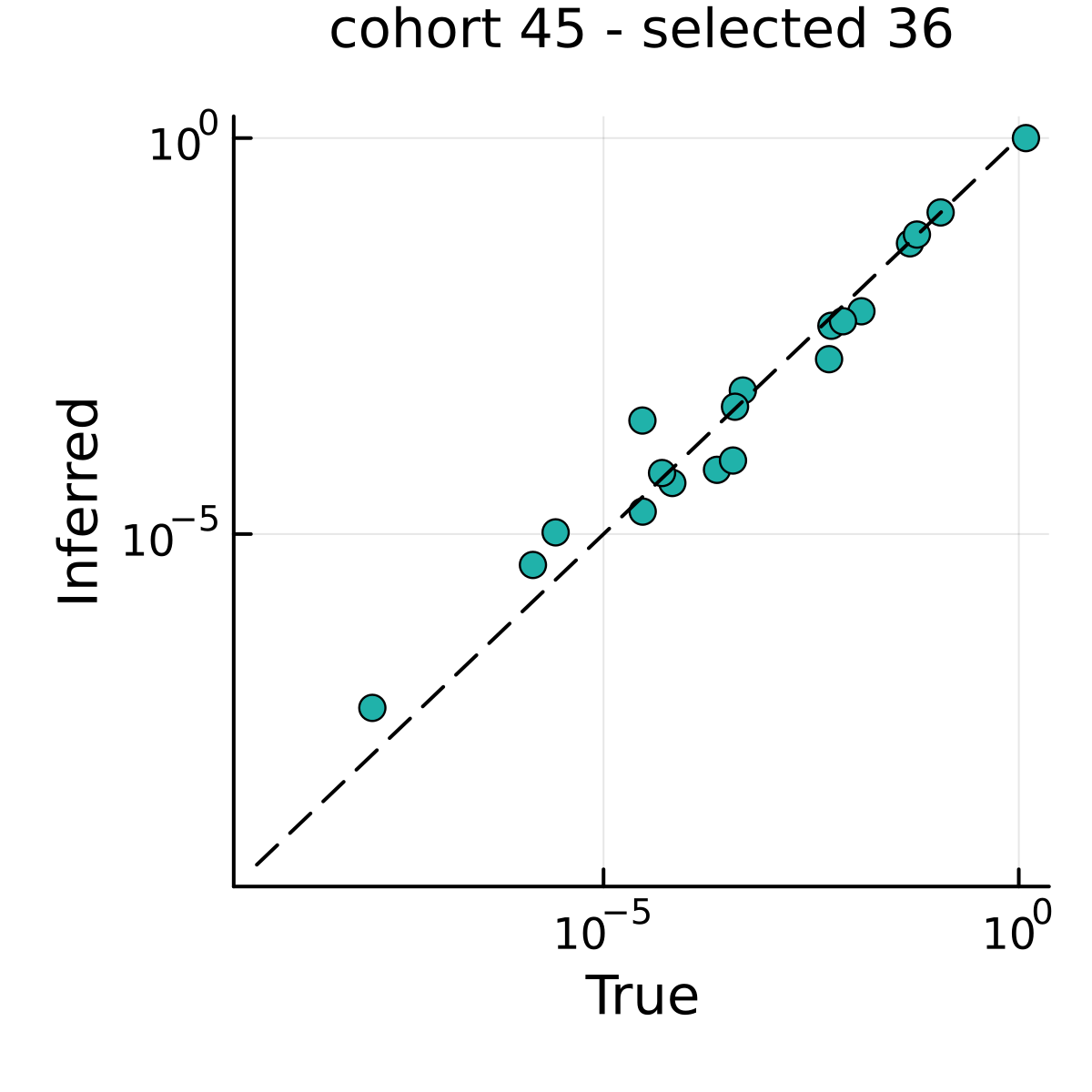}
    \end{subfigure}
     \begin{subfigure}[b]{0.22\textwidth}
        \includegraphics[width=\textwidth]{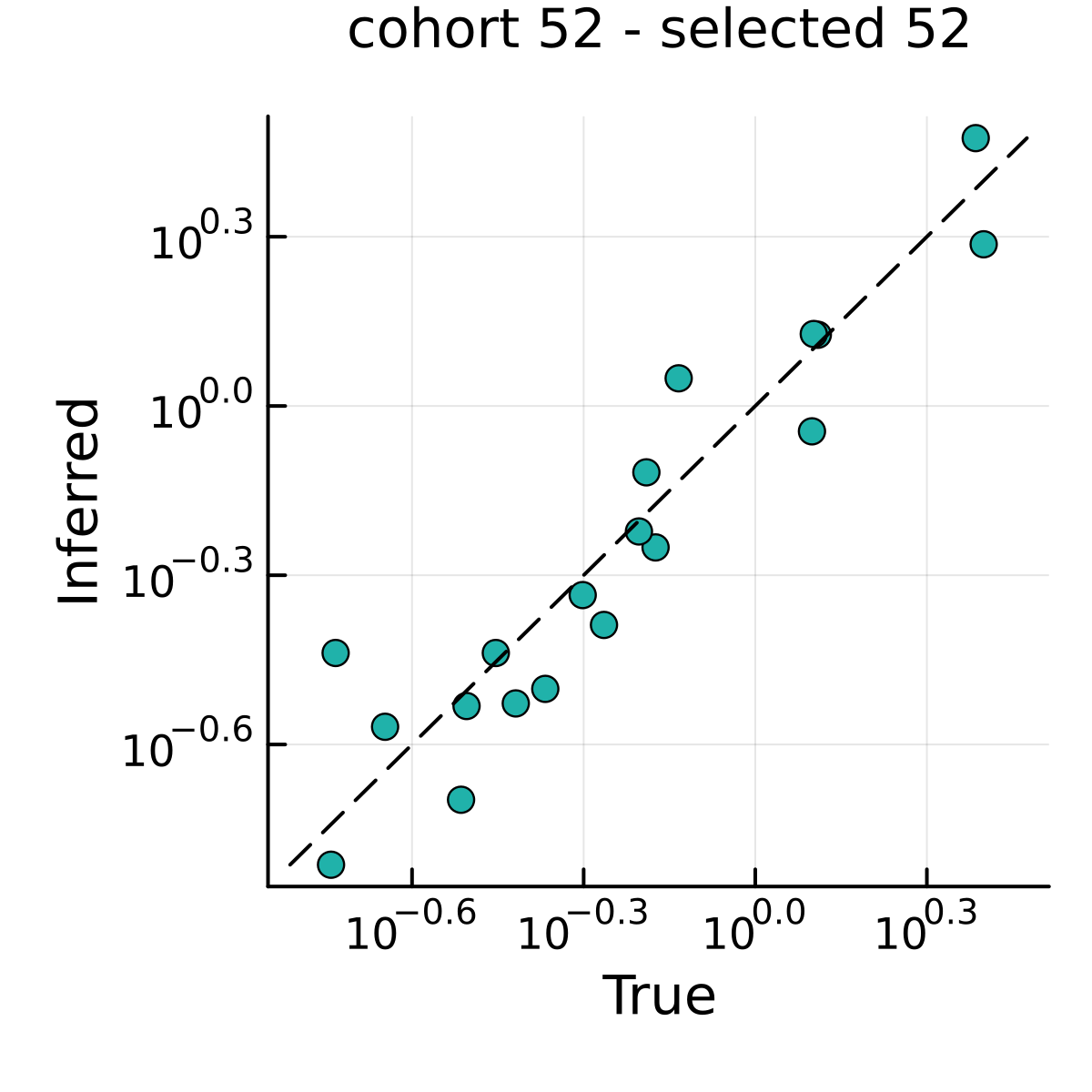}
    \end{subfigure}
     \begin{subfigure}[b]{0.22\textwidth}
        \includegraphics[width=\textwidth]{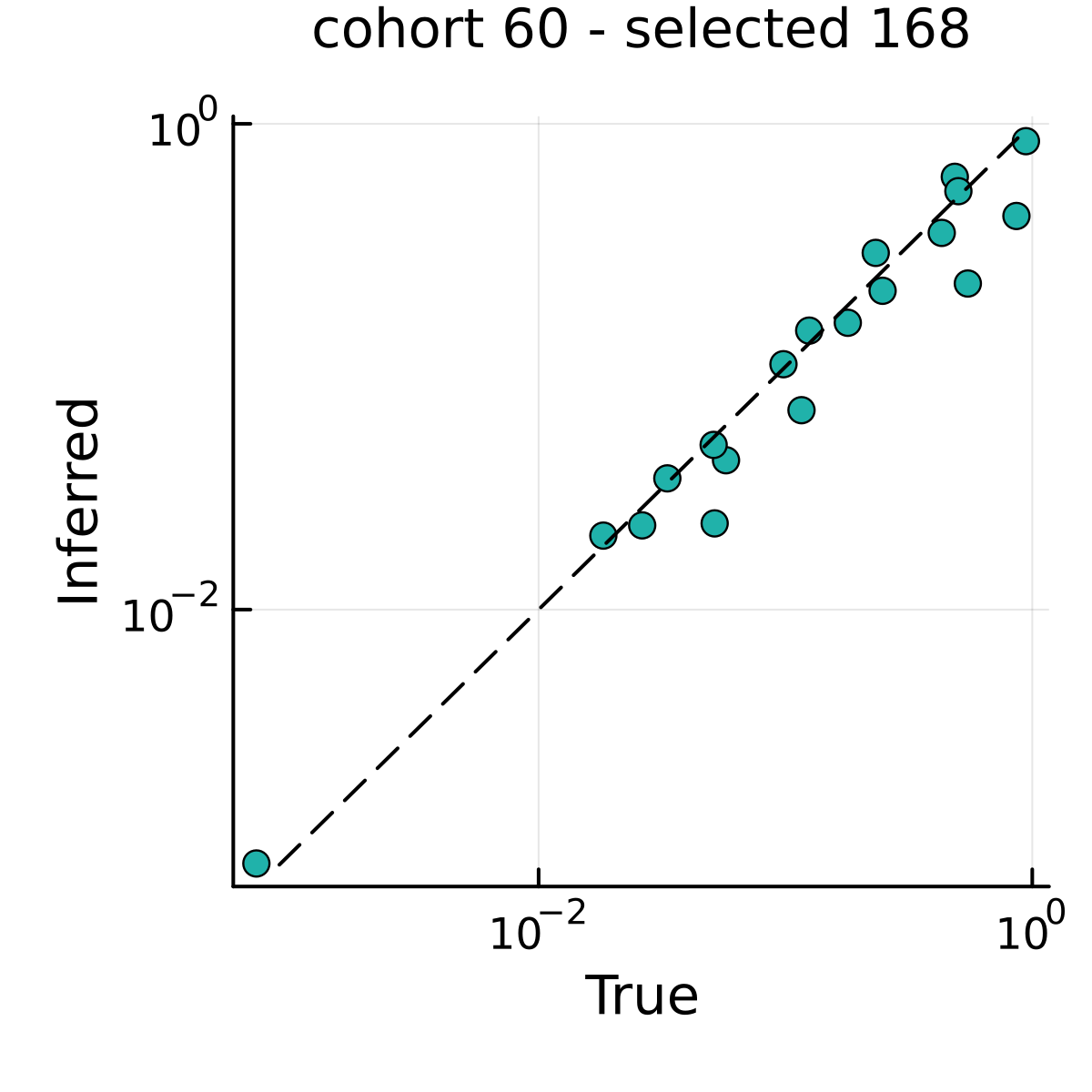}
    \end{subfigure}
     \begin{subfigure}[b]{0.22\textwidth}
        \includegraphics[width=\textwidth]{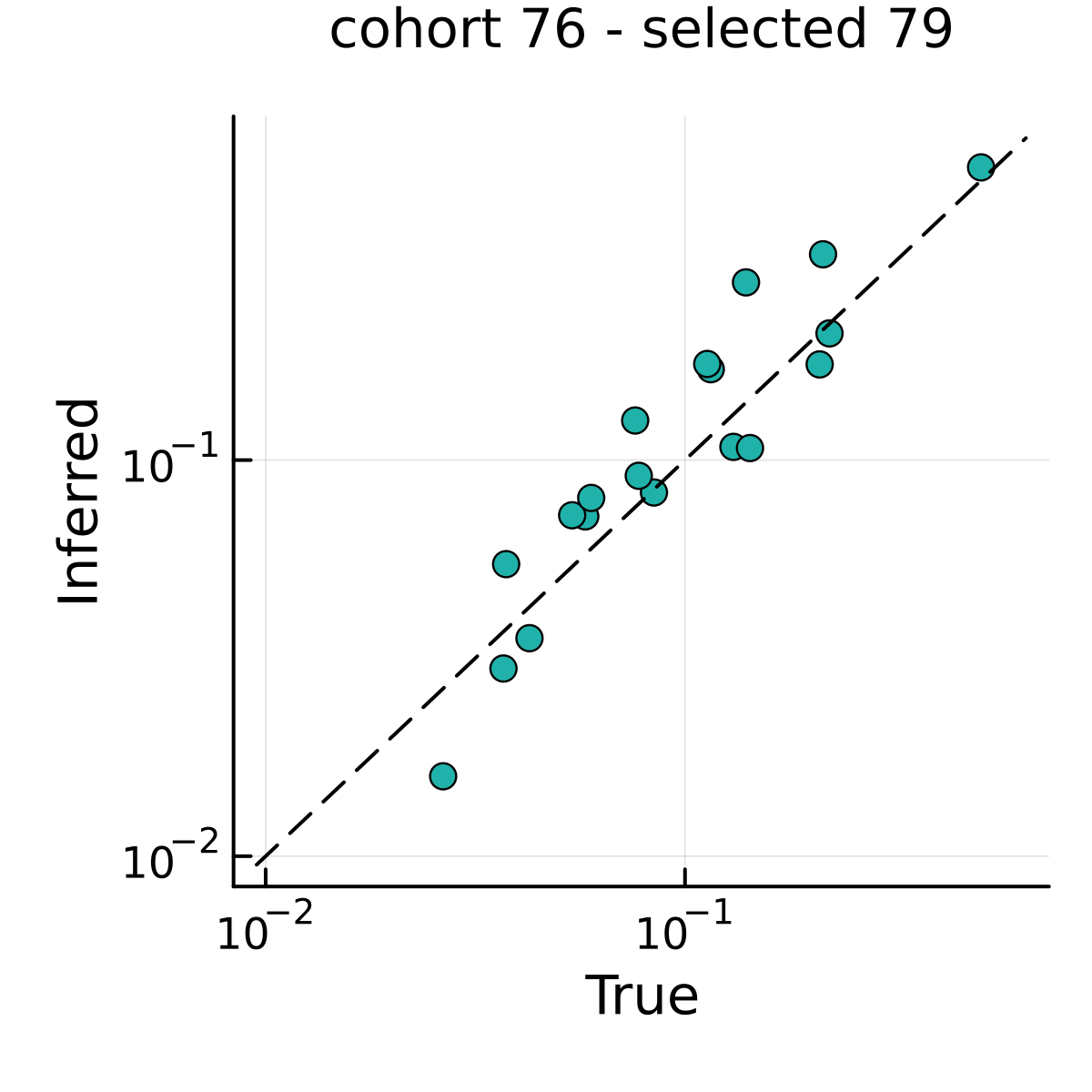}
    \end{subfigure}
     \begin{subfigure}[b]{0.22\textwidth}
        \includegraphics[width=\textwidth]{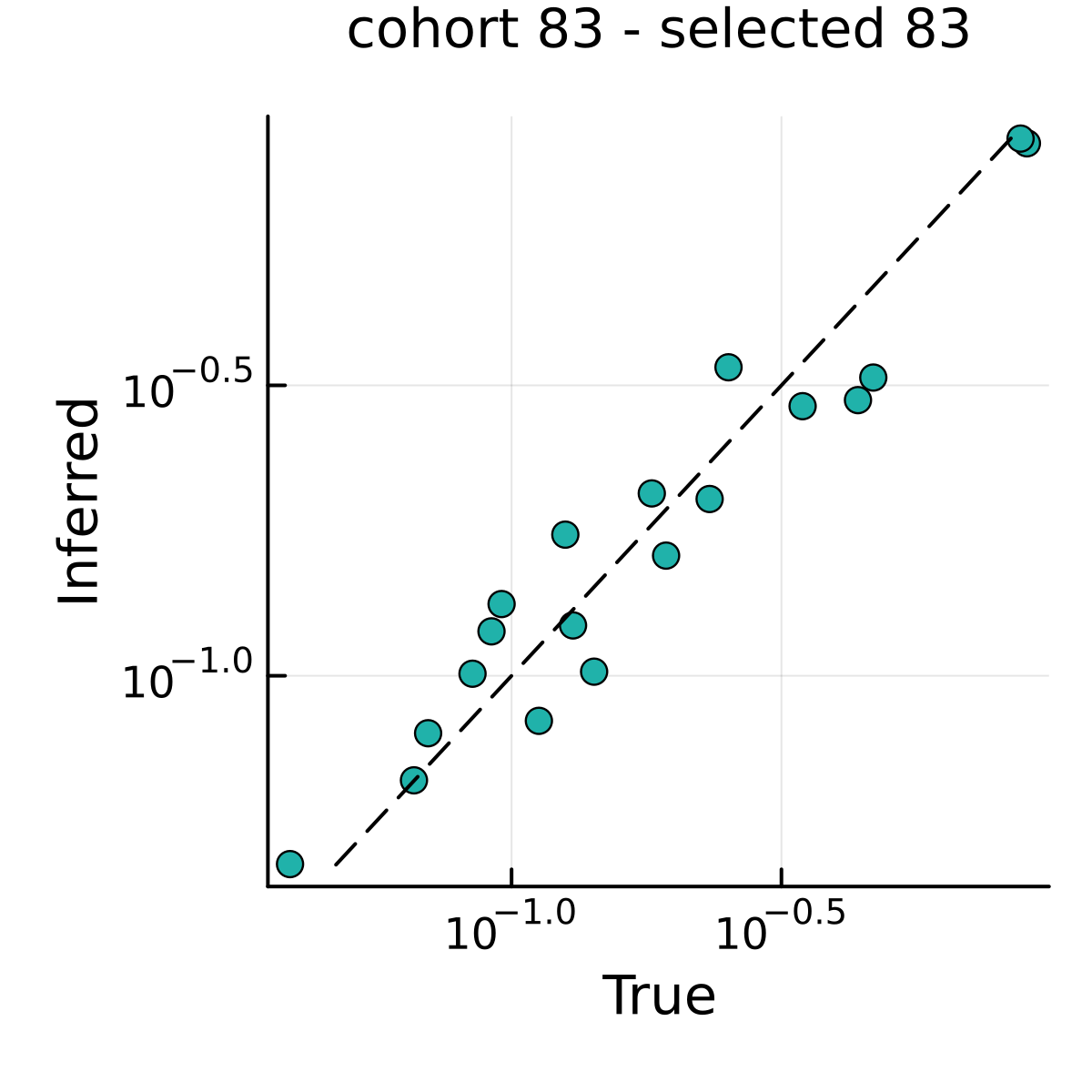}
    \end{subfigure}
     \begin{subfigure}[b]{0.22\textwidth}
        \includegraphics[width=\textwidth]{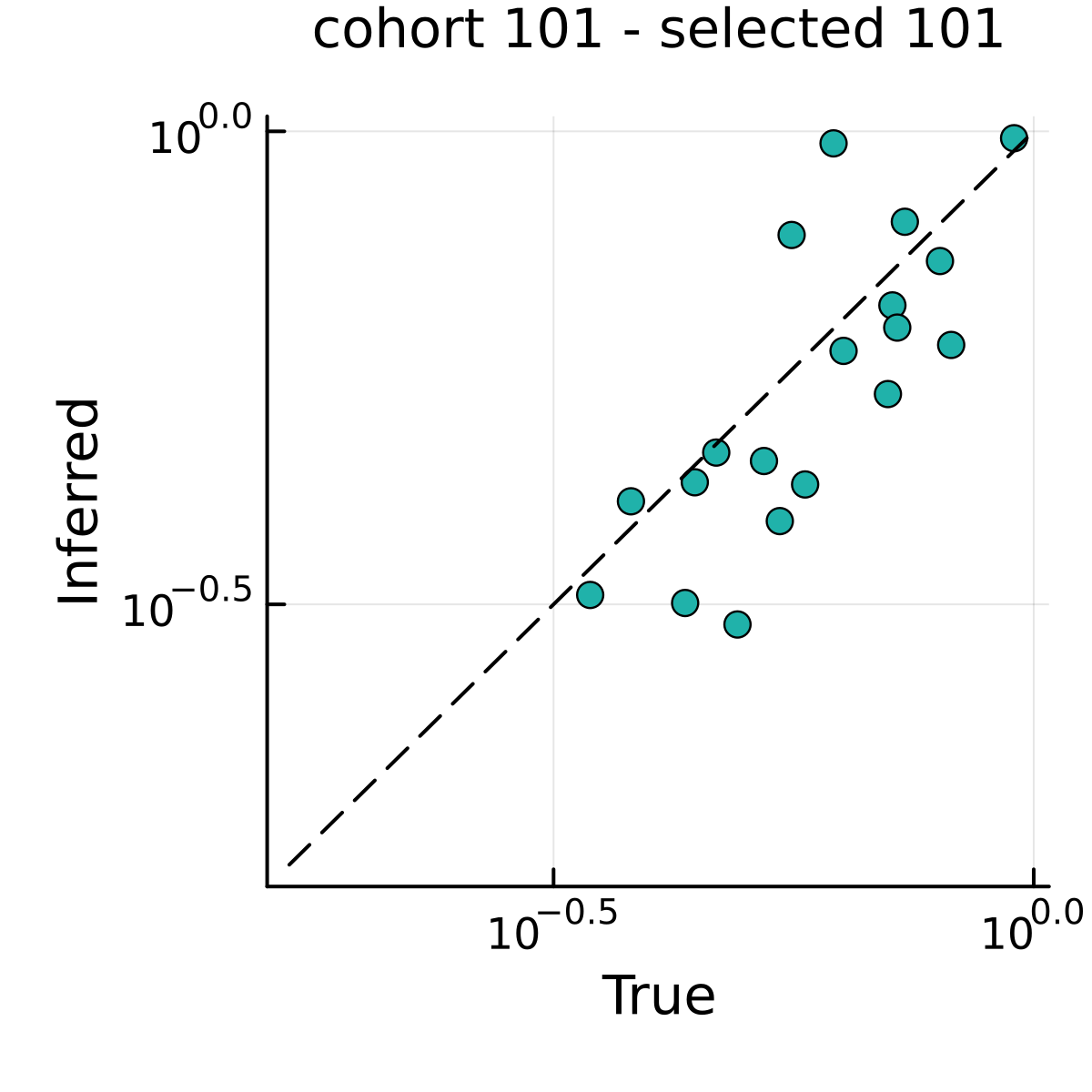}
    \end{subfigure}
     \begin{subfigure}[b]{0.22\textwidth}
        \includegraphics[width=\textwidth]{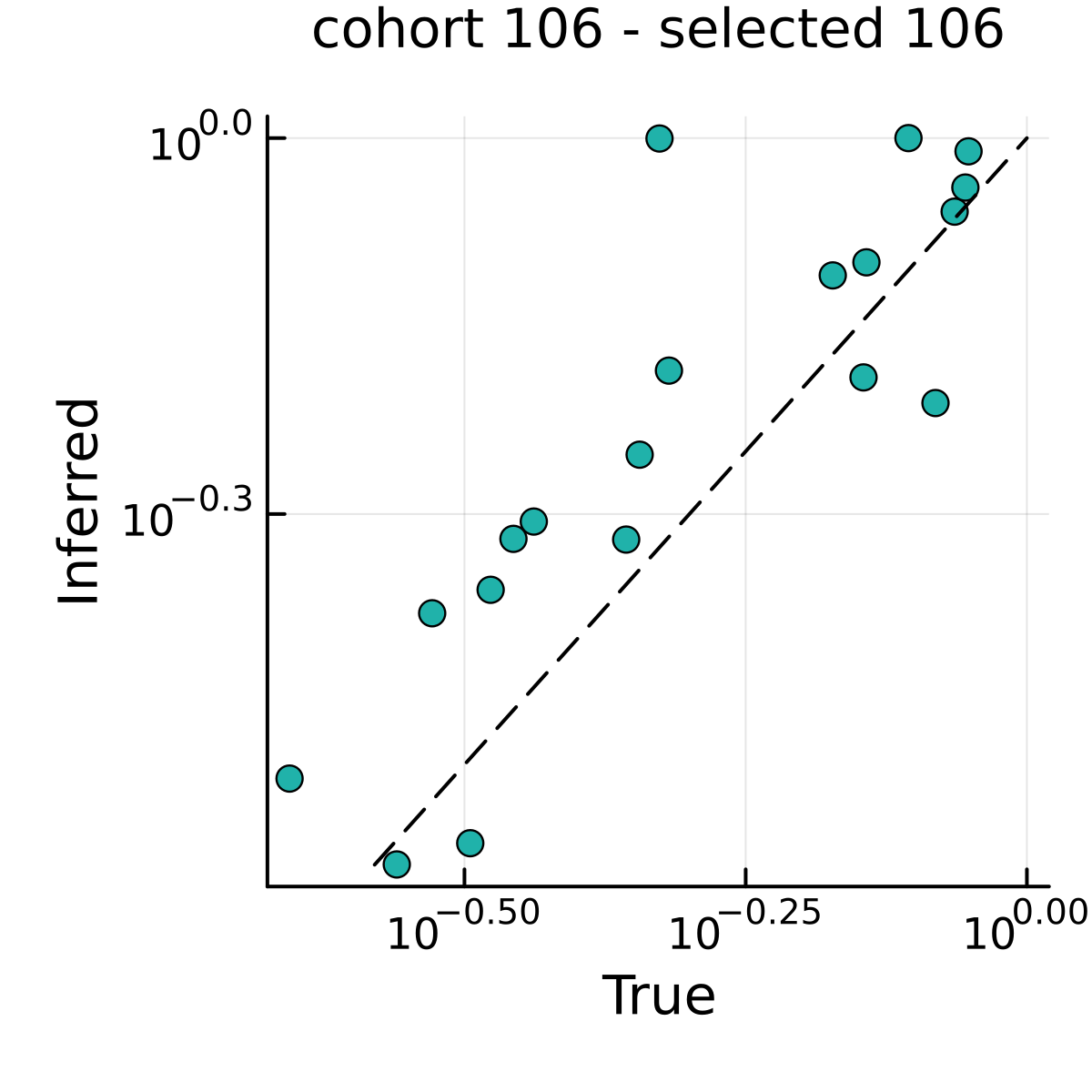}
    \end{subfigure}
     \begin{subfigure}[b]{0.22\textwidth}
        \includegraphics[width=\textwidth]{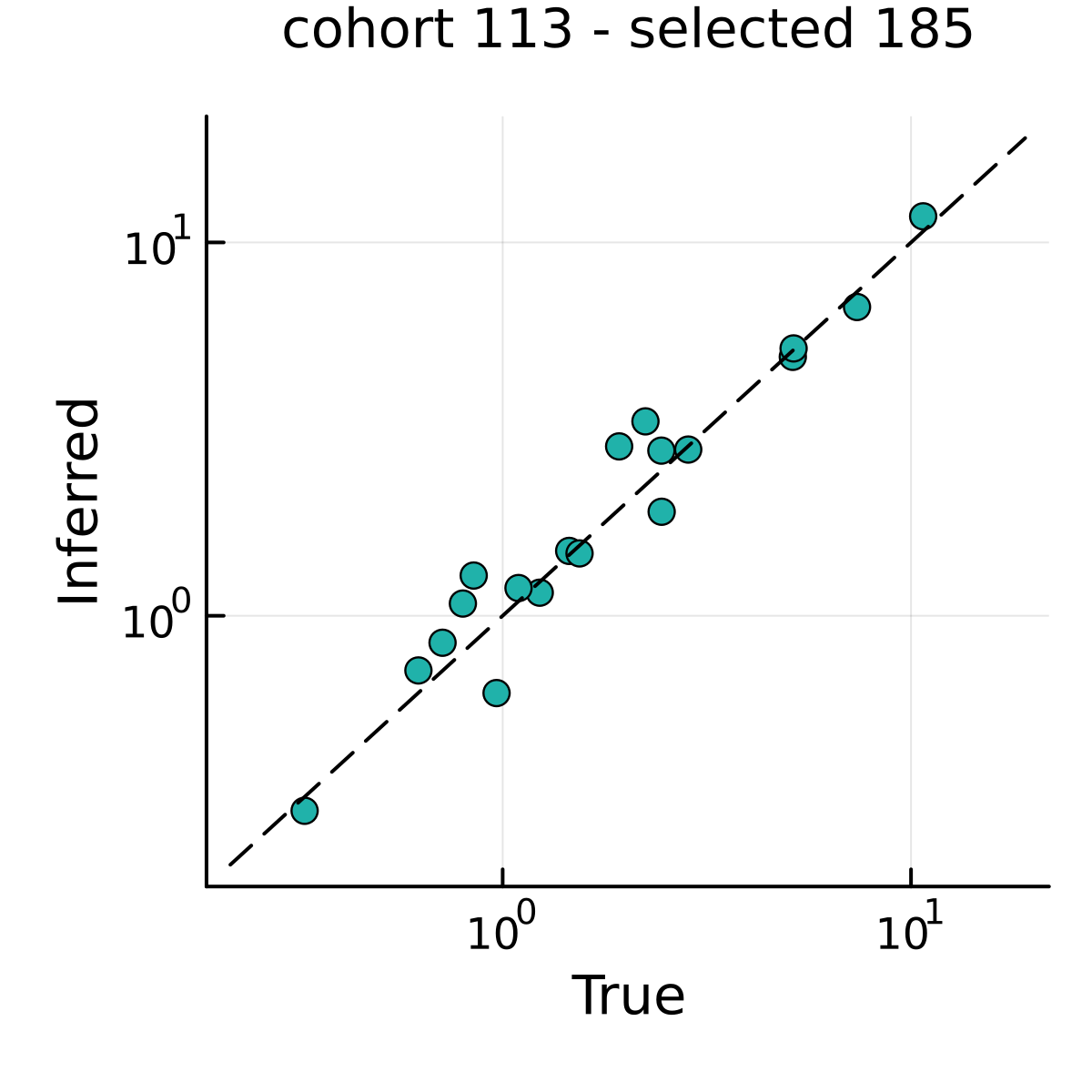}
    \end{subfigure}
     \begin{subfigure}[b]{0.22\textwidth}
        \includegraphics[width=\textwidth]{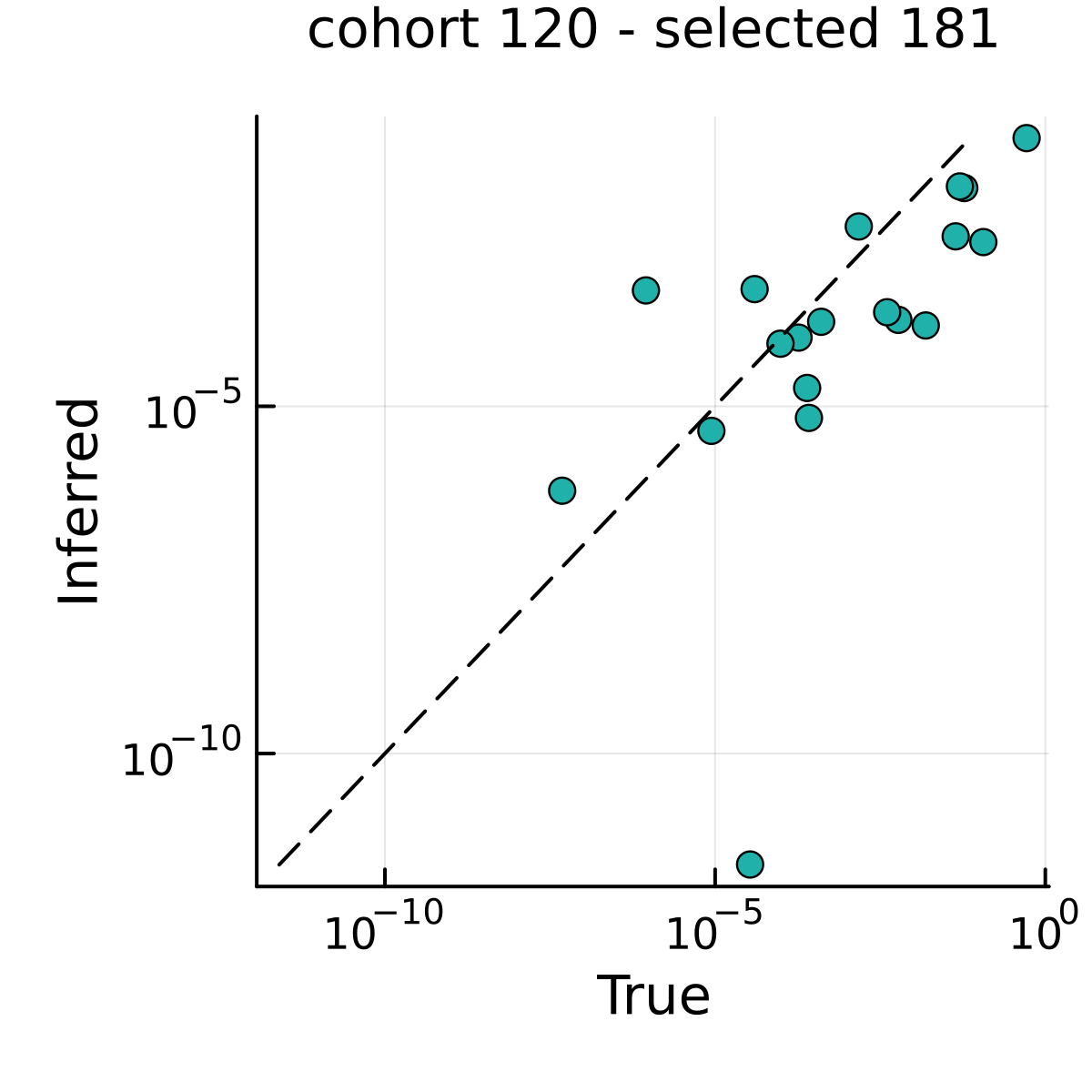}
    \end{subfigure}
     \begin{subfigure}[b]{0.22\textwidth}
        \includegraphics[width=\textwidth]{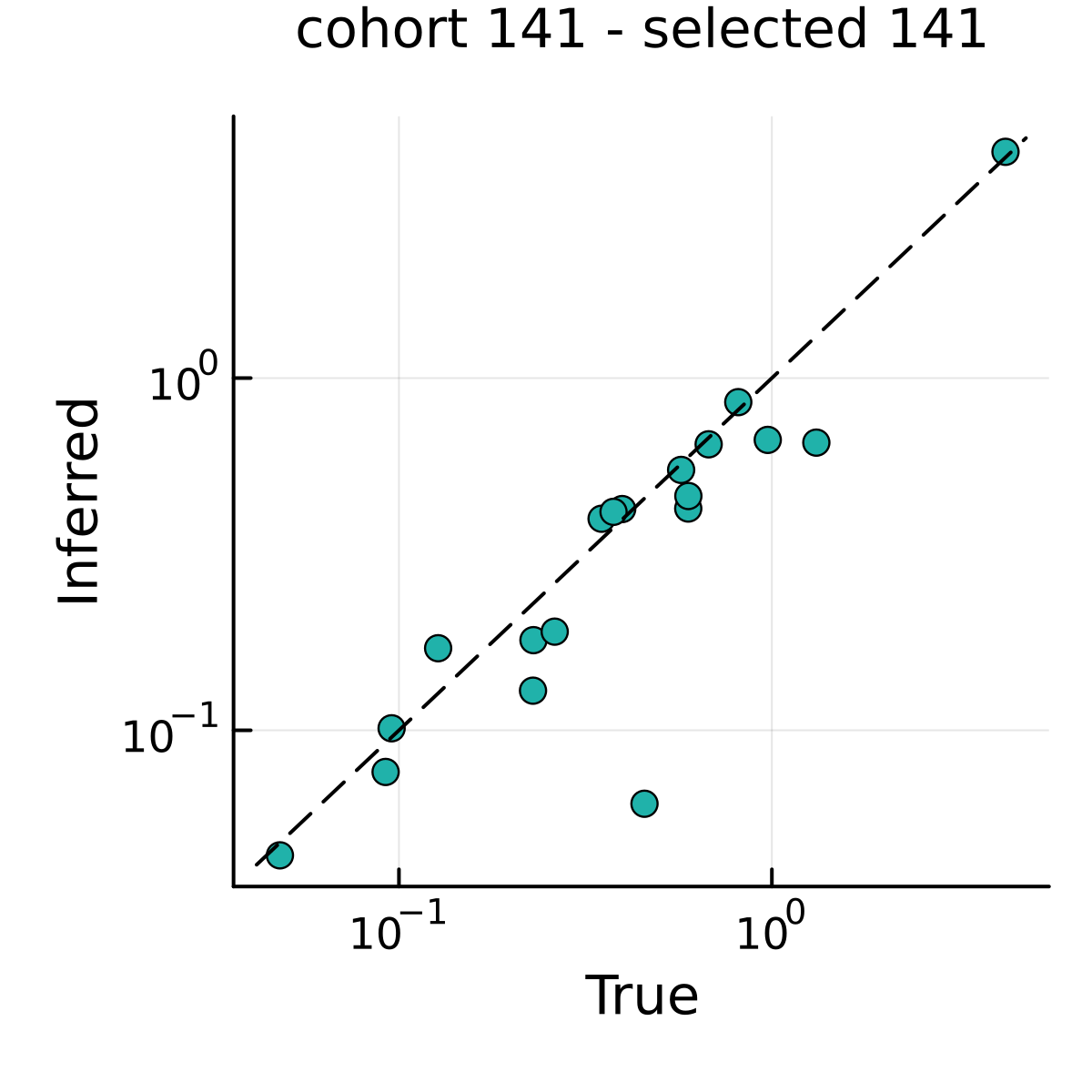}
    \end{subfigure}
     \begin{subfigure}[b]{0.22\textwidth}
        \includegraphics[width=\textwidth]{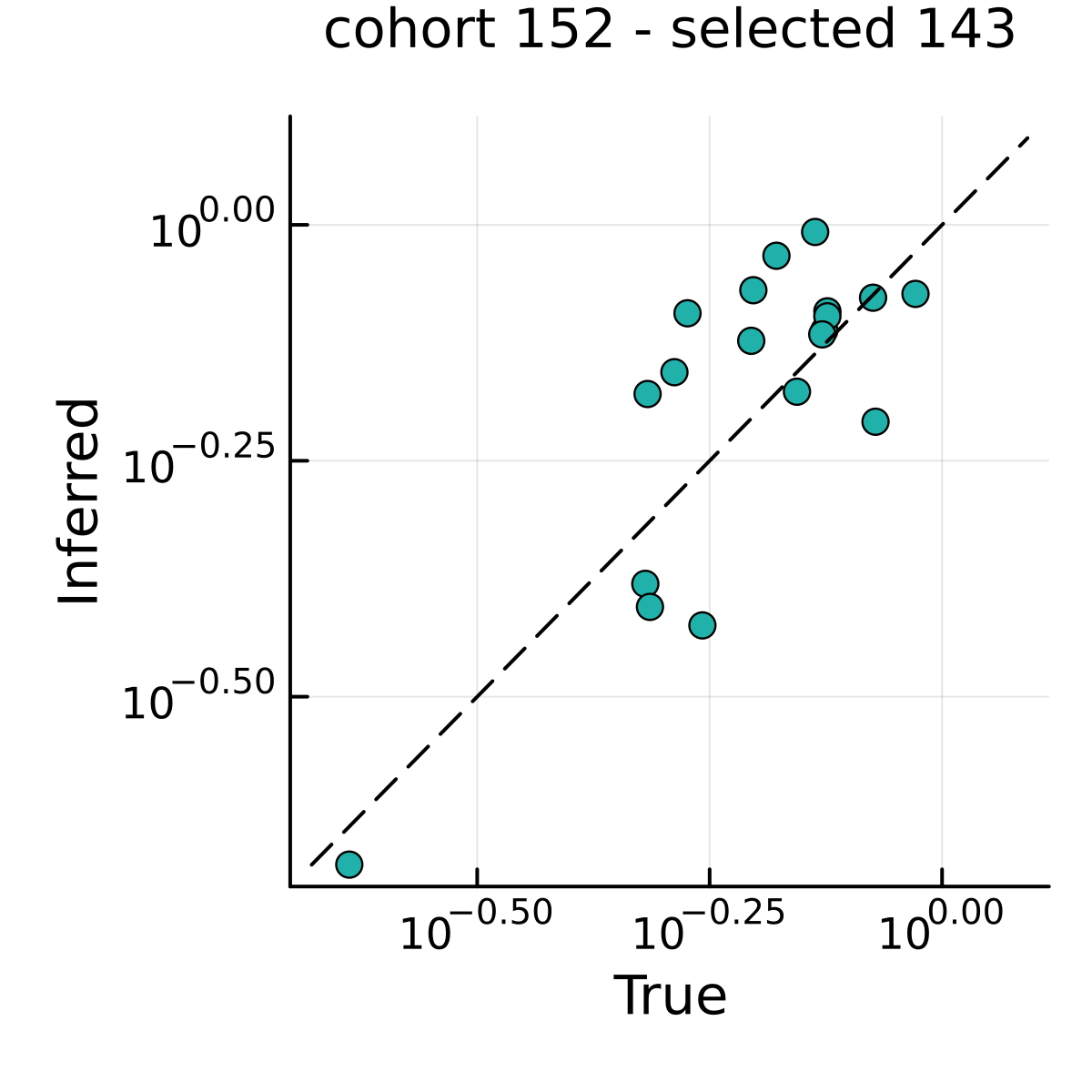}
    \end{subfigure}
     \begin{subfigure}[b]{0.22\textwidth}
        \includegraphics[width=\textwidth]{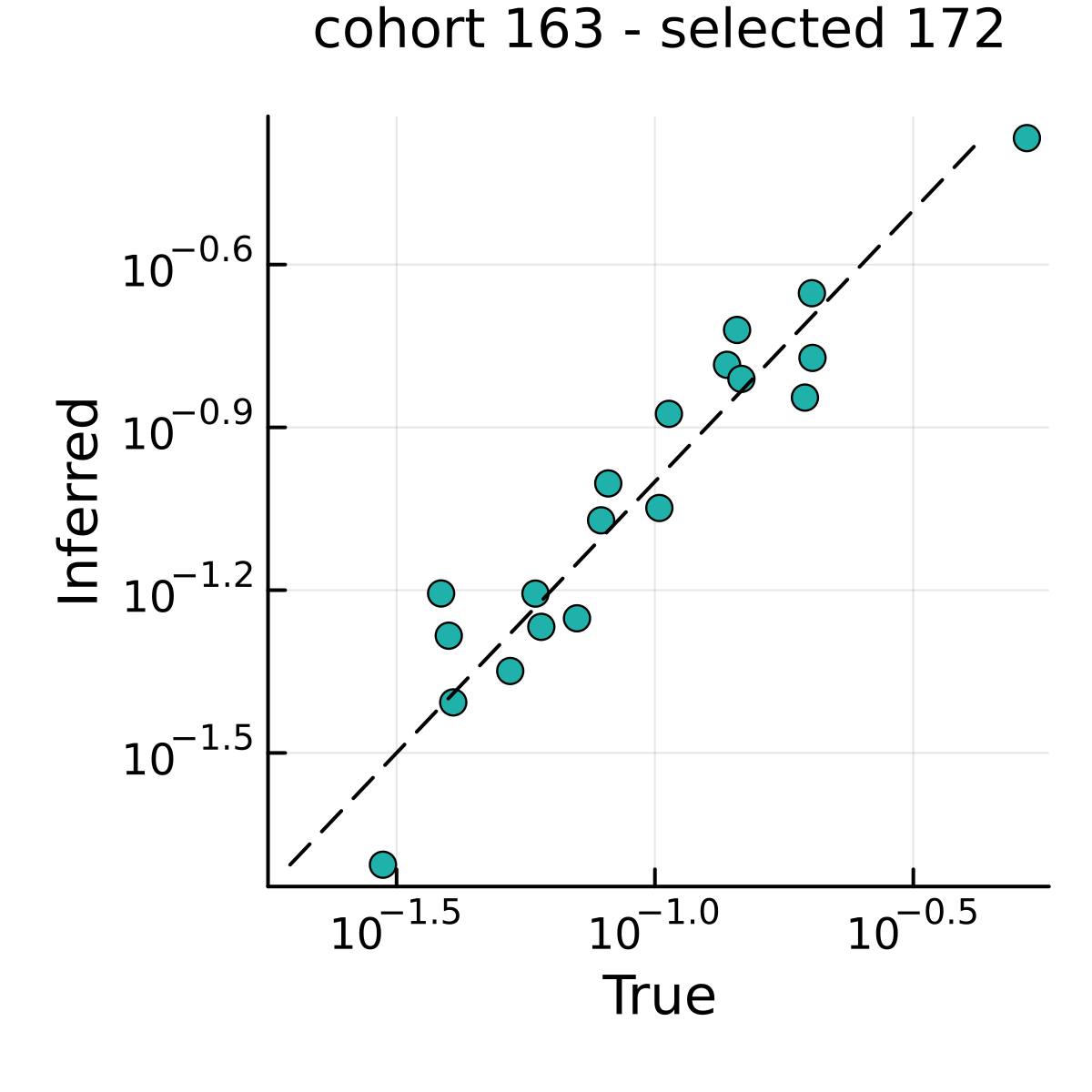}
    \end{subfigure}
     \begin{subfigure}[b]{0.22\textwidth}
        \includegraphics[width=\textwidth]{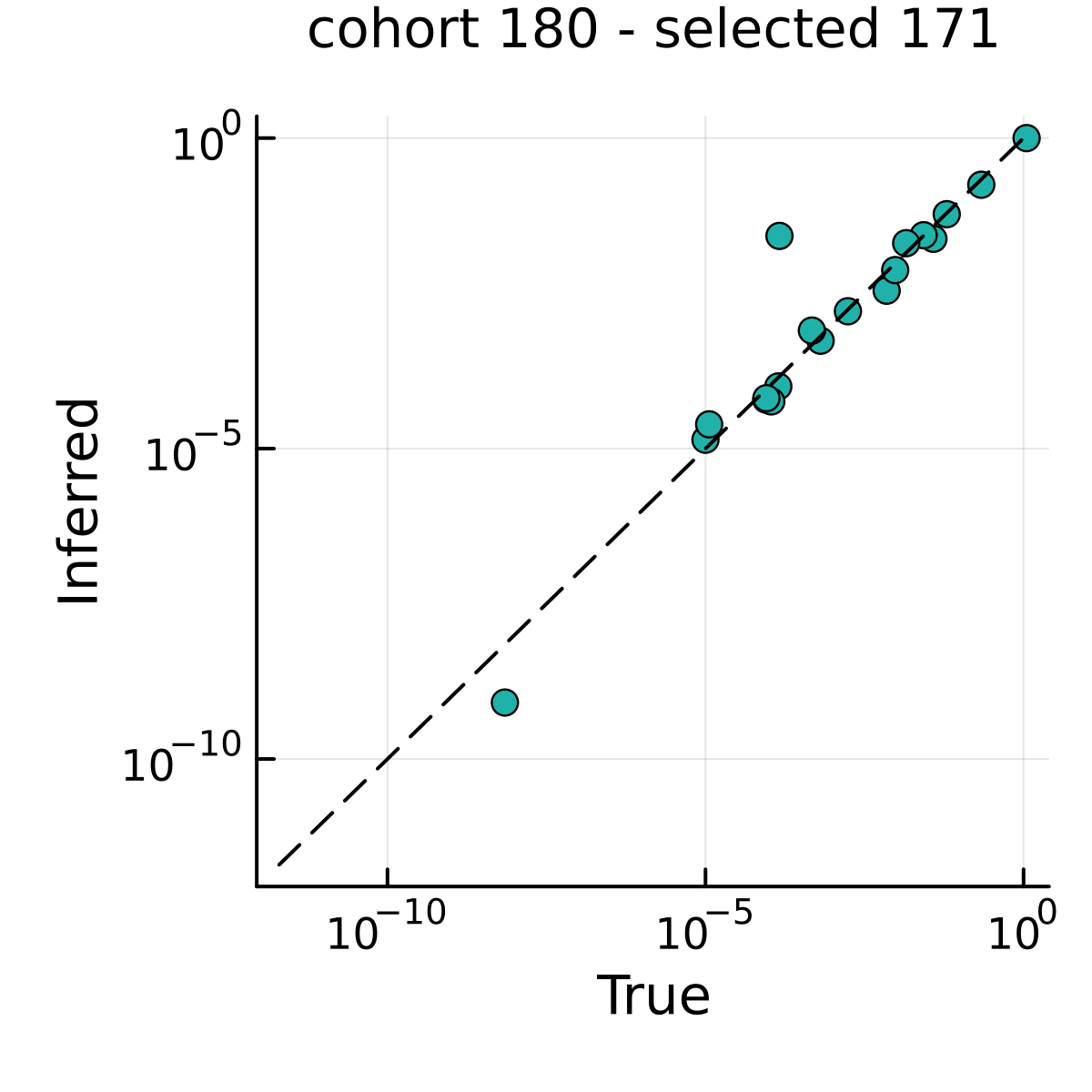}
    \end{subfigure}
     \begin{subfigure}[b]{0.22\textwidth}
        \includegraphics[width=\textwidth]{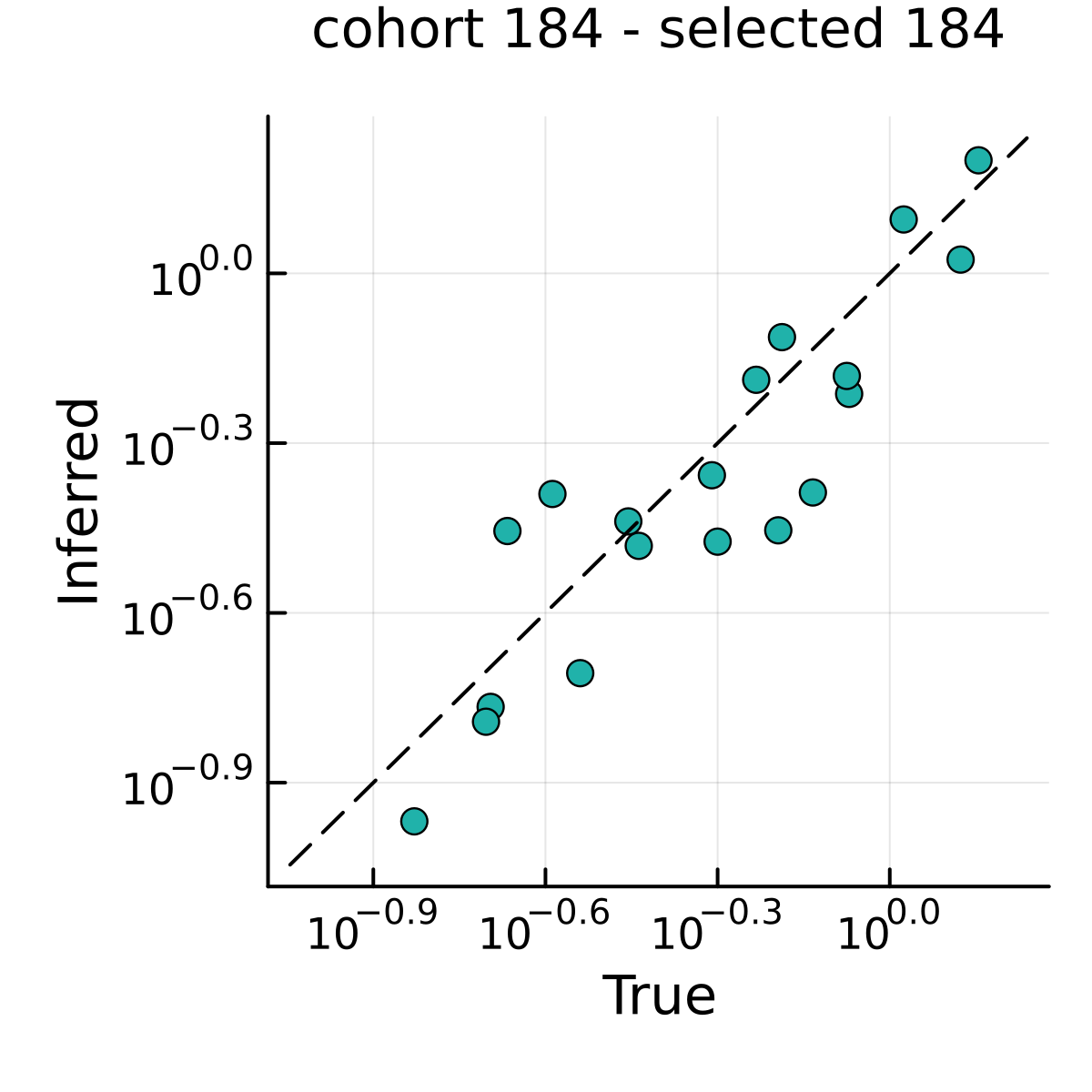}
    \end{subfigure} 
    \begin{subfigure}[b]{0.22\textwidth}
        \includegraphics[width=\textwidth]{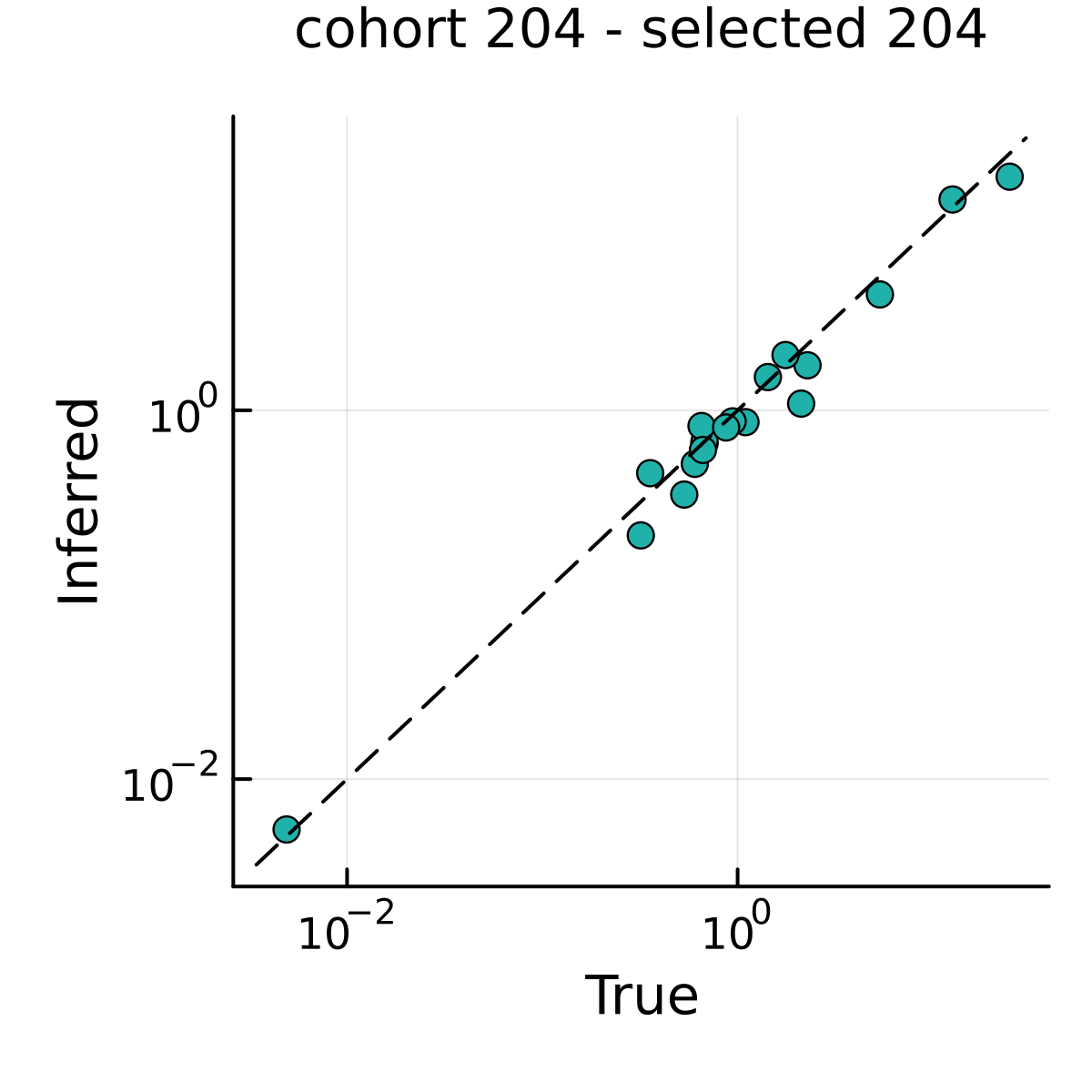}
    \end{subfigure}
     \begin{subfigure}[b]{0.22\textwidth}
        \includegraphics[width=\textwidth]{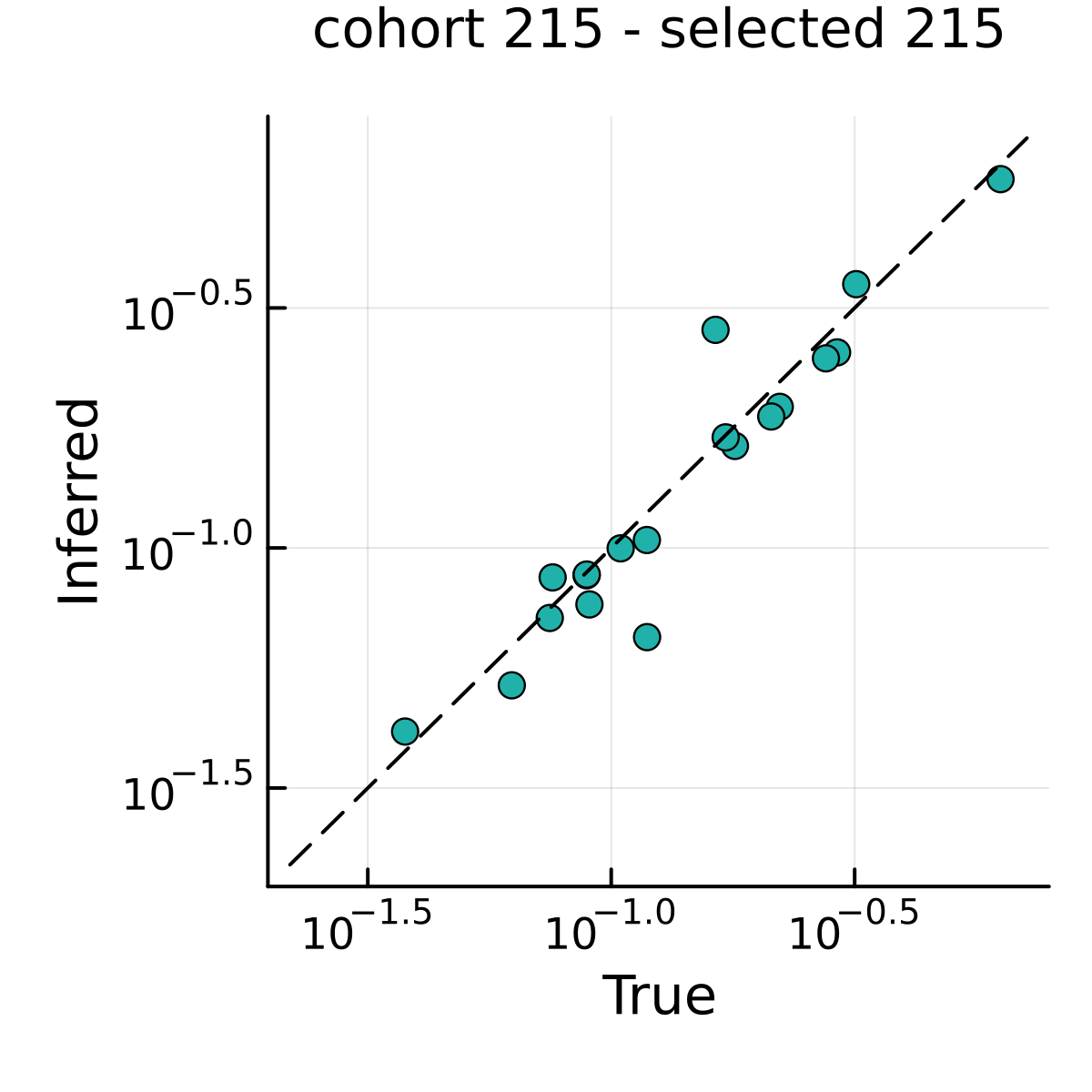}
    \end{subfigure}
     \begin{subfigure}[b]{0.22\textwidth}
        \includegraphics[width=\textwidth]{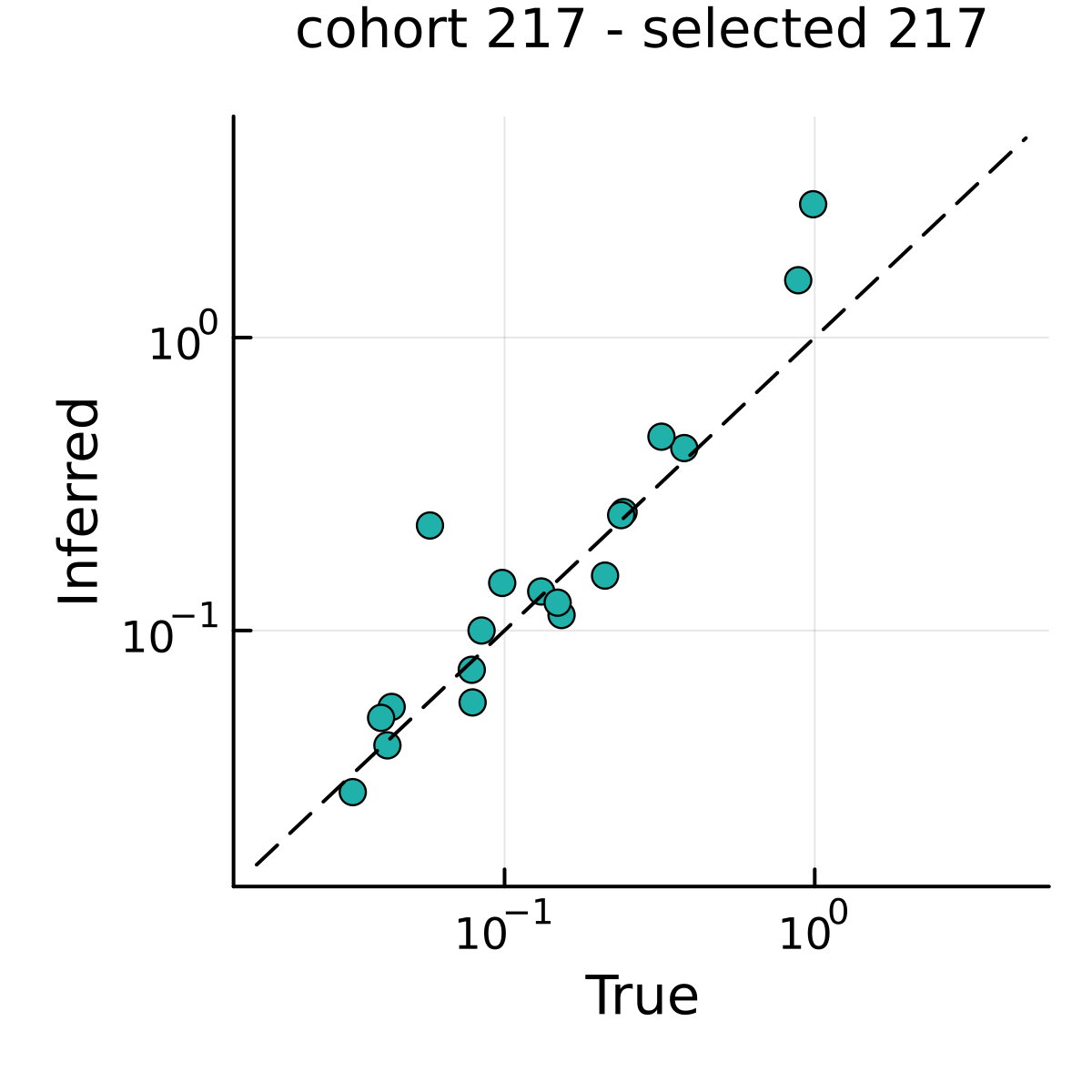}
    \end{subfigure}
    
    \caption{Comparison between the inferred (mean posterior, y-axis) and the true (x-axis) value of $R_{het}(t=1,500)$. The axes are in a log scale.}
    \label{fig:synth_R_het}
\end{figure}

\begin{figure}[h]
    \centering

    \begin{subfigure}[b]{0.22\textwidth}
        \includegraphics[width=\textwidth]{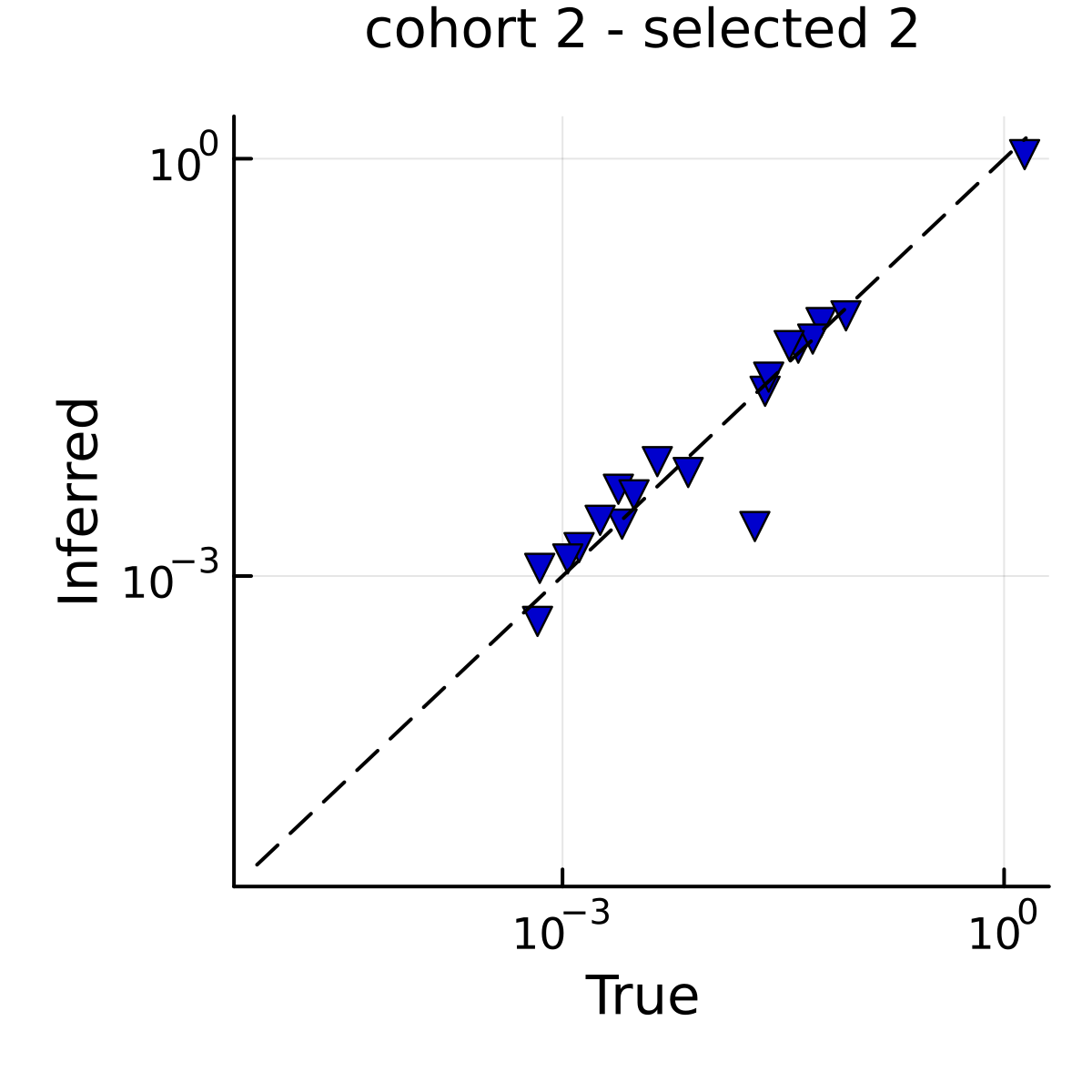}
    \end{subfigure}
     \begin{subfigure}[b]{0.22\textwidth}
        \includegraphics[width=\textwidth]{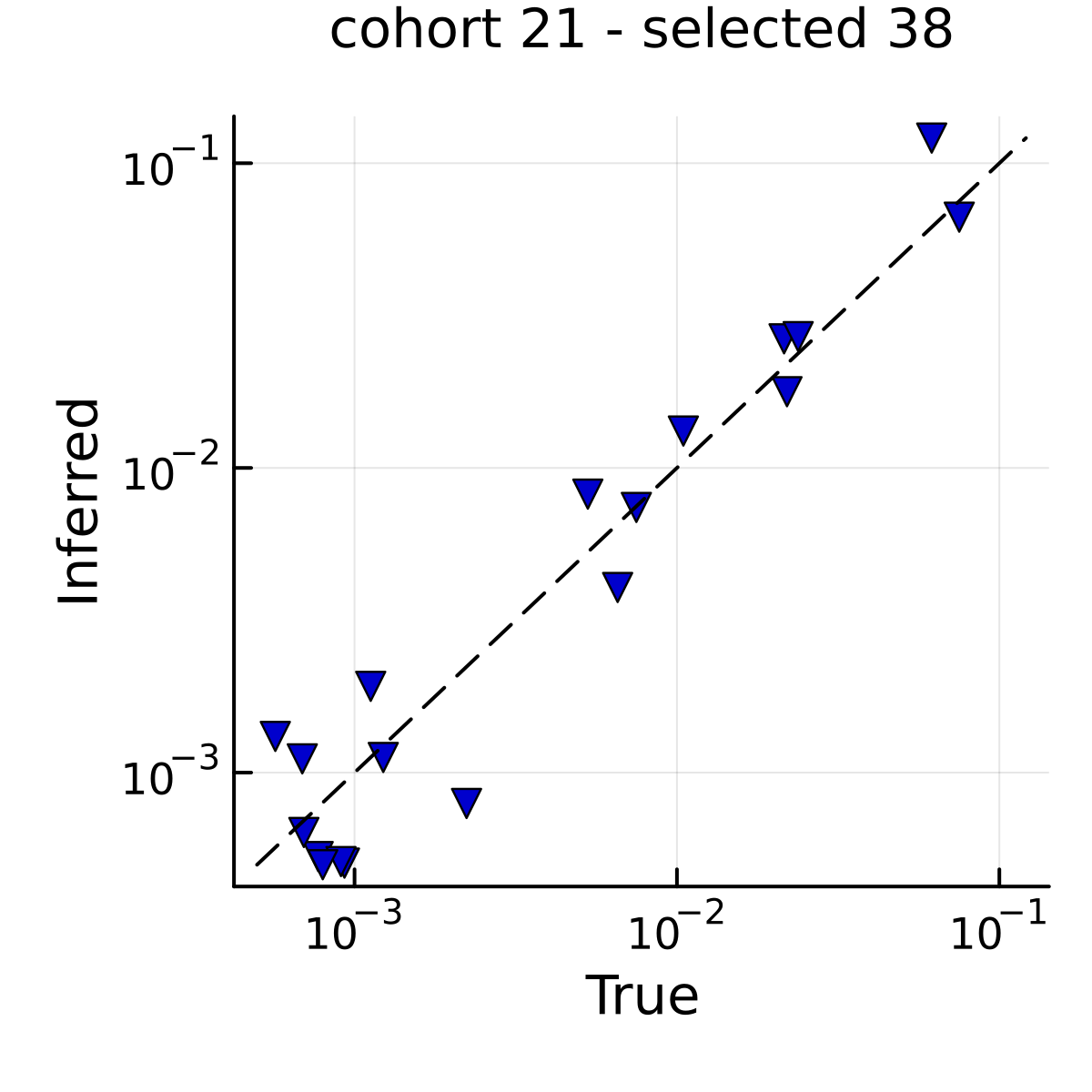}
    \end{subfigure}
     \begin{subfigure}[b]{0.22\textwidth}
        \includegraphics[width=\textwidth]{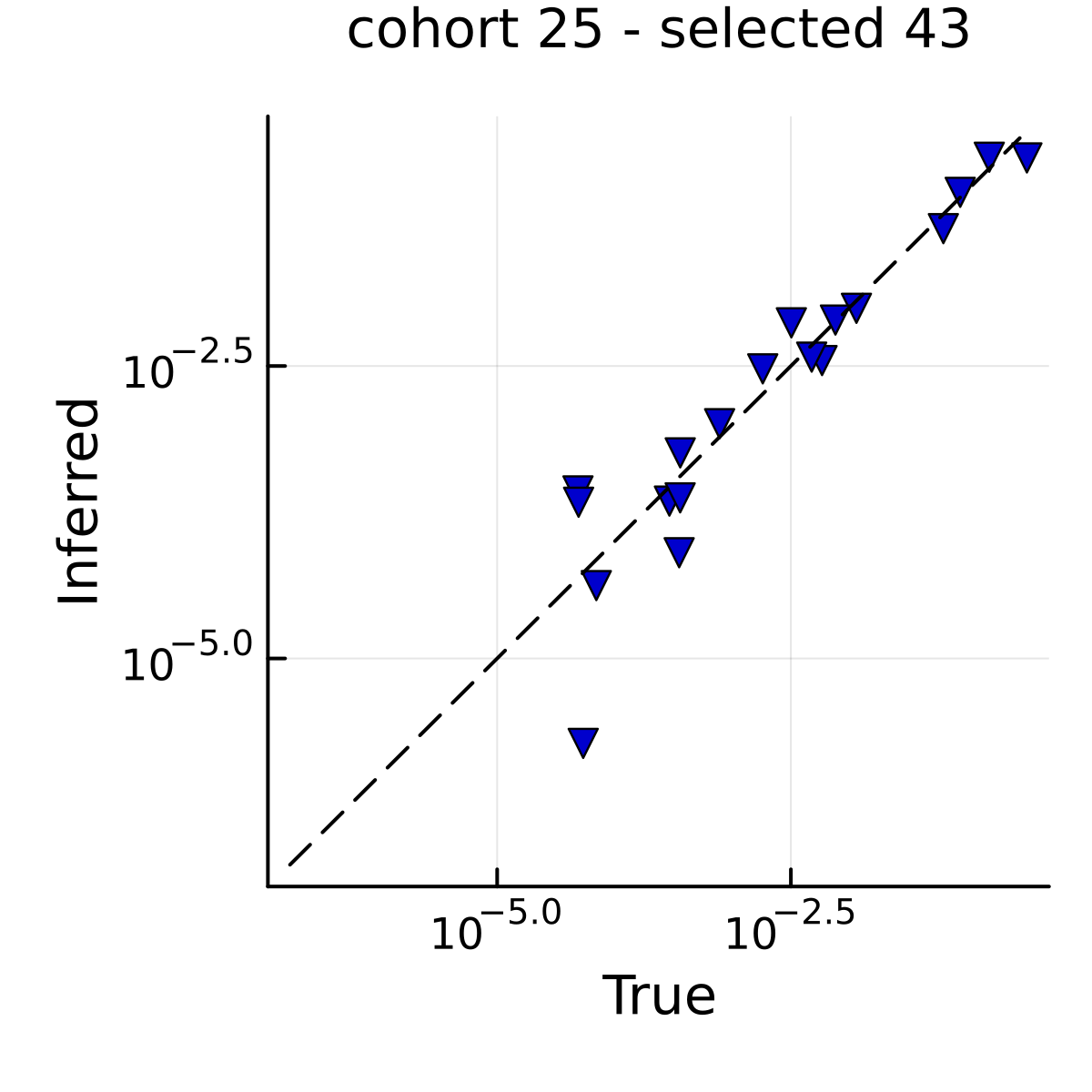}
    \end{subfigure}
     \begin{subfigure}[b]{0.22\textwidth}
        \includegraphics[width=\textwidth]{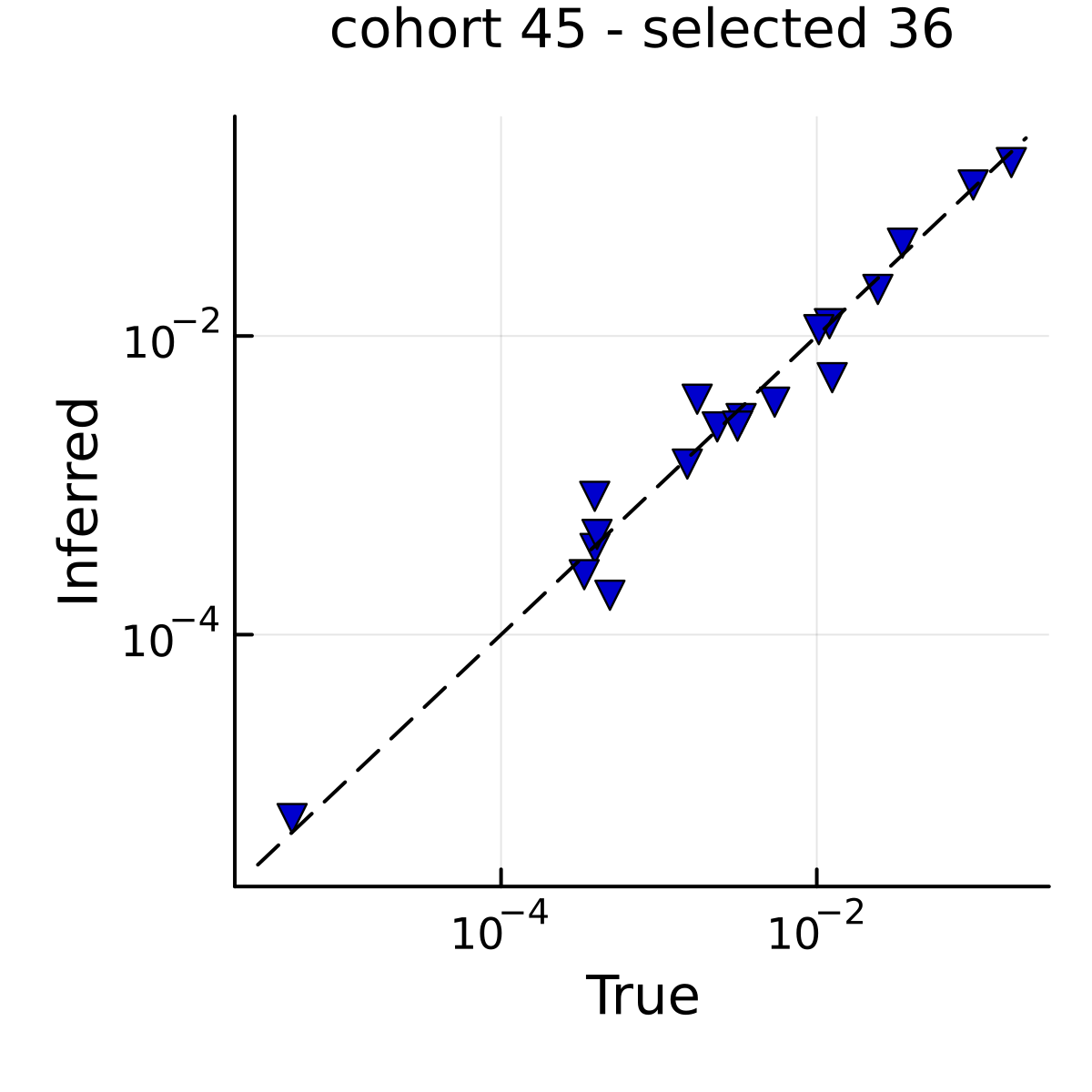}
    \end{subfigure}
     \begin{subfigure}[b]{0.22\textwidth}
        \includegraphics[width=\textwidth]{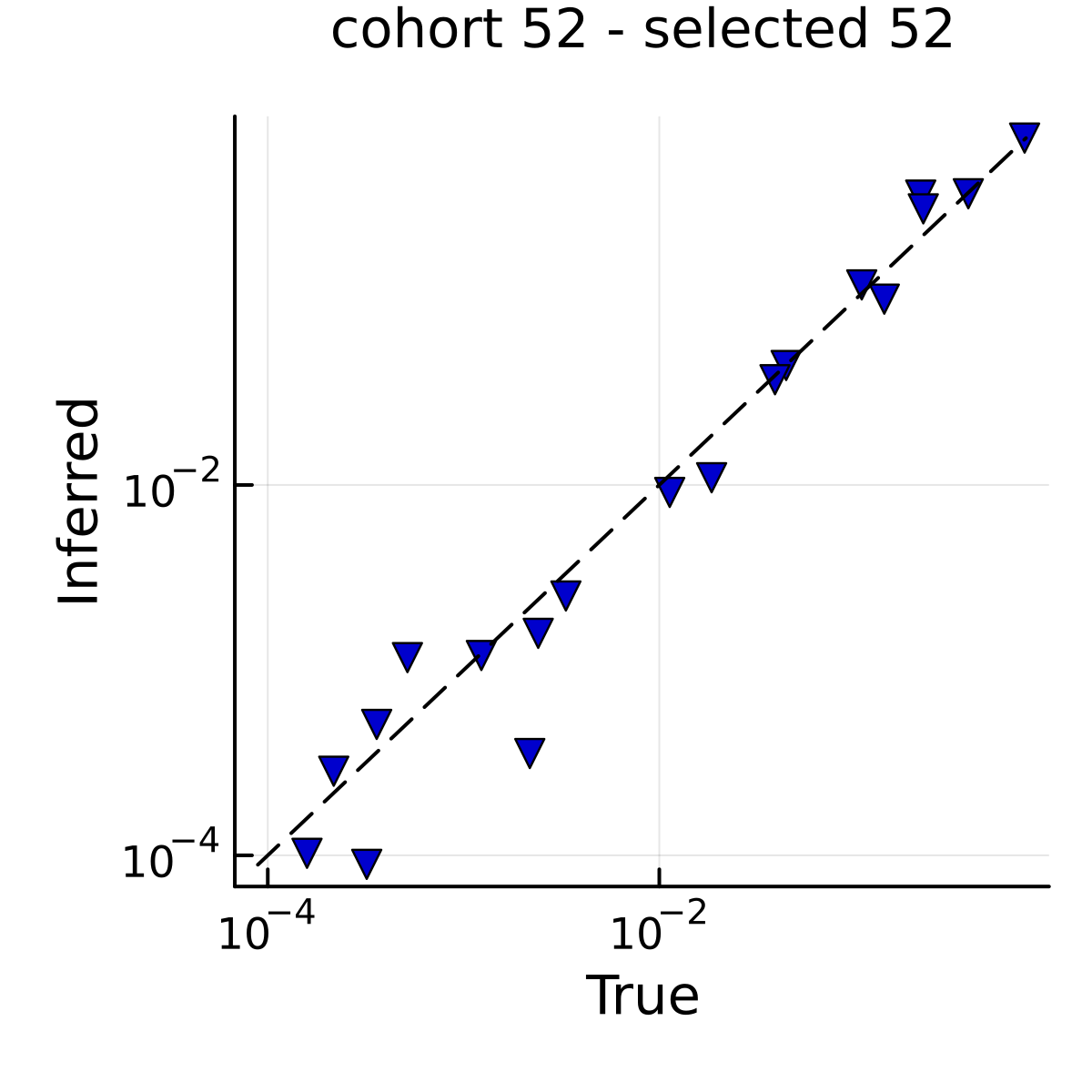}
    \end{subfigure}
     \begin{subfigure}[b]{0.22\textwidth}
        \includegraphics[width=\textwidth]{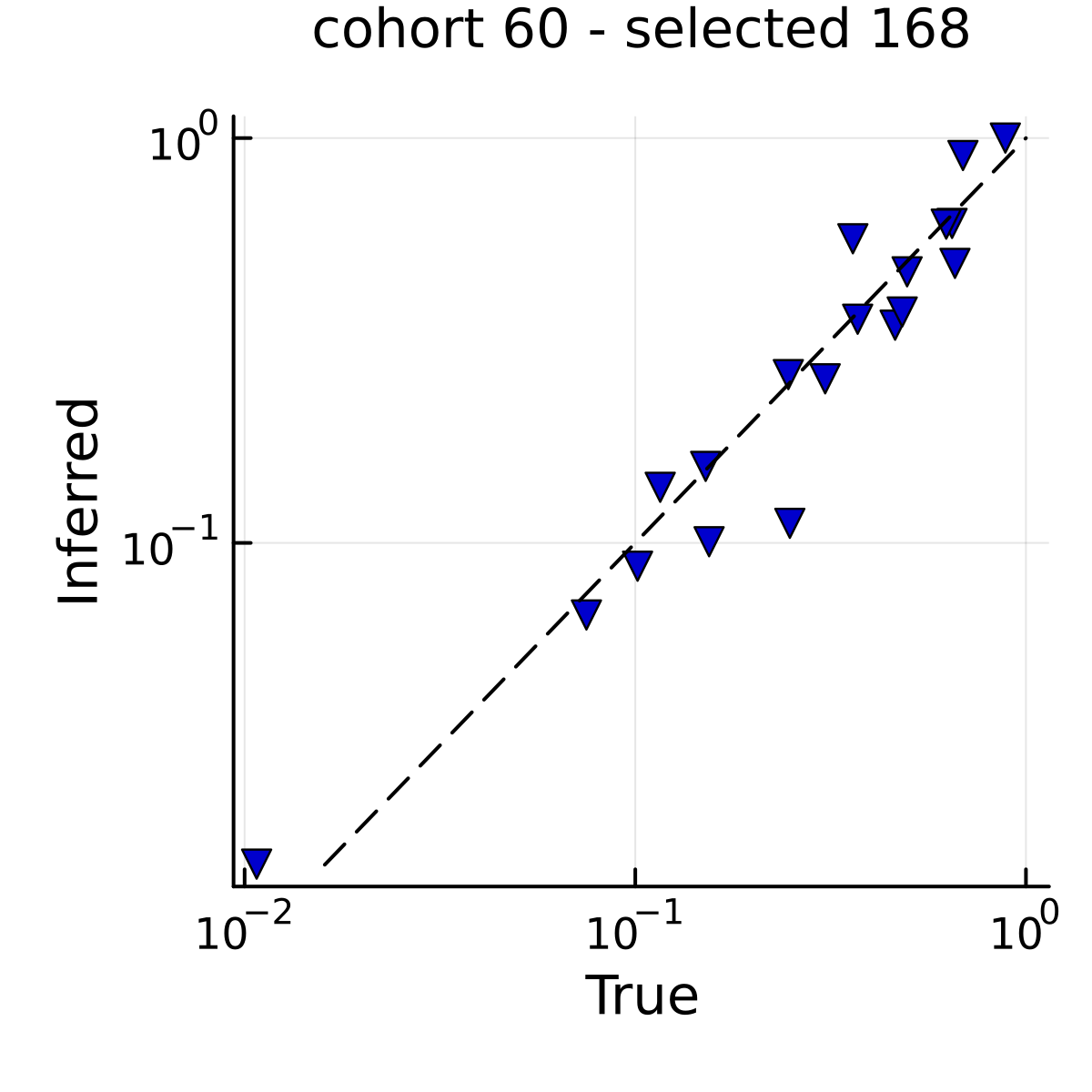}
    \end{subfigure}
     \begin{subfigure}[b]{0.22\textwidth}
        \includegraphics[width=\textwidth]{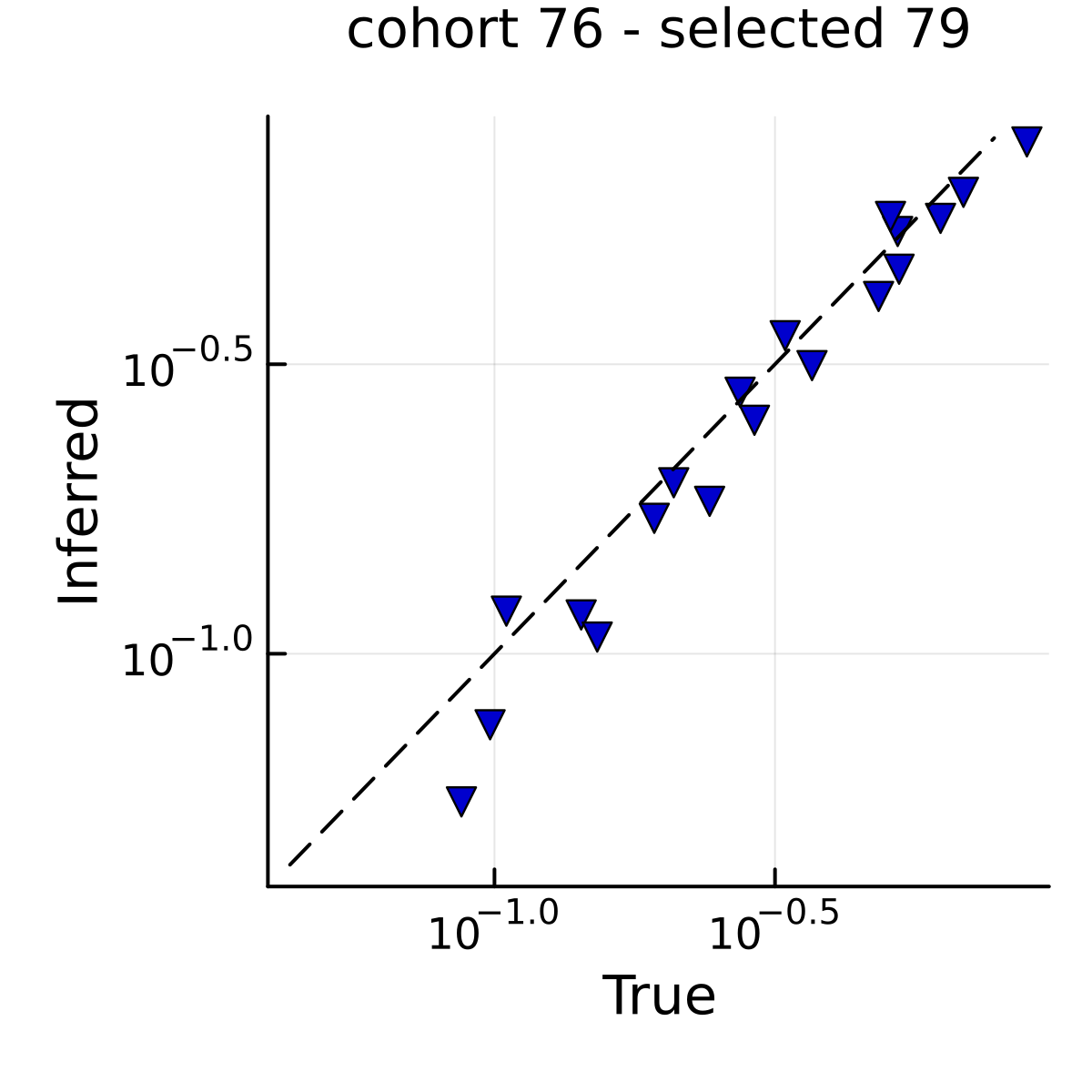}
    \end{subfigure}
     \begin{subfigure}[b]{0.22\textwidth}
        \includegraphics[width=\textwidth]{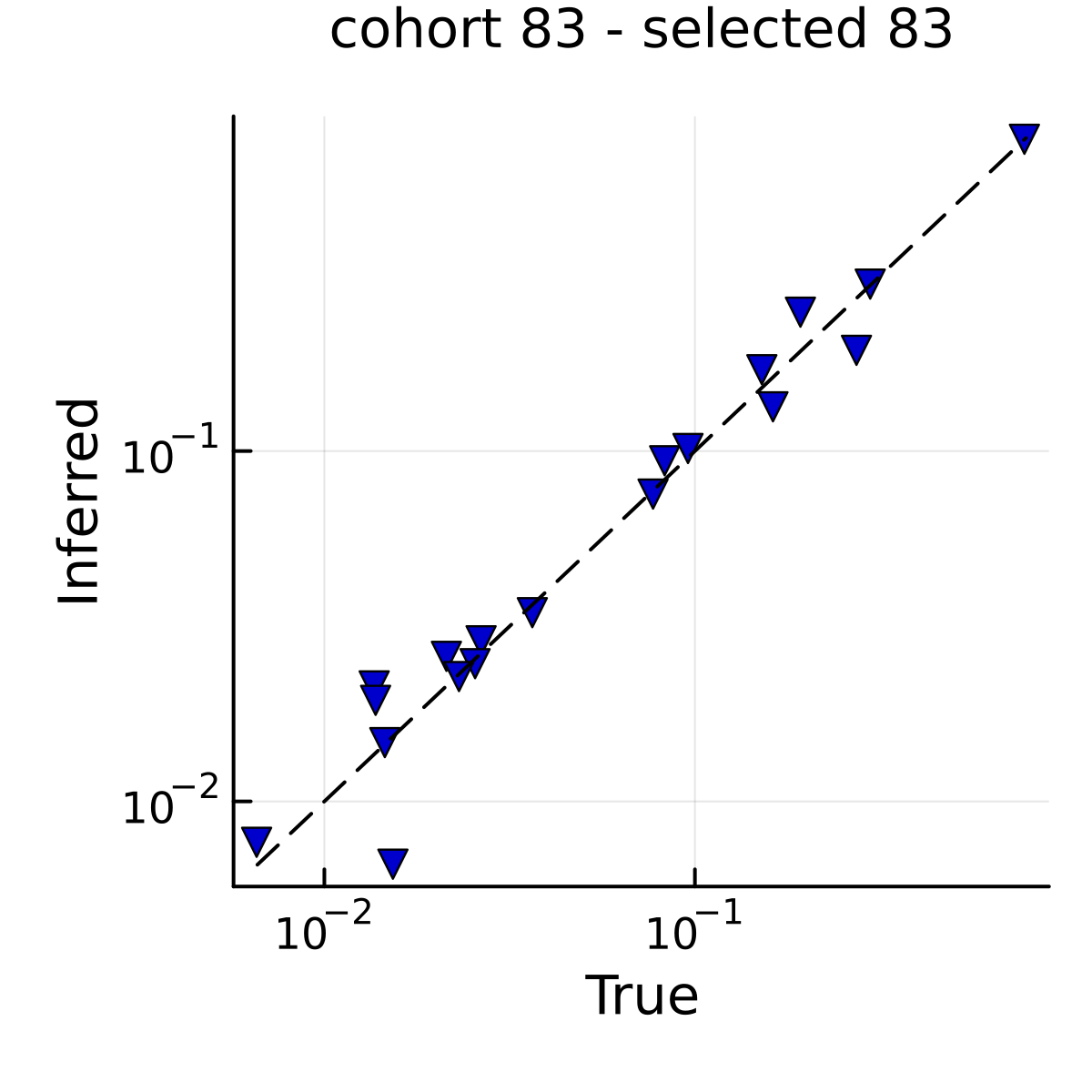}
    \end{subfigure}
     \begin{subfigure}[b]{0.22\textwidth}
        \includegraphics[width=\textwidth]{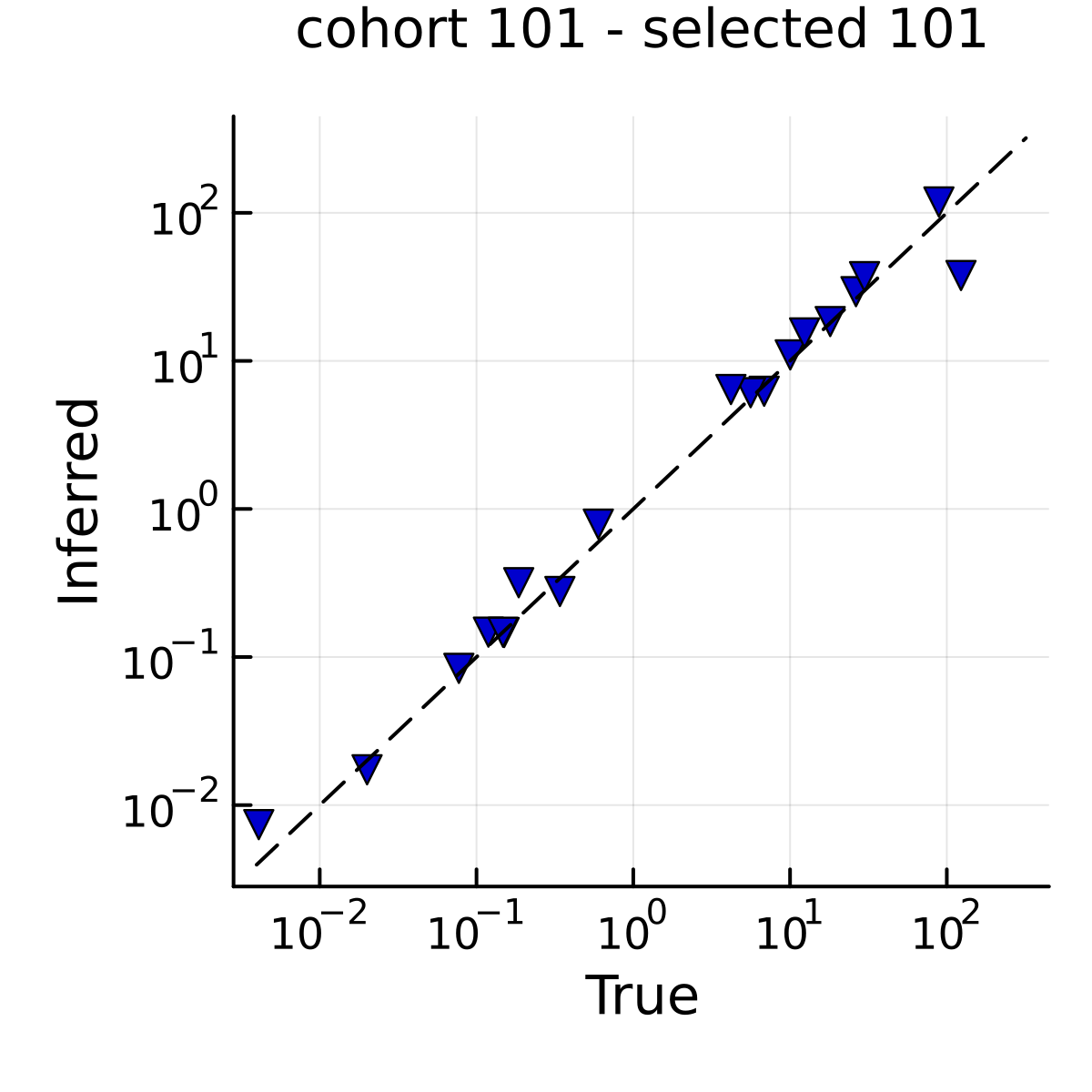}
    \end{subfigure}
     \begin{subfigure}[b]{0.22\textwidth}
        \includegraphics[width=\textwidth]{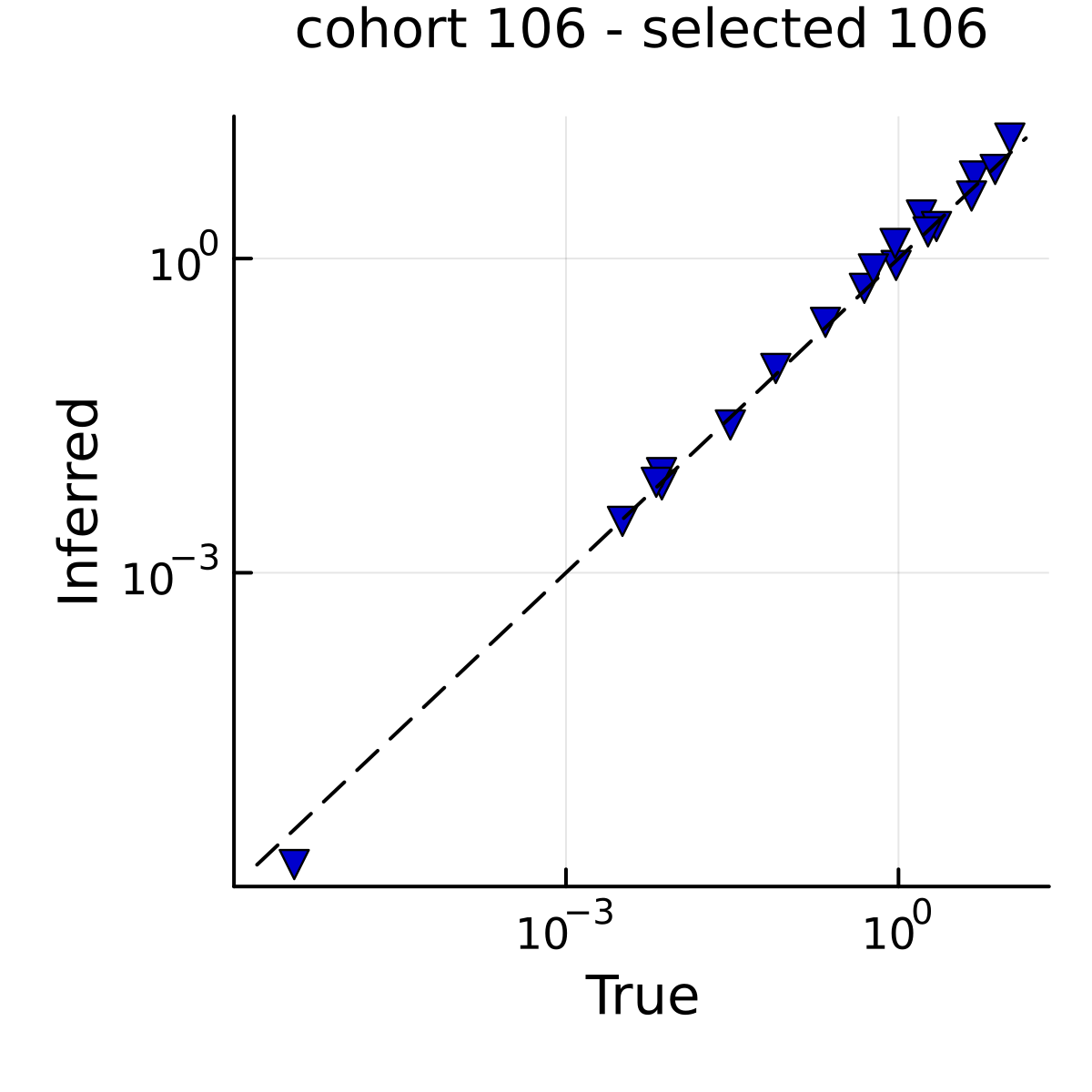}
    \end{subfigure}
     \begin{subfigure}[b]{0.22\textwidth}
        \includegraphics[width=\textwidth]{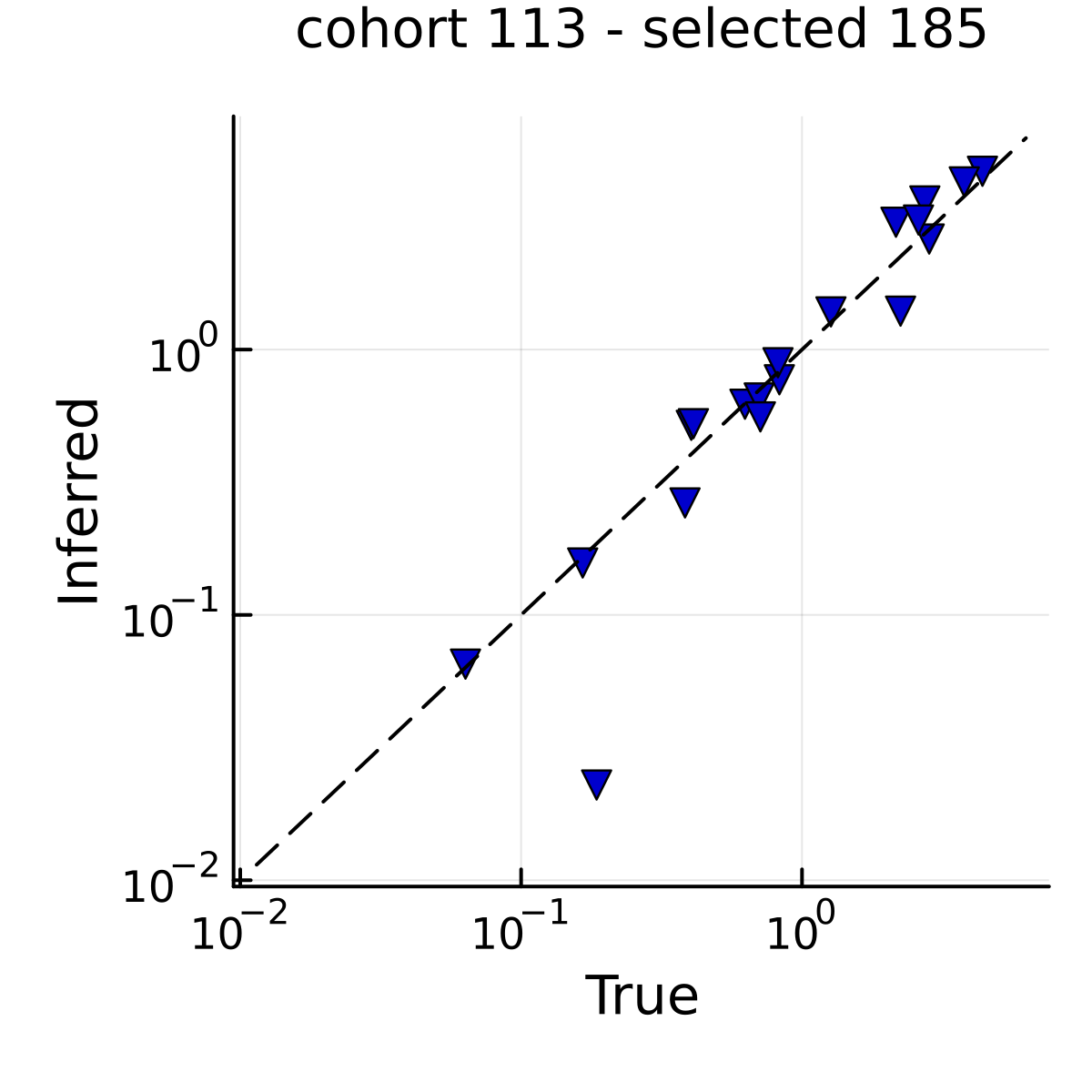}
    \end{subfigure}
     \begin{subfigure}[b]{0.22\textwidth}
        \includegraphics[width=\textwidth]{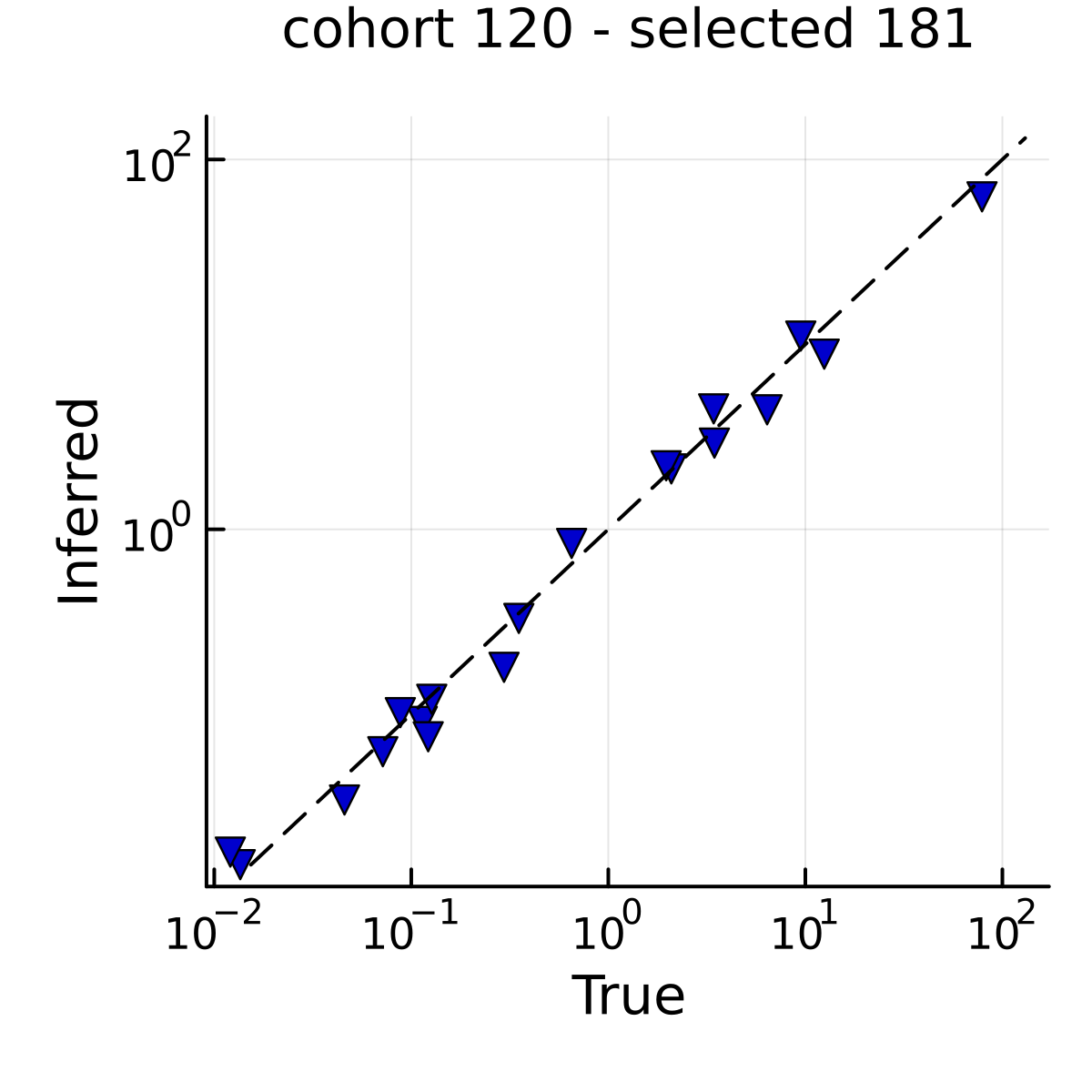}
    \end{subfigure}
     \begin{subfigure}[b]{0.22\textwidth}
        \includegraphics[width=\textwidth]{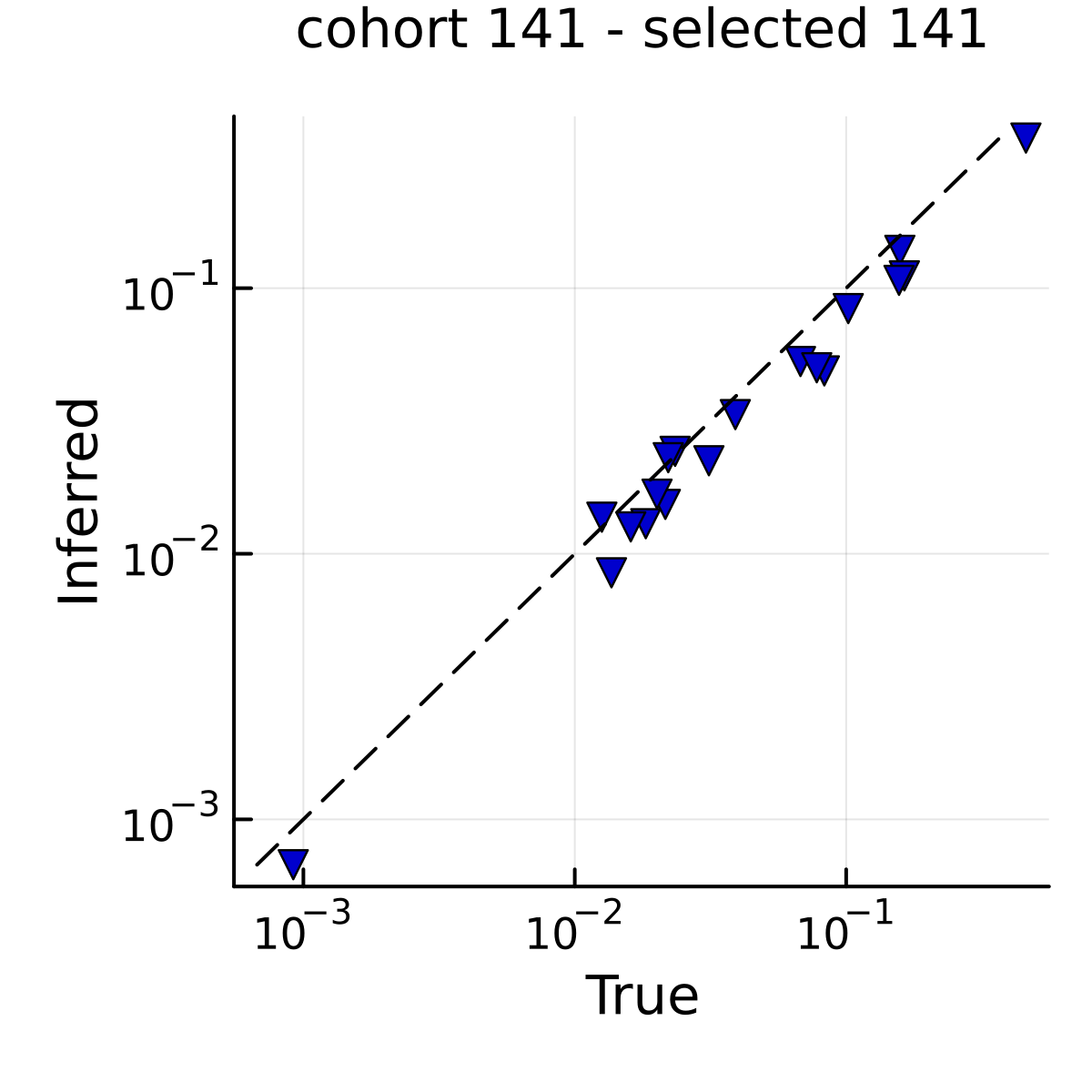}
    \end{subfigure}
     \begin{subfigure}[b]{0.22\textwidth}
        \includegraphics[width=\textwidth]{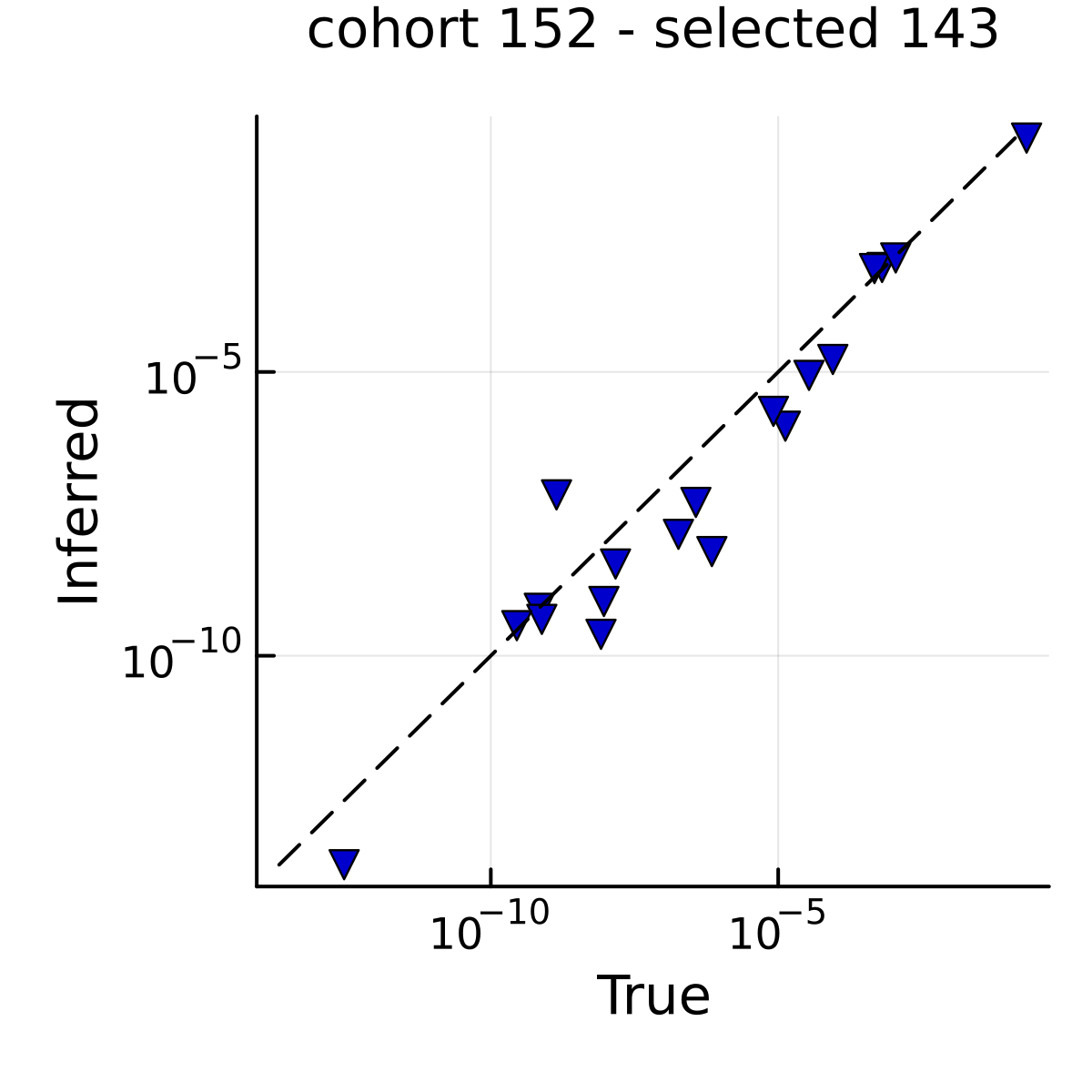}
    \end{subfigure}
     \begin{subfigure}[b]{0.22\textwidth}
        \includegraphics[width=\textwidth]{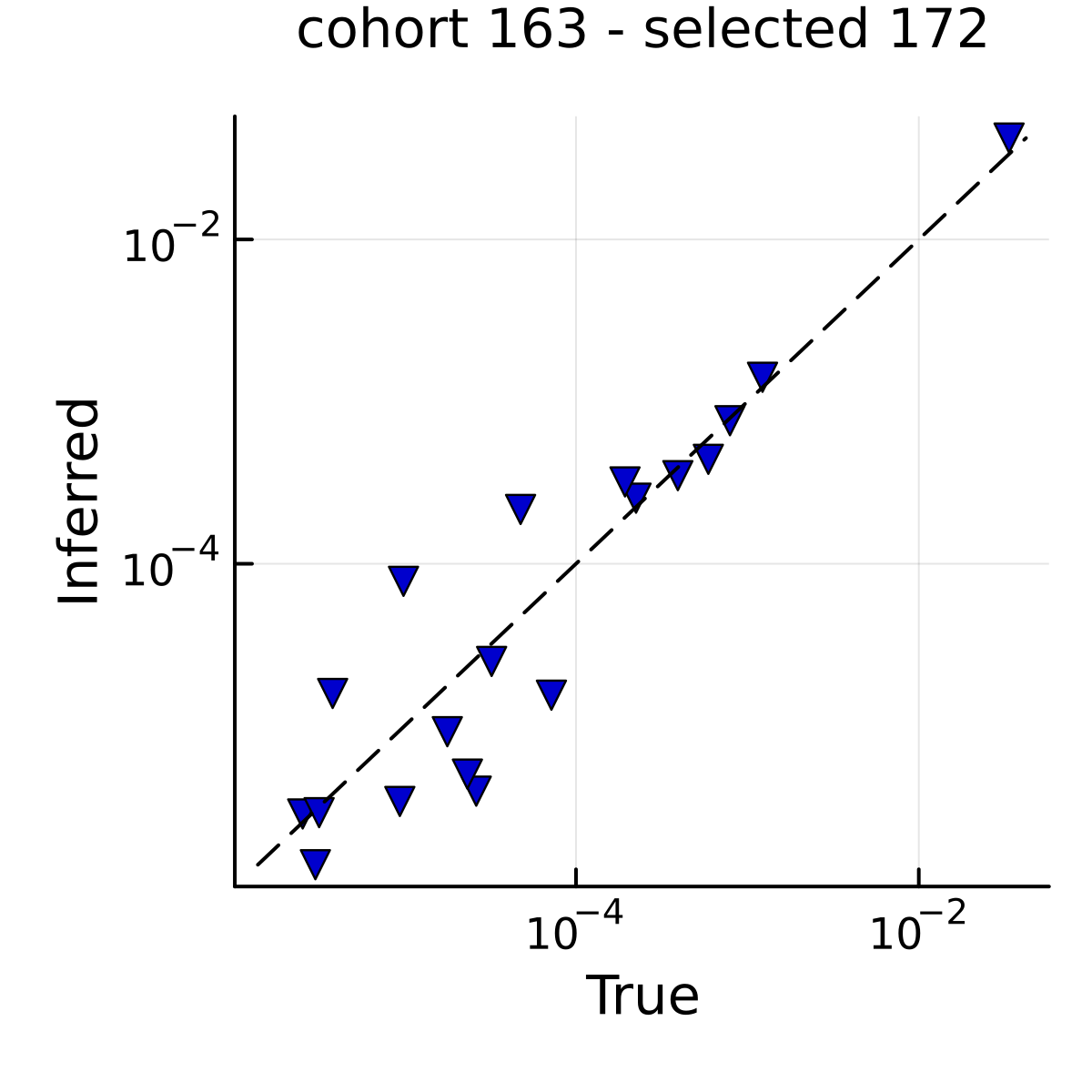}
    \end{subfigure}
     \begin{subfigure}[b]{0.22\textwidth}
        \includegraphics[width=\textwidth]{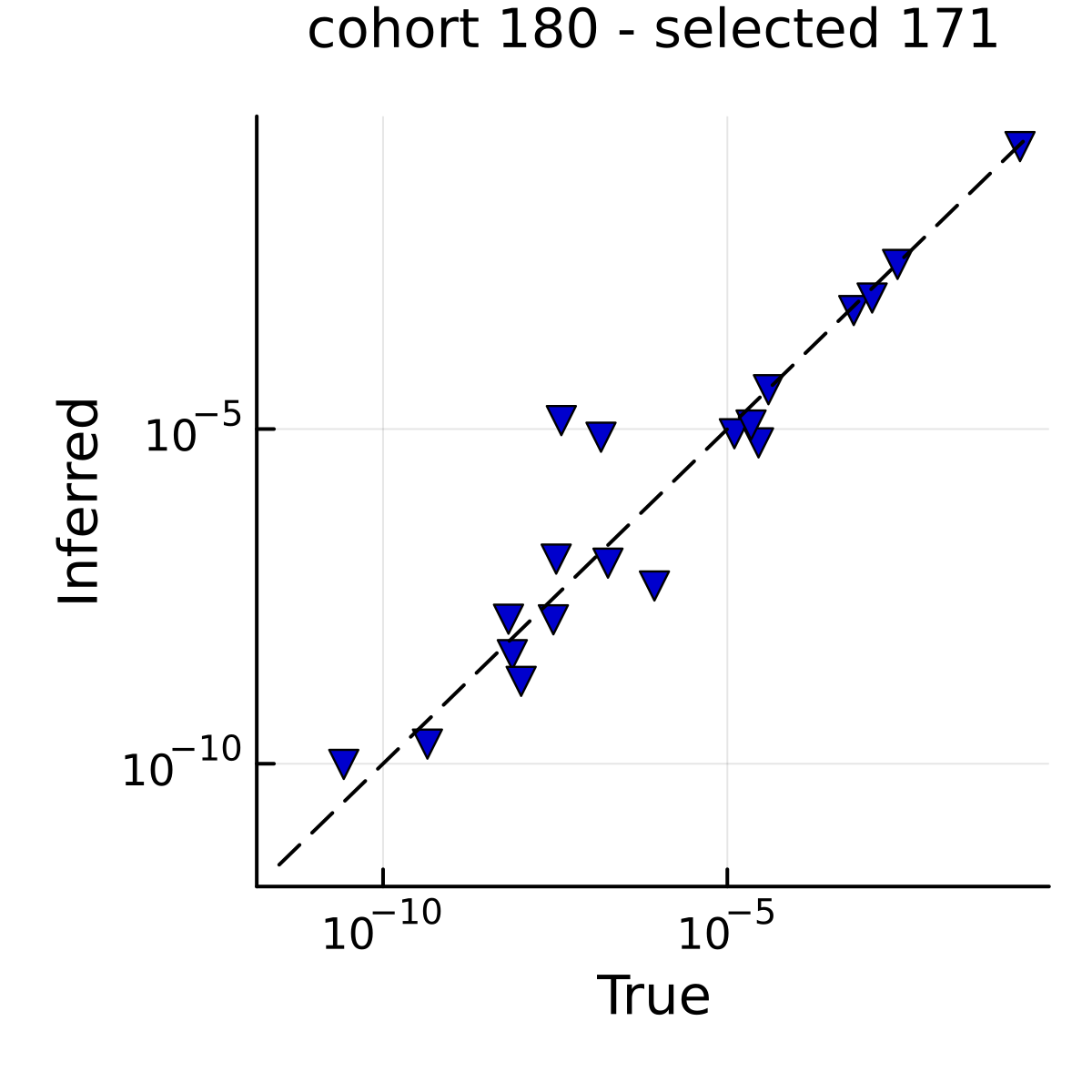}
    \end{subfigure}
     \begin{subfigure}[b]{0.22\textwidth}
        \includegraphics[width=\textwidth]{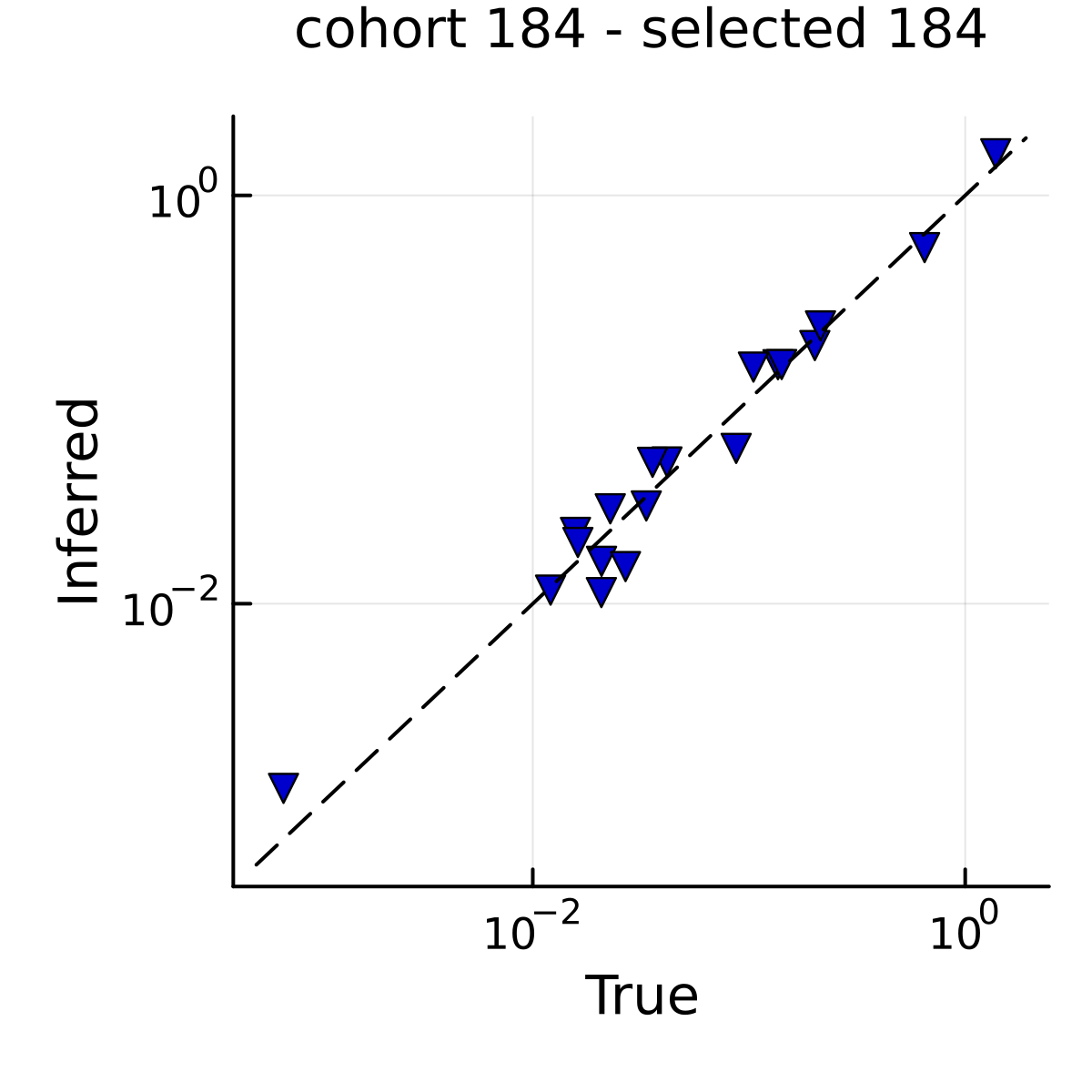}
    \end{subfigure} 
    \begin{subfigure}[b]{0.22\textwidth}
        \includegraphics[width=\textwidth]{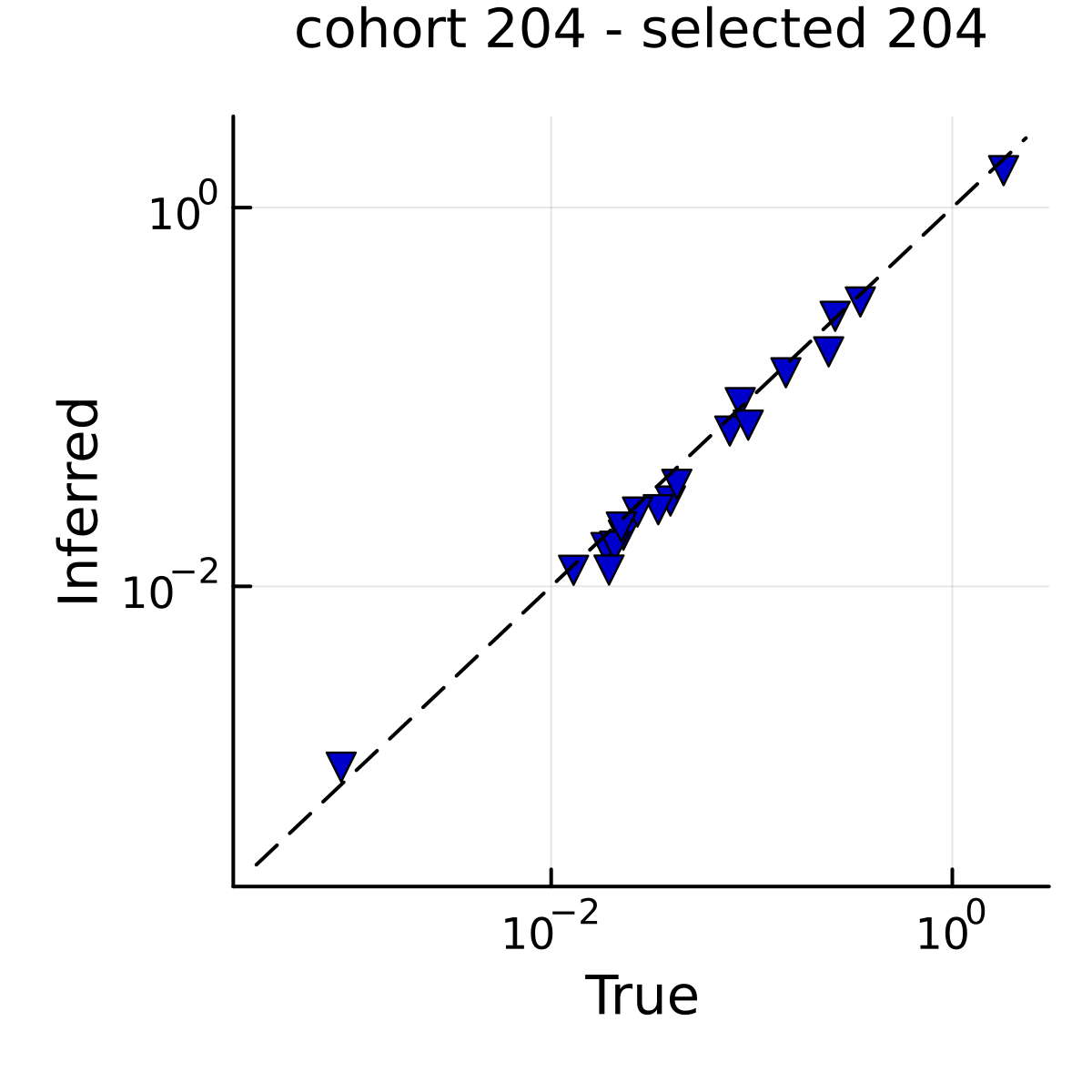}
    \end{subfigure}
     \begin{subfigure}[b]{0.22\textwidth}
        \includegraphics[width=\textwidth]{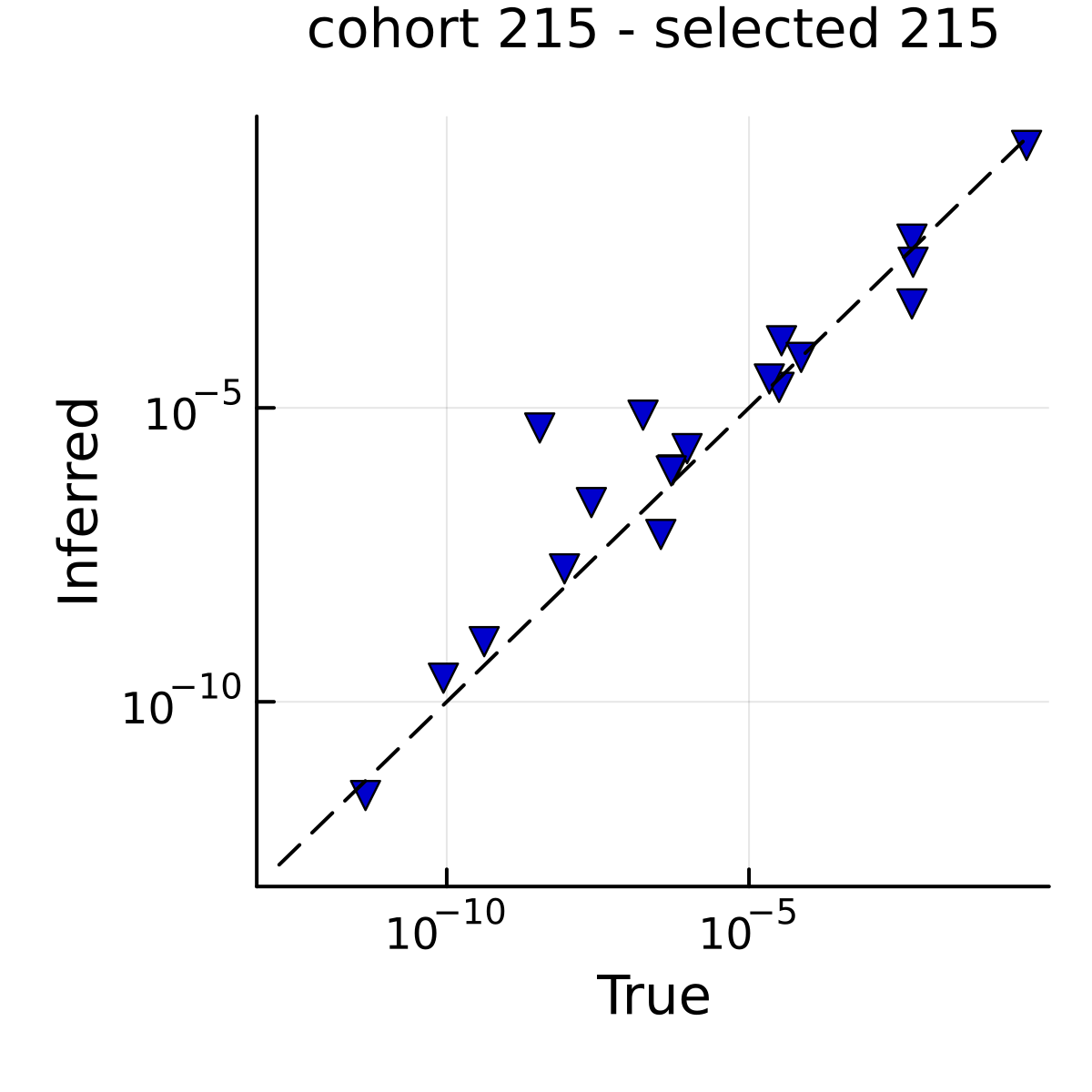}
    \end{subfigure}
     \begin{subfigure}[b]{0.22\textwidth}
        \includegraphics[width=\textwidth]{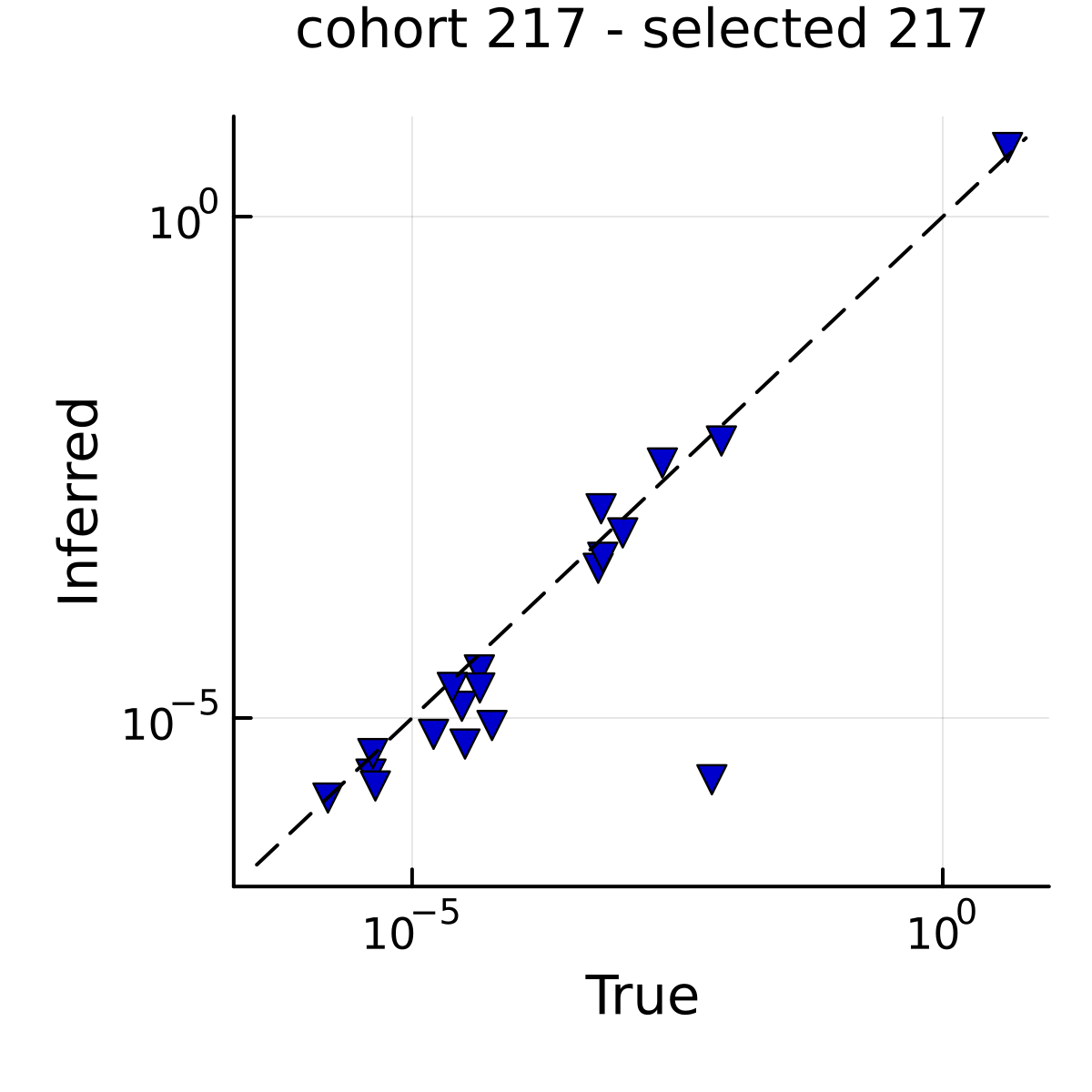}
    \end{subfigure}
    
    \caption{Comparison between the inferred (mean posterior, y-axis) and the true value of $R_{hom}(t=1,500)$. The axes are in a log scale.}
    \label{fig:synth_R_hom}
\end{figure}

Since a direct comparison between the inferred and true parameter values might not be relevant when the selected model differs from the true one, we also study another quantity that was defined by Mosca et al.~\cite{mosca2021}, the response factor ($R$-factor, see~\ref{sec:res_identif}).  
This quantity is particularly interesting since it is used to quantify the molecular response to the treatment. Thus, estimating this quantity accurately rather than the model parameters is more critical. Of course, the R-factor depends on the parameters in such a way that it could be analytically (but fastidiously) expressed.\\
In Fig.~\ref{fig:synth_R_het} and \ref{fig:synth_R_hom}, we compare the inferred (posterior mean) and true values of the heterozygous and homozygous - respectively - response factors and find that we can accurately estimate these quantities.

\FloatBarrier

\subsubsection{Dose-response relationship $\bar{\Delta}^*$}

\begin{figure}[h]
    \centering

    \begin{subfigure}[b]{0.22\textwidth}
        \includegraphics[width=\textwidth]{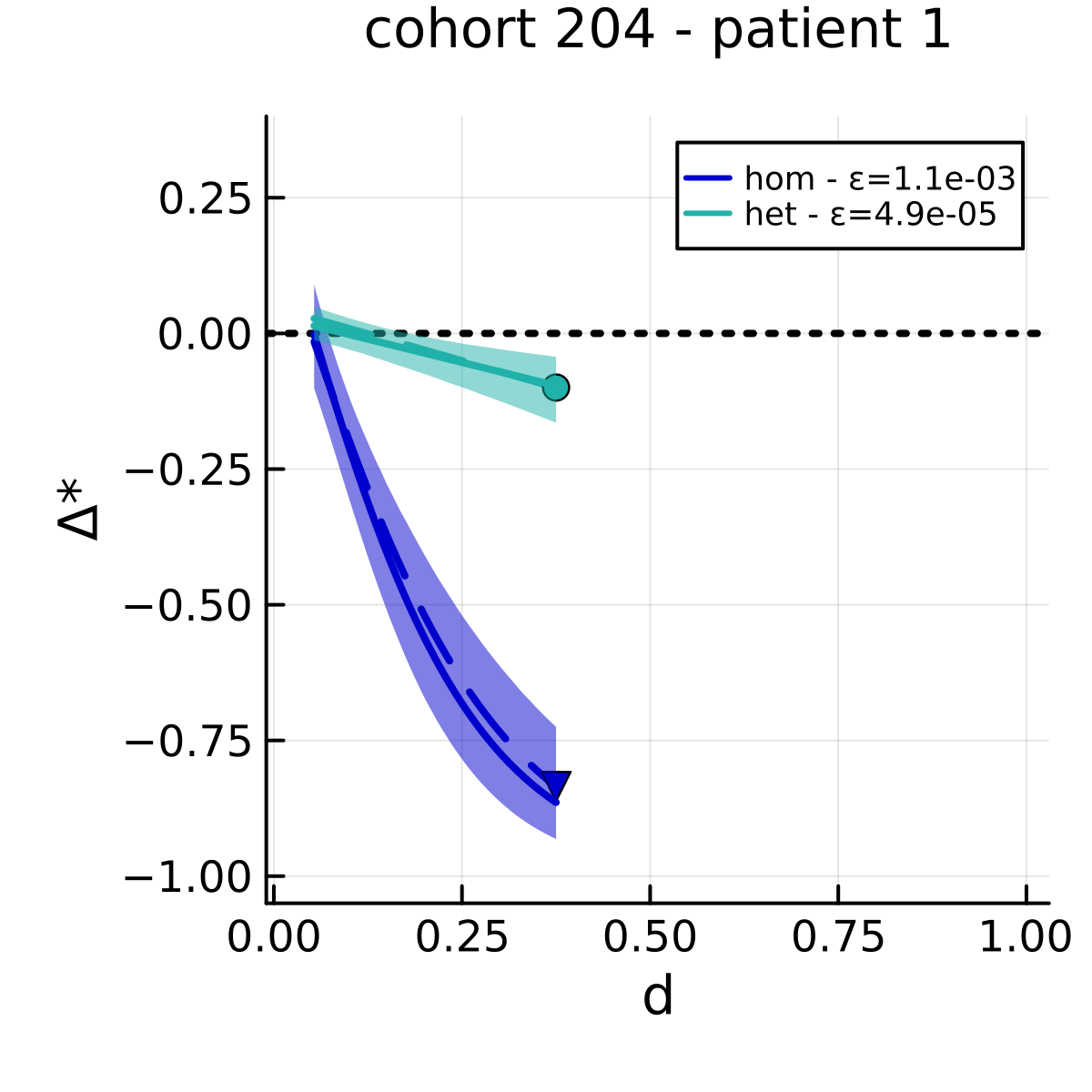}
    \end{subfigure}
    \begin{subfigure}[b]{0.22\textwidth}
        \includegraphics[width=\textwidth]{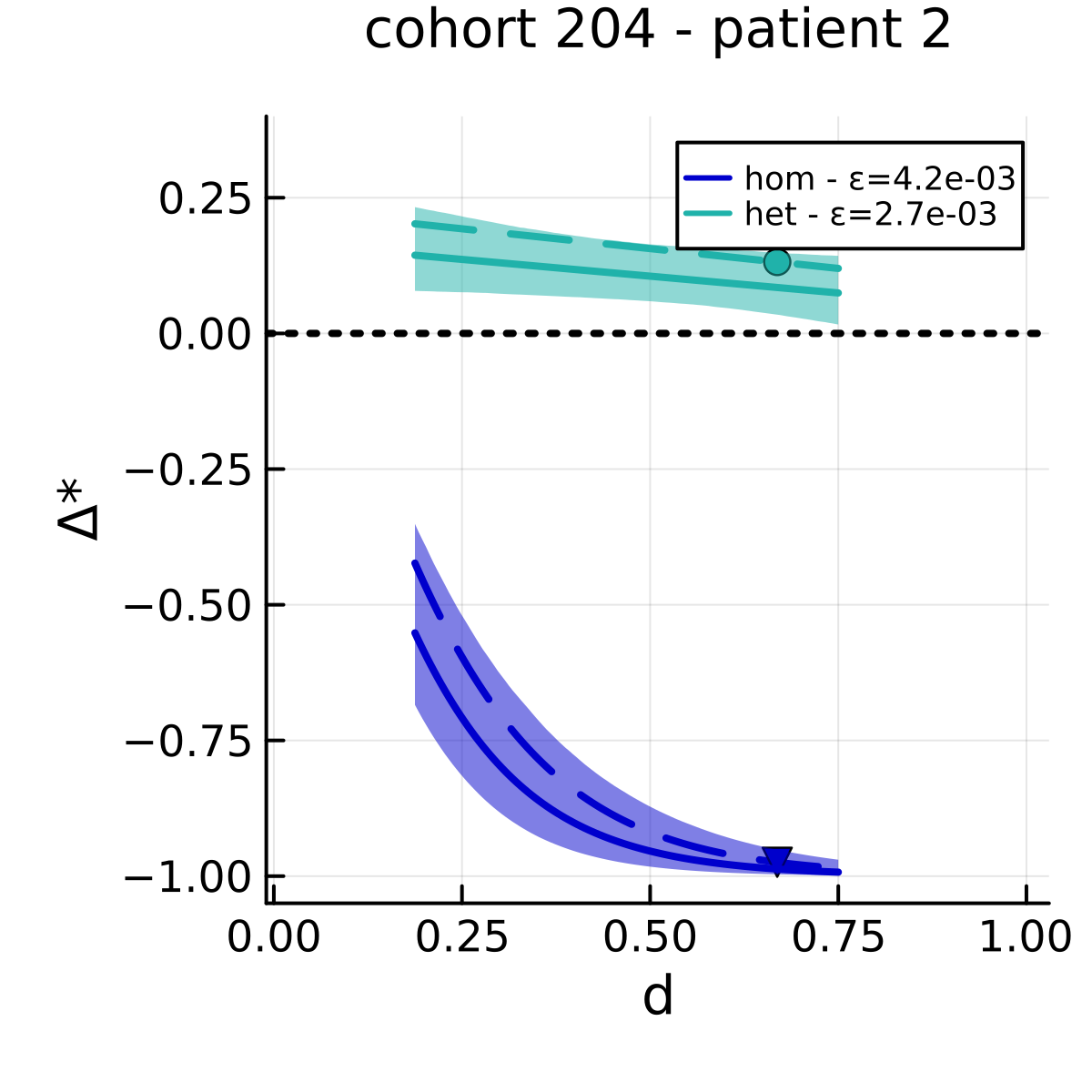}
    \end{subfigure}
    \begin{subfigure}[b]{0.22\textwidth}
        \includegraphics[width=\textwidth]{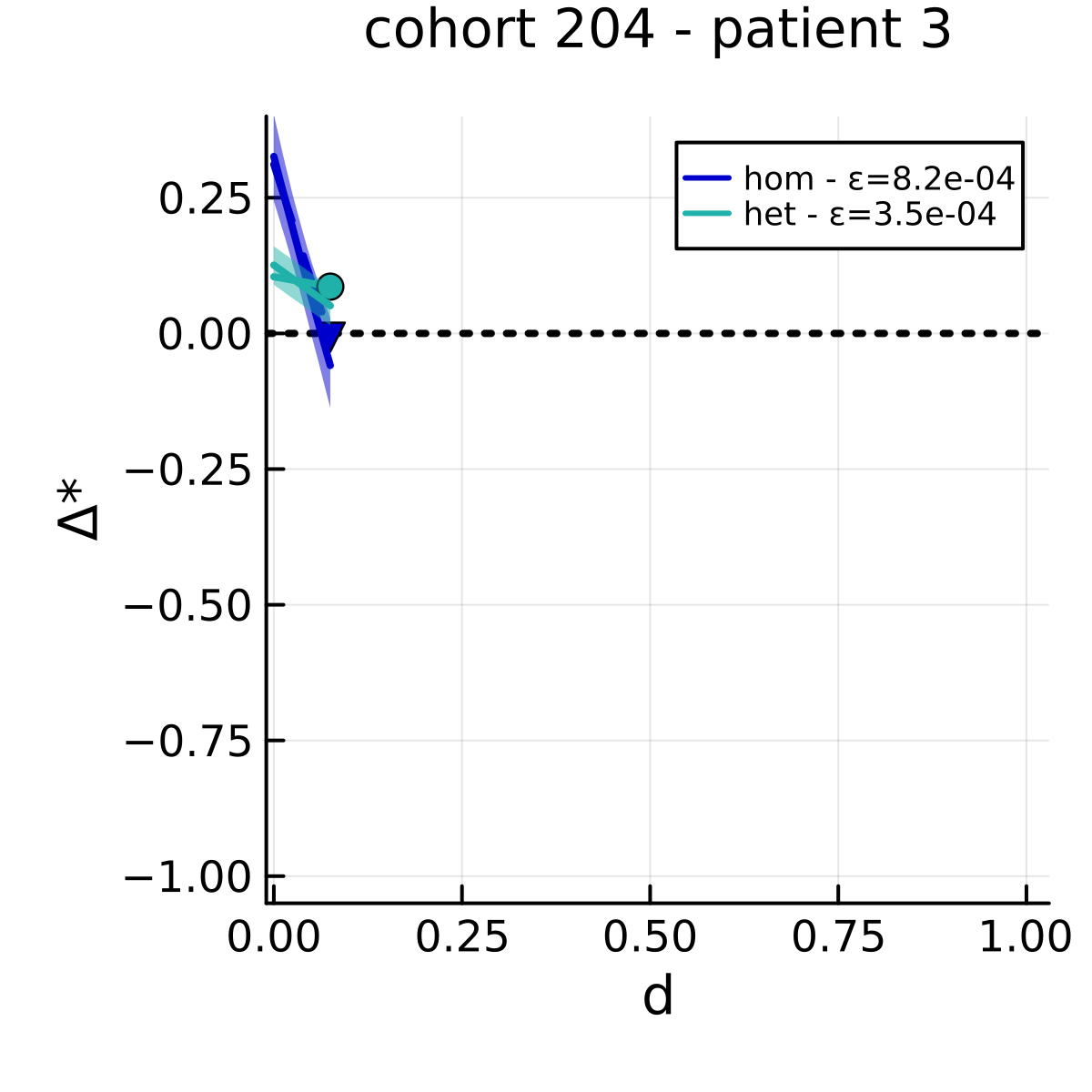}
    \end{subfigure}
    \begin{subfigure}[b]{0.22\textwidth}
        \includegraphics[width=\textwidth]{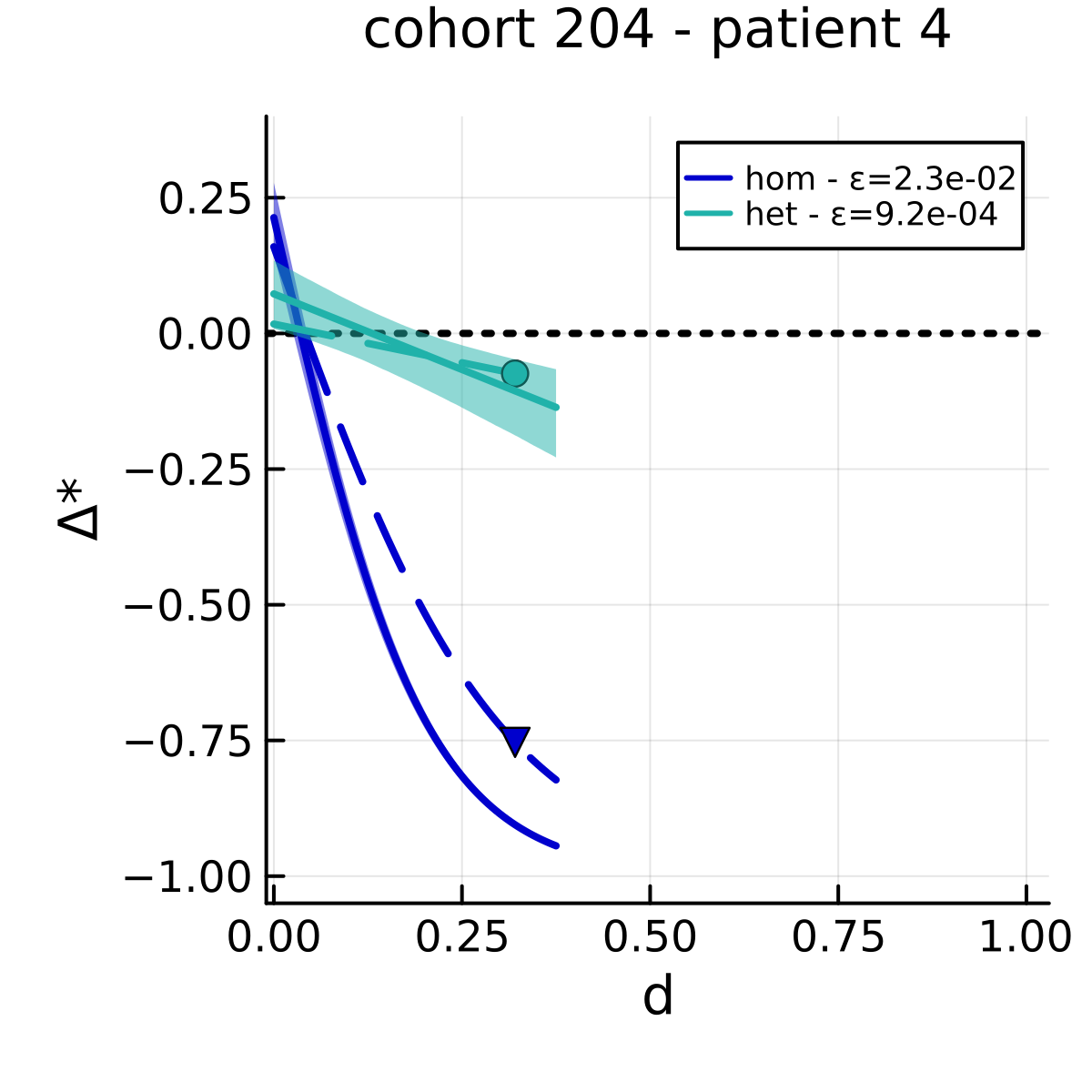}
    \end{subfigure}
    \begin{subfigure}[b]{0.22\textwidth}
        \includegraphics[width=\textwidth]{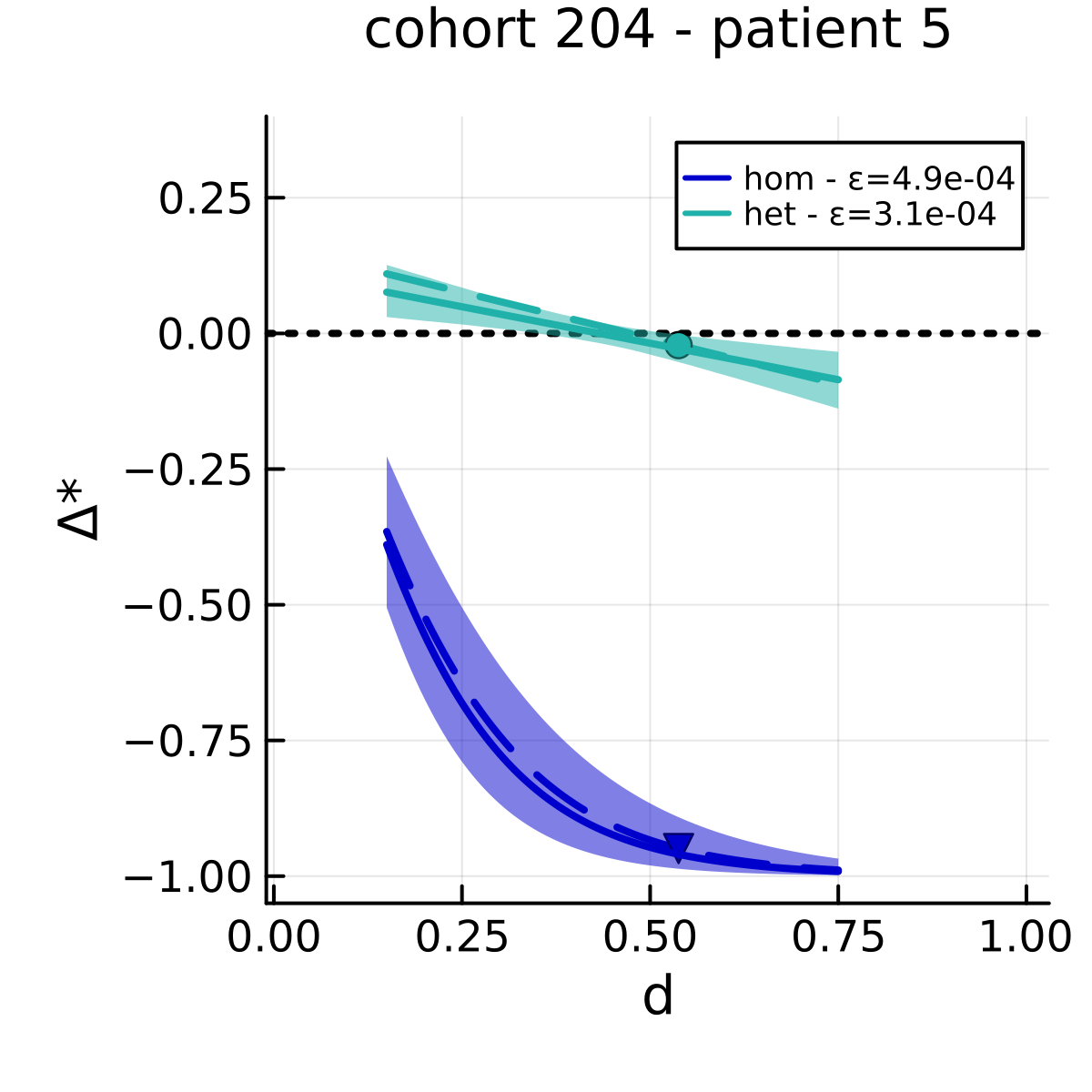}
    \end{subfigure}
    \begin{subfigure}[b]{0.22\textwidth}
        \includegraphics[width=\textwidth]{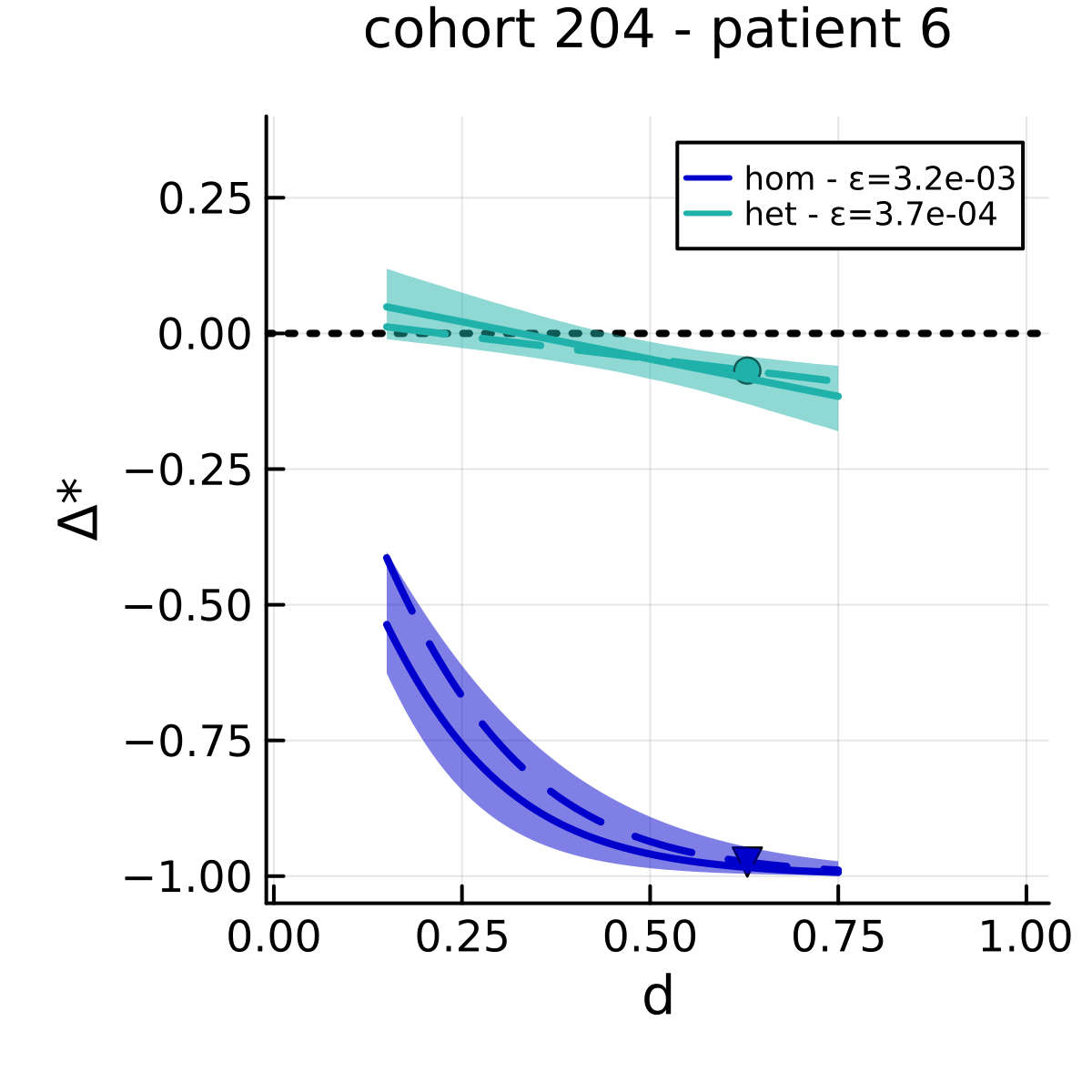}
    \end{subfigure}
    \begin{subfigure}[b]{0.22\textwidth}
        \includegraphics[width=\textwidth]{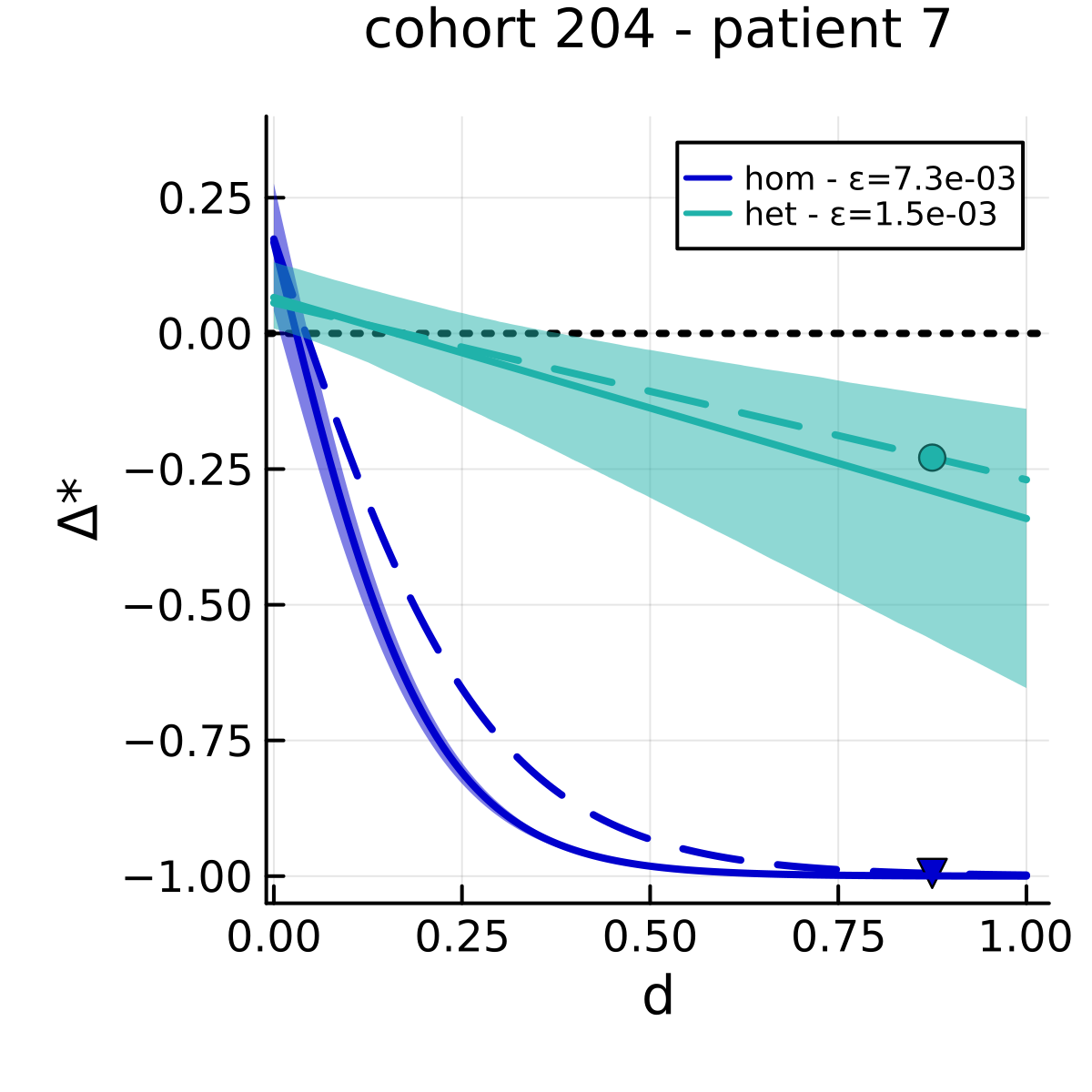}
    \end{subfigure}
    \begin{subfigure}[b]{0.22\textwidth}
        \includegraphics[width=\textwidth]{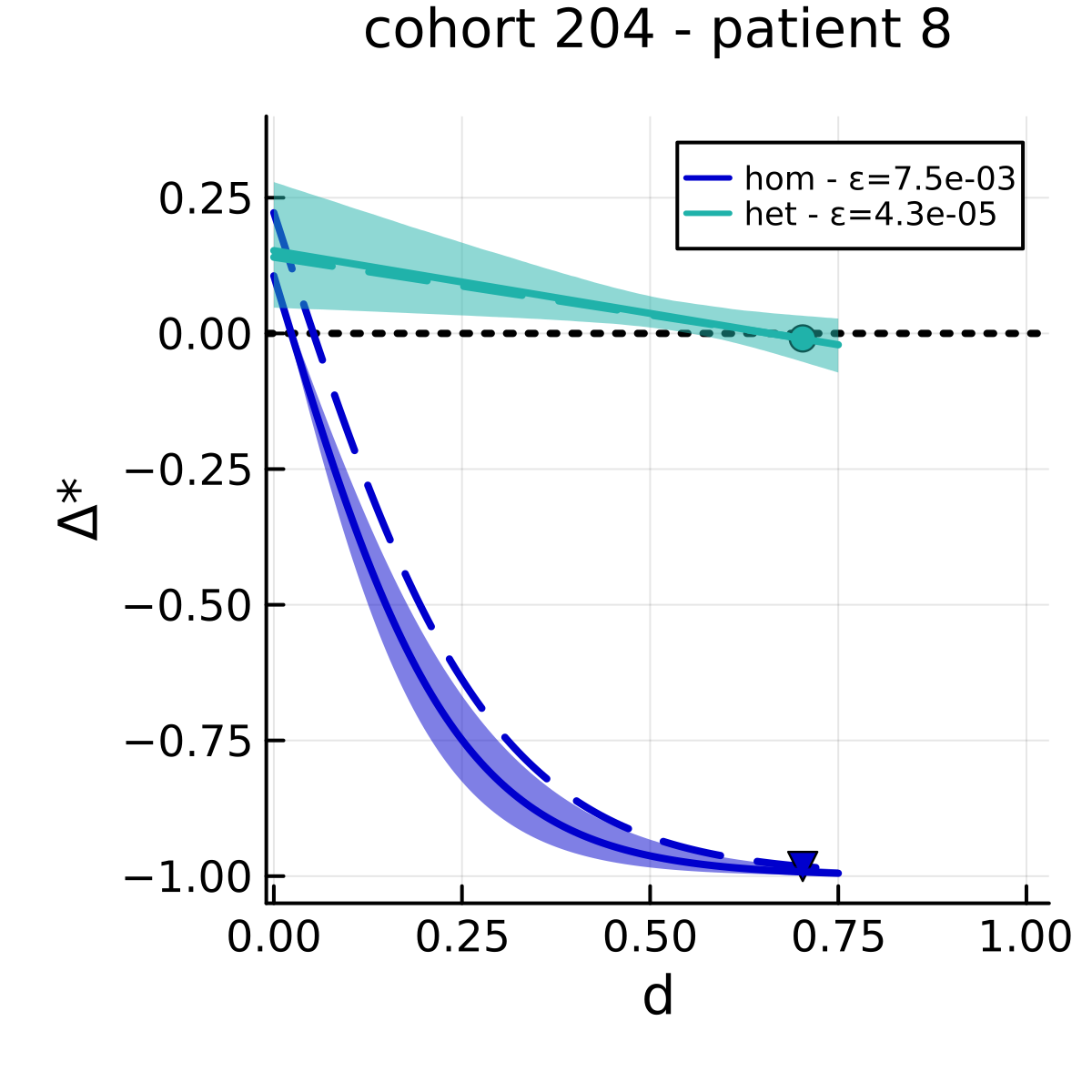}
    \end{subfigure}
    \begin{subfigure}[b]{0.22\textwidth}
        \includegraphics[width=\textwidth]{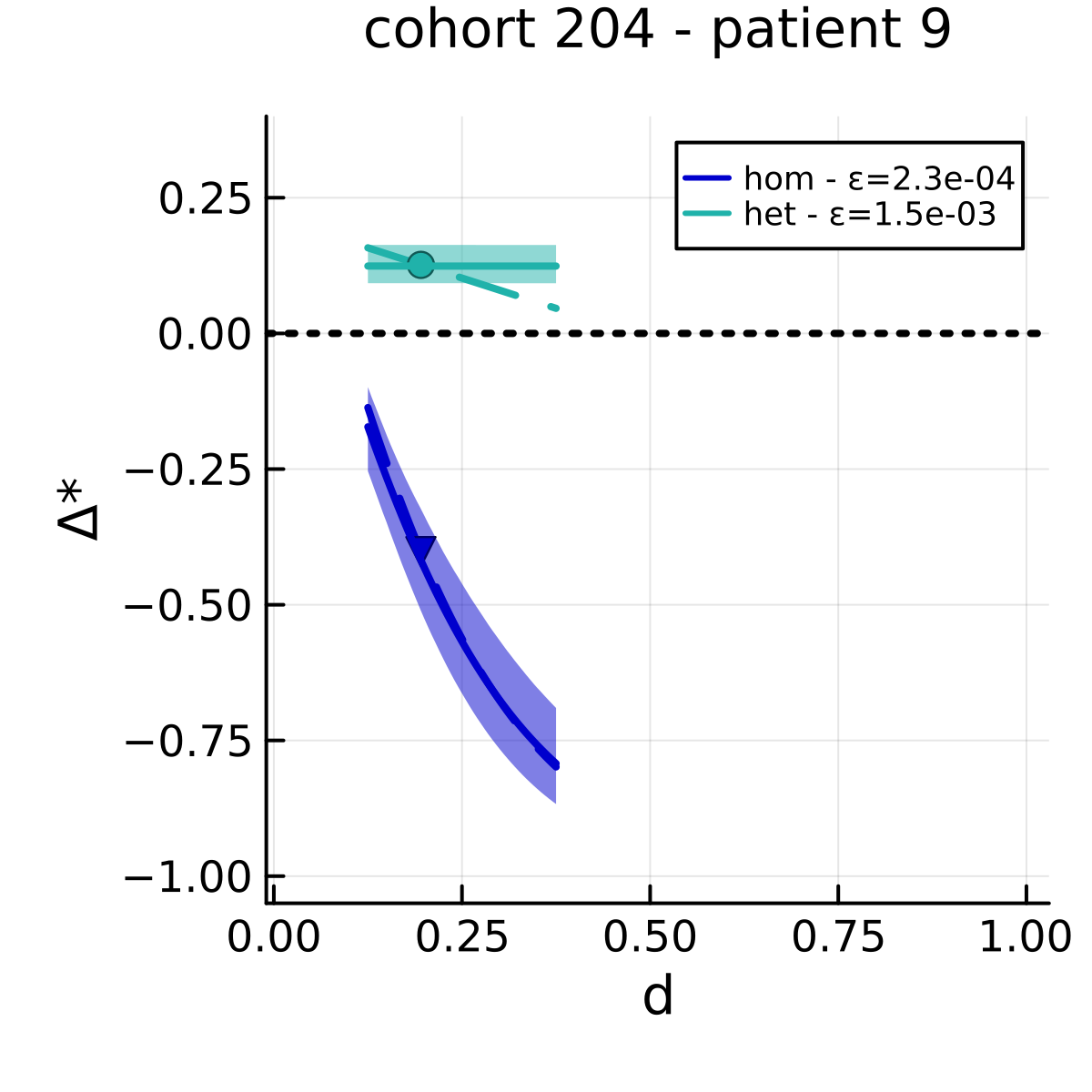}
    \end{subfigure}
    \begin{subfigure}[b]{0.22\textwidth}
        \includegraphics[width=\textwidth]{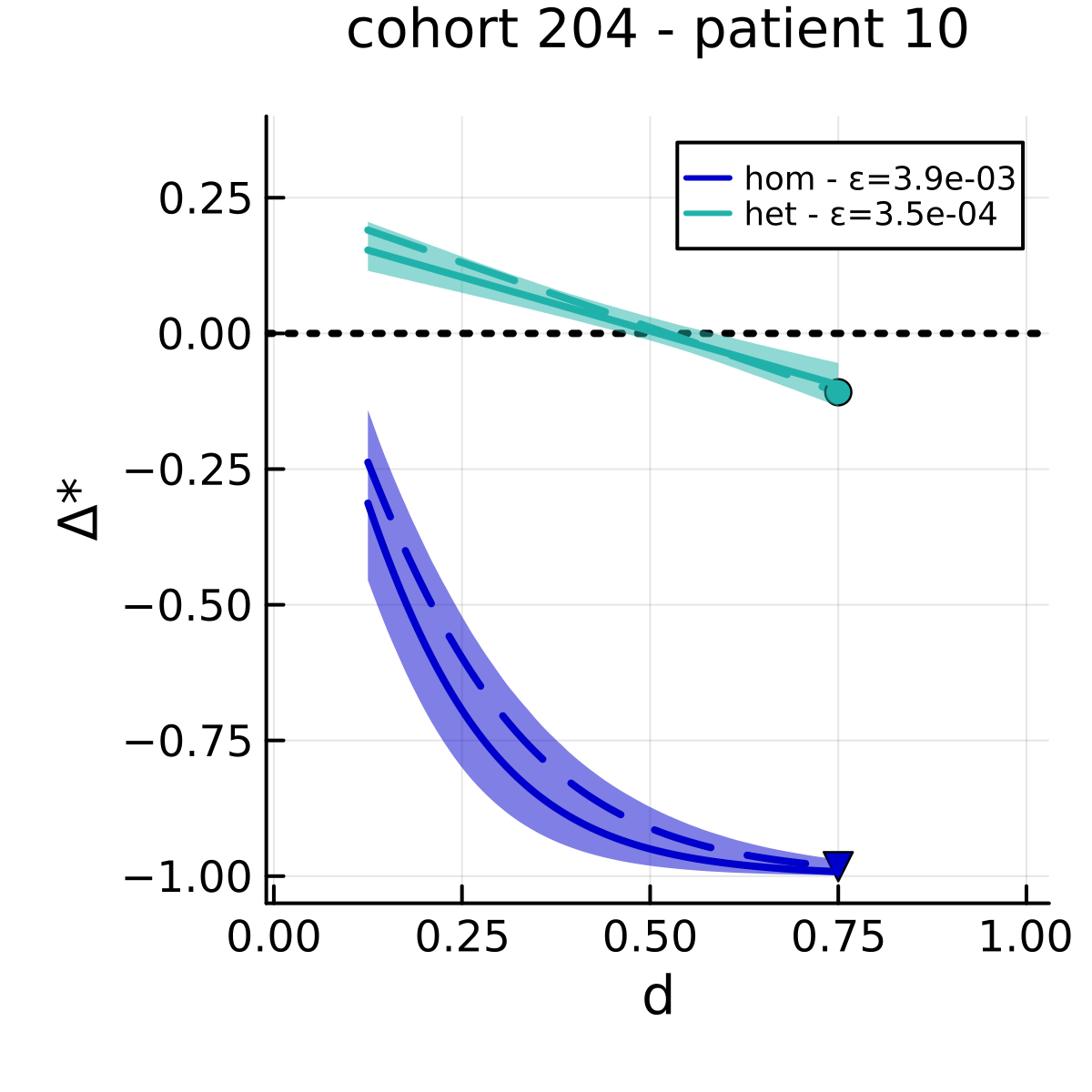}
    \end{subfigure}
    \begin{subfigure}[b]{0.22\textwidth}
        \includegraphics[width=\textwidth]{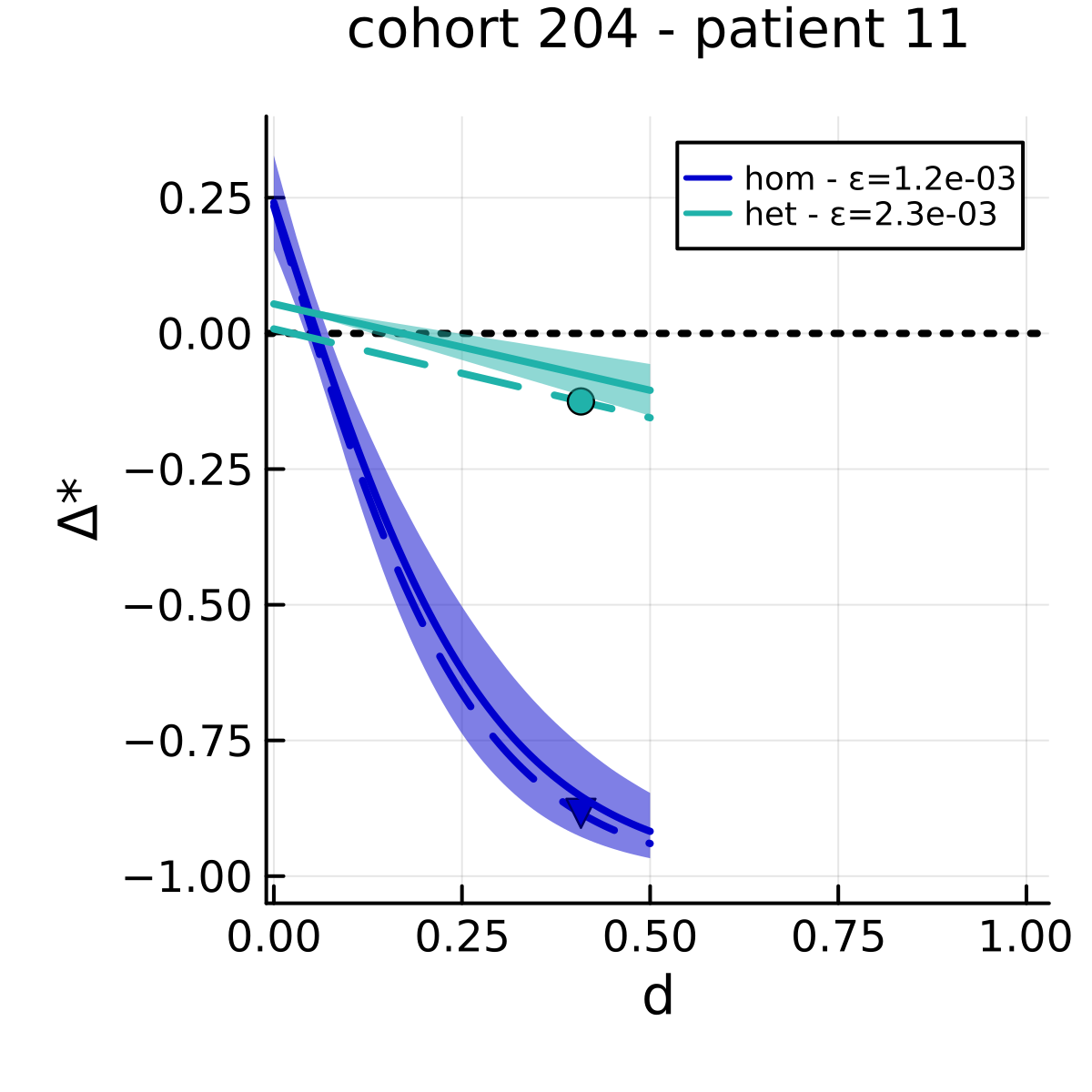}
    \end{subfigure}
    \begin{subfigure}[b]{0.22\textwidth}
        \includegraphics[width=\textwidth]{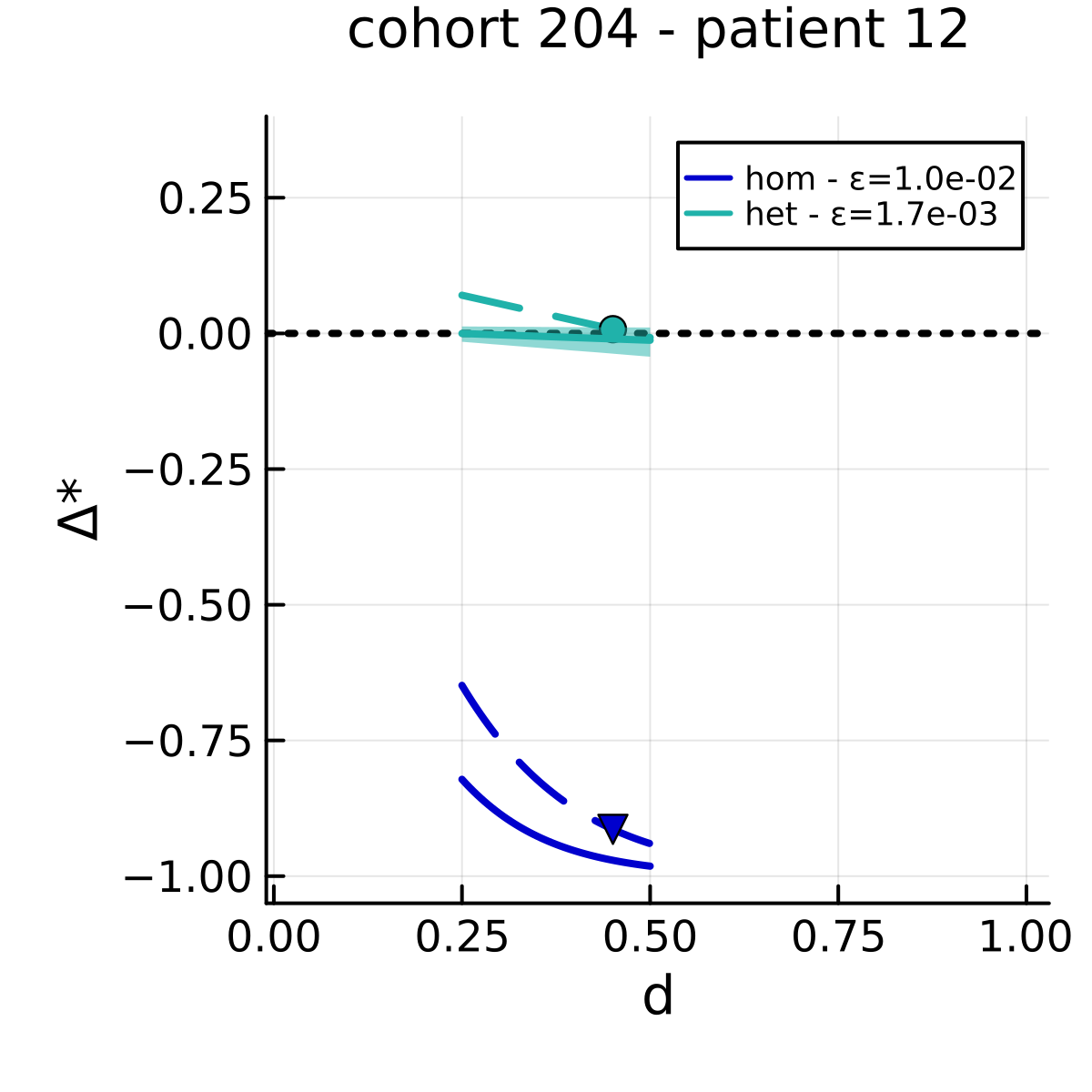}
    \end{subfigure}
    \begin{subfigure}[b]{0.22\textwidth}
        \includegraphics[width=\textwidth]{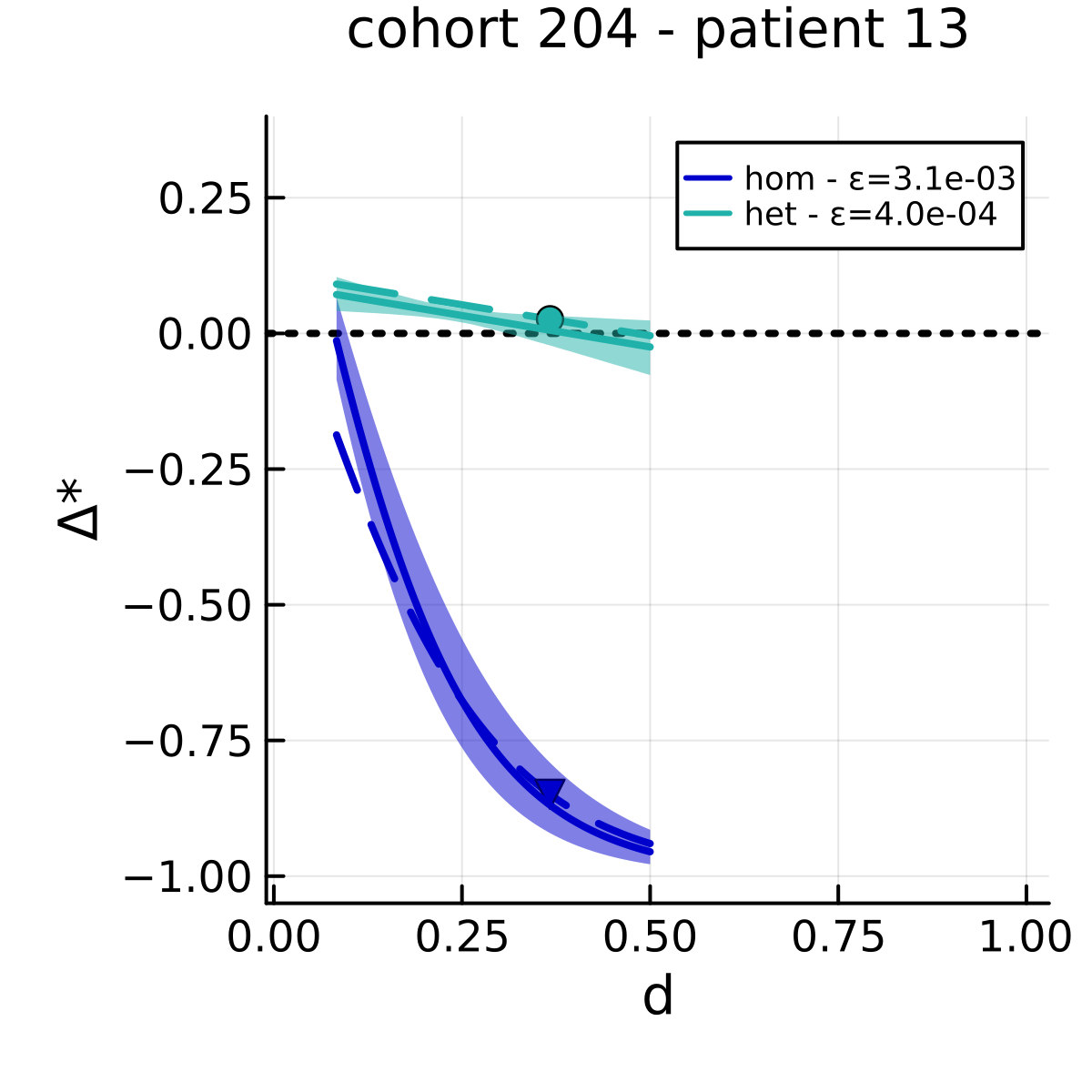}
    \end{subfigure}
    \begin{subfigure}[b]{0.22\textwidth}
        \includegraphics[width=\textwidth]{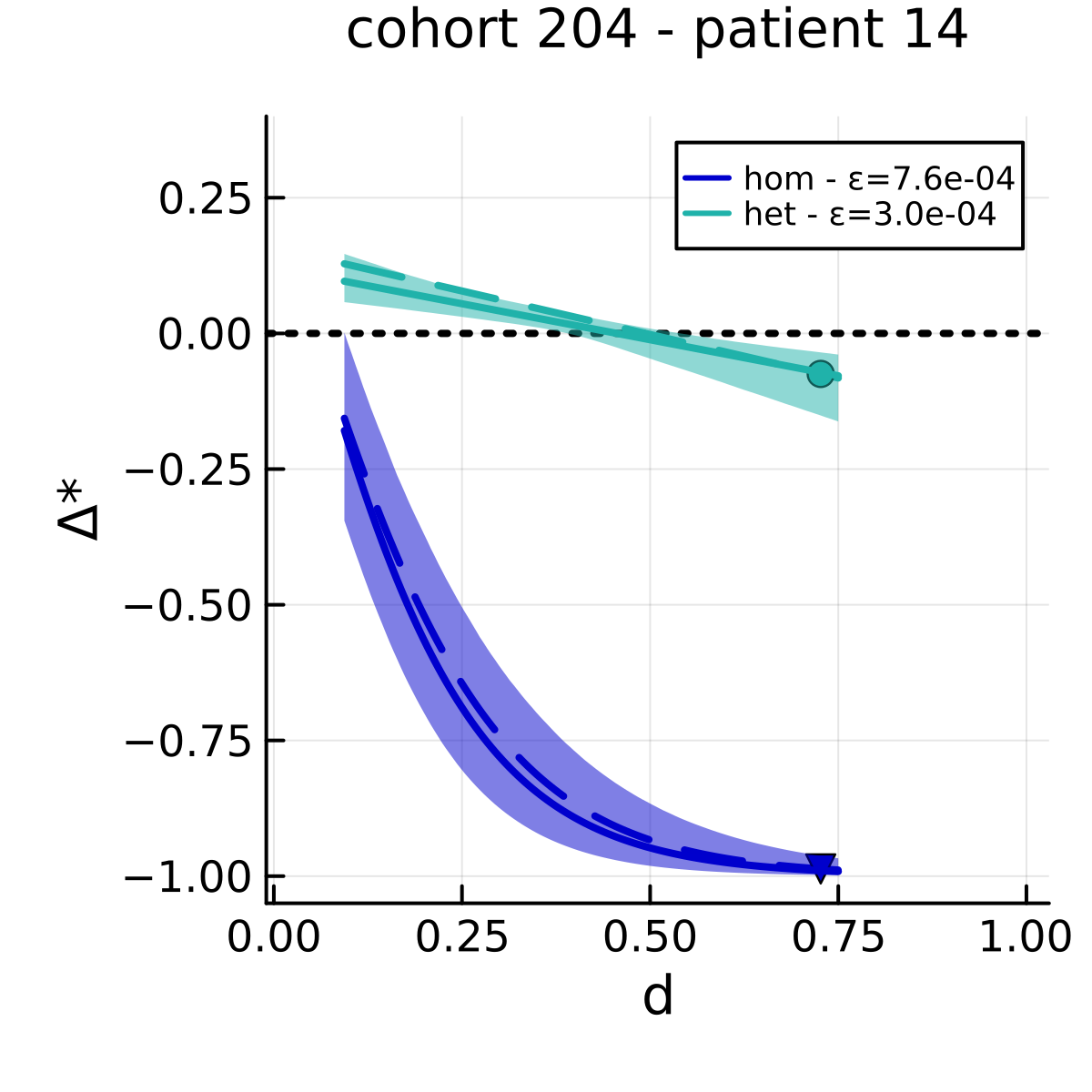}
    \end{subfigure}
    \begin{subfigure}[b]{0.22\textwidth}
        \includegraphics[width=\textwidth]{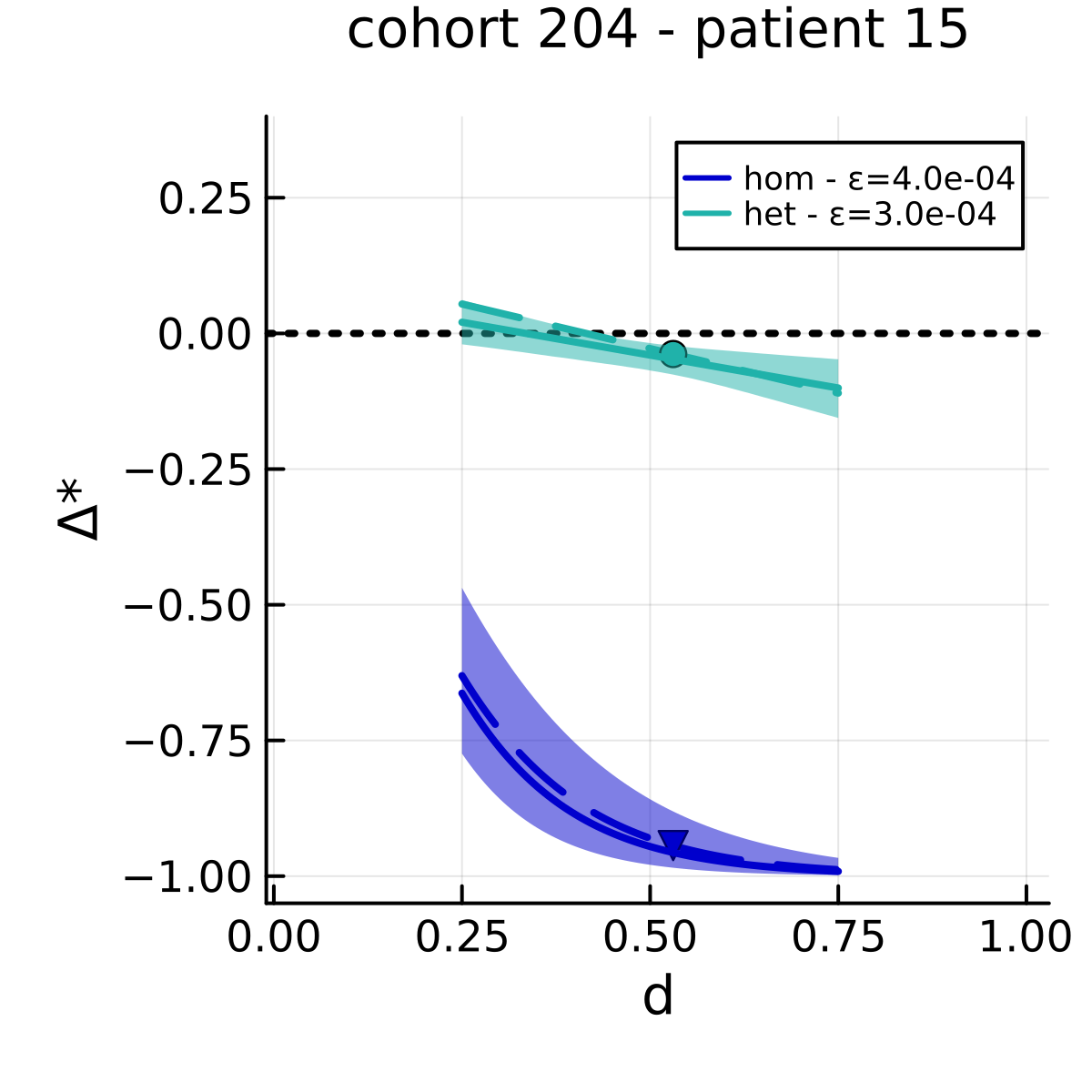}
    \end{subfigure}
    \begin{subfigure}[b]{0.22\textwidth}
        \includegraphics[width=\textwidth]{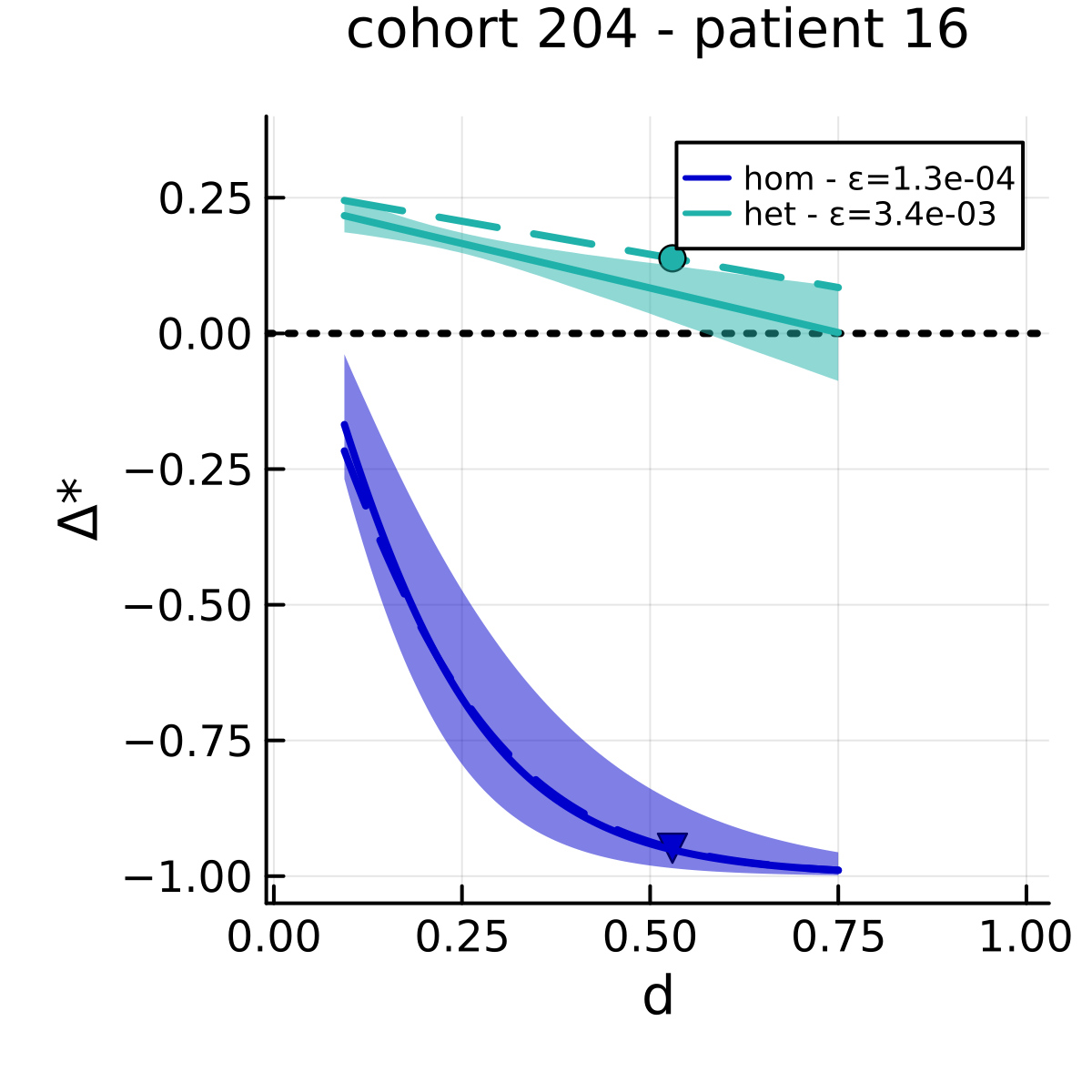}
    \end{subfigure}
    \begin{subfigure}[b]{0.22\textwidth}
        \includegraphics[width=\textwidth]{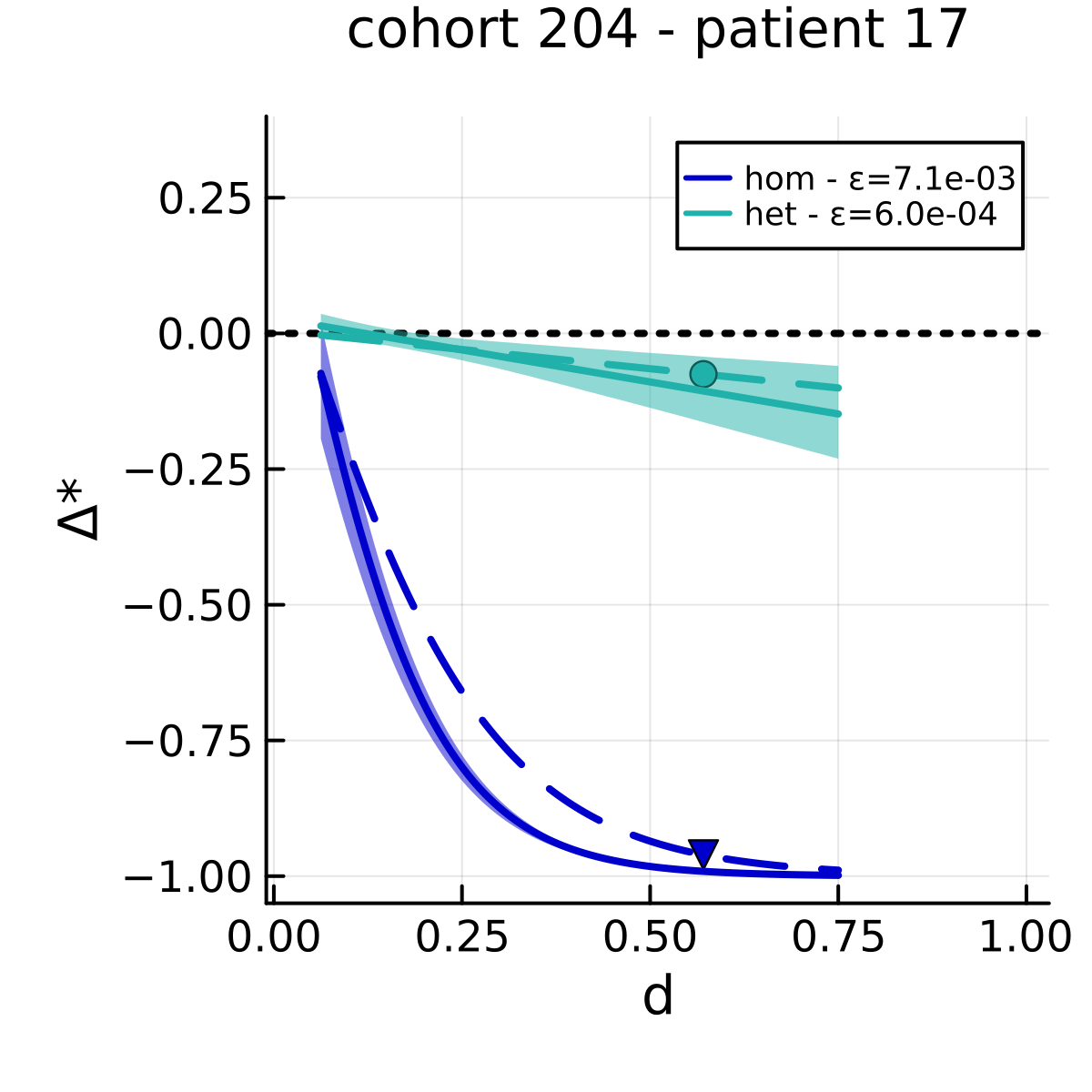}
    \end{subfigure}
    \begin{subfigure}[b]{0.22\textwidth}
        \includegraphics[width=\textwidth]{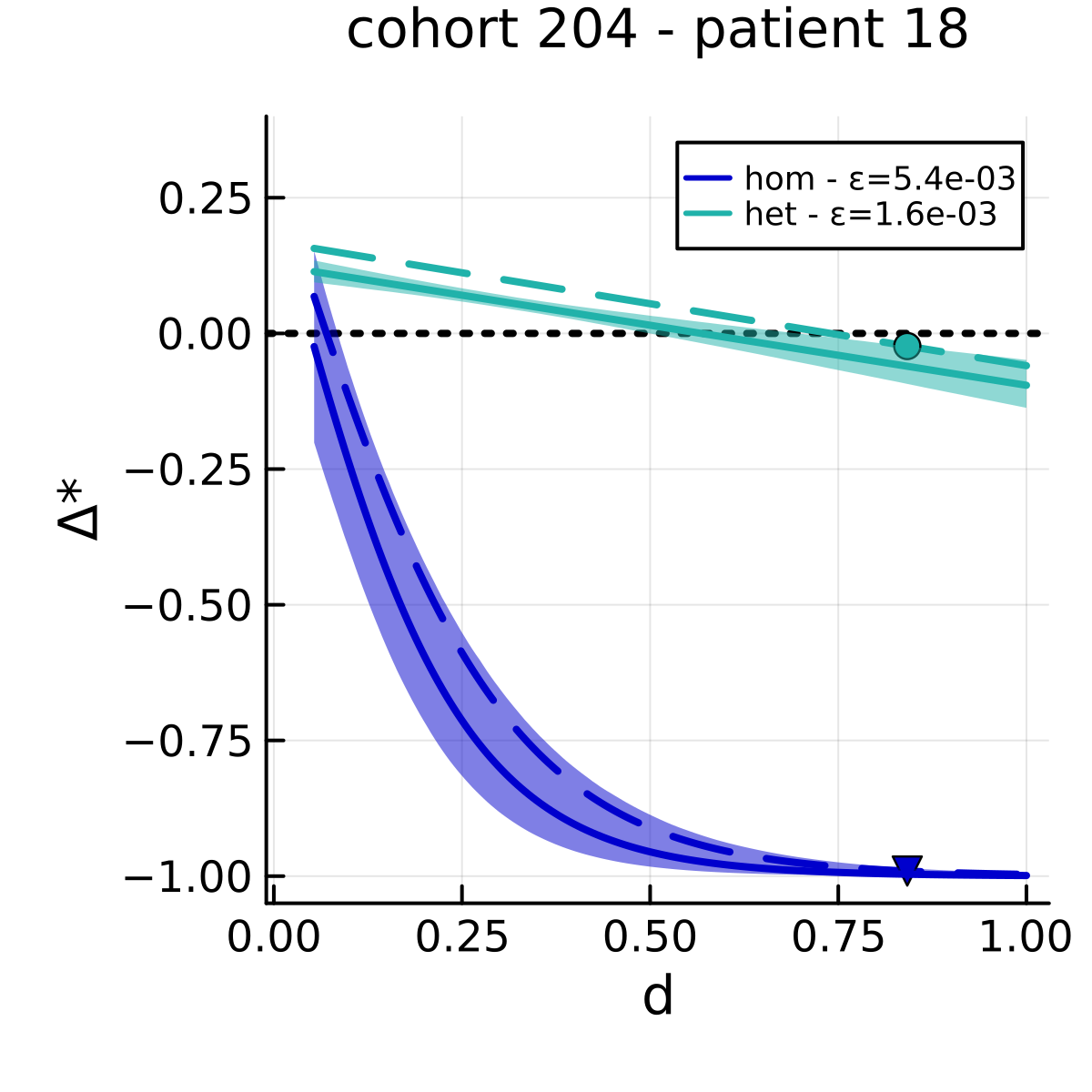}
    \end{subfigure}
    \begin{subfigure}[b]{0.22\textwidth}
        \includegraphics[width=\textwidth]{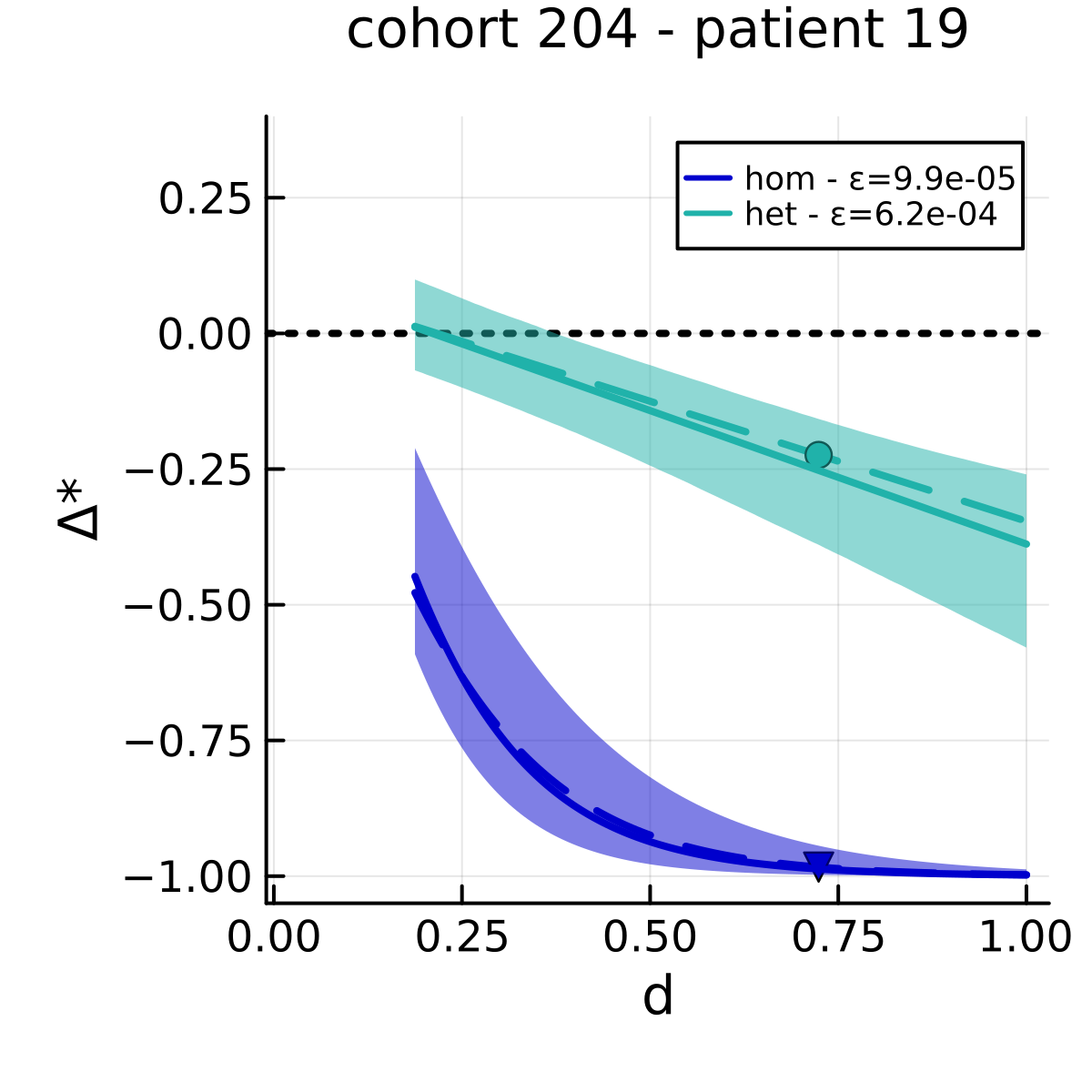}
    \end{subfigure}
    
    \caption{Comparison between the true (dash lines) and inferred (solid lines, mean posterior value) of $\bar{\Delta}^*_{het}(d)$ (green) and $\bar{\Delta}^*_{hom}(d)$ (blue), with 95\% credibility intervals, for different dose values $d$ between the minimal and maximal doses administered to each individual. The blue triangles and green circles are associated to the mean IFN$\alpha$ dose administered to the patient over the 450 first days of treatment. The horizontal dotted line corresponds to $\bar{\Delta}^*=0$. Above this line, we are in a range of values where a remission is not possible since the malignant clone continues to expand. The errors $\varepsilon$ quantify the gap between the inferred and true dose-response relationships.
 }
    \label{fig:Delta_204}
\end{figure}

\begin{figure}[h]
    \centering

\begin{subfigure}[b]{0.45\textwidth}
        \includegraphics[width=\textwidth]{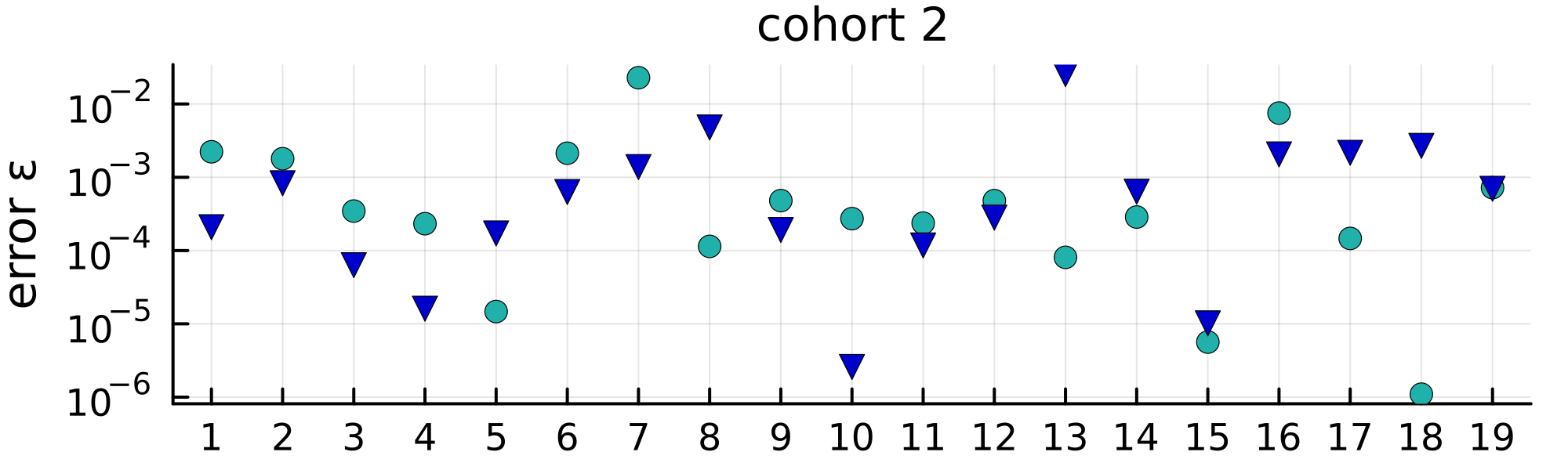}
    \end{subfigure}
    \begin{subfigure}[b]{0.45\textwidth}
        \includegraphics[width=\textwidth]{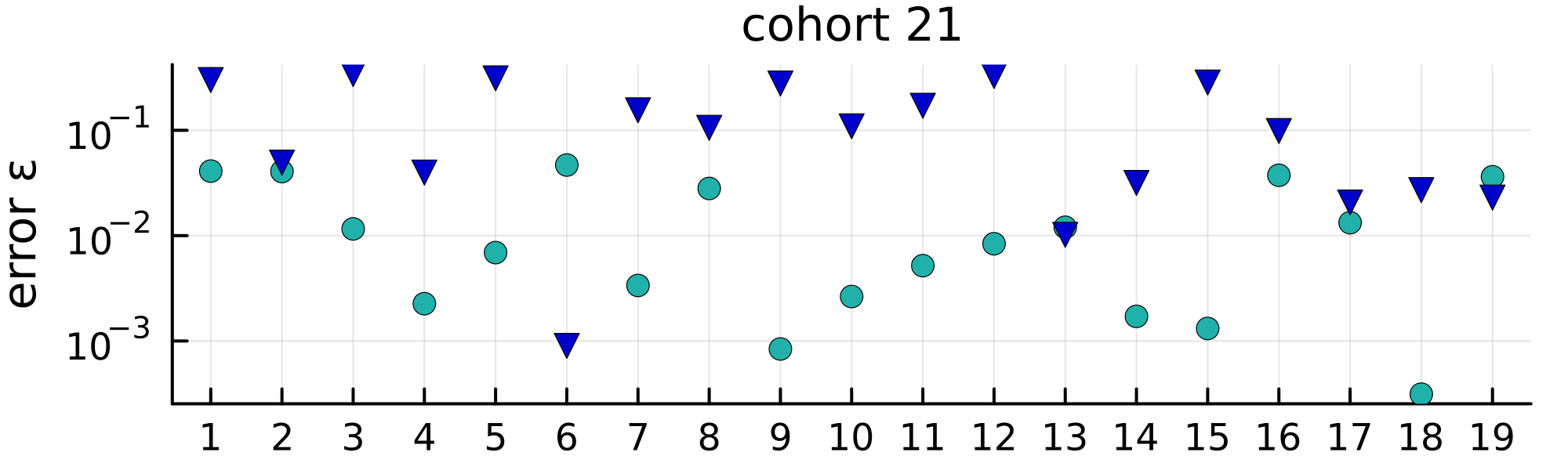}
    \end{subfigure}
    \begin{subfigure}[b]{0.45\textwidth}
        \includegraphics[width=\textwidth]{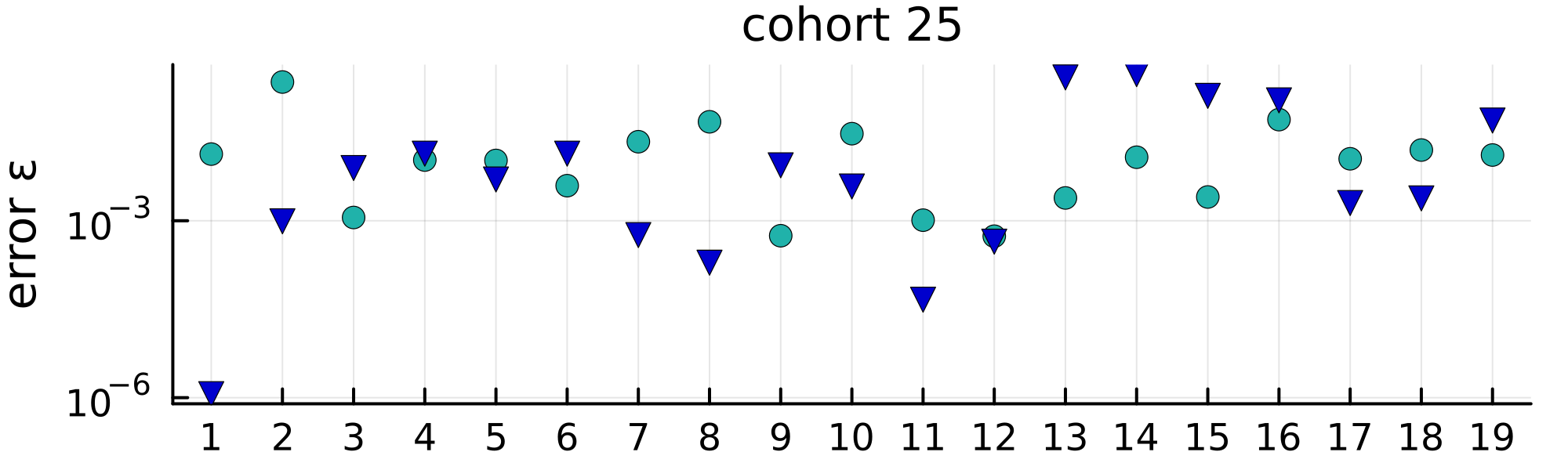}
    \end{subfigure}
    \begin{subfigure}[b]{0.45\textwidth}
        \includegraphics[width=\textwidth]{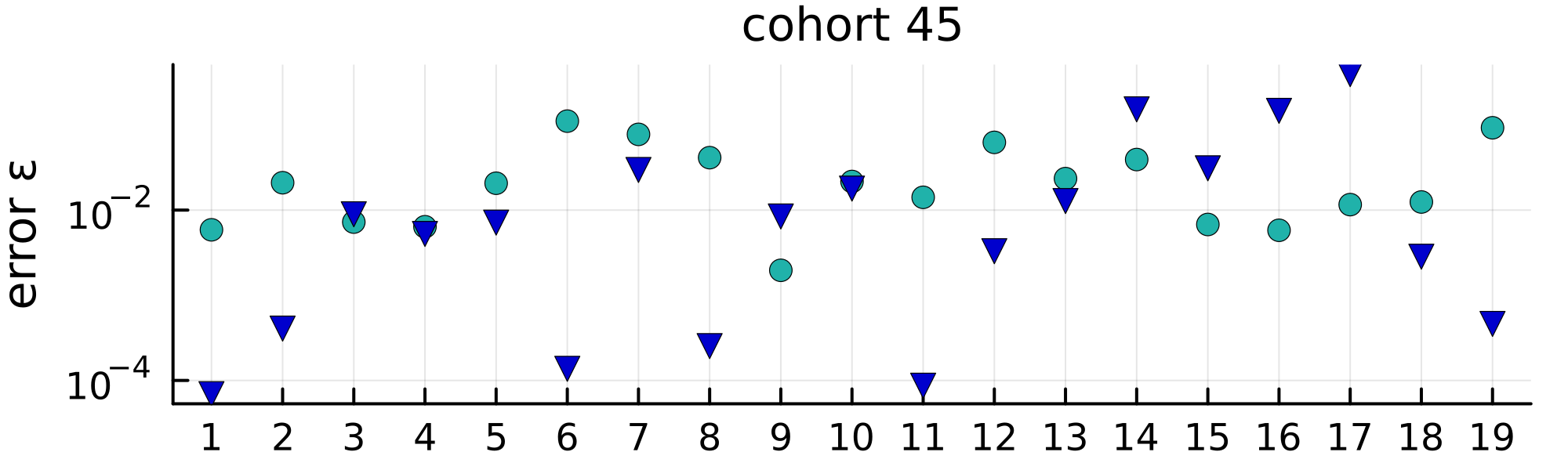}
    \end{subfigure}
    \begin{subfigure}[b]{0.45\textwidth}
        \includegraphics[width=\textwidth]{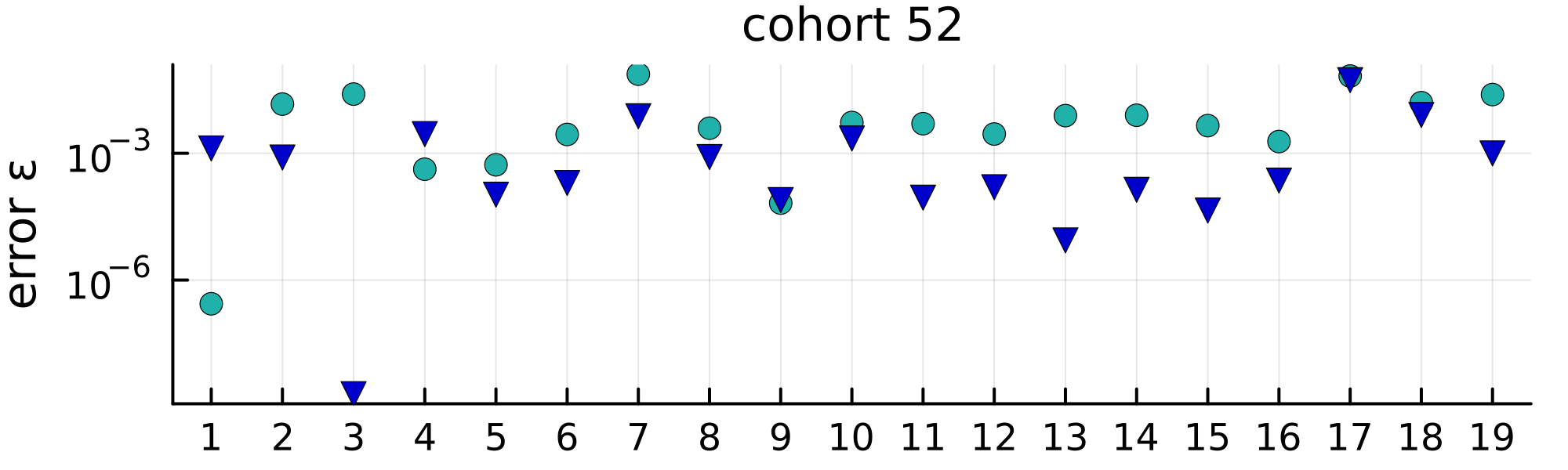}
    \end{subfigure}
    \begin{subfigure}[b]{0.45\textwidth}
        \includegraphics[width=\textwidth]{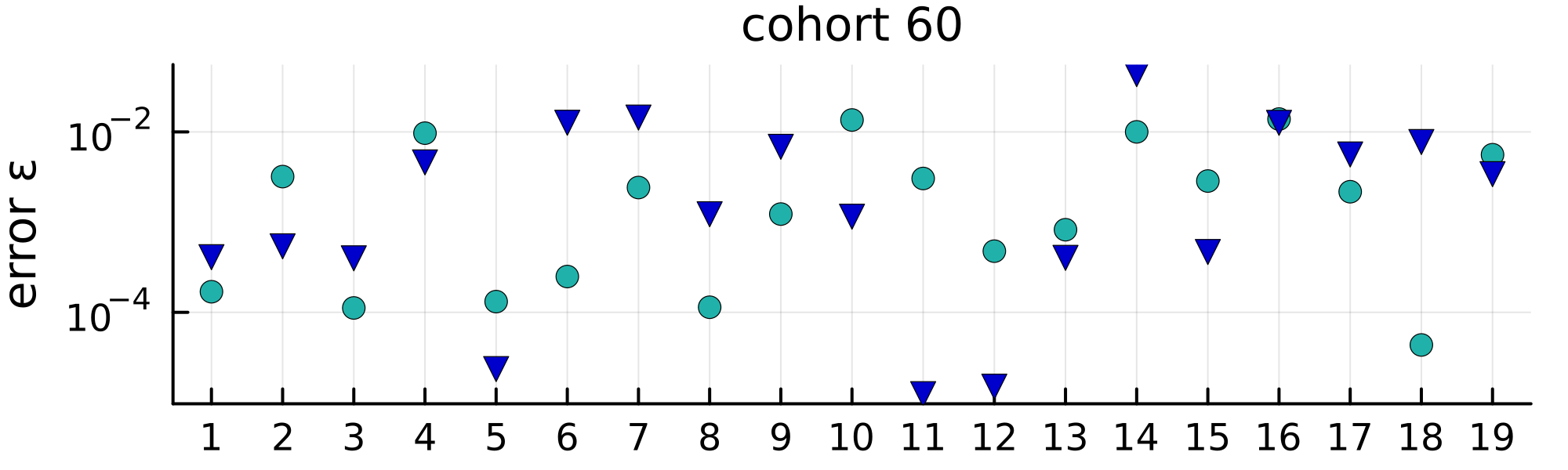}
    \end{subfigure}
    \begin{subfigure}[b]{0.45\textwidth}
        \includegraphics[width=\textwidth]{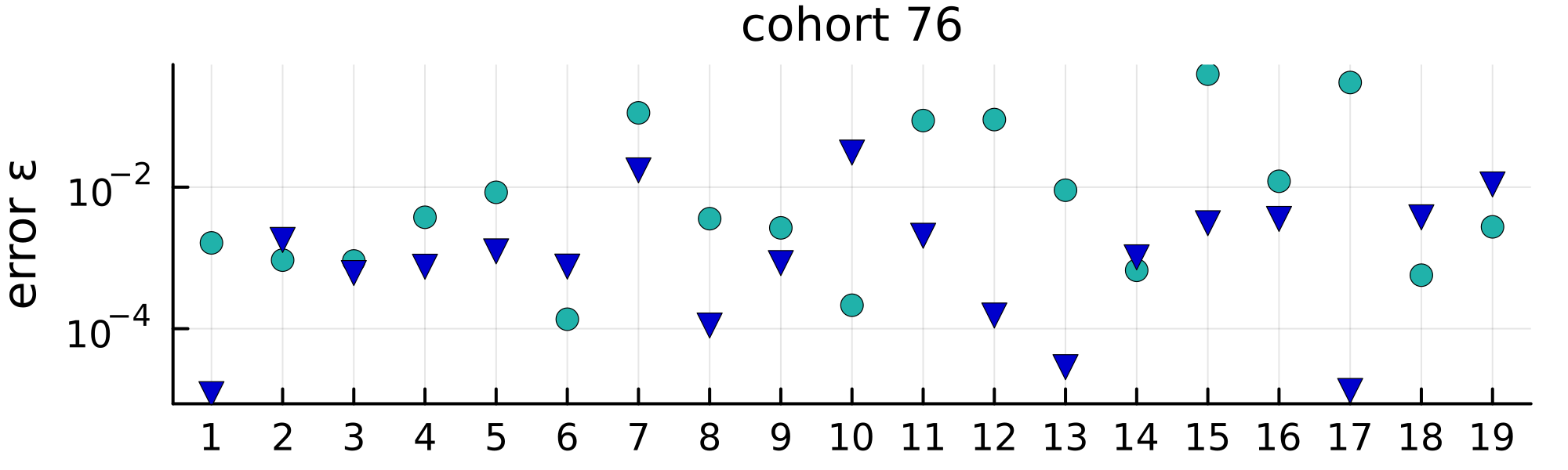}
    \end{subfigure}
    \begin{subfigure}[b]{0.45\textwidth}
        \includegraphics[width=\textwidth]{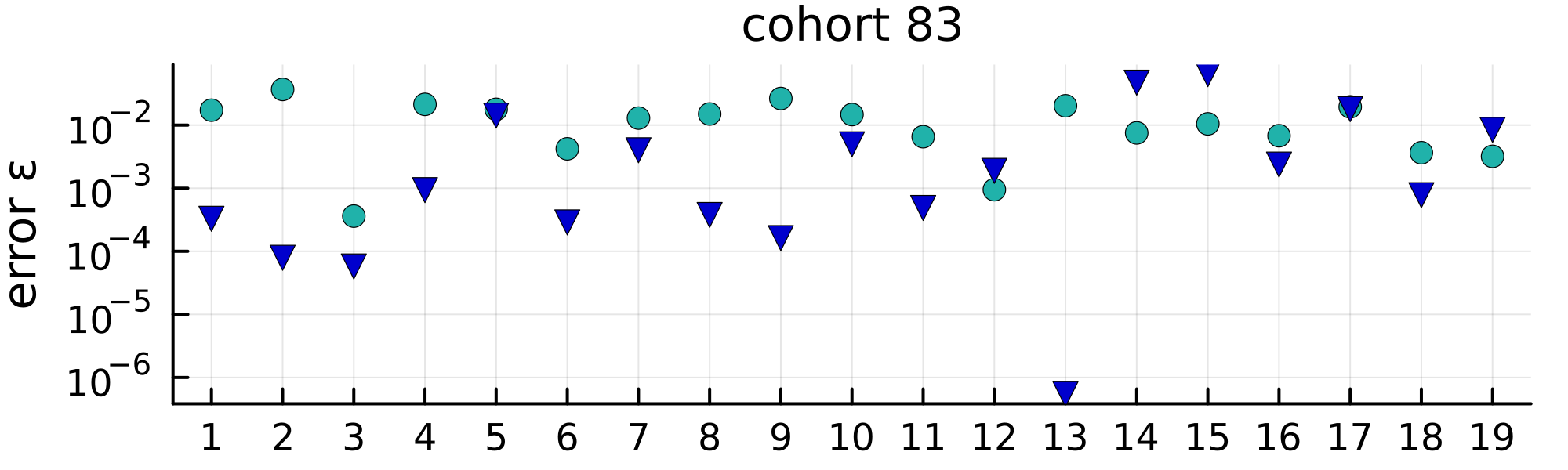}
    \end{subfigure}
    \begin{subfigure}[b]{0.45\textwidth}
        \includegraphics[width=\textwidth]{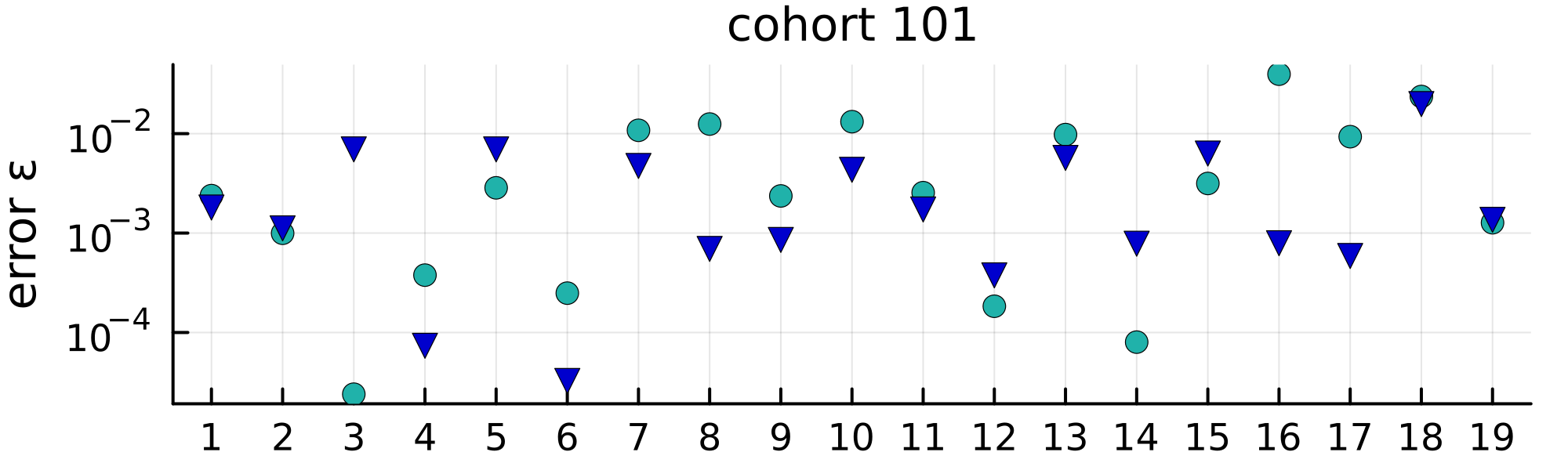}
    \end{subfigure}
    \begin{subfigure}[b]{0.45\textwidth}
        \includegraphics[width=\textwidth]{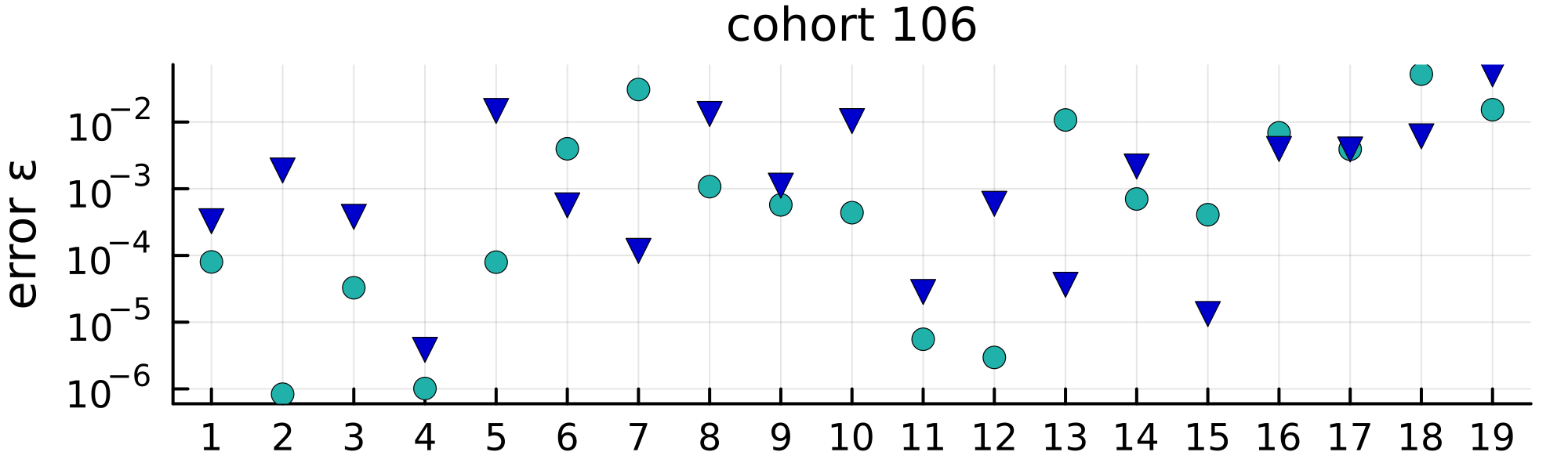}
    \end{subfigure}
    \begin{subfigure}[b]{0.45\textwidth}
        \includegraphics[width=\textwidth]{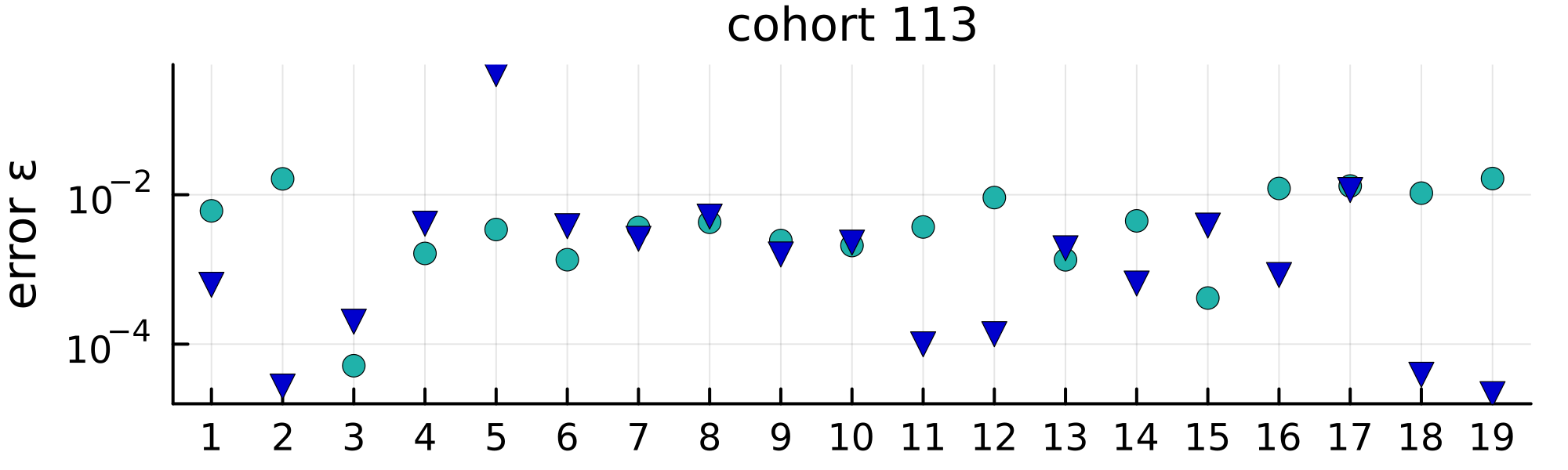}
    \end{subfigure}
    \begin{subfigure}[b]{0.45\textwidth}
        \includegraphics[width=\textwidth]{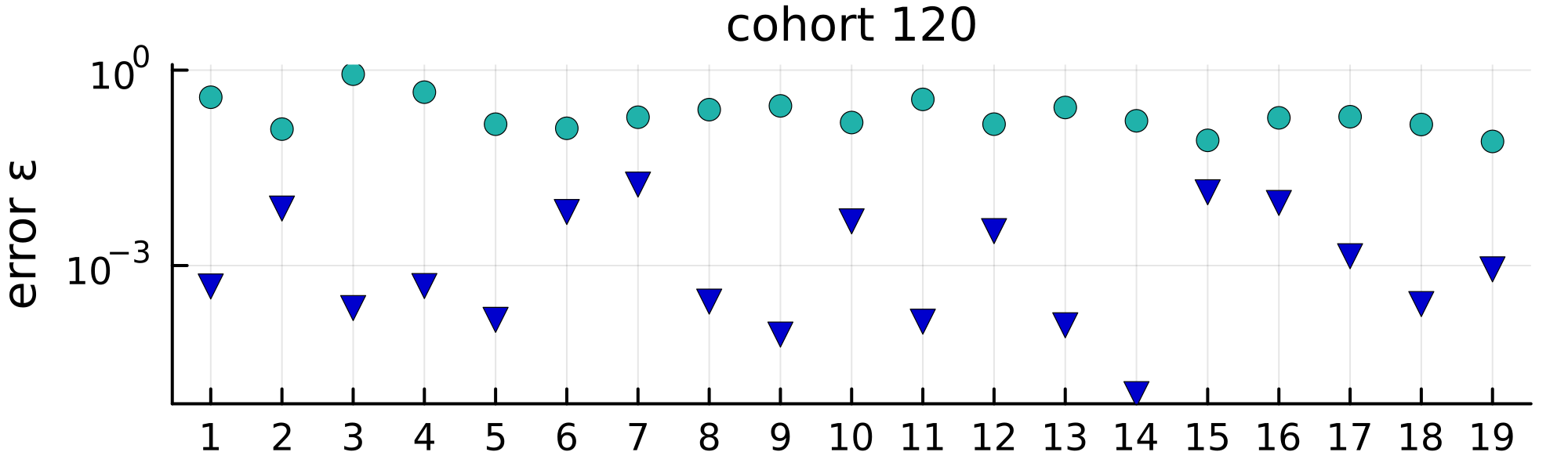}
    \end{subfigure}
    \begin{subfigure}[b]{0.45\textwidth}
        \includegraphics[width=\textwidth]{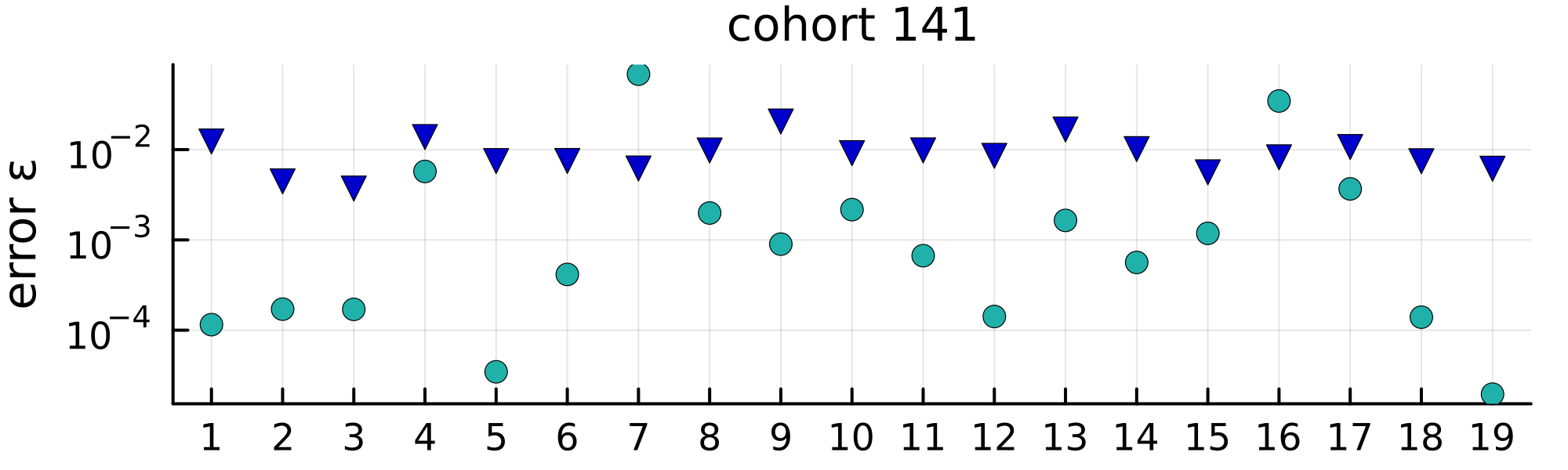}
    \end{subfigure}
    \begin{subfigure}[b]{0.45\textwidth}
        \includegraphics[width=\textwidth]{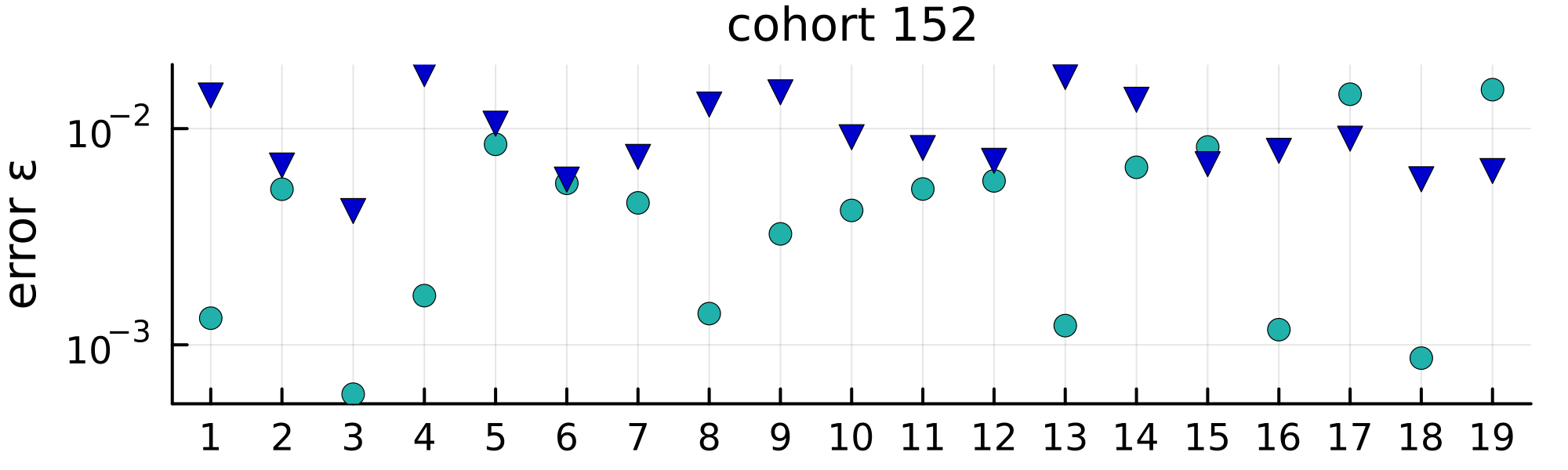}
    \end{subfigure}
    \begin{subfigure}[b]{0.45\textwidth}
        \includegraphics[width=\textwidth]{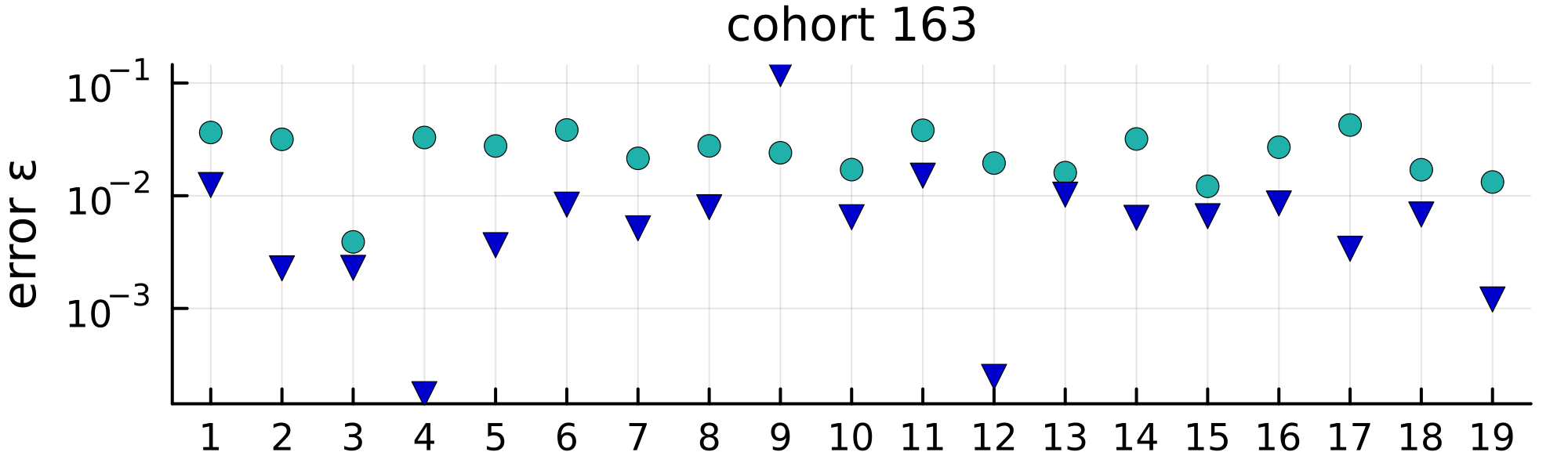}
    \end{subfigure}
    \begin{subfigure}[b]{0.45\textwidth}
        \includegraphics[width=\textwidth]{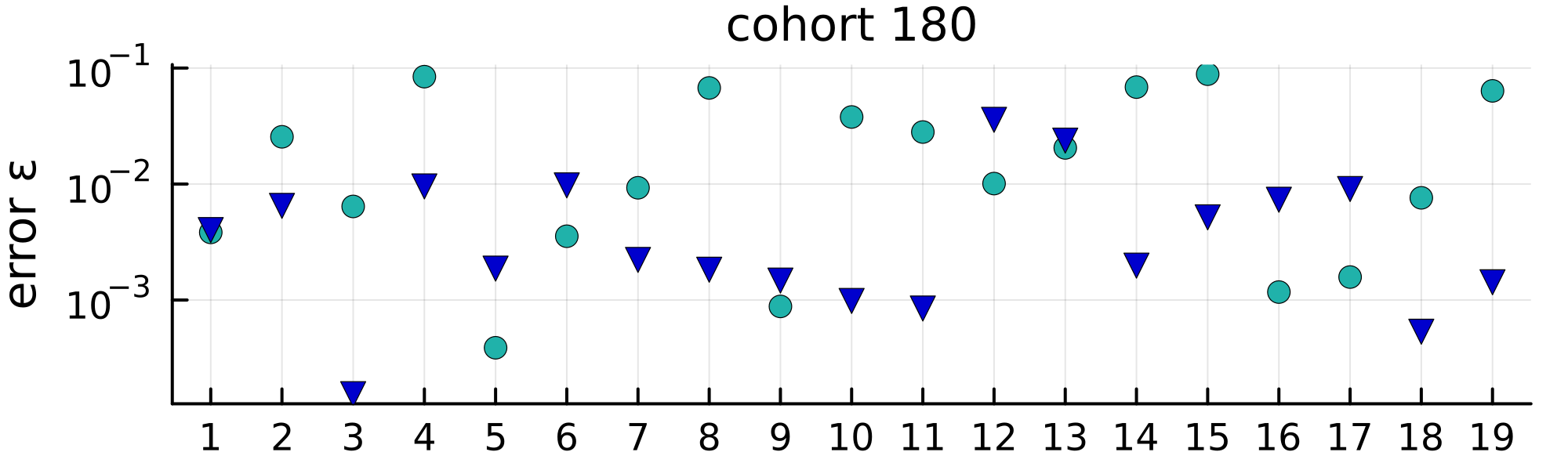}
    \end{subfigure}
    \begin{subfigure}[b]{0.45\textwidth}
        \includegraphics[width=\textwidth]{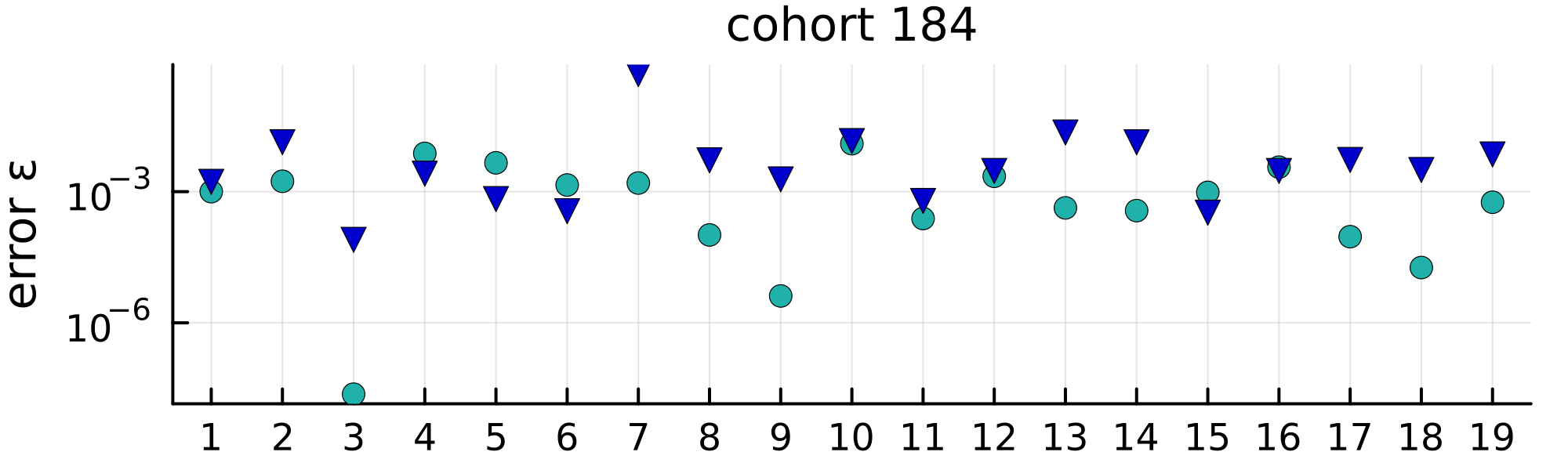}
    \end{subfigure}
    \begin{subfigure}[b]{0.45\textwidth}
        \includegraphics[width=\textwidth]{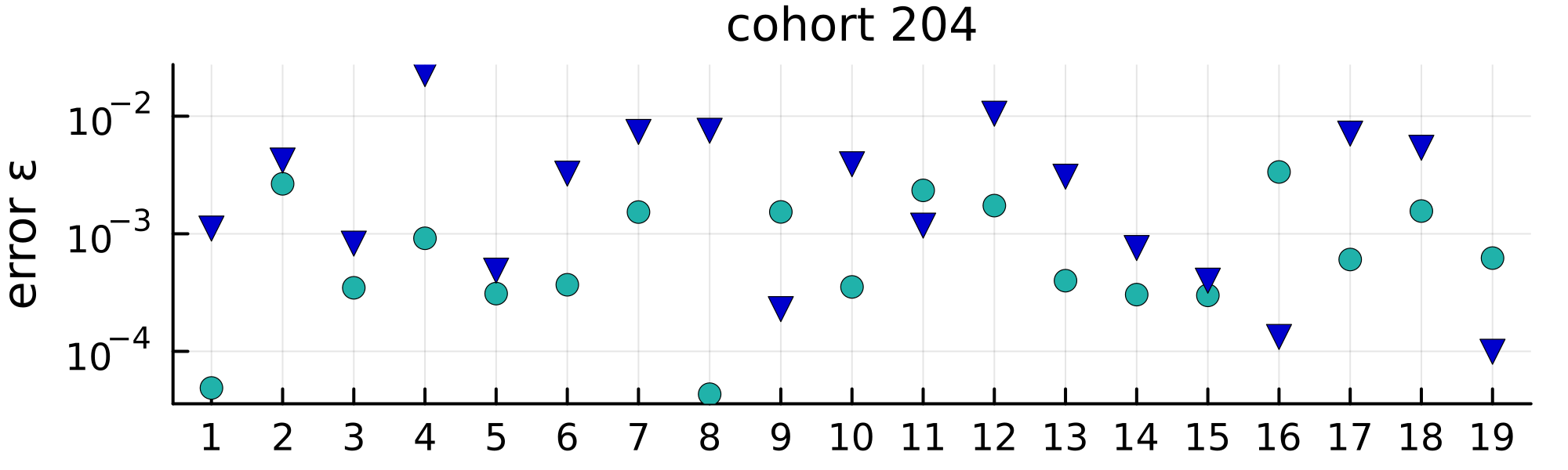}
    \end{subfigure}
    \begin{subfigure}[b]{0.45\textwidth}
        \includegraphics[width=\textwidth]{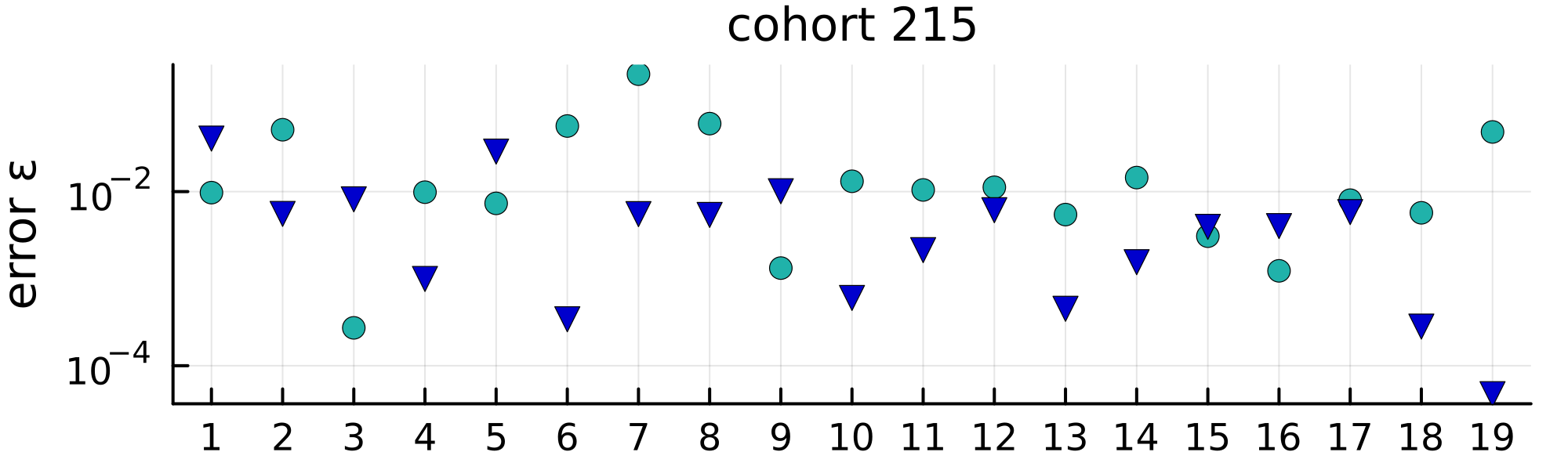}
    \end{subfigure}
    \begin{subfigure}[b]{0.45\textwidth}
        \includegraphics[width=\textwidth]{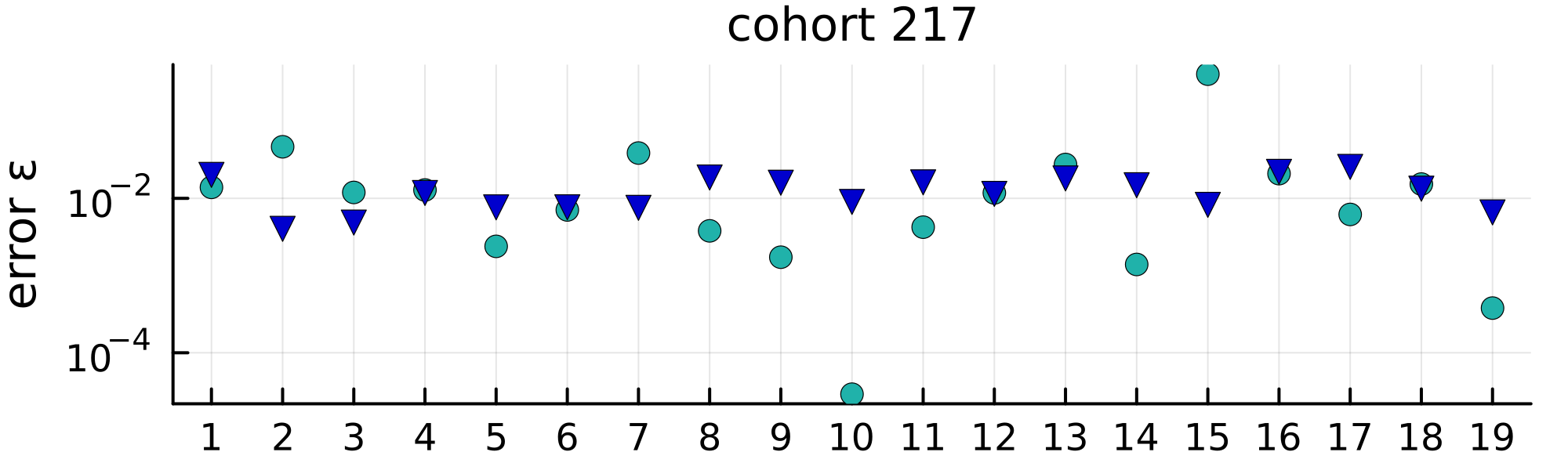}
    \end{subfigure}
    
    \caption{Errors $\varepsilon$ quantifying the difference between the true and inferred dose-response relationships $\bar{\Delta}^*_{het}$ (green) or $\bar{\Delta}^*_{hom}$ (blue) for each individual $i$ (x-axis) from each virtual cohort $m$ (subfigure).}
    \label{fig:Delta_error}
\end{figure}

Inferring how $\bar{\Delta}^*_{het}$ (or $\bar{\Delta}^*_{hom}$) evolves with the dose $d$ is crucial to understand how IFN$\alpha$ targets mutated HSCs by favouring their differentiation into progenitor cells and, therefore, understanding how the treatment might induce long-term remission. 
In particular, there might be a range of low doses for which no remission is possible since the value of $\bar{\Delta}^*$ would be strictly positive, meaning that mutated HSCs (heterozygous and/or homozygous) might still encounter more symmetrical than differentiated divisions, that is, the pool of mutated HSCs would continue to expand. 
Some of the models we study allow for such low-dose ranges without remission. Before focusing on these in the next paragraph, we can first study the inferences we made concerning the dose-response relationships of $\bar{\Delta}^*_{het}$ or $\bar{\Delta}^*_{hom}$.
For example, on cohort $m=204$ - for which the true model was correctly retrieved from our two-step model selection procedure - the comparison between the inferred and true dose-response relationships is displayed in Fig.~\ref{fig:Delta_204} for each virtual patient.
For that cohort/model, the dose-response relationship is affine for $\bar{\Delta}^*_{het}$ and affine sigmoid for $\bar{\Delta}^*_{hom}$. 
We get overall a good agreement between the inferred (mean \textit{a posteriori}) and true values of $\bar{\Delta}^*_{het}$ and $\bar{\Delta}^*_{hom}$ over a range of IFN$\alpha$ doses between the minimal and maximal ones administered to each patient (see Tab.~\ref{tab:info_patients}). Since we do not compare the value for a given dose (for example, the mean dose administered over the 450 first days of therapy, as presented in Fig.~\ref{fig:synth_Delta_het} and~\ref{fig:synth_Delta_hom}) but for a range of doses, we define the error between the inferred and true dose-response relationship, for patient $i$, by the L2-norm between the two relationships (the true and the inferred ones) normalized by the difference between the maximal and minimal IFN$\alpha$ dose administered to patient $i$. 
As we can see in Fig.~\ref{fig:Delta_204} for cohort 204, an error of about $10^{-3}$ corresponds to a good agreement when we observe some slight discrepancies already for errors of about $\sim 10^{-2}$.\\
In Fig.~\ref{fig:Delta_error}, we display these error values for each virtual patient of each synthetic cohort.

\FloatBarrier

\subsubsection{Minimal dose}
\label{sec:min_dose_synth_res}

\begin{figure}[h]
    \centering

    \begin{subfigure}[b]{0.7\textwidth}
        \includegraphics[width=\textwidth]{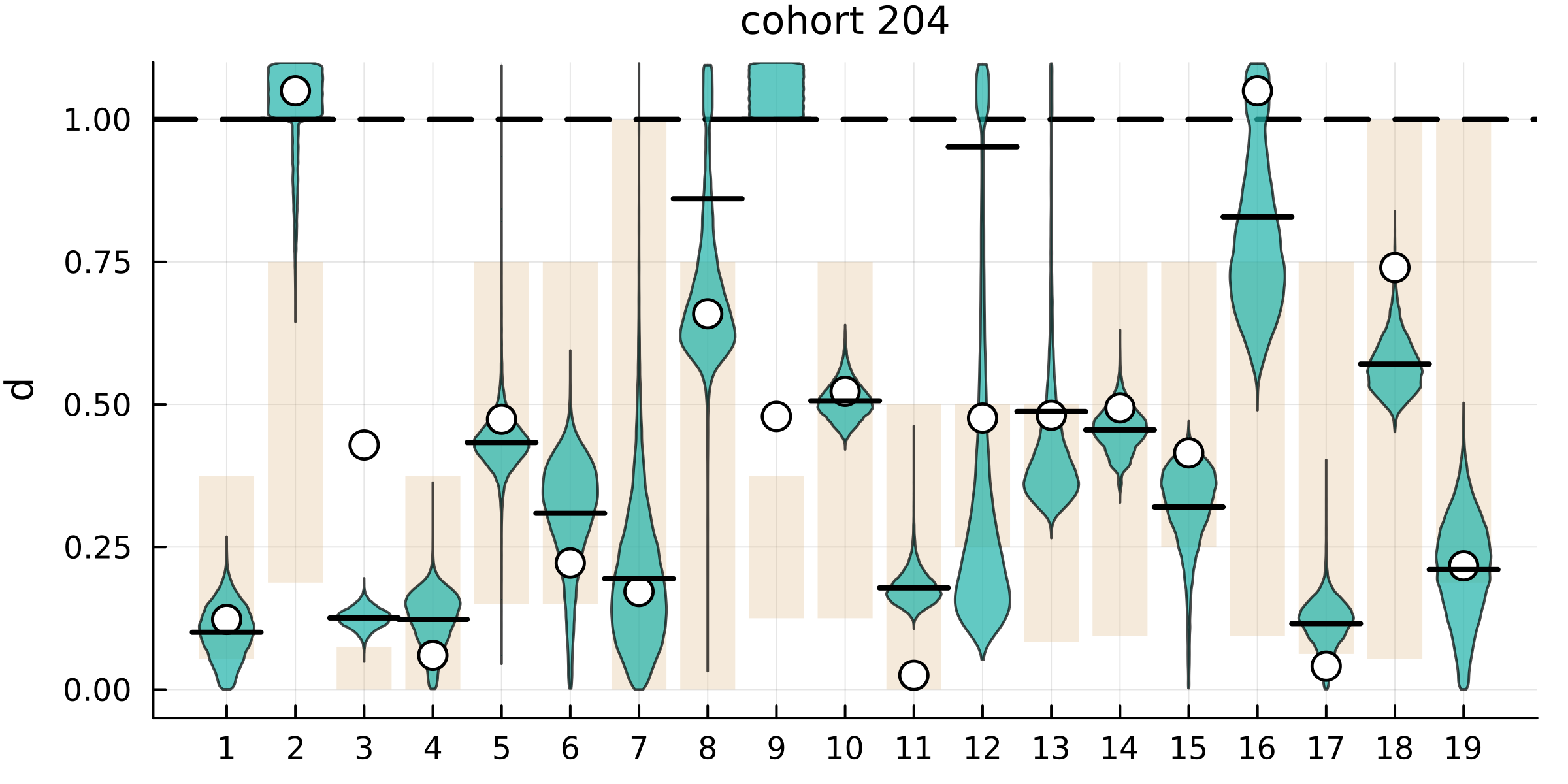}
    \end{subfigure}
    \begin{subfigure}[b]{0.7\textwidth}
        \includegraphics[width=\textwidth]{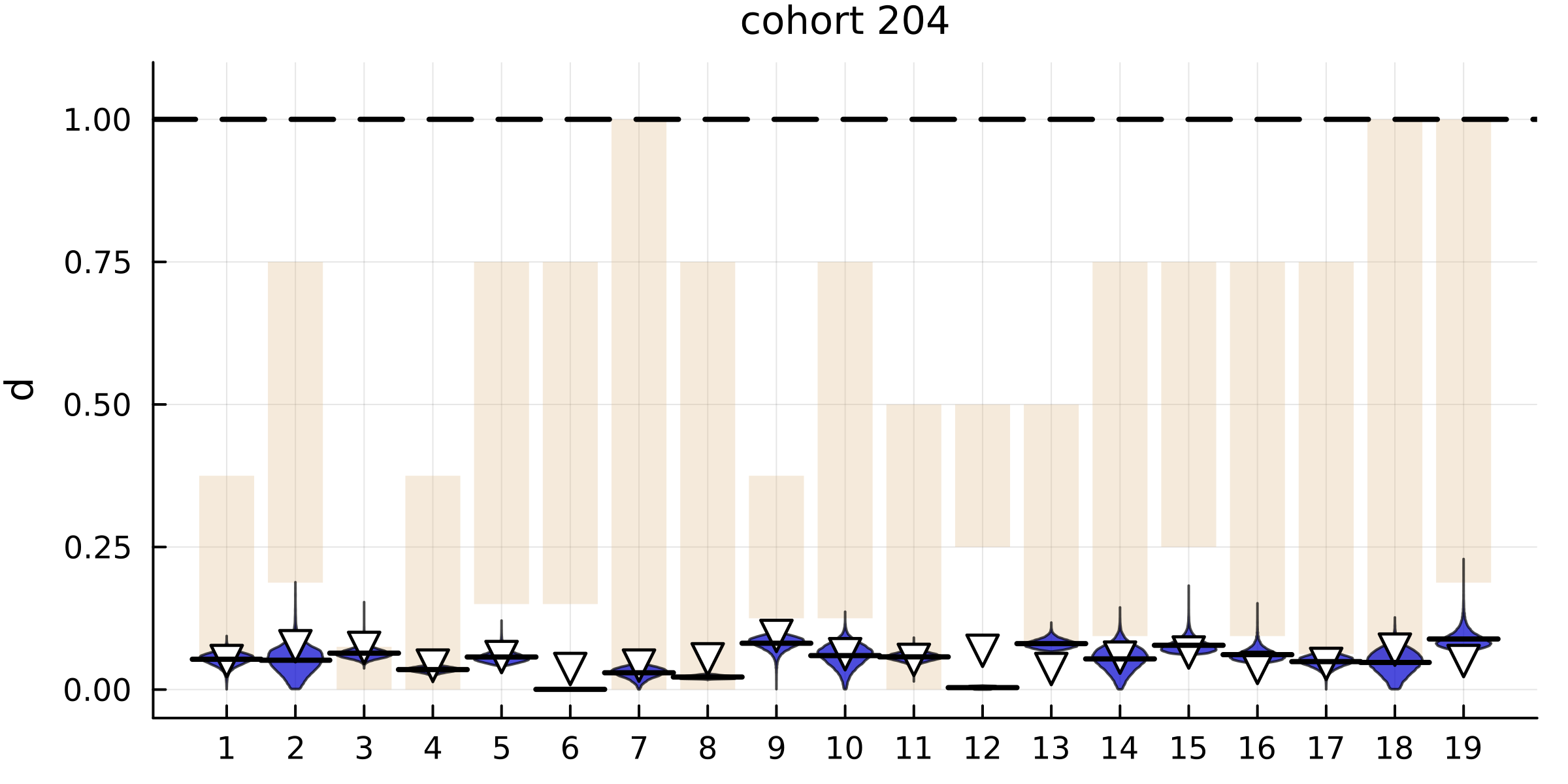}
    \end{subfigure}
    \begin{subfigure}[b]{0.7\textwidth}
        \includegraphics[width=\textwidth]{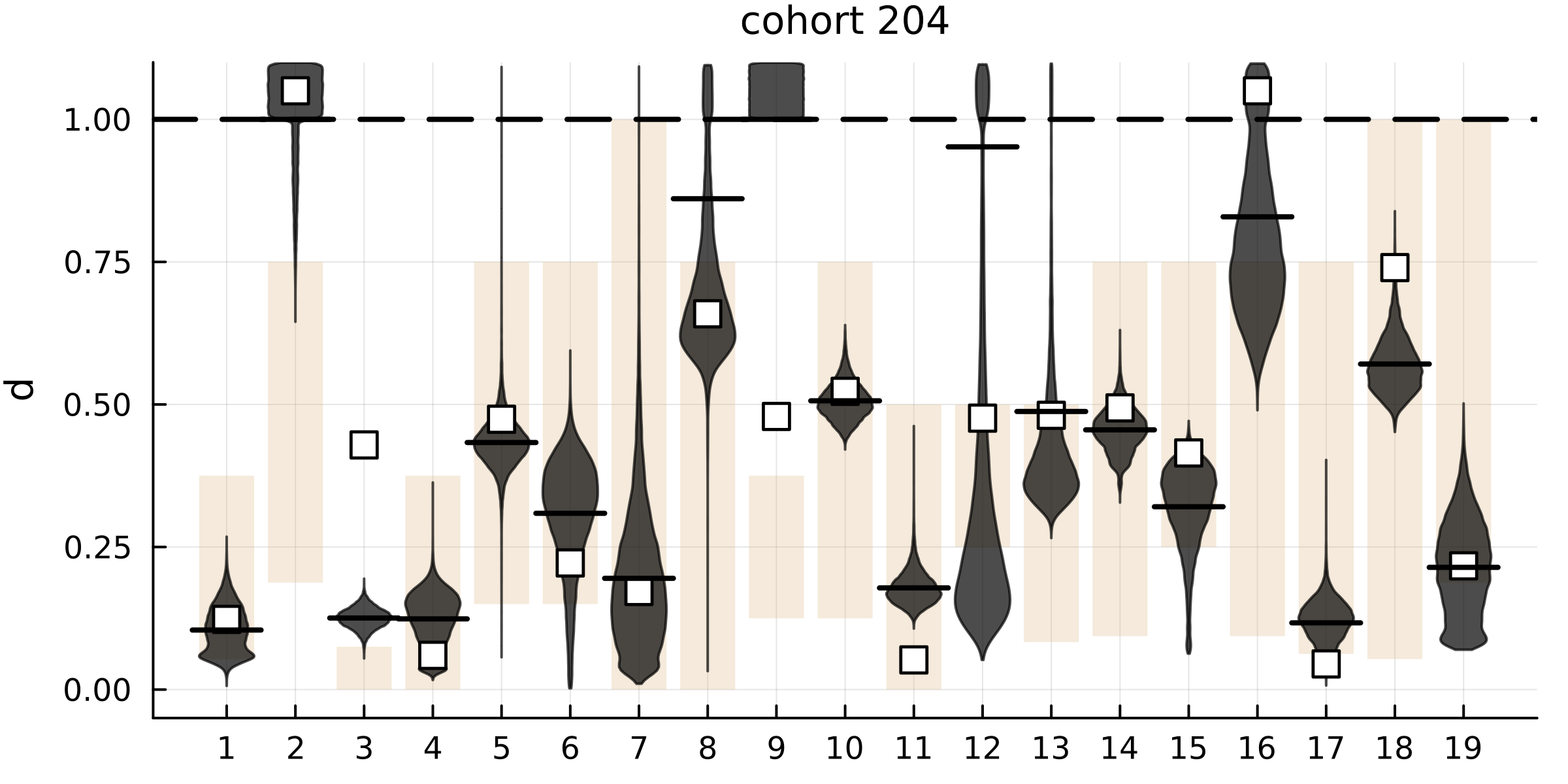}
    \end{subfigure}
    \caption{For each patient $i$ (x-axis) from the cohort $m=204$, we indicate his true heterozygous (top panel), homozygous (middle panel), or global (bottom panel) minimal dose $d^{ (i)}_{min}$ (white circles and triangles, and squares, respectively). In some cases, this minimal dose is higher than 1 (horizontal dash line), meaning that remission would not be possible in the range of IFN$\alpha$ dosages we consider. For each individual, we infer the posterior distribution of this minimal dose (violin plot). Values higher than one (horizontal dashed line) are forced to be in the range [1, 1.1] for clarity; they correspond in any case to unreachable doses. The horizontal black lines materialize the posterior mean. The objective is to estimate a minimal dose $d_{min}^{(i)}$ accounting both for the heterozygous and homozygous mutated cells (bottom panel). In that case, the minimal (global) dose is the maximum between the heterozygous and homozygous. The brown-shaded areas represent the ranges of doses administered to each patient.  }
    \label{fig:min_doses_204}
\end{figure}

Some of the models (index $m$) used to generate the data of the virtual cohorts allow the existence of an individual (index $i$) minimal dose $d_{min}^{(i,m)}$ under which no remission is possible (that is, one malignant clone, either heterozygous or homozygous, continues to expand).
When the model falls in that case, we indicated in Tab.~\ref{tab:synthetic_minimal_ind_dose} the minimal dose of each virtual patient. \\
A major part of our study, as presented in the main text, deals with estimating such an individual minimal dose (mean a posteriori).\\
Depending on the dose-response relationships we have for $\bar{\Delta}^*_{het}$ and $\bar{\Delta}^*_{hom}$, we could have both a minimal dose associated with the heterozygous clone and a minimal dose associated with the homozygous clone (as illustrated with the patients from the virtual cohort $m=204$, see Fig.~\ref{fig:min_doses_204}).
The (global) minimal dose is then defined as the maximum between both; it is the minimal dose required to induce remission for both heterozygous and homozygous HSCs.

\begin{figure}[h]
    \centering

    \begin{subfigure}[b]{0.27\textwidth}
        \includegraphics[width=\textwidth]{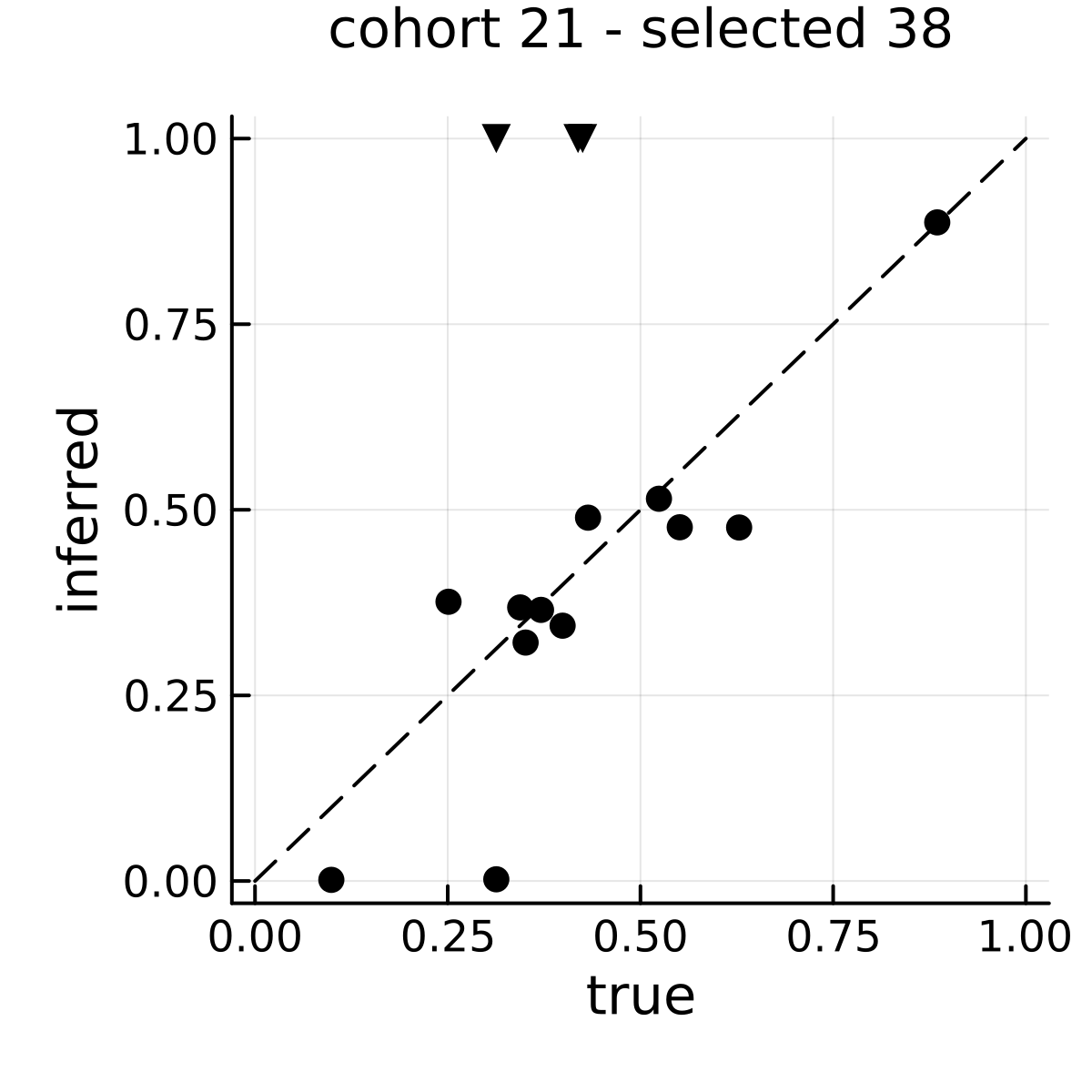}
    \end{subfigure}
    \begin{subfigure}[b]{0.27\textwidth}
        \includegraphics[width=\textwidth]{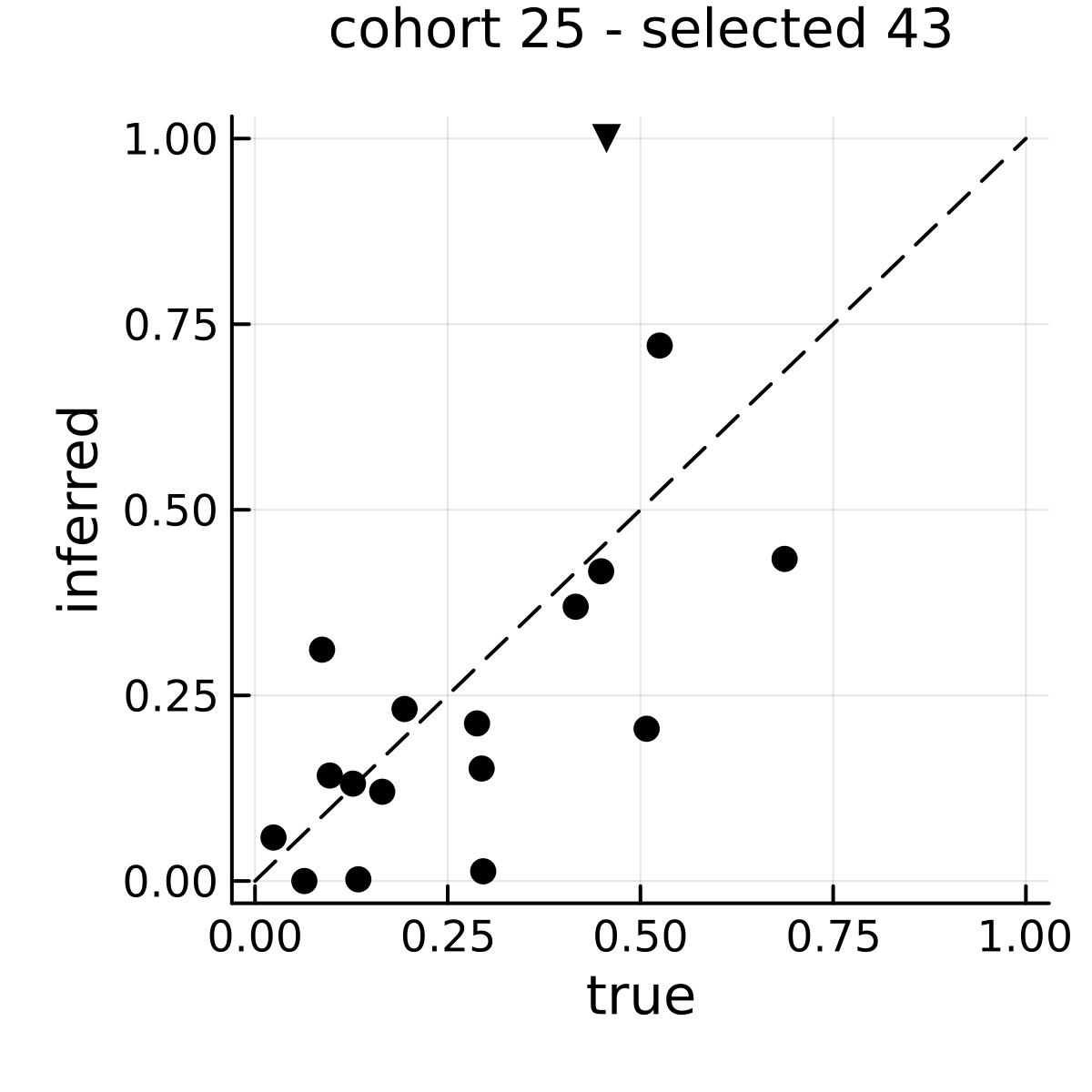}
    \end{subfigure}
    \begin{subfigure}[b]{0.27\textwidth}
        \includegraphics[width=\textwidth]{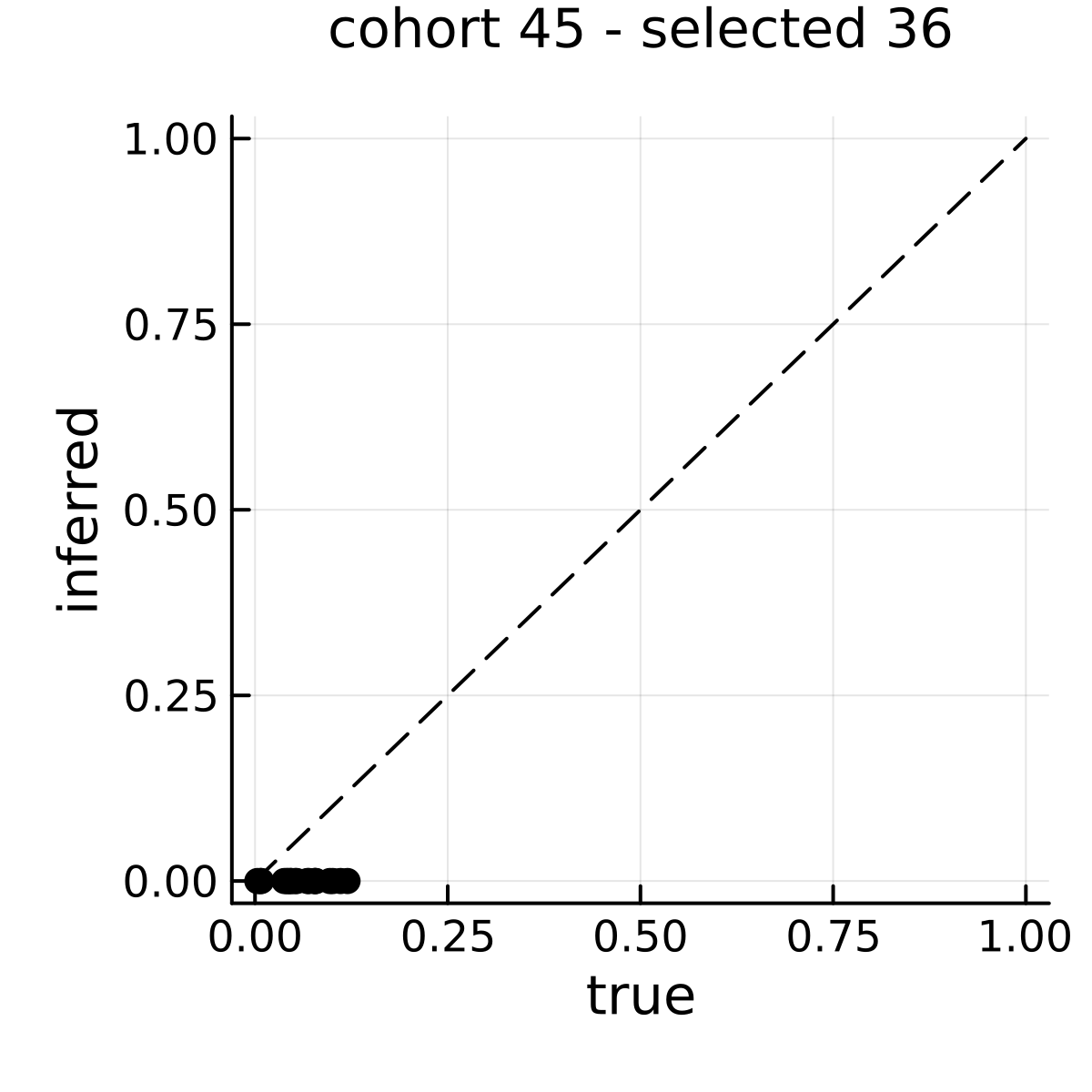}
    \end{subfigure}
    \begin{subfigure}[b]{0.27\textwidth}
        \includegraphics[width=\textwidth]{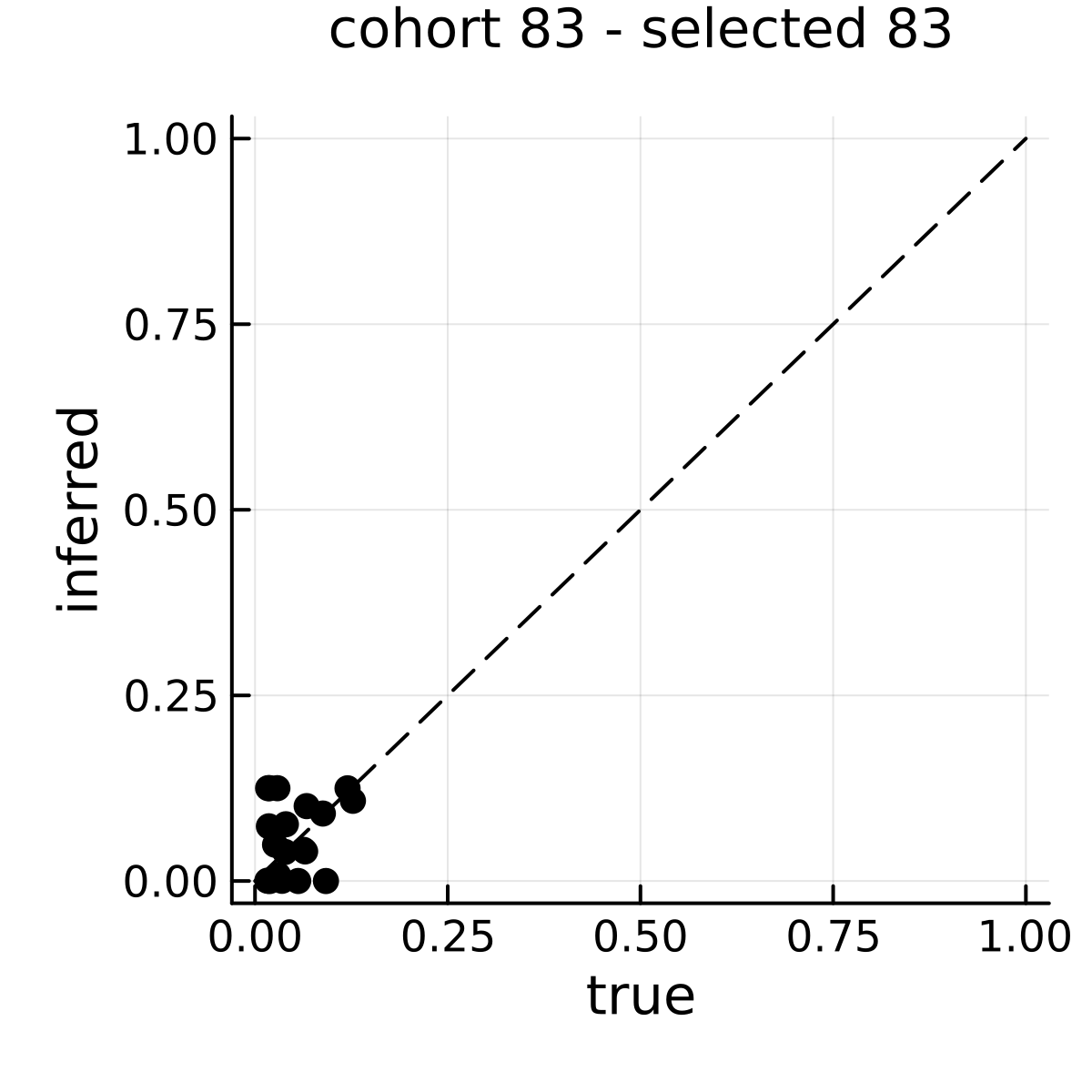}
    \end{subfigure}
    \begin{subfigure}[b]{0.27\textwidth}
        \includegraphics[width=\textwidth]{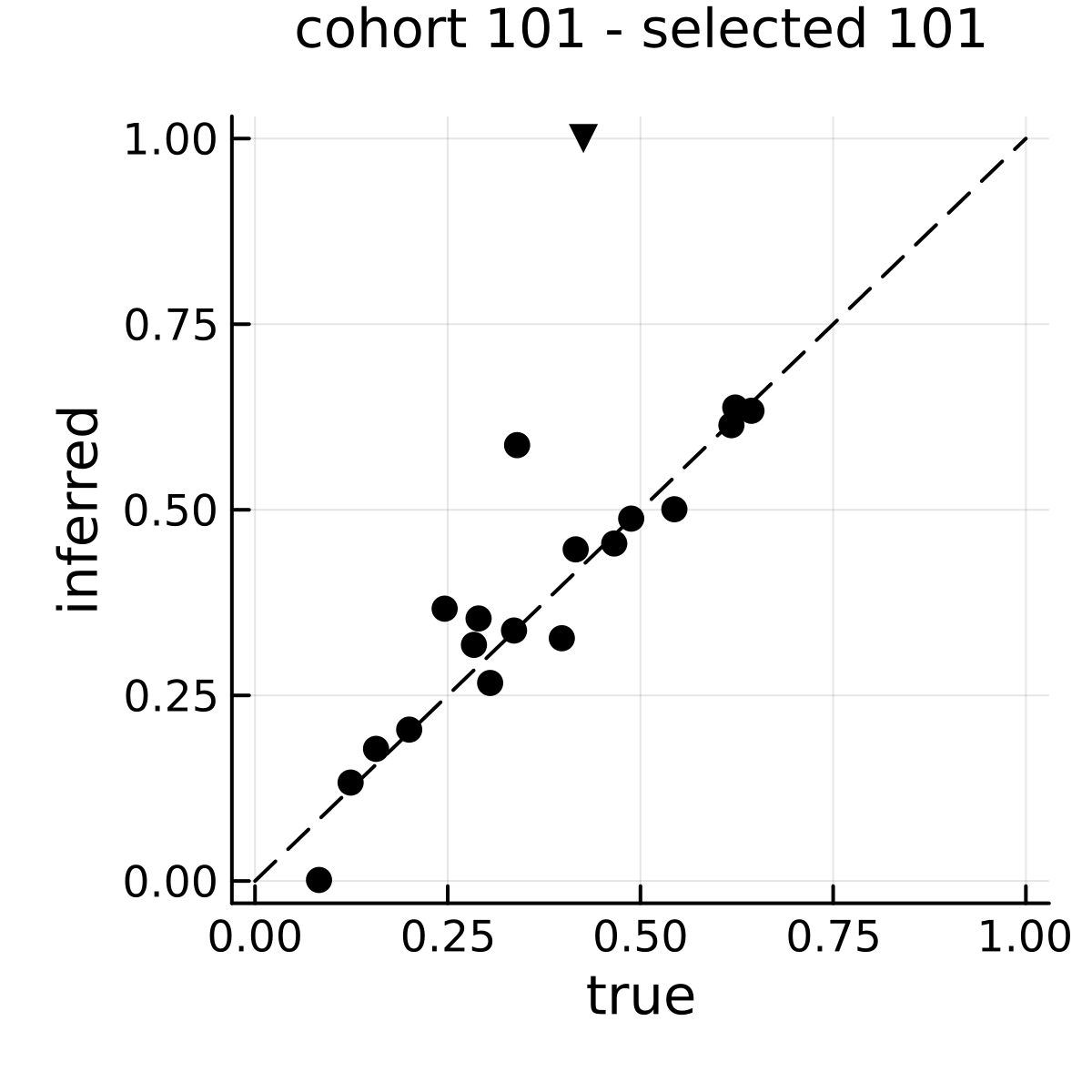}
    \end{subfigure}
    \begin{subfigure}[b]{0.27\textwidth}
        \includegraphics[width=\textwidth]{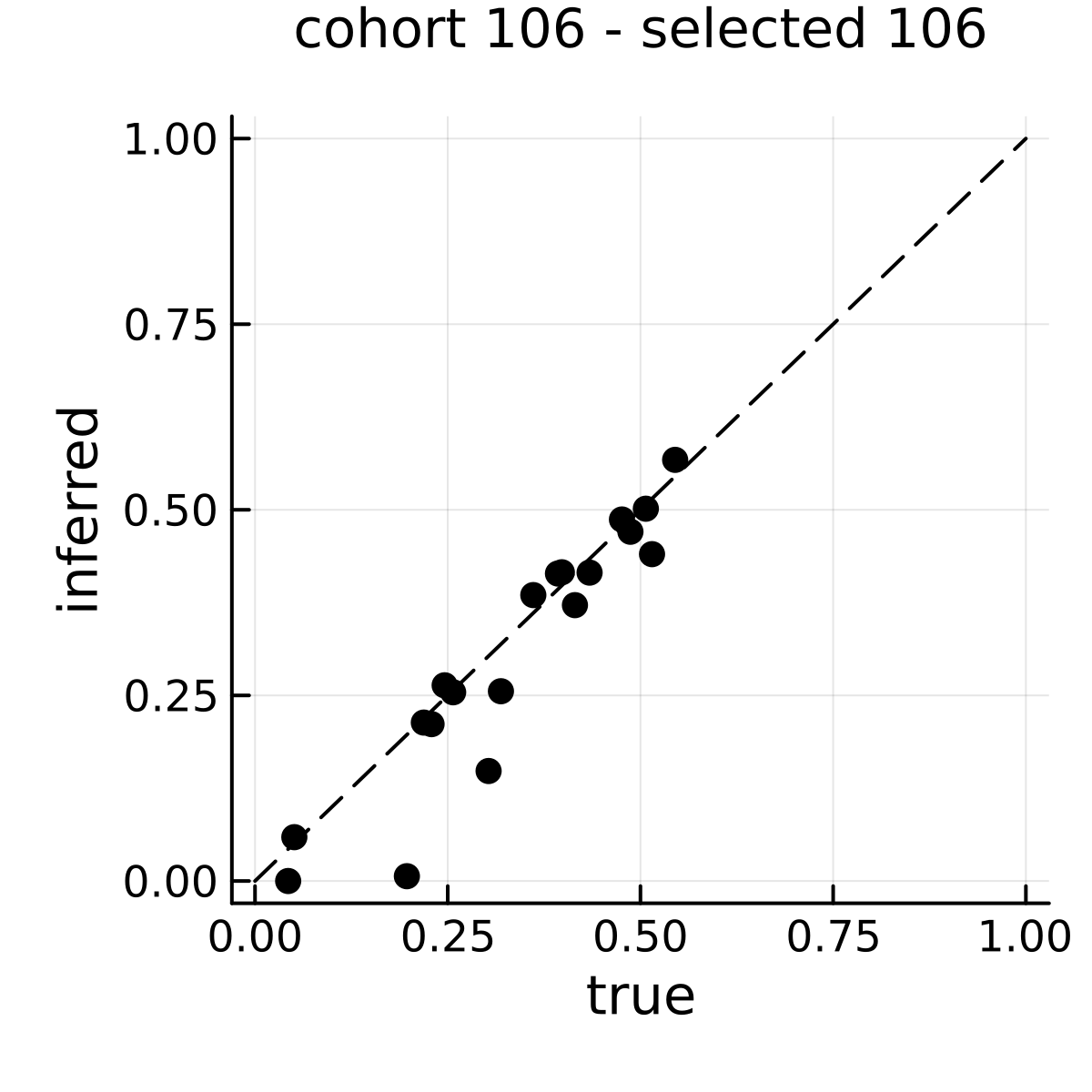}
    \end{subfigure}
    \begin{subfigure}[b]{0.27\textwidth}
        \includegraphics[width=\textwidth]{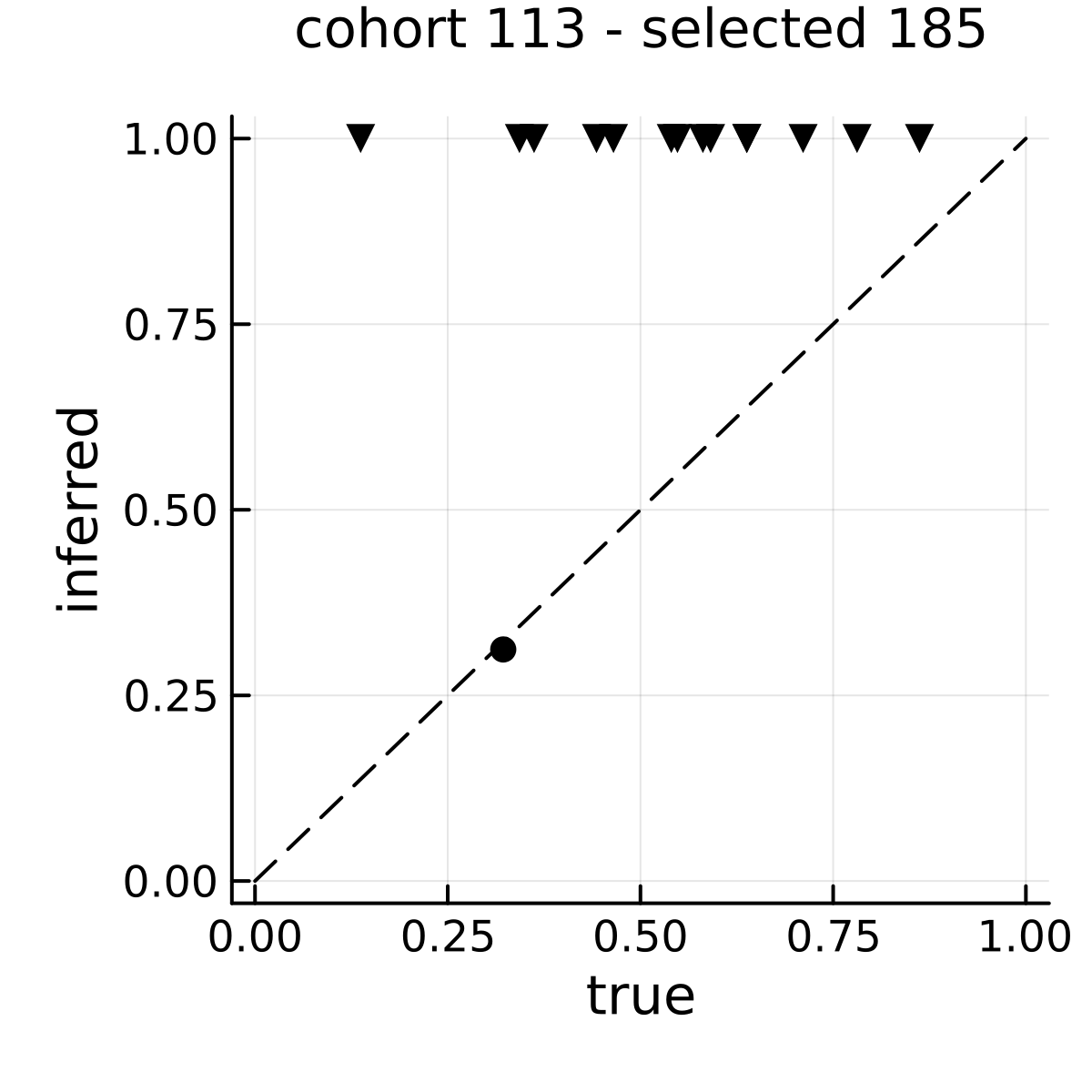}
    \end{subfigure}
    \begin{subfigure}[b]{0.27\textwidth}
        \includegraphics[width=\textwidth]{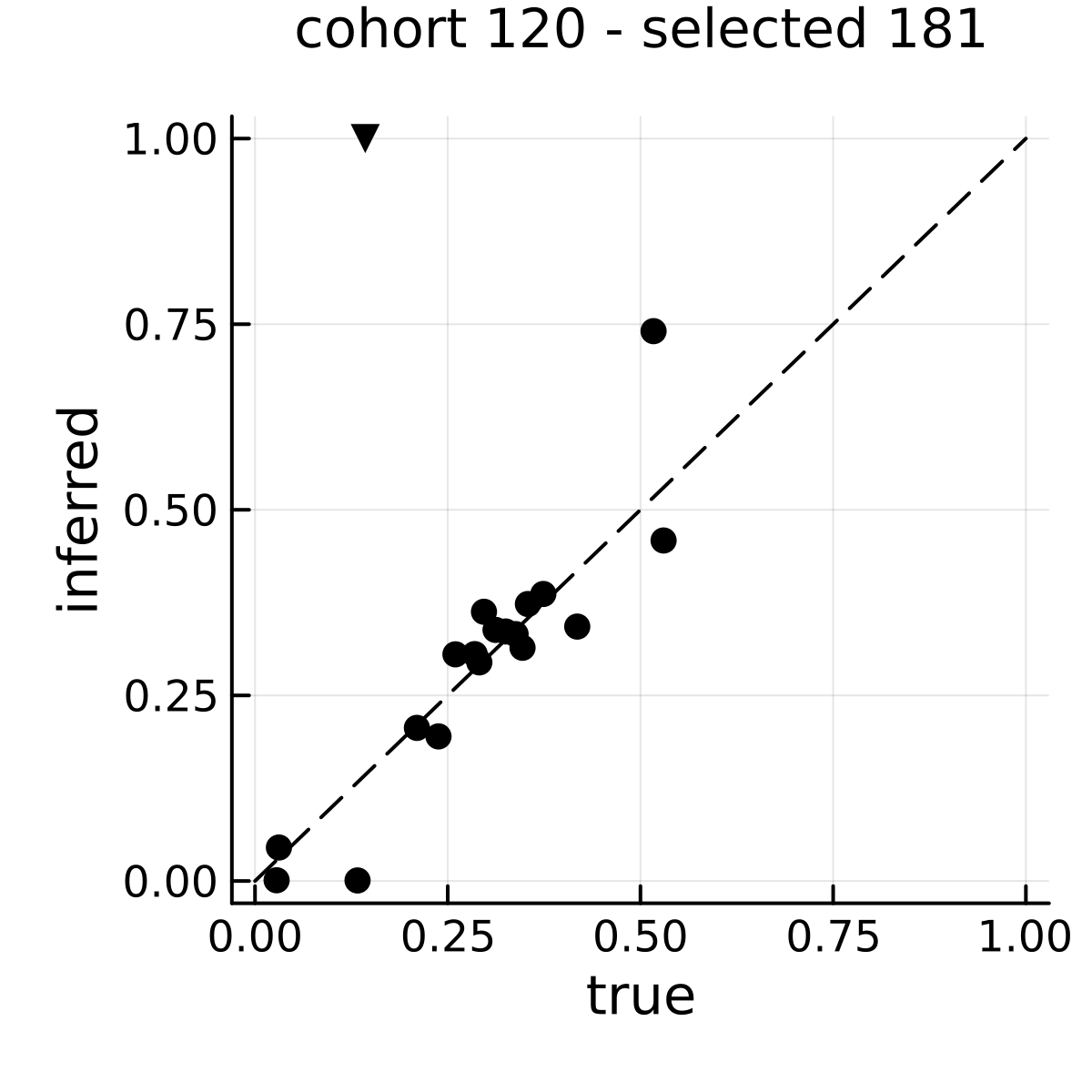}
    \end{subfigure}
    \begin{subfigure}[b]{0.27\textwidth}
        \includegraphics[width=\textwidth]{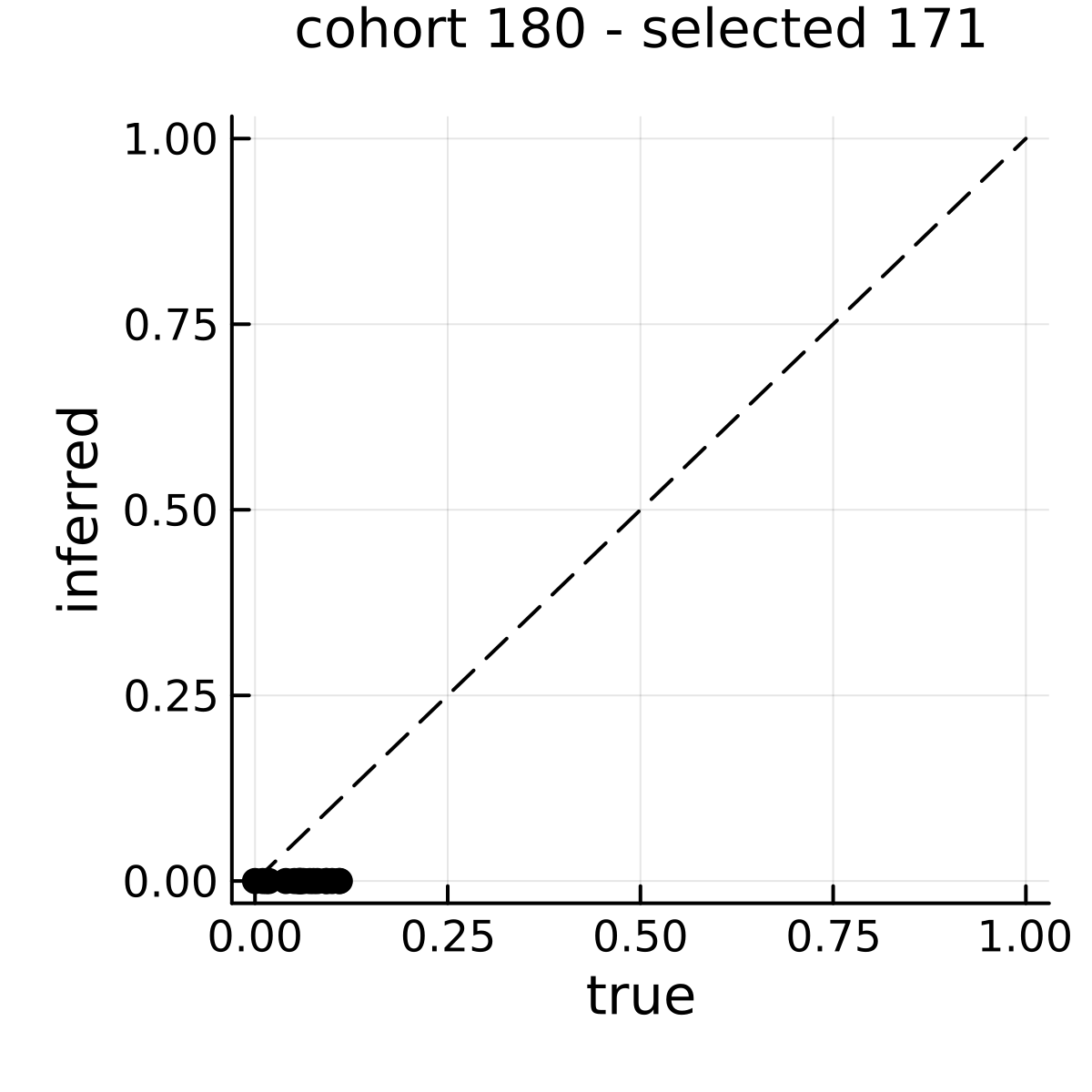}
    \end{subfigure}
    \begin{subfigure}[b]{0.27\textwidth}
        \includegraphics[width=\textwidth]{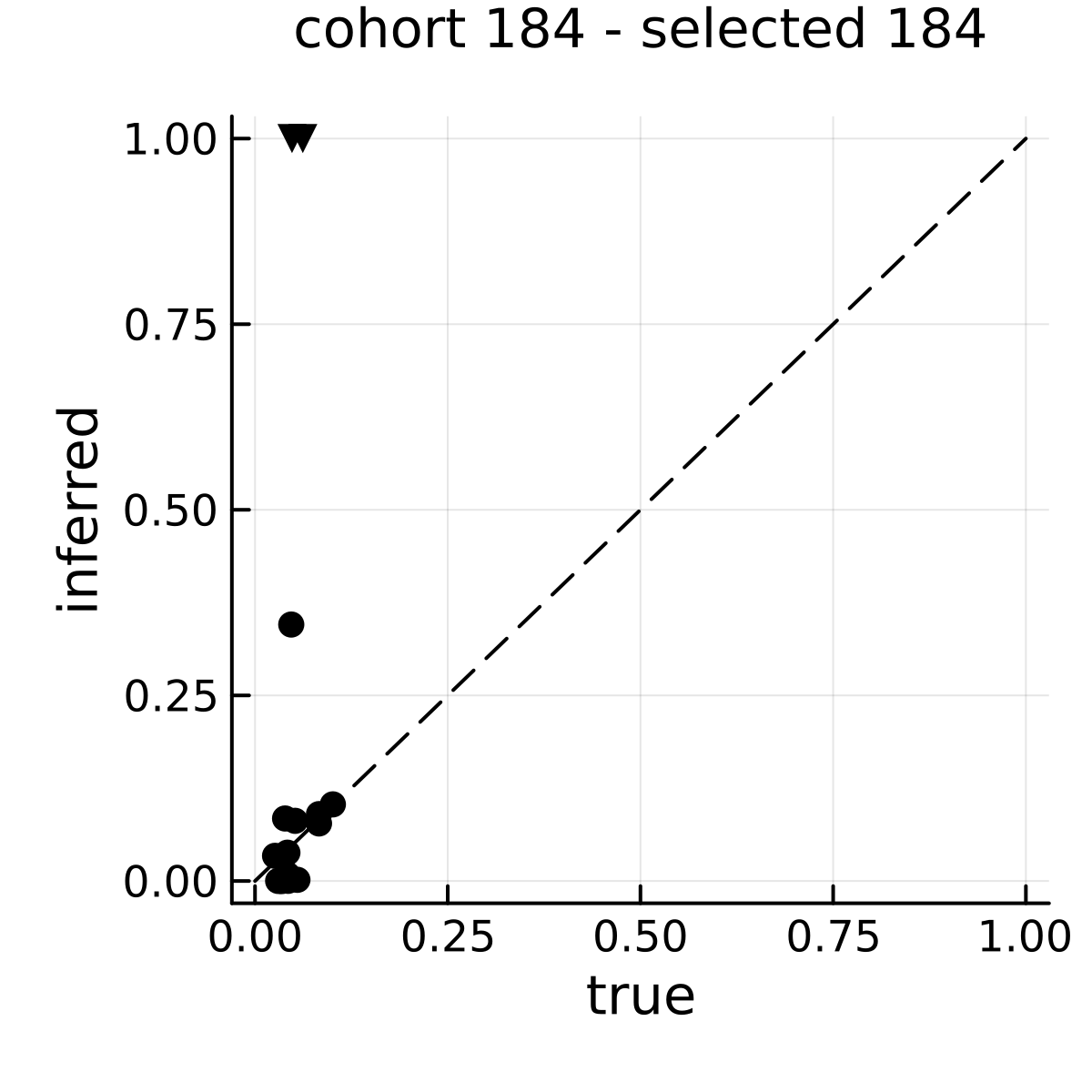}
    \end{subfigure}
    \begin{subfigure}[b]{0.27\textwidth}
        \includegraphics[width=\textwidth]{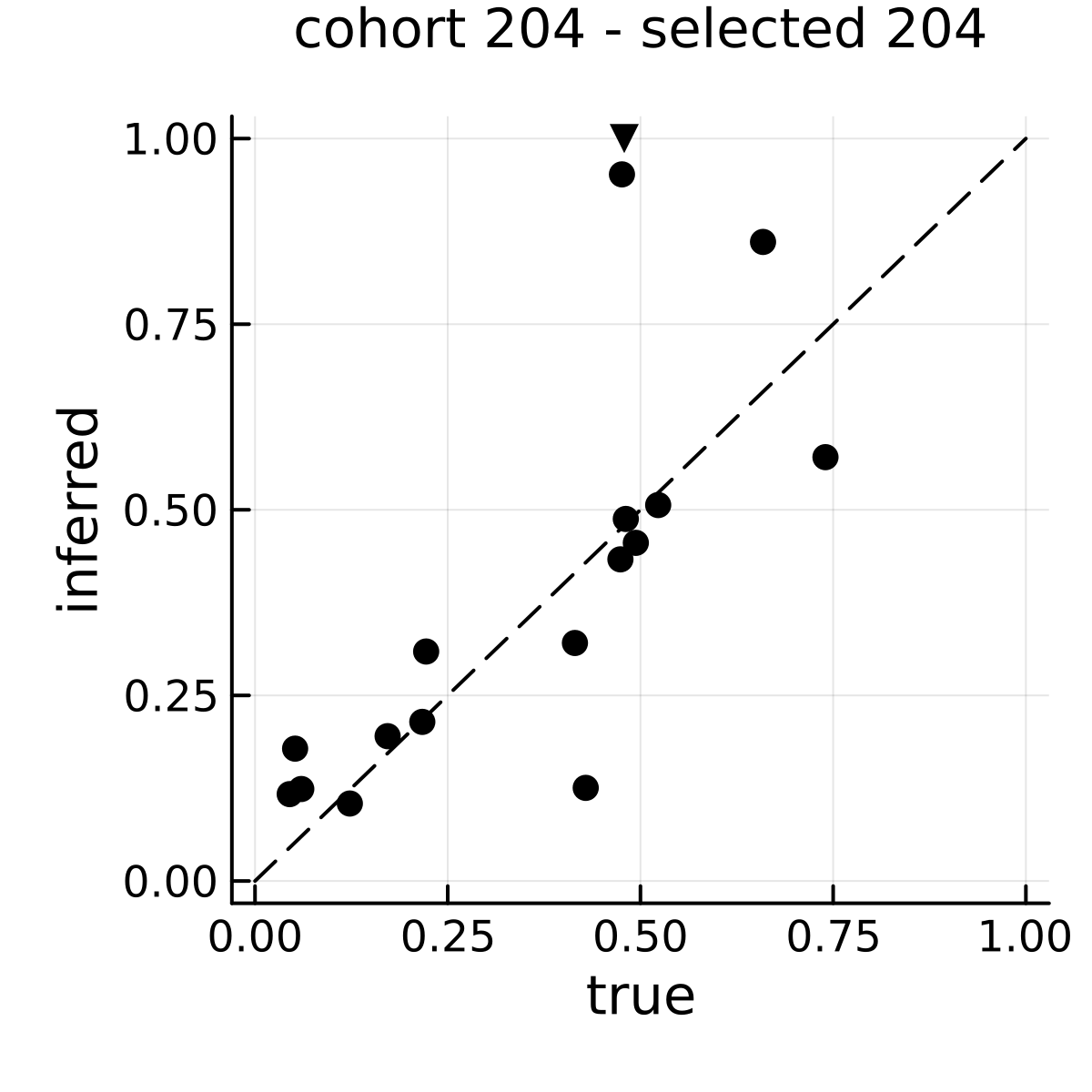}
    \end{subfigure}
    \begin{subfigure}[b]{0.27\textwidth}
        \includegraphics[width=\textwidth]{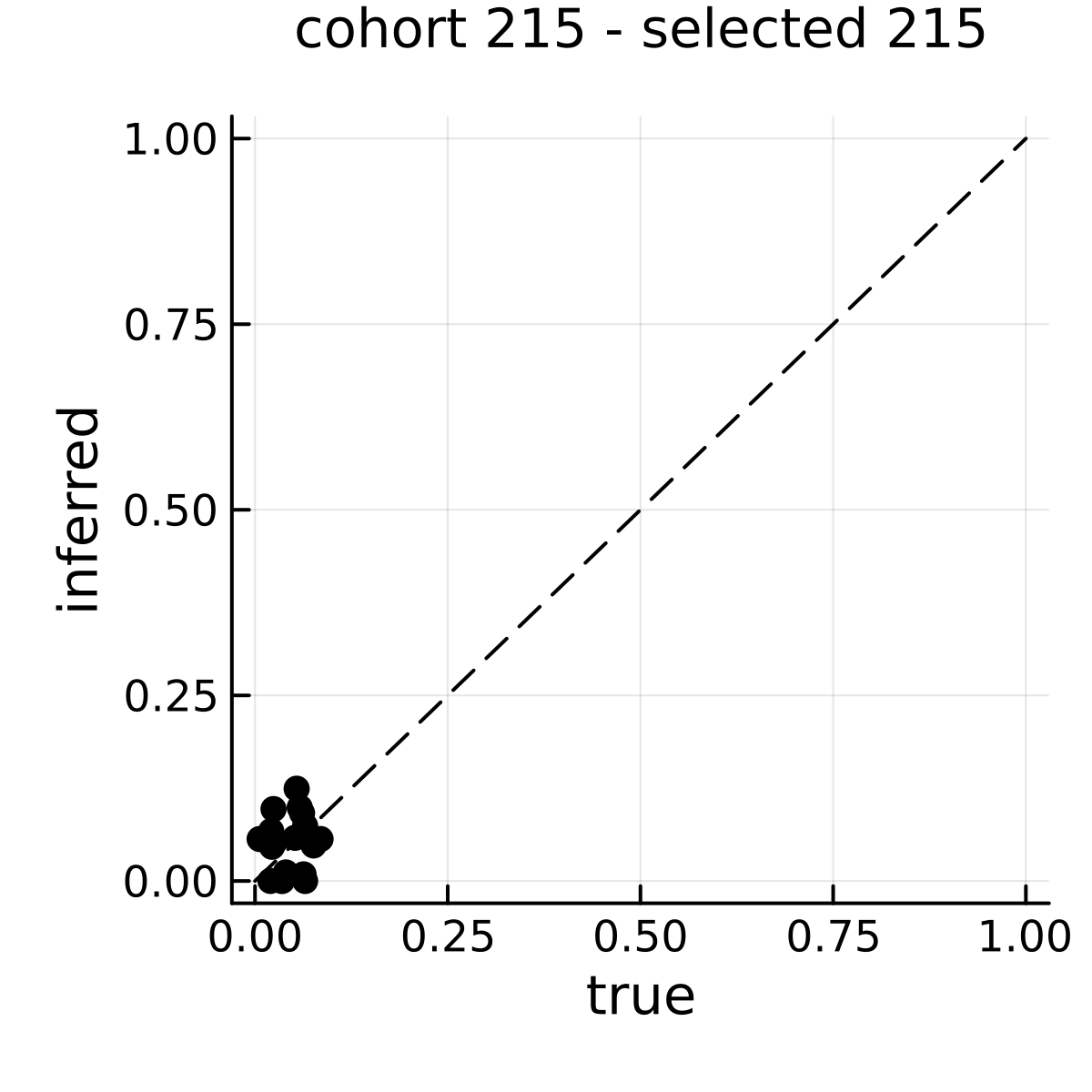}
    \end{subfigure}
    \begin{subfigure}[b]{0.27\textwidth}
        \includegraphics[width=\textwidth]{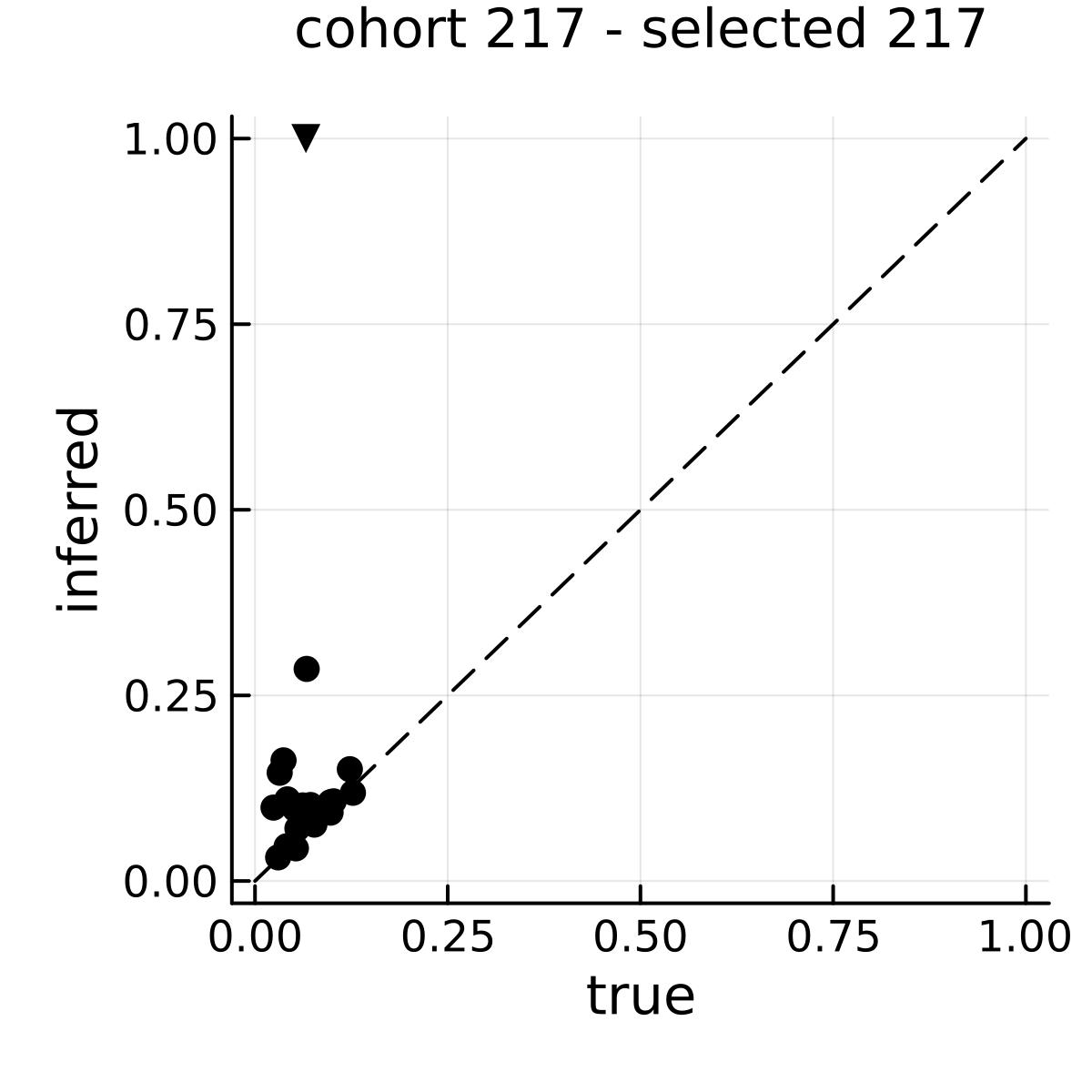}
    \end{subfigure}
    
    \caption{For each synthetic cohort $m$ (sub-panel) for which there exists a minimal dose (see Tab.~\ref{tab:synthetic_minimal_ind_dose}), we display the inferred (mean a posteriori, y-axis) vs true (x-axis) minimal dose $d_{min}^{(i,m)}$.
    Individuals for which the real minimal dose is higher than one - meaning that a global remission would not be feasible in the range of doses considered - are not displayed. 
    When a minimal dose is estimated to be higher than one, we set it to one (and represent it with a triangle) to generate the figure.
    }
    \label{fig:min_dose_inferred_vs_true}
\end{figure}

\begin{figure}[h]
    \centering

    \begin{subfigure}[b]{0.22\textwidth}
        \includegraphics[width=\textwidth]{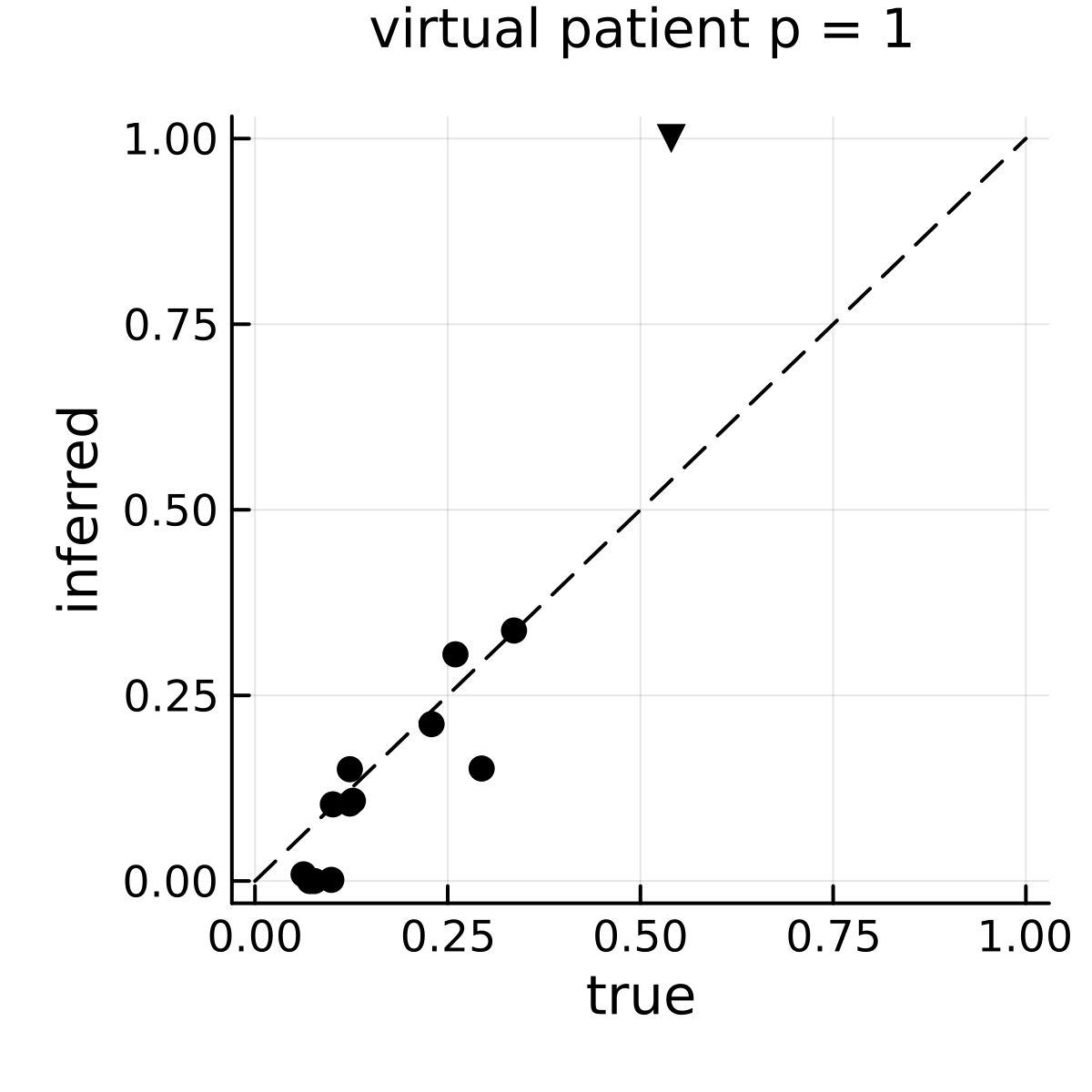}
    \end{subfigure}
    \begin{subfigure}[b]{0.22\textwidth}
        \includegraphics[width=\textwidth]{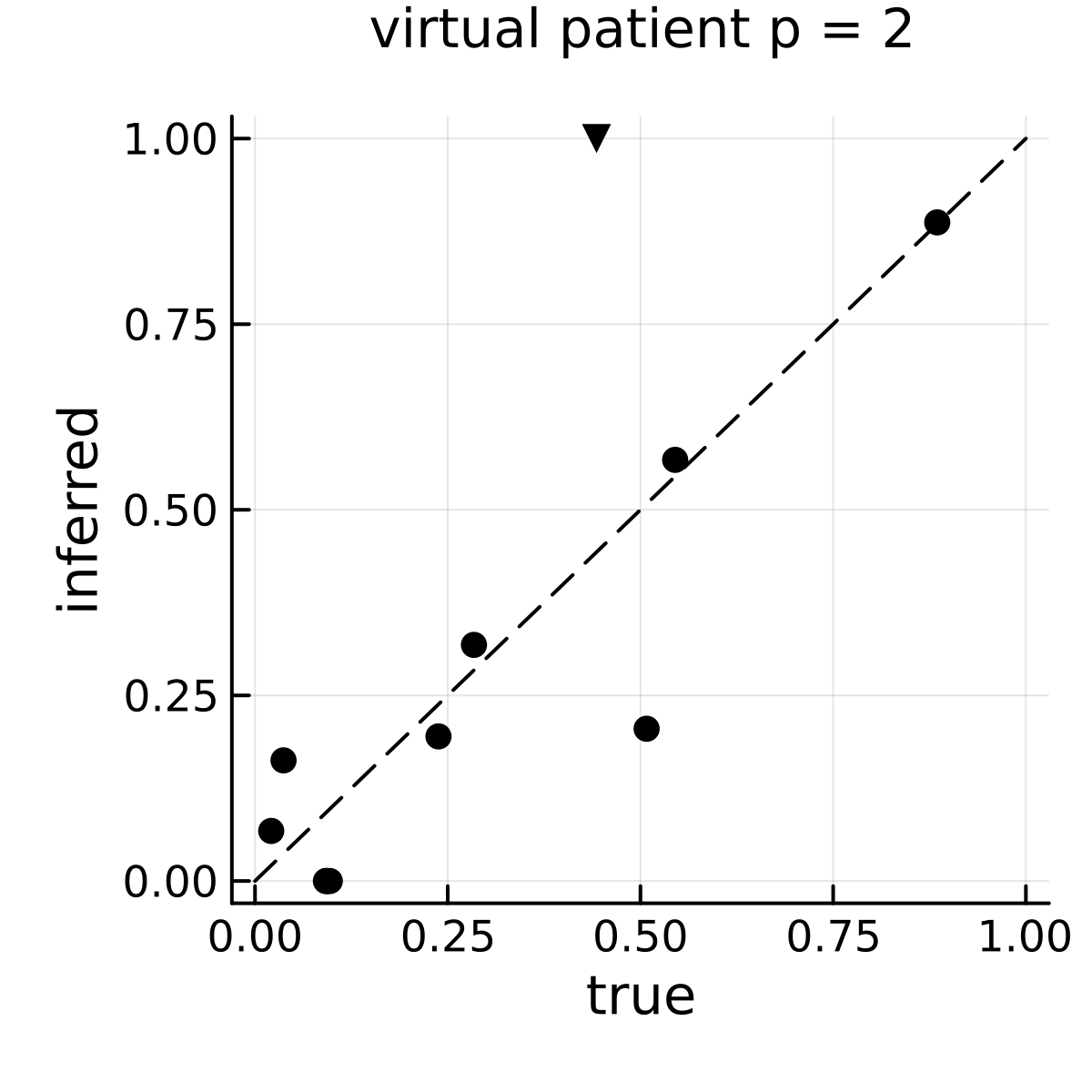}
    \end{subfigure}
    \begin{subfigure}[b]{0.22\textwidth}
        \includegraphics[width=\textwidth]{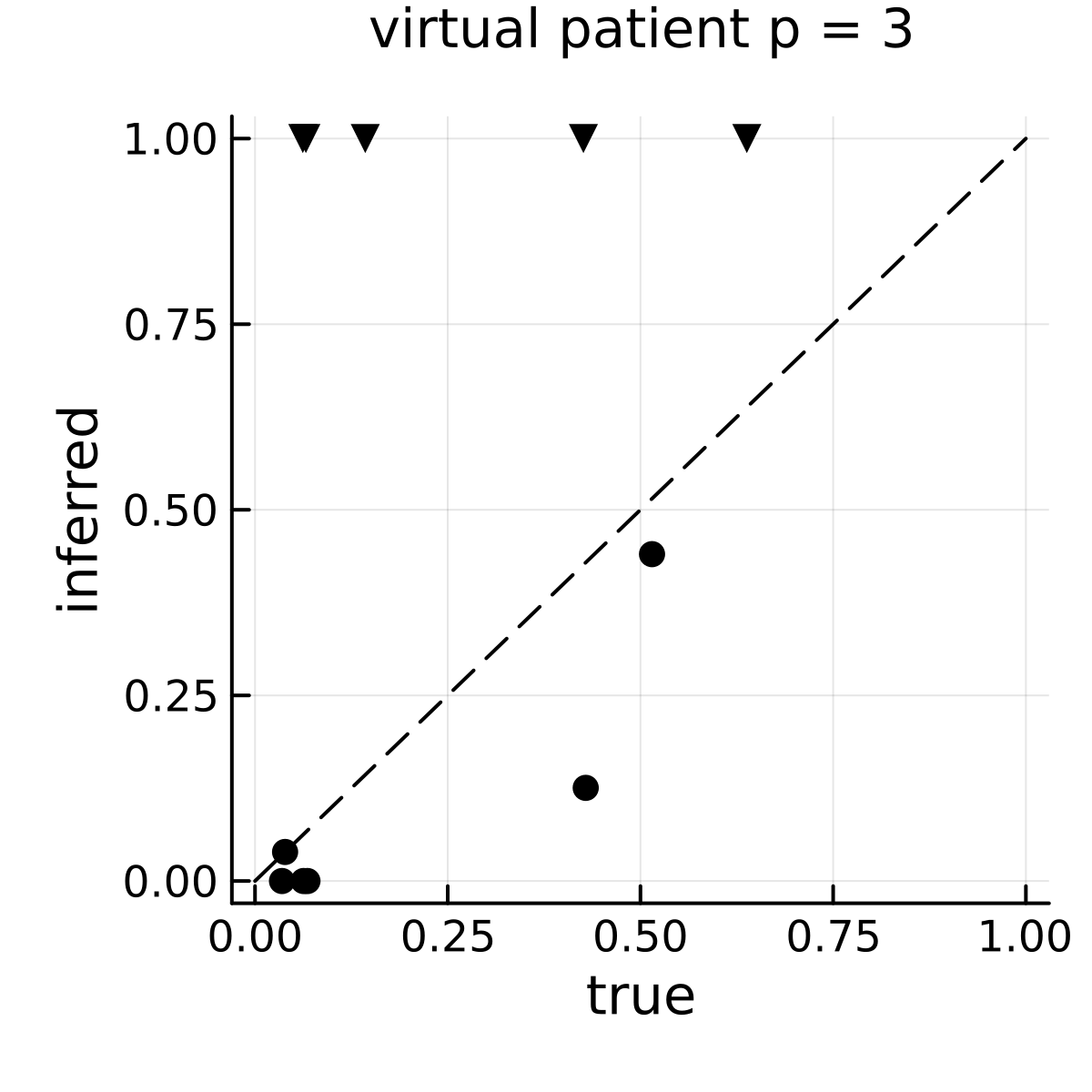}
    \end{subfigure}
    \begin{subfigure}[b]{0.22\textwidth}
        \includegraphics[width=\textwidth]{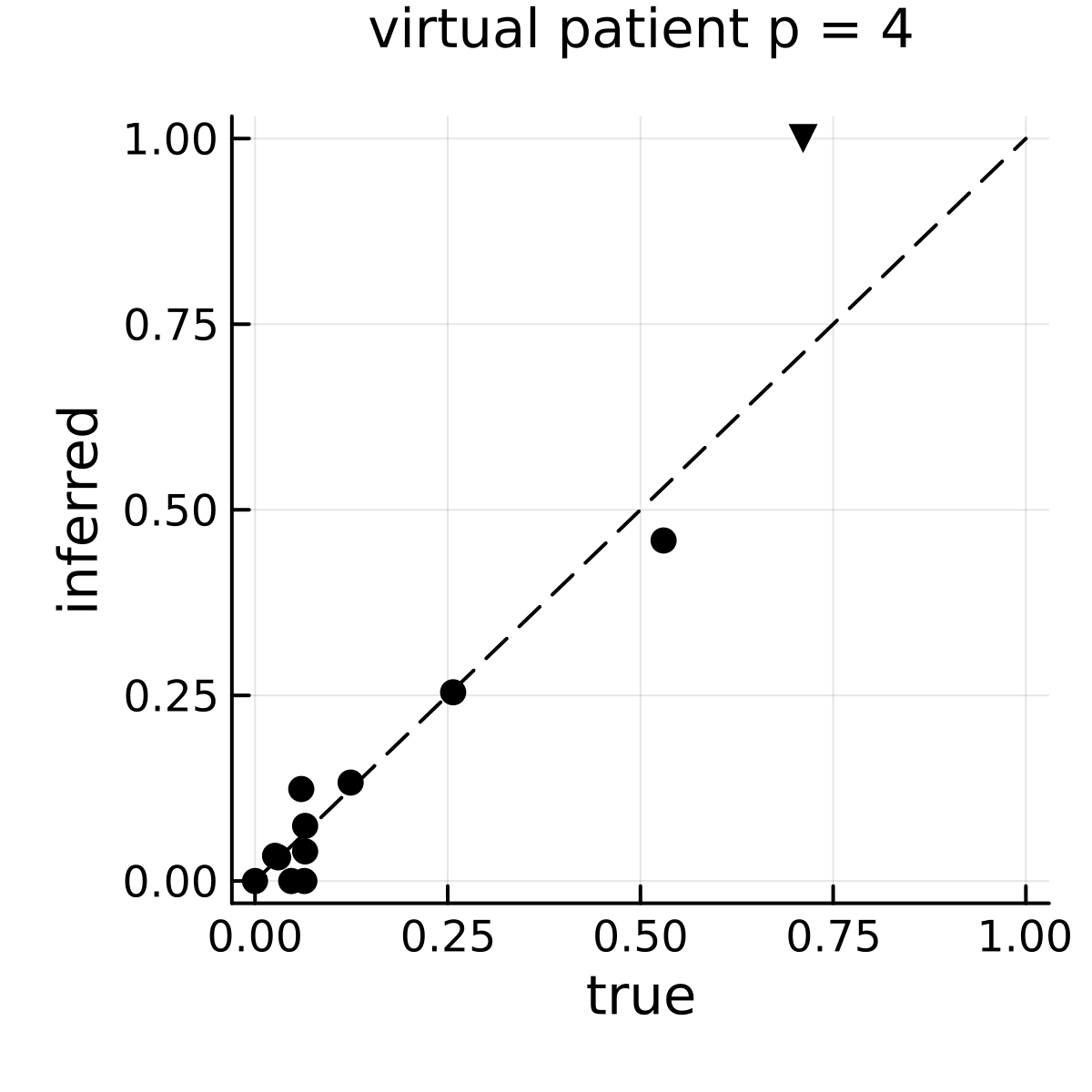}
    \end{subfigure}
    \begin{subfigure}[b]{0.22\textwidth}
        \includegraphics[width=\textwidth]{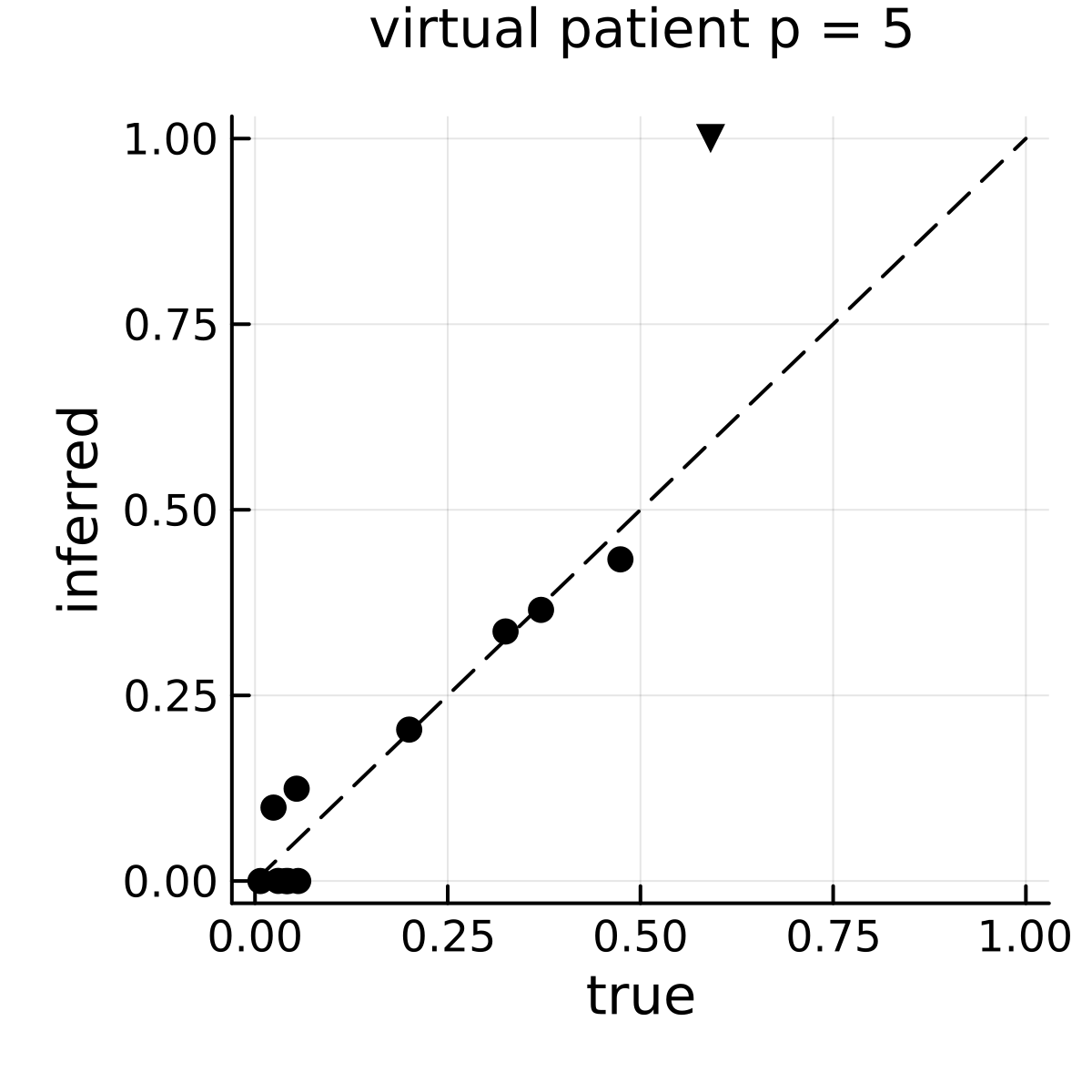}
    \end{subfigure}
    \begin{subfigure}[b]{0.22\textwidth}
        \includegraphics[width=\textwidth]{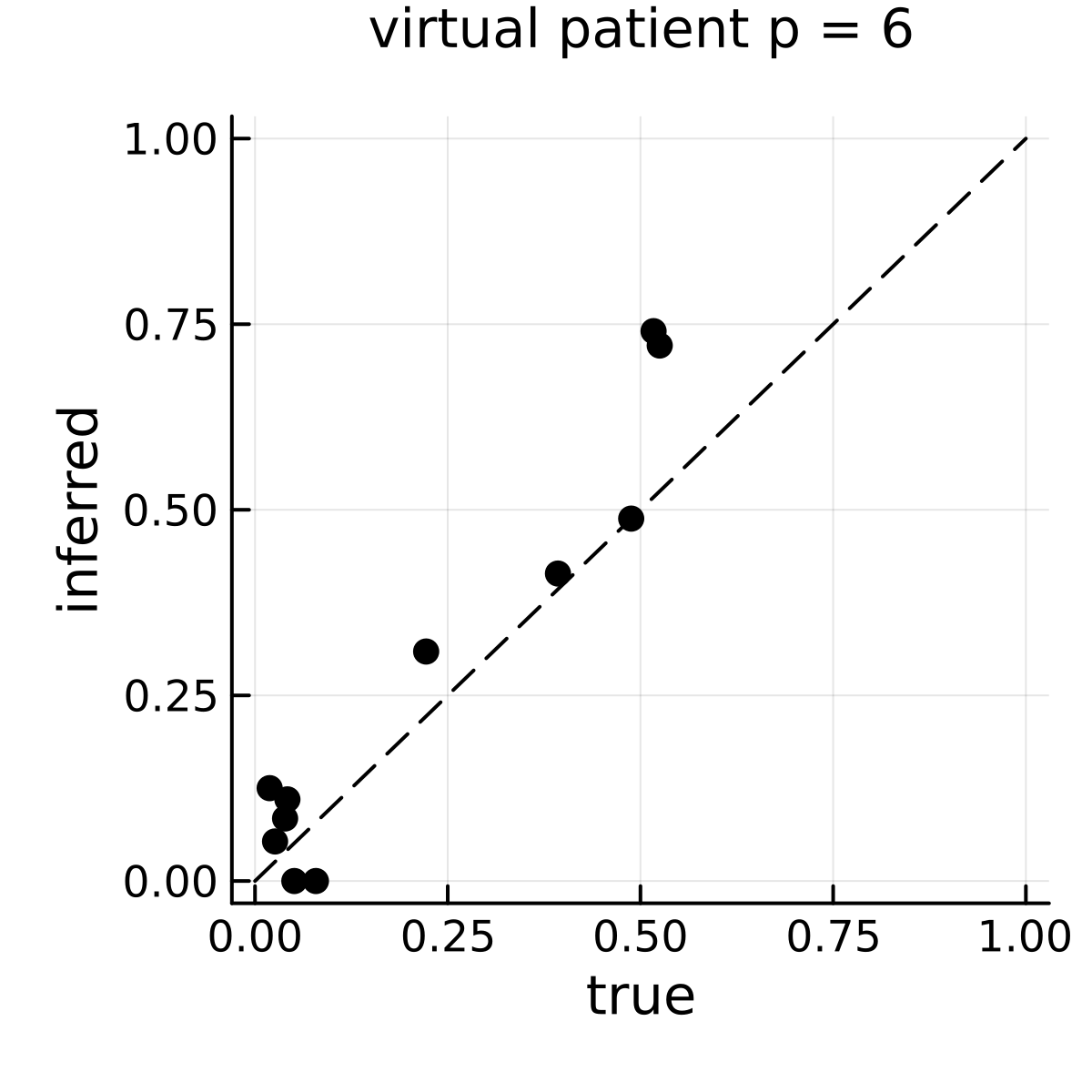}
    \end{subfigure}
    \begin{subfigure}[b]{0.22\textwidth}
        \includegraphics[width=\textwidth]{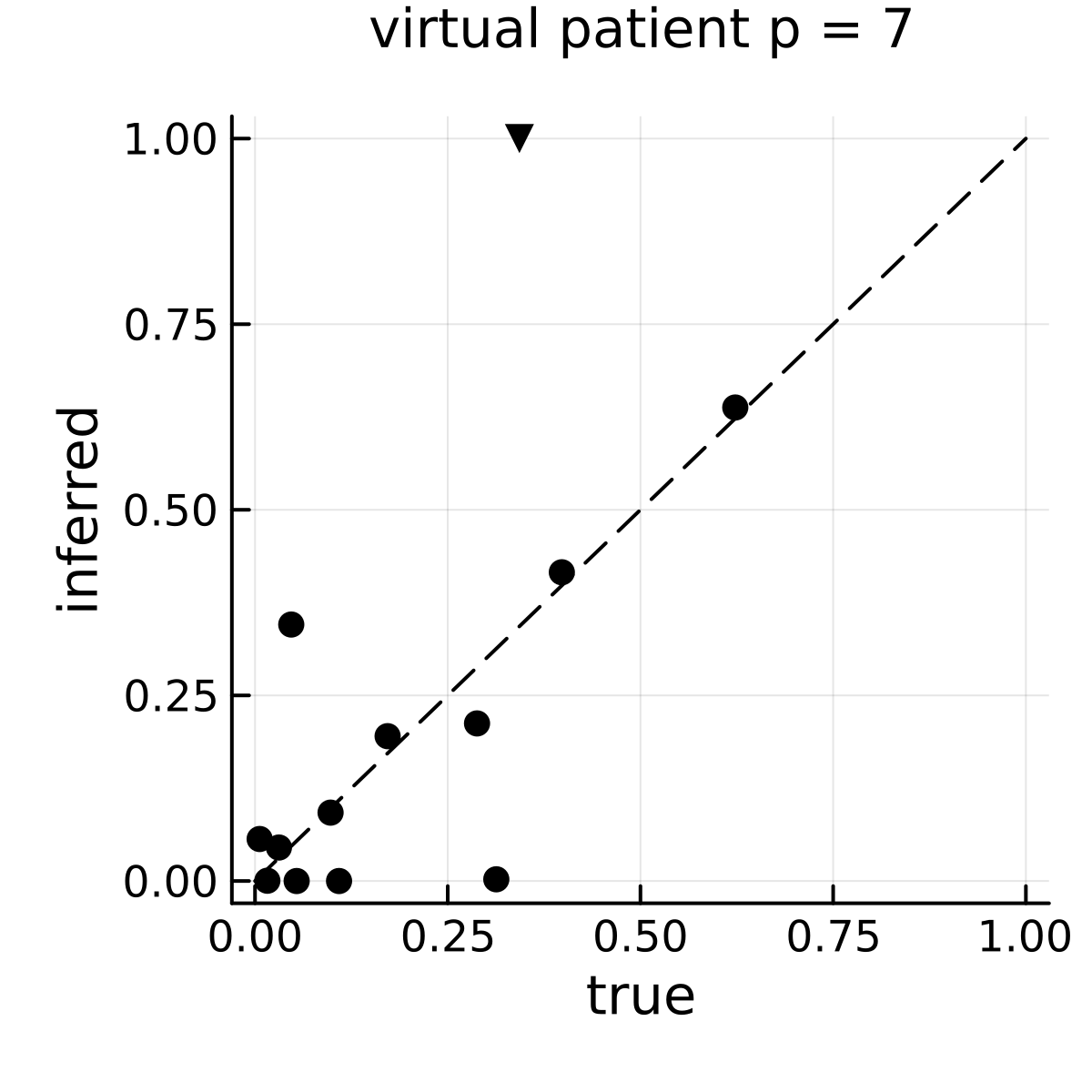}
    \end{subfigure}
    \begin{subfigure}[b]{0.22\textwidth}
        \includegraphics[width=\textwidth]{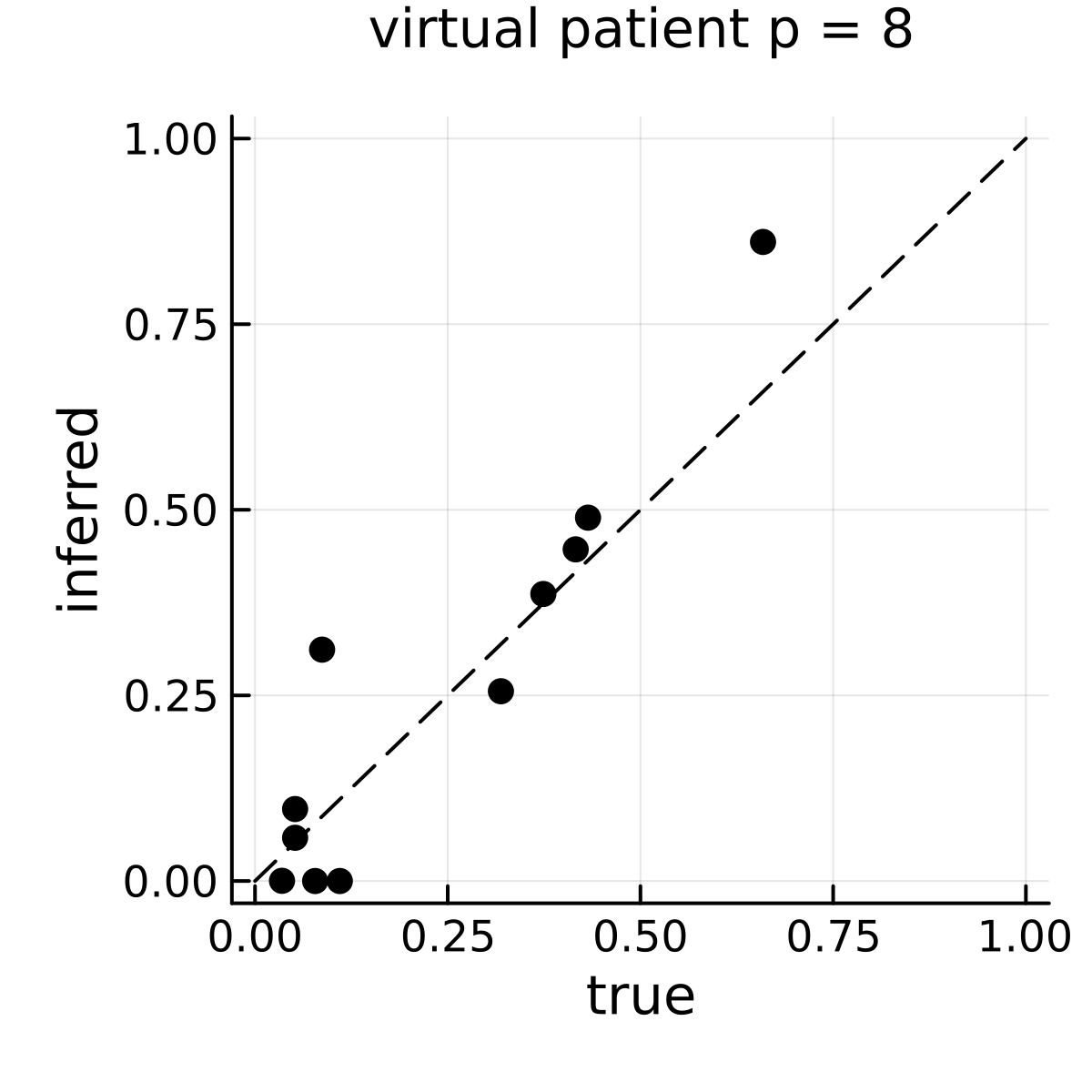}
    \end{subfigure}
    \begin{subfigure}[b]{0.22\textwidth}
        \includegraphics[width=\textwidth]{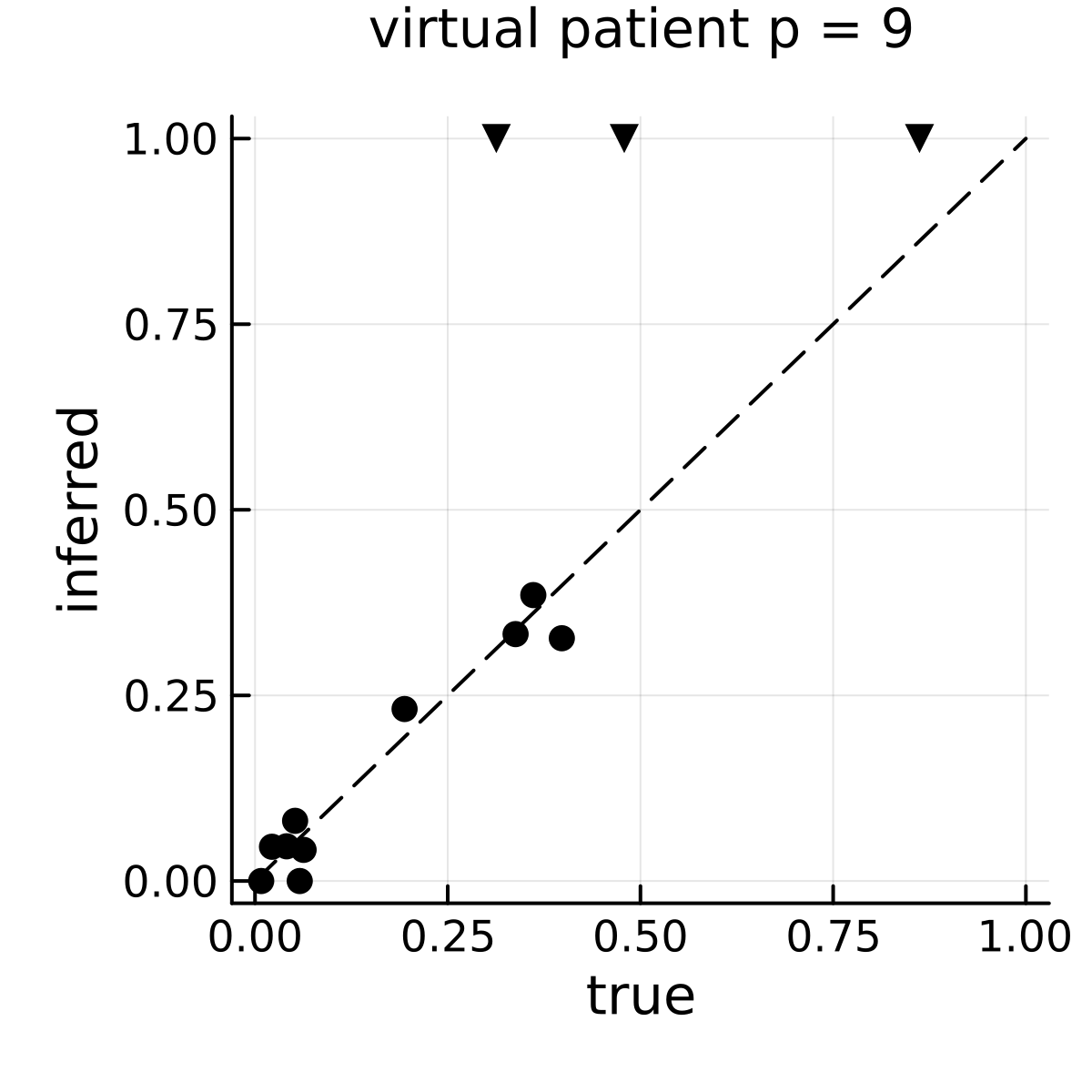}
    \end{subfigure}
    \begin{subfigure}[b]{0.22\textwidth}
        \includegraphics[width=\textwidth]{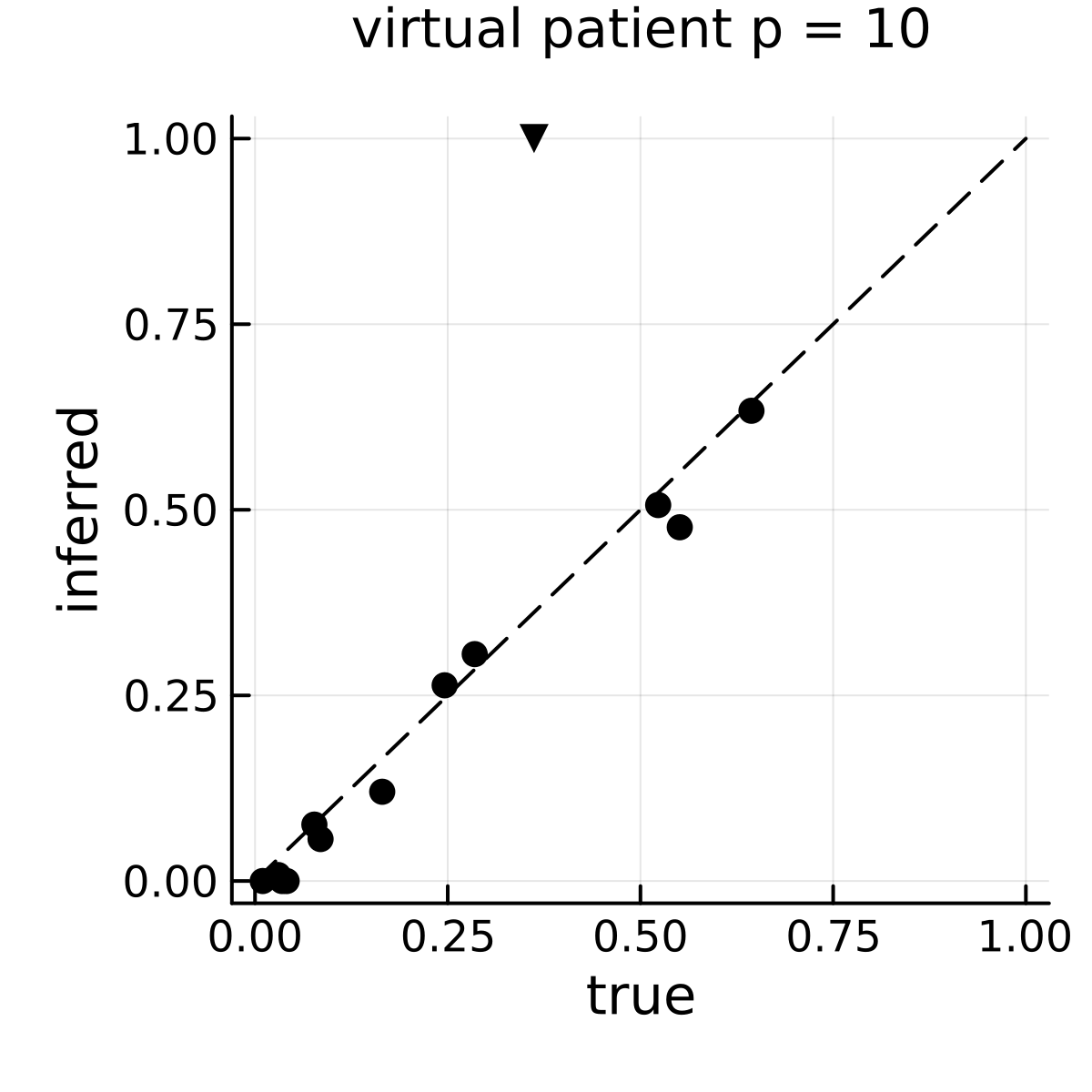}
    \end{subfigure}
    \begin{subfigure}[b]{0.22\textwidth}
        \includegraphics[width=\textwidth]{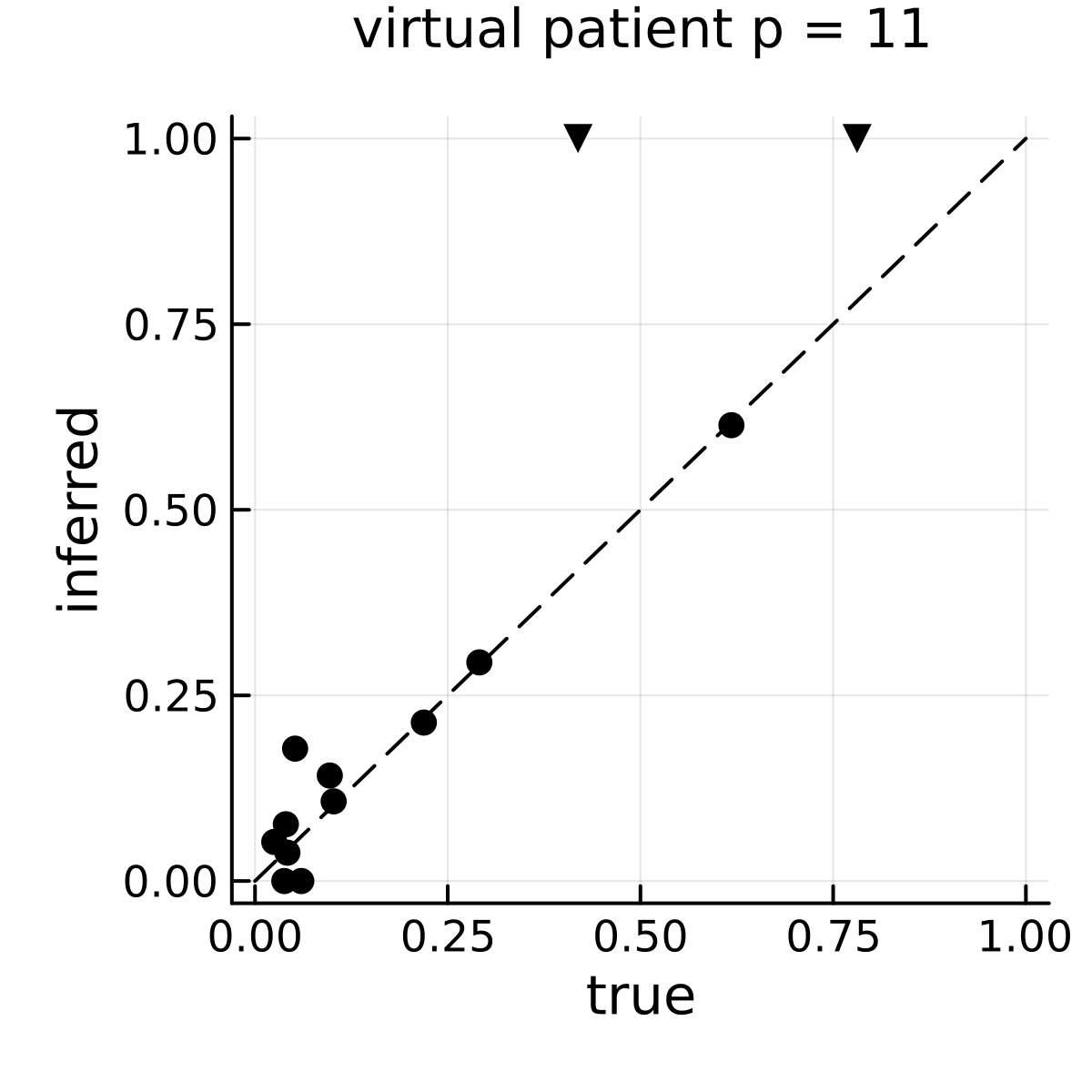}
    \end{subfigure}
    \begin{subfigure}[b]{0.22\textwidth}
        \includegraphics[width=\textwidth]{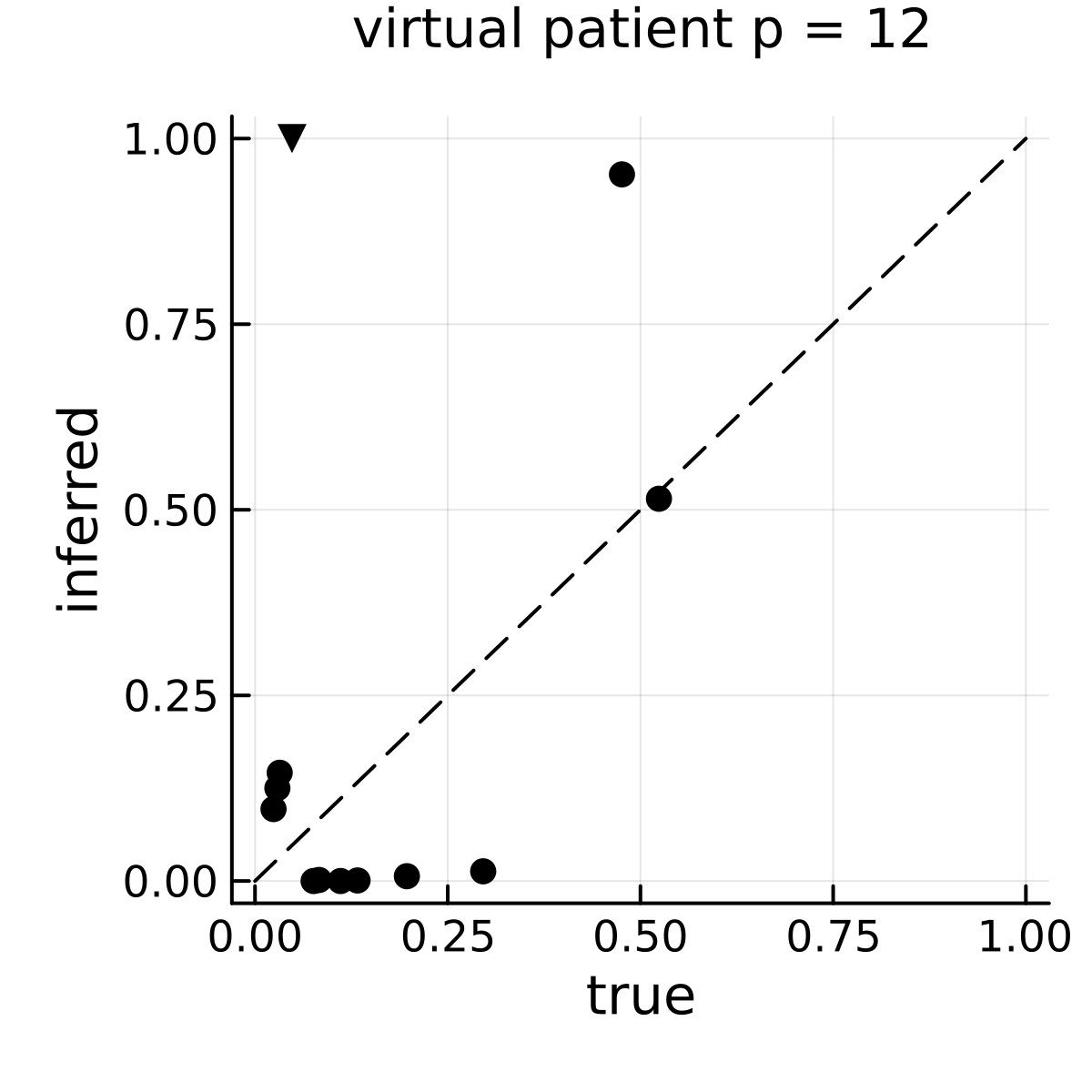}
    \end{subfigure}
    \begin{subfigure}[b]{0.22\textwidth}
        \includegraphics[width=\textwidth]{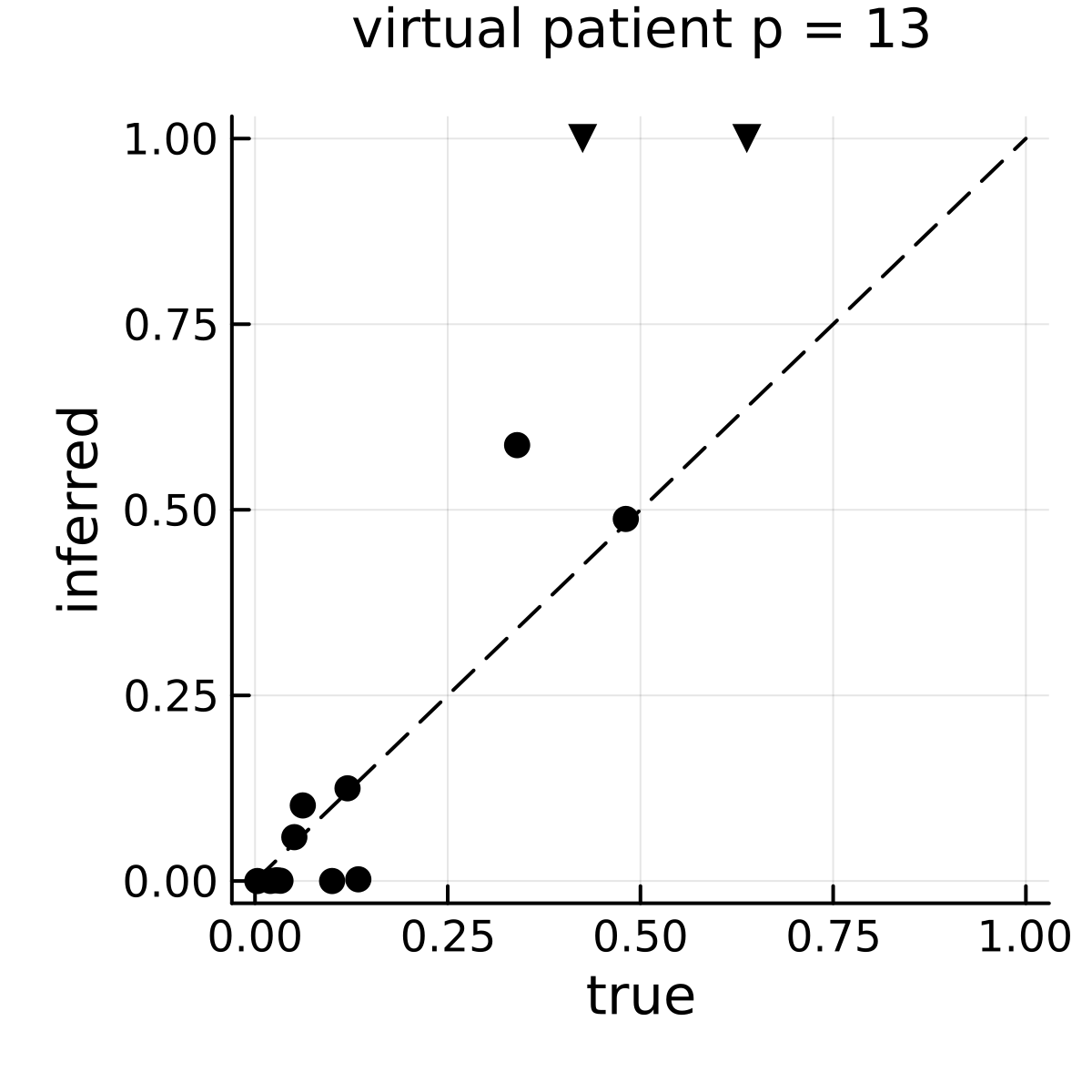}
    \end{subfigure}
    \begin{subfigure}[b]{0.22\textwidth}
        \includegraphics[width=\textwidth]{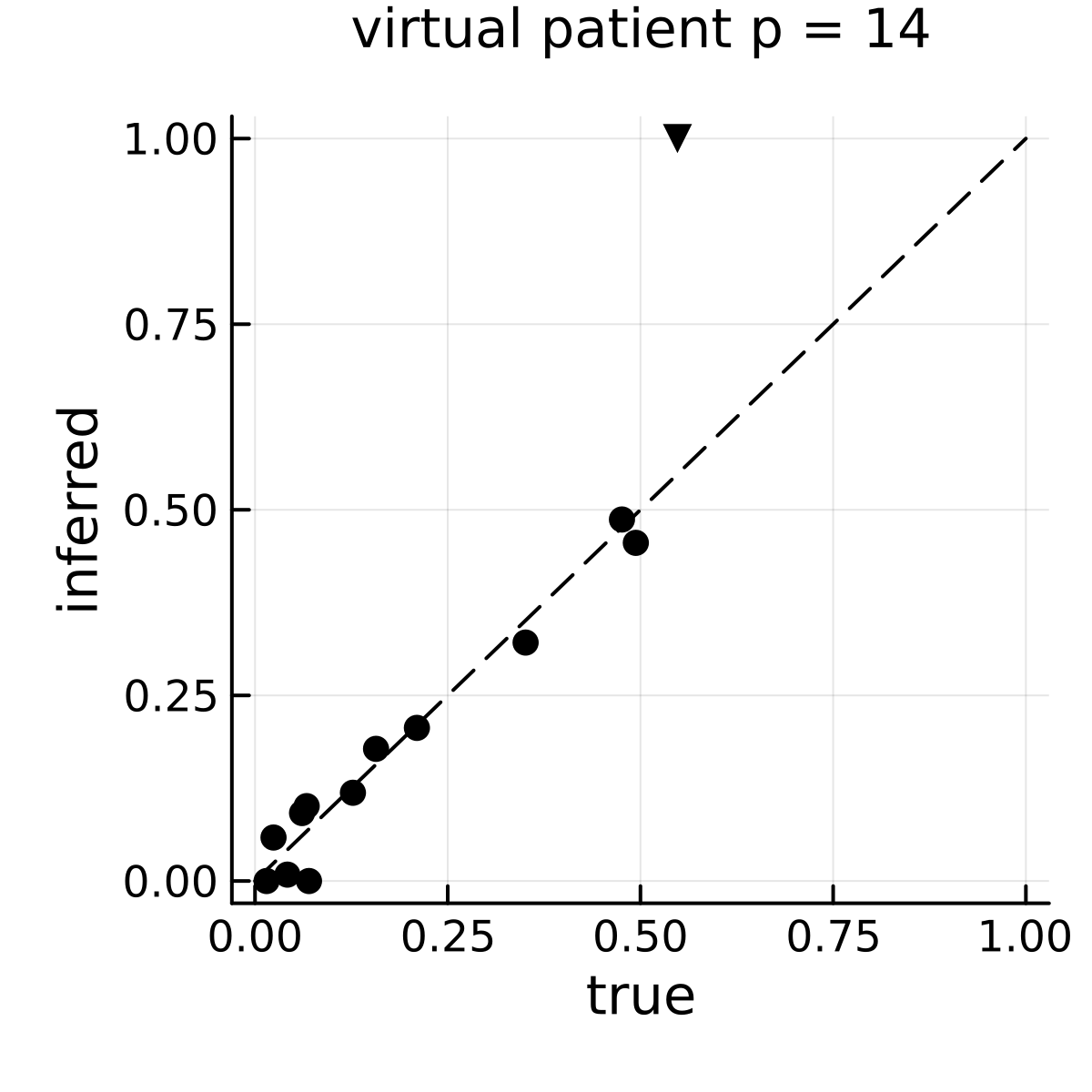}
    \end{subfigure}
    \begin{subfigure}[b]{0.22\textwidth}
        \includegraphics[width=\textwidth]{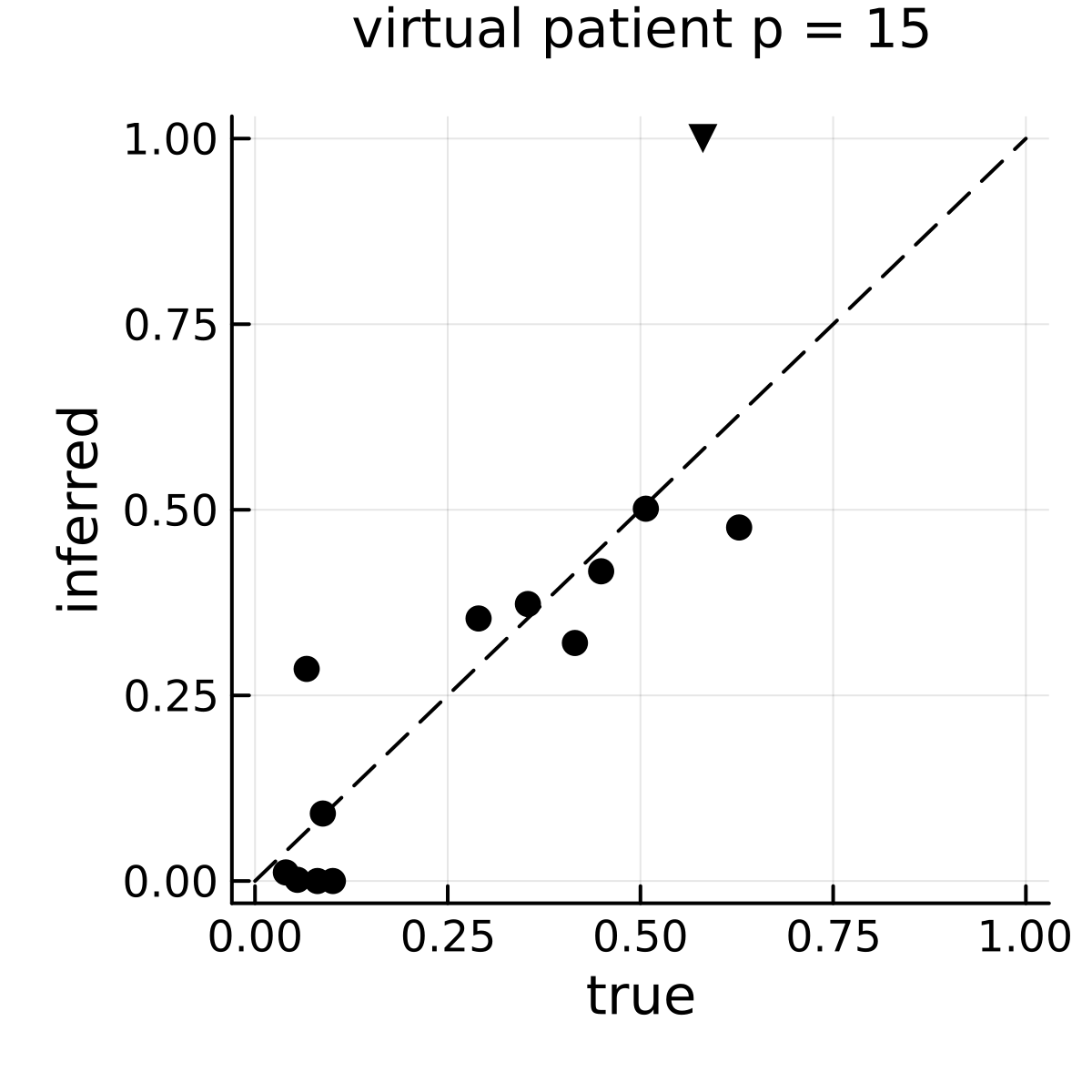}
    \end{subfigure}
    \begin{subfigure}[b]{0.22\textwidth}
        \includegraphics[width=\textwidth]{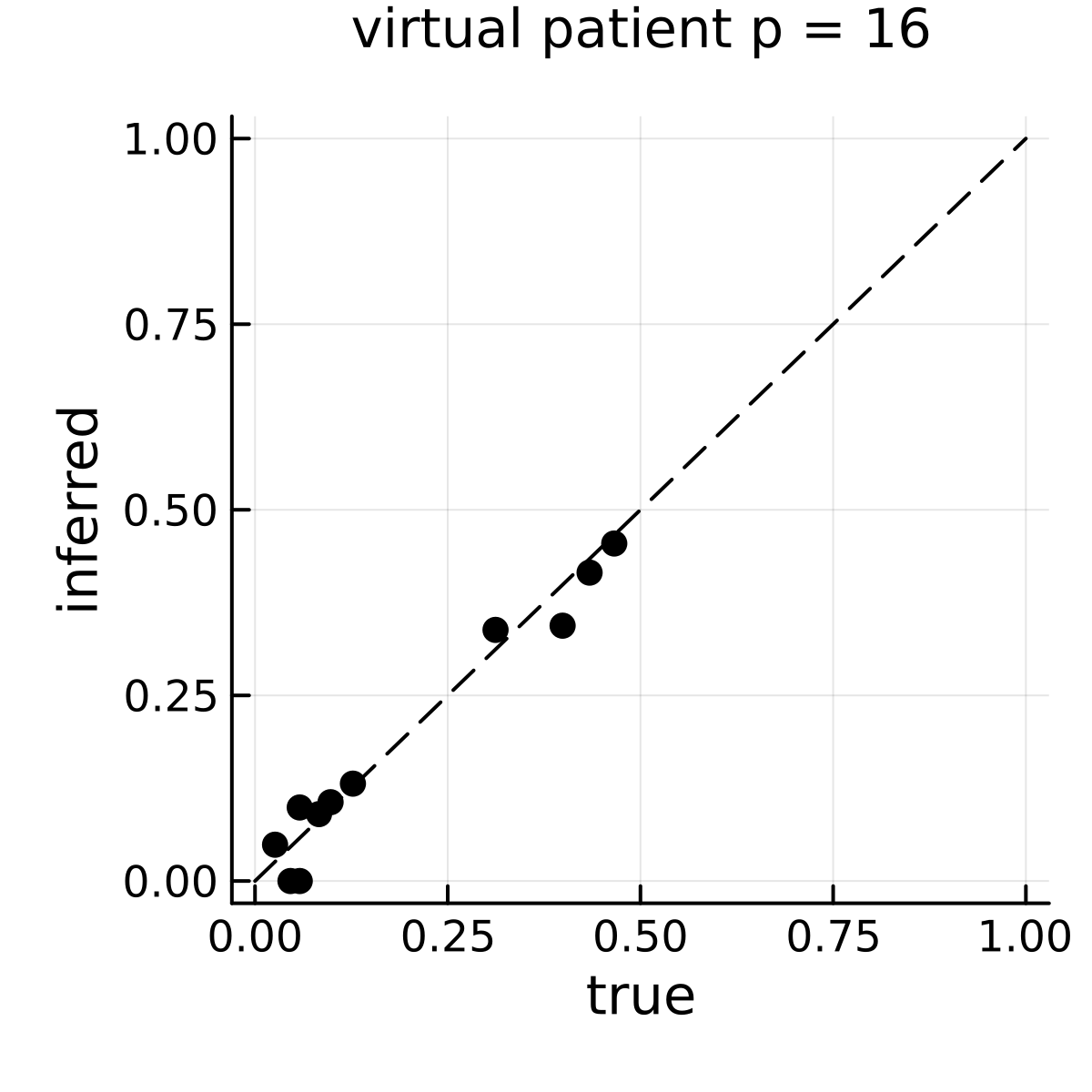}
    \end{subfigure}
    \begin{subfigure}[b]{0.22\textwidth}
        \includegraphics[width=\textwidth]{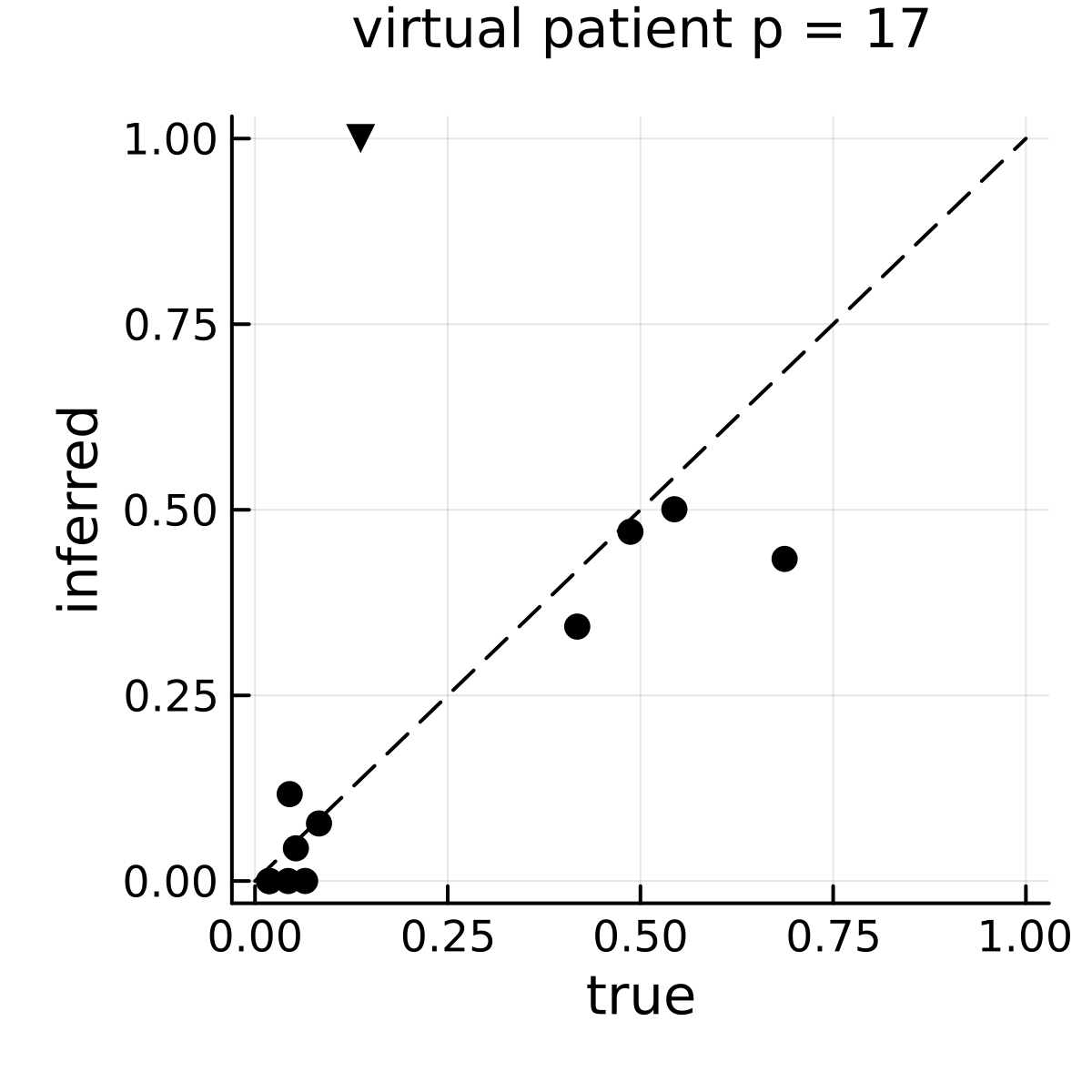}
    \end{subfigure}
    \begin{subfigure}[b]{0.22\textwidth}
        \includegraphics[width=\textwidth]{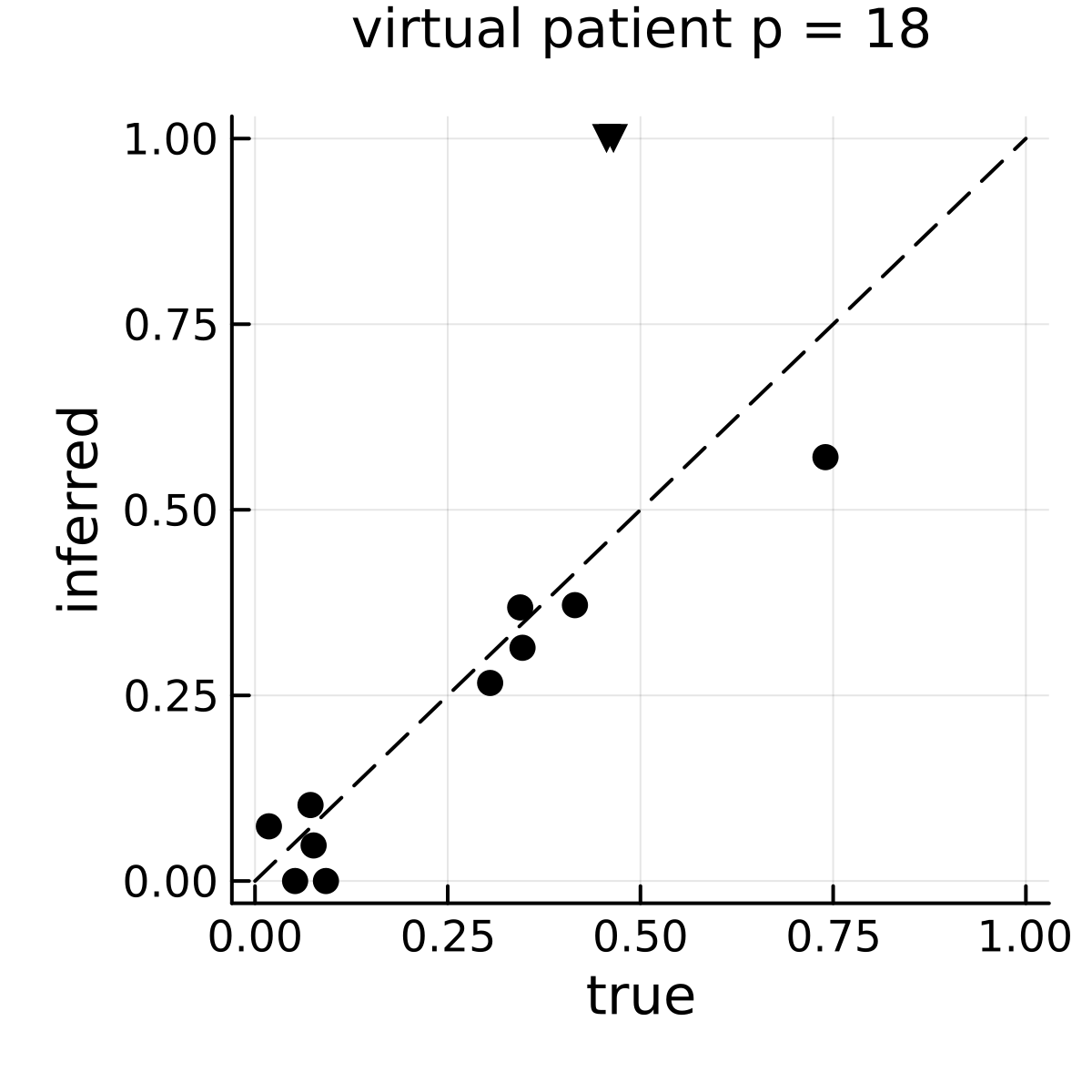}
    \end{subfigure}
    \begin{subfigure}[b]{0.22\textwidth}
        \includegraphics[width=\textwidth]{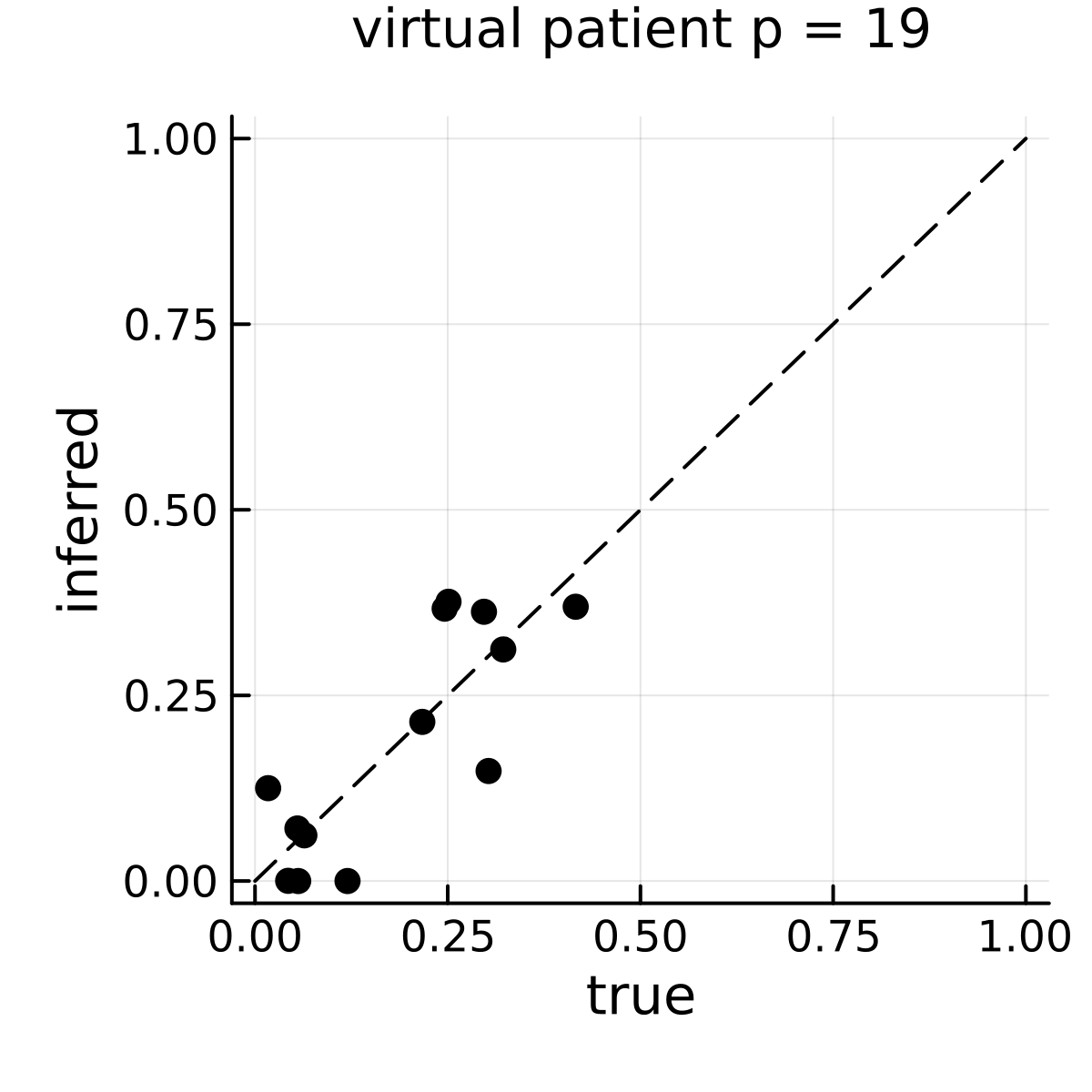}
    \end{subfigure}
    
    \caption{For each virtual patient $i$ (sub-panel), we display its inferred (y-axis, mean a posteriori) vs true minimal dose $d_{min}^{(i,m)}$ over the 13 synthetic cohorts (index $m$) that allow the existence of a minimal dose.
    Individuals for which the real minimal dose is higher than one - within a given synthetic cohort - are not displayed. 
    When a minimal dose is estimated to be higher than one, we set it to one (and represent it with a triangle) to generate the figure.}
    \label{fig:min_dose_p}
\end{figure}

In Fig.~\ref{fig:min_dose_inferred_vs_true}, we display for each of the 13 virtual cohorts that allow for the existence of a minimal dose the inferred (mean \textit{a posteriori}) vs true individual minimal doses.
In Fig.~\ref{fig:min_dose_p}, we display the same results from the $i^{th}$ virtual patient point of view. 
Over the 247 virtual patients who were presented in Tab.~\ref{tab:synthetic_minimal_ind_dose}, we find a median error (L1-norm) between the estimated and true minimal dose equal to 0.046 (i.e., equal to 8.4~$\mu$g/week).
However, the results are unequal, according to the considered virtual cohort, with some cases deserving more attention:
\begin{itemize}
    \item Concerning the virtual cohort $m=45$, the true dose-response relations were constant for $\bar{\Delta}_{hom}^*$ and affine sigmoid for $\bar{\Delta}_{het}^*$ when the selected relations were respectively constant and sigmoid (model 36). That is, there was for each individual from cohort $m=45$ a theoretic minimal dose that could not be retrieved through our selected model, where we infer a remission no matter the dose ($d_{min}=0$). However, the true minimal doses were all relatively low ($\leq 0.12$), so we did not make a significant error concerning the minimal dose. This observation also justifies why a model without a minimal dose (given that such models are more parsimonious than those with a minimal dose) was not selected.
\item We can make the same observations concerning the cohort $m=180$, for which a model without a minimal dose was selected (model 171).
\item Concerning the cohort $m=113$ (with affine dose-response relations for both $\bar{\Delta}_{hom}^*$ and $\bar{\Delta}_{het}^*$), where model 185 was selected instead (an affine sigmoid relation for $\bar{\Delta}_{hom}^*$ and a constant one for $\bar{\Delta}_{het}^*$), we estimate that remission would not be reachable for many patients for whom we infer a minimal dose higher than one. Actually, for them, we infer that $\Delta^*_{het}>0$. As shown in Fig.~\ref{fig:synth_Delta_het}, the estimations were  pretty good when considering $\Delta^{* (i,m)}_{het}(d_{450}^{(i)})$ but poorly extrapolated for the whole range of doses. \\
\end{itemize}
To sum up, we get overall accurate estimations of the minimal dose, even if the selected model is not the true one. The most problematic situations are those when a selected model involves a constant dose-response relationship, either associated with the heterozygous or homozygous malignant clone. 
We can also see that a model without a minimal dose might only be selected when the true minimal doses - over the virtual patients from the cohort - are low (or, of course, if the true model itself does not allow for the existence of a minimal dose). Conversely, when a true model does not allow for a minimal dose, neither will the selected model (except for cohort $m = 163$, where the estimated individual minimal doses were low).

\FloatBarrier
\newpage

\section{Details about the considered models}
\label{sec:list_models}

In this study, 225 models are considered. They are listed in Tab.~\ref{tab:list_models_1},~\ref{tab:list_models_2},~\ref{tab:list_models_3},~\ref{tab:list_models_4}, and~\ref{tab:list_models_5}. The differences between these 225 models stand in the way we combine the dose-response relationships for $\bar{\Delta}_{hom}^*$ (5 possibilities), $\bar{\Delta}_{het}^*$ (5 possibilities), $\bar{\gamma}_{hom}^*$ (3 possibilities), and $\bar{\gamma}_{het}^*$ (3 possibilities), resulting in $5\times 5\times 3 \times 3 = 225$ potential models. \\
Therefore, by construction, all these models are not equally distinct (or similar). The most dose-response relationships two models have in common, the closest they might be. To generalize this idea of similarity between two models, we could define a semimetric between two models as the number of dose-response relationships they do not have in common. \\

All our models are identified by an index $m$. Actually, as we presented it, our models are defined by the way we combine the dose-response relationships, that is, by a quadruplet. Let $(r_1, r_2, r_3, r_4)$ be in $\{0, 1, 2, 3, 4\}^2 \times \{0, 1, 2\}^2$ with:
\begin{itemize}
    \item $r_1$ equal to:
\begin{itemize}
\item 0 if the relation for $\bar{\Delta}^*_{hom}$ is constant, 
\item 1 if the relation for $\bar{\Delta}^*_{hom}$ is linear, 
\item 2 if the relation for $\bar{\Delta}^*_{hom}$ is affine, 
\item 3 if the relation for $\bar{\Delta}^*_{hom}$ is sigmoid, 
\item 4 if the relation for $\bar{\Delta}^*_{hom}$ is affine sigmoid;
\end{itemize}
 \item $r_2$ equal to:
\begin{itemize}
\item 0 if the relation for $\bar{\Delta}^*_{het}$ is constant, 
\item 1 if the relation for $\bar{\Delta}^*_{het}$ is linear, 
\item 2 if the relation for $\bar{\Delta}^*_{het}$ is affine, 
\item 3 if the relation for $\bar{\Delta}^*_{het}$ is sigmoid, 
\item 4 if the relation for $\bar{\Delta}^*_{het}$ is affine sigmoid;
\end{itemize}
 \item $r_3$ equal to:
\begin{itemize}
\item 0 if the relation for $\bar{\gamma}^*_{hom}$ is constant, 
\item 1 if the relation for $\bar{\gamma}^*_{hom}$ is inverse, 
\item 2 if the relation for $\bar{\gamma}^*_{hom}$ is affine, 
\end{itemize}
 \item $r_4$ equal to:
\begin{itemize}
\item 0 if the relation for $\bar{\gamma}^*_{het}$ is constant, 
\item 1 if the relation for $\bar{\gamma}^*_{het}$ is inverse, 
\item 2 if the relation for $\bar{\gamma}^*_{het}$ is affine, 
\end{itemize}
\end{itemize}
Then the model index $m$ is defined by:
$$ m =1+ r_4 + 3\times r_3 + 3^2 \times  r_2 + 5\times 3^2 \times r_1 $$

According to how $\bar{\Delta}_{het}^*$ (and $\bar{\Delta}_{hom}^*$) evolves with the dose $d$, that is, according to its dose-response relationship, there might exist a minimal dose $d_{min,het}$ above which the heterozygous (respectively, homozygous) malignant clone will be exhausted on the long-term, that is, $\bar{\Delta}_{het}(d_{min,het}) = 0$. It is the case only if the dose-response relation is affine or affine sigmoid. \\
In the first case, the minimal dose would be computed by (see eq.~(3) in the main text for the signification of the parameters):
$$d_{min,het} = - \frac{\Delta_{het}}{\Delta_{het}^*}$$
In the last case (affine sigmoid), the minimal dose would be computed by (see eq.~(5) in the main text):
$$d_{min,het} = \frac{1}{\Delta_{het}^*}\log(1+2 \Delta_{het})$$
Note that, if the dose-response relationship is linear or sigmoid, necessarily, the malignant clone will ultimately exhaust. 
If the dose-response relationship is contant, according to the sign of the parameter $\Delta_{het}^*$, either the clone will ultimately exhaust no matter the dose, either it will persist no matter the dose. If the cell population dynamics are simulated from such a model, then there might be some patients for whom a remission might be possible, and anothers for whom it would not be the case. \\
The above statements also apply to the homozygous clone. \\
Finally, we define a (global) minimal dose as the maximum between the heterozygous and homozygous minimal doses.

\begin{table}[]
    \centering

    \begin{tabular}{|c|c|c|c|c|c|c|c|c|}
\hline
Model &\multirow{2}{*}{$\bar{\Delta}_{hom}^*$} &\multirow{2}{*}{$\bar{\Delta}_{het}^*$} &\multirow{2}{*}{$\bar{\gamma}_{hom}^*$} &\multirow{2}{*}{$\bar{\gamma}_{het}^*$} & \multicolumn{3}{c|}{Existence of a minimal dose} & \multirow{2}{*}{k}\\
\cline{6-8}
$m$	 & 		 & 		 & 	 & 		& hom & het & global & \\ \hline
1	&	constant	&	constant	&	constant	&	constant	&	no	&	no	&	no	&	7	\\ \hline
2	&	constant	&	constant	&	constant	&	inverse	&	no	&	no	&	no	&	7	\\ \hline
3	&	constant	&	constant	&	constant	&	affine	&	no	&	no	&	no	&	7	\\ \hline
4	&	constant	&	constant	&	inverse	&	constant	&	no	&	no	&	no	&	7	\\ \hline
5	&	constant	&	constant	&	inverse	&	inverse	&	no	&	no	&	no	&	7	\\ \hline
6	&	constant	&	constant	&	inverse	&	affine	&	no	&	no	&	no	&	7	\\ \hline
7	&	constant	&	constant	&	affine	&	constant	&	no	&	no	&	no	&	7	\\ \hline
8	&	constant	&	constant	&	affine	&	inverse	&	no	&	no	&	no	&	7	\\ \hline
9	&	constant	&	constant	&	affine	&	affine	&	no	&	no	&	no	&	7	\\ \hline
10	&	constant	&	linear	&	constant	&	constant	&	no	&	no	&	no	&	7	\\ \hline
11	&	constant	&	linear	&	constant	&	inverse	&	no	&	no	&	no	&	7	\\ \hline
12	&	constant	&	linear	&	constant	&	affine	&	no	&	no	&	no	&	7	\\ \hline
13	&	constant	&	linear	&	inverse	&	constant	&	no	&	no	&	no	&	7	\\ \hline
14	&	constant	&	linear	&	inverse	&	inverse	&	no	&	no	&	no	&	7	\\ \hline
15	&	constant	&	linear	&	inverse	&	affine	&	no	&	no	&	no	&	7	\\ \hline
16	&	constant	&	linear	&	affine	&	constant	&	no	&	no	&	no	&	7	\\ \hline
17	&	constant	&	linear	&	affine	&	inverse	&	no	&	no	&	no	&	7	\\ \hline
18	&	constant	&	linear	&	affine	&	affine	&	no	&	no	&	no	&	7	\\ \hline
19	&	constant	&	affine	&	constant	&	constant	&	no	&	yes	&	yes	&	8	\\ \hline
20	&	constant	&	affine	&	constant	&	inverse	&	no	&	yes	&	yes	&	8	\\ \hline
21	&	constant	&	affine	&	constant	&	affine	&	no	&	yes	&	yes	&	8	\\ \hline
22	&	constant	&	affine	&	inverse	&	constant	&	no	&	yes	&	yes	&	8	\\ \hline
23	&	constant	&	affine	&	inverse	&	inverse	&	no	&	yes	&	yes	&	8	\\ \hline
24	&	constant	&	affine	&	inverse	&	affine	&	no	&	yes	&	yes	&	8	\\ \hline
25	&	constant	&	affine	&	affine	&	constant	&	no	&	yes	&	yes	&	8	\\ \hline
26	&	constant	&	affine	&	affine	&	inverse	&	no	&	yes	&	yes	&	8	\\ \hline
27	&	constant	&	affine	&	affine	&	affine	&	no	&	yes	&	yes	&	8	\\ \hline
28	&	constant	&	sigmoid	&	constant	&	constant	&	no	&	no	&	no	&	7	\\ \hline
29	&	constant	&	sigmoid	&	constant	&	inverse	&	no	&	no	&	no	&	7	\\ \hline
30	&	constant	&	sigmoid	&	constant	&	affine	&	no	&	no	&	no	&	7	\\ \hline
31	&	constant	&	sigmoid	&	inverse	&	constant	&	no	&	no	&	no	&	7	\\ \hline
32	&	constant	&	sigmoid	&	inverse	&	inverse	&	no	&	no	&	no	&	7	\\ \hline
33	&	constant	&	sigmoid	&	inverse	&	affine	&	no	&	no	&	no	&	7	\\ \hline
34	&	constant	&	sigmoid	&	affine	&	constant	&	no	&	no	&	no	&	7	\\ \hline
35	&	constant	&	sigmoid	&	affine	&	inverse	&	no	&	no	&	no	&	7	\\ \hline
36	&	constant	&	sigmoid	&	affine	&	affine	&	no	&	no	&	no	&	7	\\ \hline
37	&	constant	&	affine sigmoid	&	constant	&	constant	&	no	&	yes	&	yes	&	8	\\ \hline
38	&	constant	&	affine sigmoid	&	constant	&	inverse	&	no	&	yes	&	yes	&	8	\\ \hline
39	&	constant	&	affine sigmoid	&	constant	&	affine	&	no	&	yes	&	yes	&	8	\\ \hline
40	&	constant	&	affine sigmoid	&	inverse	&	constant	&	no	&	yes	&	yes	&	8	\\ \hline
41	&	constant	&	affine sigmoid	&	inverse	&	inverse	&	no	&	yes	&	yes	&	8	\\ \hline
42	&	constant	&	affine sigmoid	&	inverse	&	affine	&	no	&	yes	&	yes	&	8	\\ \hline
43	&	constant	&	affine sigmoid	&	affine	&	constant	&	no	&	yes	&	yes	&	8	\\ \hline
44	&	constant	&	affine sigmoid	&	affine	&	inverse	&	no	&	yes	&	yes	&	8	\\ \hline
45	&	constant	&	affine sigmoid	&	affine	&	affine	&	no	&	yes	&	yes	&	8	\\ \hline
46	&	linear	&	constant	&	constant	&	constant	&	no	&	no	&	no	&	7	\\ \hline
47	&	linear	&	constant	&	constant	&	inverse	&	no	&	no	&	no	&	7	\\ \hline
48	&	linear	&	constant	&	constant	&	affine	&	no	&	no	&	no	&	7	\\ \hline
49	&	linear	&	constant	&	inverse	&	constant	&	no	&	no	&	no	&	7	\\ \hline
50	&	linear	&	constant	&	inverse	&	inverse	&	no	&	no	&	no	&	7	\\ \hline
\end{tabular}

 \caption{List of the models (1-50 over 225) considered in the study. For each model $m$, we indicate its dose-response relationships for $\bar{\Delta}_{hom}^*$, $\bar{\Delta}_{het}^*$, $\bar{\gamma}_{hom}^*$, and $\bar{\gamma}_{het}^*$. We indicate whether there exists a minimal dose associated to the heterozygous clone (het), the homozygous one (hom), or both (global). k indicates the number of parameters to estimate for a patient whose dynamics would be described by model $m$.  }
    \label{tab:list_models_1}
\end{table}

\begin{table}[]
    \centering

    \begin{tabular}{|c|c|c|c|c|c|c|c|c|}
\hline
Model &\multirow{2}{*}{$\bar{\Delta}_{hom}^*$} &\multirow{2}{*}{$\bar{\Delta}_{het}^*$} &\multirow{2}{*}{$\bar{\gamma}_{hom}^*$} &\multirow{2}{*}{$\bar{\gamma}_{het}^*$} & \multicolumn{3}{c|}{Existence of a minimal dose} & \multirow{2}{*}{k}\\
\cline{6-8}
$m$	 & 		 & 		 & 	 & 		& hom & het & global & \\ \hline
51	&	linear	&	constant	&	inverse	&	affine	&	no	&	no	&	no	&	7	\\ \hline
52	&	linear	&	constant	&	affine	&	constant	&	no	&	no	&	no	&	7	\\ \hline
53	&	linear	&	constant	&	affine	&	inverse	&	no	&	no	&	no	&	7	\\ \hline
54	&	linear	&	constant	&	affine	&	affine	&	no	&	no	&	no	&	7	\\ \hline
55	&	linear	&	linear	&	constant	&	constant	&	no	&	no	&	no	&	7	\\ \hline
56	&	linear	&	linear	&	constant	&	inverse	&	no	&	no	&	no	&	7	\\ \hline
57	&	linear	&	linear	&	constant	&	affine	&	no	&	no	&	no	&	7	\\ \hline
58	&	linear	&	linear	&	inverse	&	constant	&	no	&	no	&	no	&	7	\\ \hline
59	&	linear	&	linear	&	inverse	&	inverse	&	no	&	no	&	no	&	7	\\ \hline
60	&	linear	&	linear	&	inverse	&	affine	&	no	&	no	&	no	&	7	\\ \hline
61	&	linear	&	linear	&	affine	&	constant	&	no	&	no	&	no	&	7	\\ \hline
62	&	linear	&	linear	&	affine	&	inverse	&	no	&	no	&	no	&	7	\\ \hline
63	&	linear	&	linear	&	affine	&	affine	&	no	&	no	&	no	&	7	\\ \hline
64	&	linear	&	affine	&	constant	&	constant	&	no	&	yes	&	yes	&	8	\\ \hline
65	&	linear	&	affine	&	constant	&	inverse	&	no	&	yes	&	yes	&	8	\\ \hline
66	&	linear	&	affine	&	constant	&	affine	&	no	&	yes	&	yes	&	8	\\ \hline
67	&	linear	&	affine	&	inverse	&	constant	&	no	&	yes	&	yes	&	8	\\ \hline
68	&	linear	&	affine	&	inverse	&	inverse	&	no	&	yes	&	yes	&	8	\\ \hline
69	&	linear	&	affine	&	inverse	&	affine	&	no	&	yes	&	yes	&	8	\\ \hline
70	&	linear	&	affine	&	affine	&	constant	&	no	&	yes	&	yes	&	8	\\ \hline
71	&	linear	&	affine	&	affine	&	inverse	&	no	&	yes	&	yes	&	8	\\ \hline
72	&	linear	&	affine	&	affine	&	affine	&	no	&	yes	&	yes	&	8	\\ \hline
73	&	linear	&	sigmoid	&	constant	&	constant	&	no	&	no	&	no	&	7	\\ \hline
74	&	linear	&	sigmoid	&	constant	&	inverse	&	no	&	no	&	no	&	7	\\ \hline
75	&	linear	&	sigmoid	&	constant	&	affine	&	no	&	no	&	no	&	7	\\ \hline
76	&	linear	&	sigmoid	&	inverse	&	constant	&	no	&	no	&	no	&	7	\\ \hline
77	&	linear	&	sigmoid	&	inverse	&	inverse	&	no	&	no	&	no	&	7	\\ \hline
78	&	linear	&	sigmoid	&	inverse	&	affine	&	no	&	no	&	no	&	7	\\ \hline
79	&	linear	&	sigmoid	&	affine	&	constant	&	no	&	no	&	no	&	7	\\ \hline
80	&	linear	&	sigmoid	&	affine	&	inverse	&	no	&	no	&	no	&	7	\\ \hline
81	&	linear	&	sigmoid	&	affine	&	affine	&	no	&	no	&	no	&	7	\\ \hline
82	&	linear	&	affine sigmoid	&	constant	&	constant	&	no	&	yes	&	yes	&	8	\\ \hline
83	&	linear	&	affine sigmoid	&	constant	&	inverse	&	no	&	yes	&	yes	&	8	\\ \hline
84	&	linear	&	affine sigmoid	&	constant	&	affine	&	no	&	yes	&	yes	&	8	\\ \hline
85	&	linear	&	affine sigmoid	&	inverse	&	constant	&	no	&	yes	&	yes	&	8	\\ \hline
86	&	linear	&	affine sigmoid	&	inverse	&	inverse	&	no	&	yes	&	yes	&	8	\\ \hline
87	&	linear	&	affine sigmoid	&	inverse	&	affine	&	no	&	yes	&	yes	&	8	\\ \hline
88	&	linear	&	affine sigmoid	&	affine	&	constant	&	no	&	yes	&	yes	&	8	\\ \hline
89	&	linear	&	affine sigmoid	&	affine	&	inverse	&	no	&	yes	&	yes	&	8	\\ \hline
90	&	linear	&	affine sigmoid	&	affine	&	affine	&	no	&	yes	&	yes	&	8	\\ \hline
91	&	affine	&	constant	&	constant	&	constant	&	yes	&	no	&	yes	&	8	\\ \hline
92	&	affine	&	constant	&	constant	&	inverse	&	yes	&	no	&	yes	&	8	\\ \hline
93	&	affine	&	constant	&	constant	&	affine	&	yes	&	no	&	yes	&	8	\\ \hline
94	&	affine	&	constant	&	inverse	&	constant	&	yes	&	no	&	yes	&	8	\\ \hline
95	&	affine	&	constant	&	inverse	&	inverse	&	yes	&	no	&	yes	&	8	\\ \hline
96	&	affine	&	constant	&	inverse	&	affine	&	yes	&	no	&	yes	&	8	\\ \hline
97	&	affine	&	constant	&	affine	&	constant	&	yes	&	no	&	yes	&	8	\\ \hline
98	&	affine	&	constant	&	affine	&	inverse	&	yes	&	no	&	yes	&	8	\\ \hline
99	&	affine	&	constant	&	affine	&	affine	&	yes	&	no	&	yes	&	8	\\ \hline
100	&	affine	&	linear	&	constant	&	constant	&	yes	&	no	&	yes	&	8	\\ \hline
\end{tabular}

 \caption{List of the models (51-100 over 225) considered in the study}
    \label{tab:list_models_2}

\end{table}

\begin{table}[]
    \centering
\begin{tabular}{|c|c|c|c|c|c|c|c|c|}
\hline
Model &\multirow{2}{*}{$\bar{\Delta}_{hom}^*$} &\multirow{2}{*}{$\bar{\Delta}_{het}^*$} &\multirow{2}{*}{$\bar{\gamma}_{hom}^*$} &\multirow{2}{*}{$\bar{\gamma}_{het}^*$} & \multicolumn{3}{c|}{Existence of a minimal dose} & \multirow{2}{*}{k}\\
\cline{6-8}
$m$	 & 		 & 		 & 	 & 		& hom & het & global & \\ \hline
101	&	affine	&	linear	&	constant	&	inverse	&	yes	&	no	&	yes	&	8	\\ \hline
102	&	affine	&	linear	&	constant	&	affine	&	yes	&	no	&	yes	&	8	\\ \hline
103	&	affine	&	linear	&	inverse	&	constant	&	yes	&	no	&	yes	&	8	\\ \hline
104	&	affine	&	linear	&	inverse	&	inverse	&	yes	&	no	&	yes	&	8	\\ \hline
105	&	affine	&	linear	&	inverse	&	affine	&	yes	&	no	&	yes	&	8	\\ \hline
106	&	affine	&	linear	&	affine	&	constant	&	yes	&	no	&	yes	&	8	\\ \hline
107	&	affine	&	linear	&	affine	&	inverse	&	yes	&	no	&	yes	&	8	\\ \hline
108	&	affine	&	linear	&	affine	&	affine	&	yes	&	no	&	yes	&	8	\\ \hline
109	&	affine	&	affine	&	constant	&	constant	&	yes	&	yes	&	yes	&	9	\\ \hline
110	&	affine	&	affine	&	constant	&	inverse	&	yes	&	yes	&	yes	&	9	\\ \hline
111	&	affine	&	affine	&	constant	&	affine	&	yes	&	yes	&	yes	&	9	\\ \hline
112	&	affine	&	affine	&	inverse	&	constant	&	yes	&	yes	&	yes	&	9	\\ \hline
113	&	affine	&	affine	&	inverse	&	inverse	&	yes	&	yes	&	yes	&	9	\\ \hline
114	&	affine	&	affine	&	inverse	&	affine	&	yes	&	yes	&	yes	&	9	\\ \hline
115	&	affine	&	affine	&	affine	&	constant	&	yes	&	yes	&	yes	&	9	\\ \hline
116	&	affine	&	affine	&	affine	&	inverse	&	yes	&	yes	&	yes	&	9	\\ \hline
117	&	affine	&	affine	&	affine	&	affine	&	yes	&	yes	&	yes	&	9	\\ \hline
118	&	affine	&	sigmoid	&	constant	&	constant	&	yes	&	no	&	yes	&	8	\\ \hline
119	&	affine	&	sigmoid	&	constant	&	inverse	&	yes	&	no	&	yes	&	8	\\ \hline
120	&	affine	&	sigmoid	&	constant	&	affine	&	yes	&	no	&	yes	&	8	\\ \hline
121	&	affine	&	sigmoid	&	inverse	&	constant	&	yes	&	no	&	yes	&	8	\\ \hline
122	&	affine	&	sigmoid	&	inverse	&	inverse	&	yes	&	no	&	yes	&	8	\\ \hline
123	&	affine	&	sigmoid	&	inverse	&	affine	&	yes	&	no	&	yes	&	8	\\ \hline
124	&	affine	&	sigmoid	&	affine	&	constant	&	yes	&	no	&	yes	&	8	\\ \hline
125	&	affine	&	sigmoid	&	affine	&	inverse	&	yes	&	no	&	yes	&	8	\\ \hline
126	&	affine	&	sigmoid	&	affine	&	affine	&	yes	&	no	&	yes	&	8	\\ \hline
127	&	affine	&	affine sigmoid	&	constant	&	constant	&	yes	&	yes	&	yes	&	9	\\ \hline
128	&	affine	&	affine sigmoid	&	constant	&	inverse	&	yes	&	yes	&	yes	&	9	\\ \hline
129	&	affine	&	affine sigmoid	&	constant	&	affine	&	yes	&	yes	&	yes	&	9	\\ \hline
130	&	affine	&	affine sigmoid	&	inverse	&	constant	&	yes	&	yes	&	yes	&	9	\\ \hline
131	&	affine	&	affine sigmoid	&	inverse	&	inverse	&	yes	&	yes	&	yes	&	9	\\ \hline
132	&	affine	&	affine sigmoid	&	inverse	&	affine	&	yes	&	yes	&	yes	&	9	\\ \hline
133	&	affine	&	affine sigmoid	&	affine	&	constant	&	yes	&	yes	&	yes	&	9	\\ \hline
134	&	affine	&	affine sigmoid	&	affine	&	inverse	&	yes	&	yes	&	yes	&	9	\\ \hline
135	&	affine	&	affine sigmoid	&	affine	&	affine	&	yes	&	yes	&	yes	&	9	\\ \hline
136	&	sigmoid	&	constant	&	constant	&	constant	&	no	&	no	&	no	&	7	\\ \hline
137	&	sigmoid	&	constant	&	constant	&	inverse	&	no	&	no	&	no	&	7	\\ \hline
138	&	sigmoid	&	constant	&	constant	&	affine	&	no	&	no	&	no	&	7	\\ \hline
139	&	sigmoid	&	constant	&	inverse	&	constant	&	no	&	no	&	no	&	7	\\ \hline
140	&	sigmoid	&	constant	&	inverse	&	inverse	&	no	&	no	&	no	&	7	\\ \hline
141	&	sigmoid	&	constant	&	inverse	&	affine	&	no	&	no	&	no	&	7	\\ \hline
142	&	sigmoid	&	constant	&	affine	&	constant	&	no	&	no	&	no	&	7	\\ \hline
143	&	sigmoid	&	constant	&	affine	&	inverse	&	no	&	no	&	no	&	7	\\ \hline
144	&	sigmoid	&	constant	&	affine	&	affine	&	no	&	no	&	no	&	7	\\ \hline
145	&	sigmoid	&	linear	&	constant	&	constant	&	no	&	no	&	no	&	7	\\ \hline
146	&	sigmoid	&	linear	&	constant	&	inverse	&	no	&	no	&	no	&	7	\\ \hline
147	&	sigmoid	&	linear	&	constant	&	affine	&	no	&	no	&	no	&	7	\\ \hline
148	&	sigmoid	&	linear	&	inverse	&	constant	&	no	&	no	&	no	&	7	\\ \hline
149	&	sigmoid	&	linear	&	inverse	&	inverse	&	no	&	no	&	no	&	7	\\ \hline
150	&	sigmoid	&	linear	&	inverse	&	affine	&	no	&	no	&	no	&	7	\\ \hline
\end{tabular}
 \caption{List of the models (101-150 over 225) considered in the study.}
    \label{tab:list_models_3}
\end{table}

\begin{table}[]
    \centering
\begin{tabular}{|c|c|c|c|c|c|c|c|c|}
\hline
Model &\multirow{2}{*}{$\bar{\Delta}_{hom}^*$} &\multirow{2}{*}{$\bar{\Delta}_{het}^*$} &\multirow{2}{*}{$\bar{\gamma}_{hom}^*$} &\multirow{2}{*}{$\bar{\gamma}_{het}^*$} & \multicolumn{3}{c|}{Existence of a minimal dose} & \multirow{2}{*}{k}\\
\cline{6-8}
$m$	 & 		 & 		 & 	 & 		& hom & het & global & \\ \hline
151	&	sigmoid	&	linear	&	affine	&	constant	&	no	&	no	&	no	&	7	\\ \hline
152	&	sigmoid	&	linear	&	affine	&	inverse	&	no	&	no	&	no	&	7	\\ \hline
153	&	sigmoid	&	linear	&	affine	&	affine	&	no	&	no	&	no	&	7	\\ \hline
154	&	sigmoid	&	affine	&	constant	&	constant	&	no	&	yes	&	yes	&	8	\\ \hline
155	&	sigmoid	&	affine	&	constant	&	inverse	&	no	&	yes	&	yes	&	8	\\ \hline
156	&	sigmoid	&	affine	&	constant	&	affine	&	no	&	yes	&	yes	&	8	\\ \hline
157	&	sigmoid	&	affine	&	inverse	&	constant	&	no	&	yes	&	yes	&	8	\\ \hline
158	&	sigmoid	&	affine	&	inverse	&	inverse	&	no	&	yes	&	yes	&	8	\\ \hline
159	&	sigmoid	&	affine	&	inverse	&	affine	&	no	&	yes	&	yes	&	8	\\ \hline
160	&	sigmoid	&	affine	&	affine	&	constant	&	no	&	yes	&	yes	&	8	\\ \hline
161	&	sigmoid	&	affine	&	affine	&	inverse	&	no	&	yes	&	yes	&	8	\\ \hline
162	&	sigmoid	&	affine	&	affine	&	affine	&	no	&	yes	&	yes	&	8	\\ \hline
163	&	sigmoid	&	sigmoid	&	constant	&	constant	&	no	&	no	&	no	&	7	\\ \hline
164	&	sigmoid	&	sigmoid	&	constant	&	inverse	&	no	&	no	&	no	&	7	\\ \hline
165	&	sigmoid	&	sigmoid	&	constant	&	affine	&	no	&	no	&	no	&	7	\\ \hline
166	&	sigmoid	&	sigmoid	&	inverse	&	constant	&	no	&	no	&	no	&	7	\\ \hline
167	&	sigmoid	&	sigmoid	&	inverse	&	inverse	&	no	&	no	&	no	&	7	\\ \hline
168	&	sigmoid	&	sigmoid	&	inverse	&	affine	&	no	&	no	&	no	&	7	\\ \hline
169	&	sigmoid	&	sigmoid	&	affine	&	constant	&	no	&	no	&	no	&	7	\\ \hline
170	&	sigmoid	&	sigmoid	&	affine	&	inverse	&	no	&	no	&	no	&	7	\\ \hline
171	&	sigmoid	&	sigmoid	&	affine	&	affine	&	no	&	no	&	no	&	7	\\ \hline
172	&	sigmoid	&	affine sigmoid	&	constant	&	constant	&	no	&	yes	&	yes	&	8	\\ \hline
173	&	sigmoid	&	affine sigmoid	&	constant	&	inverse	&	no	&	yes	&	yes	&	8	\\ \hline
174	&	sigmoid	&	affine sigmoid	&	constant	&	affine	&	no	&	yes	&	yes	&	8	\\ \hline
175	&	sigmoid	&	affine sigmoid	&	inverse	&	constant	&	no	&	yes	&	yes	&	8	\\ \hline
176	&	sigmoid	&	affine sigmoid	&	inverse	&	inverse	&	no	&	yes	&	yes	&	8	\\ \hline
177	&	sigmoid	&	affine sigmoid	&	inverse	&	affine	&	no	&	yes	&	yes	&	8	\\ \hline
178	&	sigmoid	&	affine sigmoid	&	affine	&	constant	&	no	&	yes	&	yes	&	8	\\ \hline
179	&	sigmoid	&	affine sigmoid	&	affine	&	inverse	&	no	&	yes	&	yes	&	8	\\ \hline
180	&	sigmoid	&	affine sigmoid	&	affine	&	affine	&	no	&	yes	&	yes	&	8	\\ \hline
181	&	affine sigmoid	&	constant	&	constant	&	constant	&	yes	&	no	&	yes	&	8	\\ \hline
182	&	affine sigmoid	&	constant	&	constant	&	inverse	&	yes	&	no	&	yes	&	8	\\ \hline
183	&	affine sigmoid	&	constant	&	constant	&	affine	&	yes	&	no	&	yes	&	8	\\ \hline
184	&	affine sigmoid	&	constant	&	inverse	&	constant	&	yes	&	no	&	yes	&	8	\\ \hline
185	&	affine sigmoid	&	constant	&	inverse	&	inverse	&	yes	&	no	&	yes	&	8	\\ \hline
186	&	affine sigmoid	&	constant	&	inverse	&	affine	&	yes	&	no	&	yes	&	8	\\ \hline
187	&	affine sigmoid	&	constant	&	affine	&	constant	&	yes	&	no	&	yes	&	8	\\ \hline
188	&	affine sigmoid	&	constant	&	affine	&	inverse	&	yes	&	no	&	yes	&	8	\\ \hline
189	&	affine sigmoid	&	constant	&	affine	&	affine	&	yes	&	no	&	yes	&	8	\\ \hline
190	&	affine sigmoid	&	linear	&	constant	&	constant	&	yes	&	no	&	yes	&	8	\\ \hline
191	&	affine sigmoid	&	linear	&	constant	&	inverse	&	yes	&	no	&	yes	&	8	\\ \hline
192	&	affine sigmoid	&	linear	&	constant	&	affine	&	yes	&	no	&	yes	&	8	\\ \hline
193	&	affine sigmoid	&	linear	&	inverse	&	constant	&	yes	&	no	&	yes	&	8	\\ \hline
194	&	affine sigmoid	&	linear	&	inverse	&	inverse	&	yes	&	no	&	yes	&	8	\\ \hline
195	&	affine sigmoid	&	linear	&	inverse	&	affine	&	yes	&	no	&	yes	&	8	\\ \hline
196	&	affine sigmoid	&	linear	&	affine	&	constant	&	yes	&	no	&	yes	&	8	\\ \hline
197	&	affine sigmoid	&	linear	&	affine	&	inverse	&	yes	&	no	&	yes	&	8	\\ \hline
198	&	affine sigmoid	&	linear	&	affine	&	affine	&	yes	&	no	&	yes	&	8	\\ \hline
199	&	affine sigmoid	&	affine	&	constant	&	constant	&	yes	&	yes	&	yes	&	9	\\ \hline
200	&	affine sigmoid	&	affine	&	constant	&	inverse	&	yes	&	yes	&	yes	&	9	\\ \hline
\end{tabular}

 \caption{List of the models (151-200 over 225) considered in the study.}
    \label{tab:list_models_4}
\end{table}

\begin{table}[]
    \centering
\begin{tabular}{|c|c|c|c|c|c|c|c|c|}
\hline
Model &\multirow{2}{*}{$\bar{\Delta}_{hom}^*$} &\multirow{2}{*}{$\bar{\Delta}_{het}^*$} &\multirow{2}{*}{$\bar{\gamma}_{hom}^*$} &\multirow{2}{*}{$\bar{\gamma}_{het}^*$} & \multicolumn{3}{c|}{Existence of a minimal dose} & \multirow{2}{*}{k}\\
\cline{6-8}
$m$	 & 		 & 		 & 	 & 		& hom & het & global & \\ \hline
	201	&	affine sigmoid	&	affine	&	constant	&	affine	&	yes	&	yes	&	yes	&	9	\\ \hline
202	&	affine sigmoid	&	affine	&	inverse	&	constant	&	yes	&	yes	&	yes	&	9	\\ \hline
203	&	affine sigmoid	&	affine	&	inverse	&	inverse	&	yes	&	yes	&	yes	&	9	\\ \hline
204	&	affine sigmoid	&	affine	&	inverse	&	affine	&	yes	&	yes	&	yes	&	9	\\ \hline
205	&	affine sigmoid	&	affine	&	affine	&	constant	&	yes	&	yes	&	yes	&	9	\\ \hline
206	&	affine sigmoid	&	affine	&	affine	&	inverse	&	yes	&	yes	&	yes	&	9	\\ \hline
207	&	affine sigmoid	&	affine	&	affine	&	affine	&	yes	&	yes	&	yes	&	9	\\ \hline
208	&	affine sigmoid	&	sigmoid	&	constant	&	constant	&	yes	&	no	&	yes	&	8	\\ \hline
209	&	affine sigmoid	&	sigmoid	&	constant	&	inverse	&	yes	&	no	&	yes	&	8	\\ \hline
210	&	affine sigmoid	&	sigmoid	&	constant	&	affine	&	yes	&	no	&	yes	&	8	\\ \hline
211	&	affine sigmoid	&	sigmoid	&	inverse	&	constant	&	yes	&	no	&	yes	&	8	\\ \hline
212	&	affine sigmoid	&	sigmoid	&	inverse	&	inverse	&	yes	&	no	&	yes	&	8	\\ \hline
213	&	affine sigmoid	&	sigmoid	&	inverse	&	affine	&	yes	&	no	&	yes	&	8	\\ \hline
214	&	affine sigmoid	&	sigmoid	&	affine	&	constant	&	yes	&	no	&	yes	&	8	\\ \hline
215	&	affine sigmoid	&	sigmoid	&	affine	&	inverse	&	yes	&	no	&	yes	&	8	\\ \hline
216	&	affine sigmoid	&	sigmoid	&	affine	&	affine	&	yes	&	no	&	yes	&	8	\\ \hline
217	&	affine sigmoid	&	affine sigmoid	&	constant	&	constant	&	yes	&	yes	&	yes	&	9	\\ \hline
218	&	affine sigmoid	&	affine sigmoid	&	constant	&	inverse	&	yes	&	yes	&	yes	&	9	\\ \hline
219	&	affine sigmoid	&	affine sigmoid	&	constant	&	affine	&	yes	&	yes	&	yes	&	9	\\ \hline
220	&	affine sigmoid	&	affine sigmoid	&	inverse	&	constant	&	yes	&	yes	&	yes	&	9	\\ \hline
221	&	affine sigmoid	&	affine sigmoid	&	inverse	&	inverse	&	yes	&	yes	&	yes	&	9	\\ \hline
222	&	affine sigmoid	&	affine sigmoid	&	inverse	&	affine	&	yes	&	yes	&	yes	&	9	\\ \hline
223	&	affine sigmoid	&	affine sigmoid	&	affine	&	constant	&	yes	&	yes	&	yes	&	9	\\ \hline
224	&	affine sigmoid	&	affine sigmoid	&	affine	&	inverse	&	yes	&	yes	&	yes	&	9	\\ \hline
225	&	affine sigmoid	&	affine sigmoid	&	affine	&	affine	&	yes	&	yes	&	yes	&	9	\\ \hline
\end{tabular}

 \caption{List of the models (201-225 over 225) considered in the study.}
    \label{tab:list_models_5}
\end{table}

\FloatBarrier
\newpage

\section{Pharmacokinetics of IFN$\alpha$}
\label{sec:PK}

\begin{figure}[h]
    \centering
\begin{subfigure}[b]{0.5\textwidth}
        \includegraphics[width=\textwidth]{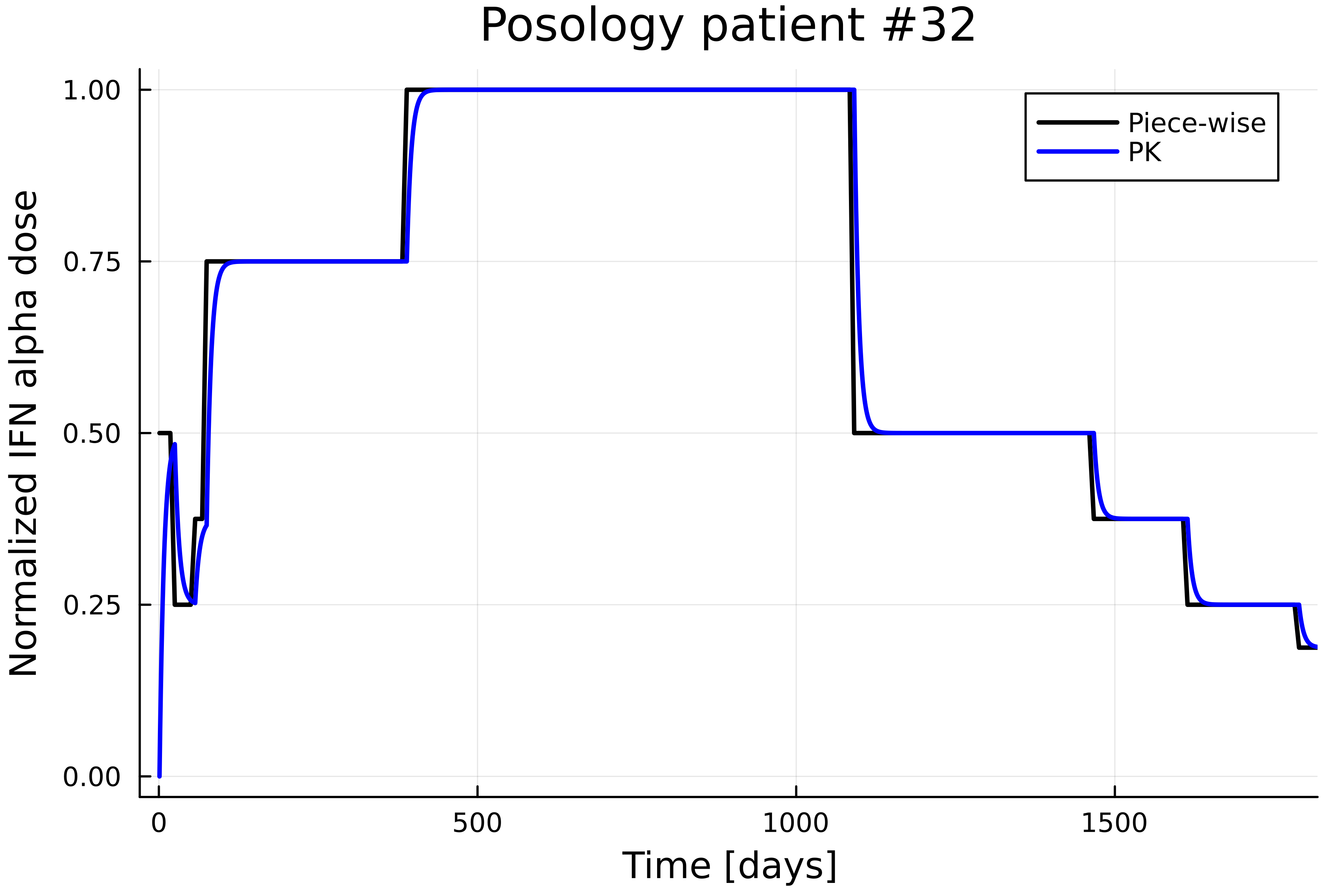}
    \end{subfigure}
    \begin{subfigure}[b]{0.5\textwidth}
        \includegraphics[width=\textwidth]{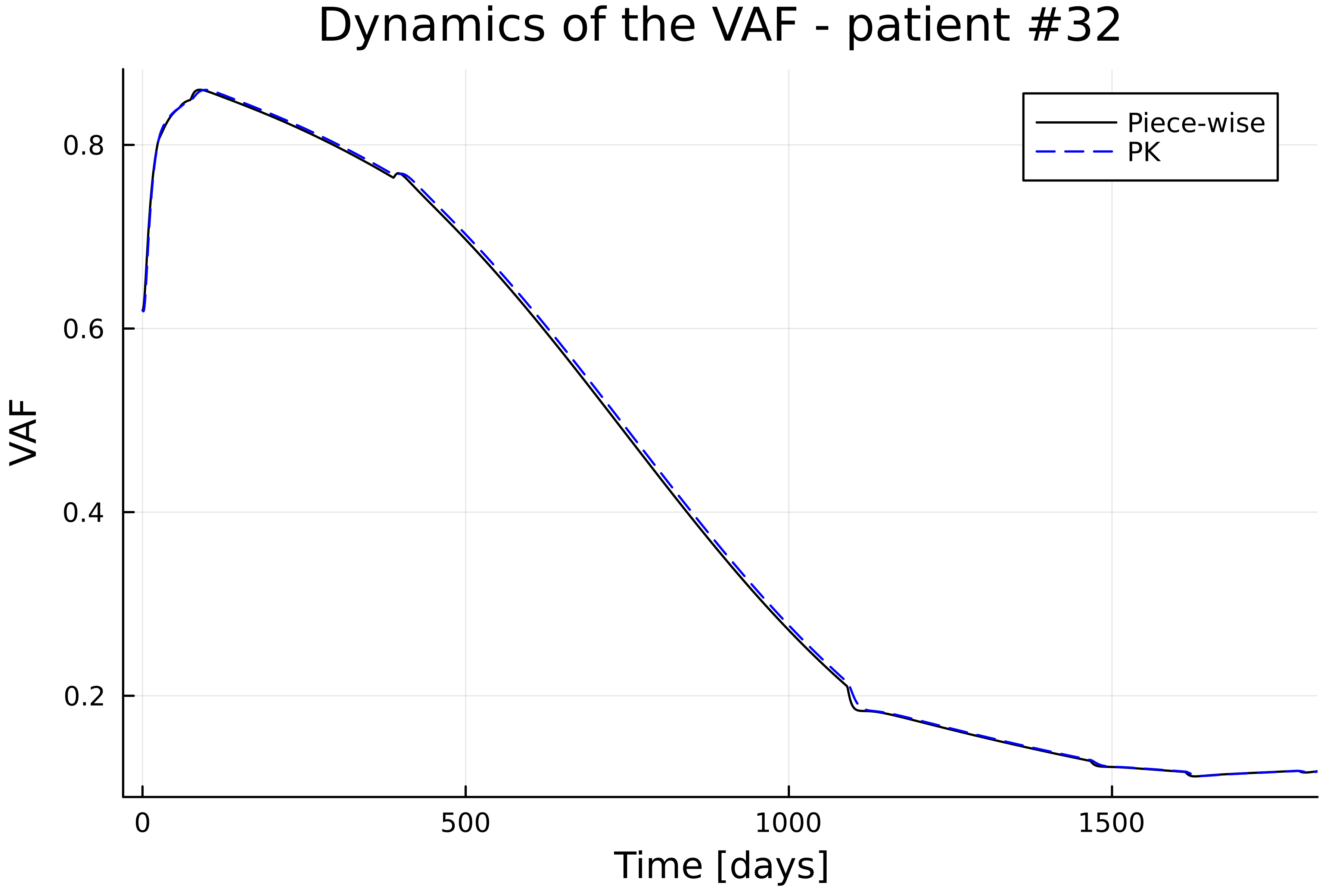}
    \end{subfigure}
    \caption{Comparison between the results we get with the PK modelling (blue line) and our model (black line) for patient \#32. Differences between both approaches are barely perceptible.}
    \label{fig:dyn_VAF_32_comp_PK}
\end{figure}

\begin{figure}[h]
    \centering

    \begin{subfigure}[b]{0.5\textwidth}
        \includegraphics[width=\textwidth]{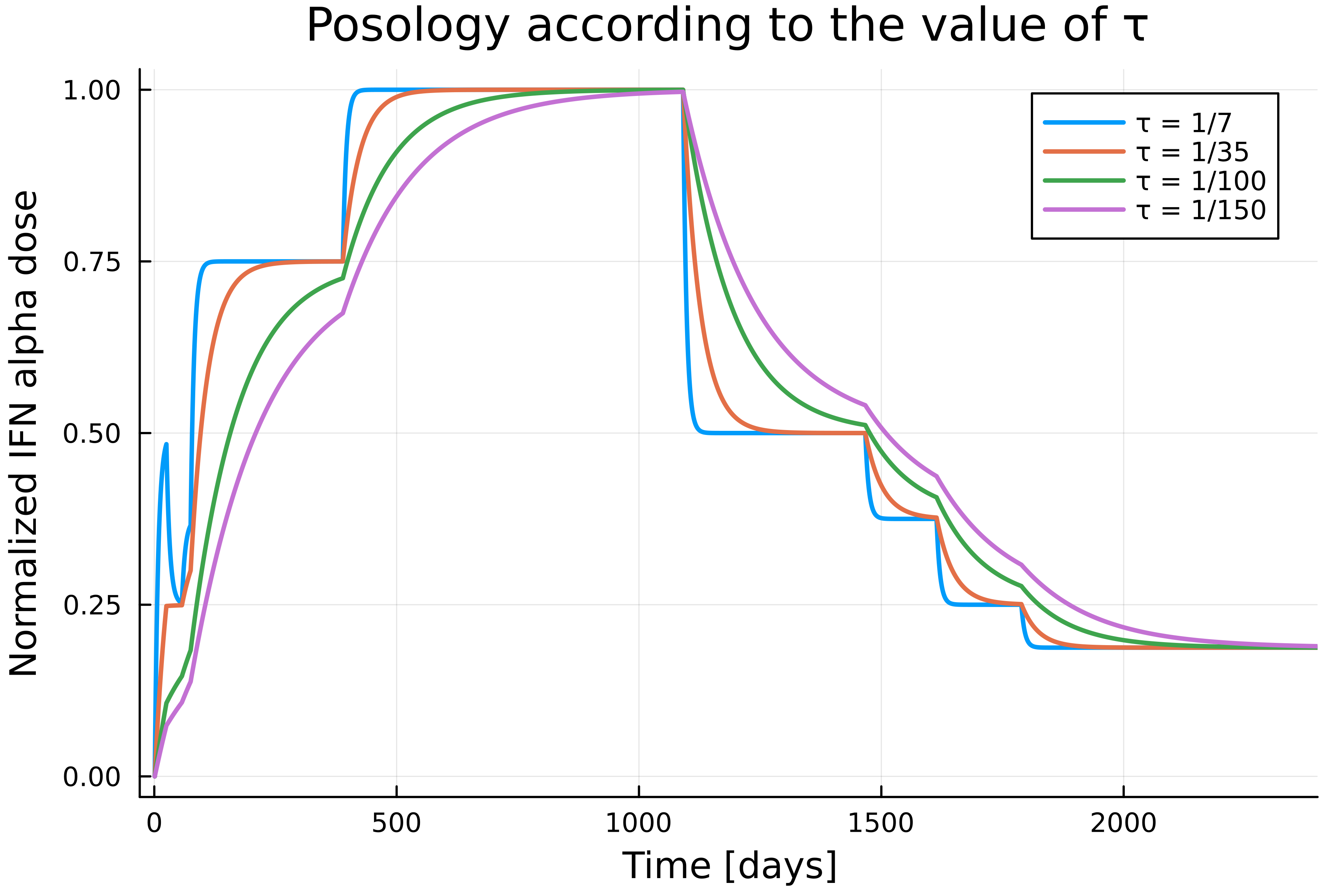}
    \end{subfigure}
   
    \begin{subfigure}[b]{0.5\textwidth}
        \includegraphics[width=\textwidth]{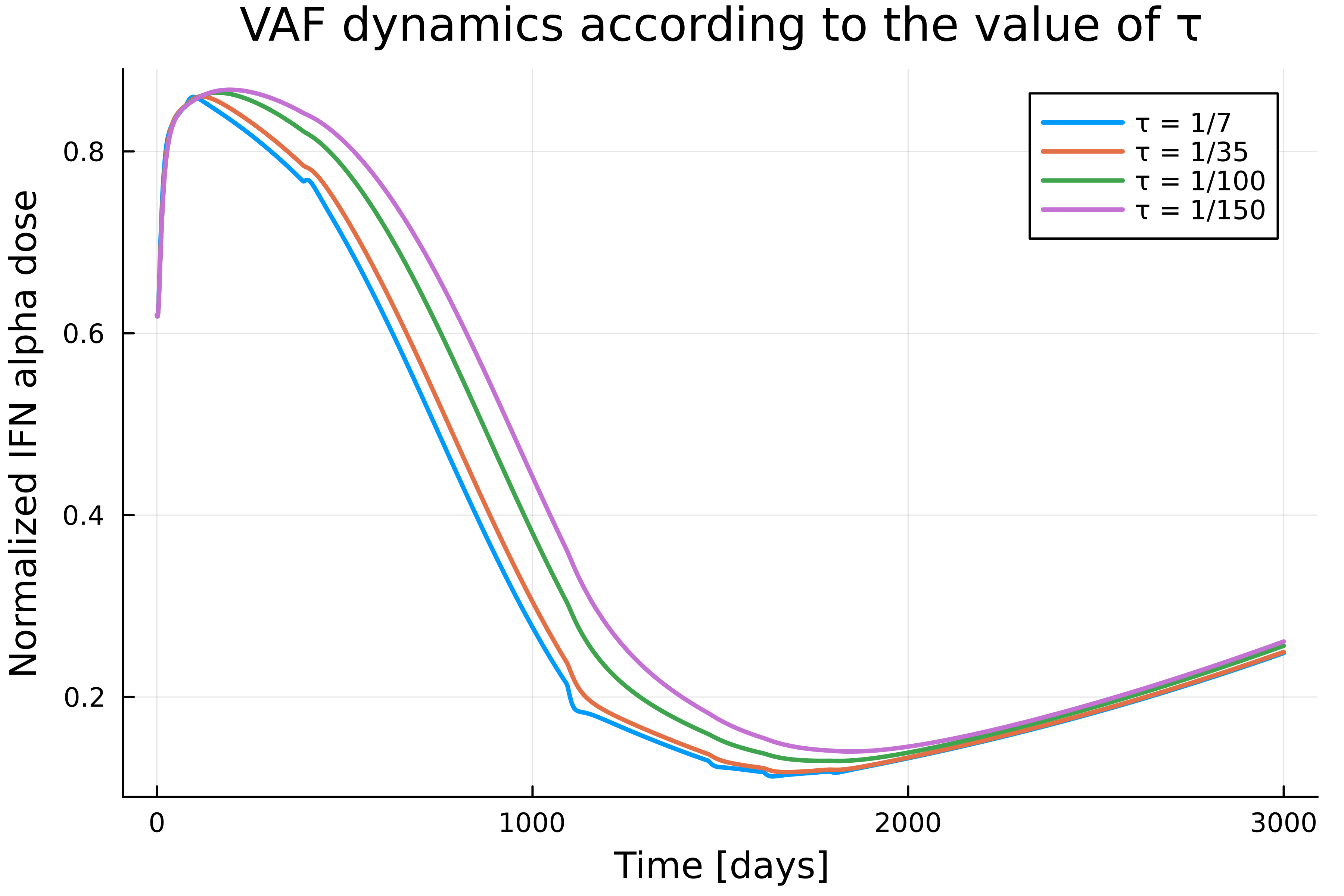}
    \end{subfigure}
    \caption{Study of the potential impact of the value of $\tau$, in eq.~\eqref{eq:PK}, on the results. Differences between two approaches (PK vs piece-wise) could only be perceptible if $\tau$ was lower than it is in reality.}
    \label{fig:poso_value_tau}
\end{figure}

In our study, we considered that some parameters were influenced by the dose intake $d(t)$, which was described as a piece-wise constant function. Therefore, we did not consider the pharmacokinetics (PK) of IFN$\alpha$. 
Considering the dose $d$ as a piece-wise function is a convenient choice; it allows to still get an analytical solution of the cell population dynamics over the therapy, and therefore to reduce the computational cost of our parameter estimation and model selection procedure. 
Here, we explore how a more complex modelization that would take into account the pharmacokinetics of IFN$\alpha$ might influence the results.
Pharmacokinetics modeling of IFN$\alpha$ was previously studied in~\cite{ottesen2020mathematical, saito2012population}.
In~\cite{pedersen2021dose}, Pedersen et al. proposed a pharmaco-kinetic modeling of IFN$\alpha$ therapy against MPN. Here, we follow their approach by considering the simplest possible equation linking the (normalized) concentration of IFN$\alpha$ in the blood $B(t)$ to the stepwise constant IFN$\alpha$ intake $d(t) \in [0,1]$:
\begin{equation}
    \label{eq:PK}
    \frac{dB(t)}{dt} = \tau(d(t)  - B(t))
\end{equation}
with $\tau=1/7$ day$^{-1}$~\cite{saito2012population}.
Pedersen et al., in their approach, make as if the patients had a daily IFN$\alpha$ administration. They acknowledge that this choice is made for simplicity, since the actual IFN$\alpha$ can be administered at different frequencies, for example every week, every two weeks, or every 10 days.  But since the exact timing of the IFN$\alpha$ administration is not known, neither for their patients, nor for the ones from the cohort of Mosca et al.~\cite{mosca2021}, the assumption is relevant and more complex models would not be suitable.\\
Then, $\bar{\Delta}_{het}^*$, $\bar{\Delta}_{hom}^*$, $\bar{\gamma}_{het}^*$ and $\bar{\gamma}_{hom}^*$, instead of explicitly depend on $d(t)$, could now depend on $B(t)$. That is, in their expressions (eq.(2)-(7) in the main text), we simply replace $d$ by $B$. 
We can not have anymore analytical expressions for the dynamics of the VAF and the CF over time and, therefore, resolve the equations numerically. 
We illustrate in Fig.~\ref{fig:dyn_VAF_32_comp_PK} how the use of a PK equation impacts our results. On the top panel, we see with this model that the differences concerning the variations of doses are only perceptible few days after there is a change of posology. These changes barely influence the dynamics of the VAF (bottom panel in Fig.~\ref{fig:dyn_VAF_32_comp_PK}), nor the dynamics of the CF (not shown). Here, we illustrated the influence of the PK-model for patient \#32, choosing for his parameter vector the posterior mean (obtained after running the hierarchical inference method for model 217). Similar results would be obtained for the other patients. The result why the differences are barely perceptible stand in the high value of $\tau$. If $\tau$ had lower values (which is not the case in reality), differences between the models with and without the PK equation would begin to be visible (Fig.~\ref{fig:poso_value_tau}).

\end{document}